\documentclass[b5paper,10pt,twoside,onecolumn,openright]{book}
\usepackage{color}
\usepackage[bookmarksnumbered=true, colorlinks=true, allcolors=blue, breaklinks=true]{hyperref}
\usepackage{vmargin}
\setpapersize{custom}{17cm}{24cm}
\setmarginsrb{1.9cm}{1.9cm}{1.9cm}{1.9cm}{0mm}{10mm}{0mm}{10mm}

\usepackage{latexsym}
\usepackage{amsmath}
\usepackage{amssymb}
\usepackage{epsfig}
\usepackage{epstopdf}
\usepackage{subfigure}
\usepackage{lscape,graphicx}
\usepackage[roman]{sublabel}
\usepackage[round]{natbib}
\usepackage{url}
\usepackage{makeidx}
\usepackage{booktabs}
\usepackage{enumitem}
\usepackage{pdfpages}
\usepackage{afterpage}
\newcommand\blankpage{%
	\null
	\thispagestyle{empty}%
	\addtocounter{page}{0}%
	\newpage}
\usepackage[T2A]{fontenc}
\usepackage[utf8]{inputenc}

\usepackage[amsmath]{ntheorem}
\theoremheaderfont{\normalfont \bfseries}
\theorembodyfont{\normalfont}
\newtheorem{algorithm}{Algorithm}[section]
\newtheorem{definition}{Definition}[section]
\newtheorem{proposition}{Proposition}[section]
\newtheorem{remark}{Remark}[section]
\newenvironment{proof}[1][Proof]{\textbf{#1. }}{\ \rule{0.5em}{0.5em}}

\renewcommand\bibname{References}

\renewcommand{\baselinestretch}{1.01}

\newcommand{\grad}{\ensuremath{^{\circ}}}
\def\sech{\mathop{\rm sech}\nolimits}
\def\tanh{\mathop{\rm tanh}\nolimits}
\def\arctan{\mathop{\rm arctan}\nolimits}

\usepackage{fancyhdr}
\renewcommand{\chaptermark}[1]{\markboth{\large \chaptername \ \thechapter. \ #1}{}}
\renewcommand{\sectionmark}[1]{\markright{\normalsize \thesection. \ #1}}
\renewcommand{\headrulewidth}{0.1pt}
\fancypagestyle{plain}{%
\fancyhead{} 
\renewcommand{\headrulewidth}{0pt} 
}

\makeatletter
\renewenvironment{thebibliography}[1]
{\section*{\Large\bibname 
\@mkboth{\MakeUppercase\bibname}{\MakeUppercase\bibname}}%
\list{\@biblabel{\@arabic\c@enumiv}}%
{\settowidth\labelwidth{\@biblabel{#1}}%
\leftmargin\labelwidth \advance\leftmargin\labelsep \@openbib@code
\usecounter{enumiv}%
\let\p@enumiv\@empty
\renewcommand\theenumiv{\@arabic\c@enumiv}}%
\sloppy \clubpenalty4000 \@clubpenalty \clubpenalty
\widowpenalty4000%
\sfcode`\.\@m%
\setlength{\parindent}{2em}}%
{\def\@noitemerr
{\@latex@warning{Empty `thebibliography' environment}}%
\endlist}
\makeatother

\makeatletter
\def\cleardoublepage{\clearpage\if@twoside\ifodd\c@page\else
  \hbox{}\thispagestyle{empty}\newpage\fi\fi}
\makeatother

\makeindex

\begin{document}
\frontmatter%
{\parindent0pt \pagestyle{empty}

\begin{figure}[h!]
\includepdf[height=\paperheight, offset=2.53cm -2.53cm]{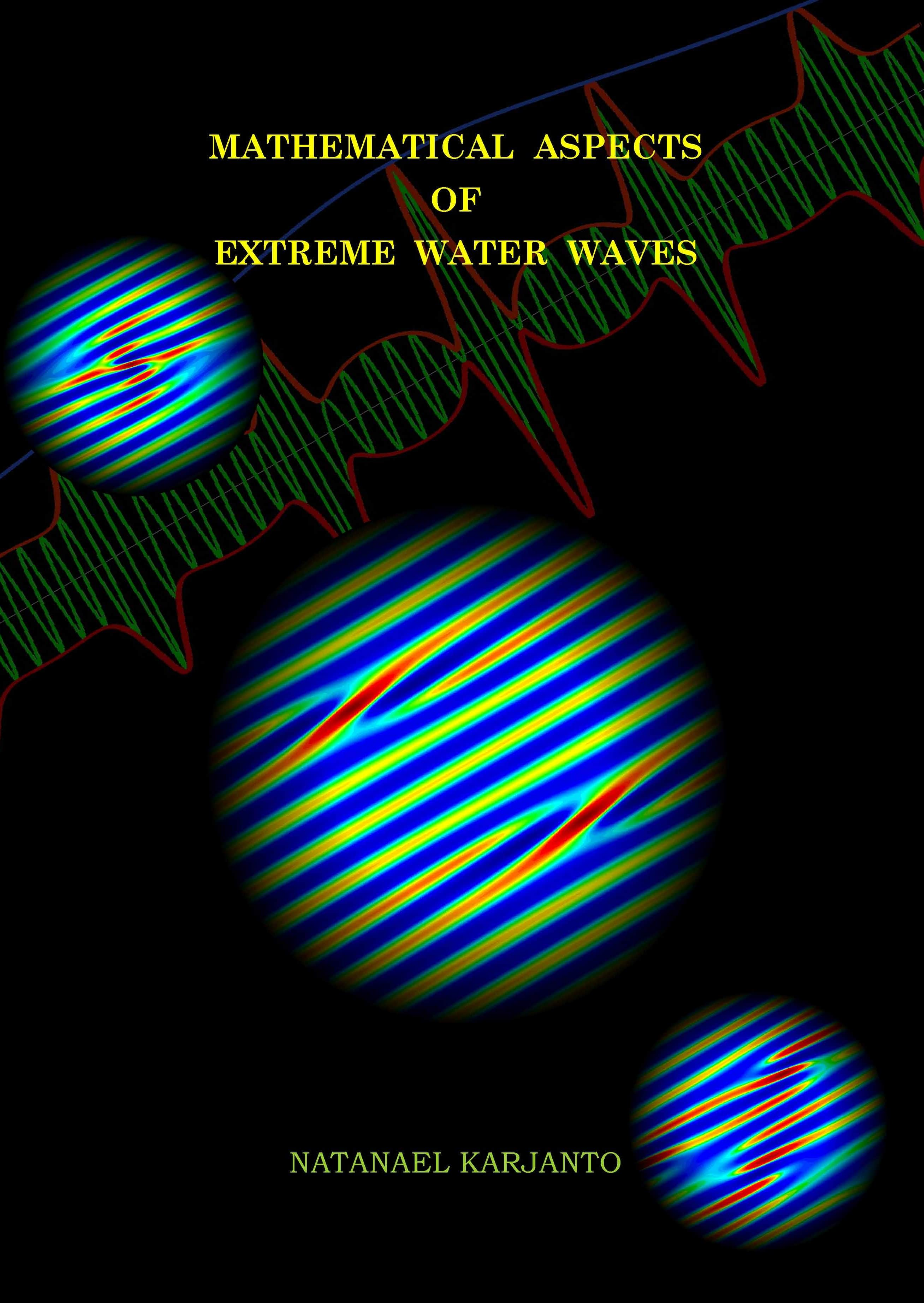}
\end{figure}
	
\afterpage{\blankpage}
	
\newpage
\begin{center}
\textbf{\Large Mathematical Aspects of Extreme Water Waves}\\
  (with Dutch and Indonesian summaries)
  \vspace*{2cm}\\
  \textbf{\large Wiskundige Aspecten van Extreme Watergolven}\\
  (met Nederlandse en Indonesische samenvattingen)
\end{center}

\newpage
{\normalsize \renewcommand{\baselinestretch}{0.99}
\begin{figure}[h]
{\footnotesize
\includegraphics[width=0.15\textwidth]{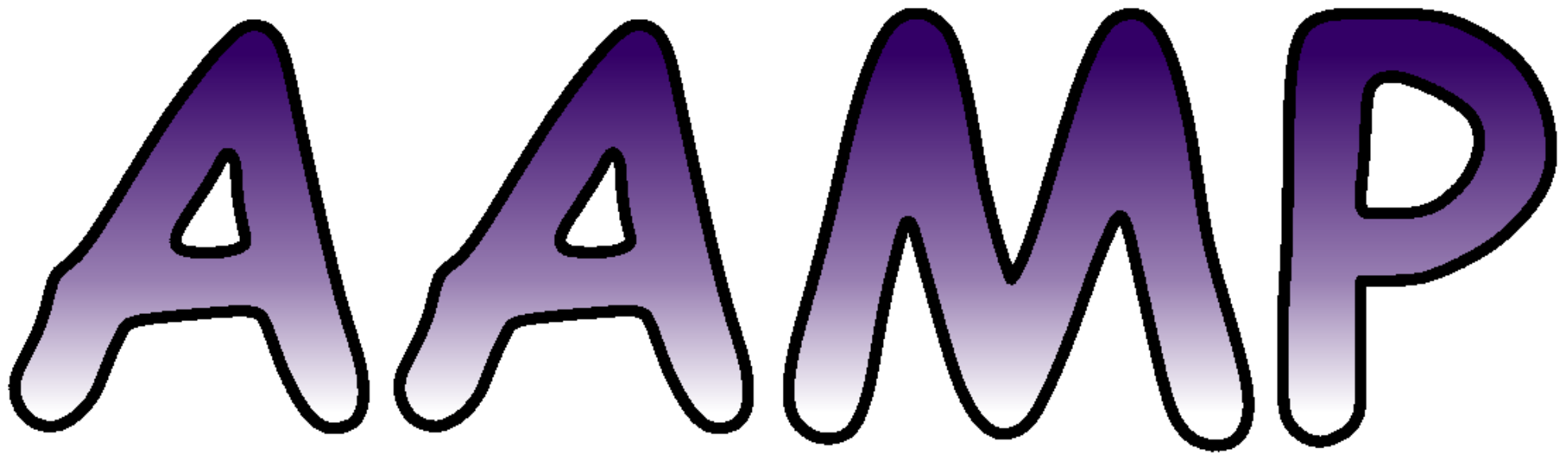}
\includegraphics[width=0.05\textwidth,viewport=259 613 353 707]{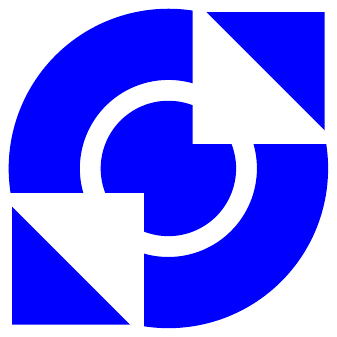} \\
Applied Analysis and Mathematical Physics  \vspace{0.5cm} \\
\includegraphics[width=0.12\textwidth]{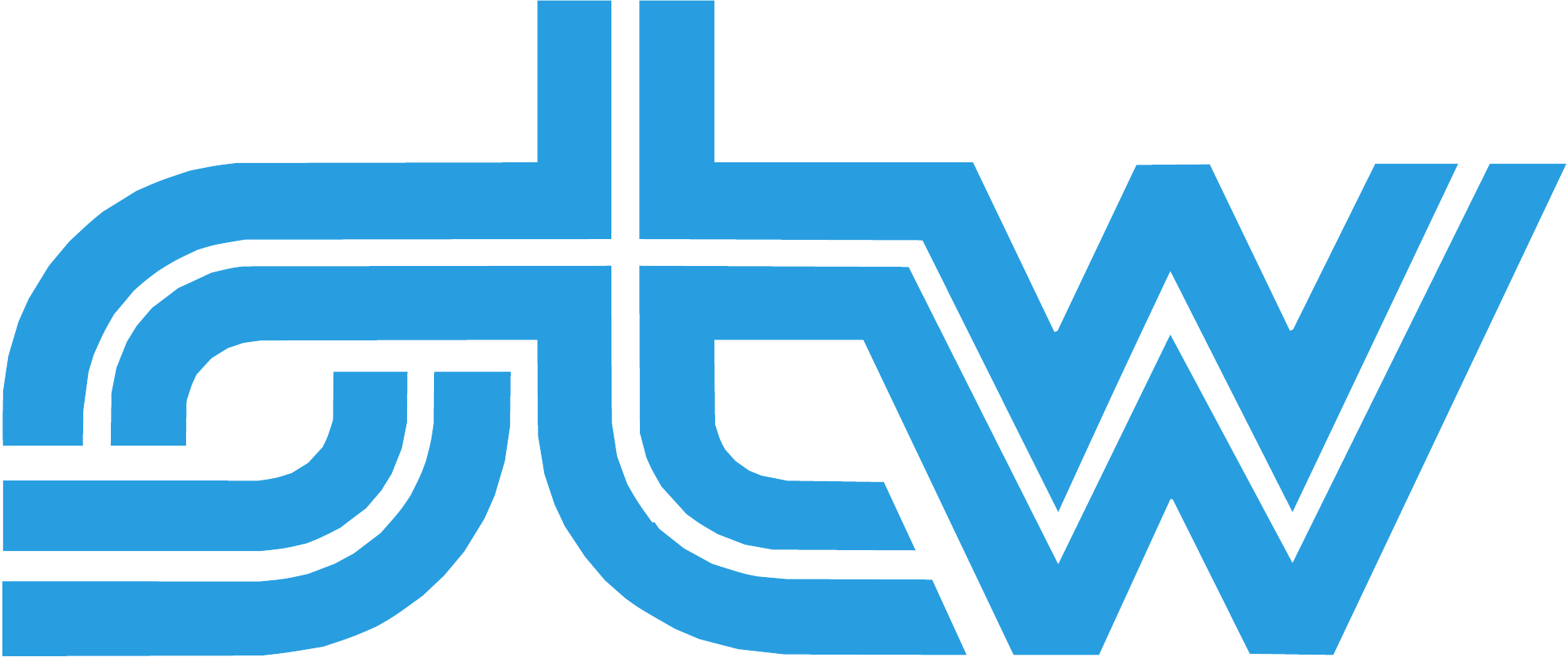}	  \\
Dutch Technology Foundation 			   \vspace{0.5cm} \\
\includegraphics[width=0.15\textwidth,viewport=247 404 348 437]{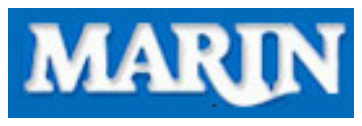}  \\
Maritime Research Institute Netherlands    \vspace{0.5cm} \\
\includegraphics[width=0.20\textwidth]{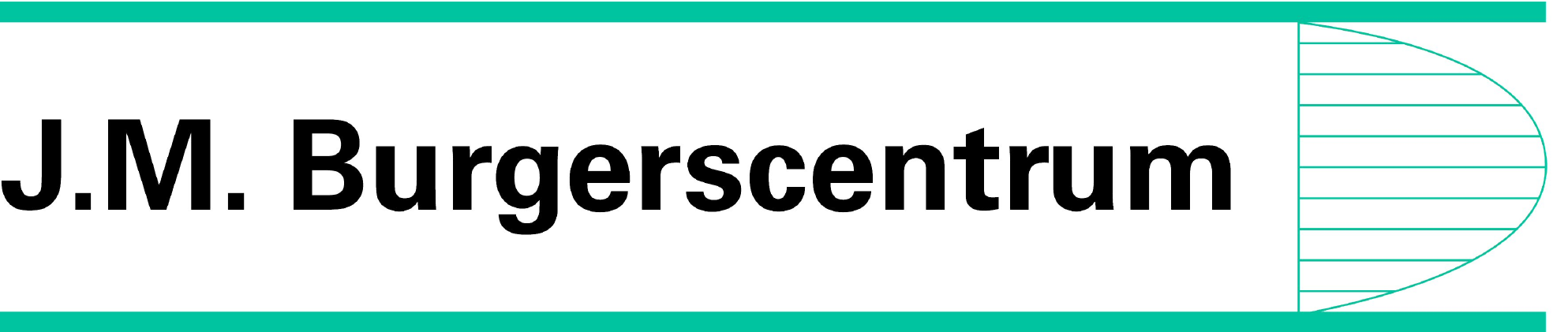}   \\
Dutch Research School for Fluid Mechanics
}
\end{figure}

\vspace*{3.5cm}

{\footnotesize The research described in this thesis was undertaken at the Department of Applied Mathematics, Faculty of Electrical Engineering, Mathematics, and Computer Science (EWI), University of Twente (UT), Enschede, The Netherlands. The research was supported by the Dutch Organization for Scientific Research (NWO), subdivision Applied Sciences the Dutch Technology Foundation (STW) under the project TWI--5374 `Prediction and Generation of Deterministic Extreme Waves in Hydrodynamic Laboratories'.\\ \par}

{\scriptsize \scshape Copyright \copyright\ 2006 by Natanael Karjanto.\\
All rights reserved. This work may not be translated or reproduced in whole or in part without the written permission from the author, except for brief excerpts in connection with reviews or scholarly analysis. Use in connection with any form of information storage and retrieval, electronic adaptation, computer software, or by similar or dissimilar methodology now known or hereafter developed is forbidden. \\ \par}

{\scriptsize 2000 Mathematics Subject Classification:\\
\textsc{76b15, 74j15, 74j30, 35q55, 35q53, 37k05, 37k40, 81u30, 76-05.}%
\\ \par}

{\scriptsize 2006 Physics and Astronomy Classification Scheme:\\
46.40.Cd, 47.35.Bb, 47.54.Bd, 05.45.Yv, 47.10.ab, 52.35.Mw, 47.54.De.%
\\ \par}

{\scriptsize {Printed in the Netherlands.\\%
This thesis is printed on acid-free paper by W\"ohrmann Print Service, Zutphen.\\
Typeset using \LaTeXe.\\
Cover design using {\sl Scribus} by Lars Pesch and Natanael Karjanto.\\
Background picture on the back cover: The Great Wave off Kanagawa by Katsushika Hokusai.\\  \par}
\texttt{ISBN-10 \qquad \ 9-03-652431-8} \\ 
\texttt{ISBN-13 \ 978-9-03-652431-5} \par}}



\newpage
\begin{center}
\vspace*{\stretch{0}}
  \textbf{\Large  MATHEMATICAL \,ASPECTS \vspace*{0.2cm} \\ {\small OF} \vspace*{0.5cm}\\ EXTREME \:WATER \:WAVES \,}

\vspace*{\stretch{6.5}}%
D\,I\,S\,S\,E\,R\,T\,A\,T\,I\,O\,N%
\vspace*{\stretch{6.5}}

to obtain\\
the doctor's degree at the University of Twente,\\
on the authority of the rector magnificus,\\
Prof. Dr. W. H. M. Zijm,\\
on account of the decision of the graduation committee,\\
to be publicly defended\\
on Friday, December 1, 2006, at 15.00\\
\vspace*{\stretch{2}}
by\\
\vspace*{\stretch{2}}
{\bf Natanael Karjanto}\\
born on April 1, 1979\\
in Bandung, West Java, Indonesia\\
\end{center}

\newpage
\begin{flushleft}
This thesis has been approved by the promotor\\
{\bf Prof. Dr. E. W. C. van Groesen}
\vspace*{0.5cm} \\
and the assistant promotor\\
{\bf Dr. Andonowati}.
\end{flushleft}

\newpage
\begin{flushleft}
{\slshape \small To my parents\\
Zakaria and Linda Karjanto\\
\includegraphics[scale = 0.10]{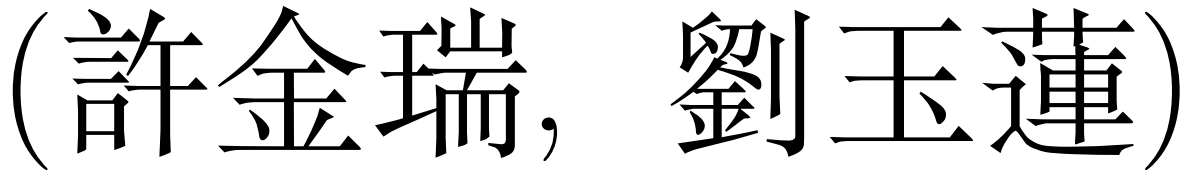} \\ \vspace*{-0.12cm}
and my sister Ferli Tiani \par}
\end{flushleft}

\newpage \thispagestyle{empty}
\begin{center}
  \textbf{\Large Composition of the Graduation Committee}
\end{center}
{\vspace*{1cm} \small
\begin{tabular}{ll}
  {\bfseries Chairperson and Secretary:} &  \\
  Prof. Dr. A. J. Mouthaan & University of Twente, The Netherlands \\
   &  \\
  {\bfseries Promotor:} &  \\
  Prof. Dr. E. W. C van Groesen \qquad & University of Twente, Applied Mathematics \\
   &  \\
  {\bfseries Assistant Promotor:} & \\
  Dr. Andonowati & University of Twente, Applied Mathematics and \\
   &  Bandung Institute of Technology, Indonesia \\
   & \\
  {\bfseries Members:} &  \\
  Prof. Dr. C. Kharif & University of Aix Marseille, France \\
  Prof. Dr. R. H. M. Huijsmans \qquad & MARIN, Wageningen, The Netherlands \\
  Prof. Dr. A. E. Mynett & WL$\mid$Delft Hydraulics, The Netherlands \\
  Prof. Dr. M. A. Peletier & Eindhoven University of Technology, The Netherlands\\
  Dr. O. Bokhove & University of Twente, Applied Mathematics \\
  Prof. Dr. S. A. van Gils & University of Twente, Applied Mathematics
\end{tabular}
}}
\newpage
\chapter*{\vspace*{-2cm} \LARGE Acknowledgement}%

{\pagestyle{plain} \small \vspace*{-0.5cm} Thank you for your interest and for taking the time to read my PhD thesis. Although only my name is displayed on the front cover of this thesis, no doubt many have contributed to the success of this project, resulting in this thesis. In the following paragraphs, I would like to express my sincere gratitude to many individuals and organizations for their generous help so that I am able to complete my academic work and to experience a memorable study period. \addcontentsline{toc}{chapter}{Acknowledgement}

The financial support from STW during this project, travel fund to attend conferences and printing this thesis is greatly appreciated. An excellent opportunity to conduct the experiments on extreme wave generation at the high-speed basin of MARIN is very much esteemed. Annual {\sl Burgersdag}, PhD courses and PhD outings organized by J. M. Burgerscentrum, as well as travel support to attend the {\sl Rogue Waves 2004} conference in Brest, France are highly regarded. The financial support during my research visit in Bandung, Indonesia from the European-Union Jakarta is highly valued. The accommodation support from the International Centre for Mechanical Sciences during a summer school in Udine, Italy is appraised. The travel support to attend the workshop in Stability and Instability of Nonlinear Waves and the conference of Nonlinear Waves and Coherent Structure in Seattle, Washington from the Society for Industrial and Applied Mathematics, the National Science Foundation, and the Pacific Institute for the Mathematical Sciences is also appreciated.

For the foremost, I am indebted to Professor E. (Brenny) van Groesen, my promotor and my daily supervisor, for granting me a privilege to be involved in the combined master-doctoral program at his chair. He has inculcated a critical attitude toward mathematical and physical problems. His patience and perseverance has helped me significantly to improve my analytical, writing, and oral skills. Above all, I thank him for reading and correcting my manuscripts meticulously, as well as improving the {\sl samenvatting} of this thesis. I express my sincere gratitude to Dr. Andonowati, my assistant promotor and also my undergraduate supervisor. She introduced me to the theory of water waves and opened up an opportunity to study abroad. I also thank her for the hospitality during my research visit in her laboratory at the Center for Mathematical Modelling and Simulation during the summer period of 2004.

I thank Professor Christian Kharif who comes all the way from Marseille, France to be a member of my graduation committee. His compliment about the content of this thesis is very encouraging. I thank Professor Ren\'e Huijsmans from Delft who formerly was in MARIN and Professor Arthur Mynett from Delft Hydraulic for the fruitful input every semester during the project meeting and their willingness to be members of my graduation committee. I thank Professor Mark Peletier from Eindhoven for some good remarks. I thank other committee members from our Mathematical Physics and Computational Mechanics (MPCM) group: Dr. Onno Bokhove for many corrections of my thesis draft and Professor Stephan van Gils for useful input in the theory of stability analysis.

Many friends and colleagues in the MPCM group have shown a great help to my academic life at campus. I thank my undergraduate friend and former office mate Hadi Susanto for countless discussions on mathematics and faith, academic advice and his hospitality during my visit to Lumajang, East Java. {\sl Matur nuwun!} I thank A. (Sena) Sopaheluwakan for many conversations, advice and helps with regard to {\sl Maple}, {\sl Matlab} and other computer-related stuff, not to forget in improving the Indonesian summary of this thesis. {\sl Hatur nuhun!} I~thank Lars Pesch for improving my academic portfolio, helping with \LaTeX \ and introducing me to {\sl Scribus} for designing my thesis cover. {\sl Danke sch\"on!} I thank Sander Rhebergen for sharing an office together, helping in translating many Dutch documents and many stories about Africa. I also thank my other cooperative office mates: Henk and Tim. I certainly enjoy many conversations during lunchtime and other occasions with the group members: Fedderik, Chris Klaij, Arek, Davit, Pablo, Alyona, Remco, Marieke, Sanne, J. (Vita) Sudirham, Milan, Jaqueline, Lie, Ivan, Mike, Joris, Dom, Vijay, Yan, Manfred, Bernard, and Bob. I thank Gerard Jeurnink for involving me in educational activities as his teaching assistant. I thank Gert Klopman for research discussions from the practical application, designing experiments, and water wave theory from a different perspective. I appreciate tireless works from Marielle Plekenpol and Diana Dalenoord in handling administrative duties, organizing nice group outings, and giving me the opportunity to write some articles in {\sl Ideaal!} I thank those who help me in translating my thesis summary into Dutch: Lars, Sander, Fedderik, Gerard, Henk, Marieke, and smoothing out by Brenny. I also thank Xu Yan for helping to insert the Chinese character of my parents' name. Certainly, I will not forget former members of the group: Frits van Beckum, Debby Lanser, Kiran Hiremath (North Carolina), Helena Margaretha (for a privilege of {\sl paranymph}), Monika Polner, Agus Suryanto, Edi Cahyono, Imelda van de Voorde (for helping in {\sl Finite Element} homework, {\sl hartelijk dank!}), Willem Visser, Christa van der Meer, Sandra Kamphuis, Renske Westerhof, Pearu Peterson (Estonia) and J. Kojo Ansong (Alberta, Canada).

I thank Johan Simonetti and Michel ten Bulte for their help in providing a nice desk and good chairs to work conveniently. I thank Henri Holtkamp and Ewout Bakker for solving computer-related problems. I thank Muharini and Marwan for many discussions on the theory of extreme water waves. I thank Tan Wooi Nee and Salemah Ismail for the hospitality during my visit to Kajang, Malaysia. I thank Diah Chaerani for the hospitality and advice during my visit to Delft. I would like to thank members of the Indonesian Students Association in Enschede as well as Indonesian Applied Mathematical Society in the Netherlands for many nice conversations: Sri Nurdiati (for hospitality in Enschede and in Bogor and countless useful counsels), Henri Uranus, Irwan Endrayanto, Eko Purnomo, Jamari, Agung Julius, Agoes Moelja, Salman, Dadan Darmana and Jenny Ngo (for helping to improve my bike-repairing skills). I also thank Jurjen Battjes, Jaap van der Vegt, Gerhard Post, Gjerrit Meinsma, Jaap Molenaar, Katarzyna Wac, Elizabeth Yasmine Wardoyo, Nail Akhmediev, Miguel Onorato, Fr\'ed\'eric Dias, Karsten Trulsen, Roger Grimshaw, Alfred Osborne, and Kristian Dysthe.

I would like to acknowledge members of the Enschede English group for kind assistance and fruitful association: Rudi and Naomi van der Moolen, Alexander and Daniela Gathier, Joop and Annie Buitenhuis, Axel and Barbara Senf, Jerry Sadler, Bas and Monique van Aken, Albert and Coby Herbeke, Michael Nwani and Marlisa ten Have, Munyaburanga, Kaba, Sikazwe, Sabaya, Nkrumah, and Robinson. Former members will never be forgotten: Anene, Dragstra, Ntela, do Santos (Angola), family van der Nent, Johnson, Losi (Canada), Jalink, family Oindo (Kenya), Chiuco (The Philippines), Sichivula, Chali and Sinyemba (Zambia), S{\o}ndergard (Denmark), Mensah, Afriyie, and Amoah (Ghana). I also thank the Enschede Oosterveld association for the friendship and organizing wholesome activities: Chin Yee Mooi, Brenda Vaartjes, and family Kruidhof. I thank Tjeerd for the tutorial on {\sl Adobe Illustrator}. \thispagestyle{plain}

Last but not least, I thank my loving parents and my dear younger sister for supporting me through prayer, as a continuous function that will never vanish to time indefinite.
\begin{flushleft}
Natanael Karjanto, \ Enschede, November 6, 2006.
\end{flushleft}
\par}
\newpage
\tableofcontents%
\addcontentsline{toc}{chapter}{Contents}
\newpage
\listoffigures%
\addcontentsline{toc}{chapter}{List of Figures}%
\newpage
{\small \parindent0pt \pagestyle{plain}
\chapter*{\vspace*{-2cm} \LARGE Abbreviations and Acronyms}
\addcontentsline{toc}{chapter}{Abbreviations and Acronyms}
\begin{center}
\begin{tabular}{ll}
  AAF & amplitude amplification factor \\
  BBC & British Broadcasting Corporation\\
      & bottom boundary condition \\
  BVP & boundary value problem \\
  CMq & Chu-Mei quotient \\
  DFSBC & dynamic free surface boundary condition \\
  DO & Dingemans-Otta \\
  DS & Davey-Stewartson \\
  IVP & initial value problem \\
  IFREMER \qquad \qquad & {\slshape Institut fran\c{c}ais de recherche pour l'exploitation de la mer} \\
          & (French Research Institute for Exploitation of the Sea) \\
  KdV & Korteweg-de Vries \\
  KFSBC & kinematic free surface boundary condition \\
  KWMBC & kinematic wavemaker boundary condition \\
  LDR & linear dispersion relation \\
  MARIN & Maritime Research Institute Netherlands \\
  MI & modulational instability \\
  mKdV & modified Korteweg-de Vries\\
  MTA &  maximum temporal amplitude \\
  NDR & nonlinear dispersion relation \\
  NLS & nonlinear Schr\"{o}dinger \\
  PV & principal value \\
  RHS & right-hand side \\
  SFB & Soliton on Finite Background
\end{tabular}
\end{center}

\clearpage \thispagestyle{empty}
\chapter*{\vspace*{-2cm} \LARGE Symbols and Notations}
\addcontentsline{toc}{chapter}{Symbols and Notations}

\begin{tabular}{lll}
  {\large \bfseries Notation} & {\large \bfseries Description} & {\large \bfseries Page} \\
    &   &   \\
  $a$, $a(x,t)$ & wave amplitude; real-valued amplitude & \pageref{wavenumber}, \pageref{spatialNLSphysical} \\
  $a_{n}(\xi)$ & amplitude spectrum, complex Fourier coefficient & \pageref{amplitude_spectrum} \\
  approx & subscript indicates an approximation of a quantity & \pageref{approx} \\
  $A_{0}(\xi)$ & plane-wave solution of the NLS equation & \pageref{planewave} \\
  $A_{j}$, $j = 1,2$ & SFB$_{1}$ and SFB$_{2}$, respectively & \pageref{SFB12} \\
  $A(\xi,\tau)$ & complex amplitude of a wave group; & \pageref{complexamplitude}\\
                & exact solutions of the NLS equation & \pageref{SFBexpression} \\
  $A^{\ast}(\xi,\tau)$ & complex conjugate of $A$ & \pageref{BR22} \\
  $\bar{A}$(x,t) & $A$ in the physical variables & \pageref{spatialNLSphysical} \\
  $\mathcal{A}(A)$ & action functional of a function $A$ & \pageref{actionfunctional} \\
  $A_\textsc{tc}(\xi,\tau)$ & complex amplitude of trichromatic waves & \pageref{Trichromatic} \\
  $\alpha_{j}$, $j = 1$--4 & coefficients of the potential function $V(F)$ & \pageref{potentialF} \\
  $\alpha_{S}(\xi)$  & position dependent constant of integration & \pageref{alphaS} \\
  $\alpha_{T}(\tau)$ & time dependent constant of integration & \pageref{alphaT} \\

  $b_{j}$, $j = -1, 0 ,1$ & coefficients of trichromatic waves & \pageref{coeffb} \\
  $B(\xi,\tau)$ & complex amplitude of $2^\textmd{nd}$-order double harmonic wave & \pageref{etaABC} \\
                & perturbation function to the plane-wave solution & \pageref{planewave} \\
  $B_{j}$, $j = 1,2$ & complex coefficient of the perturbation function $B$ & \pageref{planewave} \\
  $\beta$, $\beta_{0}$ & dispersion coefficient of the NLS equation & \pageref{beta} \\

  $c$ & nonlinear coefficient of the KdV type of equation & \pageref{KdVeqn} \\
  c\,c & complex conjugate of the preceding term(s) & \pageref{frequency} \\
  crit & subscript denotes critical value  & \pageref{crit} \\
  Crit & an optimization of a constrained variational problem & \pageref{WFB_special_ansatz} \\
  $C(\xi,\tau)$ & complex amplitude of $2^\textmd{nd}$-order nonharmonic wave & \pageref{etaABC} \\
  $C_{n}$ & conserved quantities correspond to the NLS equation & \pageref{waveenergy} \\
  $\mathbb{C}$ & the set of complex numbers & \pageref{bilcomplex} \\
\end{tabular}

\newpage
\begin{tabular}{lll}
  {\large \bfseries Notation} & {\large \bfseries Description} & {\large \bfseries Page} \\
    &   &   \\

  des & subscript denotes design parameter & \pageref{basicdesign} \\
  $\delta$ & variational or Fr\'echet derivative & \pageref{Hamiltonian}\\
  $\delta_{0}$ & convergence criterium  & \pageref{delta0}\\

  $E$ & dynamic energy, spectrum energy & \pageref{potentialenergy}, \pageref{energyspectrum} \\
  $\epsilon$ & small positive parameter & \pageref{epsilon} \\
  $\eta(x,t)$ & surface wave field, wave elevation & \pageref{Bab2NDWE} \\
  $\eta_{\textmd{c}}(x,t)$ & complexification of $\eta$ & \pageref{Hilbert} \\

  $f(z)$ & geometry of the wavemaker & \pageref{wavemakertype} \\
  $\hat{f}(\omega)$ & spectrum or Fourier transform of a function $f(t)$ & \pageref{spectrum}  \\
  $F(\xi,\tau)$ & complex amplitude of waves on finite background & \pageref{solutionA} \\
  $F_{j}$, $j = 1,2$ & SFB$_{j}$ without the plane-wave, $F_{j} = A_{j}/A_{0}$ & \pageref{SFB12} \\

  $g$ & gravitational acceleration & \pageref{rho} \\
  $G(\xi,\tau)$ & real-valued displaced amplitude & \pageref{disamp} \\
  $\gamma$, $\gamma_{0}$ & nonlinear coefficient of the NLS equation & \pageref{gamma} \\

  $h$ & water depth & \pageref{lambda} \\
  $H$ & Hamiltonian & \pageref{Hamiltoniandensity} \\
  $H_{j}$, $j = 1,2$ & the denominator of $F_{j} + 1$ & \pageref{SFB12} \\
  $\bar{H}$ & transformed Hamiltonian & \pageref{transHam} \\
  $\mathcal{H}$ & Hamiltonian density; Hilbert transform & \pageref{Hamiltoniandensity}, \pageref{Hilbert} \\

  $I$ & a value of an integral & \pageref{contour_integral2} \\
  $k_{0}$, $k(x,t)$ & wavenumber; local wavenumber & \pageref{wavenumber}, \pageref{localwavenumber} \\
  $K(\omega)$ & inverse of $\Omega(k)$; differential operator & \pageref{inverseLDR} \\
  $K_\textmd{res}(\omega)$ & residue terms of $K(\omega)$ & \pageref{epsilon} \\
  $K_\textmd{res}(-i\partial_{\tau})$ & differential operator corresponds to $K_\textmd{res}$ & \pageref{epsilon} \\

  $L$, $\mathcal{L}$ & Lagrangian, Lagrangian density & \pageref{Lagrangiandensity} \\
  $\mathcal{L}$ & linear operator & \pageref{linop} \\
  $\lambda$ & wavelength; Lagrange multiplier & \pageref{lambda}, \pageref{fullosceqn} \\

  ${\displaystyle \max_{t}}$ & maximum over time of the surface elevation $\eta$ & \pageref{MTAdefinition} \\
  $M$ & maximum amplitude at the extreme position & \pageref{designM} \\
  $\mu$, $\tilde{\mu}$   & a parameter in the Ma solution & \pageref{mu} \\
\end{tabular}

\newpage
\begin{tabular}{lll}
  {\large \bfseries Notation} & {\large \bfseries Description} & {\large \bfseries Page} \\
    &   &   \\
  $N$ & number of waves in one modulation period & \pageref{designM} \\
  $N(x)$ & number of waves of a signal in a certain interval & \pageref{number_waves} \\
  $\mathbb{N}$ & the set of natural numbers & \pageref{bilasli} \\
  $\mathbb{N}_{0}$ & the set of nonnegative whole numbers & \pageref{trigser} \\
  $\nu$, $\tilde{\nu}$  & (normalized) modulation frequency & \pageref{nu}, \pageref{nuSFB} \\

  $\mathcal{O}$ & order of a quantity & \pageref{epsilon} \\
  $\omega_{0}$, $\omega(x,t)$ & wave frequency; local frequency & \pageref{frequency}, \pageref{localfrequency} \\
  $\Omega(k)$, $\Omega(-i\partial_{x})$ & linear dispersion relation; differential operator & \pageref{lindisrel} \\%

  $P_{j}$, $j = 1,2$ & the real part of the numerator of $F_{j} + 1$ & \pageref{SFB12} \\
  $\phi$, $\phi(x,t)$ & (reduced) real-valued phase & \pageref{spatialNLSphysical} \\
  $\phi_{n}(\xi)$ & phase spectrum & \pageref{phasespectrum}\\
  $\Phi(x,t)$ & $\theta(x,t) + \phi(x,t)$, real-valued phase & \pageref{spatialNLSphysical} \\
  $\phi(\xi,\tau)$ & displaced phase & \pageref{disphase} \\
  $\phi(\xi)$      & time-independent displaced phase & \pageref{solutionF} \\
  $\phi(x,z,t)$ & velocity potential function & \pageref{phi} \\
  $Q_{j}$, $j = 1,2$ & the imaginary part of the numerator of $F_{j} + 1$ & \pageref{SFB12} \\
  ref & subscript denotes reference of a quantity & \pageref{ref} \\
  res & subscript denotes residue of a quantity & \pageref{epsilon} \\
  $2r_{0}$ & plane-wave amplitude, asymptotic amplitude & \pageref{planewavephysical} \\
  $R$ & coefficient of wave groups superposition & \pageref{Rnm} \\
  $R_{nm}$ & coefficient of $e^{im\theta}$ with order $\epsilon^{n}$ & \pageref{Rnm} \\
  $\mathbb{R}$ & the set of real numbers & \pageref{KdVeqn}\\
  $\rho$, $\tilde{\rho}$ & a quantity that depends on $\mu$ & \pageref{rho} \\

  $s$; $s_{0}$, $s_{M}$ &wave steepness; initial, extremal steepness & \pageref{designM} \\
  $S(t)$ & wavemaker motion & \pageref{wavemakertype} \\
  $\sigma$, $\tilde{\sigma}$  & (normalized) growth rate & \pageref{nu} \\
  $\sigma_{j}$, $j = 1,2$ & two growth rates in SFB$_{2}$ & \pageref{gr} \\

  $t$ & time in physical variable & \pageref{Bab2NDWE} \\
  $T_\textmd{c}$ & carrier wave period, $T_\textmd{c} = 2\pi/\omega_{0}$ & \pageref{basicdesign} \\
  $T$ & modulation period, $T = 2\pi/\nu$ & \pageref{T} \\
\end{tabular}

\newpage
\thispagestyle{plain}
\begin{tabular}{lll}
  {\large \bfseries Notation} & {\large \bfseries Description} & {\large \bfseries Page} \\
    &   &   \\
  $\tau$ & time in a moving frame of reference & \pageref{Bab2NDWE} \\
  $\theta_{j}$, $j = 1, 2$ & phase of the complex coefficients $B_{j}$, $j = 1, 2$ & \pageref{anglephases} \\
  $\theta(x,t)$ & $k_{0}x - \omega_{0}t$, phase of monochromatic wave & \pageref{etaABC} \\%

  $\mathbf{u}$ & velocity vector defined by potential function $\phi$ & \pageref{phi} \\
  $V(F)$, $V(G,\phi)$ & potential function, potential energy & \pageref{potentialenergy} \\
  $W(G,\phi)$ & normalized potential energy & \pageref{Wpotential} \\
  $x$ & position in physical variable & \pageref{Bab2NDWE} \\
  $\xi$  & position in a moving frame of reference & \pageref{Bab2NDWE} \\
  $\Xi(z,t)$ & wavemaker position & \pageref{phi} \\
  $\mathbb{Z}$ & the set of whole numbers & \pageref{amplitude_spectrum}\\
  $\mathbb{Z \setminus N}$ & the set of non positive whole numbers & \pageref{series1} \\
  $\zeta(\tau)$ & special functions related to breather solutions
  & \pageref{generalspecialWaveform}
\end{tabular}
\clearpage \thispagestyle{empty}
}

\mainmatter
\chapter{Introduction} \label{1Introduce}

\section{Background and motivation}

\subsection{Extreme wave events in the oceans}

We start to quote parts of a few reports about the occurrences of extreme waves\index{extreme waves!reports} in the oceans.

{\small
\begin{flushleft}
\begin{itemize}
\item \textsl{``In late 1942, carrying 15,000 U.S. soldiers bound for England, the {\it Queen Mary} hit a storm about 700 miles off the coast of Scotland. Without warning amid the tumult, a single, mountainous wave struck the ocean liner, rolling it over and washing water across its upper decks. Luckily, the ship managed to right itself and continue on its voyage."} ---{\it \textbf{Science News Online}, 23 November 1996.}

\item \textsl{``In the past 30 years, hundreds of ships have gone down in mysterious circumstances, taking thousands of lives with them. Naval architects now believe that a large number of these were sunk by rogue waves."} \index{rogue waves} ---{\it Monsters of the deep, \textbf{New Scientists} Magazine issue 2297, 30 June 2001.}

\item \textsl{``Since 1990, 20 vessels have been struck by waves off the South African coast that defy the linear model's predictions. And on New Year's Day, \index{New Year wave} 1995 a wave of 26m was measured hitting the {\it Draupner} \index{Draupner platform} oil rig in the North Sea off Norway."} ---{\it \textbf{BBC Horizon} Television programme, 14 November 2002.}

\item \textsl{``They are known as `rogue waves'\index{rogue waves}--the towering walls of water that, some experts suspect, sink tens of ships every year."} ---{\it \textbf{Nature}, volume \textbf{430}, 29 July 2004, page 492.}

\item \textsl{``During the last two decades, more than 200 super-tankers-ships over 200 meters (656 feet) long--have sunk beneath the waves. Rogue waves are thought to be the cause for many of these disasters, perhaps by flooding the main hold of these giant container ships. \dots\, Offshore oil rigs also get hit by rogue waves. Radar reports from the North Sea's \textit{Gorm} oil field show 466 rogue-wave encounters in the last 12 years."} ---{\it ``Monster" waves surprisingly common, satellites show, \textbf{National Geographic} news, 10 August 2004.} \index{rogue waves}

\item \textsl{``Over the last two decades more than 200 super-carriers--cargo ships over 200 m long--have been lost at sea. Eyewitness reports suggest many were sunk by high and violent walls of water that rose up out of calm seas."} ---{\it \textbf{BBC News Report} on Wave Research, 21 August 2004.}

\item \textsl{``Bad weather has sunk more than 200 supertankers and container ships during the past 20 years, and researchers believe that monster 10-story-tall ocean waves are often the culprit."} ---{\it Surf's Up-Way Up, \textbf{Scientific American}, October 2004, volume \textbf{291}, issue 4, page 38.}

\item \textsl{``Stories told by seamen about walls of water as high as 10-story buildings, waves that could destroy even big cargo ships, were treated as legends and myths. But these waves do exist. \dots\, Over two hundred ships have been their victims in the past two decades. In the North Atlantic alone, between 1995 and 1999, 27 big vessels foundered after being hit by freak waves."} ---{\it \textbf{Janson} Television programme, 28 December 2004.} \index{freak waves}
\end{itemize}
\end{flushleft}
}

These reports show that extreme waves are dangerous to merchant ships, offshore platforms, naval fleets, and other sea-going marine structures. We will take a closer look at the characteristics of these waves.

\subsection{Possible causes of extreme waves} \index{extreme waves!possible causes}

\begin{figure}[h] 		
\begin{center}
\includegraphics[scale=0.5,viewport=71 272 524 569]{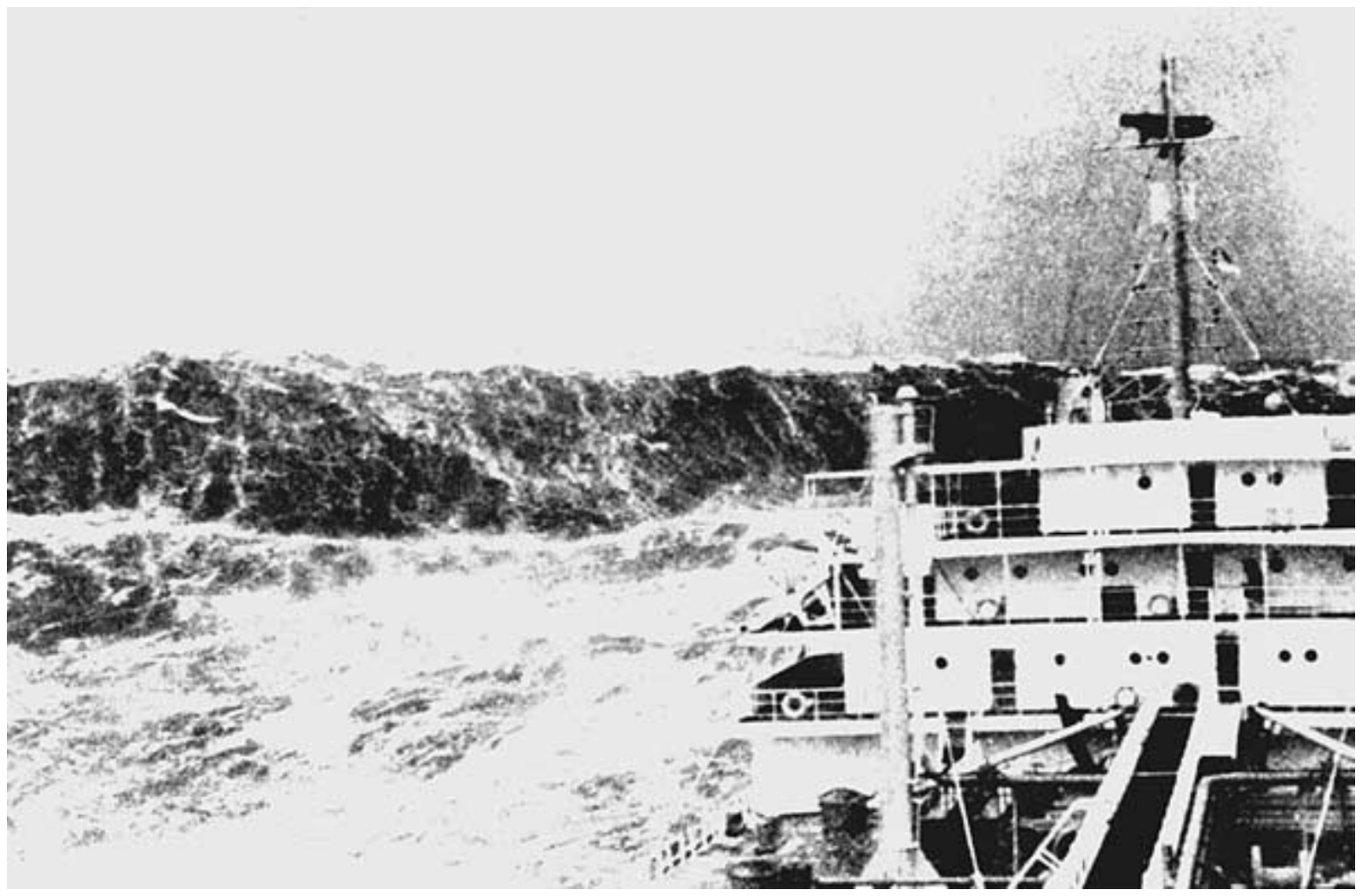}
\caption[Extreme wave photograph]{One photograph showing an extreme wave in Biscayne Bay, Florida, reproduced from \textit{Mariners Weather Log}, volume \textbf{38} no 4.} \label{freakwaves}
\end{center}
\end{figure}

As reported above, `extreme waves'\index{extreme waves!terminologies} are unusually very large waves that appear unexpectedly even under relatively calm conditions in the open ocean. They are also known as `freak waves',\index{freak waves} `rogue waves',\index{rogue waves} \index{rogue waves|see{extreme waves}} `giant waves',\index{giant waves} `monster waves',\index{monster waves} \index{monster waves|see{extreme waves}} `steep wave events',\index{steep wave events} \index{steep wave events|see{extreme waves}} `gargantuan waves',\index{gargantuan waves} \index{gargantuan waves|see{extreme waves}} `abnormal  waves',\index{abnormal waves} \index{abnormal waves|see{extreme waves}} `exceptional waves',\index{exceptional waves} \index{exceptional waves|see{extreme waves}} or `cape rollers'.\index{cape rollers} \index{cape rollers|see{extreme waves}} The wave height can reach 30 meters or more from crest to trough. They are so large that they can overwhelm and sink even the sturdiest ships. However, extreme waves can also occur in bad weather conditions when the average wave height is high and possibly several big waves come together to create a monster. See Figure~\ref{freakwaves} as an example. Extreme waves\index{extreme waves!definition} are defined as waves larger than 2.2 times the significant wave height~\citep{1Dean90}. In this context, the significant wave height is defined as the average height of the highest one-third of the waves in a long sample.

The precise causes of extreme waves are still being investigated and scientists try to find out these possible causes. These possible causes can be an internal one, which occurs in homogeneous situations, or an external one, which occurs in nonhomogeneous situations, or a combination of both causes. Internal causes of extreme waves\index{extreme waves!causes!internal} are wave focusing and nonlinear instability. Wave focusing may occur in the deep ocean, when unstable waves self-focus in bad weather conditions, creating extreme waves. Nonlinear processes may cause waves of different amplitudes and wavelengths to interact with each other and form large waves. Since the chance to generate extreme waves can be very small, many sailors claim that they seldom occur, for instance once every 10,000~years. This suggests to study the extreme waves from a statistical probability point of view. However, the latest data from satellites suggest that they take place more frequently. This is also supported by an examination of five years of wave measurements made in the South Atlantic Ocean~\citep{1Liu04}.

External causes for extreme waves\index{extreme waves!causes!external} are due to the geometric shape of coasts and the bathymetry\footnote[1]{the underwater equivalent of topography} \index{bathymetry} of a sea, interaction of wind forces and currents, and by storm systems of low pressure such as typhoons or hurricanes. An example of the geometric shape is the coast of Norway where a shallow sea bottom focuses in one spot, creating extreme waves. Examples of wind-current interaction occur in the south-eastern coast of South Africa when wind pushes against a strong Agulhas current \index{Agulhas current} flowing southward, within or near the Gulf Stream in the North Atlantic, in the south of Japan that enjoys the warm Kuroshio current \index{Kuroshio current} and the notorious seas off Cape Horn in South America. Examples of extreme waves caused by storms are typhoons in certain areas of the Pacific Ocean and Hurricane Ivan in the Gulf of Mexico. The latter one is recently reported in \textit{Science} magazine~\citep{1Wang05}.

Extreme waves\index{extreme waves!differences with tsunamis} should not be confused with tsunamis.\index{tsunamis} The latter waves are a series of traveling waves of extremely long length and period and are caused by a sudden displacement of a large volume of water. Usually, tsunamis are generated by disturbances associated with earthquakes occurring below the ocean floor, volcanic eruptions, landslides or even oceanic meteor impacts. The height of the waves in the open ocean is very small, a couple of meters at most, so they pass under ships and boats undetected. The height is growing as they reach shallow waters. Their length from one crest to the next can be up to 200 km long, and they travel in the deep ocean at speeds around 700 km/hr. Compared to tsunamis, extreme waves have very short wavelengths, less than 2000 m. Therefore, in the open ocean, tsunamis are shallow-water waves and extreme waves are deep-water waves. A wave is considered deep-water if $\lambda/h < 2$ and shallow-water if $\lambda/h > 2$, where $\lambda$ is the wavelength \label{lambda} and $h$ is the water depth. In Chapter~\ref{6Experiment}, we will report about experiments in a wave tank where the wavelength is $\lambda = 4.4$ m and the water depth $h = 3.55$ m, so $\lambda/h \approx 1.24$. Even though these waves are considered as intermediate water waves, in the wave basin they are already quite `extreme' since it is based on the theoretical model of extreme wave evolution.

\subsection{Current research on extreme waves} \index{extreme waves!current research}

The term `freak waves'\index{freak waves} \index{freak waves|see{extreme waves}} itself is already introduced as early as the 1960s by~\citet{1Draper65}. Much research has been done since then to study and to predict the occurrence of extreme waves\index{extreme waves}. Observation of extreme wave events using satellite images of the ocean surfaces has been done by the European Space Agency. One of its projects is called {\it MaxWave}, \index{extreme waves!projects!{\it MaxWave}} devoted to observe, to model, and eventually to forecast the occurrence of extreme waves~\citep{1Dankert03, 1Rosenthal05}. Another project is called {\it WaveAtlas}, \index{extreme waves!projects!{\it WaveAtlas}} which uses the images to create a worldwide atlas of extreme wave events.

Theoretical studies of extreme waves\index{extreme waves!studies!theoretical} can be divided into a deterministic approach and a statistical approach. In the deterministic approach, scientists use mathematical models to describe the evolution of extreme waves. For example, the nonlinear Schr\"{o}dinger (NLS) equation\index{NLS equation}, \index{nonlinear Schr\"{o}dinger equation|see{NLS equation}} also known as the cubic Schr\"{o}dinger equation, has been proposed as early as the 1970s to model extreme waves~\citep{1Smith76}. In his paper, the author uses the term `giant waves'.\index{giant waves} \index{giant waves|see{extreme waves}} Further in 2000,~\citet{1Osborne00} study the dynamical behavior of extreme waves in deep-water waves using solutions of the NLS equation in $1 + 1$ or $(x,t)$ and $2 + 1$ or $(x,y,t)$ dimensions. Using the inverse scattering technique, \index{inverse scattering technique} they present and discuss analytical solutions of $1 + 1$ NLS, in which one of the solutions is known as the Soliton on Finite Background (SFB). \index{SFB} \index{Soliton on Finite Background|see{SFB}} For $2 + 1$ NLS, they demonstrate numerical simulations of the equation and show the existence of unstable modes which can take the form of large amplitude extreme waves. More about the analytical and spectral study of extreme wave properties of exact solutions of the NLS equation that describe the nonlinear evolution of deep-water waves is given by~\citet{1Osborne01}. Apart from the NLS equation, some scientists use the Korteweg-de Vries (KdV) equation \index{KdV equation} \index{Korteweg-de Vries equation|see{KdV equation}} for extreme wave modeling~\citep{1Kharif00, 1Pelinovsky00, 1Kokorina02, 1Pelinovsky04}. A computation of modulated wave trains using a fully nonlinear inviscid irrotational flow model in a spatially periodic domain and its comparison with the NLS equation to produce extreme wave events is given by~\citet{1Henderson99}.

The statistical approach to study extreme wave\index{extreme waves!studies!statistical} phenomena is related to probabilistic aspects at random locations. A study of extreme wave generation in a random oceanic sea state characterized by the Joint North Wave Project spectrum can be found in~\citep{1Onorato01}. A study of extreme waves in the context of a statistical or stochastic approach in the connection with nonlinear four-wave interactions is given by~\citet{1Janssen03}. Statistics of weakly nonlinear high random waves with a derivation of new analytical models for the prediction of nonlinear extreme waves using the theory of quasideterminism\index{quasideterminism theory} of~\citet{1Boccotti00} is published recently~\citep{1Fedele05}. The statistical study of sea waves started as early as the 1950s with particular attention to the properties of wave groups in Gaussian noise. Studies of the extreme waves based on the `New Year wave' \index{New Year wave} recorded at Draupner platform\index{Draupner platform} in the North Sea are given for instance by~\citet{1Walker04, 1Gibson05}, and \citet{1Haver05}. A focusing mechanism to produce extreme waves at random locations when ocean swell traverses an area of random current is shown by~\citet{1White98}. Recently, an overview of extreme wave formation in the context of a statistical event from Longuett-Higgins and the refraction model from White and Fornberg is given by~\citet{1Heller05}. A review of the physical mechanisms of the extreme wave phenomena including both deterministic and statistical approaches is given by~\citet{1Kharif03, 1Olagnon05}, and \citet{1Grimshaw05}.

\subsection{Extreme wave modeling in a hydrodynamic laboratory}		\index{extreme waves!modelling}

The mechanisms of extreme wave generation\index{extreme waves!generation} have become an issue of interest in many projects and research topics. Maritime Research Institute Netherlands (MARIN) \index{MARIN} \index{Maritime Research Institute Netherlands|see{MARIN}} aims to generate and study large waves in one wave basin of its laboratory. Eventually, MARIN will use these large waves to test ships, offshore platforms and other floating structures in the wave basin. For extreme wave generation in a hydrodynamic laboratory, we are interested in the deterministic aspects instead of the random ones. Extreme wave generation in the entire contents of this thesis will always be deterministic. It means that we try to generate the desired extreme wave at a certain location in the wave tank. This is done by prescribing an initial signal as input to the wavemaker,\index{wavemaker} for which the wave signal is found from a mathematical model for extreme wave evolution. More details about how the experiments are conducted can be found in Chapter~\ref{6Experiment}.

In this thesis, we will concentrate on the study of a solution of a mathematical model that describes the evolution of extreme wave events. To study the theoretical aspects of extreme wave characteristics, we choose the `spatial' NLS equation\index{NLS equation!type!spatial} as a mathematical model to our problem. This type of NLS equation is suitable for the signaling problem, when we give an initial signal to the wavemaker and the model predicts the propagation in space. This is different from the `temporal' NLS equation, for which, given an initial condition, the model predicts the evolution in time. Furthermore, we have selected one family of exact solutions of the spatial NLS equation that describes extreme wave events in a wave tank. This family of exact solution is known as the SFB\index{SFB}. More information on the NLS equation will be given in Chapter~\ref{2Model} and more properties of the SFB are presented in Chapter~\ref{3Property}.

The choice to study the SFB family in great detail is motivated by the fact that the asymptotic behavior of this solution at the far-field describes (the envelope of) a modulated wave. This modulated surface wave is a solution of the linear equation obtained from a perturbation analysis of the plane-wave\index{plane-wave} solution of the NLS equation. According to the linear theory, this plane-wave solution is unstable under a long modulation, and the exponential growth is known as the linear `modulational instability',\index{modulational instability} or `sideband instability',\index{sideband instability} or `Benjamin-Feir instability' \index{Benjamin-Feir instability}~\citep{1BenjaminFeir67}. This type of instability is observed not only in water waves but also in many fields of wave propagation in nonlinear media. Therefore, in order to describe a complete evolution of a modulated wave signal in the wave tank, we choose the SFB as a nonlinear extension of the linear modulational instability into the nonlinear regime. The relation between modulational instability and the SFB solution of the NLS equation is explained in Subsection~\ref{asymptoticbehavior} of Chapter~\ref{3Property}. See also~\citep{1Akhmediev86}.

Additionally, we use the maximum temporal amplitude (MTA)\index{MTA} \index{maximum temporal amplitude|see{MTA}} of the SFB for application in wave generation. The MTA is introduced in the field of nonlinear optics~\citep{1Andonowati03}, but it has also applications in water waves. It is defined at each position as the maximum over the time variable of a wave field. From this definition, the MTA describes the largest wave amplitude that can appear in a certain position. It can also be interpreted as a stationary envelope of the wave group envelope\index{wave group(s)}. In a wave tank, the MTA has a meaningful interpretation, namely the highest points on the wall wetted after a long time of evolution. An explicit expression can be found for the MTA corresponding to the SFB. This is exceptional since an explicit expression for the MTA is not easily obtained from the mathematical model that describes the nonlinear dynamical evolution of a wave.

\begin{figure}[h]		
\begin{center}
\includegraphics[scale=0.45,angle=-90]{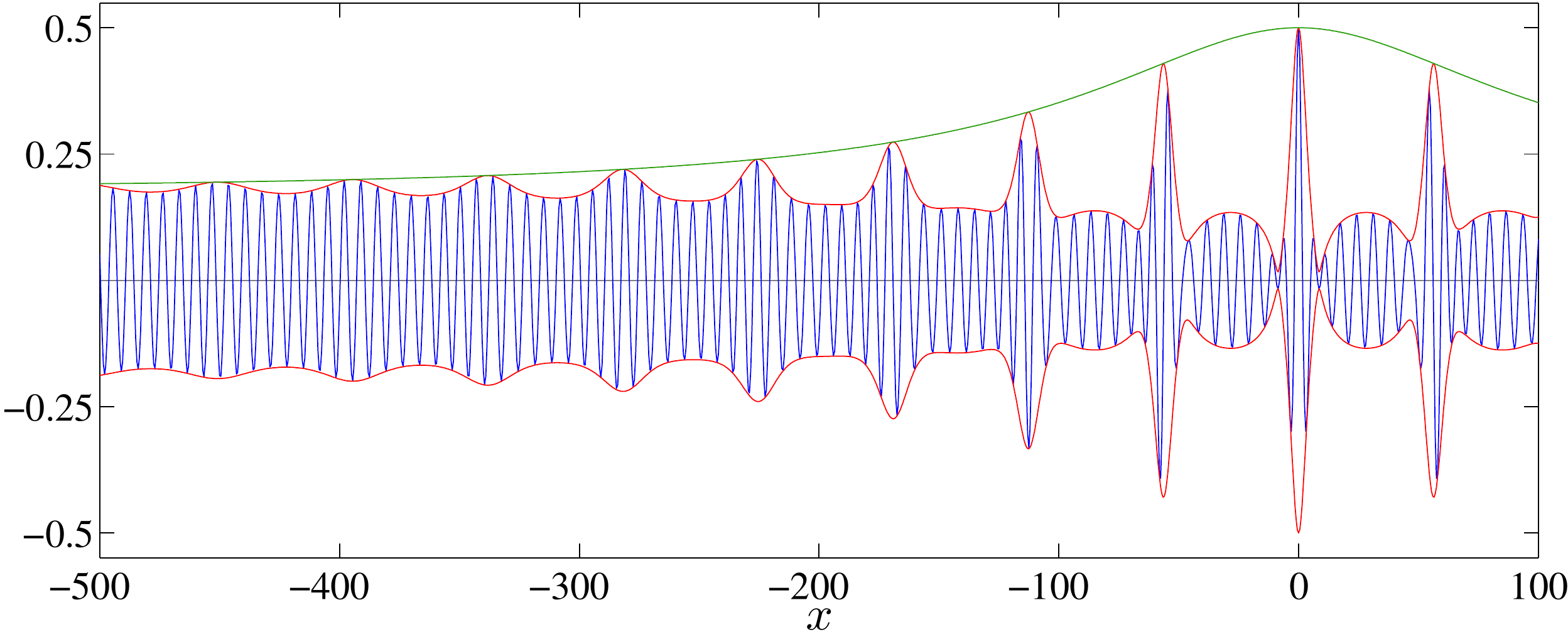}
\caption[Wave profile, envelope, and the MTA of the SFB]{It displays a plot of a wave field, its envelope, and MTA related to a characteristic member of the SFB family. At a specified time, the plot shows horizontally the distance from the position of maximal wave height at $x = 0$. The distance and the vertical elevation are measured in meters, and the waves are on a layer of water depth of 5~m.} \label{Bab1MTASFB}
\end{center}
\end{figure}

In our application, the MTA\index{MTA} is a very useful tool for designing a strategy of extreme wave generation. \index{extreme waves!generation} Using the MTA, we are able to determine the signal input for the wavemaker\index{wavemaker} that produces an extreme wave signal at the desired position. A plot of the MTA gives information about the extreme position, which is the position where the envelope of a wave signal is at its largest value. Hence, to produce this extreme wave signal at a particular position in the laboratory, the details of the initial modulated wave signal depend on the MTA plot. Therefore, it is important to know in advance the properties and the characteristics of the SFB\index{SFB} wave signal, also to avoid breaking before the wave reaches the extreme position. This breaking criterion is determined by the wave steepness\index{steepness} $a\,k \geq 0.443$, where $a$ is the wave amplitude and $k$ is the wavenumber. \label{wavenumber} A plot of the MTA together with the spatial evolution and its envelope of the SFB is given in Figure~\ref{Bab1MTASFB}. More information on the MTA of the SFB is found in Chapter~\ref{3Property}; its application to wave generation is found in Chapter~\ref{6Experiment}.

We see in Figure~\ref{Bab1MTASFB} that waves of moderate amplitude at the left propagate downstream\index{downstream} to form wave groups\index{wave group(s)}. These wave groups increase in amplitude as they propagate and reach a maximal wave height at the extreme position $x = 0$ for a specified time. We observe that close to $x = 0$, the wave groups increase in amplitude faster within the same distance than when they are far away at the left. This is due to the nonlinear effect of the Benjamin-Feir instability. At a distance sufficiently far from the extreme position, the effect is only linear due to the exponential behavior of this instability. After reaching the maximum wave height, the wave groups decrease in amplitude and return to a modulated monochromatic wave\index{monochromatic wave(s)!modulated} at the right with a phase shift compared to the monochromatic wave at the right. The MTA has a space symmetry property with respect to $x = 0$.

The NLS equation is not the only mathematical model that scientists have studied for this problem. Note that several authors prefer the modified NLS equation\index{NLS equation!modified} and other higher-order NLS equations as mathematical models to describe extreme waves\index{extreme waves} \citep{Trulsen97,Trulsen01}. In this thesis, we will concentrate on the NLS equation instead of the higher-order equations, since the NLS equation has an advantage that it possesses exact solutions that the higher-order equations do not have. Studying these families of exact solutions is very insightful from a theoretical point of view and useful from a practical point of view for an initial stage of study in extreme wave generation.

\section{Outline of the thesis}

This thesis is divided into seven chapters; the rest of the thesis is briefly summarized as follows.

\textbf{Chapter~\ref{2Model}.} In this chapter, we review known mathematical descriptions for surface wave evolution. We are interested in a nonlinear and dispersive wave equation \index{dispersive wave!nonlinear equation} that describes extreme wave phenomena. In particular, we choose the nonlinear (cubic) Schr\"{o}dinger (NLS) equation\index{NLS equation} of the spatial type instead of the temporal one since it is more suitable to model our wave evolution in a wave tank. We derive the spatial NLS equation from a KdV type equation with exact dispersion relation\index{KdV equation!exact dispersion relation} for water wave potential flow using the multiple scale method. \index{multiple scale method} We also derive the corresponding phase-amplitude equations \index{phase-amplitude equations} using the polar representation of the complex amplitude and introduce the Chu-Mei quotient from the nonlinear dispersion relation. Furthermore, we compare the approximate dispersion of both the temporal and spatial NLS equations. It is found that the spatial type approximates the dispersion relation better than the temporal one for increasing carrier wavenumber and is exact for the deep-water case. We derive the modulational instability corresponding to the NLS equation and we show that both plane-wave and single-soliton solutions of the NLS equation\index{NLS equation!solutions of} are coherent state solutions. We also present a formulation of the NLS equation from a variational principle that will be used in Chapter~\ref{3Property}.

\textbf{Chapter~\ref{3Property}.} In this chapter, we derive the properties of waves on finite background. These are exact solutions of the NLS equation. There are three special types of waves on finite background, also known as breather type of solutions, that have been proposed as models of extreme waves~\citep{1Dysthe99}. They are the Soliton on Finite Background (SFB)\index{SFB}, also known as the Akhmediev solution, the Ma breather, and the rational solution, also known as the Peregrine solution. We will concentrate on the SFB solution since it is a good model for extreme wave generation, that is practical to generate, while the Ma solution is impossible and the rational solution is difficult to generate in practice. The reason is as follows. The SFB wave signal can be generated with a moderate amplitude at the wavemaker. The Ma wave signal has necessarily the maximal amplitude already at the wavemaker. The rational wave signal is still possible to generate in principle, but it has an infinite modulation period.~\index{modulation period}

To study these waves on finite background, we will introduce a novel transformation to the displaced phase-amplitude variables with respect to a background of the monochromatic plane-wave\index{plane-wave} solution in the context of the NLS equation. The transformation of the displaced-phase is restricted to be time-independent. The change of phase with positions physically corresponds to a change of the wavelength of the carrier wave of a wave group.\index{wave group(s)} This turns out to be the only driving force responsible for the nonlinear amplitude amplification toward extreme wave events. Remarkably, the assumption that the displaced-phase is time-independent leads to the breather solutions of the NLS equation mentioned above. A particularly extensive study is devoted to the SFB solution. This solution is the unstable modulated plane-wave solution of the NLS equation, has amplitude amplification up to a maximum factor of three and is a very good candidate for extreme wave generation. The contents of this chapter are partly based on the papers by~\citet{1vanGroesen06, 1Andonowati07}.

\textbf{Chapter~\ref{4Dislocate}.} This chapter is motivated to study in detail one physical property of the SFB, namely, phase singularity\index{phase singularity} and wavefront dislocation\index{wavefront dislocation}. This phenomenon occurs at singular points of a wave field where the amplitude vanishes. We discuss it in a more general setting for surface wave fields. The phenomenon is purely linear. We introduce the Chu-Mei quotient from the (nonlinear) dispersion relation for wave groups as a result of the nonlinear transformation to phase-amplitude variables. We observe that the unboundedness of this quotient is responsible for the appearance of wavefront dislocation. The study of the physical property of the SFB and the accompanying phenomena of phase singularity and wavefront dislocation has not been discussed in the literature. The connection of the Chu-Mei quotient and wavefront dislocation also seems to be novel in the context of wave dislocation theory.

\textbf{Chapter~\ref{5HighOrder}.} This chapter deals, just as Chapter~\ref{3Property}, with special solutions of the NLS equation. We now study a family of higher-order solutions of the NLS equation that also describe modulational instability, namely SFB$_{2}$\index{SFB!SFB$_{2}$}. Similar to SFB, now denoted as SFB$_{1}$\index{SFB!SFB$_{1}$}, this class of solutions describes to the waves on finite background. Indices $1$ and $2$ refer to the number of initial sideband pairs in the spectrum domain. For a special choice of parameters, solutions from SFB$_{2}$ can have an amplitude amplification up to a factor of five. Furthermore, SFB$_{2}$ solutions can also show wavefront dislocation\index{wavefront dislocation} and phase singularity\index{phase singularity}.

\textbf{Chapter~\ref{6Experiment}.} This chapter deals with experiments on extreme wave generation\index{extreme waves!generation} that have been conducted at the high-speed basin of MARIN. The experiments are designed by Gert Klopman, Andonowati, and Natanael Karjanto. We try to investigate both qualitative and quantitative comparisons between the theoretical and the experimental signals. This kind of comparison is novel and part of the work has not been reported elsewhere in the literature. It should be noted that all experimental results show amplitude increase, non-breaking waves according to the Benjamin-Feir instability of the SFB solution of the NLS equation. We also observe that both the carrier wave frequency and the modulation period\index{modulation period} are conserved during the propagation in the wave basin. A significant difference is that the experimental wave signal does not preserve the symmetry structure as the theoretical SFB does.

The dynamic evolution in the Argand diagram \index{Argand diagram} (the complex plane) gives an understanding of the qualitative comparison between the experimental signal and a perturbed SFB signal. This comparison explains why the experimental signals have two pairs of phase singularities at two different positions and the extreme position is located in between. We also explore the concept and properties of the MTA in relation to the experimental design. We compare quantitatively the sensitivity of the MTA and its extreme position toward parameter changes and variations in a model parameter. We discover that the extreme position depends sensitively toward these changes. For SFB parameters, the extreme position is more sensitive toward the changes in the asymptotic amplitude rather than the maximum amplitude. For NLS parameters, it is essential to use an accurate value of the nonlinear coefficient for the extreme position to occur at the desired location. The investigation of these parameters using the MTA curves in the context of extreme wave generation is new and useful for understanding the experiments.

\textbf{Chapter~\ref{7Conclude}.} We present the highlights that this thesis has contributed. We end this thesis with some conclusions on extreme wave modeling. Furthermore, we give some recommendations on possible future research directions.

\newpage
{\renewcommand{\baselinestretch}{1} 
}
\setcounter{chapter}{1}
\chapter{Mathematics of water waves} 			\label{2Model}

\section{Introduction}

In this chapter, we present the mathematical description of surface water waves\index{water waves}. The theory of water waves has a long history since the time of Isaac Newton (1643-1727). For an overview of the origins and the development of water wave theory, the readers are encouraged to consult~\citep{2Craik04} and~\citep{2Craik05} for a thorough examination of Stokes' papers and letters concerning water waves as well as on how Stokes built on the earlier foundations to establish a definite theory of linear and weakly nonlinear waves.

We are interested in nonlinear and dispersive wave equations\index{dispersive wave!nonlinear equation} that describe extreme wave phenomena. In fact, the context of wave evolution in this thesis is meant to be the dispersive type of wave motion. Potential flow is considered with no shear or vertical components. Another class of wave motion is the hyperbolic type since it is formulated mathematically in terms of hyperbolic differential equations. A formal definition of dispersive waves\index{dispersive wave} is given by~\citet{2Whitham74}. Elementary solutions for linear problems are given as $\eta(x,t) = A e^{i(k x - \omega t)} + $ c.c., where c.c. denotes the complex conjugate of the preceding term; the wavenumber $k$ and the frequency $\omega$ \label{frequency} are related by the `linear dispersion relation'\index{dispersion relation!linear} $\omega = \Omega(k)$. \label{lindisrel} Dispersive means that this function has to be real, the group velocity \index{group velocity} $\Omega'(k)$ is not constant, which means that the wave velocity depends on the wavelength. Since the linear dispersion relation\index{dispersion relation!linear} for surface water waves\index{water waves} is given by $\Omega(k) = k \sqrt{(\tanh k)/k}$, this function is invertible and we will write $k = K(\omega)$, where $K = \Omega^{-1}$ is the inverse. \label{inverseLDR} The last representation is used more often in the signaling problem\index{signaling problem} while the former one is used more often in the initial value problem\index{initial value problem}. However, in this thesis, we will use both representations interchangeably.

One example of a nonlinear dispersive wave equation\index{dispersive wave!nonlinear equation} is the Korteweg-de Vries (KdV) equation\index{KdV equation}. The readers who are interested in the historical essays on the KdV equation can consult \citep{2Miles81, 2Drazin89, 2Bullough95}. For the narrow-banded spectrum\index{spectrum!narrow-banded} of wave groups, the nonlinear Schr\"{o}dinger (NLS) equation\index{NLS equation} can be derived from the KdV equation. Although both the KdV and the NLS equations are two of many mathematical models describing wave evolution, there is a significant difference. The KdV equation describes the waves themselves (the elevation of the water surface) while the NLS equation describes the envelope of wave groups. `Wave groups',\index{wave group(s)} sometimes also called `wave packets',\index{wave packet(s)} \index{wave packet(s)|see{wave group(s)}} are localized groups of waves that travel with the group velocity \index{group velocity} and naturally occur when the waves are dispersive and have a small deviation of an averaged wavelength. Hence, a wave group is composed of a superposition of a collection of waves with frequencies centered around one value.

In the following sections, we will consider various properties of the NLS equation. This chapter is organized as follows. Section~\ref{Bab2LDWE} starts with linear dispersive wave equations and the approximative description of wave groups. Section~\ref{Bab2NDWE} continues with nonlinear dispersive wave equations. This section explains the relationship between the KdV and NLS equations. Both spatial and temporal NLS equations are derived in this section, together with the corresponding phase-amplitude equations. Section~\ref{Bab2NLSDWE} deals with the property of the dispersion in the NLS equation that describes the linear modulational (Benjamin-Feir) instability\index{modulational instability}\index{Benjamin-Feir instability} and resumes the simplest coherent states of the equation. Section~\ref{varformNLS} presents the NLS equation\index{NLS equation} from a different perspective, namely from the variational formulation that will be used extensively in Chapter~\ref{3Property}.

\section{Linear dispersive wave equation} \label{Bab2LDWE}
\index{dispersive wave!linear equation}

For a linear dispersive wave equation, it is possible to write the general solution in the Fourier transform representation. Let a linear wave problem be specified by the evolution equation:
\begin{equation}
  \partial_{t} \eta + i \Omega(-i\partial_{x}) \eta = 0.
\end{equation}
Then the general solution $\eta(x,t)$ is given in the Fourier representation:
\begin{equation}
  \eta(x,t) = \frac{1}{2\pi} \int_{-\infty}^{\infty} F(\omega) e^{i(k x - \omega t)} d\omega,
\end{equation}
where $k$ and $\omega$ are related by the linear dispersion relation\index{dispersion relation!linear}. For a signaling problem, $F(\omega)$ is the Fourier transform of the initial signal $\eta(0,t)$, given as follows:
\begin{equation}
  F(\omega) = \int_{-\infty}^{\infty} \eta(0,t) e^{i \omega t} dt.
\end{equation}
The wavenumber $k$ and the frequency $\omega$ are related by the linear dispersion relation\index{dispersion relation!linear} $\omega = \Omega(k)$ or $k = K(\omega)$, where $K = \Omega^{-1}$. From the fact that $k = K[\Omega(k)]$, the relations between its derivatives can be found explicitly:
\begin{equation}
  K'(\omega_{0}) = \frac{1}{\Omega'(k_{0})} \qquad \textmd{and} \qquad
  K''(\omega_{0})= -\frac{\Omega''(k_{0})}{[\Omega'(k_{0})]^{3}}.
\end{equation}
Under the assumption that $F(\omega)$ is narrow-banded spectrum \index{spectrum!narrow-banded} around $\omega_{0}$, i.e., vanishingly small outside a small interval around $\omega_{0}$, we write the dispersion relation\index{dispersion relation} in its Taylor expansion around $\omega_{0}$:
\begin{equation}
  k = K(\omega) = \sum_{n=0}^{\infty} \frac{1}{n!} K^{(n)}(\omega_{0}) (\omega - \omega_{0})^{n}
    = k_{0} + K'(\omega_{0})(\omega - \omega_{0}) + \frac{1}{2!}K''(\omega_{0})(\omega - \omega_{0})^{2} + \dots.
\end{equation}
Therefore, we can write $\eta(x,t) = A(\xi,\tau) e^{i(k_{0}x - \omega_{0}t)}$, where $A$ is the corresponding complex-valued amplitude of the wave group:\index{wave group(s)}
\begin{equation}
  A(\xi,\tau) = \frac{1}{2\pi} \int_{-\infty}^{\infty} F(\omega_{0} + \nu) e^{-i(\tau - K_{\textmd{res}}(\omega)/\nu^{2} \xi)} d\nu,  \label{complexamplitude}
\end{equation}
where $\nu = \omega - \omega_{0} = \mathcal{O}(\epsilon)$, $\epsilon$ \label{epsilon} is a small positive parameter, $\xi = \nu^{2} x$, $\tau = \nu(t - K'(\omega_{0})x)$ and
\begin{equation}
  K_{\textmd{res}}(\omega) = K(\omega) - [k_{0} - K'(\omega_{0})\nu] = \left(\frac{1}{2!}K''(\omega_{0}) + \frac{1}{3!}K'''(\omega_{0})\nu + \dots \right)\nu^{2}.
\end{equation}
From the complex-amplitude representation~\eqref{complexamplitude}, it follows that $\nu$ is associated with the differential operator $-i \partial_{\tau}$, and so $\nu^{2} = - \partial_{\tau}^{2}$. Thus, the complex-valued amplitude $A$ satisfies
\begin{equation}
  \partial_{\xi}A + i K_{\textmd{res}}(-i\partial_{\tau}) A = 0. 	  \label{LSKres}
\end{equation}
For narrow-banded spectrum,\index{spectrum!narrow-banded} the equation~\eqref{LSKres} reduces to an approximate equation, the so-called `linear Schr\"{o}dinger' equation\index{linear Schr\"odinger equation}, by approximating $K_{\textmd{res}}$ by its lowest order term:
\begin{equation}
  \partial_{\xi}A + i \beta \partial_{\tau}^{2} A = 0,				  \label{linearSchrodinger}
\end{equation}
where the dispersion coefficient \index{dispersion!coefficient} $\beta = \frac{1}{2} K''(\omega_{0}) = -\frac{1}{2}\frac{\Omega''(k_{0})} {[\Omega'(k_{0})]^{3}}$. In the next section, we will see that a cubic term is added to equation~\eqref{linearSchrodinger} when the governing wave equation contains nonlinear terms.

\section{Nonlinear dispersive wave equation} \label{Bab2NDWE}
\index{dispersive wave!nonlinear equation}

In this section, we study the KdV-NLS relationship. We also derive the complex amplitude equation from the KdV equation with the exact dispersion relation\index{KdV equation!exact dispersion relation}. Depending on the scaling in variables, we obtain either the spatial or temporal NLS equation\index{NLS equation}.

\subsection{The KdV-NLS relationship}
\index{KdV equation!relation to the NLS equation} \index{NLS equation!relation to the KdV equation}

The relation between the KdV equation and the NLS equation has been known for many years. The NLS equation describing the slow modulation of a harmonic wave moving over a surface of a two-dimensional channel is derived using the multiple scale method~\citep{2Johnson76, 2Johnson97}. \index{multiple scale method} An analysis of the connection between the NLS and KdV periodic inverse scattering methods for wave packets is given by~\citet{2Tracy88, 2Tracy91}. \index{inverse scattering technique} Moreover, the derivation from the KdV type of equation with the exact dispersion relation\index{dispersion relation} is given by~\citet{2vanGroesen98} and~\citet{2Cahyono02}. Another reference to the derivation through perturbation theory can also be found~\citep{2Boyd01}. In addition, the latter authors also give a discussion on the KdV-induced long wave pole in the nonlinear coefficient of the NLS equation, resonance effects, and numerical illustrations. We will review again the latter derivation in this chapter and give special emphasis on the spatial NLS equation.

\subsection{Derivation of the spatial NLS equation} \label{derivespatialNLS}
\index{NLS equation!type!spatial}

In this subsection, we derive the spatial NLS equation using the `multiple time scale' method \citep{2Kevorkian61, 2Nayfeh73}. \index{multiple scale method} The KdV type of equation with the exact dispersion relation\index{KdV equation!exact dispersion relation} is given by
\begin{equation}
  \partial_{t}\eta + i \Omega(-i\partial_{x})\eta + c\,\eta \partial_{x}\eta = 0, \qquad c \in \mathbb{R}			  \label{KdVeqn}
\end{equation}
where $c$ is the coefficient in front of the nonlinear term in the KdV equation and contributes to the coefficient in front of the nonlinear term in the NLS equation as we will see; $\Omega$ is used for both the differential operator and the dispersion relation. We consider the residue of an approximation up to the third-order to find the corresponding complex-amplitude equation. An expression for wave groups\index{wave group(s)} consisting of a superposition of a first-order harmonic, a second-order double harmonic and a second-order non-harmonic long wave is given by
\begin{equation}
  \eta(x,t) = \epsilon A(\xi,\tau)e^{i\theta} + \epsilon^{2}[B(\xi,\tau)e^{2i\theta} + C(\xi,\tau)] + \textmd{c.c.}, 		  \label{etaABC}
\end{equation}
where $\epsilon$ is a small positive parameter as used commonly in perturbation theory, $\theta = k_{0}x - \omega_{0}t$, and c.c. means the complex conjugate of the preceding terms. The variables $\xi$ and $\tau$ are related to the physical variables $x$ and $t$ by
\begin{equation}
  \xi = \epsilon^{2}x \qquad \textmd{and} \qquad \tau = \epsilon(t - x/\Omega'(k_{0})). 		\label{hubungxitau}
\end{equation}
The complex amplitudes $A, B,$ and $C$ are allowed to vary slowly in the frame of reference. Substituting this Ansatz into the KdV equation~\eqref{KdVeqn} will result in an expression of the following form:
\begin{equation}
  R = \sum_{n,m} \epsilon^{n} R_{nm} e^{i m \theta} + \textmd{c.c.}, 			\label{Rnm}
\end{equation}
where $n \geq 1$, $m \geq 0$, and where the coefficients $R_{nm}$ contain expressions in $A$, $B$, and $C$ and their derivatives. In order to satisfy the KdV equation~\eqref{KdVeqn}, all coefficients of $R$ have to vanish. The first-order coefficients simply vanish: $R_{1m} = 0$, $m \geq 0$ when $k_{0}$ and $\omega_{0}$ are related by the linear dispersion relation\index{dispersion relation!linear}. The second-order coefficients read
\begin{align}
  R_{20} &= 0; \qquad
  R_{21}  = 0; \qquad  R_{2m} = 0, \quad m \geq 3; \\
  R_{22} &= i([\Omega(2k_{0}) - 2\omega_{0}]B + c\,k_{0}A^{2}).
\end{align}
Vanishing of $R_{22}$ expresses $B$ as a function of $A$:
\begin{equation}
  B(\xi,\tau) = \frac{c\,k_{0}A^{2}(\xi,\tau)}{2\omega_{0} - \Omega(2k_{0})}.		  \label{BR22}
\end{equation}
The third-order coefficients read:
\begin{align}
  R_{30} &= \left(1 - \frac{\Omega'(0)}{\Omega'(k_{0})} \right) \partial_{\tau}C - \frac{1}{2} \frac{c}{\Omega'(k_{0})} \partial_{\tau} |A|^{2} \\
  R_{31} &= \Omega'(k_{0})\partial_{\xi}A - \frac{1}{2}i \frac{\Omega''(k_{0})}{[\Omega'(k_{0})]^{2}} \partial_{\tau}^{2}A + i\, c k_{0} (A^{\ast}B + AC + A C^{\ast}) \\
  R_{32} &= -\frac{1}{\Omega'(k_{0})}\left(\Omega'(2k_{0})\partial_{\tau}B + \frac{1}{2} c \partial_{\tau} A^{2} \right) \\
  R_{33} &= 3ik_{0}c\,A B.
\end{align}
Vanishing of $R_{30}$ expresses $C(\xi,\tau)$ as a function of $A(\xi,\tau)$ and a $\xi$-dependent constant of integration $\alpha_{S}(\xi)$ for all $\theta$: \label{alphaS}
\begin{equation}
C(\xi,\tau) = \frac{1}{2} \frac{c\, |A(\xi,\tau)|^{2}} {\Omega'(k_{0}) - \Omega'(0)} + \alpha_{S}(\xi).
\end{equation}
To prevent resonance, $R_{31}$ has to vanish which leads to the dynamic evolution equation for $A$:
\begin{equation}
  \partial_{\xi}A + i \beta \partial_{\tau}^{2} A + i\gamma |A|^{2}A + \frac{2ik_{0} c}{\Omega'(k_{0})} \textmd{Re}(\alpha_{S}) A = 0.   \label{spatialNLSgauge}
\end{equation}
This equation is originally derived for the unidirectional wave propagation, in which the wave groups propagate over an uneven bottom and under the condition that reflection can be neglected. Since we consider the depth to be constant, we have Re$[\alpha_{S}(\xi)] = 0$ and both $\beta$ and $\gamma$ are constants~\citep{2Dingemans97, 2Dingemans01}. To get rid of the term Re$[\alpha_{S}(\xi)]$, we can also apply the `gauge transformation'\index{gauge transformation} \label{gaugetrans} by multiplying the evolution equation by $e^{\frac{2ik_{0}c}{\Omega'(k_{0})} \int \textmd{Re}[\alpha_{S}(\xi)] d\xi}$, see also~\citep{2Mei83}. Consequently, the new complex amplitude $\tilde{A}(\xi,\tau) = e^{\frac{2ik_{0}c}{\Omega'(k_{0})} \int \textmd{Re}[\alpha_{S}(\xi)] d\xi} A(\xi,\tau)$ now satisfies the spatial NLS equation, which after dropping the tilde reads
\begin{equation}
\partial_{\xi}A + i \beta \partial_{\tau}^{2} A + i\gamma |A|^{2}A = 0. 			\label{spatialNLS}
\end{equation}
Here the dispersion coefficient $\beta$ and the nonlinear
coefficient $\gamma$ are respectively given as follows:
\begin{align}
  \beta  &= \beta(k_{0}) =  - \frac{1}{2} \frac{\Omega''(k_{0})}{[\Omega'(k_{0})]^{3}} 			\label{beta} \\
  \gamma &= \gamma(k_{0},c^2) = \frac{k_{0}c^{2}}{\Omega'(k_{0})} \left(\frac{1}{\Omega'(k_{0}) - \Omega'(0)} + \frac{k_{0}}{2\omega_{0} - \Omega(2k_{0})} \right). 		  \label{gamma}
\end{align}
We observe that the nonlinear coefficient $\gamma$ depends quadratically on the nonlinear coefficient of the KdV equation $c$. Note also that applying the gauge transformation affects this nonlinear coefficient. In Chapter~\ref{6Experiment}, we will see that different values of $\gamma$ in the design experiments give large changes to the extreme position. For $\gamma = 0$, neglecting nonlinear effects, we see that the linear Schr\"{o}dinger equation~\eqref{linearSchrodinger} is recovered. The spatial NLS equation\index{NLS equation!type!spatial}~\eqref{spatialNLS} is the appropriate equation for the signaling problem that we consider in this thesis.

\subsection[Phase-amplitude equations]{Phase-amplitude equations for the spatial NLS equation}
\index{phase-amplitude equations!spatial NLS equation}
\label{PAspatial}

Let the complex amplitude $A$ be written in its original physical variables as $\bar{A}(x,t) = \epsilon A(\xi,\tau)$, with the transformation of variables as in~\eqref{hubungxitau}. Then the NLS equation in the physical variables is expressed as follows:
\begin{equation}
  \partial_{x}\bar{A} + \frac{1}{\Omega'(k_{0})} \partial_{t} \bar{A} + i \beta \partial_{t}^{2} \bar{A} + i \gamma |\bar{A}|^{2}\bar{A} = 0. \label{spatialNLSphysical}
\end{equation}
Now let this complex amplitude $\bar{A}$ be written in its polar form with the real-valued amplitude $a(x,t)$ and the real-valued phase $\phi(x,t)$, $\bar{A}(x,t) = a(x,t)e^{i\phi(x,t)}$. Then the phase-amplitude equations are obtained by substituting it into the NLS equation~\eqref{spatialNLSphysical}. This transformation is nonlinear and is also called `Madelung's transformation'\index{Madelung's transformation}~\citep{2Sulem99}, referring to Madelung's paper in 1927 on quantum theory~\citep{2Madelung27}. After the substitution, we remove the factor $e^{i\phi}$ and collect the real and the imaginary terms. Vanishing of both the real and the imaginary parts leads to the following coupled phase-amplitude equations in the original physical variables:
\begin{equation}
\left\{
\begin{array}{ll}
    \displaystyle{
    \partial_{x}a + \frac{\partial_{t}a}{\Omega'(k_{0})} -
    \beta \left(a \partial_{t}^{2} \phi + 2 \partial_{t}a \partial_{t}\phi \right)} &= 0 \\
    \displaystyle{
    \partial_{x}\phi + \frac{\partial_{t}\phi}{\Omega'(k_{0})} +
    \beta \left(\frac{\partial_{t}^{2}a}{a} - (\partial_{t}\phi)^{2} \right) + \gamma a^{2}} &= 0.
\end{array}
\right. \label{phaseamplitudeeqns}
\end{equation}
Let us define the local wavenumber\index{local wavenumber} $k$ and the local frequency\index{local frequency} $\omega$ as follows:
\begin{equation}
  k(x,t)      = k_{0}      + \partial_{x} \phi \qquad
  \omega(x,t) = \omega_{0} - \partial_{t} \phi.
\end{equation}
Writing the expressions in terms of local wavenumber \label{localwavenumber} and local frequency, \label{localfrequency} we can write the phase-amplitude equations in a simpler way.

The amplitude equation is the first equation of the phase-amplitude equations~\eqref{phaseamplitudeeqns}. After expressing $\partial_{t}\phi$ in terms of the local frequency and multiplying the equation with $a$, it can be written as follows:
\begin{equation}
  \frac{1}{2} \partial_{x}(a^{2}) + \frac{1}{2} \partial_{t} \left([K'(\omega_{0}) + K''(\omega_{0})(\omega - \omega_{0})] a^{2} \right) = 0.
\end{equation}
Now the amplitude equation is known as the `energy equation'\index{energy equation}. For the accuracy up to \textmd{$\cal{O}$}$(\nu^{2})$, where $\nu = \omega - \omega_{0}$ as in page~\pageref{complexamplitude}, this equation is written as follows:
\begin{equation}
  \partial_{x}(a^{2}) + \partial_{t}[K'(\omega) a^{2}] = 0, 		  \label{energyequation}
\end{equation}
which describes the conservation of energy.

The phase equation is the second equation of the phase-amplitude equations~\eqref{phaseamplitudeeqns}. Similarly, we express $\partial_{x}\phi$ and $\partial_{t}\phi$ in terms of the local wavenumber\index{local wavenumber} and the local frequency\index{local frequency}. We obtain the following relationship:
\begin{equation}
   \left[k_{0} + K'(\omega_{0})(\omega - \omega_{0}) + \frac{1}{2}K''(\omega_{0})(\omega - \omega_{0})^{2} \right] - k = \beta \frac{\partial_{t}^{2}a}{a} + \gamma\, a^{2}.
\end{equation}
This equation can now be written as the nonlinear dispersion relation\index{dispersion relation!nonlinear}. For the accuracy up to \textmd{$\cal{O}$}$(\nu^{2})$, it is given by
\begin{equation}
   K(\omega) - k = \beta \frac{\partial_{t}^{2}a}{a} + \gamma\, a^{2}.
   \label{nondisrel}
\end{equation}
This expression describes the relationship between the local wavenumber\index{local wavenumber} and the local frequency\index{local frequency} in the dispersion plane. Note that in general the right-hand side of~\eqref{nondisrel} does not vanish, and hence $(k,\omega)$ does not satisfy the linear dispersion relation\index{dispersion relation!linear}. The ratio of the second derivative of $a$ with respect of $t$ and the real amplitude $a$ itself is called the `Chu-Mei quotient'\index{Chu-Mei quotient}. Chu and Mei introduced this term for the first time when they derived the modulation equations of Whitham's theory for slowly varying Stokes waves~\citep{2Chu70, 2Chu71}. Some authors call this quotient the `Fornberg-Whitham term'~\citep{2Infeld90}, referring to~\citep{2Fornberg78}.

Note that the Chu-Mei quotient\index{Chu-Mei quotient} is an immediate consequence of the nonlinear transformation $A \longmapsto (a,\phi)$ and it is not of the nonlinearity of the evolution equation. Indeed, for the linear Schr\"{o}dinger equation ($\gamma = 0$), we find that 
\begin{equation}
  K(\omega) - k = \beta \frac{\partial_{t}^{2}a}{a},
\end{equation}
where the right-hand side vanishes for constant amplitudes, i.e., for a monochromatic mode. We will see further in Chapter~\ref{4Dislocate} that the unboundedness of Chu-Mei quotient at the vanishing amplitude is responsible for the occurrence of wavefront dislocation.~\index{wavefront dislocation}

\subsection{Derivation of the temporal NLS equation}
\index{NLS equation!type!temporal}

The same method of multiple scales \index{multiple scale method} can also be applied to derive the temporal NLS equation from the KdV equation with exact dispersion relation\index{KdV equation!exact dispersion relation}~\citep{2vanGroesen98, 2Cahyono02}.  A significant difference is the choice of variables in the moving frame of reference, which now becomes $\xi = \epsilon(x - \Omega'(k_{0})t)$ and $\tau = \epsilon^{2}t$. The first- and the second-order coefficients are the same as in the spatial case. The third-order coefficients read
\begin{align}
  R_{30} &= [\Omega'(0) - \Omega'(k_{0})]\partial_{\xi}C + \frac{1}{2}c \, \partial_{\xi}|A|^{2}; \\
  R_{31} &= \partial_{\tau}A - \frac{1}{2}i\Omega''(k_{0}) \partial_{\xi}^{2}A + i\, c k_{0} (A^{\ast} B + AC + A C^{\ast}).
\end{align}
Vanishing of $R_{30}$ expresses $C$ as a function of $A$ and a $\tau$-dependent constant of integration $\alpha_{T}$, \label{alphaT} given as follows:
\begin{equation}
  C(\xi,\tau) = \frac{1}{2} \frac{c\, |A(\xi,\tau)|^{2}}{\Omega'(k_{0}) - \Omega'(0)} + \alpha_{T}(\tau). 		\label{CR30}
\end{equation}
To prevent resonance, $R_{31}$ has to vanish which leads to an evolution equation for $A$:
\begin{equation}
  \partial_{\tau}A + i \beta_{0} \partial_{\xi}^{2} A + i\gamma_{0}|A|^{2}A + 2 i k_{0} c\, \textmd{Re}[\alpha_{T}(\tau)] A = 0.
\end{equation}
A similar assumption of unidirectional wave propagation, applying the `gauge transformation' by multiplying the evolution equation by $e^{2ik_{0}c \int \textmd{Re}[\alpha_{T}(\tau)] d\tau}$, we obtain the temporal NLS equation for $A$:
\begin{equation}
  \partial_{\tau}A + i \beta_{0} \partial_{\xi}^{2} A +  i\gamma_{0}|A|^{2}A = 0,		  \label{temporalNLS}
\end{equation}
where
\begin{align}
  \beta_{0}  &= \beta \,  [\Omega'(k_{0})]^{3} = -\frac{1}{2}\Omega''(k_{0}) \\
  \gamma_{0} &= \gamma \, \Omega'(k_{0}) = k_{0}c^{2} \left(\frac{1}{\Omega'(k_{0}) - \Omega'(0)} + \frac{k_{0}}{2\omega_{0} - \Omega(2k_{0})} \right).
\end{align}
The temporal NLS equation\index{NLS equation!type!temporal} is appropriate for the initial value problem, for example, wave evolution in the oceans, including the dynamics of extreme waves\index{extreme waves} \citep{2Osborne00, 2Onorato01}.

\subsection[Phase-amplitude equations]{Phase-amplitude equations for the temporal \\ NLS equation}
\index{NLS equation!type!temporal} \index{phase-amplitude equations!temporal NLS equation}

A similar procedure as in Subsection~\ref{PAspatial} gives the phase-amplitude equations for the temporal NLS equation. The `energy equation'\index{energy equation} is given by
\begin{equation}
  \partial_{t}(a^{2}) + \partial_{x}[\Omega'(k) a^{2}] = 0.
\end{equation}
The `nonlinear dispersion relation'\index{dispersion relation!nonlinear} is given as follows:
\begin{equation}
  \omega - \Omega(k) = \beta_{0} \frac{\partial_{x}^{2}a}{a} + \gamma_{0} a^{2}.		  \label{nondisreltem}
\end{equation}

\section[On the NLS equation as a dispersive wave equation]{On the NLS equation as a dispersive \\ wave equation}
\label{Bab2NLSDWE} \index{dispersive wave!NLS equation}

The temporal NLS equation\index{NLS equation!type!temporal}~\eqref{temporalNLS} is a nonlinear generalization of the linear equation $\partial_{\tau}A + i \beta_{0} \partial_{\xi}^{2} A = 0$. This linear equation is known as a Schr\"{o}dinger equation for the quantum mechanical probability amplitude of a particle (like an electron) moving through a region of uniform potential. Therefore, it is natural to call equation~\eqref{temporalNLS} the nonlinear Schr\"{o}dinger (NLS) equation, in this case, the temporal NLS equation. The NLS equation is also known as the `cubic Schr\"{o}dinger' equation \index{cubic Schr\"odinger equation} \index{cubic Schr\"odinger equation|see{NLS equation}} since the nonlinearity is of order three. Together with the KdV equation\index{KdV equation} and the sine-Gordon equation\index{sine-Gordon equation}, the NLS equation belongs to the classical soliton equations. Each one of these three and many more completely integrable equations possesses solutions with soliton properties. Readers interested in an overview of the classical soliton equations and the history of solitons may consult~\citep{2Scott03, 2Scott05}. A formal definition of soliton\index{soliton} is a solitary wave solution of a wave equation which asymptotically preserves its shape and velocity upon collision with other solitary waves \citep{2Scott73}. Although the term was originally applied only to solitary waves of the KdV equation, it is often used in a wider context without formal definition or verification of collision property.

As mentioned before, the NLS equation\index{NLS equation!in literature} describes the evolution of the envelope of a wave field and finds many applications. In hydrodynamics of nonlinear envelope waves, see e.g.,~\citep{2Benney67, 2Newell74, 2Whitham74, 2Yuen82}, in nonlinear optics\index{nonlinear optics}, see e.g.,~\citep{2Kelley65, 2Talanov65, 2Karpman69, 2Hasegawa73}, in nonlinear acoustics~\citep{2Tappert70}, in plasma physics, see e.g.,~\citep{2Taniuti68, 2Ichikawa72, 2Zakharov72}, and so forth. The NLS equation and its two-dimensional extension were originally derived for deep-water waves\index{deep-water waves} via a spectral method by~\citet{2Zakharov68}. Further, \citet{2Hasimoto72} and~\citet{2Davey72} derived the NLS equation for finite depth independently using multiple scale methods. \index{multiple scale method} In addition, \citet{2Yuen75} derived it using the averaged Lagrangian formulation from Whitham's theory. Moreover, a heuristic derivation of the NLS equation has been given by several authors~\citep{2Kadomtsev71, 2Karpman75, 2Jeffrey82, 2Dingemans97, 2Dingemans01}. A similar derivation of the spatial NLS equation can be found in~\citep{2Djordjevic78} under an assumption of the slowly varying bottom. The NLS equation is also a special case of the complex Ginzburg-Landau equation~\citep{2Ginzburg50, 2vanSaarloos92, 2Goldman94}. Specifically, we select the spatial type of the NLS equation instead of the temporal one since it is more suitable to model our wave evolution in a wave tank. A perspective from the variational formulation point of view will be presented below.

\subsection{Approximate dispersion of the NLS equation}
\index{NLS equation!approximate dispersion}

The linear dispersion relation\index{dispersion relation!linear} of surface water waves for finite depth is given by $\omega = \sqrt{gk\, \tanh (kh)}$. In normalized quantities, it is given by $\omega = \Omega(k) = k \sqrt{\tanh k/k}$. The NLS equation is derived under the assumption of narrow-banded spectrum. \index{spectrum!narrow-banded} In this subsection, we will compare the approximate dispersion corresponding to both the temporal and the spatial NLS equations. Let $(k_{0},\omega_{0})$ be the wavenumber and the frequency of a carrier wave of a wave group that satisfy the linear dispersion relation. We write the linear dispersion relation and its inverse in a Taylor series expansion around $k_{0}$ and $\omega_{0}$, respectively:
\begin{align}
\omega &= \omega_{0} + \Omega'(k_{0})(k - k_{0}) + \frac{1}{2}\Omega''(k_{0})(k - k_{0})^{2} + \dots \\
     k &= k_{0} + \frac{1}{\Omega'(k_{0})}(\omega - \omega_{0}) - \frac{1}{2} \frac{\Omega''(k_{0})}{[\Omega'(k_{0})]^{3}}(\omega - \omega_{0})^{2} + \dots.
\end{align}

\begin{figure}[t]		
\begin{center}
\includegraphics[width = 0.45\textwidth]{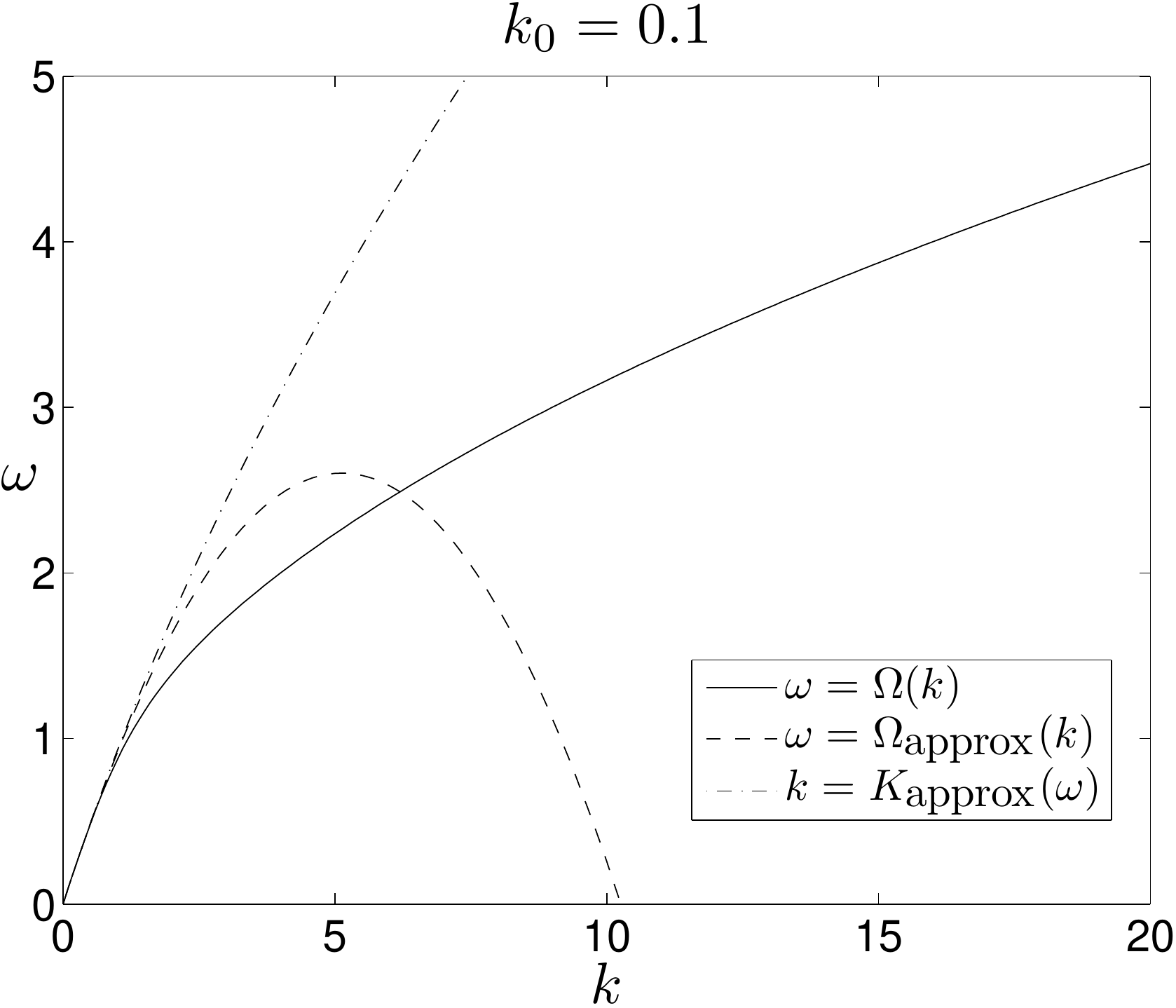} \hspace{0.75cm}
\includegraphics[width = 0.45\textwidth]{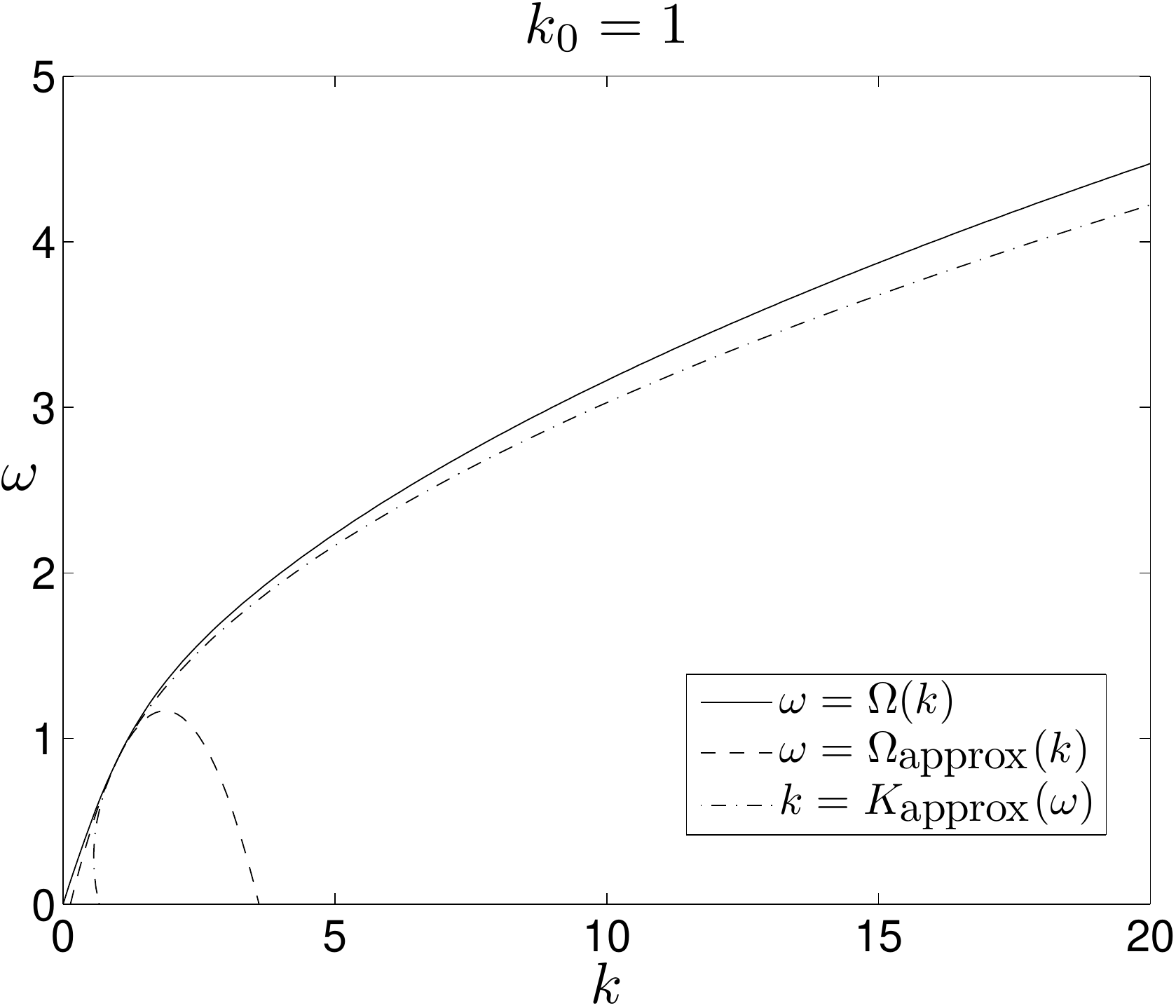}  \vspace{0.75cm} \\
\includegraphics[width = 0.45\textwidth]{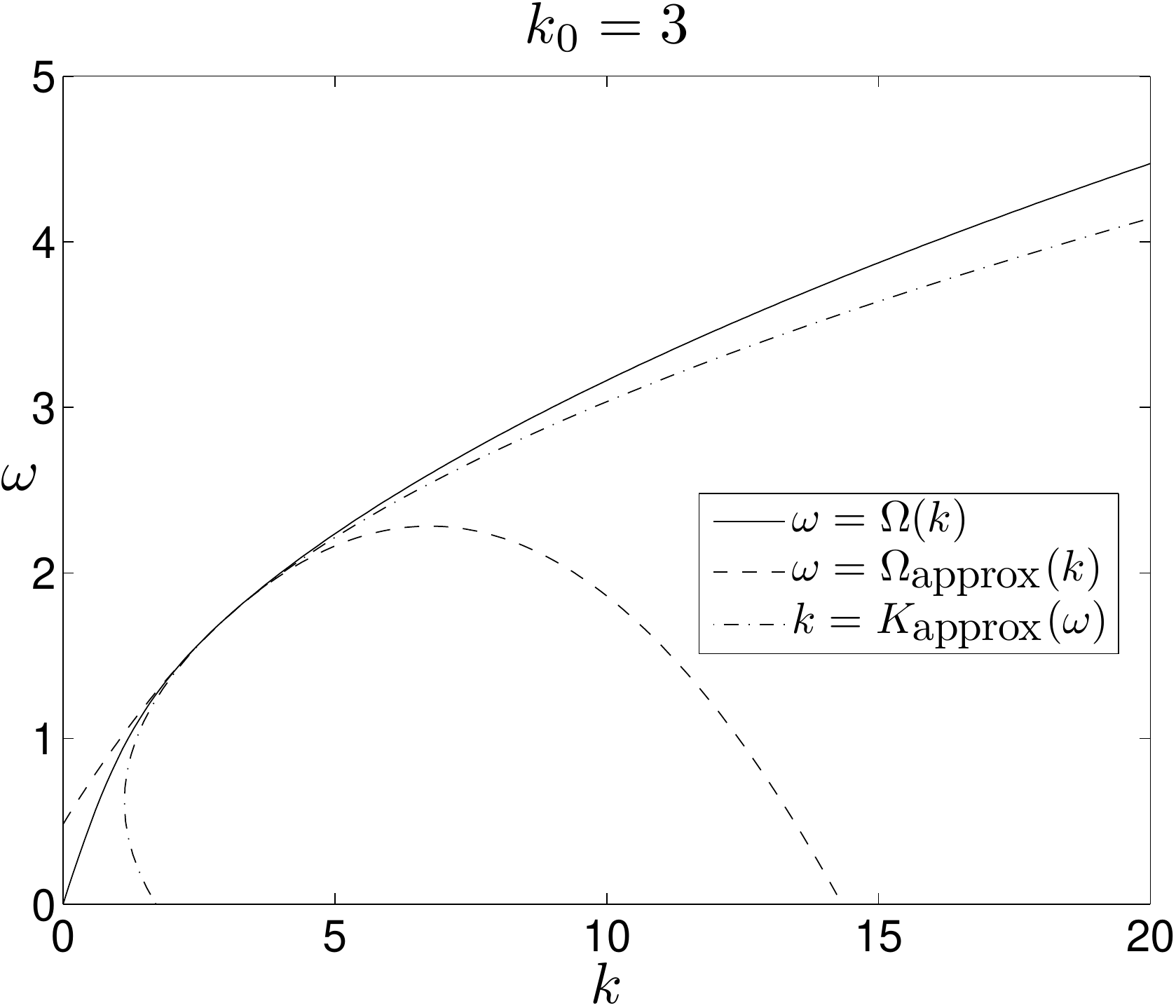}  \hspace{0.75cm}
\includegraphics[width = 0.45\textwidth]{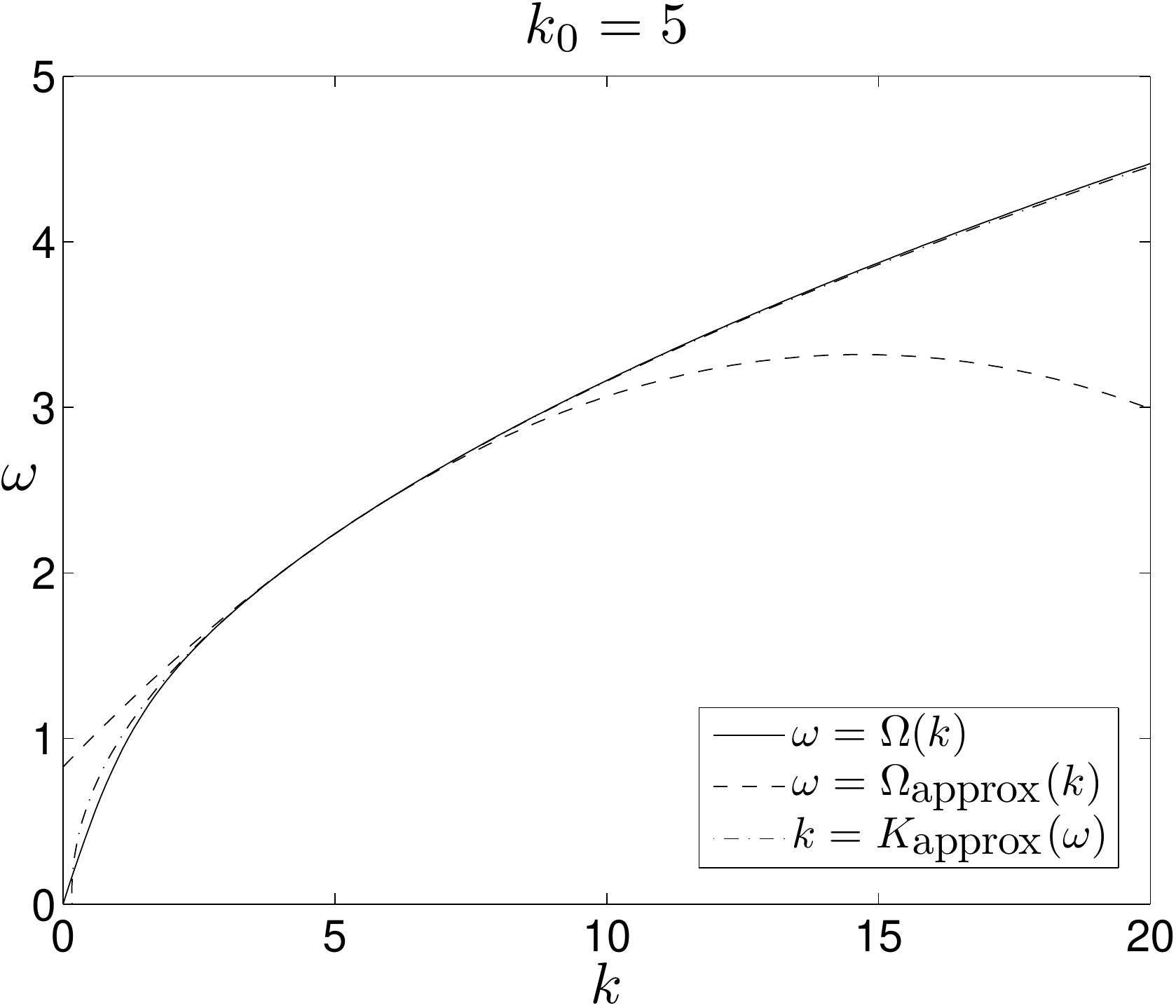}  
\caption[Linear dispersion relation and its quadratic approximations]{Plots of the linear dispersion relation $\Omega(k)$, its quadratic approximation $\Omega_{\textmd{approx}}(k)$ and the quadratic approximation of its inverse $K_{\textmd{approx}}(\omega)$ for different values of $k_{0}$. In all cases, the water depth is finite.}		    \label{LDRst}
\end{center}
\end{figure}

Let us denote the approximation up to and including quadratic terms above as $\Omega_{\textmd{approx}}(k)$ and $K_{\textmd{approx}}(\omega)$, respectively. \label{approx} Figure~\ref{LDRst} shows the plots of the linear dispersion relation and its two approximations for different values of $k_{0}$ and finite water depth. We observe that for $k_{0} \rightarrow 0$, both approximations are good but $\Omega_{\textmd{approx}}(k)$ approximates better the linear dispersion relation than $K_{\textmd{approx}}(\omega)$ for $k < k_{0}$. But for $k_{0} \rightarrow \infty$, $K_{\textmd{approx}}(\omega)$ approximates better than $\Omega_{\textmd{approx}}(k)$. This implies that the spatial NLS equation\index{NLS equation!type!spatial} is a better approximation than the temporal NLS equation\index{NLS equation!type!temporal} for a mathematical model of surface envelope water wave evolution.

For the case of deep water, the linear dispersion relation becomes $\Omega_{\textmd{deep}}(k) = \sqrt{k}$ and its inverse becomes $K_{\textmd{deep}}(\omega) = \omega^{2}$. The spatial NLS equation approximates better than the temporal NLS equation for deep-water waves\index{deep-water waves} since the former is exact\footnote[1]{Observation by Gert Klopman, 2006.} but the latter one is not, $K_{\textmd{approx,deep}}(\omega) = \omega^{2} = K_{\textmd{deep}}(\omega)$, but $\Omega_{\textmd{deep}} = \sqrt{k} \neq \Omega_{\textmd{approx,deep}}(k)$.

\subsection{Modulational instability} \label{SubsecMI}
\index{modulational instability}

Based on the value of the coefficients $\beta$ and $\gamma$, there are two types of the NLS equation. For $\beta \gamma > 0$, it is called the `focusing' type, or NLS($+$) type\index{NLS equation!type!focusing}. For $\beta \gamma < 0$, it is called the `defocusing' type, or NLS($-$) type\index{NLS equation!type!defocusing}. For the above NLS equation, since $\beta > 0$ for $k > 0$, the focusing type equation occurs for $\gamma > 0$. Positive values of $\gamma$ are found for sufficiently small wavelength: $k > k_{\textmd{crit}}$. \label{crit} The value of the critical wavenumber \index{critical wavenumber} is given by $k_{\textmd{crit}} = 1.363$ when the NLS equation is most accurately derived from the fully nonlinear water wave equation \citep{2Benjamin67}. When the NLS equation is derived from the KdV equation with exact dispersion relation, the critical value is found as $k_{\textmd{crit}} = 1.146$ \citep{2vanGroesen98}. The stability of a monochromatic wave\index{monochromatic wave(s)} depends on the wavenumber $k$: for $k < k_{\textmd{crit}}$ it is stable and for $k > k_{\textmd{crit}}$ it is unstable. The analysis of stability for finite depth is given by~\citet{2Benjamin67} and the same result is obtained independently using an average Lagrangian approach by~\citet{2Whitham67}, which is well explained in Whitham's book~\citep{2Whitham74}. \index{critical wavenumber}

As we will see in this section, a plane-wave solution of the focusing type of the NLS equation shows modulational instability if $k > k_{\textmd{crit}}$. On the other hand, a plane-wave solution with $k < k_{\textmd{crit}}$ is stable and therefore it is not so interesting to discuss in the context of modulational instability. However, this type of the NLS equation and one family of solutions the so-called `dark solitons'\index{dark soliton} received special attention from researchers in the field of nonlinear optics\index{nonlinear optics} \citep{2Kivshar98}.

In the context of this thesis, `modulational instability',\index{modulational instability} also called `sideband instability',\index{sideband instability} \index{sideband instability|see{modulational instability}} is defined as the process of growth in space of the plane-wave solution of the NLS equation as a result of a small modulation in a monochromatic wave signal. In the context of water waves\index{water waves}, the modulational instability is also known as the `Benjamin-Feir instability'\index{Benjamin-Feir instability} since~\citet{2BenjaminFeir67} investigated the phenomenon of wave trains in deep water. The same independent result is also observed in nonlinear liquids~\citep{2Bespalov66}. Apparently, the term `modulational instability' is mentioned for the first time by~\citet{2Taniuti68}. Furthermore, the instability of weakly nonlinear waves in dispersive media has been investigated by~\citet{2Lighthill65, 2Lighthill67, 2Benjamin67, 2Ostrovsky67, 2Whitham67}, and~\citet{2Zakharov67}. Much other research later on shows that modulational instability is observed in almost any field of wave propagation in nonlinear media~\citep{2Dodd82, 2Newell85, 2Lvov94, 2Remoissenet99}. For nonlinear optics\index{modulational instability!nonlinear optics}, we refer to~\citep{2Agrawal95, 2Hasegawa95}.

The references given in this paragraph refer to the temporal NLS equation\index{NLS equation!type!temporal}. The long time behavior of the modulational instability of the NLS equation is investigated by~\citet{2Janssen81}. The results are in qualitative agreement with experimental findings of~\citet{2Lake77} and the numerical computation of~\citet{2Yuen78a, 2Yuen78b}. Their works show that unstable modulations grow to a maximum limit and then subside. The energy is transferred from the primary wave to the sidebands\index{sideband(s)} for a certain period of time and then is recollected back into the primary wave mode. The long time evolution of an unstable wave train leads to a series of modulation-demodulation cycles in the absence of viscosity, known as the Fermi-Pasta-Ulam recurrence phenomenon~\citep{2Fermi55}. \index{Fermi-Pasta-Ulam recurrence phenomenon} An alternative treatment of the Benjamin-Feir instability mechanism of the two-dimensional Stokes waves on deep water was given by~\citet{2Stuart78}. Recently, \citet{2Segur05} show that any amount of a certain type of dissipation stabilizes the Benjamin-Feir instability for waves with narrow bandwidth and moderate amplitude. On the other hand, \citet{2Bridges04} show that there is an overlooked mechanism whereby the addition of dissipation leads to an enhancement of the Benjamin-Feir instability.

In the following, we derive the modulational instability \index{modulational instability!derivation} based on the NLS equation\index{NLS equation}. This stability analysis can be found also in several books~\citep{2Debnath94, 2Dingemans97}. For the readers who are interested in the history of T. B. Benjamin and his contributions to nonlinear wave theory, it is suggested to check a recent overview by~\citet{2Hunt05}.

Let us first introduce the simplest nontrivial solution of the NLS equation, called the `plane-wave'\index{plane-wave} solution. We discuss further this solution in the following subsection. Explicitly, it is given by
\begin{equation}
  A = A_{0}(\xi) = r_{0} e^{-i\gamma r_{0}^{2}\xi}. 		\label{planewave}
\end{equation}
To investigate the stability of the plane-wave\index{plane-wave!stability} solution of the NLS equation, substitute $A(\xi,\tau) = A_{0}(\xi) [1 + \epsilon B(\xi,\tau)]$ into the NLS equation~\eqref{spatialNLS}. The corresponding linearized equation reads
\begin{equation}
  \partial_{\xi}B + i \beta \partial_{\tau}^{2}B + i\gamma r_{0}^{2}(B + B^{\ast}) = 0.
\end{equation}
Substituting an Ansatz $B(\xi,\tau) = B_{1} e^{(\sigma \xi + i \nu \tau)} + B_{2} e^{(\sigma^{\ast} \xi - i \nu \tau)}$, \; $B_{1}$, $B_{2} \in \mathbb{C}$ \label{bilcomplex} into the linear equation, we get a set of two equations:
\begin{equation}
\left(
\begin{array}{cc}
  [\sigma - i(\beta \nu^{2} - \gamma r_{0}^{2})] & i\gamma r_{0}^{2}  \\
  -i\gamma r_{0}^{2} & [\sigma + i(\beta \nu^{2} - \gamma r_{0}^{2})] \\
\end{array}
\right)
\left(
\begin{array}{c}
  B_{1} \\
  B_{2}^{\ast} \\
\end{array}%
\right) = 
\left(
\begin{array}{c}
  0 \\
  0 \\
\end{array}
\right). \label{B1B2}
\end{equation}
Requiring the determinant of the matrix above to be zero, we have the condition $\sigma^{2} = 2\beta \gamma r_{0}^{2} \nu^{2} - \nu^{4}$. The coefficient $\sigma$ is real for sufficiently small $\nu$, and is then the growth rate\index{growth rate(s)} of the instability. The growth rate $\sigma > 0$ is given by $\sigma = \nu \sqrt{2 \beta \gamma r_{0}^{2} - \beta^{2}\nu^{2}}$. Defining a normalized modulation frequency\index{modulation frequency} \label{nu} $\tilde{\nu}$ by $\nu = r_{0} \sqrt{\frac{\gamma}{\beta}} \tilde{\nu}$, the growth rate $\sigma$ becomes $\sigma = \gamma r_{0}^{2} \tilde{\sigma}$, with $\tilde{\sigma} = \tilde{\nu} \sqrt{2 - \tilde{\nu}^{2}}$. Note that for instability, the normalized modulation frequency\index{modulation frequency!normalized} has to be in the instability interval $0 < \tilde{\nu} < \sqrt{2}$.

For the above two equations for $B_{1}$ and $B_{2}$~\eqref{B1B2}, we obtain $B_{1}/B_{2}^{\ast} = (\tilde{\nu}^{2} - 1) - i\tilde{\sigma}$. Taking the modulus of this ratio, we have $\left|B_{1}/ B_{2}^{\ast}\right| = 1$. Since $|B_{2}| = |B_{2}^{\ast}|$, we have also $|B_{1}| = |B_{2}|$. Therefore we write $B_{j} = |B_{j}|e^{i\theta_{j}}$, $j = 1, 2$, and we have the phase relation as follows:
\begin{equation}
  e^{i(\theta_{1} + \theta_{2})} = (\tilde{\nu}^{2} - 1) - i\tilde{\sigma}.				  \label{anglephases}
\end{equation}
The solution $A$ now reads
\begin{equation}
  A(\xi,\tau) = A_{0}(\xi) \left[1 + \epsilon |B_{1}| e^{\sigma \xi} \left(e^{i\theta_{1}} e^{i\nu \tau} + e^{-i\nu \tau} e^{i\theta_{2}} \right) \right].
\end{equation}
We see that for $\xi \rightarrow \infty$, this solution grows exponentially in space. This is the linear instability of the Benjamin-Feir modulated wave signal. Because of nonlinear effects from the cubic term nature that have been ignored in the analysis above, this growth is bounded. In the following chapter, we will see that a fully nonlinear extension of the Benjamin-Feir instability can be found, and is given by the Soliton\index{soliton!on finite background} on Finite Background (SFB)\index{SFB} and the rational solution for a very long modulation wavelength (when the modulation frequency $\nu \rightarrow 0$).

\subsection{Coherent state solutions} \label{Subseccoherent}
\index{coherent state}

\begin{figure}[h]			
\begin{center}
\subfigure[]{\includegraphics[width = 0.45\textwidth]{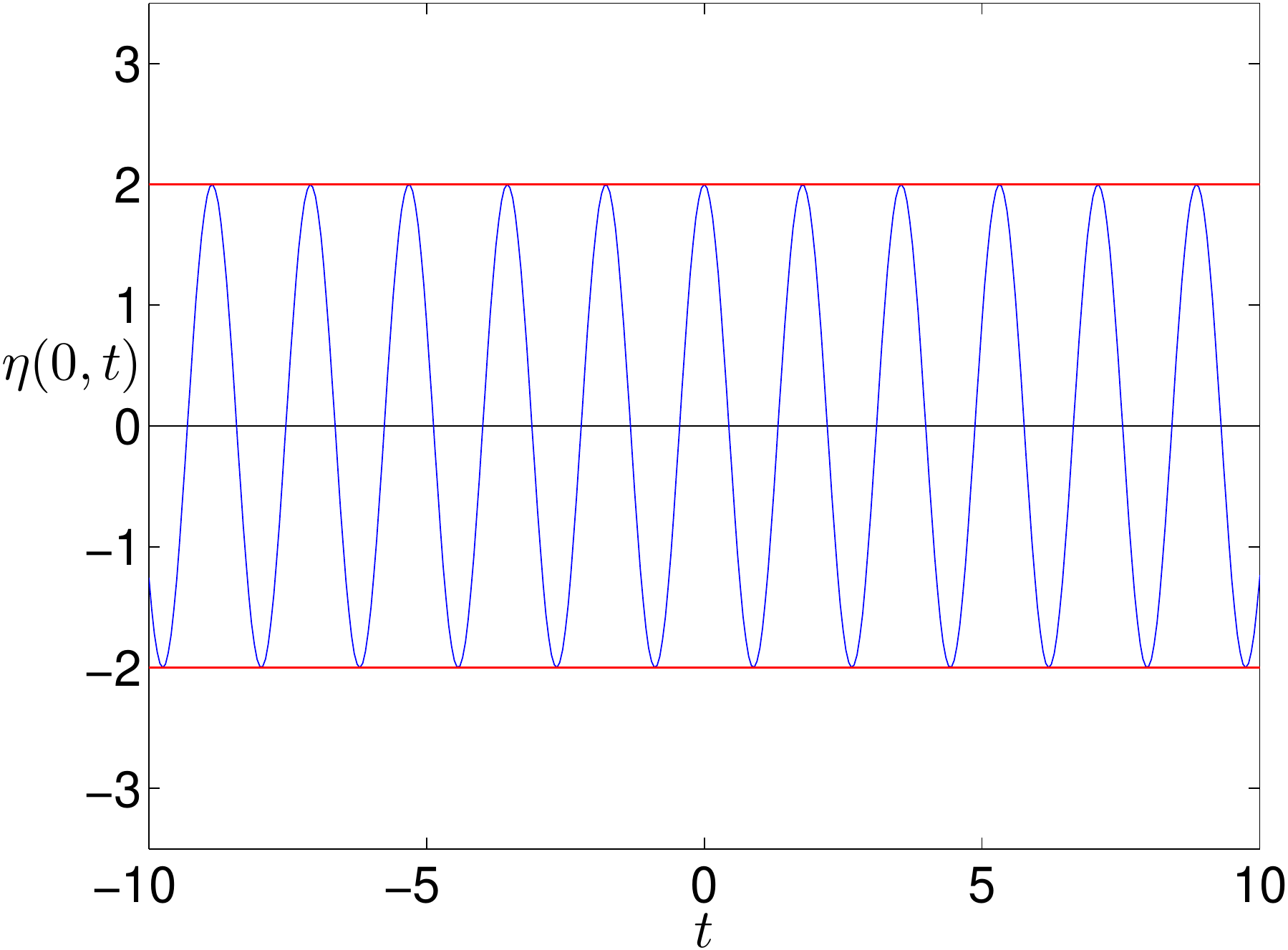}}						\hspace{0.75cm}
\subfigure[]{\includegraphics[width = 0.45\textwidth]{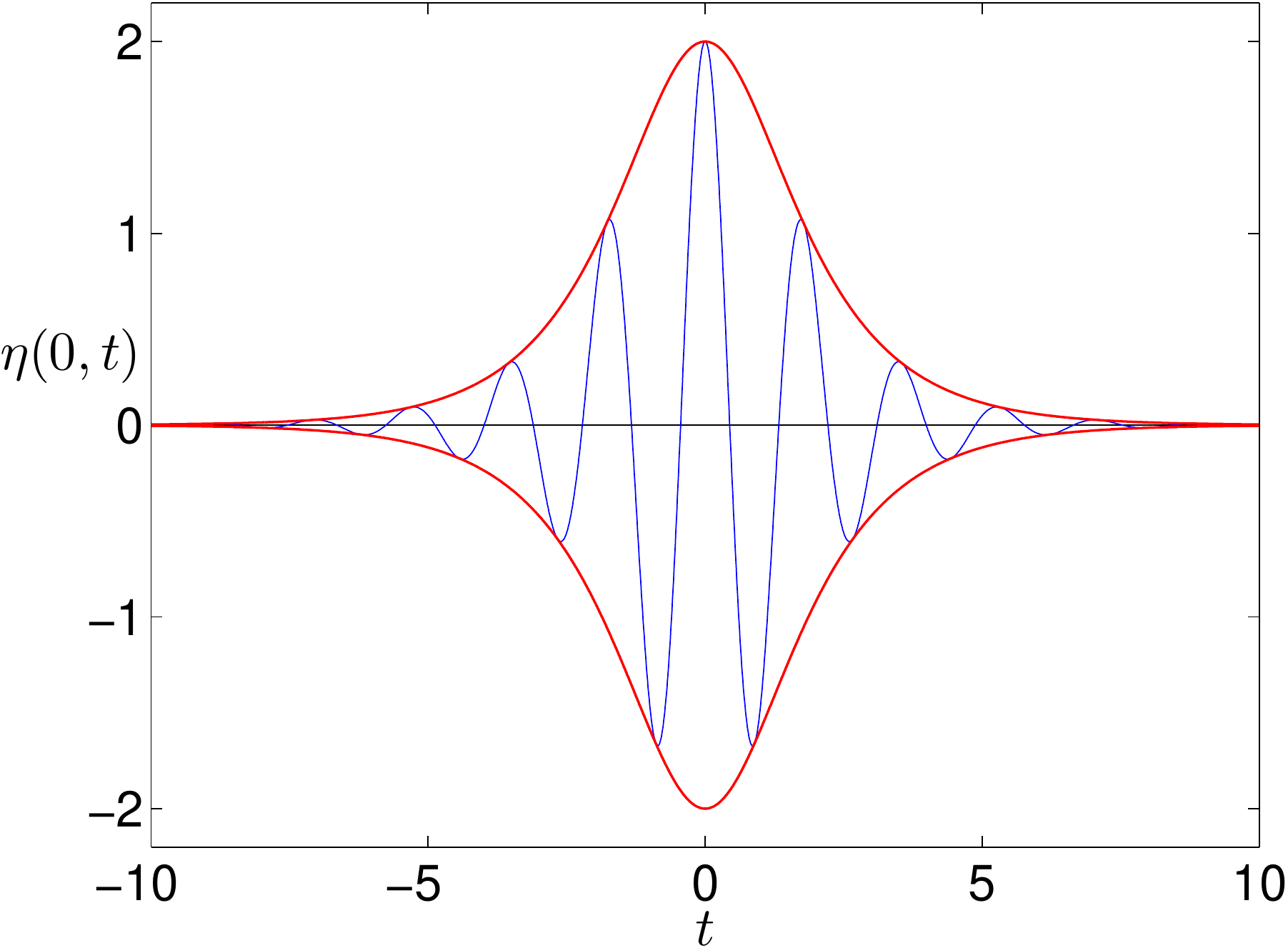}}
\caption[Wave signals of plane-wave and single soliton solutions]{Plots of the wave signal and the envelope of (a) the plane-wave solution and (b) the single soliton solution of the NLS equation. The plots are given for normalized parameters $r_{0} = \beta = \gamma = 1$.} \label{planeone}
\end{center}
\end{figure}

The NLS equation~\eqref{spatialNLS} has a number of exact solutions. The simplest nontrivial solution is the `plane-wave'\index{plane-wave!physical wave field} or the `continuous wave'\index{continuous wave} (cw) solution. It does not depend on the temporal variable $\tau$ and is given by~\eqref{planewave}, where $r_{0}$ is the plane-wave amplitude. The corresponding physical wave field in the lowest order term according to~\eqref{etaABC} is given by
\begin{equation}
  \eta(x,t) = 2 r_{0} \cos \left[(k_{0} - \gamma r_{0}^{2})x - \omega_{0} t \right]				  \label{planewavephysical}
\end{equation}
and it travels with the phase velocity of $\omega_{0}/(k_{0} - \gamma r_{0}^{2})$. Since $\gamma > 0$, this wave travels faster compared to a simple monochromatic wave\index{monochromatic wave(s)} that travels with phase velocity $\omega_{0}/k_{0}$. Note that the velocity is determined by the nonlinear dispersion relation\index{dispersion relation!nonlinear}, in agreement with~\eqref{nondisrel}. In addition, the wave has a constant amplitude $2r_{0}$. Figure~\ref{planeone}(a) shows a time signal and its envelope corresponding to the plane-wave solution at $x = 0$.

Another simple solution of the NLS equation is the `one soliton',\index{one soliton} \index{one soliton|see{single soliton}} or `single soliton',\index{single soliton} or `bright soliton'\index{bright soliton} \index{bright soliton|see{single soliton}} solution. It can be found by the inverse scattering technique~\citep{2ZakharovShabat72, 2Zakharov72}, \index{inverse scattering technique} but also much simpler by seeking a traveling-wave solution that decays at infinity. An explicit expression is given by
\begin{equation}
  A(\xi,\tau) = A_{0}(\xi)\, \sqrt{2} \sech \left(r_{0} \sqrt{\frac{\gamma}{\beta}} \tau\right),			  \label{onesoliton}
\end{equation}
where $A_{0}$ is the plane-wave solution as given above. This solution represents a solitary (envelope) wave, briefly called `envelope soliton'\index{envelope soliton}. The corresponding physical wave field of the single soliton\index{single soliton!physical wave field} for the lowest order term is given by
\begin{equation}
  \eta(x,t) = 2 r_{0} \sqrt{2} \, \sech \left(r_{0} \sqrt{\frac{\gamma}{\beta}} \left[t - x/\Omega'(k_{0}) \right] \right) \, \cos \left[(k_{0} - \gamma r_{0}^{2})x - \omega_{0} t \right].
\end{equation}
Different from the wave field corresponding to the plane-wave solution~\eqref{planewavephysical} that has a constant amplitude, this wave field of the single soliton has an amplitude that depends on time, is localized around $t = x/\Omega'(k_{0})$ and vanishes for $t \rightarrow \pm \infty$. The wave signal propagates in a nonlinear medium without spreading due to the balance of the wave dispersion and the nonlinearity of the system. Figure~\ref{planeone}(b) shows a time signal and its envelope of the single soliton solution at $x = 0$. Higher-order solutions, $N$-soliton solutions, can be constructed using the inverse scattering method of~\citep{2ZakharovShabat72}, see also~\citep{2Drazin89}. \index{inverse scattering technique}

We now remark that both these solutions have a coherent state character. Coherent state\index{coherent state} means that all Fourier components have the same phase. Generally speaking, if we write a function $f(t)$ with its Fourier transform \label{spectrum} $\hat{f}(\omega)$ as $f(t) = \int \hat{f}(\omega) e^{i \omega t} \, d\omega = \int \left|\hat{f}(\omega) \right| e^{i\vartheta(\omega)} e^{i \omega t} \, d\omega$, we will have that the phase $\vartheta(\omega)$ depends nonlinearly on $\omega$, since for $f(\omega) \in \mathbb{C}$. If this phase is constant (and then necessarily $0$ mod $\pi$ if $f$ is real), the function $f(t)$ is called `coherent'\index{coherent state}. In particular, it means that a real function $f$ has a maximal (or minimal) value at $t = 0$, since then all Fourier modes contribute optimally. The function $f(t)$ is coherent if $\vartheta(\omega) \equiv \vartheta_{0}$ for all $\omega \in \mathbb{R}$.

In order to explore more about the coherent state, we study the evolution of the spectrum of the solutions in the frequency domain. The spectrum $\hat{A}(\xi,\omega)$ of a complex amplitude function $A(\xi,\tau)$ is obtained by applying the Fourier transform with respect to the time variable $\tau$, as follows:
\begin{equation}
  \hat{A}(\xi,\omega) = \int_{-\infty}^{\infty} A(\xi,\tau) e^{-i\omega \tau} d \tau.
\end{equation}
Moreover, since the spectrum is a complex-valued function, it can be written in the polar coordinate as $\hat{A}(\xi,\omega) = |\hat{A}(\xi,\omega)| e^{i\vartheta(\xi,\omega)}$, where $|\hat{A}(\xi,\omega)|^{2}$ is called the `power spectrum'\index{spectrum!power} and $\vartheta(\xi,\omega)$ is called the `phase spectrum'\index{spectrum!phase}.

\begin{figure}[h]			
  \begin{center}
\subfigure[]{\includegraphics[width = 0.45\textwidth]{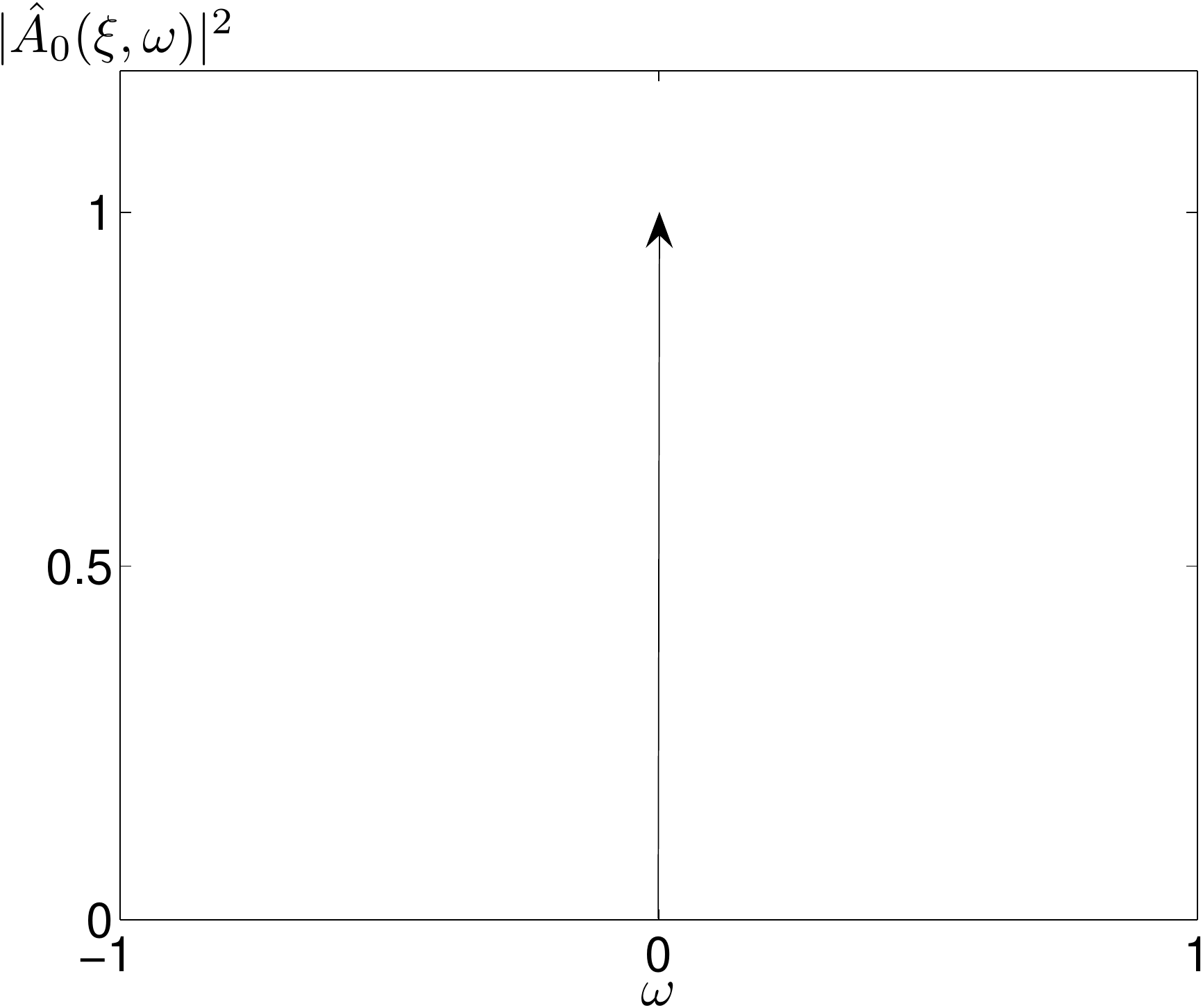}}						\hspace{0.75cm}
\subfigure[]{\includegraphics[width = 0.45\textwidth]{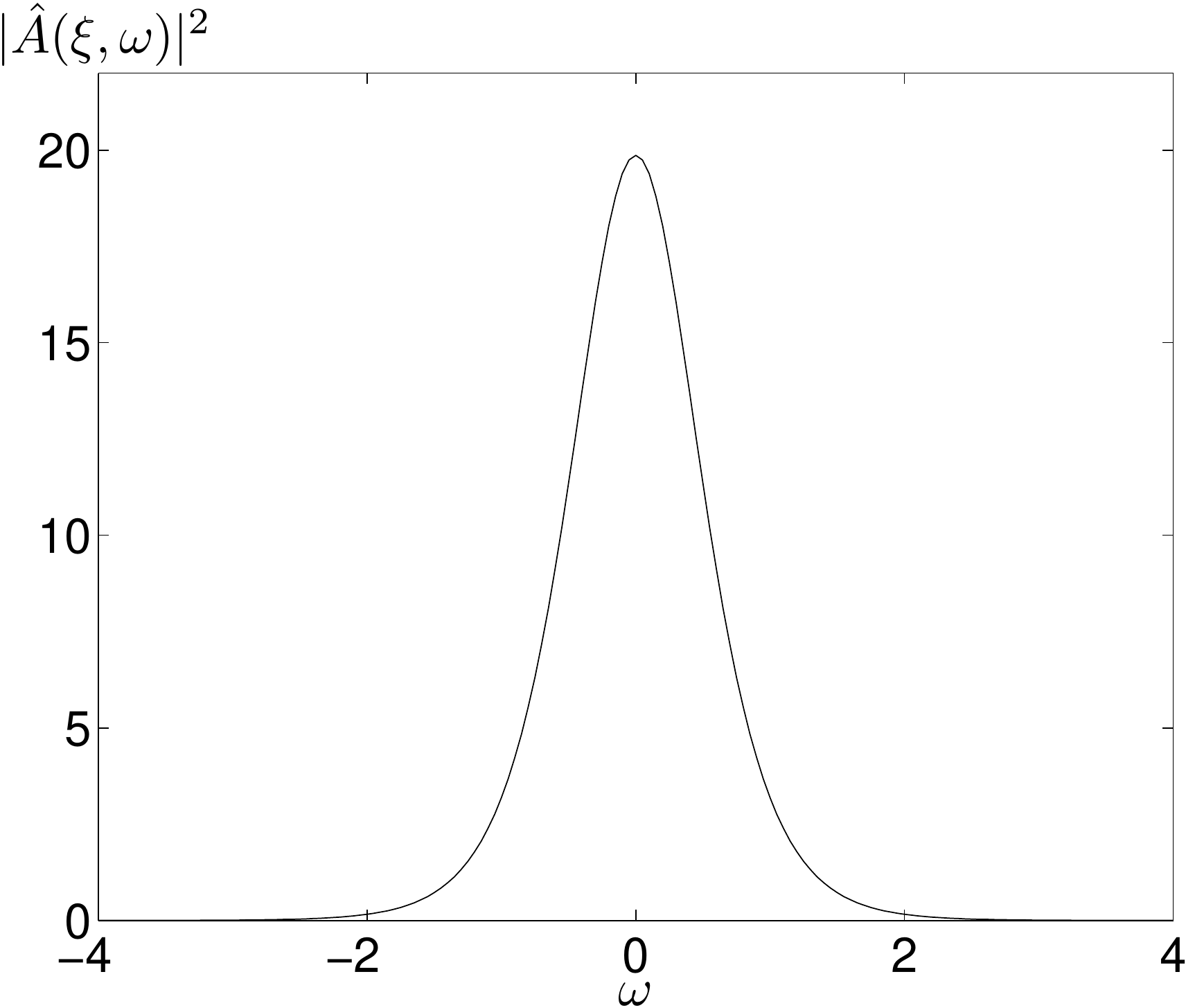}}
\caption[Power spectrum of plane-wave and single soliton solutions]{The power spectrum plots for (a) the plane-wave and (b) the single soliton solutions for normalized parameters $r_{0} = \beta = \gamma = 1$.} \label{spectra}
\end{center}
\end{figure}

The Fourier transform of the plane-wave\index{plane-wave!spectrum} is given by $\hat{A}_{0}(\xi,\omega) = A_{0}(\xi)\,\delta(\omega)$, where $\delta(\omega)$ is the Dirac's delta function. Furthermore, the Fourier transform of the single soliton\index{single soliton!spectrum} is given by
\begin{equation}
  \hat{A}(\xi,\omega) = 4 \alpha \sqrt{2} A_{0}(\xi) \sum_{m = 0}^{\infty} (-1)^{m} \frac{(2m + 1)}{\alpha^{2}(2m + 1)^{2} + \omega^{2}},	  \label{solitonspectrum}
\end{equation}
where $\alpha = r_{0} \sqrt{\gamma/\beta}$. See Appendix~\ref{Singlesolitonspectrum} on page~\pageref{Singlesolitonspectrum} for the derivation of this expression. Since both spectra have the same phase for all $\omega$, both the plane-wave and the single soliton solutions have a coherent state\index{coherent state} characteristic. Figure~\ref{spectra} shows the power spectrum for the plane-wave and the single soliton solutions. We will see in the next chapter that other solutions of the NLS equation, in particular the SFB, are not coherent.

\section[Variational formulation and conserved quantities]{Variational formulation and conserved quantities of the NLS equation} \label{varformNLS} 
\index{NLS equation!variational formulation} \index{NLS equation!conserved quantities} \index{variational formulation}

In this section, we discuss the variational formulation of the NLS equation that we will use in Chapter~\ref{3Property}. We present the Lagrangian\index{Lagrangian} and Hamiltonian\index{Hamiltonian} structures corresponding to the NLS equation~\citep{2Sulem99}.

Let $A$ denote a smooth solution of the NLS equation
\begin{equation}
  \partial_{\xi} A + i \beta \partial_{\tau}^{2} A + i \gamma |A|^{2} A = 0.%
\end{equation}
A Lagrangian density\index{Lagrangian!density} \textmd{$\cal{L}$} associated with the NLS equation can be written in terms of $A$ and is given as follows:
\begin{equation}
  \textmd{$\cal{L}$}(A) = \frac{1}{4}i(A^{\ast} \partial_{\xi}A - A \partial_{\xi}A^{\ast}) + \frac{1}{2} \beta |\partial_{\tau} A|^{2} - \frac{1}{4} \gamma |A|^{4}. 						  \label{Lagrangiandensity}
\end{equation}
Then the action functional\index{action functional} $\mathcal{A}$ is such that any solution $A$ of the NLS equation is a critical point of the corresponding action functional\index{action functional}:
\begin{equation}
  \mathcal{A}(A) = \int_{\xi_{0}}^{\xi_{1}} \int_{-\infty}^{\infty} \textmd{$\cal{L}$} \; d\tau \, d\xi.			  \label{actionfunctional}
\end{equation}
Indeed, the Euler-Lagrange equation\index{Euler-Lagrange equation} is given by
\begin{equation}
  \frac{\partial \textmd{$\cal{L}$}}{\partial A} =
  \frac{\partial}{\partial \tau} \left(\frac{\partial \textmd{$\cal{L}$}}{\partial A_{\tau}} \right) + 
  \frac{\partial}{\partial \xi } \left(\frac{\partial \textmd{$\cal{L}$}}{\partial A_{\xi }} \right),  
\end{equation}
and this equation leads to the NLS equation.

A Hamiltonian\index{Hamiltonian} structure is derived from the existence of a Lagrangian\index{Lagrangian}. The Hamiltonian density\index{Hamiltonian!density} \textmd{$\cal{H}$} corresponds to the NLS equation is given by
\begin{align}
  \textmd{$\cal{H}$}(A) &= \frac{1}{4}i(A^{\ast} \partial_{\xi}A - A \partial_{\xi}A^{\ast}) - \textmd{$\cal{L}$}(A) \\%
                        &= -\frac{1}{2} \beta |\partial_{\tau} A|^{2} + \frac{1}{4} \gamma |A|^{4}.%
                        	\label{Hamiltoniandensity}
\end{align}
The Hamiltonian $H$ is given by
\begin{equation}
  H(A) = \int_{-\infty}^{\infty} \textmd{$\cal{H}(A)$} \; d\tau.			  \label{Hamiltonian}
\end{equation}
The Hamilton equation\index{Hamilton equation} in the complex form is given by
\begin{equation}
  i\partial_{\xi}A = \delta H(A),
\end{equation}
where the right-hand side is the variational or Fr\'{e}chet derivative.\index{variational derivative}\index{Fr\'echet derivative|see{variational derivative}} This is the canonical form of the NLS equation since the right-hand side of the term reads $\delta H(A) = \beta \partial_{\tau}^{2} A + \gamma |A|^{2} A$. Thus, both the Euler-Lagrange and the Hamilton equations lead to the NLS equation.

Conservation laws are a common feature of mathematical physics, where they describe the conservation of fundamental physical quantities. It is well-known that conservation laws for many systems arise from variational principles that are invariant under transformations that belong to a continuous group. A conservation law then follows from the application of Noether's theorem\index{Noether's theorem} to the Lagrangian density~\citep{2Noether18, 2Gelfand63}. The NLS equation is a dynamical system with an infinite number of degrees of freedom and corresponds to an infinite-dimensional Hamiltonian system. It is integrable and as a consequence, it has an infinite number of conserved quantities~\citep{2Zakharov72}. These quantities have the form of an integral with respect to $\tau$ of a polynomial expression in terms of the function $A(\xi,\tau)$ and its derivatives with respect to $\tau$. The conserved quantities are given in the following form~\citep{2Lamb80}:
\begin{equation}
  (2i)^{n} C_{n} = \int_{-\infty}^{\infty} f_{n}(\tau)\, d\tau, \qquad n \geq 1,
\end{equation}
where
\begin{align}
  f_{0} 	&= 0, \qquad f_{1} = \frac{1}{2} \gamma |A|^{2} \\
  f_{n + 1} &= A \frac{d}{d\tau} \left(\frac{f_{n}}{A} \right) + \sum_{j + k = n} f_{j} f_{k}.
\end{align}
The first three conservation integrals have a simple physical meaning. The lowest order conserved quantity is called the `wave energy'\index{wave energy}, `mass'\index{mass}, `wave action'\index{wave action}, `plasmon number'\index{plasmon number}, or `wave power'\index{wave power} in optics. It is given as follows:
\begin{equation}
  C_{1} = \int_{-\infty}^{\infty} |A|^{2}\, d \tau.					  \label{waveenergy}
\end{equation}
The second-order conserved quantity is called the `(linear) momentum' and the explicit expression is given by
\begin{equation}
  C_{2} = \int_{-\infty}^{\infty} i(A \partial_{\tau}A^{\ast} - A^{\ast} \partial_{\tau}A)\, d \tau.%
\end{equation}
The third-order conserved quantity $C_{3}$ is precisely the `Hamiltonian'~\eqref{Hamiltonian} introduced above.\index{Hamiltonian}

\newpage
{\renewcommand{\baselinestretch}{1} 
}
\setcounter{chapter}{2}
\chapter{Waves on finite background} \label{3Property}
\index{waves on finite background}

\section{Introduction}

As mentioned earlier in Chapter~\ref{1Introduce}, the mathematical problem that is considered in this thesis is motivated by the following problem from the Maritime Research Institute Netherlands (MARIN)\index{MARIN}. We want to be able to generate large and non-breaking waves, known as extreme waves\index{extreme waves}, in the wave basins of MARIN. However, we do not know what kind of waves to generate. The wave generator,\index{wave generator} \index{wave generator|see{wavemaker}} or wavemaker, \index{wavemaker} has only limited capacity to make high waves directly. These high waves are already considered `extreme' if the ratio of the highest amplitude and the initial amplitude reaches a maximum value of three. Therefore, we let the `nature' to produce extreme waves in the laboratory with an input of relatively small amplitude wave signals. There are two possibilities in generating extreme waves mechanism: either linear focusing or the nonlinear modulational instability, as we have discussed in Chapter~\ref{2Model}.

We choose for modulational instability\index{modulational instability} process, in which the complete wave evolution is described by the Soliton on Finite Background (SFB)\index{SFB}, one family of solutions of the nonlinear Schr\"odinger (NLS) equation\index{NLS equation}. By providing sufficient space for the waves to develop, nature gives an amplitude amplification to these waves, leading to the formation of extreme wave events. The position where these extreme waves should occur is at a specific position in the wave basin. Practically it is desired to position the most extreme waves `halfway' of the basin, in order not to destroy the artificial beach at the end of the basin. But importantly, this extreme position should also be sufficiently far from the wavemaker so that nature has enough space in building up the extreme waves.\index{extreme waves}

Investigating extreme waves on a finite background of uniform monochromatic waves\index{monochromatic wave(s)} is related to the modulational instability of the plane-wave\index{plane-wave} solution of the NLS equation. Therefore, an initial modulated wave signal will grow exponentially in space according to the linear modulational instability but the nonlinearity will bind the growth and the wave becomes an extreme wave event. If we consider the space to be infinite, this extreme wave has the plane-wave solution as finite background. Actually, there is a family of exact solutions of the NLS equation that describes extreme wave events. This family is known as the SFB\index{SFB}~\citep{3Akhmediev87}. The SFB is not the only family of solutions with waves on finite background. There are two other solutions that belong to the type of waves on finite background, namely the Ma soliton\index{Ma soliton}~\citep{3Ma79} and the rational soliton\index{rational soliton}~\citep{3Peregrine83}. These three solutions of the NLS equation are also called breather solutions\index{breather solutions}.

The content of this chapter is a combination of two papers~\citep{3vanGroesen06, 3Andonowati07} and is organized as follows. Section~\ref{SectionWFB} will describe waves on a finite background from a variational formulation\index{variational formulation} perspective and a new displaced phase-amplitude representation is introduced. \index{displaced phase-amplitude} Sections~\ref{SectionpropertiesSFB}--\ref{SectionASE} are devoted to a special class of waves on finite background, the SFB. This extensive discussion on the SFB is the main contribution of the thesis. In Section~\ref{SectionpropertiesSFB}, we present the specifications and properties of the SFB, including several explicit expressions, the asymptotic behavior at a far distance, the physical wave field and the amplitude amplification factor. Section~\ref{SectionSFBevol} discusses the evolution of an SFB wave signal, the evolution in the Argand diagram and the phase-plane representation for the SFB envelope signal. Section~\ref{SectionASE} explains the SFB wave signal in the frequency domain and the evolution of the amplitude and phase spectrum. Section~\ref{SectionotherWFB} deals with other families of waves on finite background, namely the Ma soliton and the rational soliton. The relationship between these solutions is also explained and derived. The final section gives remarks on the content of this chapter.

\section{Description of waves on finite background} \label{SectionWFB}
\index{waves on finite background}

Waves on finite background have different characteristics from soliton type of solutions of the NLS equation, as described in Chapter~\ref{2Model} (see Subsection~\ref{Subseccoherent} on page~\pageref{Subseccoherent}). While the single-soliton vanishes at infinity, the waves on finite background are in fact nontrivial at infinity. In the case we will deal with in this chapter, the behavior at infinity will be the plane-wave solution, as defined in Chapter~\ref{2Model}, see the expression~\eqref{planewave}. In this section, the variational formulation\index{variational formulation} of the NLS equation will be used to describe such waves.

We will write the complex amplitude of the NLS equation in the form of displaced phase-amplitude\index{displaced phase-amplitude} variables, with a displacement that depends on the background; the governing equations for these variables are derived. In Subsection~\ref{pseudocoherent}, we investigate a special case by restricting the phase to depend only on space and not on time. As a consequence, the dynamics at each position is given as the motion of a nonlinear autonomous oscillator\index{nonlinear oscillator} in potential energy. The potential energy\index{potential energy} depends on the phase as a parameter and on the change of phase with the position. This change of phase with position corresponds physically to a change of the wavelength of the carrier wave and turns out to be the only driving force responsible for the nonlinear amplification towards an extreme wave. Remarkably, the assumption of the time-independent phase necessarily leads to the three families of breather solutions of the NLS equation mentioned in the preceding section.

\subsection{Displaced phase-amplitude representation}
\index{displaced phase-amplitude}

We are interested in solutions of the NLS equation~\eqref{spatialNLS} on page~\pageref{spatialNLS} \index{NLS equation!modulational instability} that are associated with modulational instability. Modulational instability\index{modulational instability}, with the Benjamin-Feir instability\index{Benjamin-Feir instability} of wave trains in surface water waves as a prime example, is commonly associated with finite-amplitude wave trains that get amplified by self-focusing processes due to modulations in the envelope amplitude~\citep{3BenjaminFeir67}. We will restrict in the following to this basic setting of a perturbation of a uniform wave train, although other `backgrounds' deserve more attention too.

\begin{figure}[h]			
\begin{center}
\includegraphics[width = 0.6\textwidth,viewport=133 433 479 697]{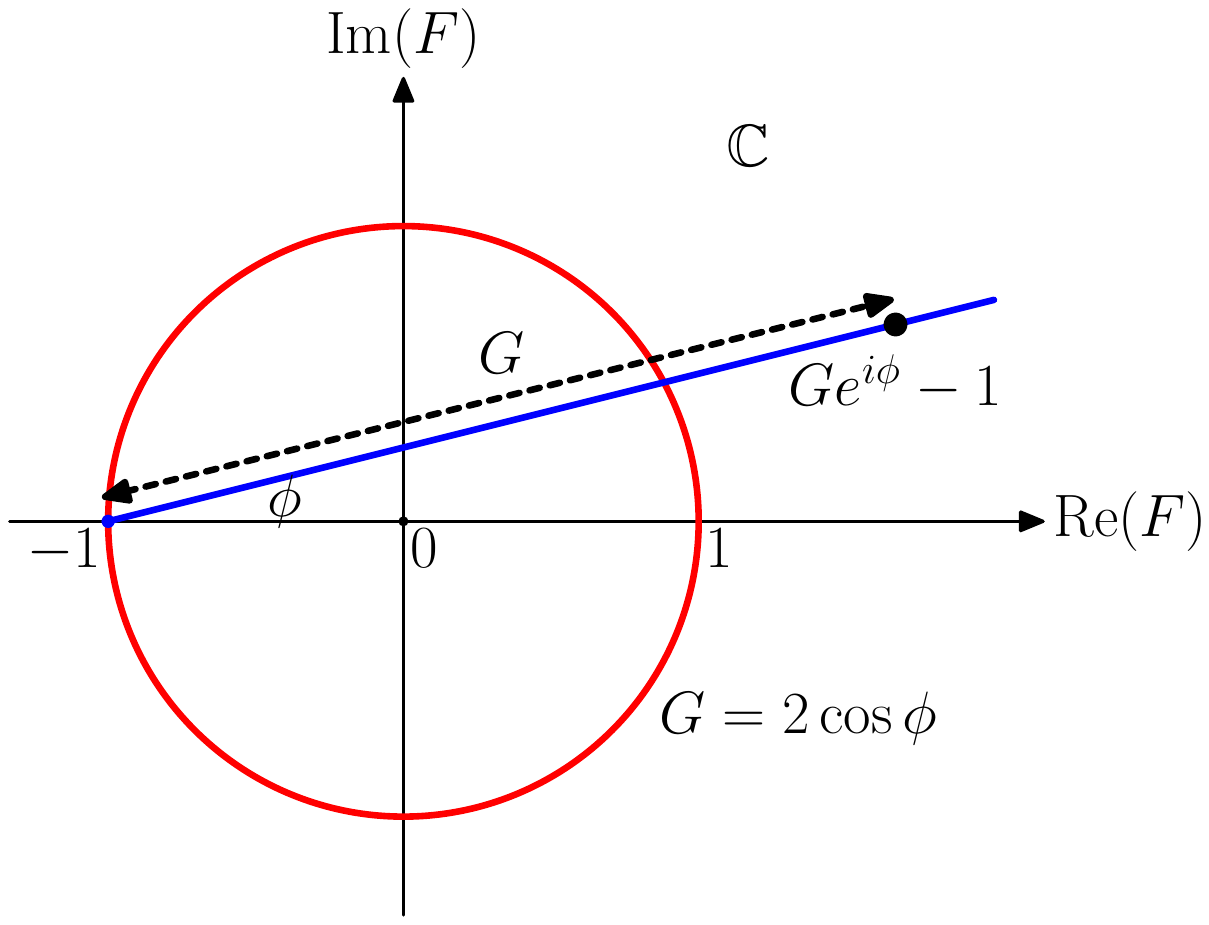}
\caption[Displaced phase-amplitude in the Argand diagram]{Indicated are the displaced phase-amplitude parameters in the Argand diagram for the amplitude of waves on finite background; the unit sphere corresponds to the set $G = 2\cos\phi$.}	    \label{sketsaargand}
\end{center}
\end{figure}

To define more precisely the class of waves we are interested in, we take as background the plane-wave solution of the NLS equation~\eqref{planewave} and will look for solutions of the form
\begin{equation}
  A(\xi,\tau) = A_{0}(\xi)\, F(\xi,\tau), \label{solutionA}
\end{equation}
where the asymptotic properties for $F$ will be specified further. For this asymptotics, we will require that except for a possible phase factor, the asymptotic value is the plane-wave solution, i.e., we require that
\begin{equation}
  |F(\xi,\tau)| \longrightarrow 1 \qquad \text{for} \qquad \xi \longrightarrow \pm \infty \quad \text{or} \quad \tau \longrightarrow \pm\infty.
\end{equation}
This asymptotic behavior\index{asymptotic behavior} motivates, without further restrictions, the introduction of a displaced phase $\phi(\xi,\tau)$ \label{disphase} and a displaced amplitude $G(\xi,\tau)$ \label{disamp} parameters as follows:
\begin{equation}
  F(\xi,\tau) = G(\xi,\tau) e^{i\phi(\xi,\tau)} - 1,   \qquad G \; \; \textmd{and} \; \; \phi \in \mathbb{R}.
\end{equation}
Since $|F| \longrightarrow 1$ at infinity, this means that $(G \cos \phi - 1)^{2} + G^2 \sin^{2} \phi = G(G - \cos 2\phi) + 1 \longrightarrow 1$. This implies that either $G \longrightarrow 0$ or $G - 2 \cos \phi \longrightarrow 0$. Since we are not interested in a trivial solution, the asymptotic requirement implies that $G - 2 \cos \phi \longrightarrow 0$. In the cases below, we will deal with solutions for which $G_{\tau}$ and $\phi_{\tau} \longrightarrow 0$. Then for some limiting phases $\phi_{\pm}$ one finds
\begin{equation}
  \phi(\xi)   \longrightarrow \phi_{\pm} \qquad \textmd{and} \qquad G(\xi,\tau) \longrightarrow 2\cos(\phi_{\pm}) \qquad \textmd{asymptotically.}
\end{equation}
In the complex plane or Argand diagram\index{Argand diagram}, these parameters are depicted in Figure~\ref{sketsaargand}.

We are interested in solutions periodic in time $\tau$ with period $T$ and an infinite range in space $\xi$. Therefore the integration boundaries of the corresponding action functional\index{action functional}, Hamiltonian or other functions will be adjusted accordingly. Consider again the action functional corresponding to the NLS equation, namely equation~\eqref{actionfunctional} on page~\pageref{actionfunctional}:
\begin{equation}
  \mathcal{A}(A) = \int_{-\infty}^{\infty} \left(\int_{-T/2}^{T/2} \frac{1}{4}i \left(A^{\ast} \partial_{\xi} A - A \partial_{\xi} A^{\ast} \right)\,d\tau - H(A) \right)\; d\xi,			  \label{3actionfunctional}
\end{equation}
where the Hamiltonian\index{Hamiltonian} $H$ is a functional of functions in $\tau$ given by equation~\eqref{Hamiltonian} on page~\pageref{Hamiltonian}:
\begin{equation}
  H(A) = \int_{-T/2}^{T/2} \left( -\frac{1}{2} \beta \left|\partial_{\tau} A \right|^{2} + \frac{1}{4} \gamma |A|^{4} \right) \; d\tau.	    \label{3hamiltonian}
\end{equation}
To arrive at the governing phase-amplitude equations, we substitute
\begin{equation}
  A(\xi,\tau) = A_{0}(\xi) \, \left[G(\xi,\tau) e^{i \phi(\xi,\tau)} - 1 \right]
\end{equation}
into the action functional~\eqref{3actionfunctional}. Then we have the following expressions:
\begin{align*}
  \frac{1}{4}i \left(A^{\ast} \partial_{\xi} A - A \partial_{\xi} A^{\ast} \right)
  &= \frac{1}{2}r_{0}^{2} \left[-G^{2} \partial_{\xi}\phi + \gamma r_{0}^{2}(G[G - 2 \cos \phi] + 1) + \partial_{\xi}(G \sin \phi) \right] \\
  \frac{1}{2} \beta |\partial_{\tau} A|^{2} 
  &= \frac{1}{2} r_{0}^{2} \beta \left[(\partial_{\tau} G)^{2} + G^{2}(\partial_{\tau} \phi)^{2} \right] \\
  \frac{1}{4}\gamma |A|^{4} &= \frac{1}{2}r_{0}^{2} \; \frac{1}{2} \gamma r_{0}^{2}(G[G - 2 \cos \phi] + 1)^{2}.
\end{align*}
The action functional now depends on $G$ and $\phi$, and since the integration of the term $\partial_{\xi}(G \sin \phi)$ gives a constant difference, we neglect it and the action functional \index{action functional} reads
\begin{equation}
  \mathcal{A}(G,\phi) = -\frac{1}{2} r_{0}^{2} \int_{-\infty}^{\infty} \left(\int_{-T/2}^{T/2} (\partial_{\xi} \phi) G^{2} \, d\tau + \bar{H}(G,\phi)\right) \; d\xi,			  \label{3actionfuncGphi}
\end{equation}
where the transformed Hamiltonian\index{Hamiltonian!transformed} $\bar{H}$ reads \label{transHam}
\begin{equation}
  \bar{H}(G,\phi) = \int_{-T/2}^{T/2} \left[-\beta (\partial_{\tau} G)^{2} - \beta (\partial_{\tau} \phi)^{2} G^{2} + W(G,\phi) \right]\;  d\tau
\end{equation}
and where we have introduced the normalized potential energy\index{potential energy} $W(G,\phi)$ as follows. The non-normalized expression for $W$ is given explicitly by
$\frac{1}{2} \gamma r_{0}^{2}(G[G - 2 \cos \phi] + 1)^{2} - \gamma r_{0}^{2}(G[G - 2 \cos \phi] + 1)^{2} = \frac{1}{2} \gamma r_{0}^{2} \left(G^{2} [G - 2 \cos \phi]^{2} - 1 \right)$. In the following, we will discard the uninteresting constant term which, moreover, gives rise to the divergence of the action functional, and take
\begin{equation}
  W(G,\phi) = \frac{1}{2} \gamma r_{0}^{2} G^{2}(G - 2 \cos \phi)^{2}. \label{Wpotential}
\end{equation}
It is to be noted that $W$ has, besides the origin $G = 0$, the unit circle in the complex plane as a set of nontrivial critical points:
\begin{equation}
  \frac{\partial W}{\partial G} = \frac{\partial W}{\partial\phi} = 0	  \qquad \textmd{for} \; \; G = 2 \cos \phi,
\end{equation}
for which $W = 0$ and $\left|Ge^{i\phi} - 1 \right| = 1$.

The governing equations follow from variations with respect to $G$ and $\phi$. Variation with respect to $\phi(\xi,\tau)$ gives the time-integrated energy equation\index{energy equation} which takes the form
\begin{equation}
  \partial_{\xi}(G^{2}) - 2 \beta \partial_{\tau}(G^{2} \partial_{\tau} \phi ) + \frac{\partial W}{\partial \phi} = 0,
\end{equation}
this energy equation shows a forcing from the dependence of $W$ on $\phi$. Variation with respect to $G(\xi,\tau)$ gives the `displaced' phase equation\index{displaced phase!equation}
\begin{equation}
  \beta \partial_{\tau}^{2} G + \left(\partial_{\xi} \phi - \beta (\partial_{\tau} \phi)^{2} \right) G - \frac{1}{2} \frac{\partial W}{\partial G} = 0,  \label{disphaseeqn}
\end{equation}
which is the transformed nonlinear dispersion relation. This equation can be interpreted as a nonlinear oscillator\index{nonlinear oscillator} equation for $G$, which now depends on $\phi$ through the dependence in $W$ and on the combination of its derivatives $\partial_{\xi} \phi - \beta (\partial_{\tau} \phi)^{2}$ that contributes to the coefficient in front of the linear term. In general, this is, therefore, a non-autonomous oscillator equation. Note that it has exactly the same form as the general phase equation~\eqref{nondisrel}, but now $W$ depends essentially on the phase $\phi$. The energy of each oscillator is not constant in general because of the dependence of the phase on $\tau$ in the expression. Indeed, if we denote the energy\index{energy} by $E$:
\begin{equation}
  E(G,\phi) = \frac{1}{2} \beta \left(\frac{\partial G}{\partial \tau}\right)^{2} + V(G,\phi),
\end{equation}
where the potential function\index{potential energy!function} \label{potentialenergy} $V(G,\phi) = \frac{1}{2}\left[\partial_{\xi}\phi - \beta (\partial_{\tau} \phi)^{2} \right]G^{2} - \frac{1}{2} W(G,\phi)$, the change of energy with respect to $\tau$ can be calculated explicitly:
\begin{equation}
\begin{aligned}
  \frac{\partial E}{\partial \tau} 
  &= \underbrace{\left( \beta \frac{\partial^{2} G}{\partial \tau^{2}} + \left[\partial_{\xi} \phi - \beta (\partial_{\tau} \phi)^{2} \right] G -  \frac{1}{2} \frac{\partial W}{\partial G} \right)}_{= \; 0} \frac{\partial G}{\partial \tau} \\ 
  & \qquad + \; \frac{1}{2} \frac{\partial}{\partial \tau} \left[\frac{\partial \phi}{\partial \xi} - \beta \left(\frac{\partial \phi}{\partial \tau} \right)^{2} \right] G^{2} - \frac{1}{2} \frac{\partial W}{\partial \phi} \frac{\partial \phi}{\partial \tau}.
\end{aligned}
\end{equation}
Vanishing of the first term is a consequence of the displaced phase equation~\eqref{disphaseeqn}. \index{displaced phase!equation} For time-periodic motions with period $T$, the time-integral over one period of $\partial_{\tau} E$ should vanish: $\int_{-T/2}^{T/2} \partial_{\tau} E(G,\phi)\, d\tau = 0$, so that there is no net energy\index{energy} input or output. In the following section, we will consider the case that the phase $\phi$ does not depend on $\tau$; then the formula above shows that the energy is of each oscillator at each fixed $\xi$ (hence fixed $\phi$) is constant.

The conservation properties for the spatial evolution are now given by the wave energy~\eqref{waveenergy}: \index{wave energy}
\begin{equation}
  \frac{\partial}{\partial \xi} \left(\int_{-T/2}^{T/2} (G^{2} - 2 G \cos \phi + 1) \, d\tau \right) = 0
\end{equation}
and the original Hamiltonian\index{Hamiltonian} $H(A)$~\eqref{3hamiltonian}:
\begin{equation}
  \frac{\partial}{\partial \xi} \left(\int_{-T/2}^{T/2} \left(
  - \frac{1}{2} \beta r_{0}^{2} \left[(\partial_{\tau}G)^{2}
  + (\partial_{\tau} \phi)^{2} G^{2} \right]
  + \frac{1}{4} \gamma r_{0}^{4} \left|G e^{i\phi} - 1 \right|^{4} \right) d\tau \right)= 0.
\end{equation}
The transformed Hamiltonian $\bar{H}(G,\phi)$ is related to the Hamiltonian\index{Hamiltonian!relation to transformed} $H(A)$ and the wave energy by the following relation:
\begin{equation}
  \frac{1}{2} r_{0}^{2} \bar{H}(G,\phi) = H(A) - \gamma \int_{-\infty}^{\infty} \left(|A|^{2} - \frac{1}{2} r_{0}^{2} \right) \, d\tau.
\end{equation}
As a consequence, the transformed Hamiltonian is also conserved: $\partial_{\xi} \bar{H}(G,\phi) = 0$. For solutions with the asymptotic behavior as above, we have $W \longrightarrow 0$ for $\xi \longrightarrow \pm \infty$. Hence it follows that $\bar{H} = 0$ asymptotically, and then the constancy in $\xi$ implies \begin{equation} 
  \bar{H}(G,\phi) = 0 \qquad \textmd{for all} \; \; \xi \in \mathbb{R}.
\end{equation}

Related to the non-autonomous character of the equation, in general, the dependence of $\phi$ on $\tau$ implies that at a fixed spatial position the motion in the Argand diagram\index{Argand diagram} is not on a straight line. In the next section, we will consider special solutions for which the motion is at each position represented by motion on a straight line, the line turning with the position.

\subsection{Pseudo-coherent wave solutions} \label{pseudocoherent}
\index{pseudo-coherent}

We now consider special solutions for which the displaced phase\index{displaced phase} $\phi$ does not depend on time $\tau$ but only depends on position $\xi$. We show that the spatial evolution is fully described by a family of constrained optimization problems for the time signals. Each optimization problem is only parameterized by the phase and the spatial dynamics comes in from the change of a multiplier $\lambda(\phi)$ \index{multiplier} with phase. Remarkably, classes of solutions of these optimization problems can be found explicitly; the corresponding solutions found in this way will then be recognized as the well-known soliton solutions of the NLS equation.

From the assumption that the phase does not depend on time
\begin{equation}
  F(\xi,\tau) = G(\xi,\tau) e^{i\phi(\xi)} - 1, 		\label{solutionF}
\end{equation}
it follows that at each position the phase is constant, and the governing oscillator equation~\eqref{disphaseeqn} for $G$ is autonomous.

Note that although the displaced phase $\phi$ is assumed to depend on $\xi$ only, if we write the complex amplitude $A = A_{0}(\xi)\,F(\xi,\tau)$ in the usual phase-amplitude form, $F(\xi,\tau) = b(\xi,\tau) e^{i\psi(\xi,\tau)}$, then the phase $\psi$ will depend on $\xi$ and $\tau$. This phase $\psi(\xi,\tau)$ is related to the displaced phase\index{displaced phase} $\phi(\xi)$ by the following relation:
\begin{equation}
  \tan \psi(\xi,\tau) = \frac{G(\xi,\tau) \sin \phi(\xi)}{G(\xi,\tau) \cos \phi(\xi) - 1}.%
\end{equation}
Therefore, the wave signal of the NLS equation itself is not coherent; we will call this a `pseudo-coherent' solution\index{pseudo-coherent}, since it is only coherent with respect to the displaced phase-amplitude variables.\index{displaced phase-amplitude}

The fact that the dependence on $\tau$ is missing in $\phi$ implies that the motions in the Argand plane are on straight lines through the point $-1$, with angle $\phi$ that depends on the position $\xi(= x)$: the solution is displaced over a distance $-1$ in the complex plane.

\begin{figure}			
\begin{center}
\includegraphics[width = 0.5\textwidth]{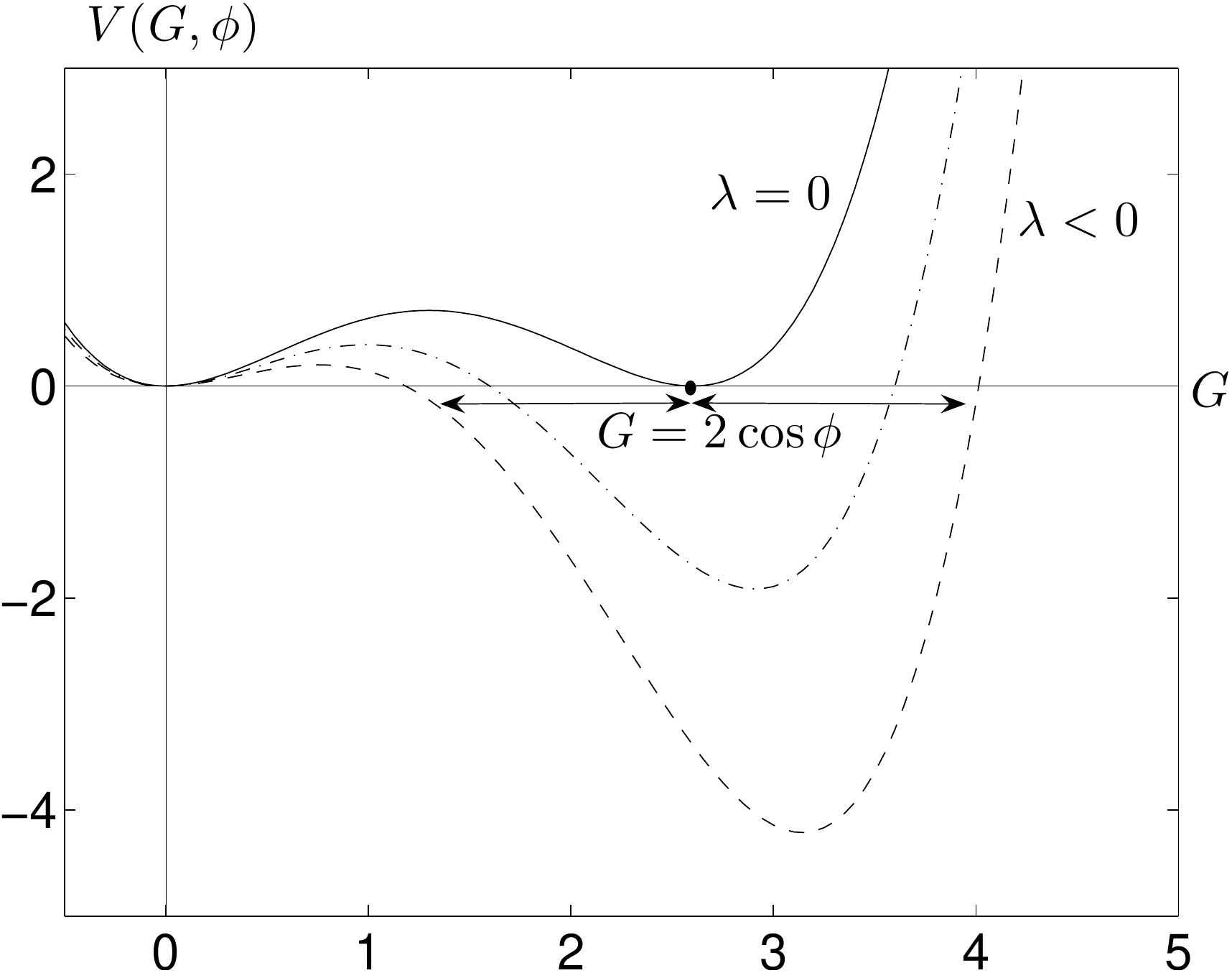}
\caption[Effective potential of the nonlinear oscillator equation]{Plot of the effective potential for various values of $\lambda < 0$. The solid line is the plot for $\lambda = 0$.} 		    \label{potential}
\end{center}
\end{figure}

The equations for $G$ and $\phi$ are a special case of the above, but it is illustrative to derive them directly from the action principle~\eqref{3actionfuncGphi}, where the Hamiltonian\index{Hamiltonian} $\bar{H}(G,\phi)$ now becomes
\begin{equation}
  \bar{H}(G,\phi) = \int_{-T/2}^{T/2} \left[-\beta (\partial_{\tau} G)^{2} + W(G,\phi) \right] \, d\tau
\end{equation}
with $W(G,\phi)$ as given in~\eqref{Wpotential}. To find the variation with respect to $\phi(\xi)$, we change the order of integration in the action functional~\eqref{3actionfuncGphi}. Consequently, we obtain the energy equation\index{energy equation} in the integral form:
\begin{equation}
  \frac{1}{T} \int_{-T/2}^{T/2} \left(-\partial_{\xi}(G^{2}) + \frac{\partial W}{\partial \phi} \right) \, d\tau = 0.
\end{equation}
This shows the presence of a forcing term from the dependence of the potential energy\index{potential energy} on the phase. Variation with respect to $G(\xi,\tau)$ gives the displaced phase equation\index{displaced phase!equation}
\begin{equation}
  (\partial_{\xi} \phi) \, G + \beta \partial_{\tau}^{2} G - \frac{1}{2} \frac{\partial W}{\partial G} = 0.				  \label{oscillatoreqn}
\end{equation}
Equation~\eqref{oscillatoreqn} shows that at fixed $\xi$ the nonlinear oscillator\index{nonlinear oscillator} equation for $G$ depends on $\phi$ and $\partial_{\xi} \phi$ but not on $\tau$, so that at each position, the equation for $G$ is autonomous. Effective potential energy\index{potential energy!effective} defined by
\begin{equation}
  V(G,\phi) = \frac{1}{2}\lambda G^{2} - W(G,\phi) = \frac{1}{2} G^{2}\left(\lambda + \gamma r_{0}^{2}(G - 2\cos \phi)^{2} \right),
\end{equation}
is the oscillator potential for $\lambda = \partial_{\xi}\phi$. For negative $\lambda$ the plots of the potential are depicted in Figure~\ref{potential}.

The conservation properties for the spatial evolution are as before: the wave energy\index{wave energy} $\int |A|^{2} d\tau$ and the Hamiltonian\index{Hamiltonian} $H(A)$; in particular, it holds that:
\begin{equation}
  \bar{H}(G,\phi) = \int_{-T/2}^{T/2} \left(-\beta (\partial_{\tau} G)^{2} + W(G,\phi) \right)\,d\tau = 0,   \qquad \textmd{for all} \;\; \xi \in \mathbb{R}.
\end{equation}
Inspection of the oscillator equation~\eqref{oscillatoreqn} shows that the solutions can be obtained in the following, somewhat surprising, variational way, the idea of which is motivated by the terms in the action functional.~\index{action functional}
\begin{proposition} \label{proposisi}
\index{constrained variational problem} Consider for each $\phi$ the constrained variational problem\footnote[1]{The notion `Crit' denotes an optimization of the corresponding constrained problem, provided that the optimization exists and the functionals are differentiable. A critical point or stationary point of the functional $\mathcal{L} = \int \frac{1}{2}\lambda G^{2}\, d\tau - \bar{H}(G,\phi)$ satisfies the condition that the first variation of $\mathcal{L}$ vanishes for all admissible variations near the critical point. \par}
\begin{equation}
  \textmd{\upshape Crit}_{G} \left\{\int_{-T/2}^{T/2} \frac{1}{2}G^{2}\, d\tau  \; \Bigg| \; \bar{H}(G,\phi) = 0 \right\}.
\end{equation}
The nontrivial solutions $\tau \longrightarrow G(\phi)$ satisfy the Lagrange multiplier\index{multiplier} \index{multiplier!Lagrange} equation for some (reciprocal) multiplier\index{multiplier!reciprocal} $\lambda(\phi)$: $\lambda G = \delta_{G}\bar{H}(G,\phi)$. Then, if $\xi \longrightarrow \phi(\xi)$ is a solution of the equation $\partial_{\xi} \phi = \lambda(\phi)$, the spatial evolution $\xi \longrightarrow G(\phi(\xi))$ leads to a solution of the NLS equation\index{NLS equation!solution in the form} given by
\begin{equation}
  A(\xi,\tau) = A_{0}(\xi)\, \left[G(\xi,\tau) e^{i\phi(\xi)} - 1 \right].			  \label{WFB_special_ansatz}
\end{equation}
\end{proposition}

The multiplier\index{multiplier} equation reads in detail
\begin{equation}
  \beta \partial_{\tau}^{2}G + \lambda G + \gamma r_{0}^{2} G(G - \cos \phi)(G - 2 \cos \phi ) = 0				  \label{fullosceqn}
\end{equation}
and has linear, quadratic and cubic nonlinear terms. It is interesting that various classes of solutions of the constrained variational problem\index{constrained variational problem} can actually be given explicitly and simply. Now solutions of the NLS equation are of the form
\begin{equation}
  G(\phi,\tau) = \frac{P(\phi)}{Q(\phi) - \zeta(\tau)}				  \label{generalspecialWaveform}
\end{equation}
where $\zeta(\tau)$ is one of three special functions, each of which corresponds to a well known special solution. For $\zeta(\tau) = \cos(\nu \tau)$, we will obtain the SFB\index{SFB} solution~\citep{3Akhmediev87}; for $\zeta(\tau) = \cosh(\mu \tau)$, it corresponds to the Ma solution\index{Ma solution}~\citep{3Ma79}; and for $\zeta(\tau) = 1 - \frac{1}{2} \nu \tau^{2}$, it is given by the rational solution\index{rational solution}~\citep{3Peregrine83}. We will specify the SFB solution in the following section and the Ma solution together with the rational solution in Section~\ref{SectionotherWFB}.
\begin{remark}
In the variational formulation\index{variational formulation} above, it is possible to take another target functional to be optimized:\footnote[1]{See the footnote of Proposition~\ref{proposisi}. \par}
\begin{equation}
  \textmd{\upshape Crit}_{G} \left\{\frac{1}{T} \int_{-T/2}^{T/2} G \, d\tau \; \Bigg| \bar{H}(G,\phi) = 0 \right\}.
\end{equation}
This is a consequence of the fact that the quadratic energy\index{energy!quadratic} is conserved, implying that 
\begin{equation}
  \frac{1}{T} \int_{-T/2}^{T/2} (G^{2} - 2 G \cos \phi)\, d\tau = \textup{constant}.
\end{equation}
Of course, the optimal solutions (and the multiplier) will be different, actually just a shift in $G$. This last formulation, for the signal at the extreme position only, was proposed by~\citet{3vanGroesen05} and \citet{3Andonowati07}.
\end{remark}

\section{Specifications and properties of the SFB} \label{SectionpropertiesSFB}

\subsection{An explicit expression}
\index{SFB!explicit expressions}

The Soliton on Finite Background (SFB) is derived by~\citet{3Akhmediev87}; the explicit expression already appeared earlier, see for instance~\citep{3Akhmediev85, 3Akhmediev86}. The authors suggested a method of obtaining exact solutions of the NLS equation by writing the complex amplitude $A$ in the form $A(\xi,\tau) = \left[p(\xi,\tau) + i q(\xi) \right] e^{i\varrho(\xi)}$ and substituting it into the NLS equation. The method consists in constructing a certain system of ordinary differential equations in $p$, the solutions of which determine the solutions of the NLS equation. After an integration procedure, the system leads to a second-order ordinary differential equation similar to~\eqref{fullosceqn}, but it has a quartic term instead of the cubic term.

We derive an explicit expression of the SFB based on the variational formulation mentioned in the previous section, where the functions $P$ and $Q$ depend on the displaced phase $\phi$. We have the following proposition:
\begin{proposition}
For any value of a modulation frequency\index{modulation frequency} $\nu$ \label{nuSFB} such that $0 < \tilde{\nu} < \sqrt{2}$, where $\tilde{\nu} = \nu/\left(r_{0} \sqrt{\gamma/\beta}\right)$ is the normalized modulation frequency\index{modulation frequency!normalized}, the solution is given by
\begin{equation}
G(\phi,\tau) = \frac{P(\phi)}{Q(\phi) - \cos (\nu \tau )}
\end{equation}
where the coefficients are given by
\begin{equation}
P(\phi) = \frac{\tilde{\nu}^{2} Q(\phi)}{\cos\phi}, \qquad \qquad Q^{2}(\phi) = \frac{2\cos^{2}\phi}{2\cos^{2}\phi - \tilde{\nu}^{2}}.
\end{equation}
and the multiplier\index{multiplier} is given by $\lambda = \gamma r_{0}^{2} (\tilde{\nu}^{2} - 2\cos^{2}\phi)$.
\end{proposition}

\begin{figure}			
\begin{center}
\includegraphics[width = 0.5\textwidth]{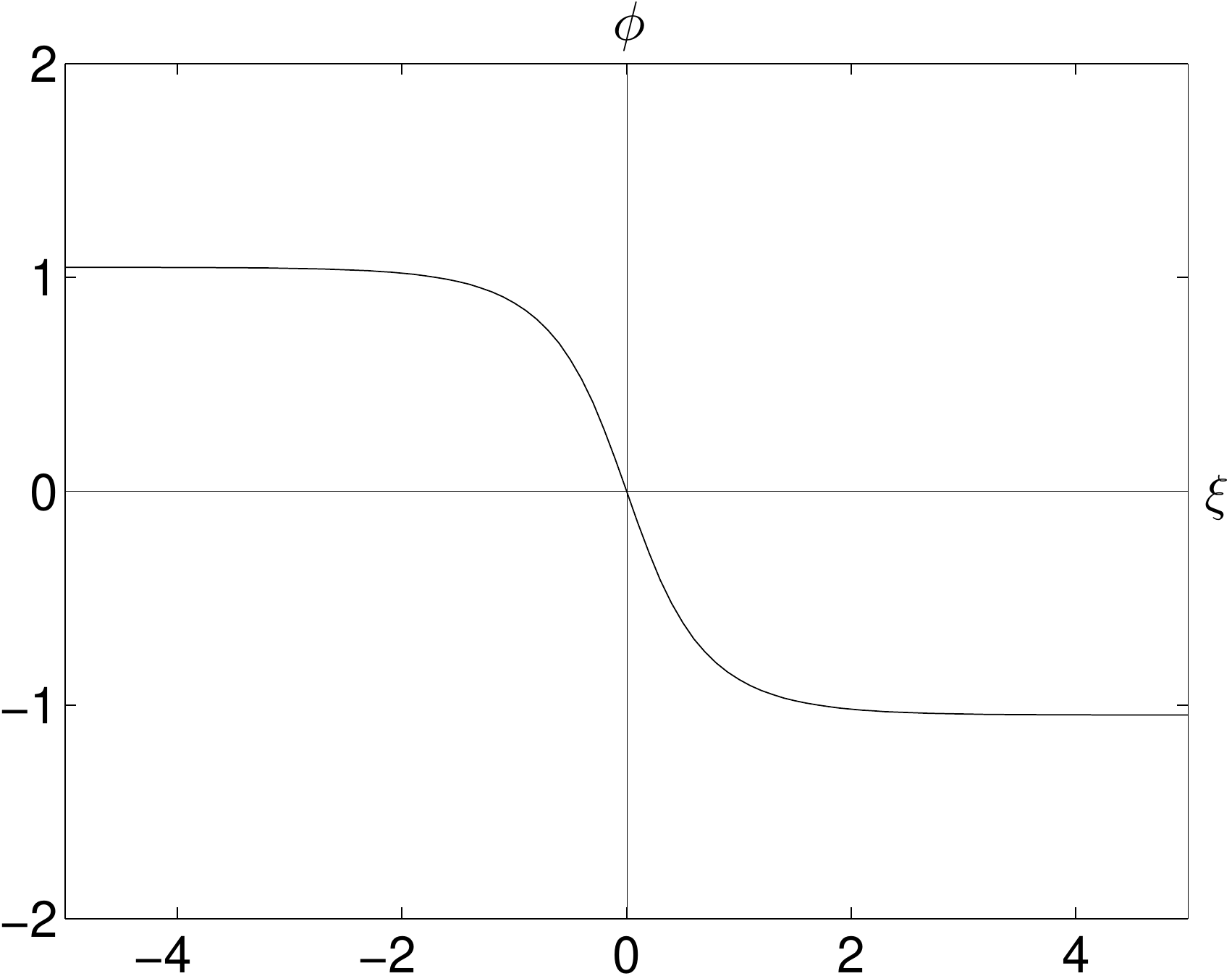}
\caption[Displaced phase of the SFB]{The plot of the displaced phase of the SFB for $\tilde{\nu} = \sqrt{1/2}$.}		    \label{reducedphase}
\end{center}
\end{figure}

The result can easily be verified by considering the function~\eqref{generalspecialWaveform} with $\zeta(\tau) = \cos (\nu\tau)$. Direct differentiation and algebraic manipulations show that this function satisfies an equation with linear, quadratic, and cubic terms, which reads
\begin{equation}
  \frac{\partial^{2}G}{\partial \tau^{2}} + \nu^{2} G - 3\nu^{2}\frac{Q}{P} G^{2} + 2\nu^{2} \frac{Q^{2} - 1}{P^{2}} G^{3} = 0.		  \label{Gpotential}
\end{equation}
Now we compare to the multiplier equation~\eqref{fullosceqn} as the governing equation for $G$ that can be written as
\begin{equation}
  \beta \partial_{\tau}^{2} G + (\lambda + 2 \gamma r_{0}^{2} \cos^{2} \phi)G -
  3 \gamma r_{0}^{2} G^{2} + \gamma r_{0}^{2} G^{3} = 0.
\end{equation}
Comparing coefficients of $G^{2}$ term, we have $Q/P = \cos \phi/\tilde{\nu}^{2}$ and from comparing coefficients $G^{3}$ term, we obtain $2\tilde{\nu}^{2} (Q^{2} - 1) = P^{2}$. Using algebraic manipulation, we arrive at the proposition above. The explicit spatial evolution for $\phi$ as a function of $\xi$ is then found by solving 
\begin{equation}
  \partial_{\xi}\phi = \lambda = \gamma r_{0}^{2} (\tilde{\nu}^{2} - 2\cos^{2}\phi), 		  \qquad 0 < \tilde{\nu}^{2} < 2.
\end{equation}
The result can be written in elementary functions, as will be given in~\eqref{displacedphase}. Note that the value of $\tilde{\nu}$ determines the asymptotic values of the phase: $\tilde{\nu} = \pm \sqrt{2}\cos\phi_{\pm}$ with $\phi_{+} > \phi_{-}$ to assure that $\phi$ is decreasing. A characteristic plot of this function is given in Figure~\ref{reducedphase}.

We write the explicit expression of the SFB in the displaced phase-amplitude\index{displaced phase-amplitude} representation given by~\eqref{solutionA} and the function $F$ with time-independent displaced phase, as in~\eqref{solutionF}. Now, expressions for the displaced phase and amplitude are given as follows:
\begin{align}
  A_{0}(\xi)  &= r_{0}e^{-i\gamma r_{0}^{2}\xi} \label{planewaveexpr}\\
  G(\xi,\tau) &= \frac{P(\xi)}{Q(\xi) - \cos[\nu (\tau - \tau_{0})]} \\
  P(\xi)      &= \tilde{\nu} \sqrt{\left(\tilde{\nu}^{2} + \sinh^{2} [\sigma (\xi - \xi_{0})]\right)/\left(1 - \tilde{\nu}^{2}/2\right)} \\
  Q(\xi)      &= \cosh [\sigma (\xi - \xi_{0})]/\sqrt{1 - \tilde{\nu}^{2}/2} \\
  \phi(\xi)   &= \arctan\left(-(\tilde{\sigma}/\tilde{\nu}^{2}) \tanh [\sigma (\xi - \xi_{0})] \right). \label{displacedphase}
\end{align}
In these expressions, $(\xi_{0},\tau_{0}) \in \mathbb{R}^{2}$ corresponds to the point where the SFB reaches a maximum value. For simplicity, we will make a shift in position and time so that the maximum value is at $(0,0)$, take $\xi_{0} = 0 = \tau_{0}$. The quantities $\nu$ and $\sigma$ characterize the SFB and they are precisely the parameters that have been introduced in Chapter~\ref{2Model} in the context of the Benjamin-Feir instability\index{Benjamin-Feir instability}. See Subsection~\ref{SubsecMI} on pages~\pageref{SubsecMI}--\pageref{Subseccoherent} for more detail.

This solution describes the modulational instability \index{modulational instability} process in the whole space from $\xi = -\infty$ to $\xi = \infty$. At the far distance $\xi \rightarrow \pm \infty$, the SFB is described by the linear theory of modulational instability. For $\xi$ in the nonlinear regime, the SFB connects an exponential behavior from two sides and binds it by large, extreme envelope waves. Note that this SFB depends on two parameters: the plane-wave amplitude $r_{0}$ and the (normalized) modulation frequency\index{modulation frequency!normalized} $\tilde{\nu}$. In order to assure that the SFB exists, $\tilde{\nu}$ has to be in the modulational instability interval $0 < \tilde{\nu} < \sqrt{2}$. The corresponding physical wave field\index{physical wave field} has the carrier wave frequency $\omega_{0}$ as an additional parameter. The physical space and time variables are scaled in a moving frame of reference by the relation $\xi = x$ and $\tau = t - x/\Omega'(k_{0})$.

We have seen an expression for the SFB, but in fact there are several different ways to present the explicit expression. The SFB\index{SFB!explicit expressions} expression~(\ref{planewaveexpr}--\ref{displacedphase}) can also be written as
\begin{equation}
  A(\xi,\tau) = A_{0}(\xi) \, \left(\frac{\tilde{\nu}^{2} \cosh(\sigma \xi) - i \tilde{\sigma} \sinh(\sigma \xi)}{\cosh(\sigma \xi) - \sqrt{1 - \frac{1}{2} \tilde{\nu}^{2}} \cos(\nu \tau)} - 1 \right).			                \label{SFBexpression}
\end{equation}
This expression is similar to the expression on page~51 of~\citep{3Akhmediev97}, with different notations $p \mapsto \nu$, $\beta \mapsto \sigma$, $a_{1} \mapsto (2 - \nu^{2})/4$ and $\psi'(\xi,\tau) \mapsto A(\xi,\tau) = \sqrt{2} \psi'(-2\xi,\tau)$.

An alternative description of the SFB is given in~\citep{3Dysthe99} and \citep{3Grimshaw01} using the transformations $\nu = \sqrt{2} \sin \varphi$, $\sigma = \sin 2\varphi$, $t \mapsto -\frac{1}{2} \xi$, $x \mapsto \frac{1}{\sqrt{2}} \tau$:
\begin{equation}
  A(\xi,\tau) = A_{0}(\xi) \frac{\cos \varphi \cos(\nu \tau) - \cosh(\sigma \xi + 2i \varphi)}{\cosh(\sigma \xi) - \cos \varphi \cos(\nu \tau)}.
\end{equation}

Another representation of the SFB is given in~\citep{3Ablowitz90} where the authors derived it using Hirota's method.\index{Hirota's method} Prescribing $\beta = -1$, $\gamma = -2$, $\nu \mapsto p$, $r_{0} \mapsto a$, and transforming the variables $\xi \mapsto t$, $\tau \mapsto x$, $A \mapsto u$ the SFB with a phase factor $-e^{2i\phi}$ can be written as:
\begin{equation}
  u(x,t) = a e^{2ia^{2}t} \, \left(\frac{1 + 2 \cos(px) e^{\sigma t + \gamma_{1} + 2i\phi} + A_{12} e^{2(\sigma t + \gamma_{1} + 2i\phi)}}{1 + 2 \cos(px) e^{\sigma t + \gamma_{1}} + A_{12} e^{2(\sigma t + \gamma_{1})}} \right)
\end{equation}
where $\sin \phi = \frac{p}{2a}$, $A_{12} = \frac{1}{\cos^{2}\phi}$, $\sigma = \pm p \sqrt{4a^{2} - p^{2}}$, and $e^{\gamma_{1}} = \pm \sqrt{1 - \frac{1}{2} \frac{p^{2}}{2a^{2}}}$.

Another explicit expression is given in~\citep{Osborne00,Osborne01} where the authors derived it using the inverse scattering technique. \index{inverse scattering technique} By prescribing $\tilde{\nu} = 1$, $\beta = -1$, $\gamma = -1$, $r_{0} \mapsto a$, changing the variables $\xi \mapsto T$, $\tau \mapsto X$, $A \mapsto u$ and dividing by the term $\cos(2 a^{2} t)$, the SFB now becomes:
\begin{equation}
  u(X,T) = a e^{2ia^{2}T} \left(\frac{\cos(a\sqrt{2}X)\sech(2a^{2}T) + i\sqrt{2}\tanh(2a^{2}T)}{\sqrt{2} - \cos(a\sqrt{2}X) \sech(2a^{2}T)} \right).
\end{equation}
This expression is the equation (4) of~\citep{3Osborne00} with a typographical error that has been corrected since it was written there $\sech(a^{2}\sqrt{2}T)$ instead of $\sech(2a^{2}T)$.

\begin{figure}[h]			
\begin{center}
\includegraphics[width = 0.55\textwidth]{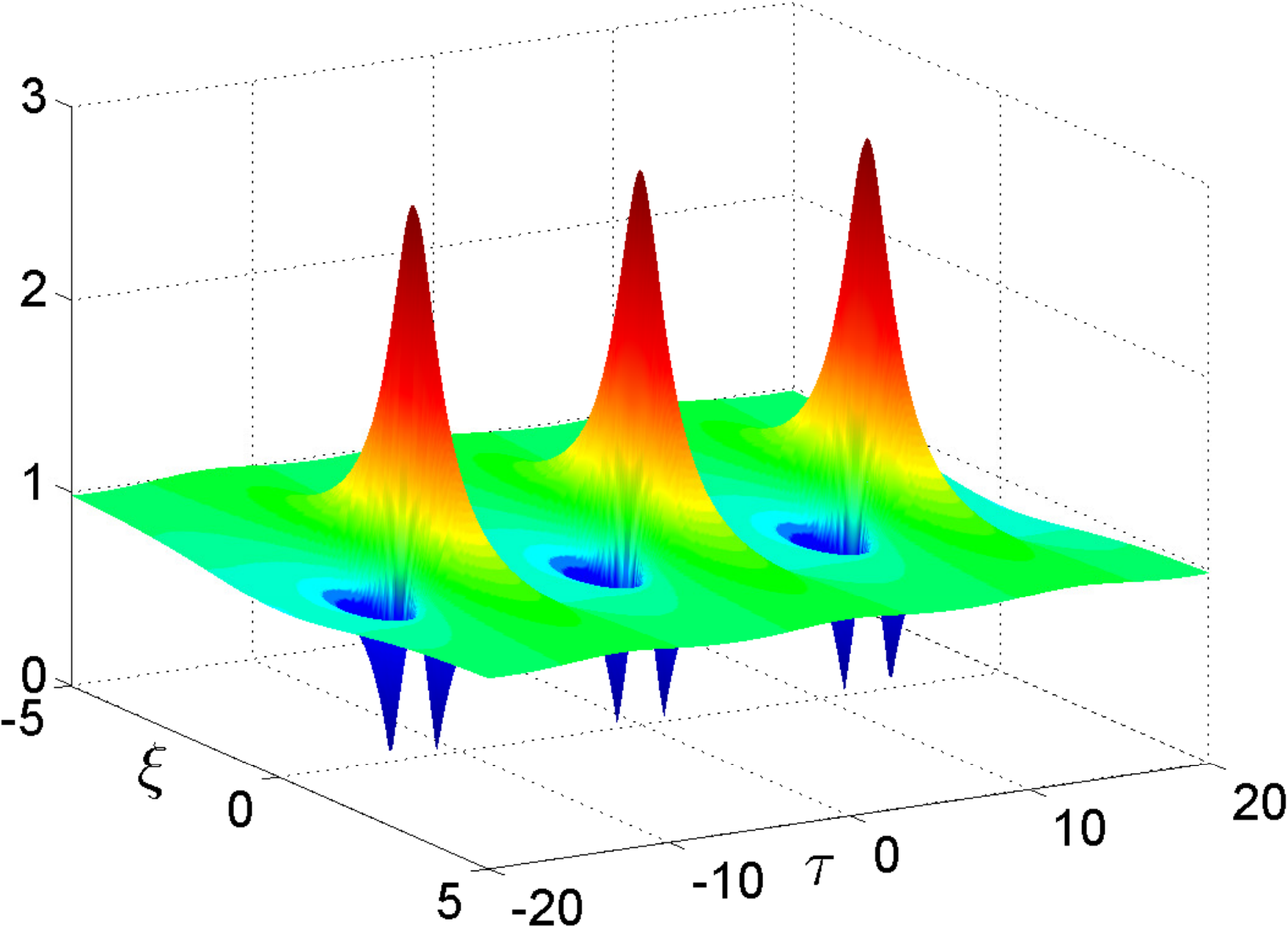}
\caption[Absolute value of the SFB]{Plot of the absolute value of the SFB for $\tilde{\nu} = 1/2$. For illustration purposes, the axes are scaled corresponding to $\beta = 1 = \gamma$.} 			\label{SFB3D}
\end{center}
\end{figure}

The name SFB\index{SFB!terminology} itself comes from the characteristics of its complex amplitude plot. At a fixed time, in particular, the times when it reaches maxima, the SFB has a soliton-like form with a finite background for $\xi \rightarrow \pm \infty$ instead of the standard soliton~\eqref{onesoliton} which has vanishing amplitude for the same limiting variable. The value of the background is always positive since it describes the envelope of the plane-wave solution in the far distance. At a fixed position, the SFB is periodic in the time variable, with the periodicity depending on the modulation frequency\index{modulation frequency} $\nu$. We denote the period of the SFB as $T = 2\pi/\nu$. A plot of the absolute value of the SFB for $r_{0} = 1$ is given in Figure~\ref{SFB3D}.
\begin{figure}[h]			
\vspace*{-0.75cm}
\begin{center}
\includegraphics[width = 0.55\textwidth]{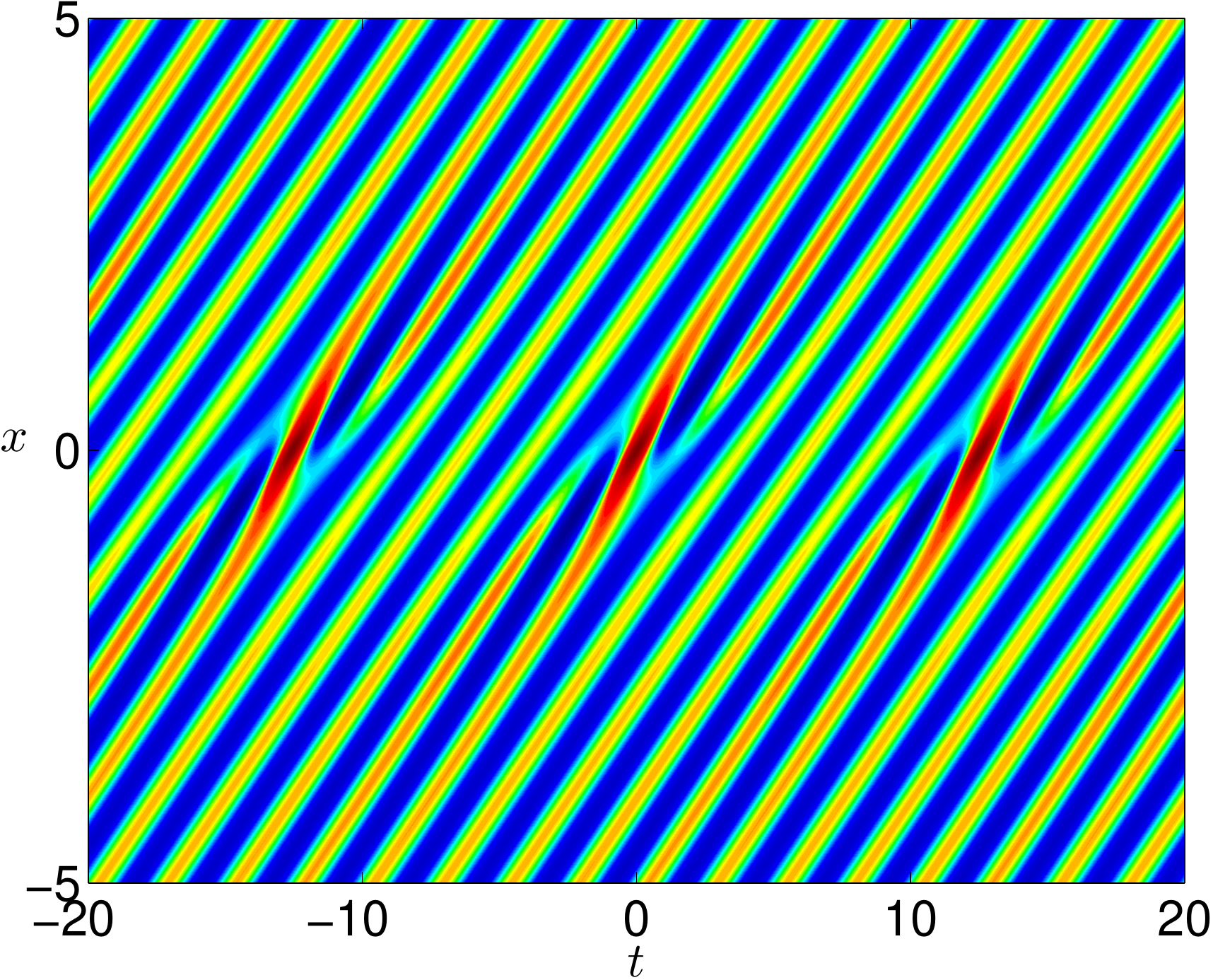}
\caption[Physical wave field of the SFB]{A density plot of the physical wave field $\eta(x,t)$ of the SFB for $\tilde{\nu} = 1/2$ corresponding to Figure~\ref{SFB3D} for $k_{0} = 2\pi$. The plot is shown in a moving frame of reference with suitably chosen velocity.}			    \label{SFBphysical}
\end{center}
\end{figure}

\subsection{The asymptotic behavior} \label{asymptoticbehavior}
\index{SFB!asymptotic behavior} \index{asymptotic behavior}

The asymptotic behavior of the SFB is given at the far distance for $\xi \rightarrow \pm \infty$. It becomes the plane-wave solution $A_{0}(\xi)$, with a phase difference of $2\phi_{0}$ from $\xi = -\infty$ to $\xi = \infty$, as we will see in this subsection. However, we want to know the linear term after the lowest order that specifies the asymptotic behavior of SFB. Thus, for the asymptotic behavior, we include both the lowest-order term and the linear term.

We only derive the result for $\xi \rightarrow -\infty$ since a similar procedure can be applied for $\xi \rightarrow \infty$. In order to find the asymptotic behavior of the SFB, we introduce a new variable $y = e^{\sigma \xi}$. (For $\xi \rightarrow \infty$, a new variable $y = e^{-\sigma \xi}$ is chosen.) The SFB that we will concentrate on is the complex function $F$ in~\eqref{solutionA}. We do not take into account the plane-wave contribution since it will give fast oscillations and distracts our analysis. The function $F$ now is given by
\begin{equation}
  F(y,\tau) = \left(\frac{(\tilde{\nu}^{2} - i \tilde{\sigma})y^2 + \tilde{\nu}^{2} + i \tilde{\sigma}}{y^{2} + 1 - \sqrt{4 - 2\tilde{\nu}^{2}}\, y \,\cos (\nu \tau)} - 1\right).
\end{equation}
Since for $\xi \rightarrow -\infty$, $y \rightarrow 0$, we develop the function $F(y,\tau)$ in a Taylor expansion around $y = 0$:
\begin{equation}
  F(y,\tau) = F(0,\tau) + \partial_{y}F(0,\tau)y + \frac{1}{2} \partial_{y}^{2}F(0,\tau)y^{2} + \dots \, .
\end{equation}
After some simple manipulations we arrive at
\begin{align}
  F(0,\tau) 			&= \tilde{\nu}^{2} - 1 + i \tilde{\sigma}, \\
  \partial_{y}F(0,\tau) &= \sqrt{4 - 2\tilde{\nu}^{2}}(\tilde{\nu}^{2} + i\tilde{\sigma})\cos(\nu \tau).
\end{align}
Therefore, the asymptotic behavior for the SFB at $\xi \rightarrow \mp \infty$ is given by\index{asymptotic behavior!SFB}
\begin{equation}
  A(\xi,\tau) \approx A_{0}(\xi) \, \left[e^{i\phi_{0}} + \sqrt{4 - 2\tilde{\nu}^{2}} e^{i\phi_{1}} e^{\pm \sigma \xi} \cos (\nu \tau) \right],  \label{SFB_asymp}
\end{equation}
where
\begin{equation}
  \tan \phi_{0} = \mp \frac{\tilde{\sigma}}{\tilde{\nu}^{2} - 1}
  \qquad \textmd{and} \qquad
  \tan \phi_{1} = \mp \frac{\tilde{\sigma}}{\tilde{\nu}^{2}}.
  \label{phaseshift}
\end{equation}
Notice that $\tan 2 \phi_{1} = \tan \phi_{0}$ and the phases corresponding to the central frequency and the first sideband\index{sideband(s)} experience phase shifts of $2\phi_{0}$ and $2\phi_{1}$, respectively. The former one describes precisely the phase shift of the SFB.

This asymptotic behavior of the SFB corresponds to the linear modulational (Ben-jamin-Feir) instability as mentioned already in Subsection~\ref{SubsecMI} of the previous chapter. A similar relation between the phases $\theta_{1}$ and $\theta_{2}$ is also obtained~\eqref{anglephases} that $\tan(\theta_{1} + \theta_{2}) = \tan \phi_{0}$ by the transformation in the time variable $\tau \mapsto \tau - \phi_{0}$. This instability is based on the linear theory but the SFB describes a complete nonlinear evolution of a modulated wave signal. The relation between modulational instability\index{modulational instability!relation to asymptotic behavior} and the asymptotic behavior\index{asymptotic behavior!relation to modulational instability} of the SFB is known and given by~\citet{3Akhmediev86}.

\subsection{Physical wave field}
\index{SFB!physical wave field} \index{physical wave field}

The SFB family, just like all solutions of the NLS equation, describe the envelopes of physical wave fields, in our case, the surface water waves. In order to study the corresponding physical wave field, we use the notion of the physical wave field $\eta(x,t)$ as a wave group\index{wave group(s)} with the complex amplitude $A$ and the wave group frequency $\omega_{0}$, as introduced in Section~\ref{Bab2LDWE} and also Section~\ref{Bab2NDWE}. We consider only the lowest-order contribution and we rewrite again the physical wave field $\eta$ as follows:
\begin{equation}
  \eta(x,t) = A(\xi,\tau)\,e^{i(k_{0}x - \omega_{0}t)} + \textmd{c.c}.			  \label{physicalwavefield}
\end{equation}

The corresponding physical wave field\index{physical wave field!SFB} of the SFB now depends on three essential parameters, namely the wave group frequency $\omega_{0}$, the (normalized) modulation frequency\index{modulation frequency!normalized} $\tilde{\nu}$, and the plane-wave amplitude $r_{0}$. We called them `essential' since we do not consider two other parameters $(\xi_{0},\tau_{0})$. By shifting these parameters we can always obtain the maximum amplitude at the desired location and time, as we did earlier by prescribing $\xi_{0} = 0 = \tau_{0}$. Both the dispersion coefficient $\beta$ and the nonlinear transfer coefficient $\gamma$ in the NLS equation depend on $\omega_{0}$. The SFB physical wave field is periodic in time with the modulation period\index{modulation period} of $T = 2\pi/\nu$. \label{T} During the downstream evolution, it experiences a phase shift of $2\phi_{0}$, where $\phi_{0}$ is given by~\eqref{phaseshift}.

We often write the complex amplitude $A$ in its polar form: $A = a e^{i\phi}$, where $a$ and $\phi$ are the real-valued amplitude and the real-valued phase, respectively. The SFB physical wave field then becomes
\begin{equation}
  \eta(x,t) = 2 a(x,t)\, \cos \Phi(x,t), \qquad \textmd{where} \qquad \Phi(x,t) = k_{0}x - \omega_{0} t + \phi(x,t). 		\label{physicaleta}
\end{equation}
We use this expression to give a graphical illustration of the SFB wave field. Using the shifting mentioned above, the SFB reaches the highest amplitude at $x = \xi = 0$ and we call it the `extreme position'\index{extreme position}. By taking only the lowest-order contribution, we neglect the Stokes effect from higher-order terms. Consequently, the wave elevation from~\eqref{physicaleta} is not completely realistic. Figure~\ref{SFBphysical} shows the corresponding physical wave field of the SFB in scaled parameters $\beta = \gamma = r_{0} = 1$. Notice that the wave field shows `wavefront dislocation'\index{wavefront dislocation} at the extreme position for the interval $0 < \tilde{\nu} < \sqrt{3/2}$, where waves are merging or splitting~\citep{3NyeBerry74}. This phenomenon will be explained in more detail in Chapter~\ref{4Dislocate}.

For both the theory and the experiment of extreme wave generation, \index{extreme waves!generation} we are interested in the ratio of the maximum amplitude and the value of the background. So, we have the following definition. 
\begin{definition}[Amplitude amplification factor, AAF] \label{AAFdefinition}
  \index{AAF} \index{AAF!definition} \index{amplitude amplification factor|see{AAF}}
  The AAF of a wave field is the ratio of the maximum amplitude and the value of its background.
\end{definition}
For the SFB, the AAF depends only on the modulation frequency and does not depend on the two other essential parameters. Explicitly, it is given by~\citet{3Onorato00}:
\begin{equation}
  \textmd{AAF}(\tilde{\nu}) = 1 + \sqrt{4 - 2\tilde{\nu}^{2}}, \qquad 0 < \tilde{\nu} < \sqrt{2}.
\end{equation}
This result is easily obtained from substituting the point where the SFB amplitude reaches its maximum $(x,t) = (0,0)$ and taking the ratio of the background. In \citep{3Onorato00}, the AAF for the SFB is given as a function of the wave steepness\index{steepness} and the number of waves under the modulation.\index{AAF!SFB}

\begin{figure}[h]		
\begin{center}
\includegraphics[width = 0.5\textwidth]{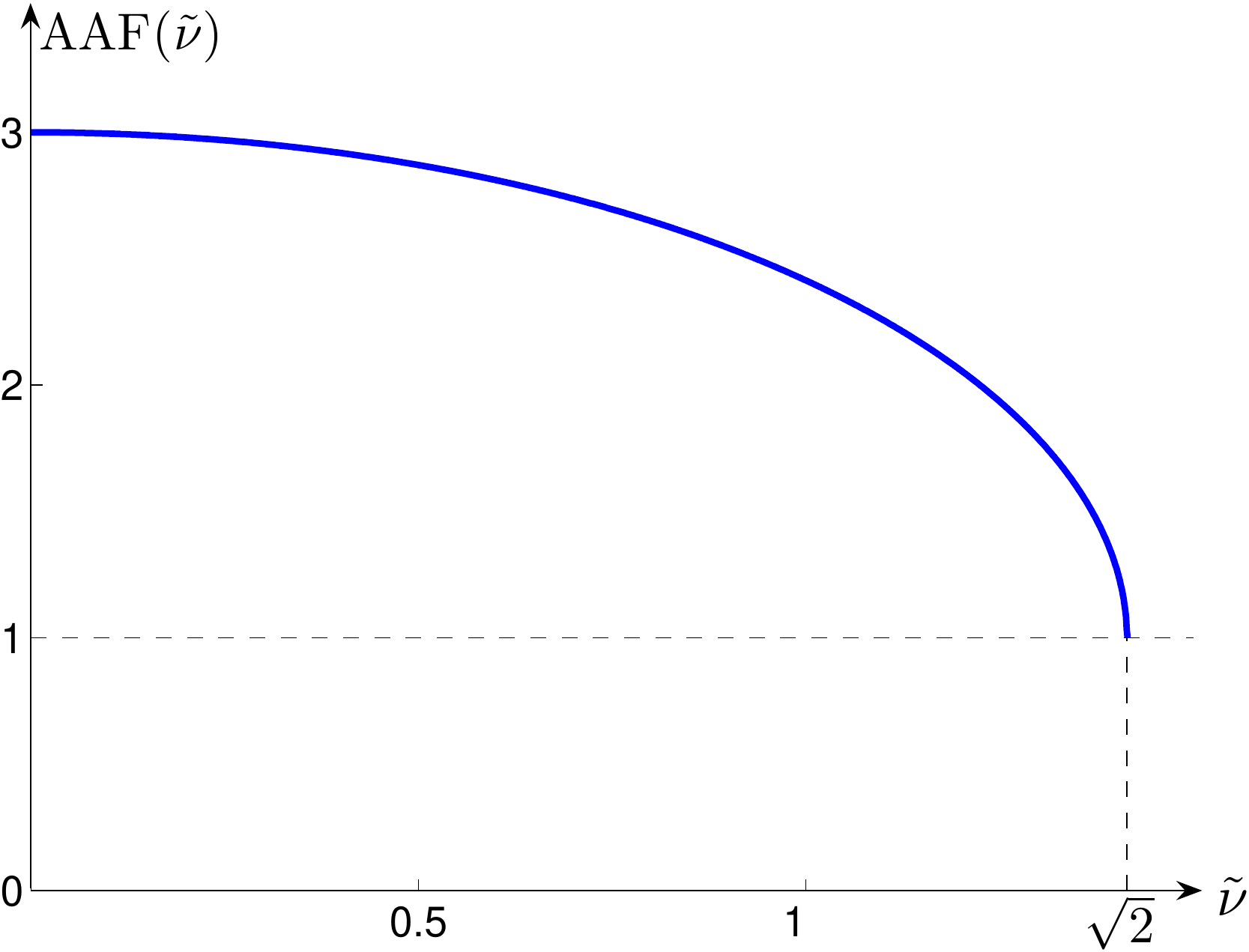}
\caption[Amplitude amplification factor of the SFB]{The plot of the amplitude amplification factor of the SFB as a function of the modulation frequency $\tilde{\nu}$.}				    \label{AAF_SFB}
\end{center}
\end{figure}

For a very long modulated wave signal, i.e., when the modulation frequency $\nu \rightarrow 0$, the AAF reaches a limiting value of 3. For $\tilde{\nu} \geq \sqrt{2}$, the SFB does not exist anymore and the AAF $= 1$ since the wave signal is just the stable plane-wave signal. The AAF decreases monotonically for increasing modulation frequency. Figure~\ref{AAF_SFB} shows the AAF plot of the SFB.

\subsection{Maximum temporal amplitude} \label{MTAsubsection}
\index{SFB!MTA}

In this subsection, we introduce the concept of maximum temporal amplitude (MTA) which turns out to be a useful tool in the process to generate extreme wave events at a specific position in a wave basin. We begin with a formal definition of the maximum temporal amplitude. 
\begin{definition}[Maximum temporal amplitude, MTA] \label{MTAdefinition}
  \index{MTA!definition}
  The MTA is defined as the maximum over the time variable of a physical wave field $\eta(x,t)$:
  \begin{equation}
    \textmd{\upshape MTA}(x) = \max_{t} \; \eta(x,t).
  \end{equation}
\end{definition}
This concept of MTA is introduced by~\citet{3Andonowati03} when discussing optical pulse deformation in nonlinear media. However, this quantity has also applications in water wave theory, in particular, it is useful in the process of wave generation. From the definition above, the MTA describes the largest wave amplitude that can appear in a certain position. It can also be interpreted as a stationary envelope of the wave groups envelope.\index{wave group(s)!envelope}
Since we have an explicit expression for the SFB, we can find an explicit formulation for the MTA, given as follows~\citep{3Andonowati07}:
\begin{equation}
  \textmd{MTA}(x) = 2 r_{0} \sqrt{1 + \frac{\tilde{\nu}^{2} \sqrt{4 - 2\tilde{\nu}^{2}}}
  {\cosh(\sigma x) - \sqrt{1 - \frac{1}{2}\tilde{\nu}^{2}}}}.
  \label{MTAexpression}
\end{equation}
This explicit expression is found from the real amplitude of the SFB physical wave field, given by
\begin{equation}
  a^{2}(x,t) = 4 r_{0}^{2} \frac{\left((\tilde{\nu}^{2} - 1) \cosh(\sigma x) - \sqrt{1 - \frac{1}{2}\tilde{\nu}^{2}} \cos\left[\nu\left(t - x/V_{0} \right)\right] \right)^{2} + \tilde{\sigma}^{2} \sinh^{2}(\sigma x)}{\left(\cosh(\sigma x) - \sqrt{1 - \frac{1}{2}\tilde{\nu}^{2}} \cos\left[\nu\left(t - x/V_{0} \right)\right] \right)^{2}}.
\end{equation}
We observe that $\cos\left[\nu\left(t - x/V_{0} \right)\right] = 1$ for maximum over time. The rest is a simple algebraic calculation to arrive at the MTA expression for the SFB~\eqref{MTAexpression}. In general, however, an explicit expression for the MTA is not easy to find.

\begin{figure}[h]		
\begin{center}
\includegraphics[width = 0.4\textwidth, angle = -90]{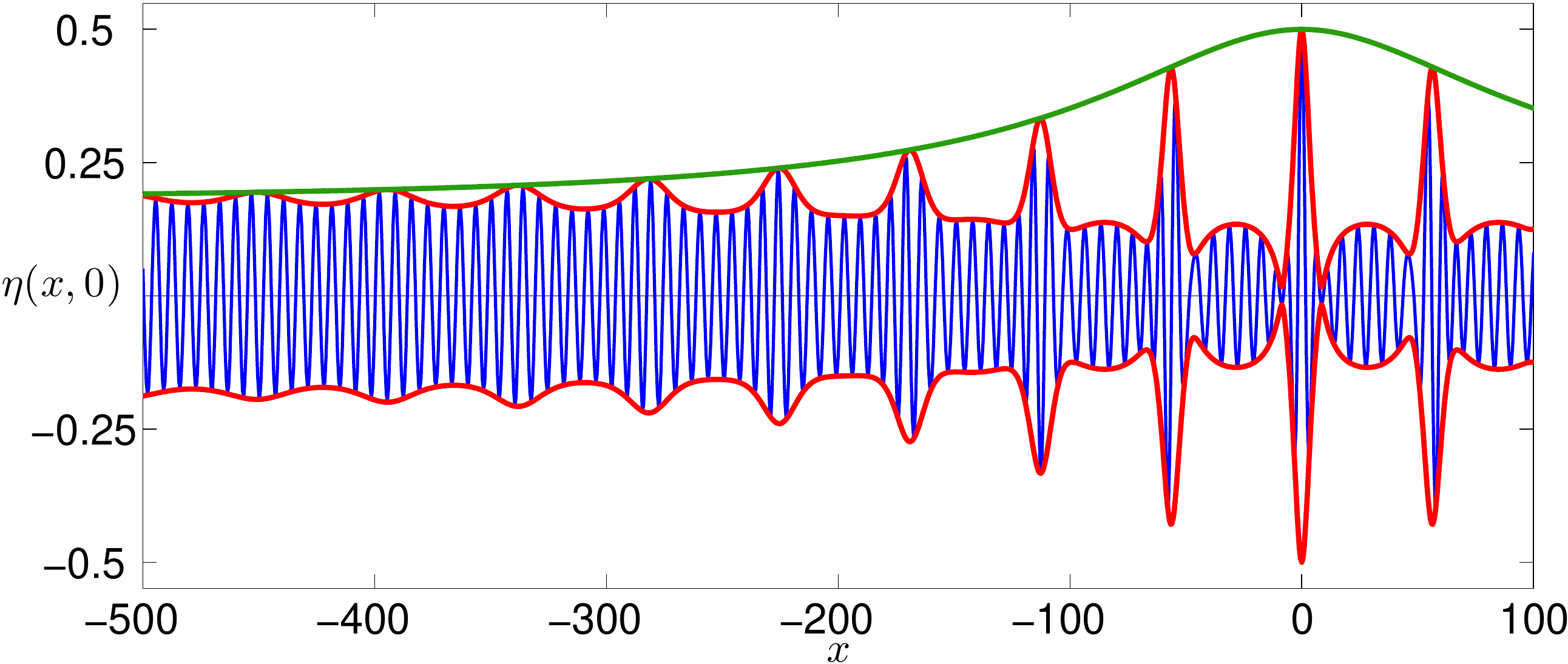}
\caption[Wave profile, envelope and the MTA of the SFB]{A plot of individual waves, the corresponding wave envelope and the time-independent MTA of the SFB. The horizontal axis is the distance from the extreme position and the vertical axis is the surface elevation. Units along the axes are in meters and the water depth of layer is 5~m. See also Figure~\ref{Bab1MTASFB} in Chapter~\ref{1Introduce}.} 	\label{MTASFB}
\end{center}
\end{figure}

In our application, the MTA is a very useful tool for designing a strategy of extreme wave generation. \index{extreme waves!generation} Using the MTA, we are able to determine the signal input to give to the wavemaker in order to produce an extreme wave signal at a specific desired position. We assume that the wave will not break during its downstream\index{downstream} evolution. An MTA plot gives information about the extreme position, namely the position where the envelope of a wave signal is at its largest value. If we desire to produce this extreme wave signal at a particular position, to be denoted as $x = 0$, taking the signal at $x = - x_{0} < 0$ as an initial signal will then produce this maximal wave at $x = 0$. Hence, the signal at $x = - x_{wm}$, where $x_{wm}$ is the distance to the wavemaker from the extreme position, will be the signal to be fed to the wavemaker.\index{wavemaker} This given signal as input to the wavemaker will lead to becoming the extreme wave signal at the desired location, assuming that the NLS equation is an accurate description of the reality, there is no reflection and the domain is infinite. A plot of the MTA together with the spatial evolution and its envelope of the SFB is given in Figure~\ref{MTASFB}.

\section{Spatial evolution of the SFB wave signal} \label{SectionSFBevol}
\index{SFB!wave signal evolution}

In the preceding section, we have seen in Figure~\ref{MTASFB} a snapshot of the SFB wave profile and see the downstream\index{downstream} evolution of a small amplitude modulated wave into a large one during propagation over a sufficiently large distance. In this section, we will illustrate in more detail the spatial evolution of the signal. Additionally, a dynamic evolution model of the envelope signal\index{envelope signal} in the Argand diagram\index{Argand diagram} gives an illustration of the properties of the SFB. Particularly interesting is the wave signal at the extreme position\index{extreme position} $\xi = x = 0$. Only at this position, the SFB is a real-valued function, while at $x \neq 0$, the SFB is a complex-valued function. Furthermore, the amplitude will vanish only at this position for the interval $0 < \tilde{\nu} < \sqrt{3/2}$ and leads to phase singularity\index{phase singularity}. More properties of the signal at this position will be described in this section, where we will discuss the SFB wave signal evolution, the evolution in the Argand diagram and the phase plane representation of the SFB envelope signal at the extreme position.

\subsection{Wave signal evolution}

We have seen that the SFB solution behaves asymptotically as a plane-wave solution. The corresponding wave signal is a monochromatic wave\index{monochromatic wave(s)} with constant envelope $2r_{0}$, frequency $\omega_{0}$ and travels with the phase velocity $\omega_{0}/(k_{0} - \gamma r_{0}^{2})$. However, the modulated wave signal will grow in space during its evolution according to the linear modulational (Benjamin-Feir) instability. The SFB describes the complete nonlinear evolution of the modulated wave signal for all $x$. The SFB wave signal reaches an extreme amplitude at the extreme position, after which it returns to its initial state. The modulation period\index{modulation period} $T$ is maintained during its evolution and depends on the modulation frequency\index{modulation frequency} $\nu$, $T = 2\pi/\nu$.

\begin{figure}[!h]			
\begin{center}
\includegraphics[width = 0.6\textwidth]{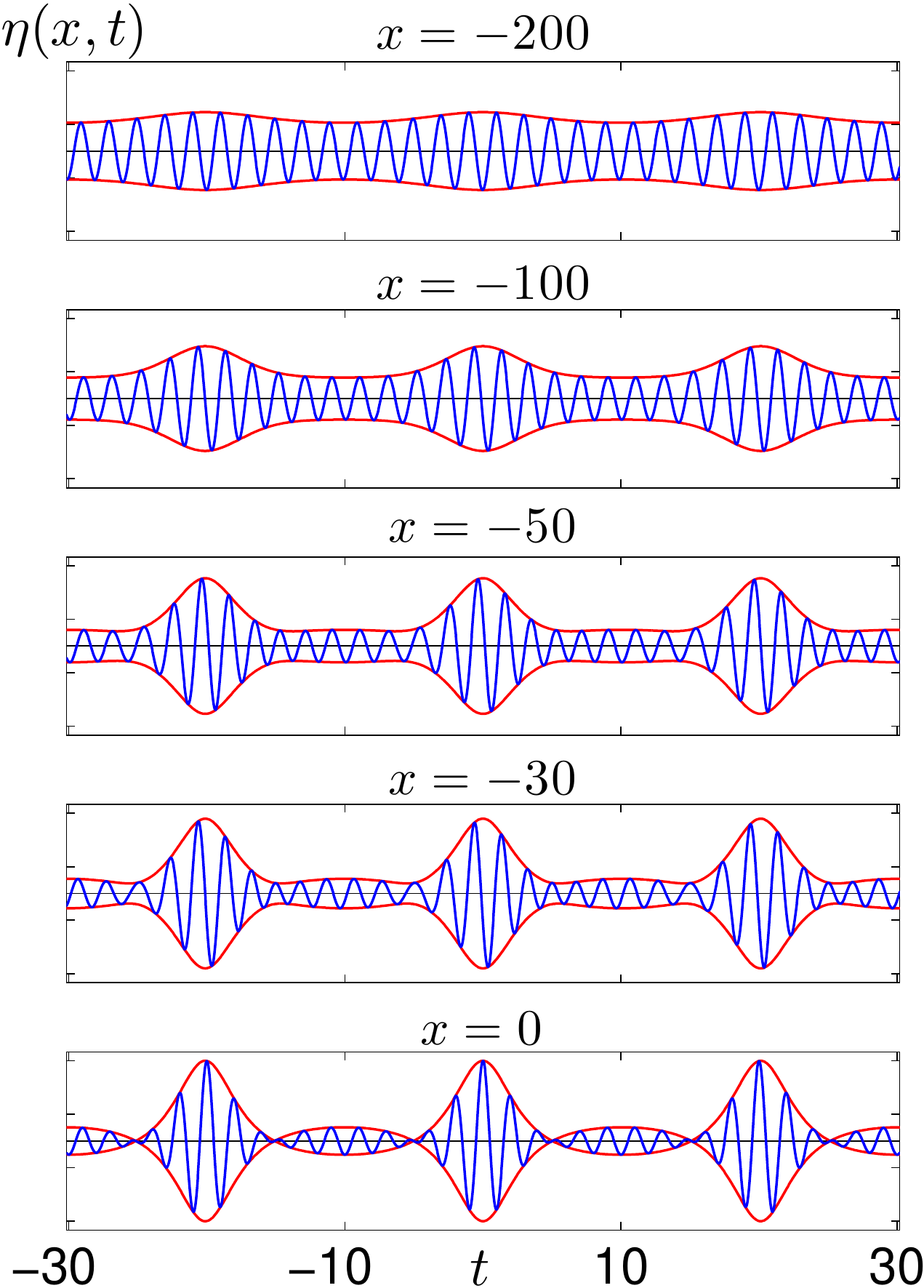}
\caption[Downstream evolution of the SFB wave signal]{The downstream\index{downstream} evolution of the SFB for $\tilde{\nu} = 1$ from a modulated wave signal into the extreme wave signal. The signals are taken at different positions: $x = -200$, $x = -100$, $x = -50$, $x = -30$ and $x = 0$, respectively. The time axis is in second and the position is measured in meter.}			  \label{SFB_evolution}
\end{center}
\end{figure}

Figure~\ref{SFB_evolution} shows the spatial evolution of the SFB wave signal from a modulated wave signal into the extreme signal at the extreme position. In this figure, we see that parts of a wave group signal grow but other parts decrease in amplitude as the signal propagates toward the extreme position. At this extreme position, a pair of `phase singularities' may occur in one modulation period. More explanation about `phase singularity'\index{phase singularity} will be given in Chapter~\ref{4Dislocate}.

\subsection{Evolution in the Argand diagram} \label{evolargand}
\index{Argand diagram} \index{SFB!Argand diagram}

In Subsection~\ref{SectionWFB}, we have introduced waves on the finite background using the displaced phase-amplitude representation.\index{displaced phase-amplitude} Particularly, in Subsection~\ref{pseudocoherent}, we restricted the displaced phase to be time-independent and remarkably the SFB can be written in this representation. In this subsection, we will see that all evolution curves of the SFB in the Argand diagram are centered in $(-1,0)$ and the angle with the real axis depends on the value of the displaced phase, i.e., on the position $\xi$. \index{displaced phase}

Since the SFB is a complex-valued function, we can depict its evolution in the complex plane, the Argand diagram, with axes the real and the imaginary parts of the complex-amplitude $A$. This complex-amplitude $A$ itself depends on the two variables space $\xi$ and time $\tau$. Hence, the representation of the SFB will be a four-dimensional manifold. Here, we will fix one variable and use the other one to plot the evolution for different values in the Argand diagram. The dynamic evolution of the SFB for different $\tau$ parameterized in $\xi$ is given in~\citep{3Akhmediev97}, and for different $\xi$ parameterized in $\tau$ in~\citep{3vanGroesen06}. We are more interested in the latter case since it is a significant help to understand the experimental results to be presented in Chapter~\ref{6Experiment}.

\begin{figure}[h]			
\begin{center}
\includegraphics[width = 0.45\textwidth, viewport= 36 211 529 612]{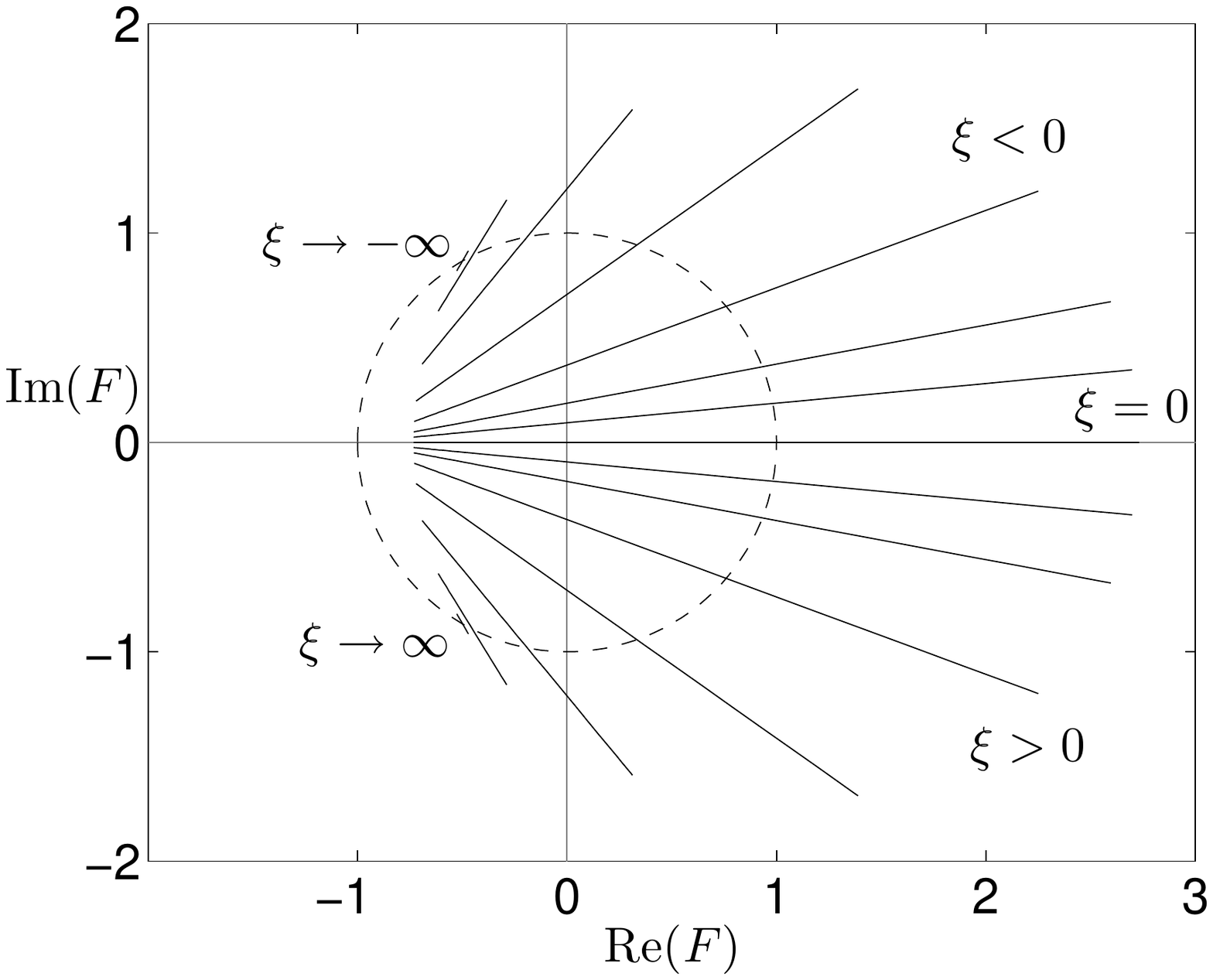}  \hspace{0.5cm}
\includegraphics[width = 0.45\textwidth, viewport= 35 211 529 612]{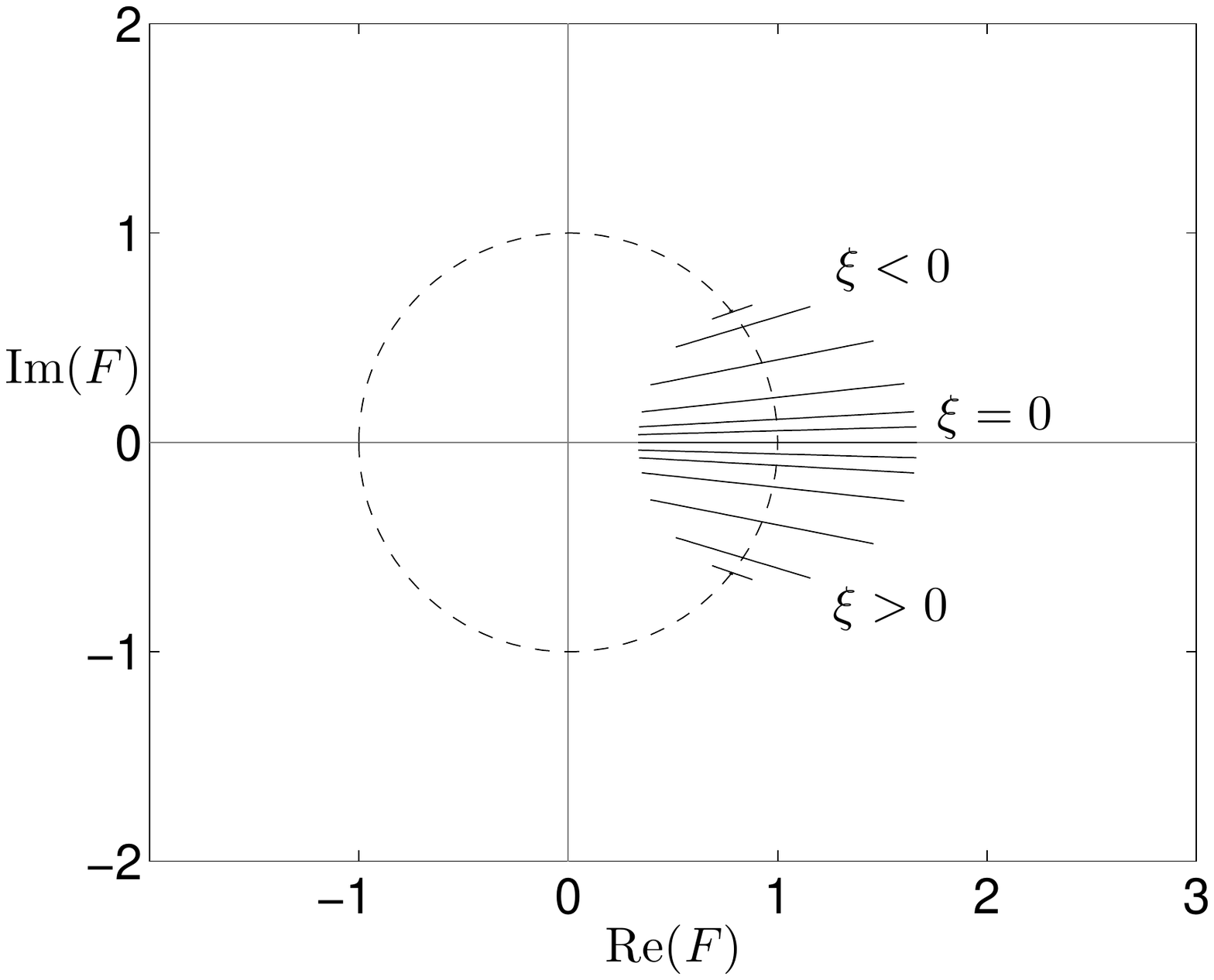}
\caption[Dynamic evolution of the SFB in the Argand diagram]{The dynamic evolution of the SFB parameterized in $\tau$ at different $\xi$ for $\tilde{\nu} = \sqrt{1/2}$ (left) and $\tilde{\nu} = 4/3$ (right).} \label{ArgandSFB}
\end{center}
\end{figure}

In the Argand diagram\index{Argand diagram}, the time trajectories lie on straight lines. These lines have an angle that depends on the `background' and the angle decreases as the position increases. At a fixed position, for one modulation period, a specific part of the lines is passed twice. All lines are centered in $(-1,0)$ due to the displaced phase-amplitude representation. The angle that each line makes with the real axis is determined by the time-independent displaced phase $\phi(\xi)$. For $\xi < 0$, the evolution curves are above the real axis and below it for $\xi > 0$. For $\xi \rightarrow \pm \infty$, the trajectory reduces to a point. For $\xi = 0$, the curve is the longest and lies on the real axis. For the interval $0 < \tilde{\nu} < \sqrt{3/2}$, it also passes the origin twice for one modulation period. Therefore, this is the only position where we have (one pair of) phase singularities in each modulation period. Figure~\ref{ArgandSFB} shows the dynamic evolution of the SFB for different positions $\xi$ and parameterized in time $\tau$.

\subsection{Phase plane representation of envelope signal}
\index{SFB!phase plane representation} \index{envelope signal}	\index{phase plane}

We want to investigate a dynamic property of the SFB envelope signal in the phase plane. \index{phase plane!SFB} Since the dynamic evolution of the displaced amplitude\index{displaced amplitude} $G$ is described by a second-order differential equation~\eqref{fullosceqn}, we can study its dynamics by investigating `phase curves'\index{phase curves} in the phase plane. The SFB envelope signal is given by twice the magnitude of its complex-valued amplitude $2|A|$. Phase curves for this quantity remain positive due to the absolute value. A better representation would be a four-dimensional manifold, presenting the real and the imaginary parts of the complex amplitude separately. However, such a representation is rather difficult to interpret the dynamics.

Nevertheless, the SFB envelope signal is particularly interesting at the extreme position $x = 0$ since there $A$ is a real-valued function. At this position, the physical time coincides with the time in the moving frame of reference, $\tau = t$, so there is no effect of spatial displacement. The `extreme signal'\index{extreme signal} mentioned here refers to the SFB wave signal at the extreme position. We write the extreme signal as a function that depends on the parameter $\tilde{\nu}$. The parameter $r_{0}$ is less interesting and therefore we take the normalized value $r_{0} = 1$. The SFB extreme signal is given as follows:
\begin{equation}
  \eta(0,t; \tilde{\nu}) = 2|F(t;\tilde{\nu})| \cos
  \left[\omega_{0}t + \frac{1}{2}\pi(1 - \textmd{sign}[F(t;\tilde{\nu})]) \right],
\end{equation}
where sign$[F(t;\tilde{\nu})]$ is given by $1$ when $F(t;\tilde{\nu}) > 0$ and $-1$ when $F(t;\tilde{\nu}) < 0$; and $F(t;\tilde{\nu})$ is the envelope of the extreme signal, the `extreme signal envelope'.\index{extreme signal envelope} Note that generally, the term `envelope' always refers to a nonnegative quantity. In our context, however, we allow the envelope to be negative too. We know that since the wave signal itself depends on the modulation frequency $\nu$, then certainly the extreme envelope does too. The smaller the modulation frequency, the larger the period will be, and the longer the time needed to repeat that periodic pattern.

\begin{figure}[h]			
\begin{center}
\includegraphics[width = 0.5\textwidth, viewport = 58 211 534 628]{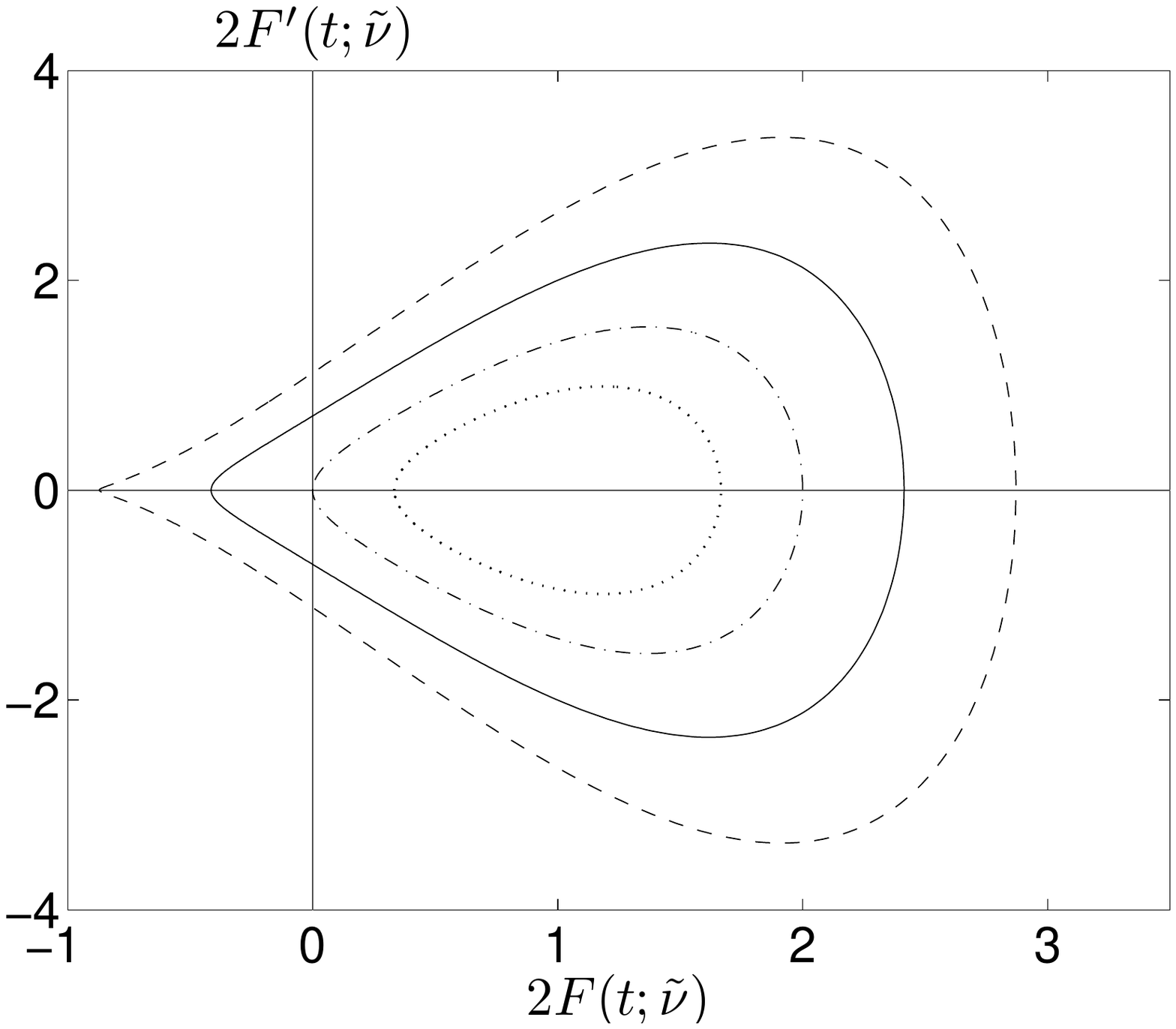}
\caption[Phase curves of the SFB wave envelope]{Phase curves in the phase plane of the corresponding SFB wave envelope at the extreme position. The curves are taken for different values of $\tilde{\nu}$: $\tilde{\nu} = 1/2$ (dashed curve), $\tilde{\nu} = 1$ (solid curve), $\tilde{\nu} = \sqrt{3/2}$ (dash-dot curve) and $\tilde{\nu} = 4/3$ (dotted curve).} 			\label{phaseplane}
\end{center}
\end{figure}

We examine this extreme signal envelope by studying the phase curves of the envelope signal\index{envelope signal} in a phase plane. \index{phase plane} The horizontal axis of the phase plane denotes twice extreme envelope $2F(t;\tilde{\nu})$ and the vertical axis denotes twice the derivative of the extreme envelope $2F'(t;\tilde{\nu})$. The motion along the phase curve as a function of time is clockwise as time increases. The closed curve is transferred precisely once in one modulation period. Pairs of phase singularities are visible if the phase curve crosses the vertical axis, which is the case for $0 < \tilde{\nu} < \sqrt{3/2}$. Figure~\ref{phaseplane} shows phase plots for different values of the modulation frequency $\tilde{\nu}$.

In Section~\ref{SectionWFB}, the displaced amplitude\index{displaced amplitude} $G$ was represented by the second-order differential equation for all~$x$~\eqref{fullosceqn}. Yet, it is difficult to write such representation for the complex amplitude without the plane-wave contribution $F$ for $x \neq 0$. However, thanks to the fact that $F$ is a real function at $x = 0$, we can write a second-order differential equation for $F$ as follows:
\begin{equation}
  \frac{d^{2}F}{dt^{2}} + \frac{dV}{dF} = 0. \label{FNewton}
\end{equation}
This equation describes a particle moving in a nonlinear oscillator\index{nonlinear oscillator} with potential function $V(F)$. It has a mechanical analogy with Newton's second law of motion \index{Newton's second law of motion} in classical mechanics. The potential function is given explicitly by
\begin{equation}
  V(F) = \alpha_{1}F + \frac{1}{2} \alpha_{2}F^{2} + \frac{1}{3} \alpha_{3} F^{3} + \frac{1}{4} \alpha_{4}F^{4},			  \label{potentialF}
\end{equation}
where $\alpha_{1} = \frac{\gamma}{\beta} r_{0}^{2}(\tilde{\nu}^{2} - 2)$, $\alpha_{2} = \frac{\gamma}{\beta} r_{0}^{2}(\tilde{\nu}^{2} - 3)$, $\alpha_{3} = 0$ and $\alpha_{4} = \frac{\gamma}{\beta} r_{0}^{2}$. Furthermore, the energy conservation\index{energy!conservation} equation for $F$ is given by
\begin{equation}
  E(F) = \frac{1}{2}\left(\frac{dF}{dt}\right)^{2} + V(F)
       = \frac{1}{4} \frac{\gamma}{\beta} r_{0}^{2} (3 - 2\tilde{\nu}^{2}).
\end{equation}

The dynamic evolution for the displaced amplitude $G$ at each position is governed by~\eqref{Gpotential}. At $x = 0$, we can also write a nonlinear oscillator equation for $G$ similar to~\eqref{FNewton}:
\begin{equation}
  \frac{d^{2}G}{dt^{2}} + \frac{dV}{dG} = 0, \label{GNewton}
\end{equation}
with the potential function $V(G)$ is given explicitly by
\begin{equation}
  V(G) = \frac{1}{2} \tilde{\alpha}_{2}G^{2} + \frac{1}{3} \tilde{\alpha}_{3}G^{3} + \frac{1}{4} \tilde{\alpha}_{4}G^{4},
  \label{potentialG}
\end{equation}
where $\tilde{\alpha}_{2} = \frac{\gamma}{\beta} r_{0}^{2} \tilde{\nu}^{2}$, $\tilde{\alpha}_{3} = -3 \frac{\gamma}{\beta} r_{0}^{2} $ and $\tilde{\alpha}_{4} = \frac{\gamma}{\beta} r_{0}^{2}$. At $x = 0$, the relation between $F$ and $G$ becomes $G = F + 1$. Substituting this into the potential function of $G$~\eqref{potentialG}, we obtain the relation of the potential functions $V(G) = V(F) + E(F)$. Using these both relations, we confirm the fact that if $G$ satisfies the oscillator equation~\eqref{GNewton}, then $F$ satisfies the oscillator equation~\eqref{FNewton}. Moreover, the energy conservation equation for the displaced amplitude $G$ at all position vanishes: 
\begin{equation}
  E(G) = \frac{1}{2}\left(\frac{dG}{d\tau}\right)^{2} + V(G) = 0.%
\end{equation}

\section{Amplitude spectrum evolution} \label{SectionASE}
\index{SFB!spectrum} \index{spectrum}

So far we have considered the properties of the SFB in the time domain by studying the signal and its propagation. There are interesting properties that we can explore if we study this solution in the frequency domain. To accomplish this, we need the Fourier transform that maps wave signals from the time domain into the frequency domain. Using this transformation, we obtain the corresponding `spectrum evolution' of a wave signal. Furthermore, we can extract from this spectrum the `absolute amplitude spectrum evolution'\index{spectrum!amplitude} and a `phase spectrum evolution'.~\index{spectrum!phase}

Since the complex-valued amplitude $A(\xi,\tau)$ is a periodic function in $\tau$ with period of $T = 2\pi/\nu$, we write it in the Fourier series representation:
\begin{equation}
  A(\xi,\tau) = \sum_{n = -\infty}^{\infty} a_{n}(\xi) e^{i n \nu \tau}.
\end{equation}
Now, the spatial-dependent complex-valued Fourier coefficient is given by
\begin{equation}
  a_{n}(\xi) = \frac{1}{T}\int_{-\frac{T}{2}}^{\frac{T}{2}} A(\xi,\tau) e^{-i n \nu \tau} d\tau, \quad n \in \mathbb{Z}. 		\label{amplitude_spectrum}
\end{equation}
We introduce a dummy variable $\tilde{\tau} = \nu \tau$ and we apply the fact that $A(\xi,\tau)$ is an even function with respect to variable $\tau$, and therefore $\int_{-T/2}^{T/2} A(\xi,\tau) \sin (n \nu \tau) \, d\tau = 0$. After dropping the tilde, the expression~\eqref{amplitude_spectrum} becomes
\begin{equation}
  a_{n}(\xi) = \frac{1}{2\pi} \int_{-\pi}^{\pi} A\left(\xi,\frac{\tau}{\nu}\right) \cos (n \tau)\, d \tau,  \qquad n \in \mathbb{Z}.	  \label{spectrumSFB}
\end{equation}
From this, it follows that $a_{n} = a_{-n}$ and that $|a_{n}| = |a_{-n}|$.

The complex Fourier coefficients $a_{n}(\xi)$ form an infinite sequence $\{a_{n}(\xi)\}_{n=-\infty}^{\infty}$, which is called the `spectrum'\index{spectrum} of the complex function $A(\xi,\tau)$. By expressing $a_{n}$ in polar form with its magnitude and argument, namely, $a_{n}(\xi) = |a_{n}(\xi)|e^{\,i\varphi_{n}(\xi)}$, the sequence of the real-valued absolute amplitude $\{|a_{n}(\xi)|\}_ {n=-\infty}^{\infty}$ is called the `absolute amplitude spectrum'\index{spectrum!amplitude} and the sequence of real-valued phases $\{\varphi_{n}(\xi)\} _{n=-\infty}^{\infty}$ is called the `phase spectrum'\index{spectrum!phase} of the function $A(\xi,\tau)$. The coefficient $a_{n}$, $n \in \mathbb{Z}$ corresponds to the frequency $\omega_{0} + n \nu$, $n \in \mathbb{Z}$ of the $n^{\textmd{th}}$ complex Fourier mode of the physical wave field $\eta = A e^{i(k_{0}x - \omega_{0} t)} + $ c.c. The carrier wave frequency $\omega_{0}$ is called the central frequency and $\omega_{0} \pm n \nu$ is called the $n^{\textup{th}}$ `sideband'.\index{sideband(s)}

The explicit expression for $a_{n}$, $n \in \mathbb{Z}$ is given without derivation in the book of \citet{3Akhmediev97}; we write their result again here according to our notation. For $n = 0$:
\begin{equation}
  a_{0}(\xi) = A_{0}(\xi) \left(\frac{\tilde{\nu}^{2}\cosh(\sigma \xi) - i \tilde{\sigma} \sinh(\sigma \xi)} {\sqrt{\cosh^{2}(\sigma \xi) - \left( 1 - \frac{1}{2}\tilde{\nu}^{2}\right)}} - 1 \right),		  \label{magnitude0spectrum}
\end{equation}
and for $n \in \mathbb{Z}$, $n \neq 0$:
\begin{equation}
  a_{n}(\xi) = A_{0}(\xi) \frac{\tilde{\nu}^{2}\cosh (\sigma \xi) - i \tilde{\sigma} \sinh(\sigma \xi)}{\sqrt{\cosh^{2}(\sigma \xi) - \left(1 - \frac{1}{2}\tilde{\nu}^{2} \right)}}\left(\frac{\cosh (\sigma \xi) - \sqrt{\cosh^{2}(\sigma \xi) - \left(1 - \frac{1}{2}\tilde{\nu}^{2} \right)}}{\sqrt{1 - \frac{1}{2}\tilde{\nu}^{2}}} \right)^{\!\!n}.			  \label{magnitudespectrum}
\end{equation}
The proof of this result is given in Appendix~\ref{SFBspectrum} on page~\pageref{SFBspectrum}.

For $\xi = \mp \infty$, the absolute of all amplitudes $|a_{n}(\xi)|$ vanish, except for $|a_{0}(\xi)|$ which approaches $r_{0}$. The asymptotic behavior of the amplitude spectrum\index{spectrum!asymptotic behavior} for $\xi \rightarrow \mp \infty$ is given in more detail by
\begin{align}
  |a_{0}(\xi)| &= r_{0} \left[1 - \tilde{\nu}^{2}(2 - \tilde{\nu}^{2}) e^{\pm 2 \sigma \xi} \right], \\
  |a_{n}(\xi)| &= r_{0} \tilde{\nu} \sqrt{2} \left(\sqrt{1 - \frac{1}{2}\tilde{\nu}^{2}} e^{\pm \sigma \xi} \right)^{|n|},   \; n \in \mathbb{Z}, \,n \neq 0.
\end{align}
Conservation of energy\index{energy!conservation}, i.e., $1/T \int_{-T/2}^{T/2} |A|^{2}\, d\tau =$ constant $= r_{0}^{2}$, using the property that $|a_{n}| = |a_{-n}|$, results into the fact that for the spectrum we have
\begin{equation}
  E = |a_{0}(\xi)|^{2} + 2 \sum_{n = 1}^{\infty} |a_{n}(\xi)|^{2} = r_{0}^{2}.		  \label{energyspectrum}
\end{equation}

\begin{figure}[h]			
\begin{center}
\includegraphics[width = 0.45\textwidth]{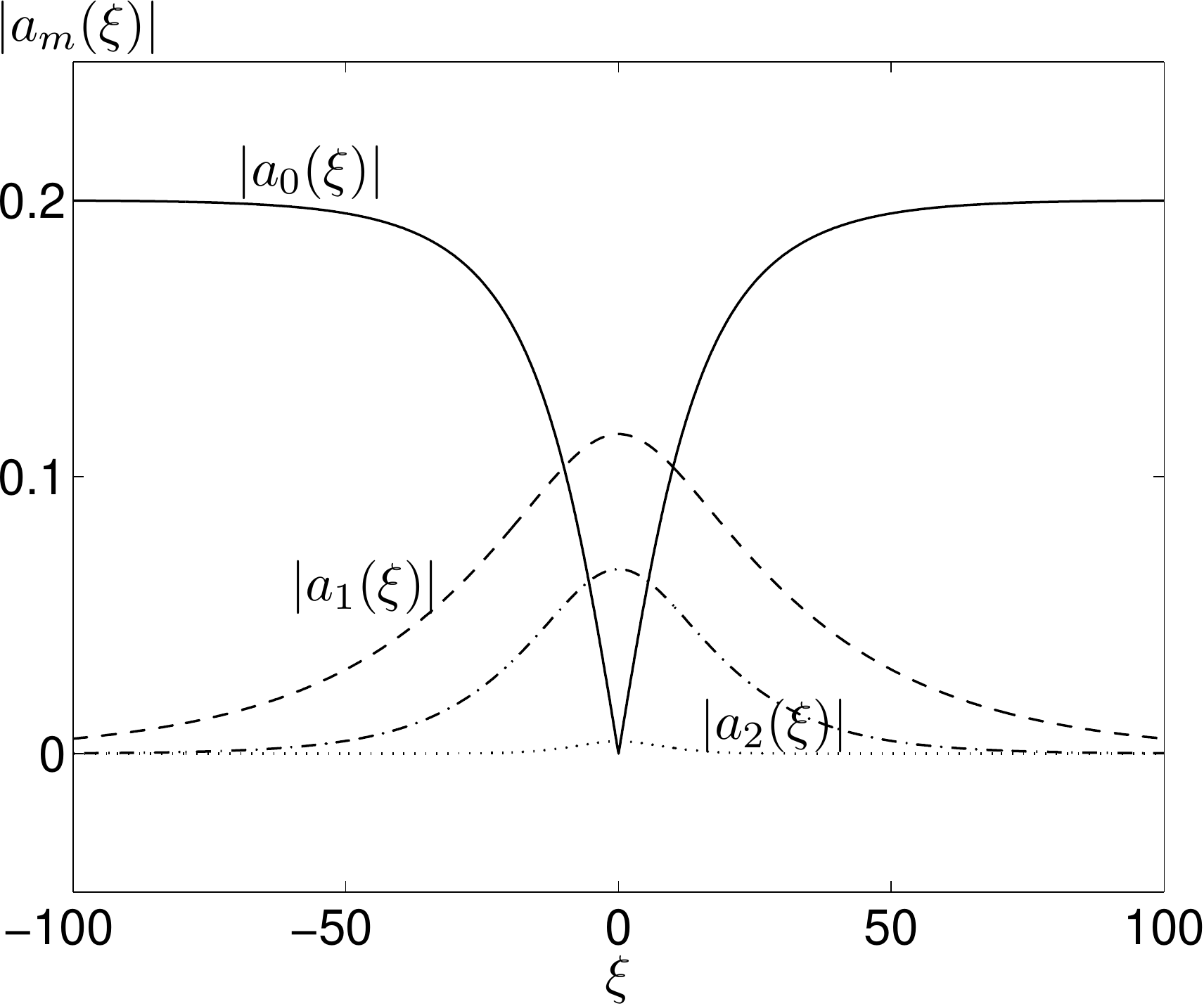} 	\hspace{0.75cm}
\includegraphics[width = 0.45\textwidth]{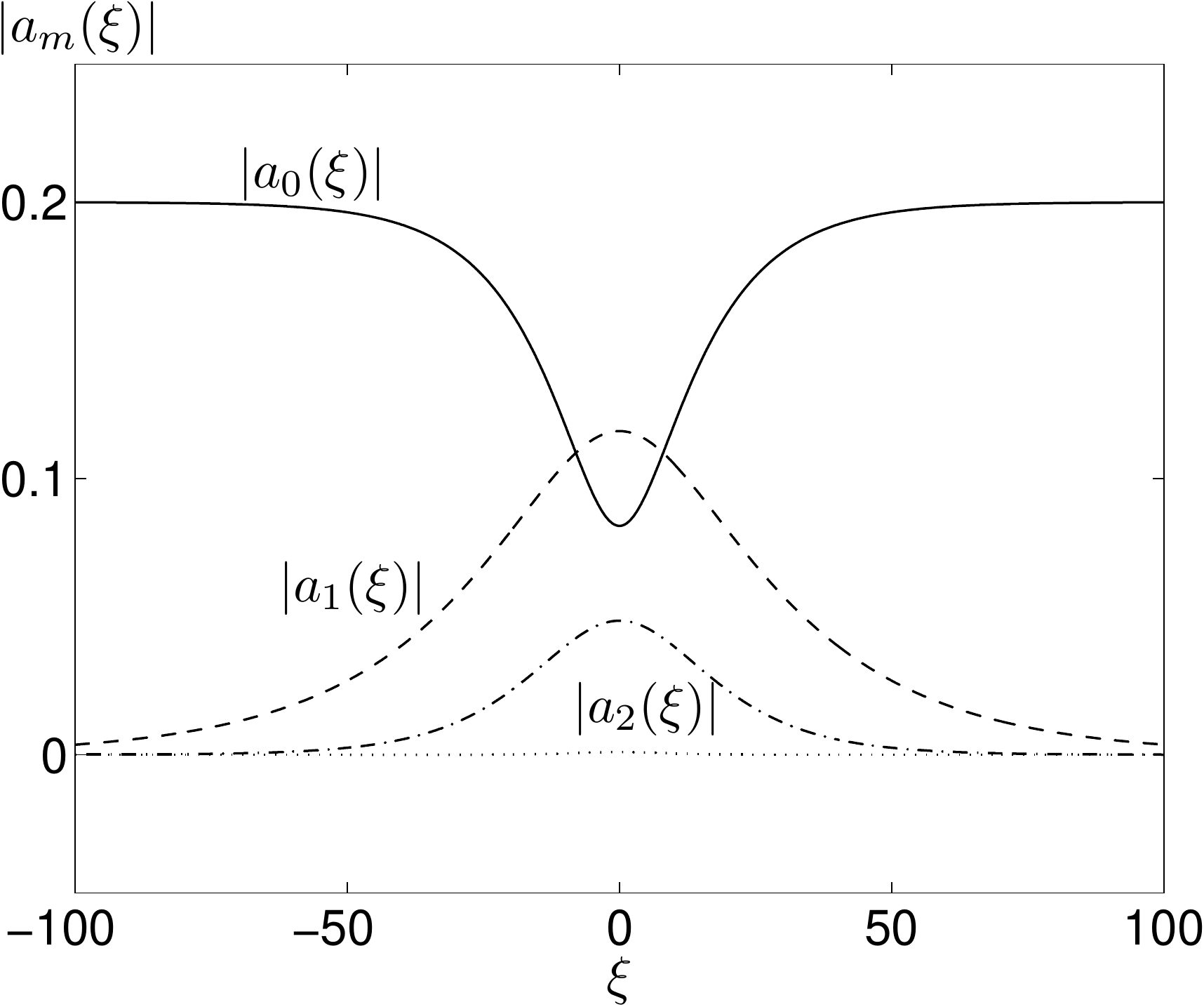}
\caption[Absolute amplitude spectrum of the SFB]{Plots of the absolute amplitude spectrum of the SFB corresponding to the central frequency (solid curve), the first sideband (dashed curve), the second sideband (dash-dotted curve) and the remaining energy in the higher-order sidebands, defined as $2 \sum_{n = 3}^{\infty} |a_{n}(\xi)|^{2}$, (dotted curve) for two values of the modulation frequency $\tilde{\nu}$: $\tilde{\nu} = \sqrt{1/2}$ (left), and $\tilde{\nu} = 1$ (right). For $\tilde{\nu} = \sqrt{1/2}$ the central amplitude vanishes at the extreme position, which means that all energy is transferred to the sidebands.} \label{ASE_SFB1}
\end{center}
\end{figure}

Figure~\ref{ASE_SFB1} shows plots of the amplitude spectrum corresponding to the central frequency, the first sideband, the second sideband, and the remaining energy in the higher-order sidebands.\index{sideband(s)} For all $\tilde{\nu}$ in the instability interval, we observe that the first sideband dominates the second sideband during the whole evolution. That is due to the fact that the term $|a_{n}|$ in~\eqref{magnitudespectrum} decreases in amplitude as $n$ increases. There is a special case when the central frequency vanishes at the extreme position, namely for $\tilde{\nu} = \sqrt{1/2}$. This means that at the extreme position, the energy from the central frequency has been transferred during the spatial evolution completely to its sidebands. In other cases, the energy from the central frequency is only partly distributed to its sidebands.\index{sideband(s)}

By writing the spectrum in the polar form, we can obtain information about the phase spectrum\index{spectrum!phase} evolution $\phi_{n}(\xi)$, $n \in \mathbb{Z}$:
\begin{equation}
\tan \phi_{n}(\xi) = \left\{
\begin{array}{ll}
{\displaystyle \frac{-\tilde{\sigma} \sinh(\sigma \xi)}{\tilde{\nu}^2 \cosh(\sigma \xi) - \sqrt{\cosh^{2}(\sigma \xi) - \left(1 - \frac{1}{2}\tilde{\nu}^{2}\right) }}},         							 & \; \hbox{for $n = 0$;} \\
{\displaystyle -\frac{\tilde{\sigma}}{\tilde{\nu}^2} \tanh(\sigma \xi)}, & \; \hbox{for $n \in \mathbb{Z}, \: n \neq 0$.}
\end{array}
\right. \label{phasespectrum}
\end{equation}
The asymptotic behavior of the phase for $\xi \rightarrow \mp \infty$ is given by
\begin{equation}
\tan \phi_{n}^{\mp} = \left\{
\begin{array}{ll}
    {\displaystyle \pm \frac{\tilde{\sigma}}{\tilde{\nu}^{2} - 1}}, & \; \hbox{for $n = 0$;} \\
    {\displaystyle \pm \frac{\tilde{\sigma}}{\tilde{\nu}^{2}}},     & \; \hbox{for $n \in \mathbb{Z}, \: n \neq 0$.}
\end{array}
\right.
\end{equation}
Since not all Fourier components have the same phase, the SFB is not coherent, which is different from the plane-wave\index{plane-wave} solution~\eqref{planewave} and the single soliton\index{single soliton} solution~\eqref{onesoliton}, as has been discussed in Subsection~\ref{Subseccoherent} on page~\pageref{Subseccoherent}. The result above agrees with the phase information when we look at the asymptotic behavior of the SFB~\eqref{SFB_asymp}. It is remarkable that the phases for all sidebands are equal since they do not depend on $n$. 

The phase shift experienced by the central frequency is equal to the phase shift of the physical wave packet profile.

\section{Other soliton waves on finite background} \label{SectionotherWFB}
\index{waves on finite background}

In this section, we derive two other exact solutions of the NLS equation. These exact solutions are known in the literature, together with the SFB treated above, as `breather' solutions\index{breather solutions} of the NLS equation~\citep{3Dysthe99, 3Dysthe00, 3Grimshaw01}. The name `breather' reflects the behavior of the profile which is periodic in time or space and localized in space or time. The breather concept was introduced by~\citet{3Ablowitz74} in the context of the sine-Gordon partial differential equation. \index{sine-Gordon equation}  Breather solutions\index{breather solutions} are also found in other equations, for instance in the Davey-Stewartson (DS) equation \index{Davey-Stewartson equation}~\citep{3Tajiri00} and modified Korteweg-de Vries (mKdV) equation~\citep{3Drazin89}. \index{KdV equation!modified}

The breather solutions are found from our analysis in Subsection~\ref{pseudocoherent} in the form~\eqref{generalspecialWaveform}:
\begin{equation}
  G(\phi,\tau) = \frac{P(\phi)}{Q(\phi) - \zeta(\tau)}
\end{equation}
where $P$ and $Q$ depend on phase. We will see that different functions $\zeta(\tau)$  will lead to different breather solutions of the NLS equation.
\begin{figure}[h]			
\begin{center}
\includegraphics[width = 0.45\textwidth]{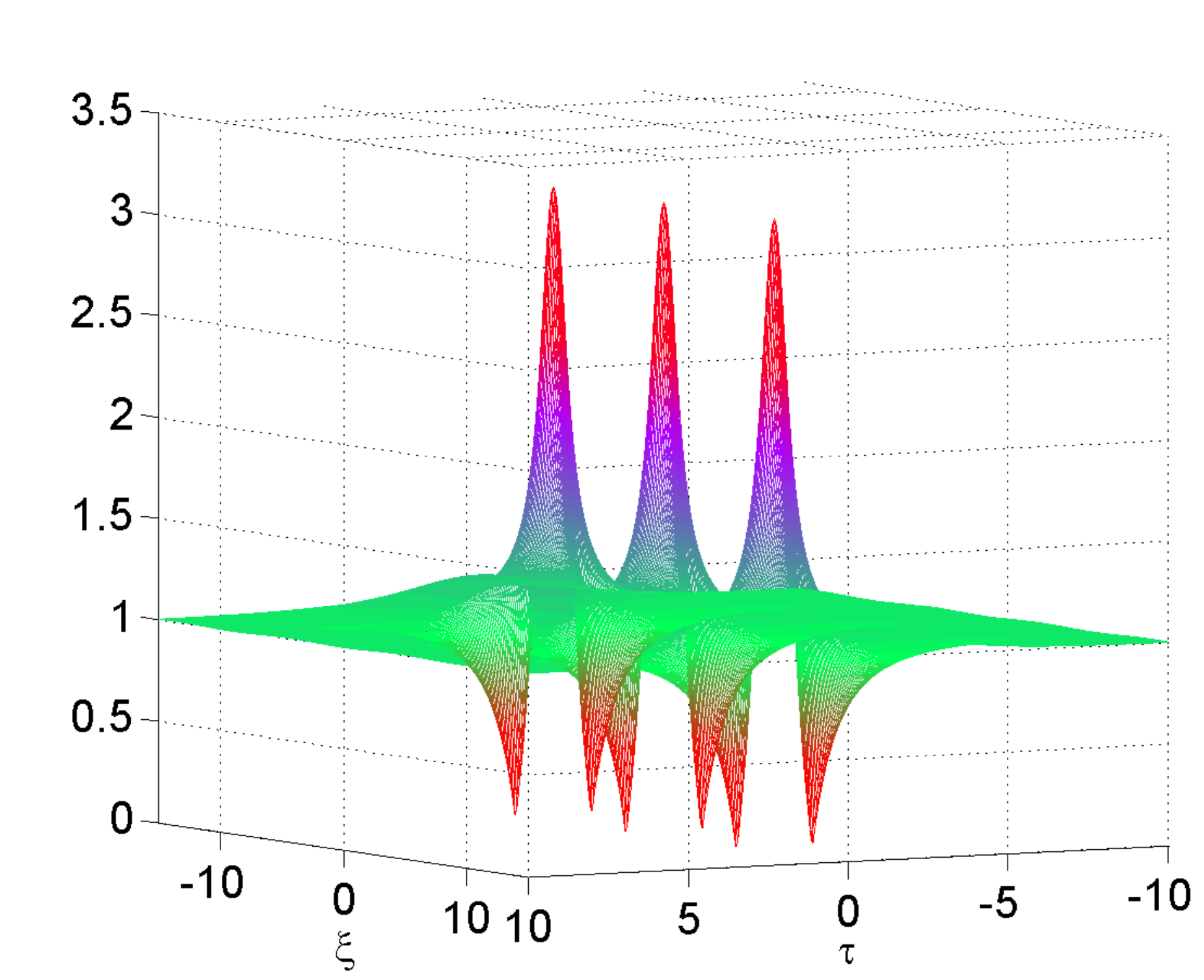}			\hspace{0.75cm}
\includegraphics[width = 0.45\textwidth]{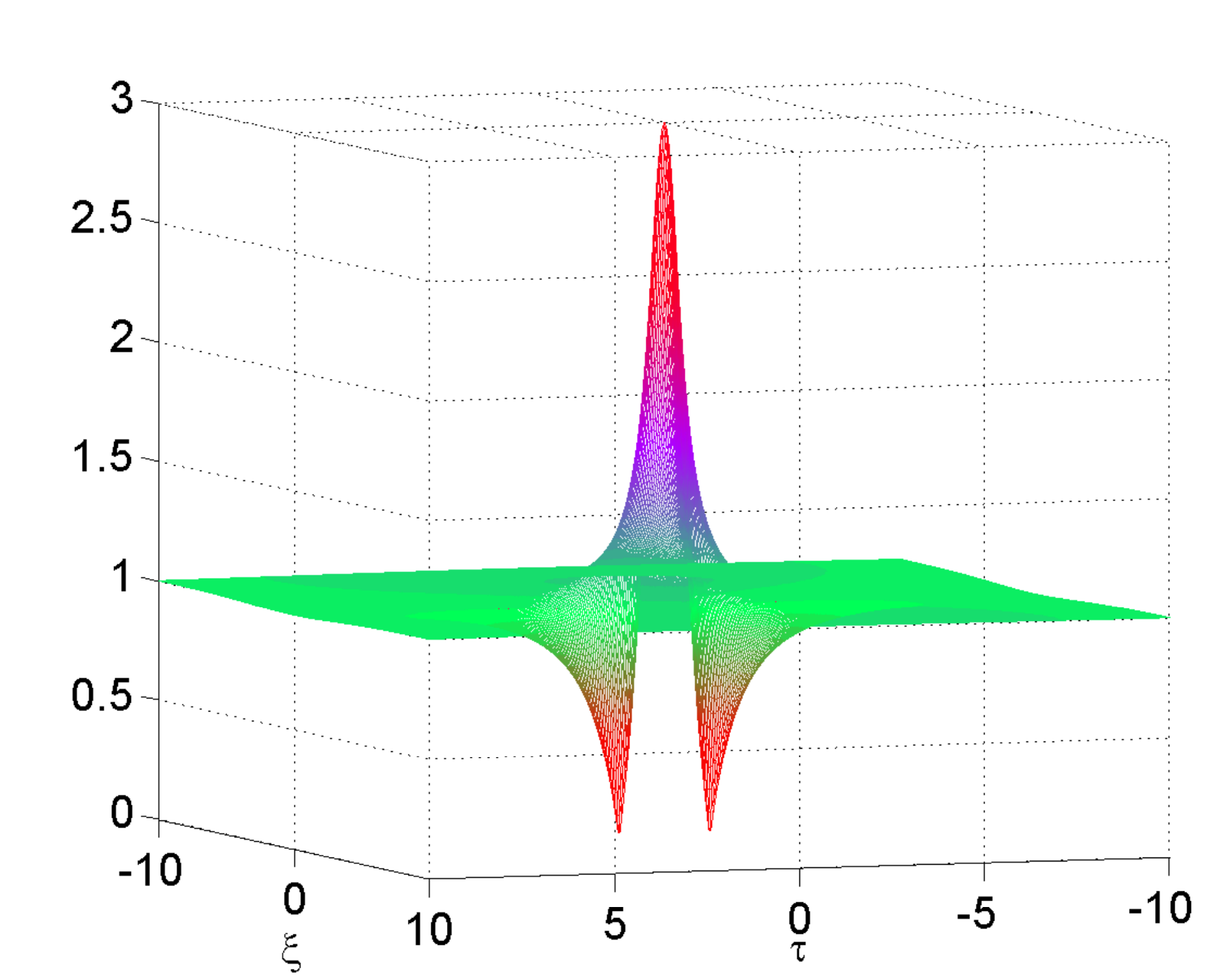}
\caption[Absolute value of the Ma and rational breathers]{Plots of the absolute value of the Ma breather for $\tilde{\mu} = 0.4713$ (left) and the rational breather (right), for $r_{0} = 1$ in all cases. For illustration purposes, the axes are scaled corresponding to $\beta = 1 = \gamma$.} 		\label{MaRa3D}
\end{center}
\end{figure}
\begin{figure}[h!]			
\begin{center}
\subfigure[]{\includegraphics[width = 0.3\textwidth]{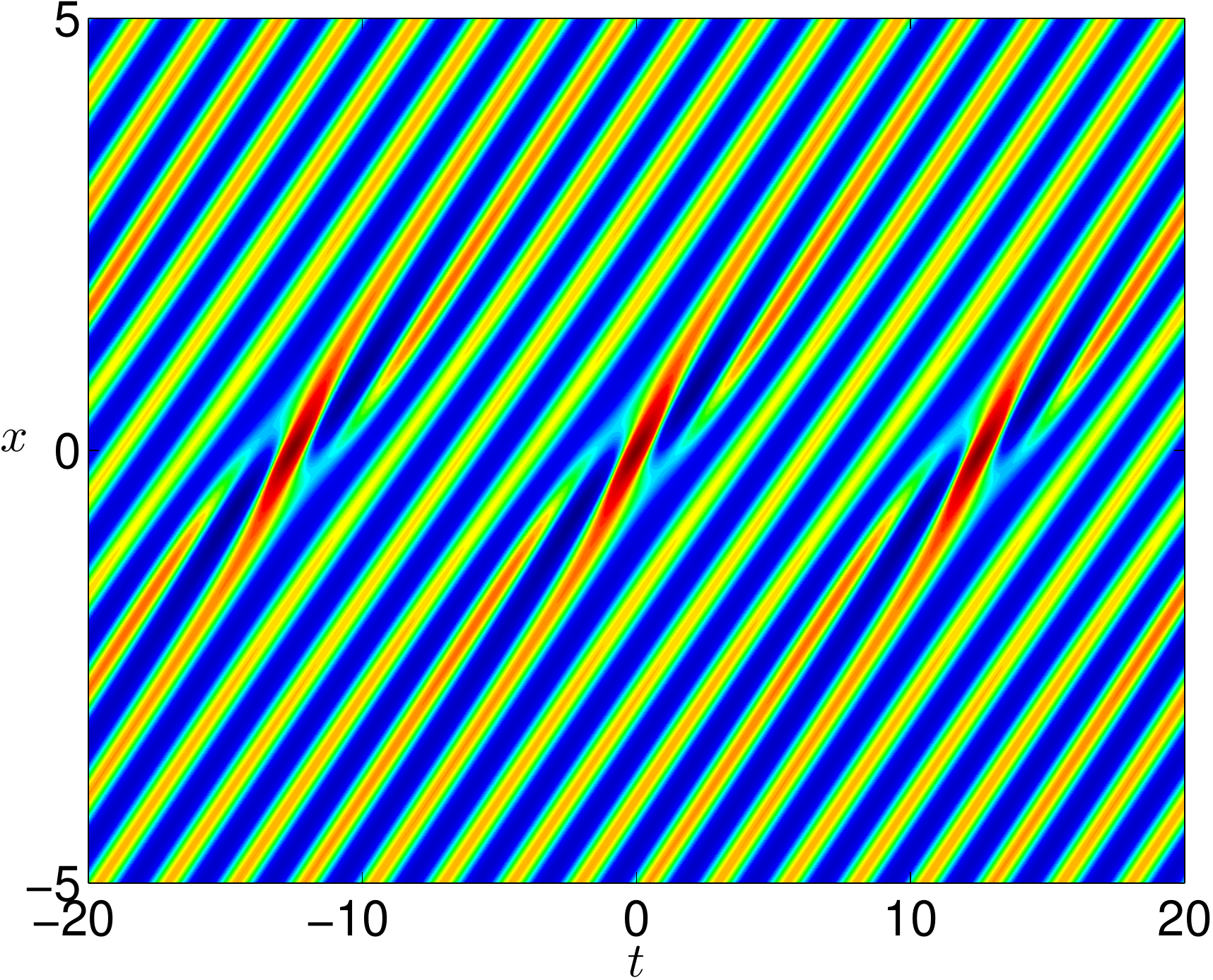}}			\hspace{0.45cm}
\subfigure[]{\includegraphics[width = 0.3\textwidth]{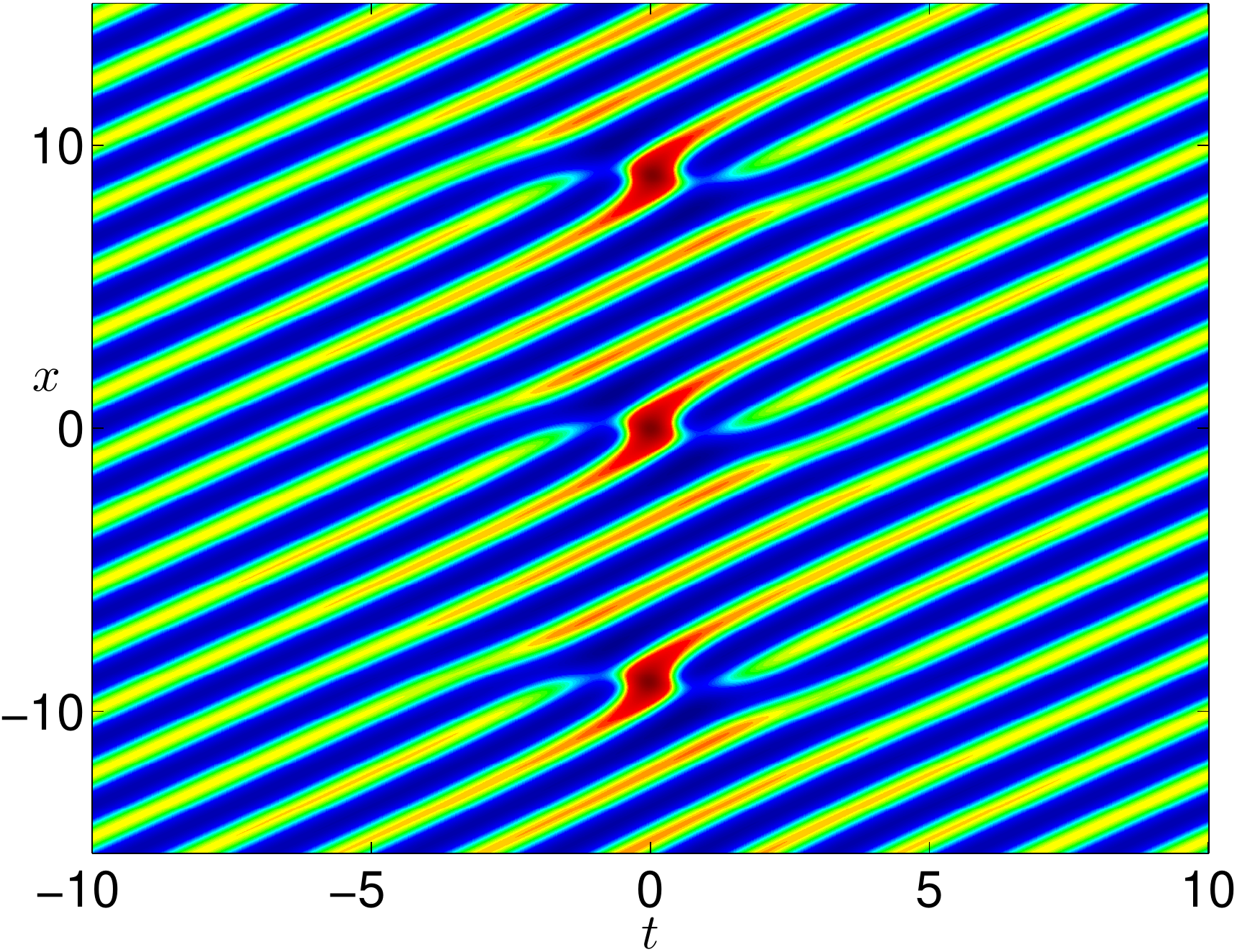}}		\hspace{0.45cm}
\subfigure[]{\includegraphics[width = 0.3\textwidth]{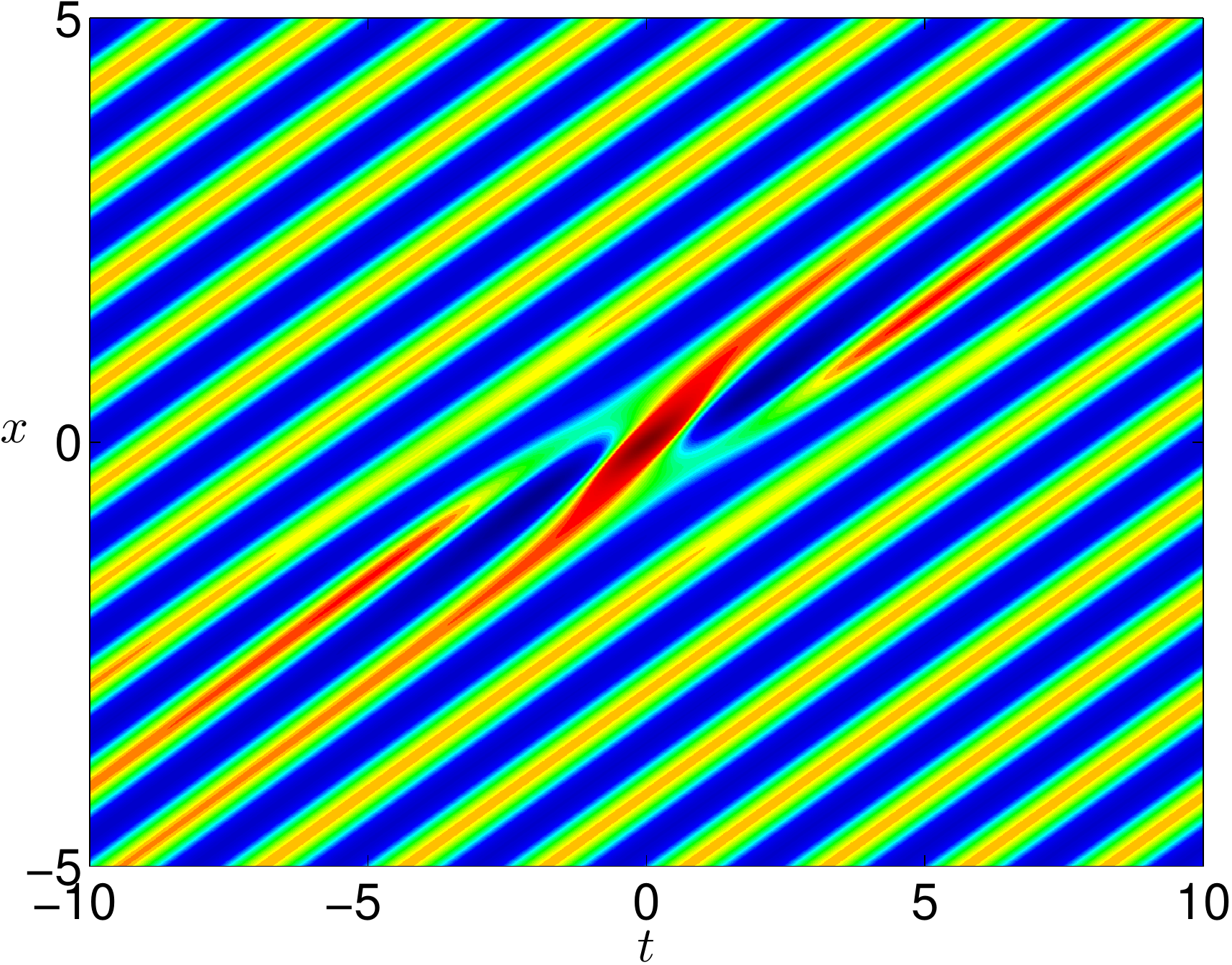}}
\caption[Physical wave field of the breather solutions]{Density plot of the SFB for $\tilde{\nu} = 1/2$ (a), the Ma breather for $\tilde{\mu} = 0.4713$ (b) and the rational breather (c). For all cases $r_{0} = 1$ and $k_{0} = 2\pi$. The plots are shown in a moving frame of reference with suitably chosen velocity.} 		\label{physical}
\end{center}
\end{figure}

\subsection{Breather solutions of the NLS equation}
\index{breather solutions} \index{NLS equation!breather solutions} \index{SFB}

Firstly, for $\zeta(\tau) = \cos(\nu \tau)$, we obtain the SFB which we have discussed in the preceding sections. Secondly, we take $\zeta(\tau) = \cosh(\mu \tau)$, where $\mu = r_{0} \sqrt{\gamma/\beta} \tilde{\mu}$. \label{mu} The function $G$ satisfies the following differential equation:
\begin{equation}
  \partial_{\tau}^{2}G = \mu^{2} G - 3\mu^{2}\frac{Q}{P} G^{2} - 2\mu^{2} \frac{1 - Q^{2}}{P^{2}} G^{3}.
\end{equation}
Comparing with~\eqref{fullosceqn}, the solution~\eqref{generalspecialWaveform} is obtained with $P(\phi) = -\tilde{\mu}^{2} Q(\phi)/$ $\cos \phi$, $Q^{2}(\phi) = 2 \cos^{2}\phi/(2 \cos^{2}\phi + \tilde{\mu}^{2})$ and the displaced phase\index{displaced phase} satisfies $\tan \phi(\xi) = - \frac{\tilde{\rho}}{\tilde{\mu}^{2}} \tan(\rho \xi)$, where $\tilde{\rho} = \tilde{\mu} \sqrt{2 + \tilde{\mu}^{2}}$ and $\rho = \gamma r_{0}^{2} \tilde{\rho}$. \label{rho} The corresponding solution of the NLS equation~\eqref{WFB_special_ansatz} can then be written after some manipulations as
\begin{equation}
  A(\xi,\tau) = A_{0}(\xi) \left(\frac{-\tilde{\mu}^{2} \cos(\rho \xi) + i \tilde{\rho} \sin(\rho \xi)}{\cos(\rho \xi) \pm \sqrt{1 + \frac{1}{2} \tilde{\mu}^{2}} \cosh(\mu \tau)} - 1 \right).
\end{equation}
This is the Ma solution\index{Ma solution} or Ma breather\index{Ma breather}~\citep{3Ma79}.

Thirdly, we take $\zeta(\tau) = 1 - \frac{1}{2} \nu^{2} \tau^{2}$. (The same solution can also be obtained by substituting $\zeta(\tau) = 1 + \frac{1}{2} \mu^{2} \tau^{2}$.) Then, the function $G$ satisfies the differential equation
\begin{equation}
  \partial_{\tau}^{2}G = \frac{3\nu^{2}}{P} G^{2} - 4\nu^{2} \frac{Q - 1}{P^{2}} G^{3}.
\end{equation}
Comparing with~\eqref{fullosceqn}, we have $P(\phi) = \tilde{\nu}^{2}/\cos \phi$, $Q(\phi) = 1 + P^{2}/(4\tilde{\nu}^{2})$ and the displaced phase\index{displaced phase} satisfies $\tan \phi(\xi) = - 2\gamma r_{0}^{2} \xi$. Substituting into the Ansatz~\eqref{generalspecialWaveform}, we obtain
\begin{equation}
  A(\xi,\tau) = A_{0}(\xi) \left(\frac{4 (1 - 2 i \gamma r_{0}^{2} \xi)}{1 + 4 (\gamma r_{0}^{2} \xi)^{2} - 2 \frac{\gamma}{\beta} r_{0}^{2} \tau^{2}} - 1 \right).
\end{equation}
This is the rational solution or rational breather, \index{rational breather|see{rational solution}} also called the algebraic solution\index{algebraic solution} \citep{3Peregrine83}. Plots of the absolute value of the Ma soliton\index{Ma soliton} and the rational soliton\index{rational soliton} for $r_{0} = 1$ are given in Figure~\ref{MaRa3D}.

\subsection{Relation between the breather solutions}
\index{breather solutions!relation between} \index{SFB!relation with other solutions}

As shown by~\citet{3Dysthe99}, there is a relationship among the breather solutions of the NLS equation. Different parameters are used for explicit expressions of the SFB and the Ma breather in \citep{3Dysthe99} as well as in \citep{3Grimshaw01}. Introducing new parameters $\varphi$ and $\vartheta \in \mathbb{R}$ by the following relations: $\tilde{\nu} = \sqrt{2} \sin \varphi$, $\tilde{\sigma} = \sin 2\varphi$, $\tilde{\mu} = \sqrt{2} \sinh \vartheta$ and $\tilde{\rho} = \sinh 2\vartheta$, we can write our expressions similar to the ones in these papers.

\begin{figure}[h]		
\begin{center}
\includegraphics[width = 0.5\textwidth, viewport = 73 597 445 815]{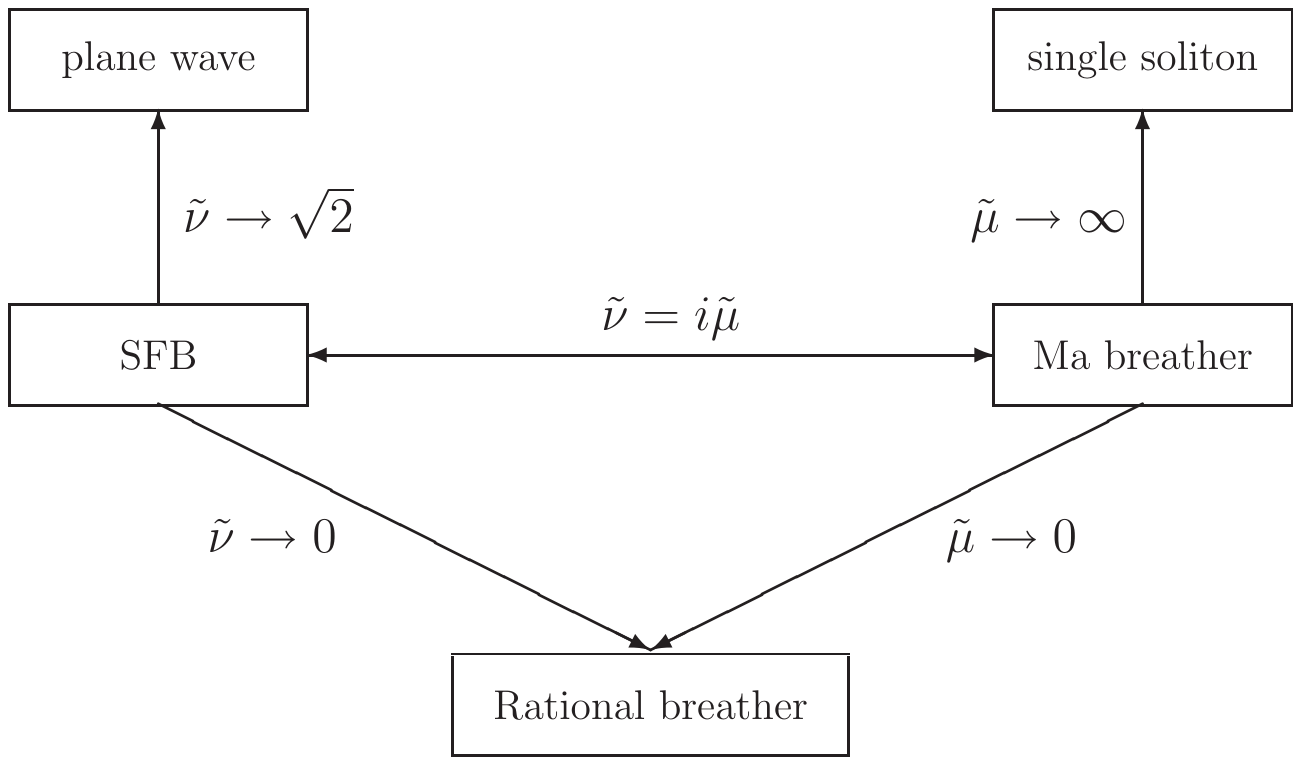}
\caption[Diagram of the breather solutions relation]{The schematic diagram for the derivation of the Ma breather and the rational breather from the SFB.} \label{segitiga}
\end{center}
\end{figure}

For the expressions given above for these solutions, we observe that the SFB becomes the Ma breather if we substitute $\nu = i \mu$ and it becomes the rational breather if $\nu \rightarrow 0$. Similarly, the Ma breather becomes the rational breather for $\mu \rightarrow 0$. The Ma breather becomes the single soliton\index{single soliton} solution for $\mu \rightarrow \infty$ \citep{3Akhmediev97}. Figure~\ref{segitiga} explains the schematic diagram of these relations. These relations can also be seen from the expression in the previous section. For the SFB, we take $\zeta(\tau) = \cos (\nu \tau)$, which leads to the Ma breather if we substitute $\nu = i \mu$, so that $\zeta(\tau) = \cos (i \mu \tau) = \cosh (\mu \tau)$. Taking the Taylor expansion either around $\nu = 0$ or $\mu = 0$, we get $\zeta(\tau) = 1 - \frac{1}{2} \nu^{2} \tau^{2}$ or $\zeta(\tau) = 1 + \frac{1}{2} \mu^{2} \tau^{2}$, and both cases will lead to the rational breather.

The relations have also consequences for the amplitude amplification factor (AAF)\index{AAF}, defined as the ratio between the maximum amplitude and the value of its background. The expressions for the breather solutions as given above were shifted such that the maximum amplitude is at $(\xi,\tau) = (0,0)$ and the value of the background is $r_{0}$.

\begin{figure}[h]		
\begin{center}
\includegraphics[width = 0.6\textwidth]{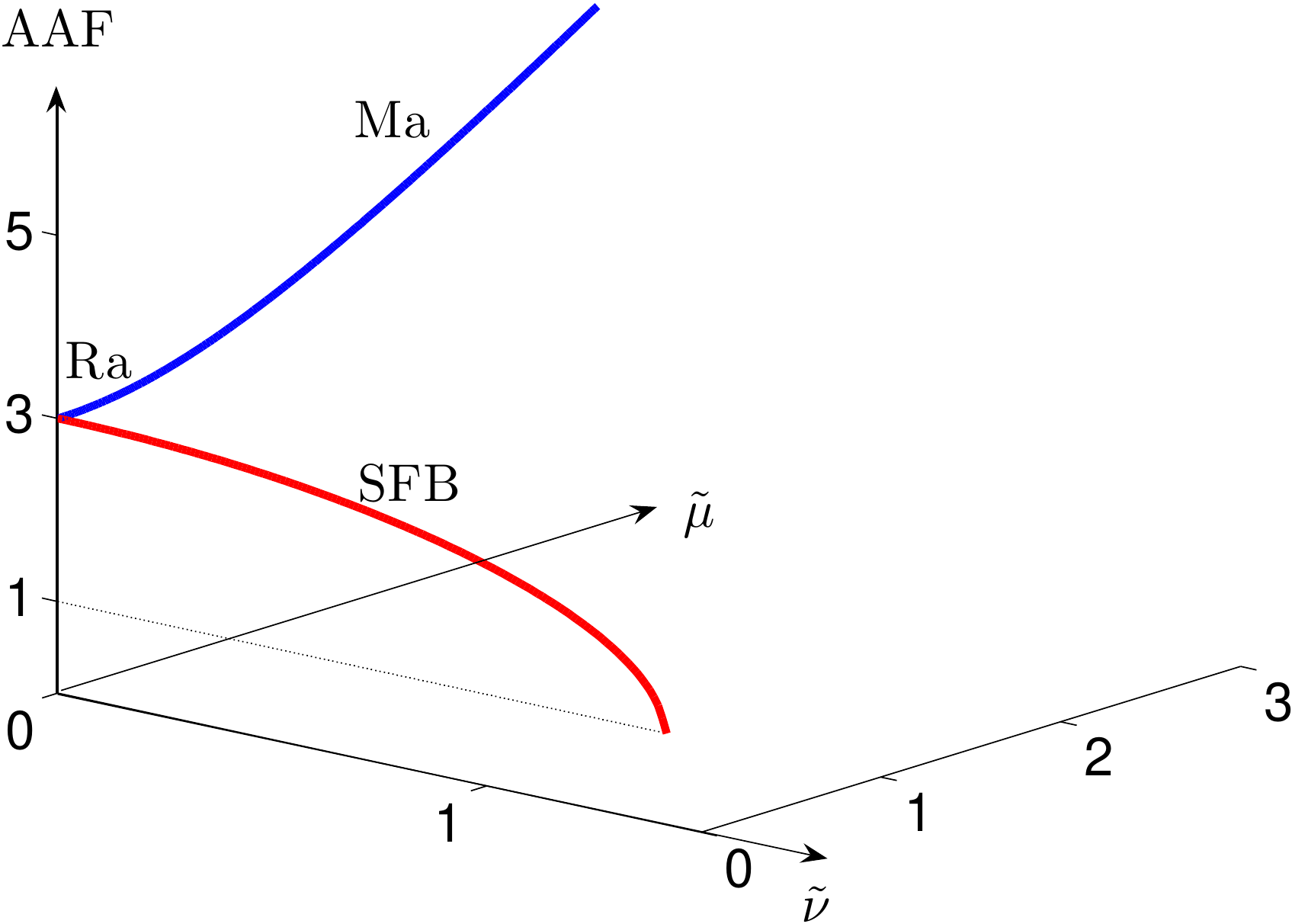}
\caption[Amplitude amplification factor of the breather solutions]{The plot of the amplitude amplification factor for the SFB and the Ma breather. The meeting point of the two curves is the AAF of the rational breather.} 		\label{AAF_plot}
\end{center}
\end{figure}

For the SFB, the amplification is given by $\textmd{AAF}_{\textmd{\tiny S}}(\tilde{\nu}) = 1 + \sqrt{4 - 2\tilde{\nu}^{2}}$. For $0 < \tilde{\nu} < \sqrt{2}$, the amplification is bounded and $1 < \textmd{AAF}_{\textmd{\tiny S}}(\tilde{\nu}) < 3$. The AAF for the Ma breather\index{AAF!Ma solution} is given by $\textmd{AAF}_{\textmd{\tiny Ma}}(\tilde{\mu}) = 1 + \sqrt{4 + 2\tilde{\mu}^{2}}$. Hence, for $\tilde{\mu} > 0$, we have $\textmd{AAF}_{\textmd{\tiny Ma}} > 3$. The AAF for the rational breather is exactly $3$\index{AAF!rational solution}, which follows by letting either $\tilde{\nu}$ or $\tilde{\mu}$ go to zero:
\begin{equation}
\textmd{AAF}_{\textmd{\tiny Ra}} =  \lim_{\tilde{\nu} \rightarrow 0} \textmd{AAF}_{\textmd{\tiny S}} = 3 = \lim_{\tilde{\mu} \rightarrow 0} \textmd{AAF}_{\textmd{\tiny Ma}}.
\end{equation}
The plot of the AAF for all the three breather solutions is given in Figure~\ref{AAF_plot}. For $\tilde{\mu} \rightarrow \infty$ in the Ma breather, the single-soliton solution~\eqref{onesoliton} is obtained, see page~61--64 of~\citet{3Akhmediev97}.

\subsection{Physical wave fields}
\index{breather solutions!physical wave fields} \index{physical wave field} \index{physical wave field!Ma solution} \index{physical wave field!rational solution} \index{Ma solution!physical wave field} \index{rational solution!physical wave field}

The corresponding physical wave fields of the waves on the finite background are particularly interesting to investigate. Considering only the first-order contribution, the physical wave field $\eta$ is given by~\eqref{physicalwavefield}. We apply the same moving frame of reference as in the case of SFB. Density plots for three breather solutions are given in Figure~\ref{physical}. For illustration, we choose $r_{0} = 1$, $k_{0} = 2\pi$ in all cases, modulation frequency $\tilde{\nu} = 1/2$ for the SFB and $\tilde{\mu} = 0.4713$ for the Ma breather. For better representation, the physical wave fields are shown in a moving frame of reference with suitably chosen velocity.

The SFB has extreme values at $x = \xi = 0$ and is periodic in time. The Ma breather has extreme values at $\tau = 0$ and is periodic in space. The rational breather is neither periodic in time nor in space, but decaying asymptotically in space and time to the plane-wave\index{plane-wave} solution~\eqref{planewave} with its maximum at $(x,t) = (0,0)$. Despite some differences, the breather solutions show `wavefront dislocation'\index{wavefront dislocation}, when splitting or merging of waves occurs \citep{3NyeBerry74}. For the SFB, it occurs at the interval $0 < \tilde{\nu} < \sqrt{3/2}$, for the Ma soliton\index{Ma soliton} for all $\mu > 0$ and also for the rational soliton.\index{rational soliton}

\section{Conclusion}

In this chapter, we have discussed the description of waves on the finite background and its application to extreme water wave generation in the hydrodynamic laboratory. We introduced a transformation to displaced phase-amplitude\index{displaced phase-amplitude} variables with respect to a background of monochromatic waves and viewed it from the variational formulation\index{variational formulation} perspective. The dynamical evolution of these waves is governed by a nonlinear oscillator\index{nonlinear oscillator} equation for the displaced amplitude\index{displaced amplitude} with potential energy that depends on the displaced phase. \index{displaced phase} The displaced phase-amplitude transformation and its physical interpretation are new contributions of this thesis.

Furthermore, by restricting to a special case that the displaced phase is time-independent, the nonlinear oscillator equation for the wave signal at each position becomes autonomous. We observe that the change of the displaced phase with respect to the position is the only driving force for a spatial evolution toward extreme wave events. The restriction led to three exact solutions of the NLS equation for waves on finite background, also known as the breather solutions: the SFB solution\index{SFB}, the Ma solution\index{Ma solution}, and the rational solution.~\index{rational solution}

We studied extensively many properties of the SFB since this solution describes the spatial evolution of the envelope wave signal from a slightly modulated wave into a large extreme wave and returns into the initial signal with a different phase. The asymptotic behavior\index{asymptotic behavior} of the SFB in the far distances is described by the linear modulational (Benjamin-Feir) instability. The SFB that binds the exponential growth of this instability is the extension in the nonlinear regime and therefore describes a complete evolution of the modulational instability process. \index{modulational instability} \index{Benjamin-Feir instability}

Particularly interesting is that we use the SFB in a real-life application for a model of extreme wave generation\index{extreme waves!generation} in the wave basin of a hydrodynamic laboratory. We use the concept of MTA to quantify the spatial evolution from a moderate amplitude into a large wave. This knowledge of MTA is used for the design of the wave generation. Since the initial SFB wave signal has a moderate amplitude, it is possible to generate it by the wavemaker. This is not the case for the initial wave signal that corresponds to the Ma solution which requires the highest amplitude waves at the wavemaker. Theoretically, the limiting case of the SFB for infinitely long modulation can reach an amplitude amplification of factor three, provided that the wave basin has enough space. This limiting case is precisely the wave signal that corresponds to the rational solution in which the wave signal has infinite periodicity and an exponentially confined extreme wave.

We also observed that the corresponding physical wave field of the SFB solution (for $0 < \tilde{\nu} < \sqrt{3/2}$), the Ma solution (for $\mu > 0$) and the rational solution show splitting and merging waves. The following chapter will discuss these phenomena in more detail.

\newpage
{\renewcommand{\baselinestretch}{1} 
}

\setcounter{chapter}{3}
\chapter{Wavefront dislocation in surface water waves} \label{4Dislocate}

\section{Introduction}

We have seen in the previous chapter that the physical wave field of the Soliton on Finite Background (SFB) shows wavefront dislocation which happens for modulation frequency $\tilde{\nu} < \sqrt{3/2}$. This phenomenon occurs when the real amplitude of the SFB vanishes and then the phase is undefined, which is often called a phase singularity\index{phase singularity}. In this chapter, we will investigate in more detail the phenomena of wavefront dislocation and phase singularity in the field of water waves. Many studies in the literature are dedicated to phenomena related to `phase singularity' and `wavefront dislocation'.~\index{wavefront dislocation} Since generally both phenomena occur simultaneously, the term phase singularity is more often used in physical optics to describe what we will refer to as wavefront dislocation, for instance in~\citep{4Balistreri00}. Phase singularities are also called `intensity zeros',\index{intensity zeros} \index{intensity zeros|see{phase singularity}} `topological charges',\index{topological charges} \index{topological charges|see{phase singularity}} or `optical vortices' \index{optical vortices} \index{optical charges|see{phase singularity}}~\citep{4Coullet89, 4Nye99, 4Berry00, 4Berry01, 4Soskin01}. In water waves, the `disappearance of waves'\index{waves disappearance} in a modulated train of surface gravity waves was described by~\citet{4Tanaka95}.

`Dislocation'\index{dislocation} \index{dislocation!wavefront|see{wavefront dislocation}} is known for a long time in the field of material science. There it is used to describe an irregularity within a crystal structure, often responsible for the plastic deformation of metals and other crystalline solids. The concept was introduced as early as 1934 and proposed independently by several authors~\citep{4Orowan34, 4Polanyi34, 4Taylor34}. Other references to dislocation in crystals\index{dislocation!crystals} are~\citep{4Read53, 4Honeycombe84, 4Dieter88, 4Hull01}. The above phenomena are also found in several other branches of physics. A simple example of phase singularity\index{phase singularity!north pole} is the singular time zone at the north pole~\citep{4Dennis01}. See Figure~\ref{kutubutara}(b) for a visualization. In~\citep{4Soskin01}, `singular optics'\index{singular optics} is discussed as a new branch of modern physical optics that deals with a wide class of effects associated with phase singularities in wave fields and with the topology of wavefronts. An introduction and a summary of a workshop on singular optics about recent progress in the field is given by~\citet{4Berry04}.

Extensive references to many topics related to dislocations from theoretical to experimental observations and applications can be found in \citep{4Nabarro7904}. Experimental observations in a neon discharge in two-dimensional space-time are reported by~\citet{4Krasa81}. A study of it in the Aharonov-Bohm effect is done by~\citet{4Berry80}. An analysis for constructing a theory of wavefront dislocation using catastrophe theory is developed by~\citet{4Wright79}. A study of the phenomenon in optics, particularly in monochromatic light waves is reported by~\citet{4Basistiy95}. Line singularities in vector and electromagnetic waves, including the paraxial case, when waves propagate in a certain direction and the general case, when waves propagate in all directions, is discussed by~\citet{4Nye97}. A theoretical framework for understanding the local phase structure and the motion of the most general type of dislocation in a scalar wave, how this dislocation may be categorized and how its structure in space and time is related, has been studied by~\citet{4Nye81}. Statistical calculations associated with dislocations for isotropically random Gaussian ensembles, that is, superpositions of plane-waves equidistributed in direction but with random phases, are given by~\citet{4Berry00}. Knotted and linked phase singularities in monochromatic waves \index{monochromatic wave(s)} by constructing exact solutions of the Helmholtz equation are given by~\citet{4Berry01}.

Apparently, the same phenomenon is also observed in 3D surfaces of constant phase (wavefronts) of a wave field. A new concept of `wavefront dislocation'\index{wavefront dislocation!concept introduced} was introduced in 1974 by~\citet{4Nye&Berry74} and is used to explain the experimentally observed appearance and disappearance of crest or trough pairs in a wave field. The authors show that wavefront dislocation appears as a result of partial spatial overlapping of two quasi-monochromatic pulses. In their paper, the examples are given in two and three space dimensions plus time. They also predict the gliding and climbing of wavefront dislocation on the pulse envelope. Other terminologies that are also often used to describe the phenomenon are death and birth of waves\index{wavefront dislocation!other terminologies!death and birth of waves}, and annihilation and creation of waves\index{wavefront dislocation!other terminologies!annihilation and creation of waves}. When dealing with waves, \citet{4Nye&Berry74}~showed that dispersion is not really involved when wavefront dislocation occurs, while, on the other hand, \citet{4Trulsen98} explained that wavefront dislocation is a consequence of linear dispersion alone and predicted by the linear Schr\"{o}dinger equation\index{linear Schr\"odinger equation}, an example of the paramount importance of a linear dispersive wave equation.\index{dispersive wave!linear equation}

\begin{figure}[h]			
\begin{center}
\subfigure[]{\includegraphics[width = 0.25\textwidth]{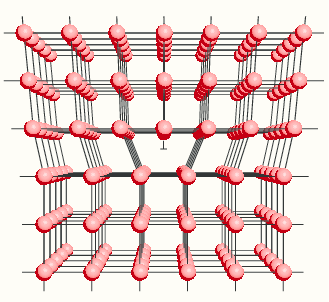}} 			\hspace{1cm}
\subfigure[]{\includegraphics[width = 0.24\textwidth]{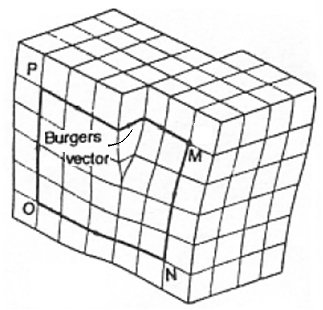}} 				\hspace{1cm}
\subfigure[]{\includegraphics[width = 0.32\textwidth]{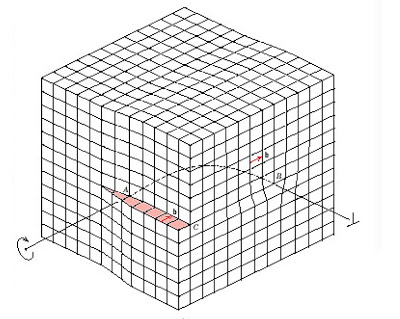}}
\caption[Types of dislocation]{Schematic diagrams showing several types of dislocation line: pure edge dislocation (a), pure screw dislocation (b) and mixed edge-screw dislocation (c).} \label{edgescrew}
\end{center}
\end{figure}

In $(1 + 1)$D, dislocation happens at a set of points in $(x,t)$ space. In higher dimensions, however, for example in $(2 + 1)$D or $(3 + 1)$D, dislocation is a set of lines in $(x,y,t)$ space or a set of planes in $(x,y,z,t)$ space. There are several types of dislocation lines in higher-order dimensions. But before that, let us introduce the analog of the `Burgers vector'\index{Burgers vector} in the crystal according to~\citet{4Read53}. It is a vector perpendicular to the wavefronts of length equal to the wavelength. If the Burgers vector is perpendicular to the dislocation line, it is called `pure edge'\index{dislocation!pure edge} type. If the Burgers vector is parallel or antiparallel to the dislocation line, it is called `pure screw'\index{dislocation!pure screw} type. If both types are observed, it is called `mixed edge-screw'\index{dislocation!mixed edge-screw} dislocation. In this thesis, since we consider $(1 + 1)$D case, we observe only the pure edge type of dislocation point. Dislocations in crystals are shown in Figure~\ref{edgescrew}.

In this chapter, we restrict to wave fields with one spatial and one temporal variable. Even for this simplest case, we sensed some confusion in the cited references above about the equivalence of the possibly different phenomena of phase singularity\index{phase singularity} and wavefront dislocation.~\index{wavefront dislocation} Moreover, it was not very clear if these phenomena are exceptional, rare events should be expected at any point of vanishing amplitude.~\index{vanishing amplitude} From a more practical point of view, we wanted to use the appearance of wavefront dislocations that we had found in the theoretical expression of the SFB as a check-in measured signals of waves that were generated in a hydrodynamic laboratory. The robustness of such a phenomenon for perturbations of various kinds is then required, a result that was not found in the cited references.

This chapter is organized as follows. In Section~\ref{Preliminaries}, we present the basic notions of phase singularity, wavefront dislocation, and introduce the Chu-Mei quotient that will be used in this chapter. Further, we give the most trivial examples of surface wave fields, namely superpositions of two and three monochromatic waves. With these examples, we will show that already for trichromatic waves phase singularity and wavefront dislocation will be generic properties, but also that phase singularity is not necessarily accompanied by wavefront dislocation. In Section~\ref{WDwavegroups} we study these aspects for wave groups. We will show that the unboundedness of the Chu-Mei quotient is a necessary condition for the occurrence of wavefront dislocation. A perturbation analysis is done to show that the boundedness of the Chu-Mei quotient is an exceptional case. Although all these phenomena are essentially linear, in Subsection~\ref{SubsectionSFB} we investigate these phenomena for the special solution of the NLS equation, the SFB. We give a novel contribution to the theory of wave dislocation by introducing the Chu-Mei quotient and relating it with wavefront dislocation. The final section concludes the chapter with some conclusions and remarks.

\section{Preliminaries} \label{Preliminaries}

This section is devoted to collect preliminary definitions that will be used in this chapter to study wavefront dislocation. We also illustrate degenerate and generic cases of the phenomena by using simple wave fields that consist of a superposition of a few harmonic modes.

\subsection{Basic Notions}

Let $\eta(x,t)$  be a real-valued function that describes a surface wave field in one space variable $x$ and time $t$. The complexification of $\eta$ is defined with the Hilbert transform\index{Hilbert transform} $\mathcal{H}$$[\eta]$, given by $\eta_{\textmd{c}}(x,t) = \eta(x,t) + i \mathcal{H}[\eta(x,t)]$. The Hilbert transform of a function $f$ is denoted as $\mathcal{H}[f(t)]$ and is given by an improper integral:
\begin{equation}
  \textmd{$\cal{H}$}[f(t)] = \frac{1}{\pi} \textsc{pv} \int_{-\infty}^{\infty} \frac{f(\tau) \, d\tau}{\tau - t}   		\label{Hilbert}
\end{equation}
where \textsc{pv} denotes the Cauchy principal value of the corresponding integral. Note that our wave field $\eta$ depends on $x$ and $t$ and the Hilbert transform is applied with respect to the time variable $t$. Nonetheless, we can also apply the transformation with respect to the space variable $x$. Written in polar form with real-valued phase and amplitude variables we get $\eta_{\textmd{c}}(x,t) = a(x,t) e^{i\Phi(x,t)}$. The {\slshape local wavenumber}\index{local wavenumber} and {\slshape local frequency}\index{local frequency} are defined respectively as $k(x,t) = \partial_{x}\Phi$ and $\omega(x,t) = -\partial_{t} \Phi$.

\begin{figure}			
\begin{center}
\subfigure[]{\includegraphics[width = 0.55\textwidth,viewport = 164 506 447 696]{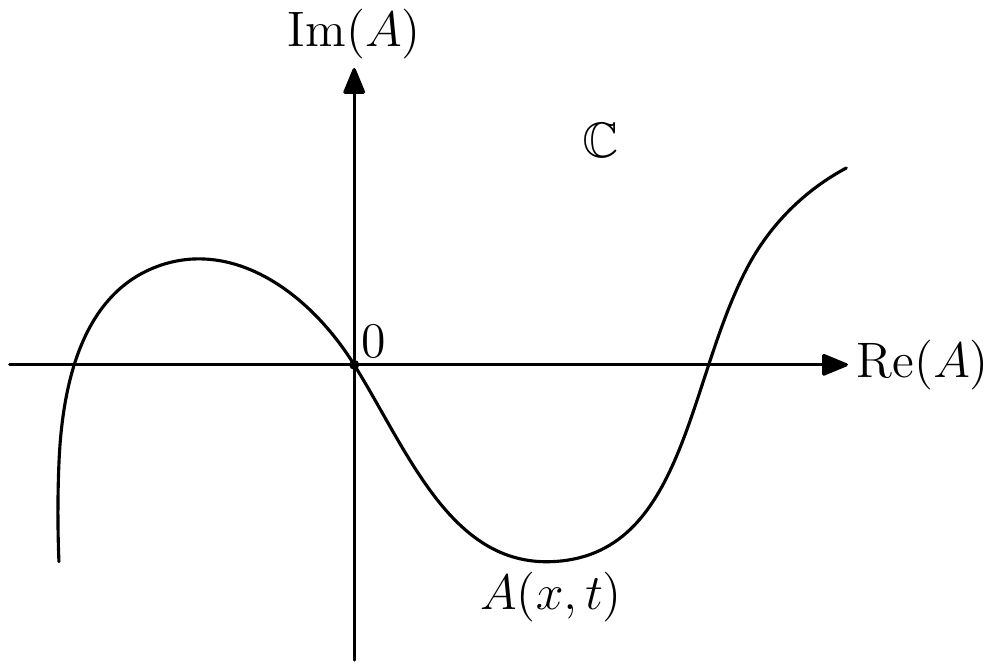}}    \hspace{0.75cm}
\subfigure[]{\includegraphics[width = 0.35\textwidth,viewport = 189 452 423 706]{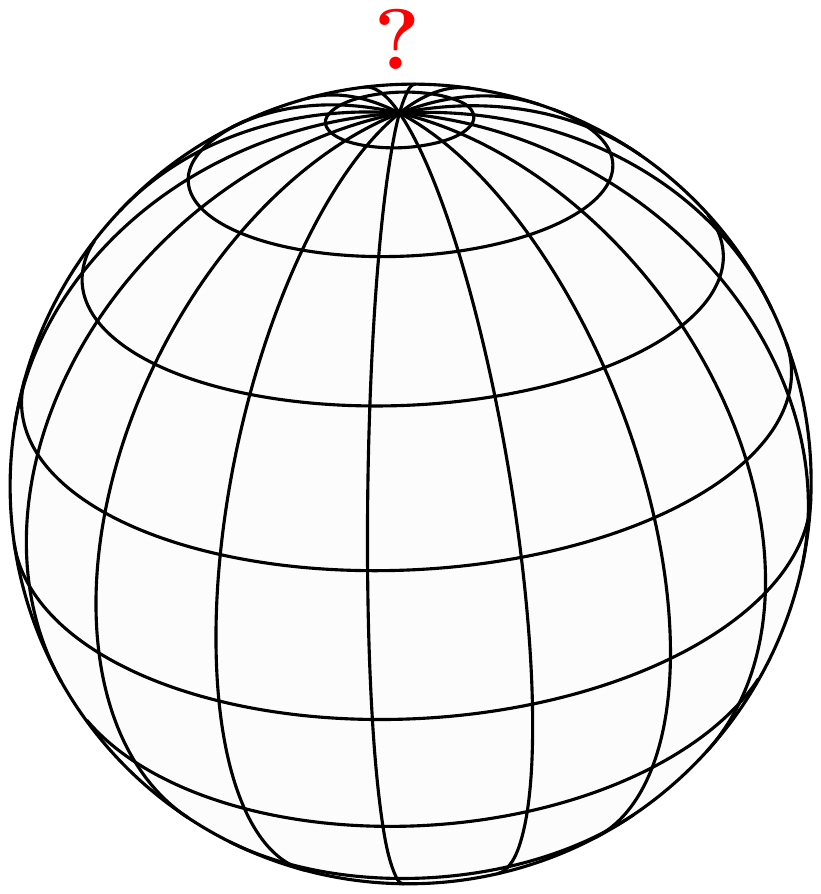}}
\caption[Phase singularity illustrations]{(a) A sketch of a trajectory of the complex amplitude in the Argand diagram when a singularity appears where the trajectory crosses the origin. Typically the phase will experience a $\pi/2$ jump. (b) An example of singularity is the singular time zone at the North Pole.}	    \label{kutubutara}
\end{center}
\end{figure}

The phase $\Phi$ is uniquely defined for smooth functions $\eta$ for all $(x,t) \in \mathbb{R}^{2}$ for which the amplitude does not vanish. When the wave field has {\slshape vanishing amplitude}\index{vanishing amplitude}, i.e., if $a(\hat{x},\hat{t}) = 0$, we call $(\hat{x},\hat{t}) \in \mathbb{R}^{2}$ a {\slshape singular point}\index{singular point(s)}. In the Argand diagram\index{Argand diagram} (the complex plane), the time signal at a fixed position corresponds to an evolution curve $t \mapsto \eta_{\textmd{c}}(x,t)$; a singular point $(\hat{x},\hat{t})$ lies on an evolution curve that is at the origin of the complex plane. Figure~\ref{kutubutara}(a) shows a sketch of a complex-amplitude $A$ that passes the origin and thus experiences singularity.

For nonzero amplitude, the phase of $\eta_{\textmd{c}}$ has a well-defined value, but at a singular point, the phase $\Phi$ is undetermined or may even become singular. We will say that the wave field $\eta(x,t)$ has a {\slshape phase singularity}\index{phase singularity} at the singular point\index{singular point(s)} $(\hat{x},\hat{t})$ if $\Phi(x,t)$ is not continuous. As is clear from the interpretation in the Argand diagram, in most cases the trajectory will cross the origin and the phase will be discontinuous and have a $\pi$-jump. Only in the case where the origin acts as a reflection point, the phase will be continuous. Examples are easily constructed for both cases by superposition of just a few waves, as we will show further on in Subsections~\ref{Bichromatic} and~\ref{Trichromatic}.

Wavefront dislocation\index{wavefront dislocation} is observed when waves at a certain point and time merge or split. Necessarily this can happen only at a singular point, as we will see. Formally we will define that the wave field $\eta(x,t)$ has {\slshape wavefront dislocation} of strength\index{wavefront dislocation!strength} $n \neq 0$ in the area of the $(x,t)$ plane that is enclosed by a contour if the following contour integral has the given integer multiple of $2\pi$~\citep{4Nye&Berry74, 4Berry81, 4Berry98}:
\begin{equation}
\oint d\Phi = \oint (k \mathop{dx} - \omega \mathop{dt}) = \iint \left(\frac{\partial \omega}{\partial x} + \frac{\partial k}{\partial t} \right) \mathop{dx} \mathop{dt} =  2n\pi, \qquad n \neq 0.					    \label{contour_integral1}
\end{equation}

Instead of taking an arbitrary closed curve, it is also possible to investigate the property of a given singular point. Then, by taking a circle of radius $\epsilon$, and allowing the radius to shrink to zero, the strength of the singular point\index{singular point(s)!strength} is found from 
\begin{equation}
I = \lim_{\epsilon \rightarrow 0} \oint_{C(\epsilon)} \mathop{d\Phi}
  = \lim_{\epsilon \rightarrow 0} \int_{0}^{2\pi} \frac{d\Phi}{d\theta} \mathop{d\theta}				    \label{contour_integral2}
\end{equation}
where $\theta$ is the angle variable describing the circle. If $I = 0$ there is no wavefront dislocation, while if $I = \pm 2\pi$ there is wavefront dislocation. More specifically, splitting of waves for progressing time will occur if for increasing $x$ the value of $I = -2\pi$; for $I = 2\pi$, merging of waves will happen for increasing $x$. Indeed, the formula to measure the number of waves of a signal in a certain time interval $(0,T)$ at a fixed position $x$ is given by
\begin{equation}
N(x) = \frac{1}{2\pi} \int_{0}^{T} \omega(x,t) \mathop{dt} 			\label{number_waves}.
\end{equation}
This shows that when crossing the singular position $\hat{x}$, the number of waves may change if the time interval contains singular times. When the time interval is just around one singular time instant, we get that the change is equal to the strength of the singular point as defined above.

\index{Fornberg-Whitham term|see{Chu-Mei quotient}} For a full description of the appearance of wavefront dislocation, it turns out to be useful to introduce another quantity. This is the so-called\index{Chu-Mei quotient} `Chu-Mei quotient'\footnote[1]{Some authors call this quotient the `Fornberg-Whitham term'\index{Fornberg-Whitham term}~\citep{4Infeld90}, referring to~\citep{4Fornberg78}. However, throughout this chapter, we call it the `Chu-Mei quotient', since they introduced it for the first time~\citep{4Chu70, 4Chu71} when they derived the modulation equations of Whitham's theory~\citep{4Whitham67} for slowly varying Stokes waves. However, the quotient already appeared earlier in~\citep{4Karpman67, 4Karpman69} when they consider modulated waves in nonlinear media. \par} defined for signaling problems by
\begin{equation}
\textmd{CMq} = \frac{\partial^{2}_{t}a}{a};
\end{equation}
for initial value problems, the derivative with respect to $x$ is used instead of to $t$. See equations~\eqref{nondisrel} and~\eqref{nondisreltem} on pages~\pageref{nondisrel} and~\pageref{nondisreltem}, respectively. This quotient appears in the dispersion relation\index{dispersion relation} for both linear and nonlinear dispersive wave equations\index{dispersive wave} and has a clear interpretation in this context, as we will show in Section~\ref{WDwavegroups}.

In the remainder of this chapter, we will show the following inclusive relations between the concepts introduced above. In Section~\ref{WDwavegroups} we will show that unboundedness of the Chu-Mei quotient is a necessary condition for wavefront dislocation to occur in wave groups. Further, wavefront dislocation at a point implies that there is phase singularity and phase singularity can only occur at singular points.~\index{singular point(s)} Although we give examples that the reversed implications are not valid, this will only happen for degenerate cases. \emph{Generically} it will be the case that at a singular point there is phase singularity\index{phase singularity}, wavefront dislocation, and the unboundedness of the Chu-Mei quotient.~\index{Chu-Mei quotient!unboundedness}

A monochromatic wave\index{monochromatic wave(s)} will not have any singular point,~\index{singular point(s)!monochromatic wave} and therefore it is not interesting for our investigations. A bichromatic wave can have singular points, which may have phase singularity. In the latter case, there will be no wavefront dislocation and the Chu-Mei quotient will be finite. A combination of three monochromatic waves can show all the phenomena; we will briefly describe these illustrative cases in the following subsections.

\subsection{Bichromatic wave field} \label{Bichromatic}
\index{bichromatic waves}

Consider the superposition of two monochromatic waves,~\index{monochromatic wave(s)!superposition of two} known as the bichromatic waves. A complexified form is
\begin{equation}
\eta_{\textmd{c}}(x,t) = |A_+| \exp(i\theta_+) + |A_-|\exp(i\theta_-)
\end{equation}
where $A_\pm = |A_\pm|e^{i\phi_\pm}$ are the complex-valued amplitudes and where $\theta_\pm = k_\pm x - \omega_\pm t + \phi_\pm$ are the phases of the constituent monochromatic waves. Assume that the waves have different phase velocities. Inspection of the real amplitude of the superposition, shows that singular points can only, and will, happen for $|A_+| = |A_-|$ and then for $\theta_+ -\theta_- = \frac{1}{2}n \pi$, $n \in \mathbb{Z}$. We consider the bichromatic with constituent waves of the same amplitude (taken to be unity for simplicity). Hence, we consider \index{bichromatic waves}
\begin{equation}
\begin{aligned}
\eta(x,t) &= \sin[(k_{0} + \kappa)x - (\omega_{0} + \nu)t] - \sin[(k_{0} - \kappa)x - (\omega_{0} - \nu)t] \\ 
          &= \frac{1}{2}\sin(\kappa x - \nu t) e^{i(k_{0}x - \omega_{0}t)} +  \textmd{c.c.} 			      
\end{aligned}          
\end{equation}
where both $\kappa$ and $\nu$ are small quantities, denote the modulation wavenumber and the modulation frequency, respectively. Indeed, the spatial period is given by $\lambda_{s} = 2\pi/\kappa$ and the temporal period is $T = 2\pi/\nu$. This wave field has a degeneracy of singular points:~\index{singular point(s)!bichromatic waves} the amplitude vanishes on straight lines in the $(x,t)$ plane that satisfy $\kappa \hat{x} - \nu \hat{t} = n\pi$, $n \in \mathbb{Z}$. At each point on such a line, there is a jump of $\pi$ in the phase, a phase singularity.~\index{phase singularity!bichromatic waves} \index{bichromatic waves}
\begin{figure}[h]			
\begin{center}
\subfigure[]{\includegraphics[width = 0.47\textwidth]{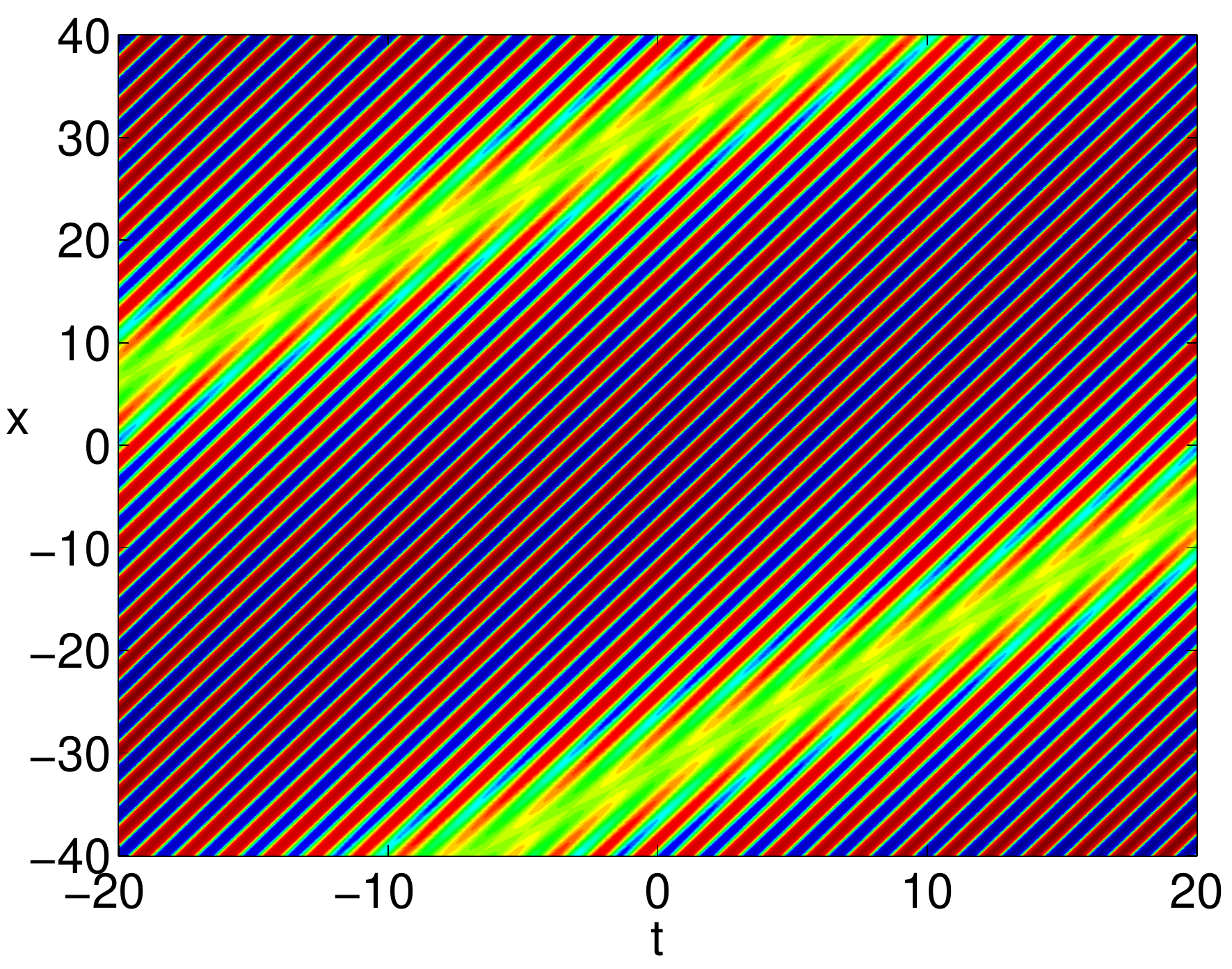}}	    \hspace{0.75cm}
\subfigure[]{\includegraphics[width = 0.43\textwidth]{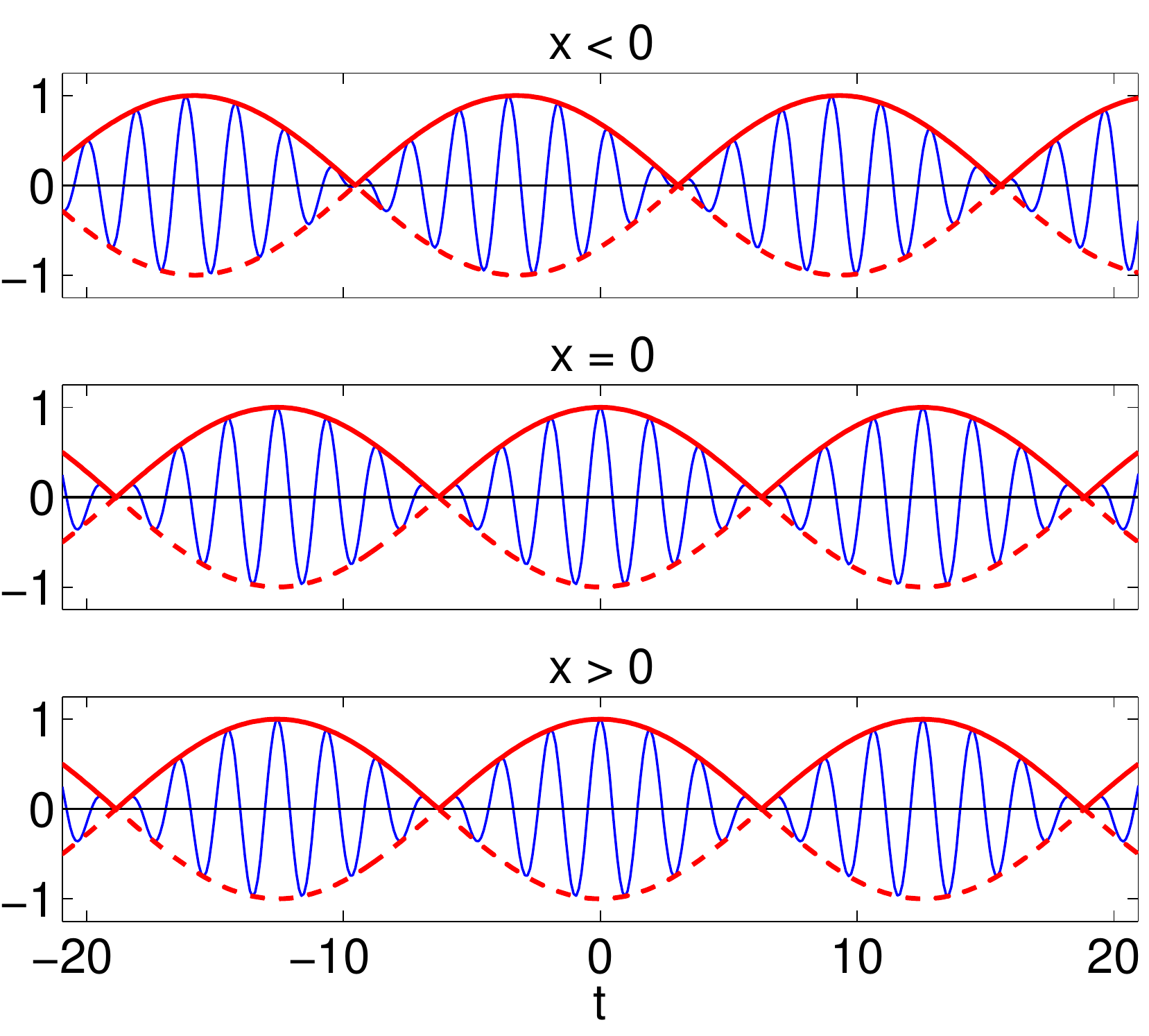}}
\caption[Bichromatic waves]{A density plot of the bichromatic wave field (a) and the corresponding wave signals for three different positions (b). The number of waves near the singular points does not change, there is no wavefront dislocation in this wave field.}			    \label{gambarbichromatic}
\end{center}
\end{figure}

The Chu-Mei quotient\index{Chu-Mei quotient!bichromatic waves} can be calculated explicitly and is bounded at a singular point: $\textmd{CMq} = \lim_{t \rightarrow \hat{t}} \frac{\partial_{t}^{2}a}{a}(x = \hat{x},t) = - \nu^{2}$. Using Proposition~\ref{ChuMei2WD} below, this implies that this wave field does not have wavefront dislocation.~\index{wavefront dislocation!bichromatic waves} Without using that result, this can also be shown in a direct way by calculating the strength at each point. Calculating the contour integral~\eqref{contour_integral2} around any singular point $(\hat{x},\hat{t})$ it is found that indeed $I(\hat{x},\hat{t}) = 0$. Figure~\ref{gambarbichromatic} shows the density plot of the bichromatic wave field. No wavefront dislocation is visible but the waves do show phase singularity\index{phase singularity}: the crests become troughs and vice versa. The evolution in the Argand diagram\index{Argand diagram!bichromatic waves} for this simple solution is on the real axis, crossing the origin twice in each period. \index{bichromatic waves}

\subsection{Trichromatic wave field} \label{Trichromatic}
\index{trichromatic waves}

Consider the superposition of three monochromatic waves. \index{monochromatic wave(s)!superposition of three} Already in this case, generically phase singularity\index{phase singularity} and wavefront dislocation\index{wavefront dislocation} will occur whenever the amplitude vanishes. An example is a solution of the linear version of the NLS equation~\eqref{NLSE}:
\begin{equation}
\partial_{\xi} A + i \beta \partial_{\tau}^{2} A = 0.
\end{equation}
Actually, almost any combination of three monochromatic waves can be used as an example. This solution is expressed as $\eta(x,t) = A_{\textsc{tc}}(\xi,\tau) e^{i(k_{0}x - \omega_{0}t)} + $ c.c., where \index{trichromatic waves}
\begin{equation}
A_{\textsc{tc}}(\xi,\tau) = \sum_{n = -1}^{1} b_{n} e^{i(\kappa_{n}\xi - \nu_{n}\tau)},
\end{equation}
with $b_{n} \neq 0$, $\kappa_{0} = 0 = \nu_{0}$, $\nu_{1} = \nu = -\nu_{-1}$ and $\kappa_{1} = \nu^{2} = \kappa_{-1}$. 	\index{trichromatic waves}

This wave field is periodic in both spatial and temporal variables $(\xi,\tau)$. If the coefficients are such that there are singular points, the phenomena described above can be investigated. To start with, a simple example of a degenerate case\index{trichromatic waves!degenerate} is $\eta_{\textsc{degenerate}} = \frac{1}{2}[1 + \sin(\kappa x - \nu t)]e^{i(k_{0}x - \omega_{0}t)}$. Clearly, there is a singular point, but there is no phase singularity and wavefront dislocation: the evolution in the Argand diagram\index{Argand diagram!degenerate trichromatic waves} is a straight line on the nonnegative real axis with the origin as a reflection point. \index{trichromatic waves}

Phase singularity\index{phase singularity!trichromatic waves} occurs when the coefficients of the trichromatic wave satisfy the following condition:
\begin{eqnarray}
(b_{1} - b_{-1})^{2} \leq b_{0}^{2} \leq (b_{1} + b_{-1})^{2},   \qquad &\textmd{if}& \; b_{1}b_{-1} > 0, \\
(b_{1} + b_{-1})^{2} \leq b_{0}^{2} \leq (b_{1} - b_{-1})^{2},   \qquad &\textmd{if}& \; b_{1}b_{-1} < 0. 	\label{coeffb}
\end{eqnarray}
If the coefficients do not satisfy this condition, the real-valued amplitude remains positive definite and therefore there is no phase singularity. The singular points~\index{singular point(s)!trichromatic waves} are the set of points $(\xi,\tau)$ that satisfy the following conditions:
\begin{align}
\cos(2\nu^{2}\xi) &= \frac{(b_{1} + b_{-1})^{2} [b_{0}^{2} - (b_{1} - b_{-1})^{2}]}{2 b_{0}^{2} b_{1} b_{-1}} - 1 \\
\cos(2\nu \tau)   &= \frac{b_{0}^{2} - (b_{1} - b_{-1})^{2}}{2 b_{1} b_{-1}} - 1.
\end{align}
There are four singular points within one spatial period and one singular point within one temporal period. \index{trichromatic waves}

\begin{figure}[h]			
\begin{center}
\subfigure[]{\includegraphics[width = 0.4\textwidth,viewport = 81 211 481 601]{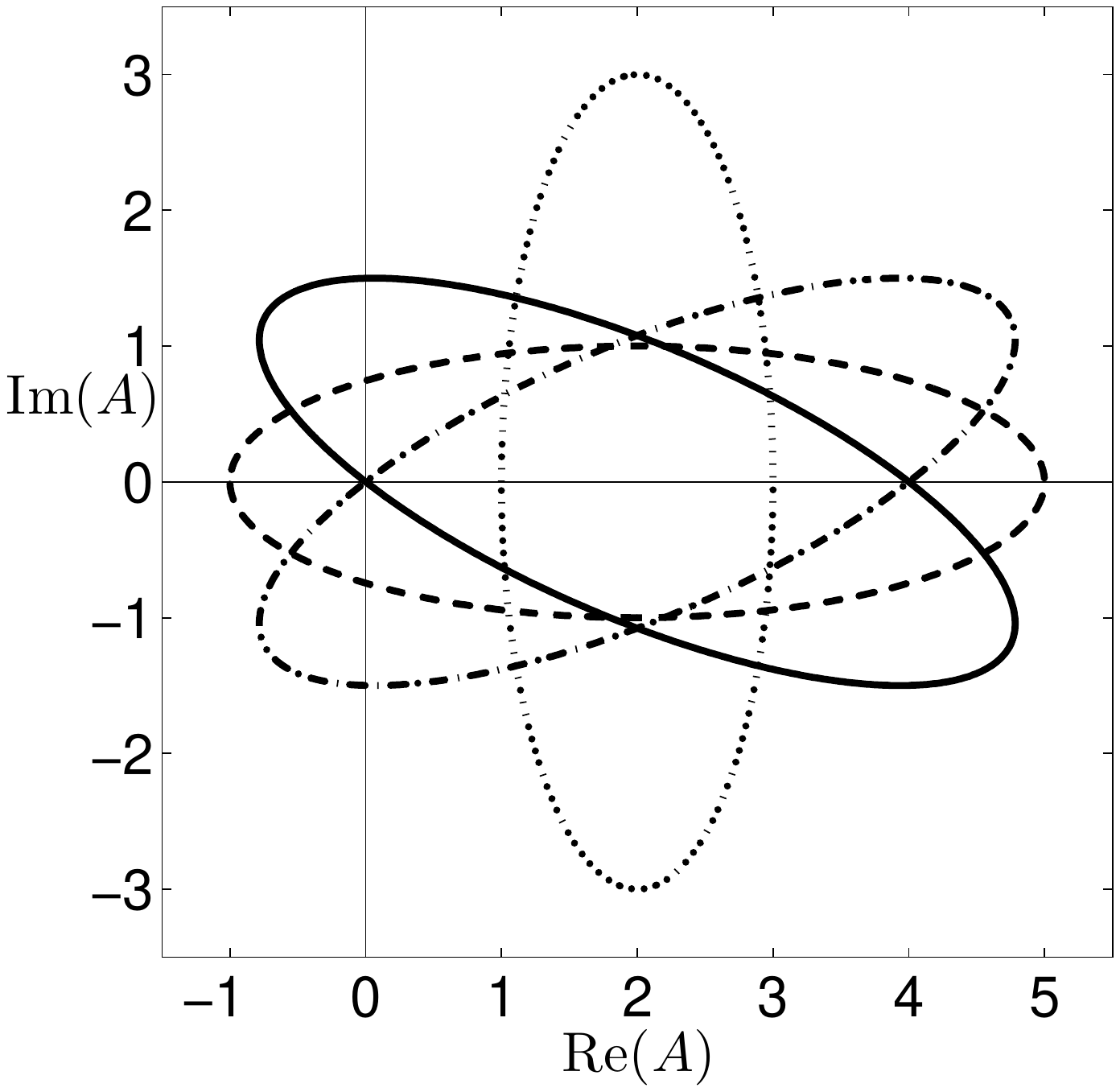}}	    \hspace{0.75cm}
\subfigure[]{\includegraphics[width = 0.5\textwidth]{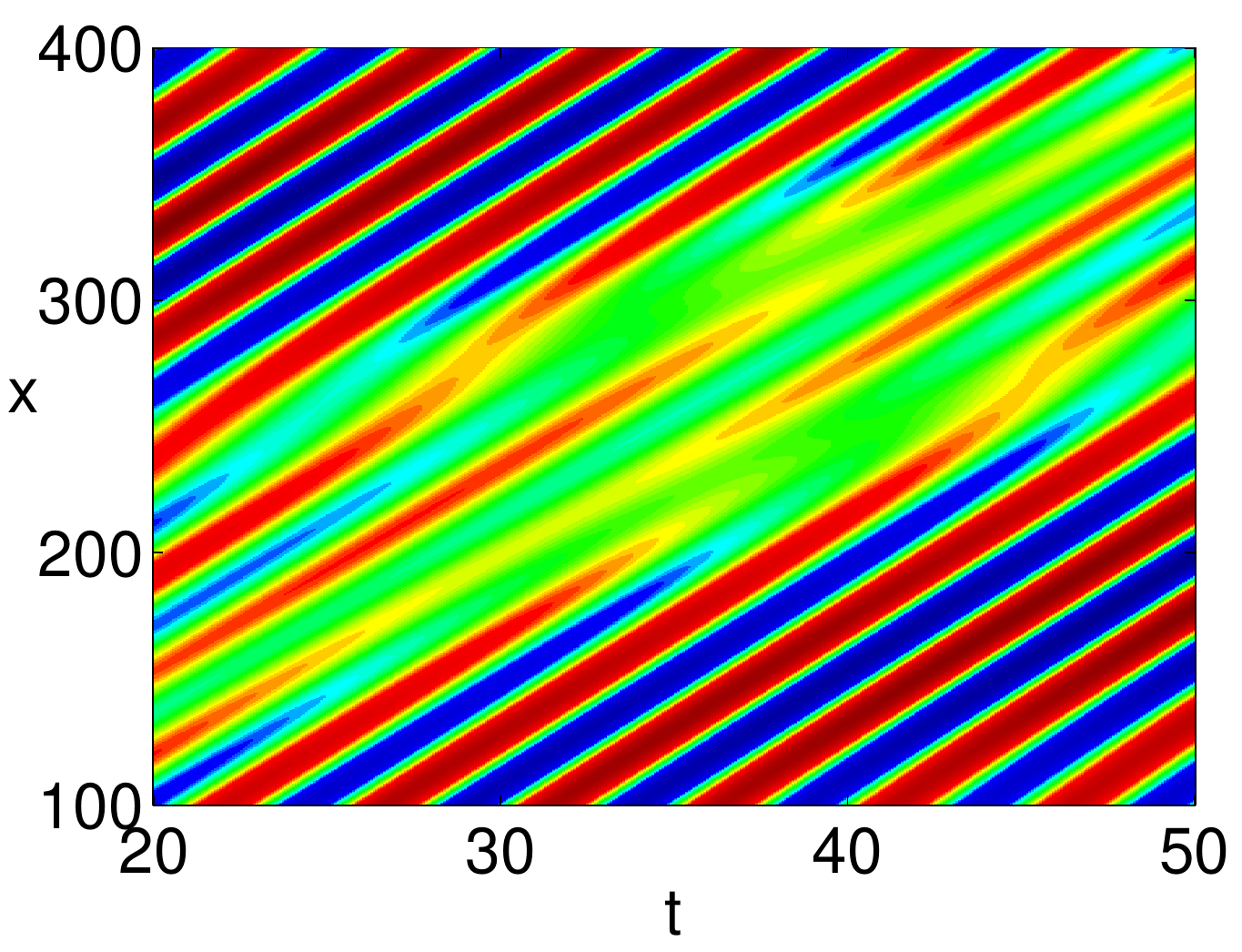}}
\caption[Argand diagram of trichromatic waves]{\index{Argand diagram!trichromatic waves}(a) The evolution in the Argand diagram is shown for the trichromatic wave parameterized by $\tau$ and plots are given for different values of $\xi$: $\xi = 0$ (dotted); $\xi = \hat{\xi}_{1}$ (solid); $\xi = \pi/(2\nu^{2})$ (dashed); $\xi = \hat{\xi}_{2}$ (dash-dot). The evolution curves are counterclockwise ellipses and follow a clockwise direction for increasing $\tau$. When the ellipse crosses the origin, a phase singularity and wavefront dislocation occur. (b) The density plot is shown of the trichromatic wave near a phase singularity where splitting and merging of waves can be seen.}		    \label{linearsolution} \index{trichromatic waves}
\end{center}
\end{figure}

\begin{figure}[h!]			
\begin{center}
\subfigure[]{\includegraphics[width = 0.45\textwidth]{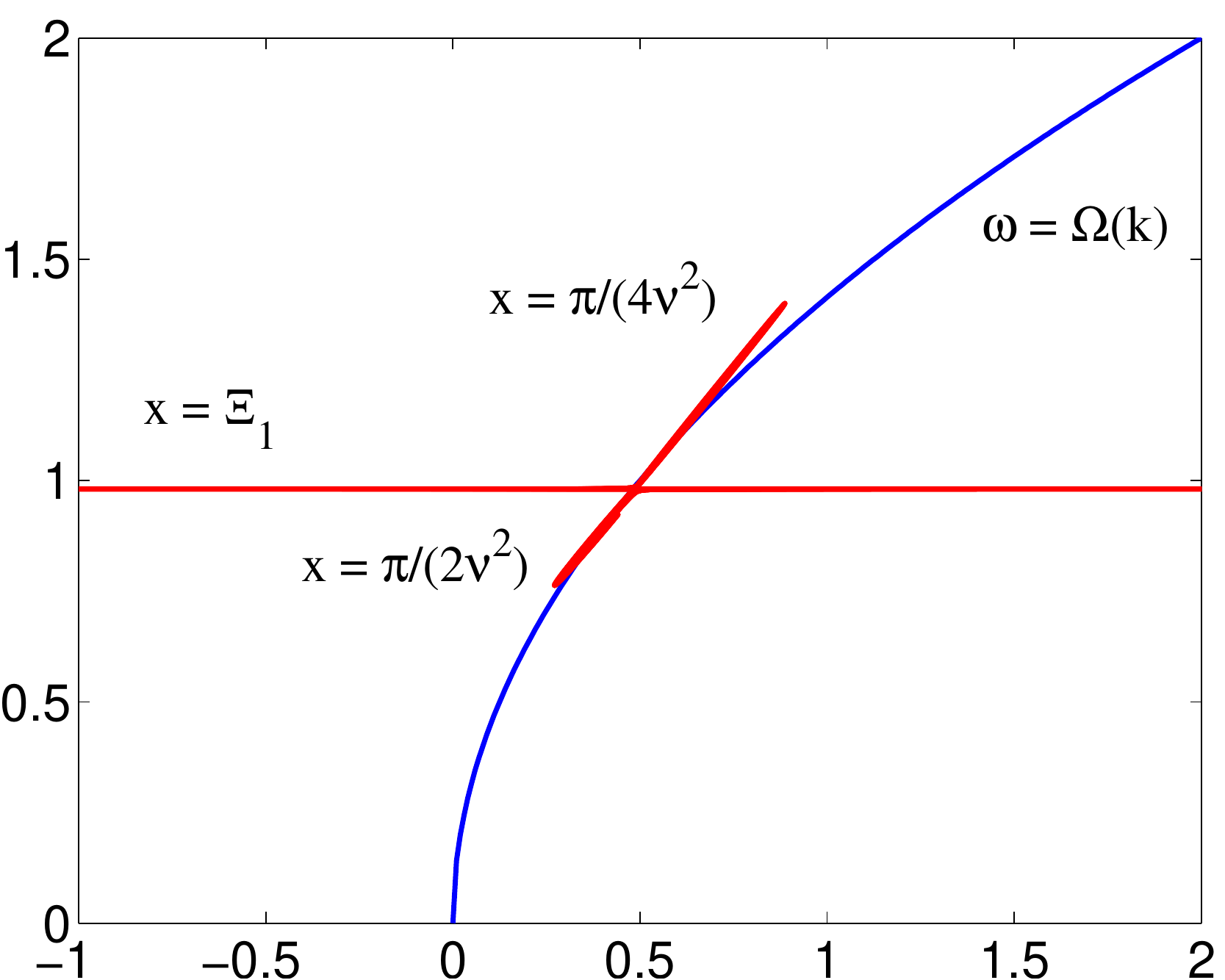}}	    \hspace{0.75cm}
\subfigure[]{\includegraphics[width = 0.45\textwidth]{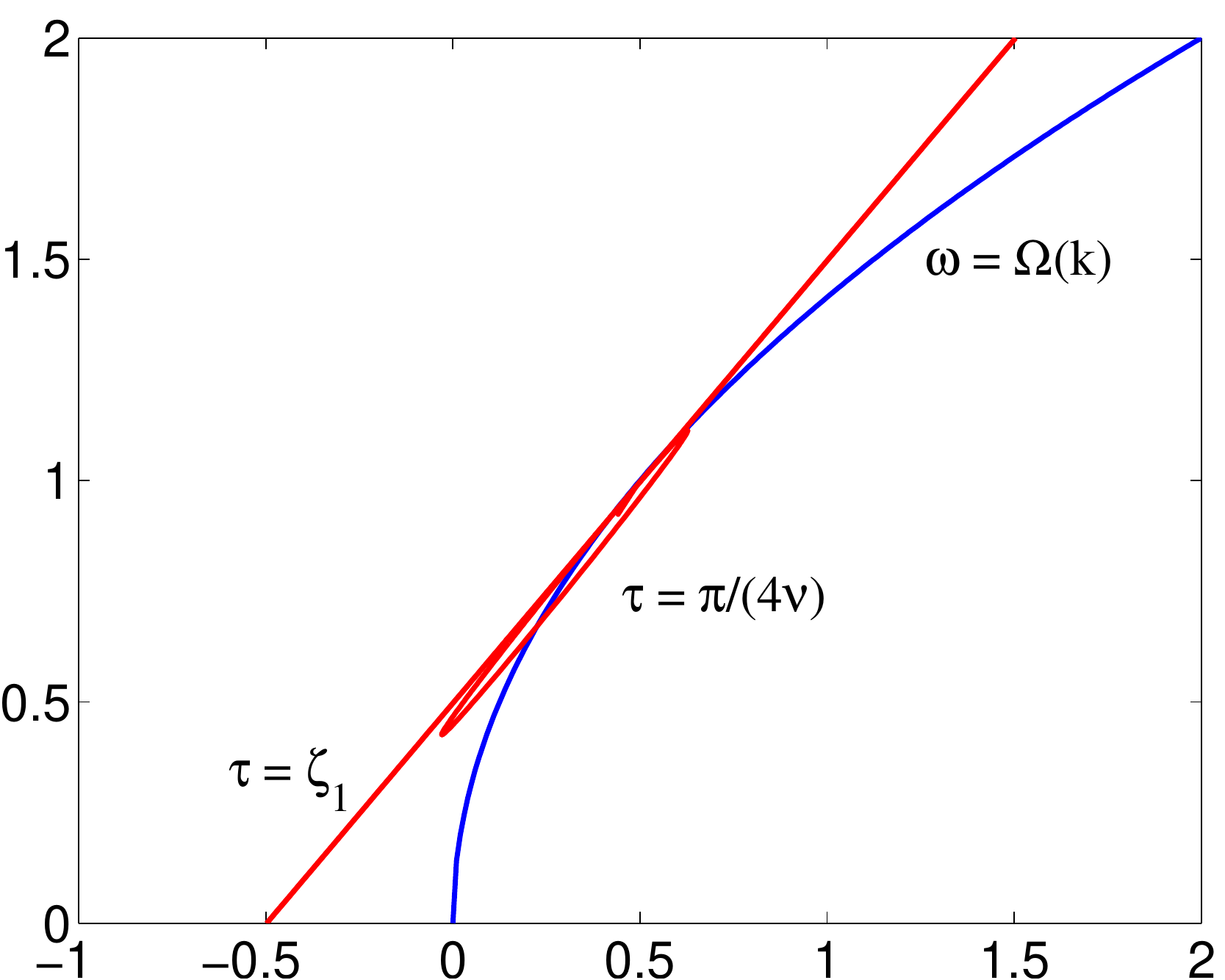}}
\caption[Local wavenumber and local frequency of trichromatic waves]{Plots of the local wavenumber $k$ (horizontal axis) and the local frequency $\omega$ (vertical axis) in the dispersion plane of the considered trichromatic wave.~\index{trichromatic waves} (a) For a fixed position. some trajectories are shown parameterized by the time, showing that the local wavenumber\index{local wavenumber} becomes unbounded at the instance of singularity; similarly, (b) for a fixed time with trajectories parameterized by position, showing the local frequency\index{local frequency} becoming unbounded at the singular position.} \label{Linear_LWLF}
\end{center}
\end{figure}

In general, as said before, at each singular point there will be a phase singularity and wavefront dislocation\index{wavefront dislocation!trichromatic waves}, and an unbounded Chu-Mei quotient.~\index{Chu-Mei quotient!trichromatic waves} We illustrate some aspects for the case studied by~\citet{4Trulsen98}, for which $b_{0} = 2$, $b_{1} = -2$, and $b_{2} = 1$, $\nu = 1/13$ and $\omega_{0} = 1$. The motion of the amplitude in the complex plane, shown in Figure~\ref{linearsolution}(a), makes it clear that there are singular points with phase singularity. The appearance of wavefront dislocation is shown in the density plot in Figure~\ref{linearsolution}(b) and can be investigated in detail by counting the number of waves in one period. The Chu-Mei quotient is unbounded at the singular points.~\index{singular point(s)} This is related to the fact that the local frequency\index{local frequency} and local wavenumber\index{local wavenumber} become unbounded, as shown in Figure~\ref{Linear_LWLF}. \index{trichromatic waves}
\index{trichromatic waves}

\section{Wavefront dislocations in wave groups}
\label{WDwavegroups} \index{wavefront dislocation!in wave groups}
\index{wave group(s)!wavefront dislocation}

In the previous section, we showed that already a superposition of three monochromatic waves can show wavefront dislocation at singular points. In this section, we will consider linear and nonlinear dispersive wave equations\index{dispersive wave!linear equation} \index{dispersive wave!nonlinear equation} and show that a necessary condition for a wave group to have a wavefront dislocation is that the Chu-Mei quotient is unbounded. Moreover, we will also show that the unboundedness of this term is a generic property: if it is bounded at a singular point for an exceptional case, any perturbation of the waves will result in an unbounded Chu-Mei quotient.~\index{Chu-Mei quotient!unboundedness}

\subsection{Linear and nonlinear dispersive wave equations}
\index{dispersive wave!linear equation} \index{dispersive wave!nonlinear equation}

We consider a linear or nonlinear dispersive wave equation. As a model for mainly unidirectional propagation, we can take an evolution equation of the KdV type:
\begin{equation}
\partial_{t}\eta + i \Omega(-i\partial_{x})\eta + \partial_{x} N(\eta) = 0. 		\label{LDWE}
\end{equation}
Here $k \mapsto \Omega(k)$ determines the dispersion relation\index{dispersion relation} and the inverse will be denoted by $K$: $K = \Omega^{-1}$. The weak nonlinearity is given by $N(\eta) = a \eta^{2} + b \eta^{3}$, but is of little relevance for the following discussion about wavefront dislocation as we shall see, so taking a linear equation for which $a = 0 = b$ is possible.

When looking for a wave group\index{wave group(s)} with carrier frequency $\omega_{0}$, the evolution is described with a complex amplitude $A$ and is then given in the lowest-order by
\begin{equation*}
\eta(x,t) = \epsilon A(\xi,\tau)e^{i\theta_{0}} + \textmd{c.c.}
\end{equation*}
where $\theta_{0} = k_{0}x - \omega_{0}t$, with $k_{0} = K(\omega_{0})$ and c.c.~denotes the complex conjugate of the preceding term. The amplitude is described in a time-delayed coordinate system: $\xi = x$ and $\tau = t - x/V_{0}$ where $V_{0} = \Omega'(k_{0}) = 1/K'(\omega_{0})$. This transformation is suitable for studying the evolution in space, for the signaling problem. The resulting equation for $A$ is then the spatial nonlinear Schr\"{o}dinger (NLS) equation\index{NLS equation!type!spatial}, given by
\begin{equation}
\partial_{\xi}A + i \beta \partial_{\tau}^{2} A + i\gamma |A|^{2}A = 0. 	\label{NLSE}
\end{equation}
Here $\beta = - \Omega''(k_{0})/(2[\Omega'(k_{0})]^{3})$ is related to the group velocity dispersion\index{dispersion!group velocity}, while $\gamma$ is a transfer coefficient from the nonlinearity ($\gamma = 0$ for the linear equation).

By writing $A$ in its polar form with the real-valued amplitude $a$ and the real-valued phase $\phi$, $A = a(x,t)e^{i\phi(x,t)}$, and substituting into~\eqref{NLSE}, we obtain the coupled phase-amplitude equations. In the original physical variables, the amplitude equation is known as the `energy\index{energy equation} equation'~\eqref{energyequation} on page~\pageref{energyequation}, and the phase equation can be written as the `nonlinear dispersion relation'~\eqref{nondisrel}\index{dispersion relation!nonlinear} on page~\pageref{nondisrel}. Even for a linear equation, this phase equation contains an additional nonlinear term which results from the fact that the transformation $A \mapsto (a,\phi)$, $A = a e^{i\phi}$ itself is nonlinear.

At vanishing amplitude,~\index{vanishing amplitude} the nonlinear term of the equation vanishes, $\gamma a^{2} = 0$, which shows that the nonlinearity does not play an important role at vanishing amplitude and hence for the phenomena to follow. Only the Chu-Mei quotient plays a significant role in understanding phase singularity and wavefront dislocation phenomena. Unboundedness of the Chu-Mei quotient implies that $K(\omega) - k$ becomes unbounded, and hence that the local wavenumber\index{local wavenumber} or the local frequency\index{local wavenumber} become unbounded.

Now we will show that the Chu-Mei quotient is unbounded if the wavefront dislocation occurs.\index{wavefront dislocation!Chu-Mei quotient} \index{Chu-Mei quotient!unboundedness} The following proposition and its proof have not been mentioned in the literature before.
\begin{proposition}
A necessary condition for a wave field to have a wavefront dislocation is that the Chu-Mei quotient is unbounded at a singular point.		  \label{ChuMei2WD}
\end{proposition}
\begin{proof}
  The proposition means that if the contour integral $d\phi$ is nonzero, then the Chu-Mei quotient is unbounded at the singular points. We will show its contraposition, namely if the Chu-Mei quotient is bounded at singular points,~\index{singular point(s)} then the contour integral vanishes and there is no wavefront dislocation. The Chu-Mei quotient being bounded at a singular point means that either both local wavenumber\index{local wavenumber} and local frequency \index{local wavenumber} are bounded at the singular point or that the local wavenumber or local frequency is unbounded, but $|K(\omega(x,t)) - k(x,t)| < \infty$. For the first case, since both quantities are bounded, the integrand in the contour integral~\eqref{contour_integral1} is bounded, and hence vanishes in the limit for vanishing contour around the point. For the latter case, it means that there exists a positive constant $M$ such that $K(\omega(x,t)) - M \leq k(x,t) \leq K(\omega(x,t)) + M$. Hence, since $K(\omega) \rightarrow \pm \infty$ if and only if $\omega(x,t) \rightarrow \pm \infty$, both wavenumber and frequency have to be unbounded. For evaluating the contour integral~\eqref{contour_integral1}, observe that
  \begin{equation*}
    \oint(K(\omega) \mathop{dx} - \omega \mathop{dt})  - \oint M \mathop{dx} \leq \oint (k \mathop{dx} - \omega \mathop{dt}) \leq \oint(K(\omega) \mathop{dx} - \omega \mathop{dt}) + \oint M \mathop{dx}.
  \end{equation*}
  The contribution $\oint M \mathop{dx}$ vanishes in the limit for shrinking contour, and the same holds for the integral $\oint(K(\omega) \mathop{dx} - \omega \mathop{dt})$ by selecting a limiting contour such as a rectangle for which the length of the sides are chosen appropriately, for instance, $dx = \mathcal{O}(\omega/K(\omega)) \mathop{dt}$. Thus, also in this case, the contour integral~\eqref{contour_integral1} vanishes, and there is no wavefront dislocation.
\end{proof}

\subsection{The Chu-Mei quotient under perturbation} \label{ChuMei}
\index{Chu-Mei quotient!perturbation}

We will now show that the boundedness of the Chu-Mei quotient at a singular point is exceptional: almost any perturbation of the wave field will make the quotient to become unbounded. This is intuitively clear by looking at the trajectory in the Argand diagram\index{Argand diagram}: at a singular point, the trajectory crosses the origin, $a = 0$, and it will be exceptional if it does this with vanishing `acceleration' $\partial_{t}^{2}a = 0$.

The translation of this result to complex-valued functions will give the required statement. Indeed, let $F: \mathbb{R}^{2}\rightarrow \mathbb{C}$, and denote by $F^{\prime}$ and $F^{\prime\prime}$ respectively the first and second derivative with respect to the parameter $t$ or, actually, in any direction. Then defining the amplitude $a$ as $a^{2} = |F|^{2}$, after some manipulations we get
\begin{equation*}
\frac{\partial_{t}^{2}a}{a} = \frac{\textmd{Re} \left(F^{\prime\prime } \cdot F^{\ast}\right)} {\left\vert F\right \vert^{2}} + \frac{\left[\textmd{Im}  \left(F^{\prime} \cdot F^{\ast}\right)\right]^{2}} {\left\vert F \right\vert^{4}},
\end{equation*}
where all quantities at the right-hand side should be evaluated at a singular point for which $a = \left\vert F\right\vert = 0$. Boundedness of this expression is highly exceptional, and a generic perturbation of a function for which it is bounded will lead to unboundedness.

\subsection{SFB wave field} \label{SubsectionSFB}
\index{SFB}

The NLS equation has many interesting special solutions. One family is the so-called SFB that has been studied extensively in the previous chapter. The SFB physical wave field depends on three parameters: the carrier wave frequency $\omega_{0}$, the plane-wave amplitude $r_{0}$ and the normalized modulation frequency $\tilde{\nu}$, where $0 < \tilde{\nu} < \sqrt{2}$. Singular points of the SFB can be found by requiring the real-amplitude to vanish. This implies that both the real and the imaginary parts of the SFB complex amplitude vanish. Alternatively, we can also derive this using an argument of the displaced phase-amplitude representation~\eqref{WFB_special_ansatz}.

\begin{figure}[h]			
\begin{center}
\subfigure[]{\includegraphics[width = 0.45\textwidth]{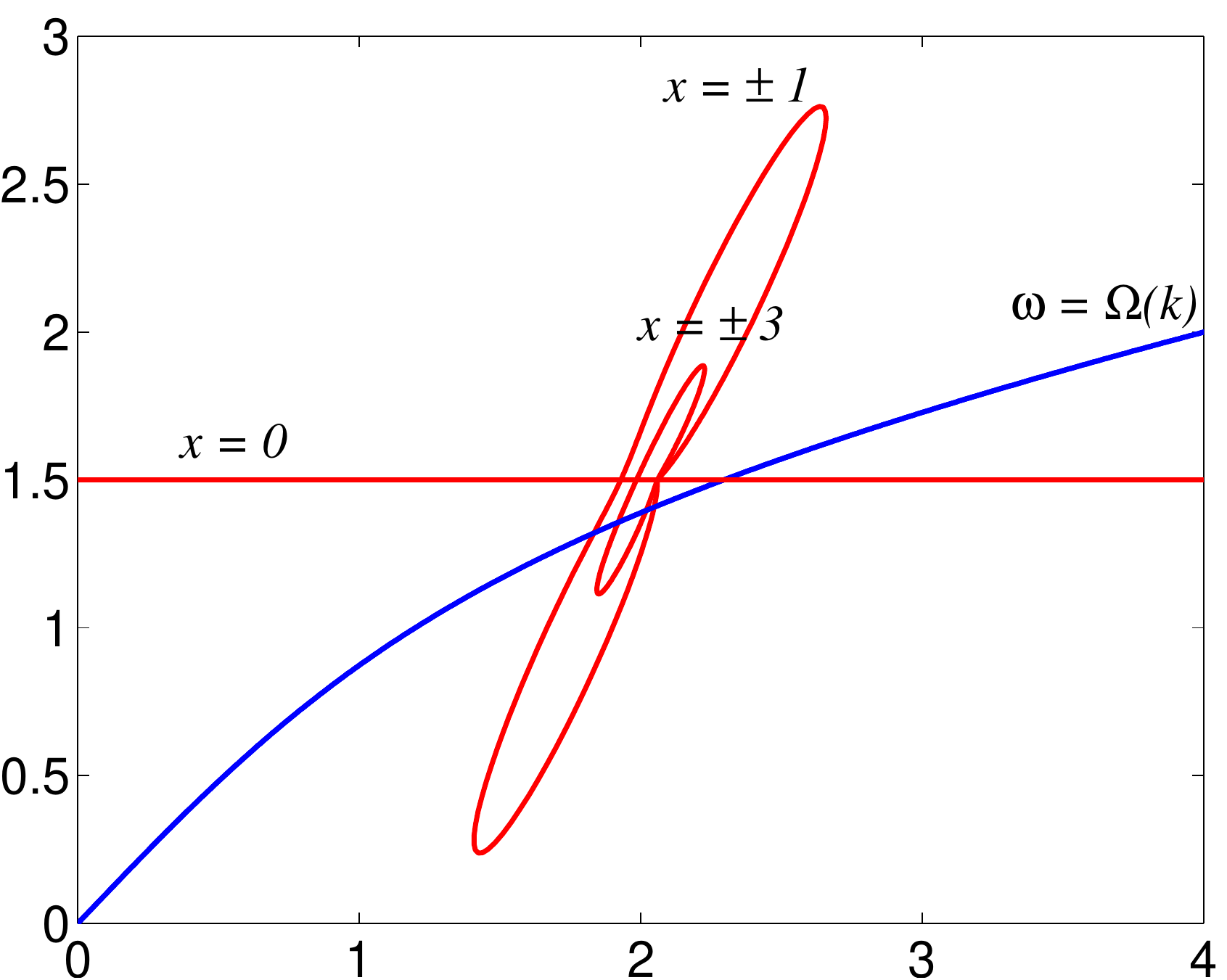}}    		\hspace{0.75cm}
\subfigure[]{\includegraphics[width = 0.45\textwidth]{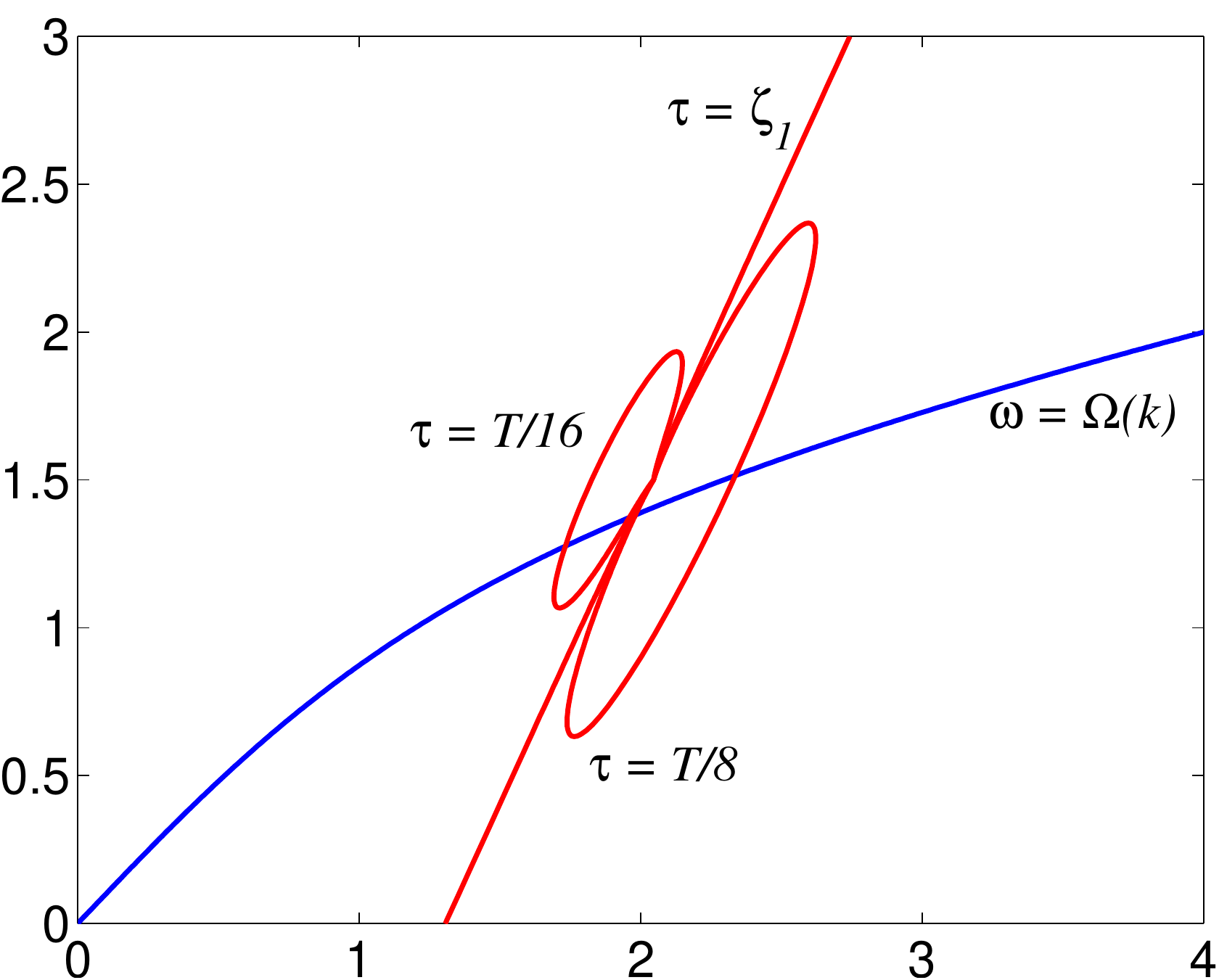}}
\caption[Local wavenumber and local frequency of the SFB]{Plots of the local wavenumber $k$ (horizontal axis) and the local frequency $\omega$ (vertical axis) in the dispersion plane for $\nu = \frac{1}{2}$. At $x = 0$, the local wavenumber becomes unbounded (a), and at $\tau = \zeta_{1}$, the local frequency becomes unbounded (b).}		    \label{SFB_LWLF}
\end{center}
\vspace*{-0.5cm}
\end{figure}

The fact that $\phi$ is independent of $\tau$ means that at each position the trajectory in the Argand diagram\index{Argand diagram!SFB} is on a straight line through the point $-1$ under an angle $\phi$. Hence, only when $\phi = 0$, which means at $\xi = 0$, there can be a singular point. At that position, singular points will occur if\index{singular point(s)!SFB}
\begin{equation}
\cos(\nu \tau) = \frac{2(1 - \tilde{\nu}^{2})}{\sqrt{4 - 2\tilde{\nu}^{2}}}.
\end{equation}
Vanishing amplitude\index{vanishing amplitude!SFB} occurs for $0 < \tilde{\nu} \leq \sqrt{3/2}$ for which there is phase singularity\index{phase singularity!SFB wave}. For $\sqrt{3/2} < \tilde{\nu} < \sqrt{2}$, the amplitude remains positive definite and there is no phase singularity. Such phase singularity occurs at $\xi = 0$ for two instants in each temporal period. At the phase singularities, the local wavenumber\index{local wavenumber} and local frequency\index{local wavenumber} become unbounded, as shown in Figure~\ref{SFB_LWLF}; this confirms the fact that the Chu-Mei quotient is unbounded at the singular points.

There are wavefront dislocations at the singular points.~\index{wavefront dislocation!SFB} Let $(0,\zeta_{1})$ and $(0,\zeta_{2})$ be singular points of the SFB in one modulation period. We use the contour integral~\eqref{contour_integral2} to calculate the strength of the singular points. The contour integral around the first singular point is given by
\begin{equation}
I(0,\zeta_{1}) =  \lim_{\epsilon \rightarrow 0} \left(\phi(2\pi) - \lim_{\theta \rightarrow \pi/2^{+}} \phi(\theta) +  \lim_{\theta \rightarrow \pi/2^{-}} \phi(\theta) - \phi(0) \right) = -2\pi.
\end{equation}
Similarly, the contour integral around the second singular point is found to be $I(0,\zeta_{2}) = 2\pi$. These calculations show that during each modulation period, the splitting of waves occurs at $(0,\zeta_{1})$ and the merging of waves occurs at $(0,\zeta_{2})$ for increasing space.

\begin{figure}[h!]			
\begin{center}
\subfigure[]{\includegraphics[width = 0.5\textwidth]{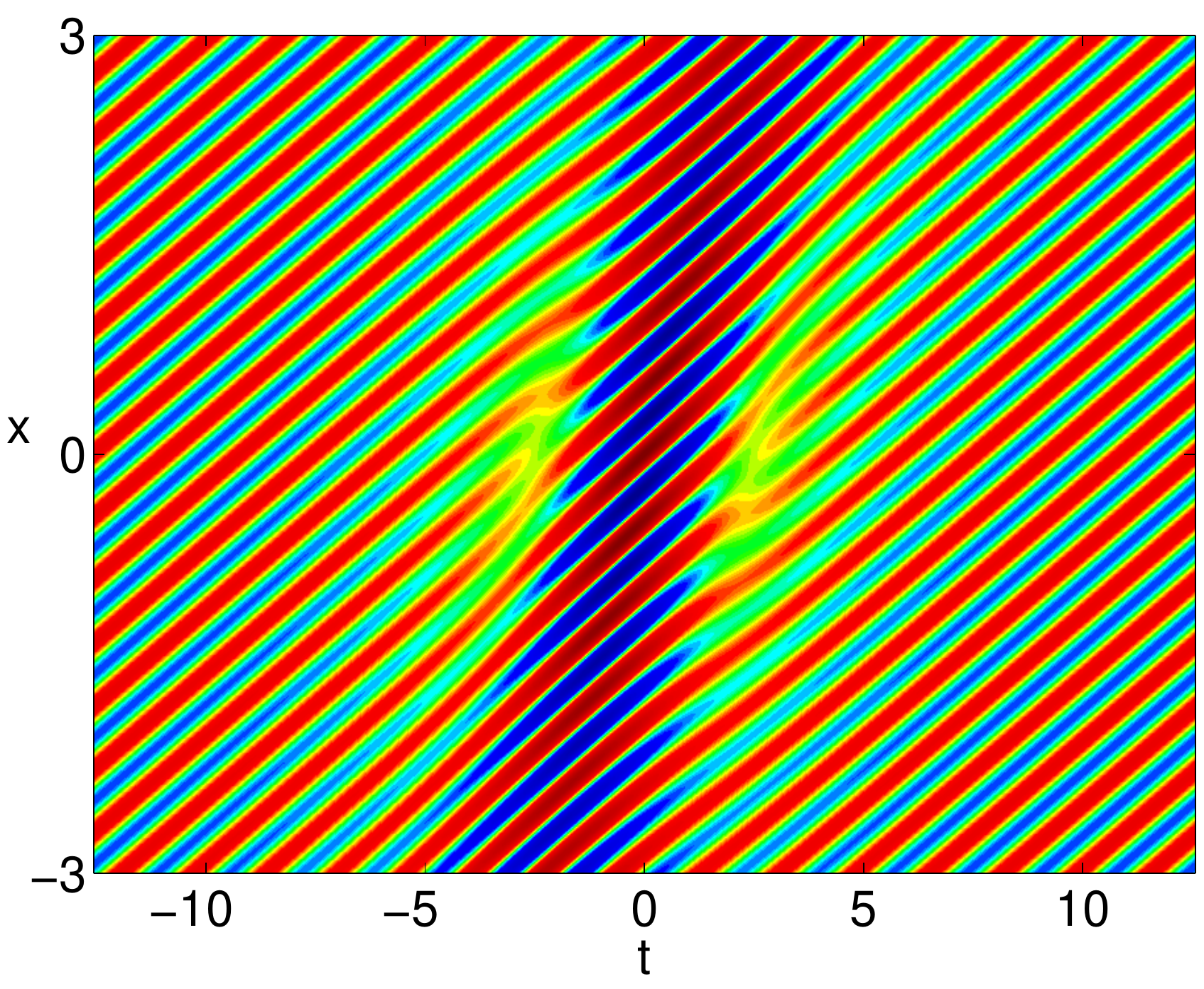}}    	\hspace{0.75cm}
\subfigure[]{\includegraphics[width = 0.4\textwidth]{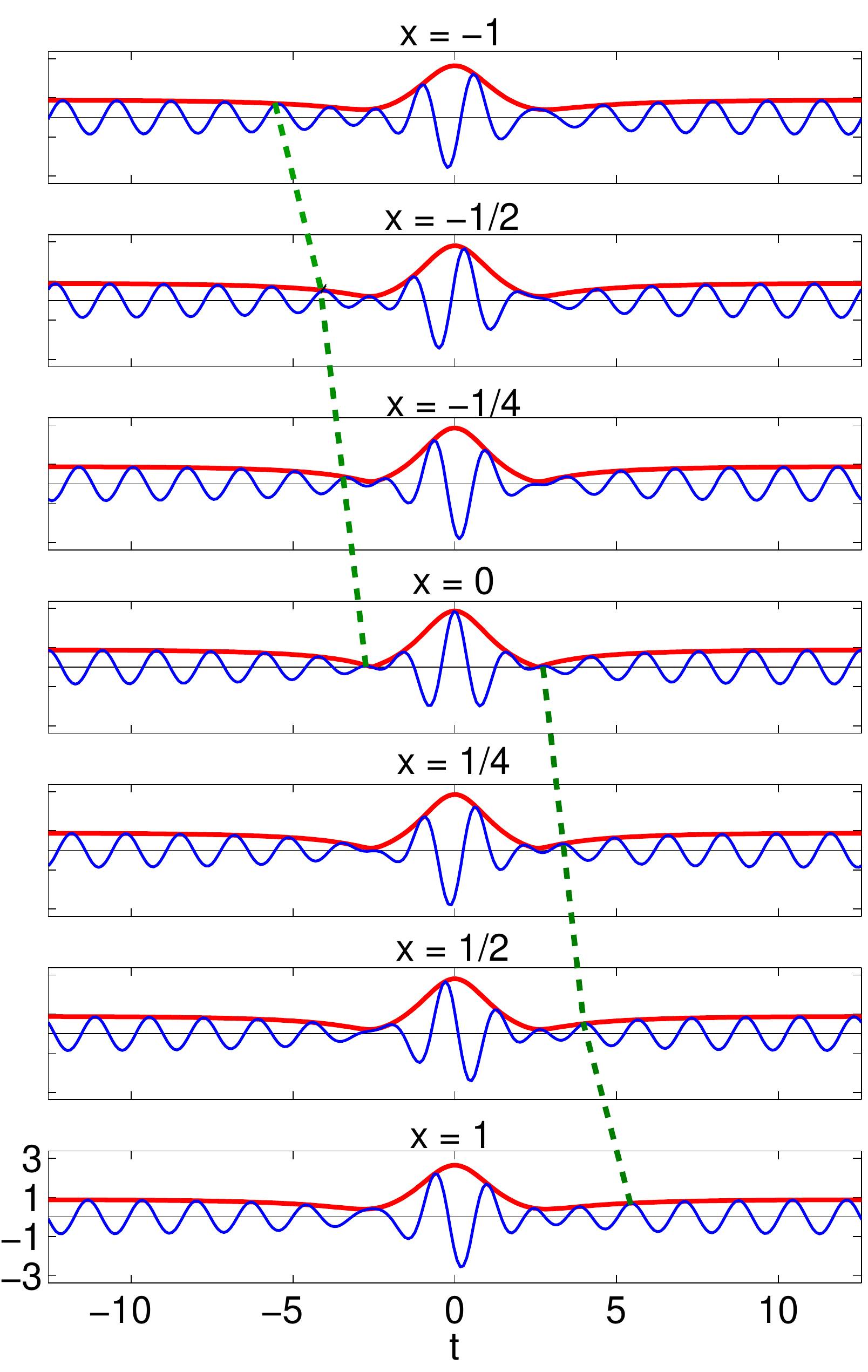}}
\caption[Wavefront dislocations in the SFB]{Density plot of the SFB wave field with wavefront dislocations (a), and the corresponding wave signals for different positions in a moving frame of reference (b) for $\nu = \frac{1}{2}$. The number of waves decreases from 8 to 7 for the half period $t \in \left[-\frac{1}{2}T, 0 \right]$ and it increases from 7 to 8 for the next half period $t \in \left[0, \frac{1}{2}T \right]$.}  		\label{SFB}
\end{center}
\end{figure}

Figure~\ref{SFB} shows a density plot of the SFB wave field around two phase singularities. We observe the splitting and merging waves in pairs. In plots of the time signal at different positions, we see the splitting and merging in more detail. In this example, for a half modulation period $t \in \left[-\frac{1}{2}T, 0 \right]$, the number of waves decreases from 8 to 7, indicating that waves are merging when passing the singularity. At another half modulation period $t \in \left[0, \frac{1}{2}T \right]$, it increases from 7 to 8, which indicates that waves are splitting when passing the singularity. However, the number of waves in one modulation period for $x \rightarrow \pm \infty$ remains the same before and after undergoing the singularity, namely $\omega_{0}/\nu$.

Observation and investigation of wavefront dislocation in modulated surface water waves\index{wavefront dislocation!modulated water waves} have been done by~\citet{4Tanaka95}. His investigation is based on the modulated gravity waves corresponding to Benjamin-Feir instability and is done numerically. By taking an analogy to our signaling problem, the corresponding envelope function experiences vanishing amplitude at two different positions. He observed that between these two vanishing amplitudes, the wave crests `disappear', as is confirmed by the decrease in the number of waves.~\index{waves disappearance} \index{waves disappearance|see{wavefront dislocation}}

\section{Conclusion}

We discussed the phenomena of phase singularity and wavefront dislocation that can happen at singular points of a wave field where the amplitude vanishes. We investigated for wave fields in one spatial dimension the appearance of these essentially linear phenomena. We used simple examples of trichromatic waves to see the relationship between these concepts. We introduced the Chu-Mei quotient\index{Chu-Mei quotient} as it is known to appear in the `nonlinear dispersion relation'\index{dispersion relation!nonlinear} for wave groups as a consequence of the nonlinear transformation of the complex amplitude to real phase-amplitude variables. We also linked the unboundedness of this quotient to the unboundedness of the local wavenumber and frequency at singular points. This unboundedness is a generic property and is necessary for the occurrence of phase singularity\index{phase singularity} and wavefront dislocation. Viewing the phenomena from this angle is novel and never been considered in the literature. It is important to stress again that the phenomena are essentially linear since nonlinear terms in the equation are of higher-order at a singular point. We showed that for an interesting class of solutions of the NLS equation, the Solitons on Finite Background, wavefront dislocations occur there too.

\newpage
{\renewcommand{\baselinestretch}{1} 
}
\setcounter{chapter}{4}
\chapter{Higher-order waves on the finite background} \label{5HighOrder}
\index{waves on finite background!higher-order}

\section{Introduction}

This chapter deals with one of the special solutions of the NLS equation, but a different class than has been studied in Chapter~\ref{3Property}. We will study a family of higher-order solutions of the NLS equation that also describes modulational instability\index{modulational instability} and this study can also be seen as being motivated by the problem of extreme wave generation in the wave basin of a hydrodynamic laboratory. To distinguish from the previous Soliton on Finite Background (SFB)\index{SFB}, we give an index to the special classes of solutions that are based on the number of pairs of initial sidebands.\index{sideband(s)} In the context of this chapter, the SFB that we have discussed extensively in Chapter~\ref{3Property} is denoted as SFB$_{1}$\index{SFB!SFB$_{1}$} and its modulation frequency as $\nu_{1}$. The higher-order SFBs are denoted as SFB$_{2}$, SFB$_{3}$, and so on. In this chapter, we only concentrate on SFB$_{2}$.\index{SFB!SFB$_{2}$}

In the spectral domain, the initial Benjamin-Feir spectrum is represented by one central frequency and one pair of sidebands\index{sideband(s)} inside the instability interval. However, if there is more than one pair of sidebands as the initial condition of modulation, and all these are inside the instability interval, then, due to the four-wave mixing process, all will be amplified. However, since this process is a nonlinear superposition of the elementary processes of the Benjamin-Feir instability\index{Benjamin-Feir instability} with one pair of initial sidebands, the total effect can no longer be described by the SFB$_{1}$. Higher-order solutions of the NLS equation can then be constructed using a Darboux transformation. \index{Darboux transformation} Readers who are interested in the topic of Darboux transformation may consult~\citep{5Matveev91, 5Rogers02}.

Similar to SFB$_{1}$, SFB$_{2}$ also describes waves on the finite background\index{waves on finite background}, and these solutions were found by~ \citet{5Akhmediev85}. SFB$_{2}$ in this thesis refers to the dynamic evolution in which the modulation frequencies are $\nu_{2}$ and $2\nu_{2}$, where $\nu_{2} = \frac{1}{2}\nu_{1}$. Note that this choice of modulation frequencies is a special case, which is chosen since then the explicit expressions are less complicated than an arbitrary combination of modulation frequencies. For a special choice of parameters, solutions from SFB$_{2}$ can have an amplitude amplification up to a factor of 5. Therefore, by providing sufficient space for waves to propagate downstream, \index{downstream} extreme wave events with an even larger amplitude than waves from SFB$_{1}$ can be generated. Another similarity with SFB$_{1}$ is that SFB$_{2}$ solutions also show wavefront dislocation and phase singularity.

The aim of this chapter is to present some aspects of the SFB$_{2}$\index{SFB!SFB$_{2}$} which have not been discussed in the literature; the chapter is organized as follows. After this introduction, Section~\ref{specification} discusses the specifications and properties of SFB$_{2}$, including the explicit expression, asymptotic behavior, physical wave field, and maximum temporal amplitude. Section~\ref{spatialevol2} explains the spatial evolution of the SFB$_{2}$ signal. We will also see phase singularity and give phase plane representation of the solutions. Section~\ref{SectionASE2} presents the SFB$_{2}$ spectrum and its evolution. Finally, Section~\ref{SR2} gives conclusions and remarks about this chapter.

\section{Specifications and properties of SFB$_{2}$}		\label{specification}

\subsection{An explicit expression}
\index{SFB!SFB$_{2}$!explicit expression}

We have seen in Subsection~\ref{pseudocoherent} on page~\pageref{pseudocoherent} that SFB$_{1}$ can be written in a displaced phase-amplitude expression as given by $A(\xi,\tau) = A_{0}(\xi)\,F(\xi,\tau)$, where $F(\xi,\tau)$ is given by~\eqref{solutionF}. SFB$_{2}$ cannot be expressed in a similar way with these displaced phase-amplitude variables, since the evolution in the Argand diagram is not a set of straight lines, as we will see in Subsection~\ref{SFB2Argand}.

An explicit expression for SFB$_{2}$ can be found in~\citep{5Akhmediev97}. Another expression can also be found in~\citep{5Ablowitz90, 5Calini02}. The authors derived it using Hirota's method\index{Hirota's method} and associated it with the dark-hole soliton\index{dark soliton} solutions of the defocusing NLS equation. Explicitly, both SFB$_{1}$ and SFB$_{2}$ can be given by the following expression:
\begin{equation}
A_{j}(\xi,\tau;\tilde{\nu}_{j}) = A_{0}(\xi) \left(\frac{P_{j}(\xi,\tau) + i Q_{j}(\xi,\tau)}{H_{j}(\xi,\tau)} - 1 \right), \qquad j = 1,2,		  \label{SFB12}
\end{equation}
where the indices $j = 1,2$ corresponds to SFB$_{1}$ and SFB$_{2}$, respectively. For both SFBs, $A_{0}(\xi)$ denotes the plane-wave\index{plane-wave} solution of the NLS equation, as given by~\eqref{planewave} on page~\pageref{planewave}. Expressions for $P_{j}$, $Q_{j}$, and $H_{j}$ are given as follows:
\begin{align}
P_{1}(\xi,\tau) &= \tilde{\nu}_{1}^{2}\cosh [\sigma (\xi - \xi_{0})]\\
Q_{1}(\xi,\tau) &= - \tilde{\sigma}   \sinh [\sigma (\xi - \xi_{0})]\\
H_{1}(\xi,\tau) &= \cosh [\sigma (\xi - \xi_{0})] - \sqrt{1 - \frac{1}{2} \tilde{\nu}_{1}^{2}} \cos[\nu_{1} (\tau - \tau_{0})]
\end{align}
{\small
\begin{align}
  P_{2}(\xi,\tau) = \frac{3}{2\tilde{\nu}_{2}\sqrt{2}} \left[\cos \nu_{2}(\tau + \tau_{01} - 2\tau_{02}) +
                    \frac{1}{9}\cos \nu_{2}(3\tau - 2\tau_{02} - \tau_{01}) \right] \nonumber \\
               + \, \frac{1}{2 \tilde{\sigma}_{1}}(2 - \tilde{\nu_{2}}^{2}) \cosh \sigma_{1}(\xi - \xi_{01}) \cos 2\nu_{2}(\tau - \tau_{02}) \nonumber \\
               + \, \frac{2}{\tilde{\sigma}_{2}}  (1 - 2\tilde{\nu}_{2}^{2}) \cosh \sigma_{2}(\xi - \xi_{02}) \cos \nu_{2}(\tau - \tau_{01}) \nonumber \\
               - \, \frac{4}{3\tilde{\nu}_{2}\sqrt{2}}\left[\frac{\tilde{\nu}_{2}^{2}(17\tilde{\nu}_{2}^{2} - 10)}
                    {2 \tilde{\sigma}_{1}\,\tilde{\sigma}_{2}} \cosh \sigma_{1}(\xi - \xi_{01})\,\cosh \sigma_{2}(\xi - \xi_{02}) \right. \nonumber \\
        \left. + \, \sinh \sigma_{1}(\xi - \xi_{01})\,\sinh \sigma_{2}(\xi - \xi_{02}) \right] \\
  Q_{2}(\xi,\tau) = \frac{1}{2} \sinh \sigma_{1}(\xi - \xi_{01})\,\cos 2\nu_{2}(\tau - \tau_{02}) \nonumber \\
               + \, \sinh \sigma_{2}(\xi - \xi_{02})\,\cos \nu_{2}(\tau - \tau_{01}) \nonumber \\
               + \, \frac{\tilde{\nu}_{2}\sqrt{2}}{\tilde{\sigma}_{1}\,\tilde{\sigma}_{2}}
                    \left[\tilde{\sigma}_{2}\cosh \sigma_{1}(\xi - \xi_{01})\,\sinh \sigma_{2}(\xi - \xi_{02}) \right. \nonumber \\
        \left. - \, \tilde{\sigma}_{1}\sinh \sigma_{1}(\xi - \xi_{01})\,\cosh \sigma_{2}(\xi - \xi_{02}) \right]\\
  H_{2}(\xi,\tau) = \frac{3}{4\tilde{\nu}_{2}\sqrt{2}}\left[\cos \nu_{2}(\tau + \tau_{01} - 2\tau_{02}) +
                    \frac{1}{9}\cos \nu_{2}(3\tau - 2\tau_{02} - \tau_{01}) \right] \nonumber \\
               + \, \frac{1}{2 \tilde{\sigma}_{1}} \cosh \sigma_{1}(\xi - \xi_{01})\,\cos 2\nu_{2}(\tau - \tau_{02})
               + \, \frac{1}{\tilde{\sigma}_{2}} \cosh \sigma_{2}(\xi - \xi_{02})\,\cos \nu_{2}(\tau - \tau_{01}) \nonumber \\
               - \, \frac{2}{3\tilde{\nu}_{2}\sqrt{2}}\left[\frac{\tilde{\nu}_{2}^{2}(4\tilde{\nu}_{2}^{2} - 5)}
                    {\tilde{\sigma}_{1}\, \tilde{\sigma}_{2}} \cosh \sigma_{1}(\xi - \xi_{01})\,\cosh \sigma_{2}(\xi - \xi_{02}) \right. \nonumber \\
        \left. + \, \sinh \sigma_{1}(\xi - \xi_{01})\,\sinh \sigma_{2}(\xi - \xi_{02}) \right].%
\end{align}
}
To guarantee the existence of the SFBs, the modulation frequencies $\nu_{j}$, $j = 1,2$ have to be in the instability interval, $0 < \tilde{\nu}_{1} < \sqrt{2}$ and $0 < \tilde{\nu}_{2} < \sqrt{1/2}$, where $\tilde{\nu}_{j} = \nu_{j}/\left(r_{0} \sqrt{\gamma/\beta}\right)$. For SFB$_{1}$, $\sigma$ is the growth rate\index{growth rate(s)} that corresponds to the Benjamin-Feir instability\index{Benjamin-Feir instability}, given as $\sigma = \gamma r_{0}^{2} \tilde{\sigma}$, where $\tilde{\sigma} = \tilde{\nu}_{1} \sqrt{2 - \tilde{\nu}_{1}^{2}}$. For simplicity, we take $(\xi_{0},\tau_{0}) = (0,0)$, so that SFB$_{1}$ reaches its maximum at $(\xi,\tau) = \left(0, 2n\pi/\nu_{1} \right)$, $n \in \mathbb{Z}$.

For SFB$_{2}$\index{SFB!SFB$_{2}$}, $\sigma_{1}$ and $\sigma_{2}$ \label{gr} are two growth rates\index{growth rate(s)} in the modulational instability\index{modulational instability} process that correspond to the first and second sideband, respectively. They are given as $\sigma_{1} = \sigma(\tilde{\nu}_{2}) = \gamma r_{0}^{2} \tilde{\sigma}_{1}$, where $\tilde{\sigma}_{1} = \tilde{\nu}_{2} \sqrt{2 - \tilde{\nu}_{2}^{2}}$ and $\sigma_{2} = \sigma(2 \tilde{\nu}_{2}) = \gamma r_{0}^{2} \tilde{\sigma}_{2}$, where $\tilde{\sigma}_{2} = 2 \tilde{\nu}_{2} \sqrt{2 - 4 \tilde{\nu}_{2}^{2}}$. In the following subsection, we will see that the asymptotic behavior of SFB$_{2}$\index{asymptotic behavior!SFB$_{2}$} depends on these growth rates. The parameters $\xi_{01}$, $\xi_{02}$, $\tau_{01}$ and $\tau_{02}$ are arbitrary real and determine the position and time lag of SFB$_{2}$. Let us now define $\Delta \xi := |\xi_{01} - \xi_{02}|$ and $\Delta \tau := |\tau_{01} - \tau_{02}|$ as parameters related to space and time variables, respectively. For definiteness, we will consider SFB$_{2}$ with maxima at $(\xi,\tau) = \left(0, 2n\pi/\nu_{2} \right)$, $n \in \mathbb{Z}$. This is obtained by choosing $\Delta \xi = 0 = \Delta \tau$ or by taking $\xi_{01} = 0 = \xi_{02}$ and $\tau_{01} = \pi/\nu_{2} = \tau_{02}$. With this choice of parameters, SFB$_{2}$ reduces to SFB$_{1}$ for $\tilde{\nu}_{2} \rightarrow \sqrt{1/2}$:
\begin{equation}
\lim_{\tilde{\nu}_{2}\rightarrow \sqrt{1/2}} A_{2}(\xi,\tau;\tilde{\nu}_{2}) = A_{1}\left(\xi,\tau;\tilde{\nu}_{1} = \sqrt{1/2} \right).
\end{equation}
\index{SFB!SFB$_{2}$!relation to SFB$_{1}$}
\begin{figure}[h]			
\begin{center}
\includegraphics[width = 0.45\textwidth]{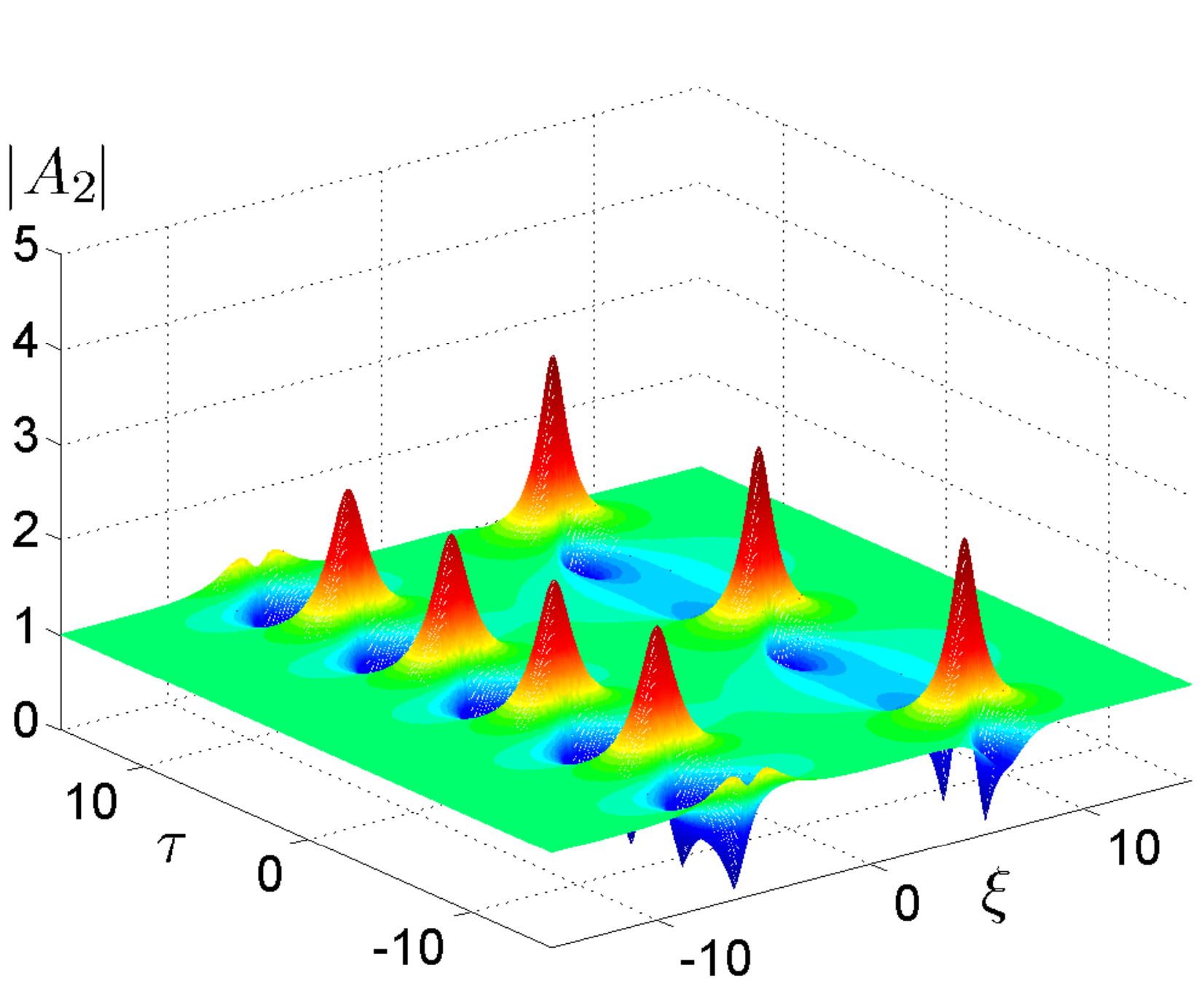}    \hspace*{0.75cm}
\includegraphics[width = 0.45\textwidth]{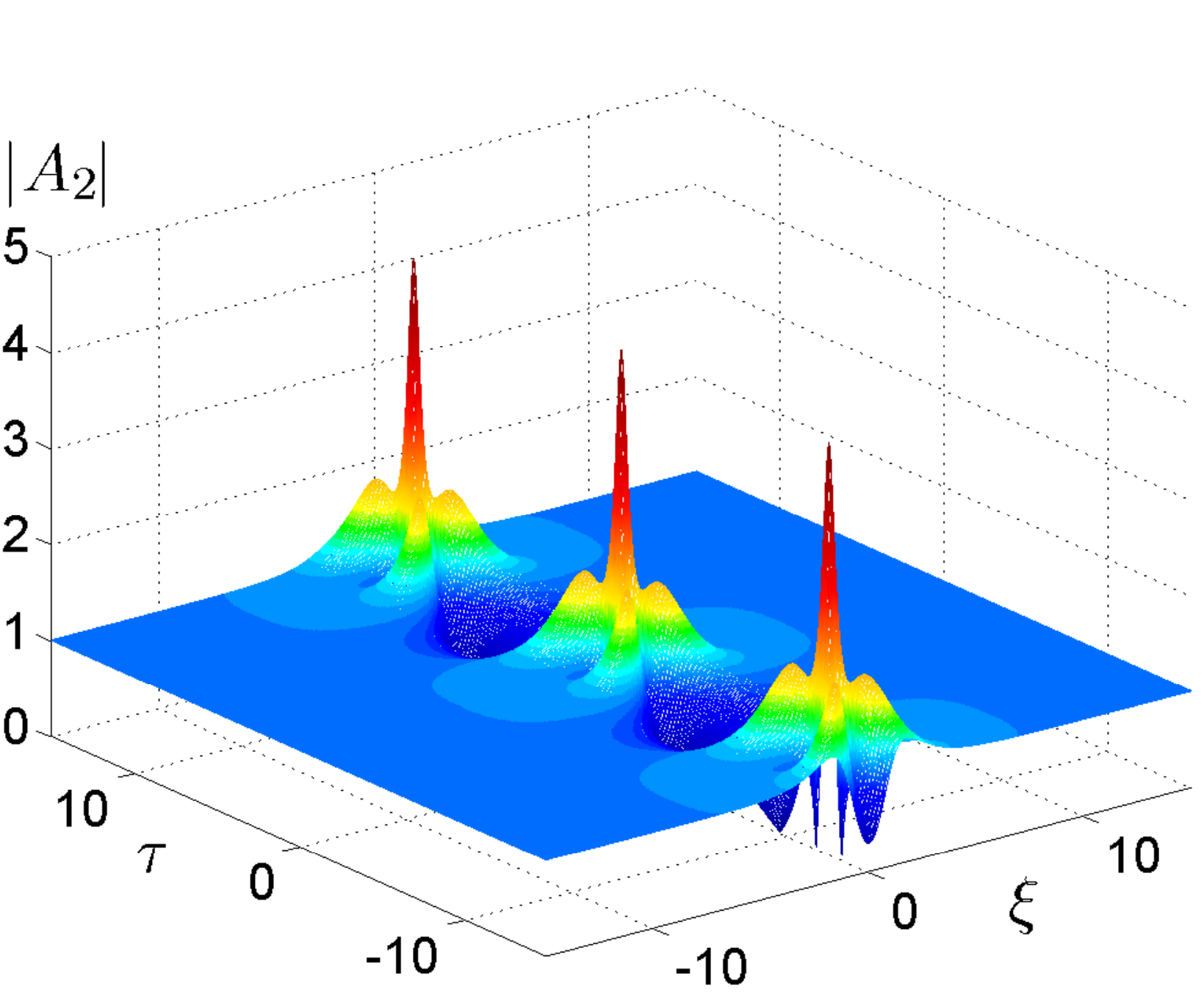}
\caption[Absolute value of SFB$_{2}$]{Three-dimensional plots of the absolute value of SFB$_{2}$ for $\tilde{\nu}_{2} = \frac{1}{2}$, $\Delta \tau = 0$, $\Delta \xi = 10$ (left), and $\Delta \xi = 0$ (right), which show an interaction of two SFB$_{1}$s. For illustration purposes, the axes are scaled corresponding to $\beta = 1 = \gamma$ in the NLS equation.} 		\label{SFBinteract}
\end{center}
\end{figure}
\begin{figure}[h!]			
\begin{center}
\includegraphics[width = 0.54\textwidth]{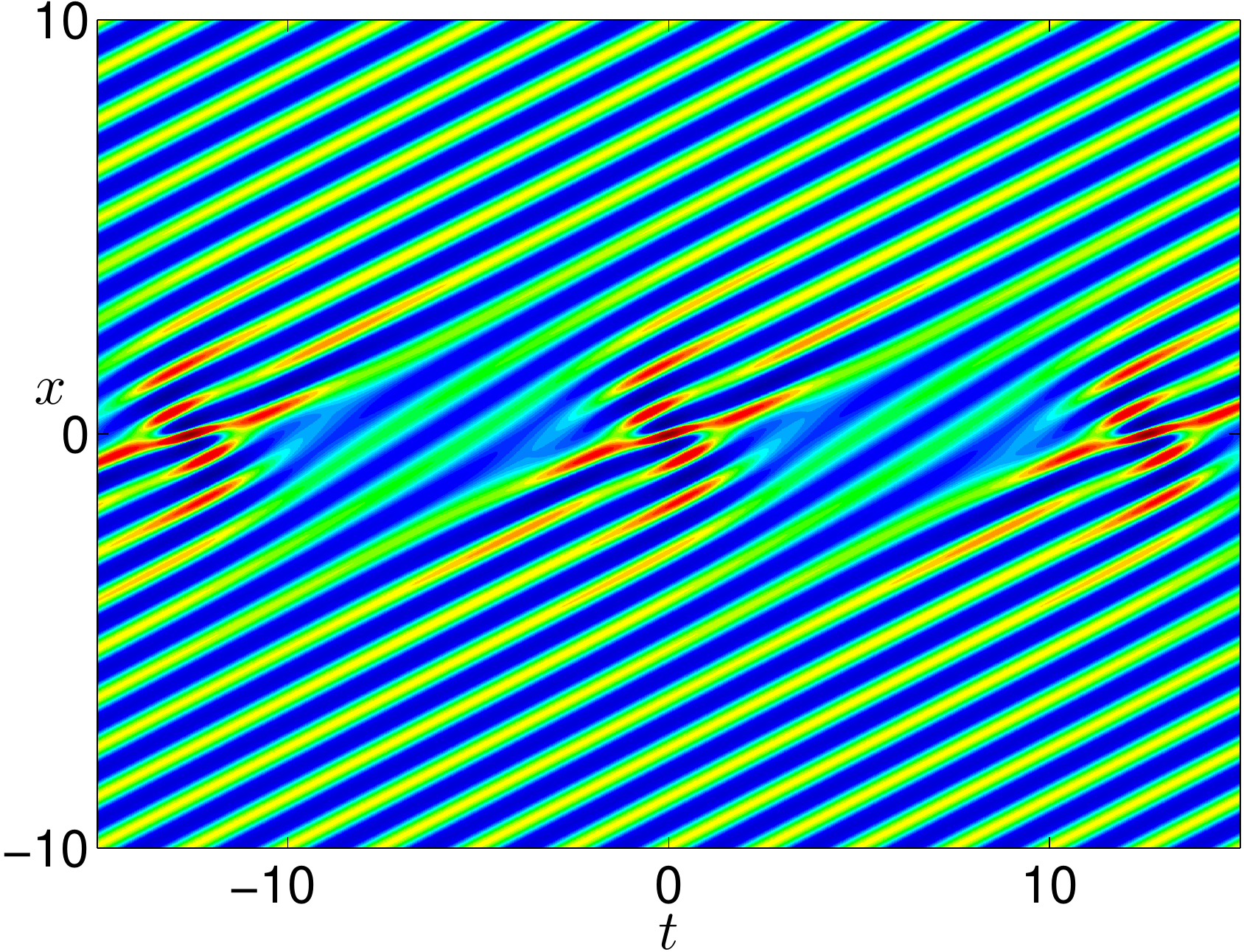}
\caption[Physical wave field of the SFB$_{2}$]{A density plot of SFB$_{2}$ physical wave field for $\tilde{\nu}_{2} = \frac{1}{2}$.} \label{SFB2physical}
\end{center}
\end{figure}

\vspace*{-0.75cm} The plots of the absolute value of SFB$_{2}$ for different values of parameters show that there are interactions between two SFB$_{1}$s with different modulation period. Figure~\ref{SFBinteract} shows the plots of SFB$_{2}$ for both $\Delta \xi \neq 0$ and $\Delta \xi = 0$. For nonzero $\Delta \xi$, the plot shows two SFB$_{1}$s which are separated by a peak distance of approximately $\Delta \xi$. They have modulation frequencies of $\frac{1}{2}\nu_{1}$ and $\nu_{1}$ and modulation period of $4\pi/\nu_{1}$ and $2\pi/\nu_{1}$, respectively. By allowing $\Delta \xi$ to decrease to zero, an interaction occurs between these two SFB$_{1}$s and a change in the modulation period takes place. For $\Delta \xi = 0$, we see the result of the SFB$_{1}$ interaction, which has a higher maximum amplitude than the individual SFB$_{1}$s. The modulation period of this SFB now becomes $4\pi/\nu_{1} = 2\pi/\nu_{2}$.

\subsection{The asymptotic behavior} \label{SFB2asbe}
\index{SFB!SFB$_{2}$!asymptotic behavior} \index{asymptotic behavior!SFB$_{2}$}

Similar to SFB$_{1}$, the asymptotic behavior of SFB$_{2}$ at the far distance $\xi \rightarrow \pm \infty$ is given by the plane-wave\index{plane-wave} solution $A_{0}(\xi)$ of the NLS equation with a phase difference of $2 \phi_{20}$ from $\xi = -\infty$ to $\xi = \infty$, where $\phi_{20}$ is given in Appendix~\ref{ApendiksSFB2} on page~\pageref{ApendiksSFB2}. However, we can apply the same procedure as in Subsection~\ref{asymptoticbehavior} on page~\pageref{asymptoticbehavior} to find the asymptotic behavior that includes the lowest order term and the linear terms. Since SFB$_{2}$ has two growth  rates\index{growth rate(s)}, the asymptotic behavior will then depend on the value of its modulation frequency $\tilde{\nu}_{2}$. Different values of $\tilde{\nu}_{2}$ give different growth rate ratio\index{growth rate(s)!ratio}. Here we present some selected asymptotic expressions of SFB$_{2}$ based on different growth rate ratio \index{growth rate(s)!ratio} for $\xi \rightarrow \mp \infty$:\\ 
$\bullet$ $\sigma_{2}/\sigma_{1} \ll 1$: 
{\small
\begin{equation}
A_{2}(\xi,\tau) \approx A_{0}(\xi) \left[e^{i\phi_{20}} + (e^{\sigma_{1} \xi})^{2}\, w_{23} + e^{\sigma_{1} \xi} \, \tilde{w}_{21} \, \cos(\nu_{2} \tau) + (e^{\sigma_{1} \xi})^{2} \, \tilde{w}_{22} \,\cos(2\nu_{2} \tau) \right]
\end{equation}
} 
$\bullet$ $\sigma_{2}/\sigma_{1} \approx 1$: 
{\small
\begin{equation}
A_{2}(\xi,\tau) \approx A_{0}(\xi) \left[ e^{i\phi_{20}} + e^{\sigma_{1} \xi}\,w_{21}\,\cos(\nu_{2} \tau) + e^{\sigma_{2} \xi} \,w_{22}\,\cos(2 \nu_{2} \tau) \right]
\end{equation}
}
$\bullet$ $\sigma_{2}/\sigma_{1} \approx 2$: 
{\small
\begin{equation}
A_{2}(\xi,\tau) \approx A(\xi) \left[ e^{i\phi_{20}} + e^{\sigma_{2} \xi} \, w_{23} + e^{\sigma_{1} \xi} \, w_{21}\,\cos(\nu_{2} \tau) + e^{\sigma_{2} \xi} \bar{w}_{22} \, \cos(2\nu_{2} \tau) \right].
\end{equation}
}
The quantities $w_{2l}$, $l = 1, 2, 3$, $\bar{w}_{22}$, $\tilde{w}_{22} \in \mathbb{C}$ and are given in Appendix~~\ref{ApendiksSFB2} on page~\pageref{ApendiksSFB2}. From the above expressions, we can see that each sideband carries different phase information due to different growth rate ratio.

\subsection{Physical wave field}
\index{SFB!SFB$_{2}$!physical wave field} \index{physical wave field!SFB$_{2}$}

We also use the expression~\eqref{physicaleta} on page~\pageref{physicaleta} to give an illustration of the physical wave field corresponding to SFB$_{2}$. Again, in this case, we consider only the lowest order contribution of the physical wave field. Figure~\ref{SFB2physical} shows a density plot of SFB$_{2}$  physical wave field for scaled parameters $\beta = 1 = \gamma = r_{0}$. We observe a similar phenomenon of wavefront dislocation in SFB$_{2}$ as we have in SFB$_{1}$, but now with a varied pattern. In SFB$_{1}$, wavefront dislocation occurs at $x = 0$ and periodically in time but only for $0 < \tilde{\nu}_{1} \leq \sqrt{3/2}$. In SFB$_{2}$, wavefront dislocation also occurs at $x = 0$ and is periodic in time but now for all $0 < \tilde{\nu}_{2} < \sqrt{1/2}$. Depending on the value of $\nu_{2}$, there can be either one or two pairs of wavefront dislocations during one modulation period. For $0 < \tilde{\nu}_{2} < \tilde{\nu}_{2}^{\ast}$, SFB$_{2}$ has two pairs of wavefront dislocations and for $\tilde{\nu}_{2}^{\ast} < \tilde{\nu}_{2} < \sqrt{1/2}$, it has only one pair during one modulation period, where $\tilde{\nu}_{2}^{\ast} = \frac{1}{6}\sqrt{8\sqrt{10} - 10}$. Like SFB$_{1}$, in between one such pair of wavefront dislocations, the real-valued amplitude reaches its extreme value.

\begin{figure}[h]			
\begin{center}
\includegraphics[width = 0.6\textwidth]{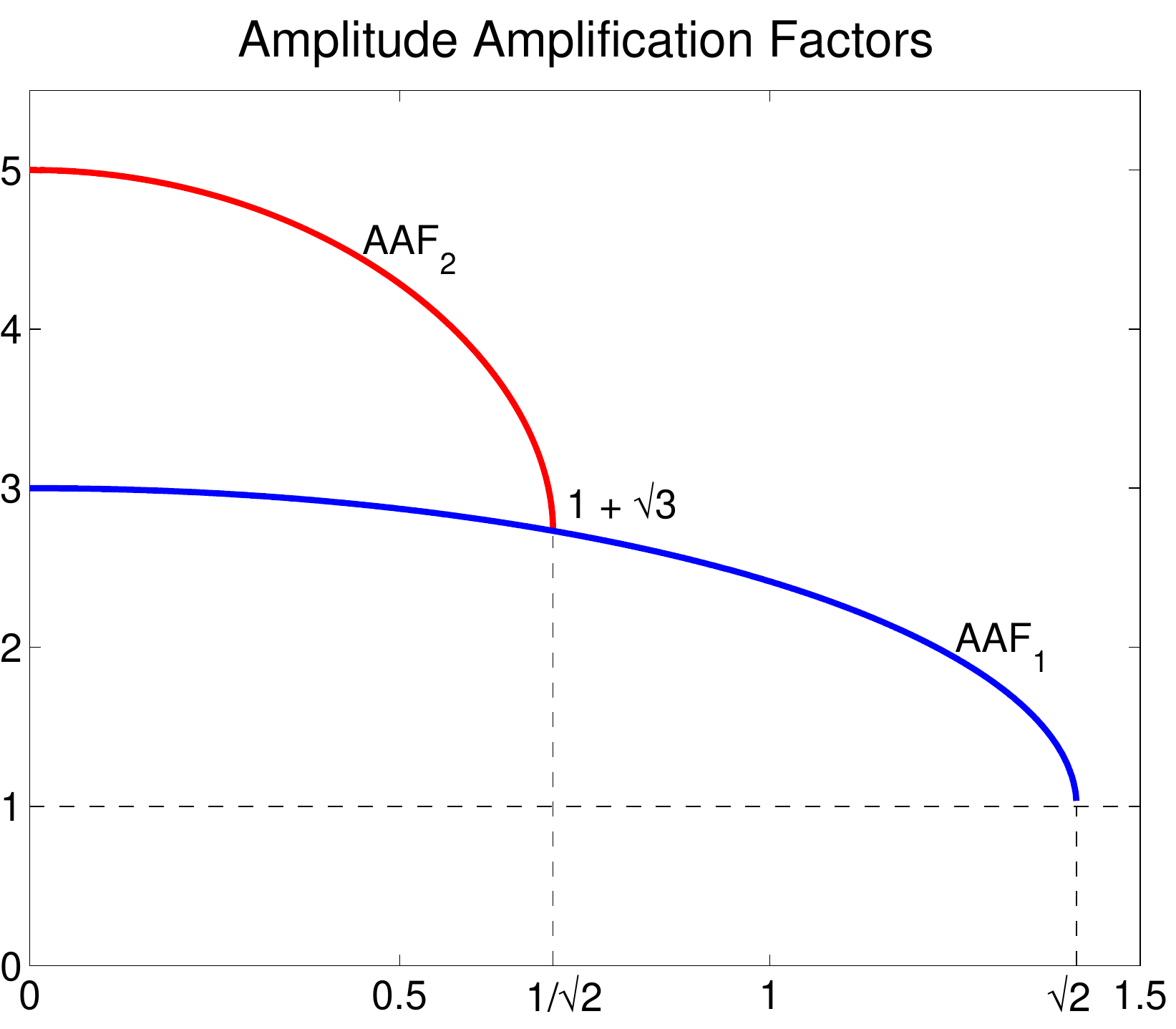}
\caption[Amplitude amplification factors of SFB$_{1}$ and SFB$_{2}$]{Plots of amplification amplitude factors of SFB$_{1}$ and SFB$_{2}$. Both AAF plots are monotonically decreasing for increasing modulation frequency. The same value of AAF $= 1 + \sqrt{3}$ is reached when $\tilde{\nu}_{2} = \sqrt{1/2}$ and $\tilde{\nu}_{1} = \sqrt{1/2}$, because the limiting value of $\tilde{\nu}_{2} \rightarrow \sqrt{1/2}$ leads SFB$_{2}$ to SFB$_{1}$ with $\tilde{\nu}_{1} = \sqrt{1/2}$.} 		\label{AAFs}
\end{center}
\end{figure}

For both theory and experiment on extreme wave generation\index{extreme waves!generation} of SFB$_{2}$ wave field, we are also interested in the amplitude amplification factor, denoted as AAF and defined in Definition~\ref{AAFdefinition} on page~\pageref{AAFdefinition}. The AAF of SFB$_{1}$ and SFB$_{2}$ are denoted as AAF$_{1}$ and AAF$_{2}$, respectively. We know that the AAF of SFB$_{1}$ is given by
\begin{equation}
\textmd{AAF}_{1}(\tilde{\nu}_{1}) = 1 + \sqrt{4 - 2\tilde{\nu}_{1}^{2}}
\end{equation}
for $0 < \tilde{\nu}_{1} < \sqrt{2}$. For a very long modulation, when the (normalized) modulation frequency $\tilde{\nu}_{1} \rightarrow 0$, it can reach up to a maximum value of 3. It decreases monotonically for increasing values of modulation frequency. The AAF of SFB$_{2}$\index{SFB!SFB$_{2}$!AAF} is given by
\begin{equation}
\textmd{AAF}_{2}(\tilde{\nu}_{2}) = 1 + \sqrt{4 - 2\tilde{\nu}_{2}^{2}} + 2 \sqrt{1 - 2\tilde{\nu}_{2}^{2}}.
\end{equation}
Interestingly, there exists a relationship between AAF$_{1}$ and AAF$_{2}$ and it can be written explicitly as a simple linear relation:
\begin{equation}
  \textmd{AAF}_{2}(\tilde{\nu}_{2}) = 1 + \left(\textmd{AAF}_{1}(\tilde{\nu}_{2}) - 1 \right) +
    \left(\textmd{AAF}_{1}(2\tilde{\nu}_{2}) - 1 \right).
\end{equation}
For a very long modulation, AAF$_{2}$ can reach up to a maximum value of 5 for $\nu_{2} \rightarrow 0$. For the whole range of frequencies $0 < \tilde{\nu}_{2} < \sqrt{1/2}$, the boundary values are $1 + \sqrt{3} < \textmd{AAF}_{2} < 5$. Figure~\ref{AAFs} shows the plot of the amplification amplitude factors.

\subsection{Maximum temporal amplitude}
\index{SFB!SFB$_{2}$!MTA} \index{MTA!SFB$_{2}$}

The concept of the maximum temporal amplitude (MTA) is introduced in Subsection~\ref{MTAsubsection} and the definition is given in Definition~\ref{MTAdefinition} on page~\pageref{MTAdefinition}. The MTA is useful for the generation of extreme waves\index{extreme waves!generation} in the wave basin, particularly to determine a signal input for the wavemaker so that the extreme wave signal is produced at a certain location.
\begin{figure}[h]			
\begin{center}
\includegraphics[width = 0.4\textwidth, angle = -90]{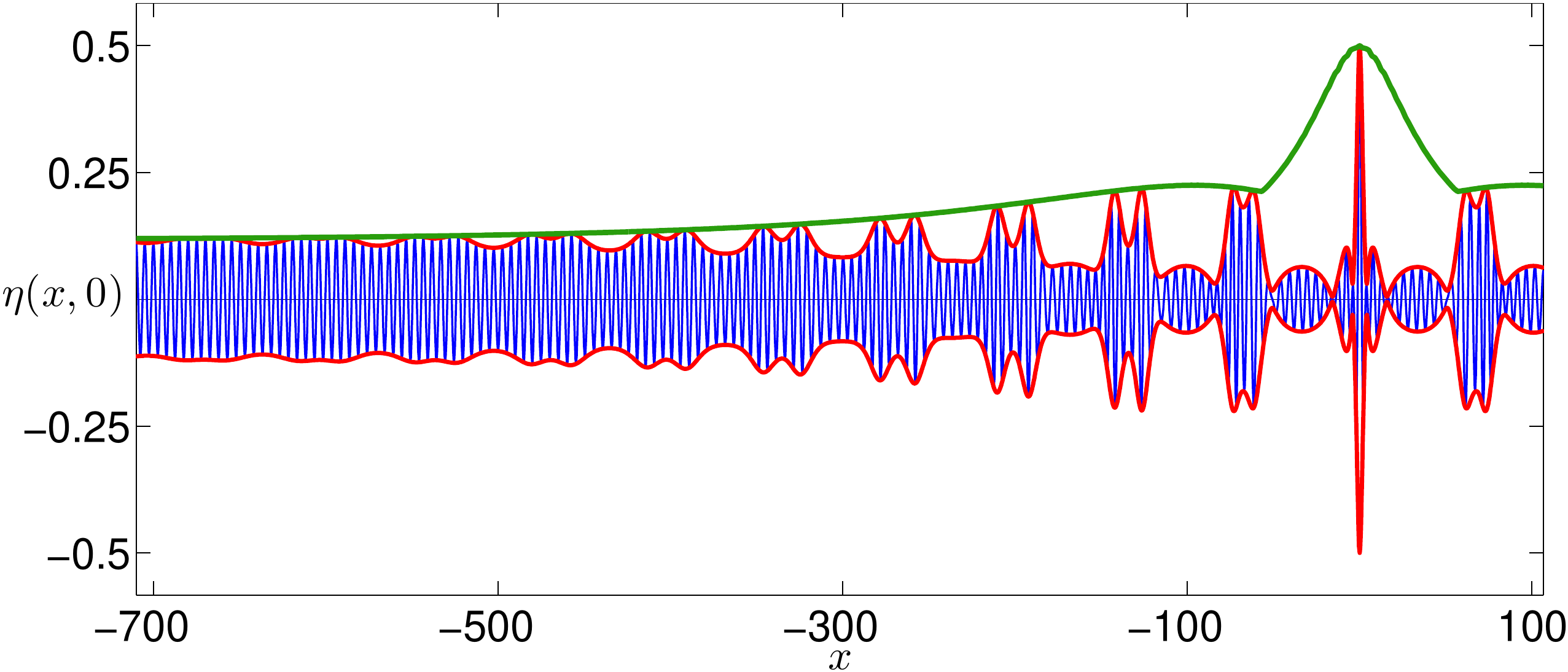}
\caption[Wave profile, envelope, and the MTA of SFB$_{2}$]{A plot of individual waves, the corresponding wave envelope, and the time-independent MTA of SFB$_{2}$ for $\tilde{\nu}_{2} = \frac{1}{2}$. The horizontal axis is the distance from the extreme position and the vertical axis is the surface elevation. Both axes are measured in meter.} 	\label{MTA2plot}
\end{center}
\end{figure}

The MTA plot of SFB$_{2}$ is illustrated in Figure~\ref{MTA2plot}. This figure shows the propagation of waves of varying amplitude after they are being generated by the wavemaker at the left-hand side. In this example, the water depth is $3.55$ m, the wavelength is approximately $6.2$ m and both horizontal and vertical distances are in meters. The wave signal is generated on the left-hand side, and waves propagate to the right and reach an extreme condition at $x = 0$. The shown SFB$_{2}$ wave field has an initial signal of $15$~cm amplitude and it reaches an extreme amplitude of $45$ cm, an amplitude amplification of factor three. The MTA plot for SFB$_{2}$ shows a somewhat different gradual pattern evolution than MTA for SFB$_{1}$, see Figure~\ref{MTASFB} on page~\pageref{MTASFB} for a comparison. For SFB$_{2}$ the MTA increases monotonically and very slowly until 200~m from the extreme position, then decreases slightly, and after that, it starts to grow significantly faster than before it reaches the area of the extreme position. Afterward, it returns to the initial behavior in the reverse order.

The double modulated wave train envelope travels downstream\index{downstream} with increasing amplitude for SFB$_{2}$. Wave groups\index{wave group(s)} are deformed and increase in amplitude until they reach an extreme position. For SFB$_{2}$, the visible pattern is a result of wave packet interaction and the effect becomes more significant and tangible as the wave profile approaches the extreme position. This focusing of the two constituent wave packages is the explanation of the fast growth from a certain position relatively close to the extreme position. It is also very significant that close to the extreme position, the amplitude can grow to very large values within a short distance.

\section{Spatial evolution of SFB$_{2}$ wave signal} \label{spatialevol2}

We have observed in the previous section that double modulated wave groups\index{wave group(s)} with small amplitude evolve into large-amplitude waves over a sufficiently long distance. These wave groups\index{wave group(s)!interaction} interact during the downstream\index{downstream} evolution toward the extreme position. In this section, we will illustrate in more detail the spatial evolution of the signal. We pay more attention to the wave signal at $x = 0$ since at this position the complex amplitude SFB$_{2}$ becomes a real-valued function. Also at this position, the amplitude vanishes and leads to phase singularity. \index{phase singularity!SFB$_{2}$} Illustrations in the dispersion plane are presented to see that indeed the Chu-Mei quotient\index{Chu-Mei quotient} is unbounded at such instances. We also present the dynamic evolution of the envelope signal of SFB$_{2}$ in the Argand diagram and the phase plane at the extreme position.

\subsection{Wave signal evolution}
\index{SFB!SFB$_{2}$!wave signal evolution}

\begin{figure}[h!]			
\begin{center}
\includegraphics[width = 0.5\textwidth]{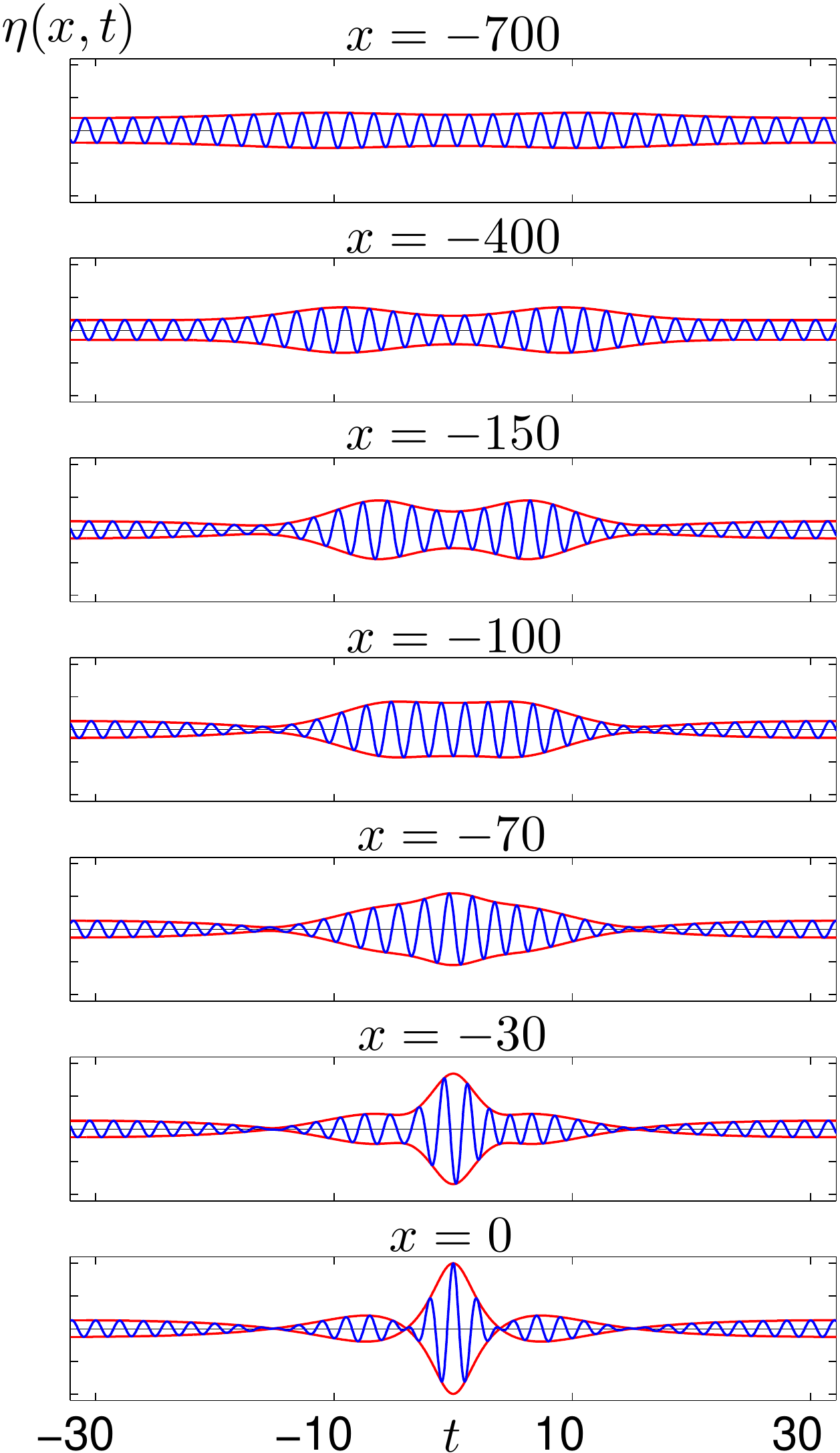}
\caption[Evolution of SFB$_{2}$ wave signal]{The evolution of the time signal of SFB$_{2}$ for $\tilde{\nu}_{2} = \frac{1}{2}$ from a modulated wave signal into an extreme signal.} 	    \label{SFB2evolution}
\end{center}
\end{figure}

We have understood that SFBs behaves asymptotically as a plane-wave solution. Similar to the case of SFB$_{1}$, a modulated wave signal with two different modulation frequencies from SFB$_{2}$ will grow in space during its evolution according to the modulational instability\index{modulational instability}. SFB$_{2}$ describes the complete nonlinear evolution of this modulated wave signal. The modulation period $T_{2}$ is preserved during the evolution and is determined by the modulation frequency $\nu_{2}$, $T_{2} = 2\pi/\nu_{2}$. Figure~\ref{SFB2evolution} shows the evolution of SFB$_{2}$ from a modulated monochromatic wave signal into an extreme condition. We see in this figure that in one modulation period, two parts of a wave group\index{wave group(s)!interaction} signal grow and interact as the wave signal evolve in space toward the extreme signal. The signal at the extreme position, or the extreme signal, of SFB$_{2}$ shows a different pattern from the SFB$_{1}$ extreme signal. It has two pairs of phase singularities for $0 < \tilde{\nu}_{2} < \tilde{\nu}_{2}^{\ast}$ and one pair of singularities for $\tilde{\nu}_{2}^{\ast} < \tilde{\nu}_{2} < \sqrt{1/2}$, as we shall see in the next subsection.

\subsection{Phase singularity} \label{SFB2PS}
\index{SFB!SFB$_{2}$!phase singularity} \index{phase singularity!SFB$_{2}$}

We have understood in Chapter~\ref{4Dislocate} that phase singularity and wavefront dislocation are generically a result of vanishing amplitude. We have also understood that the unboundedness of the Chu-Mei quotient\index{Chu-Mei quotient} is responsible for the occurrence of these phenomena. In this subsection, we explain that the same phenomena also occur in the SFB$_{2}$ physical wave field. We have shown that for SFB$_{1}$, phase singularity occurs for a certain value of modulation frequency $\nu_{1}$, namely $0 < \tilde{\nu}_{1} \leq \sqrt{3/2}$. However, SFB$_{2}$ always shows phase singularity and wavefront dislocation for any $\tilde{\nu}_{2}$ with $0 < \tilde{\nu}_{2} < \sqrt{1/2}$. Different from SFB$_{1}$ which has only one pair of singularities, SFB$_{2}$ can have two pairs of singularities for $0 < \tilde{\nu}_{2} < \tilde{\nu}_{2}^{\ast}$, where $\tilde{\nu}_{2}^{\ast} = \frac{1}{6} \sqrt{8\sqrt{10} - 10}$.
\begin{figure}[h]			
\begin{center}
\subfigure[]{\includegraphics[width = 0.45\textwidth]{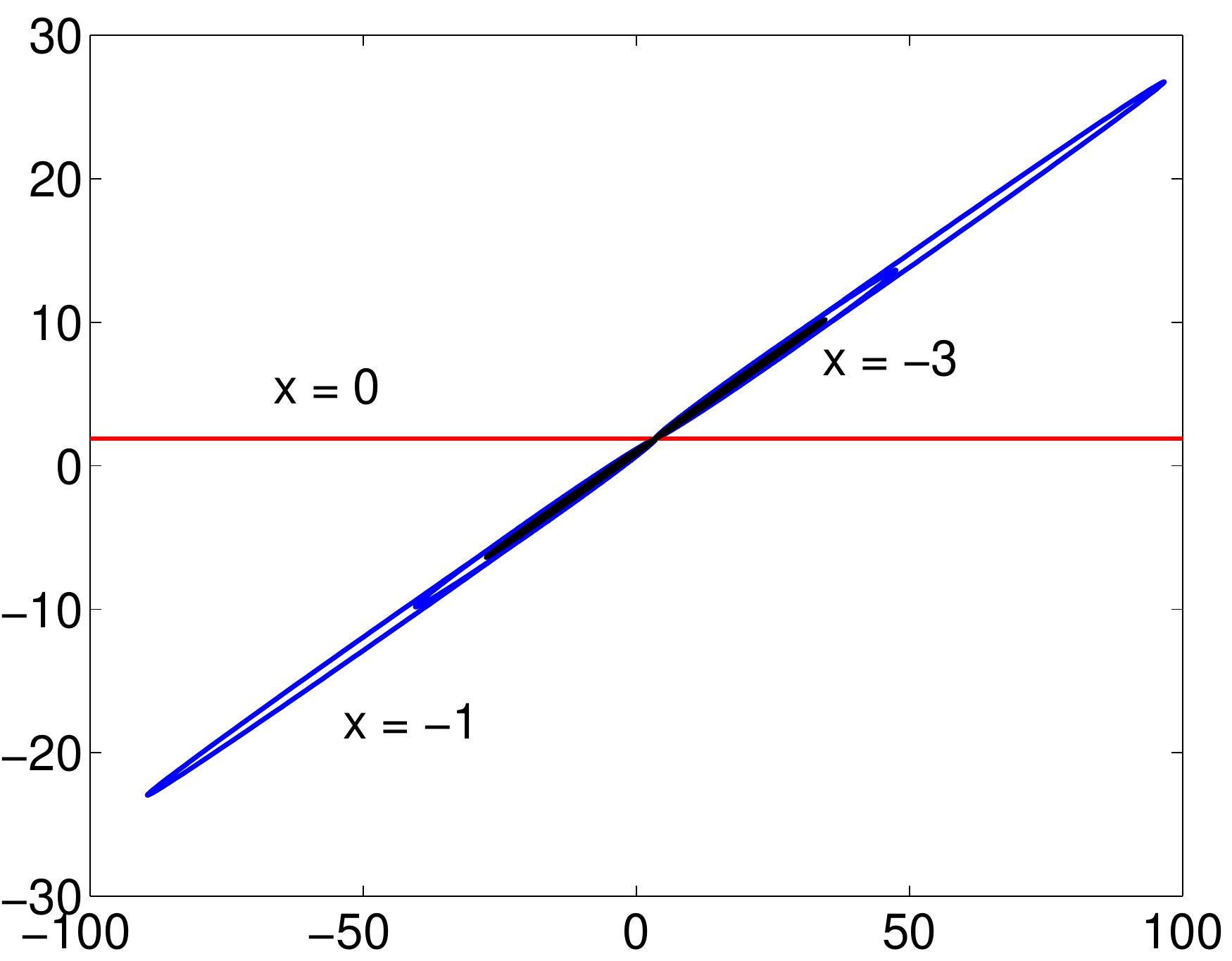}}	    \hspace{0.75cm}
\subfigure[]{\includegraphics[width = 0.45\textwidth]{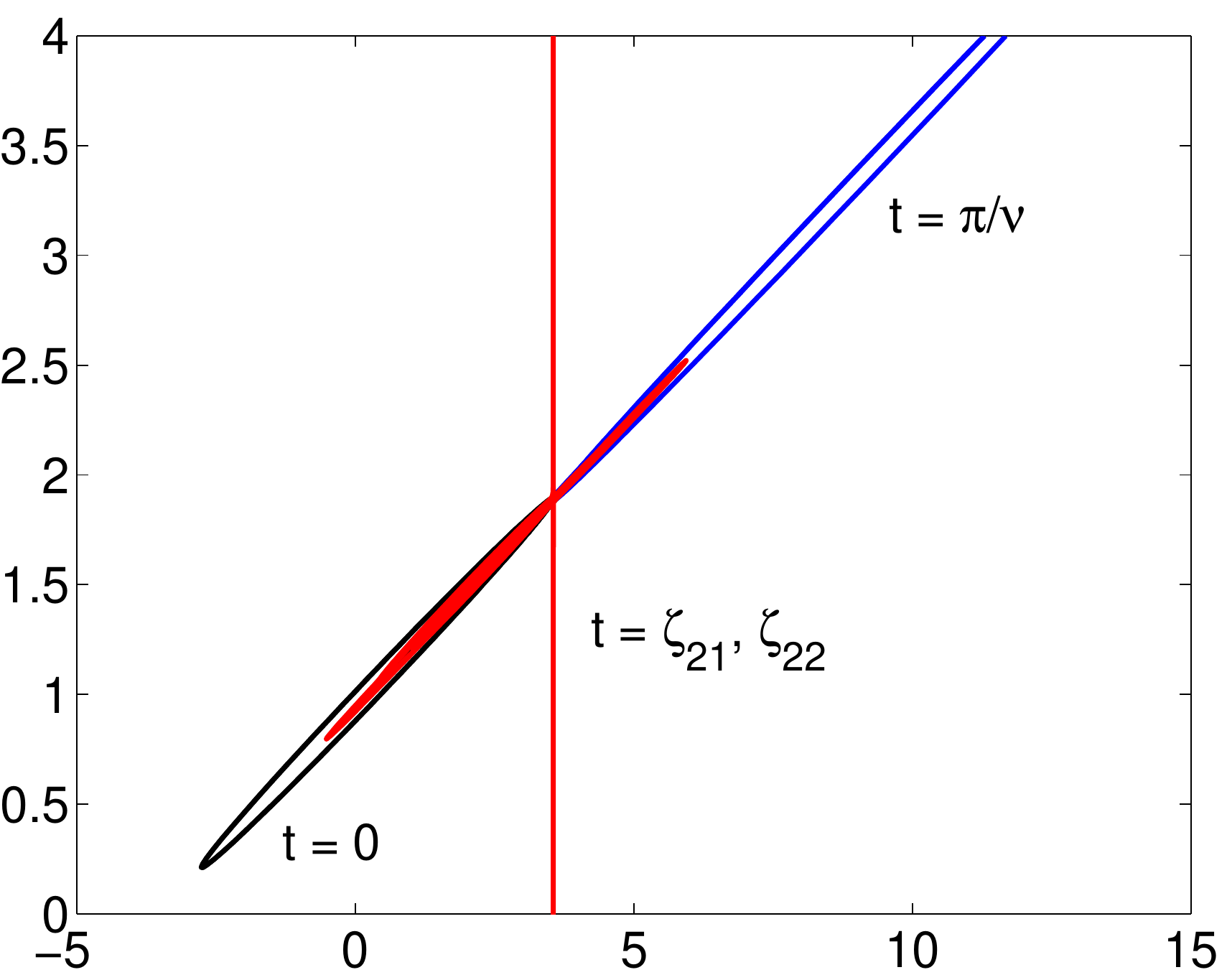}}
\caption[Local wavenumber and local frequency of SFB$_{2}$]{Plots of local wavenumber $k$ (horizontal axis) and local frequency $\omega$ (vertical axis) in the dispersion plane for $\tilde{\nu}_{2} = \frac{1}{2}$. At $x = 0$, the local wavenumber is unbounded (a) and at $\tau = \zeta_{21}$ or $\zeta_{22}$, the local frequency is unbounded (b).} 		\label{LWLF2}
\end{center}
\end{figure}

For SFB$_{2}$, a phase singularity occurs when $\cos(\nu_{2}\zeta_{2})$ satisfies the following cubic equation:
\begin{equation}
\cos^{3}(\nu_{2} \zeta_{2}) + C_{2} \cos^{2}(\nu_{2} \zeta_{2}) + C_{1} \cos(\nu_{2} \zeta_{2}) + C_{0} =  0, 		\label{cubic}
\end{equation}
where $C_{0}$, $C_{1}$, and $C_{2}$ are coefficients depending on $\tilde{\nu}_{2}$. Explicit expressions of these coefficients are given in Appendix~\ref{ApendiksSFB2} on page~\pageref{SFB2cubicPS}. By defining the intermediate variables $Q = \frac{1}{9}(3C_{1} - C_{2}^{2})$ and $R = \frac{1}{54}(9C_{1}C_{2} - 27C_{0} - 2C_{2}^{3})$, as well as the discriminant of the cubic equation $D = Q^{3} + R^{2}$, we find that for normalized quantities $r_{0}$, $\beta$, and $\gamma$ for each $\tilde{\nu}_{2}$ with $0 < \tilde{\nu}_{2} < \sqrt{1/2}$, the discriminant is negative $D < 0$. Therefore, all roots of the cubic equation~\eqref{cubic} are all real and unequal. However, when solving for $\zeta_{2}$, not all solutions are real. Two real solutions that correspond to the time when the envelope vanishes are given by: 
{\small
\begin{align}
\zeta_{21} &= \frac{\sqrt{\beta/\gamma}}{r_{0}\tilde{\nu}_{2}} \arccos\left[-\frac{C_{2}}{3} + 2\sqrt{-Q} \cos\left(\frac{\theta}{3} + \frac{4\pi}{3} \right)  \right], \qquad \; 0 < \tilde{\nu}_{2} < \sqrt{\frac{1}{2}},\\
\zeta_{22} &= \frac{\sqrt{\beta/\gamma}}{r_{0}\tilde{\nu}_{2}} \arccos\left[-\frac{C_{2}}{3} + 2\sqrt{-Q} \cos\left(\frac{\theta}{3} + \frac{2\pi}{3} \right)  \right], \qquad \; 0 < \tilde{\nu}_{2} < \tilde{\nu}_{2}^{\ast}, \;
\end{align}
} where $\cos \theta = R/\sqrt{-Q^{3}}$.

Another way to confirm the occurrence of phase singularity\index{phase singularity} is by showing that the Chu-Mei quotient\index{Chu-Mei quotient} is unbounded at the singular points. This can be shown by plotting the local wavenumber\index{local wavenumber} and local frequency\index{local frequency}, as we also have shown for SFB$_{1}$ in Subsection~\ref{SubsectionSFB} on page~\pageref{SubsectionSFB}. Figure~\ref{LWLF2} shows the plot of trajectories in the dispersion plane when local wavenumber and local frequency are unbounded.

\subsection{Argand diagram} \label{SFB2Argand}
\index{SFB!SFB$_{2}$!Argand diagram}

\begin{figure}[h]			
\begin{center}
\includegraphics[width = 0.5\textwidth]{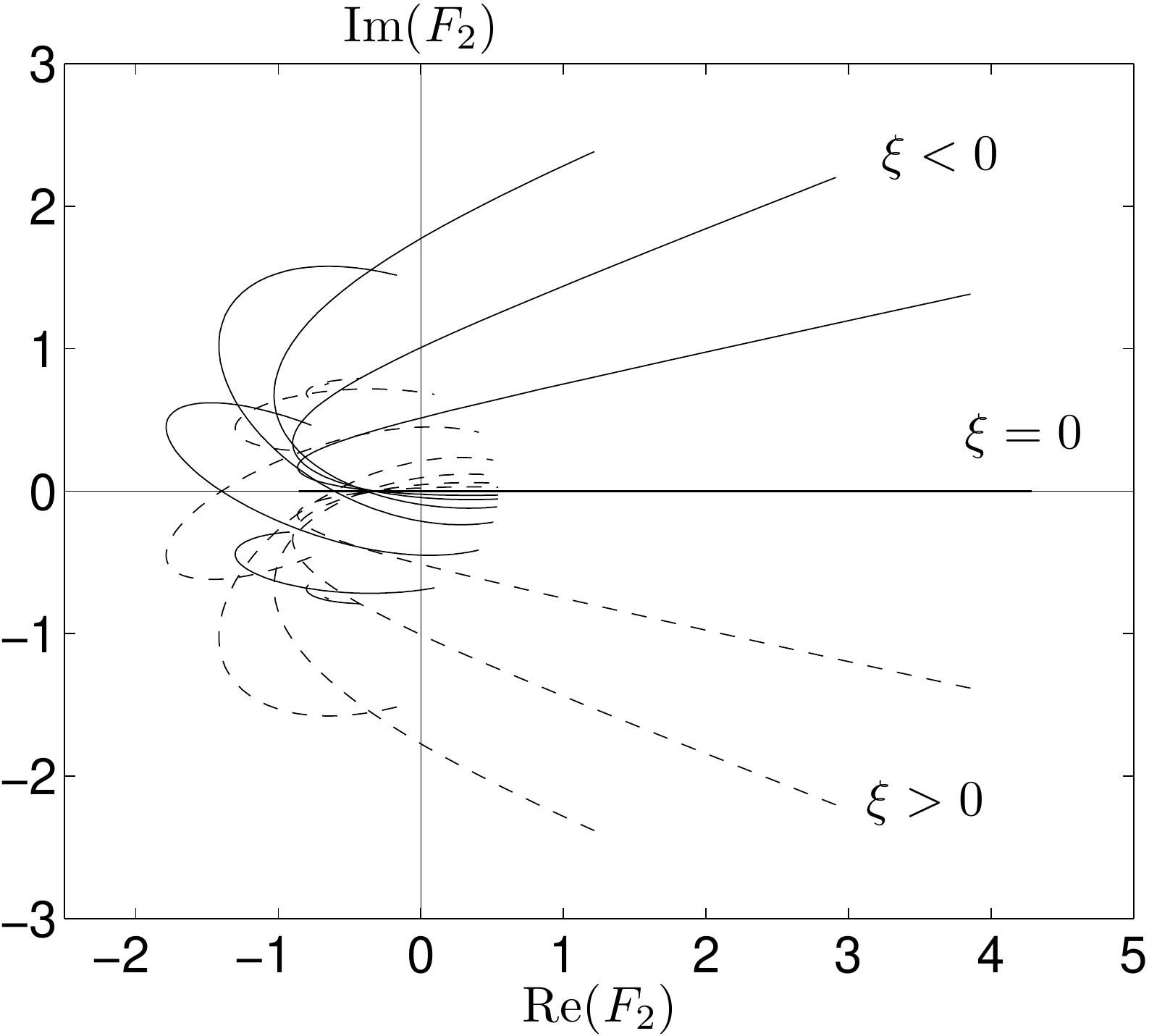}
\caption[Evolution in the Argand diagram of SFB$_{2}$]{The dynamic evolution of SFB$_{2}$ parameterized in time at different positions for $\tilde{\nu}_{2} = \frac{1}{2}$. At $\xi = 0$, the curve lies on the real axis; solid curves indicate $\xi < 0$; and dashed curves indicate $\xi > 0$.}	    \label{SFB2_Argand}
\end{center}
\end{figure}

The evolution in the Argand diagram for SFB$_{2}$ is a collection of curves in a complex plane parameterized in time or space. The horizontal and the vertical axes are the real and the imaginary parts of SFB$_{2}$ after removing the plane-wave solution, respectively. The space trajectories of the evolution in the Argand diagram for SFB$_{2}$ can be found in~\citep{5Akhmediev97}. For time trajectories, the evolution in the Argand diagram for SFB$_{2}$ shows a different pattern than for SFB$_{1}$. Instead of a collection of straight lines, it is a collection of curves. Figure~\ref{SFB2_Argand} shows the time trajectories in the Argand diagram of SFB$_{2}$ for $\tilde{\nu}_{2} = \frac{1}{2}$.

\subsection{Phase plane representation}
\index{SFB!SFB$_{2}$!phase plane} \index{phase plane!SFB$_{2}$}

\begin{figure}[h]			
\begin{center}
\includegraphics[width = 0.5\textwidth]{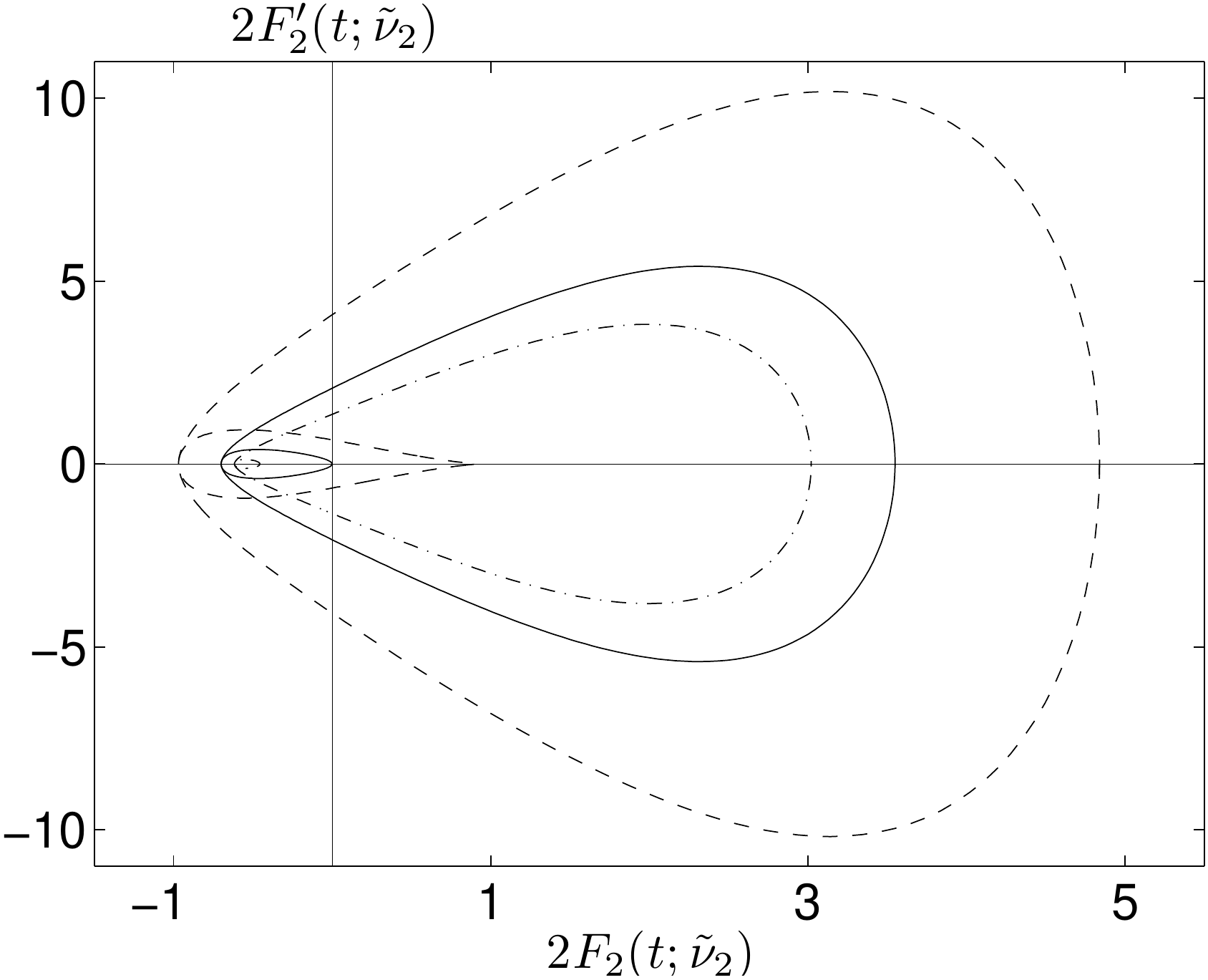}
\caption[Phase curves of SFB$_{2}$ wave envelope]{Plots of phase curves in the phase plane SFB$_{2}$ at the extreme position. The curves correspond to different values of $\tilde{\nu}_{2}$: $\tilde{\nu}_{2} = \frac{1}{4}$ (dashed), $\tilde{\nu}_{2} = \tilde{\nu}_{2}^{\ast}$ (solid) and $\tilde{\nu}_{2} = 0.7$ (dash-dot). Note that for each value of $\tilde{\nu}_{2}$, the phase plot crosses $F_{2} = 0$, and for $\tilde{\nu}_{2} < \tilde{\nu}_{2}^{\ast}$ it crosses this axis twice, corresponding to one and two pairs of phase singularities.}		    \label{phase_plane_envelope}
\end{center}
\end{figure}

In this subsection, we investigate a dynamic property of SFB$_{2}$, namely the envelope signal in the phase plane. At the extreme position, SFB$_{2}$ is a real-valued function. We study the extreme envelope by plotting the corresponding phase curves in a phase plane. Figure~\ref{phase_plane_envelope} shows phase plots of SFB$_{2}$ extreme envelope for different values of $\tilde{\nu}_{2}$. The horizontal axis denotes twice the extreme envelope $2 F_{2}(t;\nu_{2})$ and the vertical axis denotes twice the derivative of the extreme envelope $2 F'(t;\nu_{2})$. For an increasing time, the motion along the phase curve is in the clockwise direction.

\section{Amplitude spectrum evolution} \label{SectionASE2}
\index{SFB!SFB$_{2}$!spectrum evolution}	\index{spectrum!amplitude}

In the preceding sections, we have considered the characteristics and properties of SFB$_{2}$ in the time domain. In this section, we study the corresponding characteristics in the frequency domain. The changes of the spectrum of a wave signal during the evolution reflect the changes of the signal itself during downstream\index{downstream} evolution. The SFB$_{2}$ spectrum has two pairs of sidebands\index{sideband(s)} at the initial state. During the evolution, the spectrum develops into more pairs of sidebands and eventually returns again to the initial state. This energy distribution and recollection from and to the central frequency is caused by the nonlinear effects of the modulational instability\index{modulational instability}, which is also observed in SFB$_{1}$. For SFB$_{1}$, the amplitude spectrum corresponding to the first sideband always decays more slowly than the second sideband for $\xi \rightarrow \pm \infty$. However, for SFB$_{2}$, this is not always the case due to the two growth rates\index{growth rate(s)}. Depending on the modulation frequency $\tilde{\nu}_{2}$, which in turns gives different growth rate ratio, the first sideband\index{sideband(s)} decays faster for $\xi \rightarrow \pm \infty$ if $\tilde{\nu}_{2}^{\ast} < \tilde{\nu}_{2} < \sqrt{1/2}$.

\begin{figure}[h!]			
\begin{center}
\subfigure[]{\includegraphics[width=0.45\textwidth]{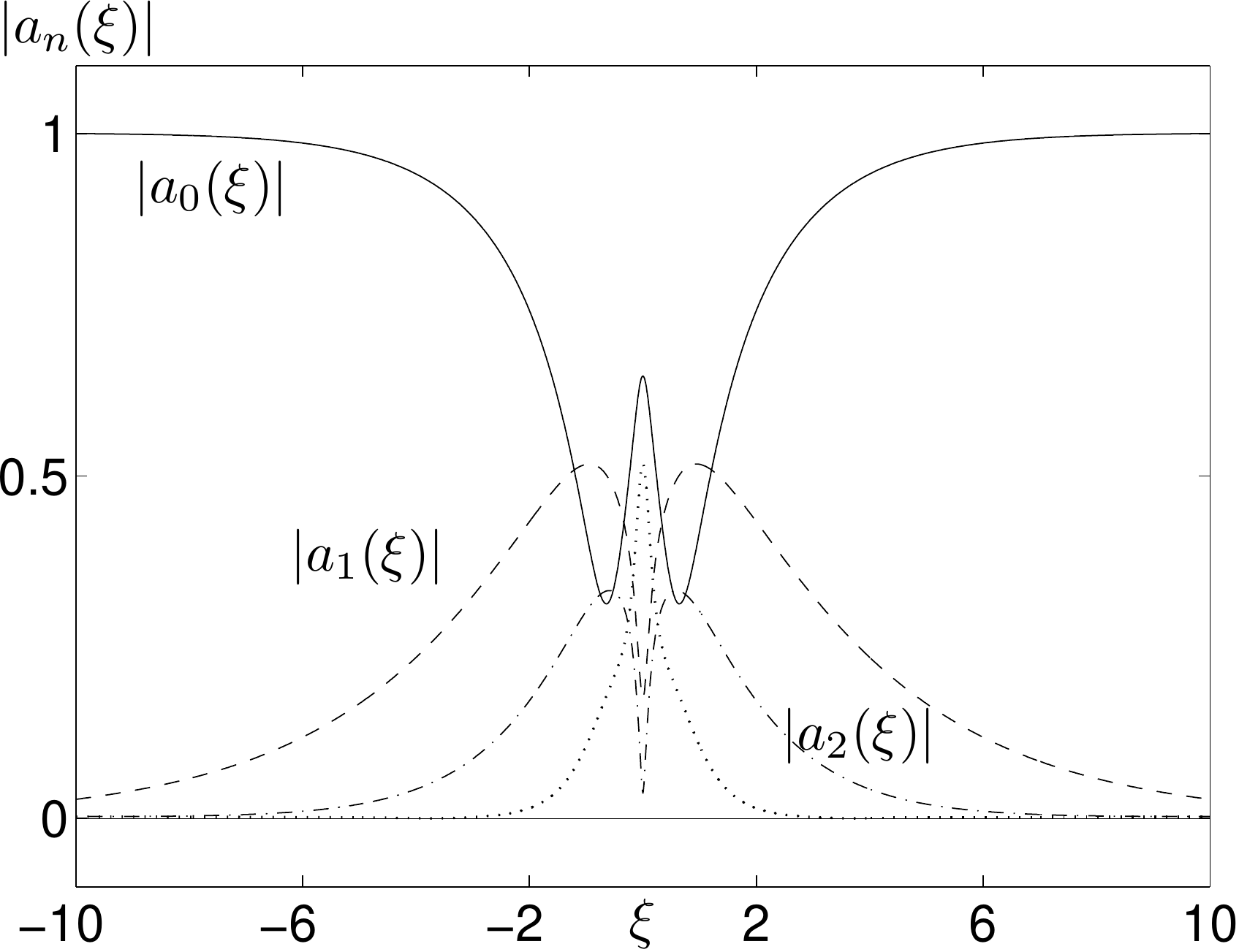}} 		\hspace{0.75cm}
\subfigure[]{\includegraphics[width=0.45\textwidth]{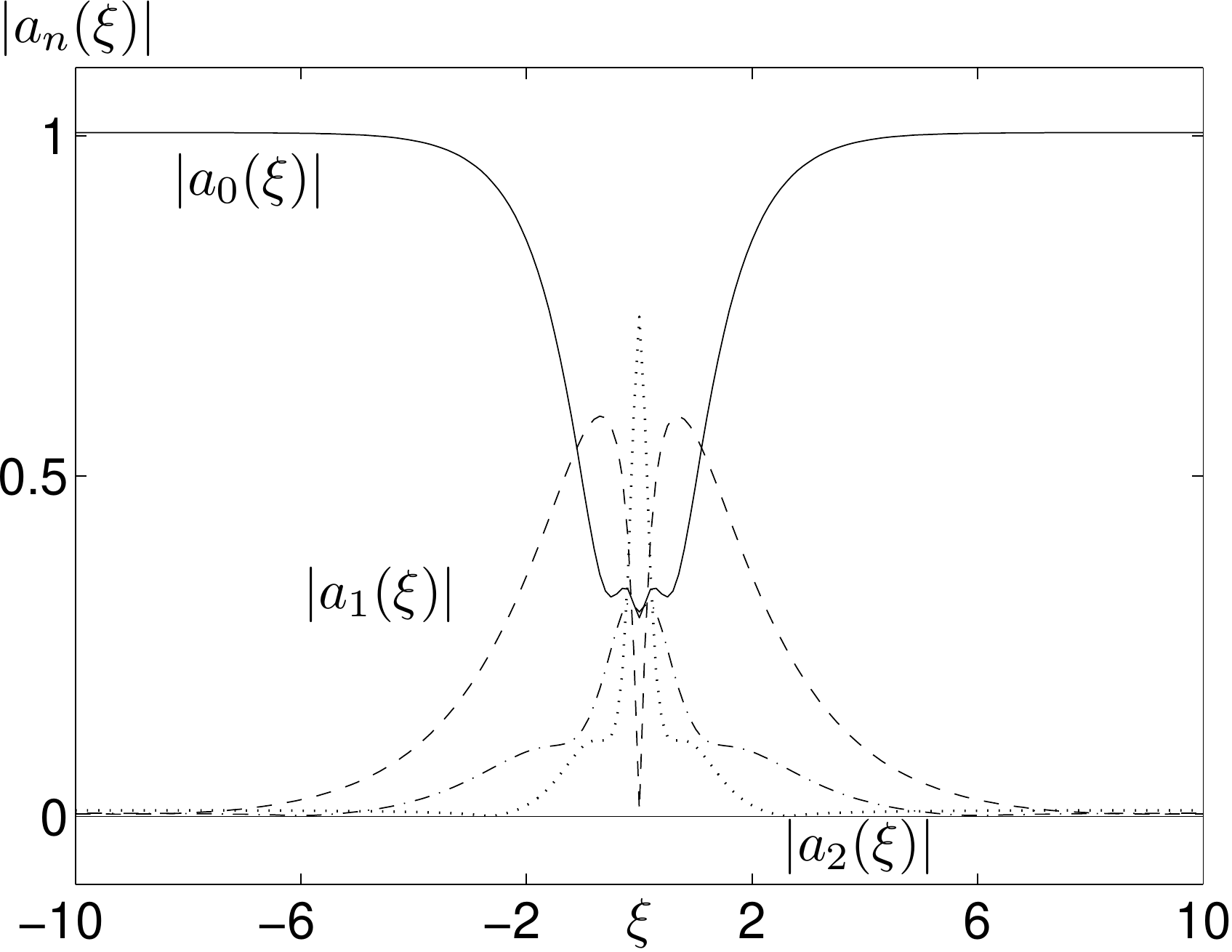}} 		\vspace{0.10cm}
\subfigure[]{\includegraphics[width=0.45\textwidth]{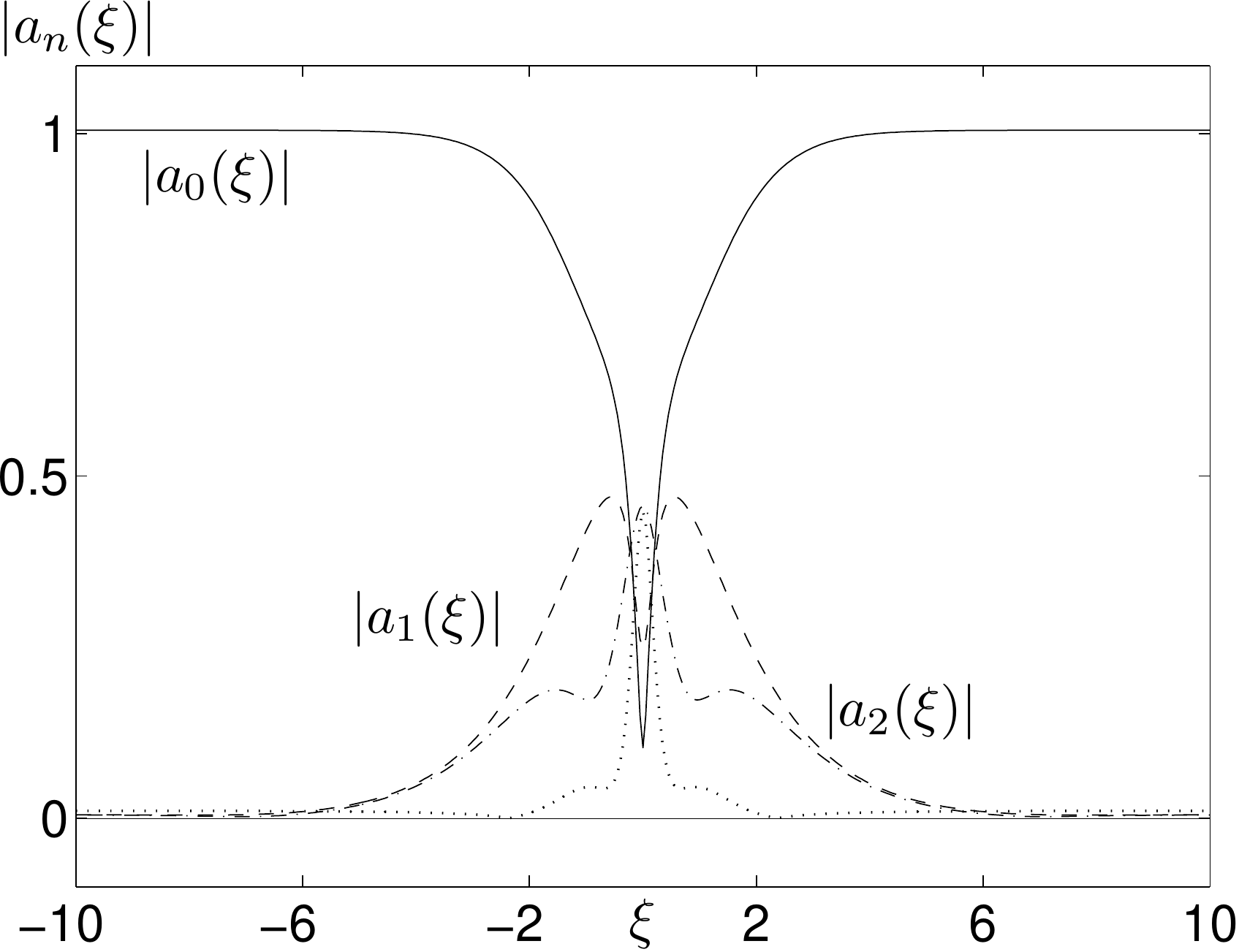}} 		\hspace{0.75cm}
\subfigure[]{\includegraphics[width=0.45\textwidth]{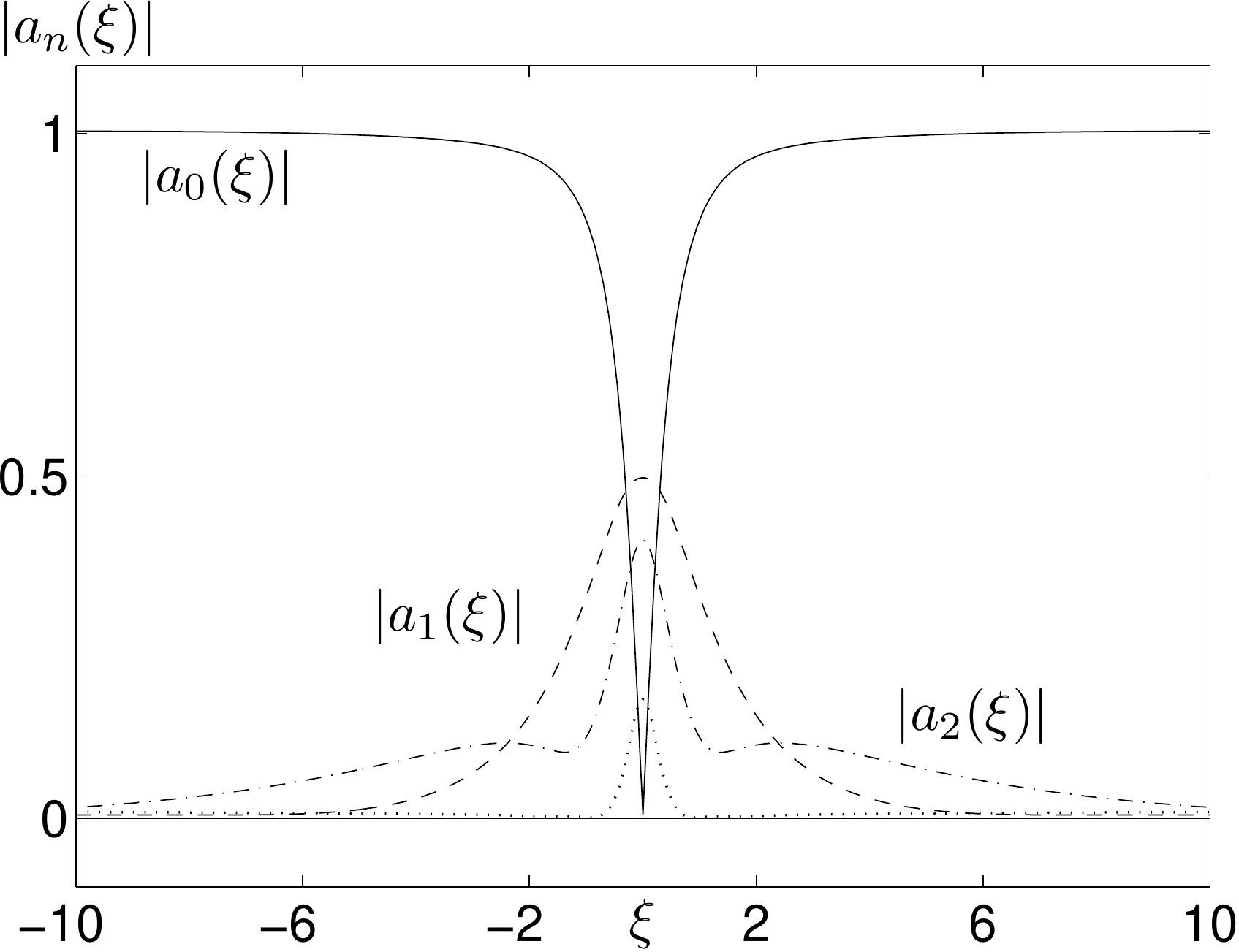}}
\caption[Absolute amplitude spectrum of SFB$_{2}$]{Plots of the absolute amplitude spectrum SFB$_{2}$ corresponding to the central frequency, the first sideband, and the second sideband. The plots are given for different values of modulation frequency $\tilde{\nu}_{2}$: (a) $\tilde{\nu}_{2} = 1/4$, (b) $\tilde{\nu}_{2} = 1/2$, (c) $\tilde{\nu}_{2} = \sqrt{2/5}$, and (d) $\tilde{\nu}_{2} = 0.7$.} \label{ASESFB2}
\end{center}
\end{figure}

We have been able to derive the spectrum corresponding of SFB$_{1}$ as presented in Appendix~\ref{SFBspectrum} on page~\pageref{SFBspectrum}. Deriving an exact expression for the spectrum of SFB$_{2}$ is complicated. However, the corresponding absolute amplitude spectrum can be calculated numerically. This spectrum contains an integral expression and the numerical approximation to this integral is calculated by applying the trapezoidal rule. Figure~\ref{ASESFB2} shows the plots of this absolute amplitude spectrum for four different values of the modulation frequency $\tilde{\nu}_{2}$. The different behavior between the first sideband\index{sideband(s)} and the second sideband is caused by the different ratio of the growth rates. \index{growth rate(s)!ratio}

For $0 < \tilde{\nu}_{2} < \sqrt{2/5}$, the first growth rate\index{growth rate(s)} has a lower value than the second one, $\sigma_{1} < \sigma_{2}$. Consequently, the first sideband dominates the second sideband\index{sideband(s)} if the position far enough from the extreme position. For $\tilde{\nu}_{2} = \sqrt{2/5}$, both growth rates have the same value, $\sigma_{1} = \sigma_{2}$, and therefore both amplitude spectra have the same growth and decay rates when $\xi \rightarrow \pm \infty$. But for $\sqrt{2/5} < \tilde{\nu}_{2} < \sqrt{1/2}$, the first growth rate is larger than the second one, $\sigma_{1} > \sigma_{2}$. Consequently, the first sideband is dominated by the second sideband for $\xi \rightarrow \pm \infty$, but as they evolve toward the extreme position, the first sideband takes over and starts to dominate the second sideband.\index{sideband(s)}

\section{Conclusion and remark} \label{SR2}

In this chapter, we studied some characteristics of higher-order waves on finite background\index{waves on finite background!higher-order}, particularly the SFB$_{2}$ solution of the NLS equation that describes the modulational instability\index{modulational instability} with two pairs of initial sidebands.\index{sideband(s)} We also made some comparisons with SFB$_{1}$ that has been discussed extensively in Chapter~\ref{3Property}. Since the initial evolution of SFB$_{2}$ has two different modulation wavelengths, an interaction of two wave groups\index{wave group(s)!interaction} takes place as they propagate downstream, see Figure~\ref{MTA2plot}. Due to this wave group interaction, we observed that SFB$_{2}$ can have a higher amplitude amplification than SFB$_{1}$, up to a maximum factor of 5. This factor depends strongly on the modulation frequency and an explicit expression between amplitude amplification factors of SFB$_{1}$ and SFB$_{2}$ has been presented. \index{SFB!SFB$_{2}$}

The corresponding physical wave field of SFB$_{2}$\index{physical wave field!SFB$_{2}$} shows an interesting pattern of wavefront dislocation. Depending on the value of modulation frequency, either one pair or two pairs of wavefront dislocations in one modulation period are observed. For $0 < \tilde{\nu}_{2} < \tilde{\nu}_{2}^{\ast}$, we have two pairs of wavefront dislocations, while for $\tilde{\nu}_{2}^{\ast} < \tilde{\nu}_{2} < \sqrt{1/2}$, only one pair of wavefront dislocations is observed, where $\tilde{\nu}_{2}^{\ast} = \frac{1}{6}\sqrt{8 \sqrt{10} - 10}$. Wavefront dislocation occurs at the extreme position $x = 0$ and periodically in time. In fact, at these instances, the corresponding amplitude of the SFB$_{2}$ wave field vanishes and the corresponding phase becomes singular, the phase singularity. \index{phase singularity} Additionally, the Chu-Mei quotient\index{Chu-Mei quotient} becomes unbounded, as shown by the dispersion plot in Figure~\ref{LWLF2}, when the local wavenumber and the local frequency become unbounded.

We also mentioned that the asymptotic behavior\index{asymptotic behavior!SFB$_{2}$} for $\xi \rightarrow \pm \infty$ and the amplitude spectrum\index{spectrum!amplitude} evolution of SFB$_{2}$ depend on the growth rate ratio\index{growth rate(s)!ratio}, which is determined by the modulation frequency $\nu_{2}$. Some illustrative plots are presented in Figure~\ref{ASESFB2}. We also observed that during the spectrum evolution, the energy from the central frequency is distributed into its sidebands\index{sideband(s)} and collected again to the initial state. This process is similar to the extreme wave formation from a single modulated wave and how that returns to its initial condition.

{\renewcommand{\baselinestretch}{1} 
}
\setcounter{chapter}{5}
\chapter[Experimental results]{Experimental results on extreme wave generation}		\label{6Experiment}
\index{extreme waves!experiments} \index{extreme waves!generation}

\section{Introduction}
\index{extreme waves!experiments}

In the previous chapters, we have discussed the theoretical aspects of extreme wave generation. As explained in the introductory chapter, the original aim of this investigation is the wish of MARIN (Maritime Research Institute Netherlands) \index{MARIN} to generate large amplitude, non-breaking waves in its wave basin.\index{wave basin} These large waves will be used for testing model ships and offshore constructions in extreme wave conditions as occur in the open oceans. In this chapter, we will discuss some results from the experiments on extreme waves\index{extreme waves!experiments} that have been conducted at MARIN during the summer period of 2004.

Based on the modulational instability\index{modulational instability} of a modulated wave signal that has been discussed in Chapter~\ref{2Model}, we know theoretically that the SFB family studied in Chapter~\ref{3Property} (and~\ref{5HighOrder}) is a potentially good class of extreme waves. This chapter describes to what extent our aim to generate large waves has been achieved. We present comparisons between the experimental results from MARIN and our theoretical analysis based on the SFB. Another comparison of the experimental results and wave signals calculated by the nonlinear wave model {\it HUBRIS} has been presented in {\itshape Rogue Waves 2004} conference held by IFREMER in Brest, France \citep{6Huijsmans05}. \index{extreme waves!experiments}

Choosing the SFB\index{SFB} as a family for generating extreme waves implies that we want to exploit modulational instability as the process to make extreme waves with the aid of natural processes. This is a highly nonlinear process, very different from pseudo-linear dispersive wave focusing, as in the linear frequency focusing wave \index{linear frequency focusing wave}~\citep{6LonguetHiggins74, 6Chaplin96}. From the start, it is not clear whether the asymptotic description using the NLS equation\index{NLS equation} is robust enough to describe the complicated dynamic reality. In principle, higher-order nonlinear effects, combined with dispersion, could counterbalance the nonlinear amplification predicted by SFB. Robustness of the process would mean that the SFB modulational instability\index{modulational instability} would survive these and other effects which are not accounted for by the NLS model. In the literature, except for our paper~\citep{6Huijsmans05}, no experiments are described that relate in detail observed waves with the SFB solution. It was mentioned by~\citet{6Onorato04} that SFB waves were generated in a laboratory, but a precise account of the results has not been published. Personal contacts with~\citet{6Onorato04} revealed that only the development into large waves was observed, but that no detailed analysis was performed. It is our aim to provide such a detailed analysis of the MARIN experiments, and the description here serves as an initiation thereof. \index{extreme waves!experiments}

\section{Experimental setting}
\index{extreme waves!experiments} \index{experiments!setting}

\subsection{Wave basin aspects}
\index{extreme waves!experiments} \index{wave basin}

\begin{figure}[h]			
\begin{center}
\includegraphics[angle = 0, width = 0.9\textwidth]{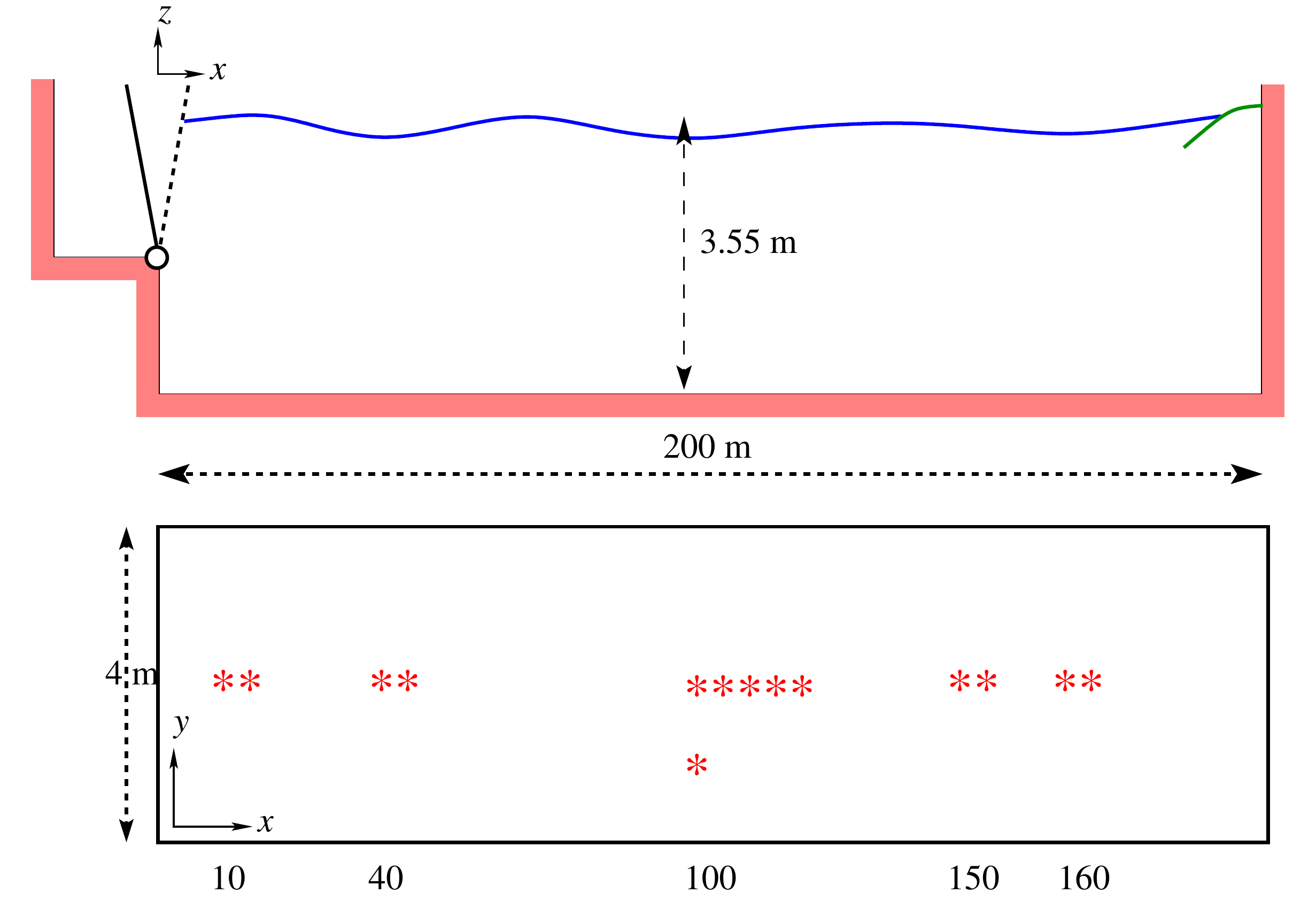}
\caption[Wave basin and wave gauge positions]{Sketch of the wave basin and the positions of the wave gauge.}			  \label{wavebasin}
\end{center}
\end{figure}

The experiment was conducted at the `high-speed basin'\index{wave basin!high speed} of MARIN.\index{MARIN} This basin has a dimension of 200~m long, 4~m wide and the water depth is 3.55~m. It is used for wave generation of regular and irregular waves in the longitudinal direction. The main application of this basin is for resistance, propulsion, and seakeeping tests for high-speed vessels. See the website of MARIN for more information about the facilities. Unidirectional waves are generated by a flap-type wavemaker\index{wavemaker} at one side and the waves are absorbed on the other side by an artificial beach. The hinge of the flap-type wavemaker is at 1.27 m above the basin floor. A predefined wave board signal was given to the hydraulic wavemaker and for each experiment, the stroke of the wave flap is measured. Figure~\ref{wavebasin} shows a simple sketch of the wave basin and the position of wave gauges\index{wave gauge} that were used during the experiments. \index{extreme waves!experiments}
\index{experiments!setting}

The electronic wave gauges measure the wave signal in the vertical ($z$) direction. It is important to note that only point measurements at discrete positions can be obtained. There are 14 gauges in total. Except for one, all gauges are installed in the middle of the lateral ($y$) direction. To check the unidirectionality of the waves, one gauge is placed 1~m from the flume centerline, as indicated in Figure~\ref{wavebasin}. Two gauges are installed close to the wavemaker at $x = 10$~m in order to compare the initial stage of the wave signal evolution. Two gauges are installed around $x = 40$~m, $x = 150$~m  and $x = 160$~m. Six gauges are installed around $x = 100$~m since in the initial design of the experiment this is the extreme position. However, another set of designs are proposed, to avoid initial breaking, for which the extreme position is at $x = 150$~m from the wavemaker. Because of technical reasons, the wave gauges were not replaced and the positioning of Figure~\ref{wavebasin} was kept. With two measurements at around both 150~m and 160~m, we expected to catch sufficient useful information on extreme wave\index{extreme waves} characteristics including large non-breaking waves and phase singularities. \index{extreme waves!experiments} \index{experiments!setting}

\subsection{Removing second-order effects} 
\index{extreme waves!experiments} \index{experiments!setting} \index{experiments!removing second-order effects} \index{second-order effects}

At each wave gauge\index{wave gauge}, the actual wave height is measured with respect to time. Therefore we obtain a time signal of the measured wave height. This measured wave signal contains higher-order effects. We know that the first-order physical wave field constructed with the theoretical SFB wave signal that we have discussed in Chapter~\ref{3Property} does not possess these higher-order contributions. Since we want to make both qualitative and quantitative comparisons with this theoretical signal, we would like to remove the higher-order effects from the measured wave signal. In the following, we describe how to remove the second-order contributions, since the high-frequency components of this contribution perturb the signal with rapid oscillations without essentially changing its behavior at the carrier frequency or the modulation frequency of the wave groups.\index{wave group(s)} Furthermore, in order to prevent that the second-order free waves are generated by the wavemaker, we have applied the second-order steering of the wavemaker. The theory of wave generation based on the fully nonlinear water wave equation with its boundary conditions including the second-order steering is presented in Appendix~\ref{wavegeneration}. \index{experiments!setting} \index{extreme waves!experiments}

Now we discuss in some detail the effect of second-order contributions that are generated by nature from the primary, first-order waves. Let $\eta_{1}$ denote a measured wave signal \index{experiments!measured signal} at the position $x = x_{1}$ m, where $x_{1}$ is one of the positions of the wave gauges. We assume that the measured wave signal $\eta_{1}$ consists of first-order (linear) harmonic waves, second-order double harmonic waves, and second-order non-harmonic long waves, given as follows:
\begin{equation}
\eta_{1}(t) = A_{1}(t) e^{-i\omega_{0}t} + \epsilon[B_{1}(t) e^{-2i\omega_{0} t} +  C_{1}(t)] + \textmd{c.c}.
\end{equation}
Note that $A_{1}$ in this expression is at the order of $\epsilon$ compared to $A$ in expression~\eqref{etaABC}. The second-order terms from the above expression appear from a generation through the nonlinear mode. This means that for a linear wave with frequency $\omega_{0}$ and complex amplitude $A_{1}$, the nonlinear effect will generate frequency $2\omega_{0}$ and zero frequency, with complex amplitudes $B_{1}$ and $C_{1}$, respectively. These complex amplitudes will be of the order of $A_{1}^{2}$. \index{experiments!setting} \index{extreme waves!experiments}

We now want to remove the second-order terms from the measured wave signal $\eta_{1}$ using an iteration method. After $N$ number of iterations, we obtain a reduced signal without the second-order contribution, denoted as $\eta_{N + 1}$. So, we write $\eta_{N + 1}(t) = A_{N + 1}(t) e^{-i \omega_{0} t} + \textmd{c.c}$. In the following, we will call the measured signal in which the second-order contribution has been removed the `experimental signal'. \index{experiments!experimental signal} \index{experiments!setting} \index{extreme waves!experiments}

The first step is to find the complexification of the measured wave signal $\eta_{1}$ by applying the Hilbert transform\index{Hilbert transform} $\mathcal{H}$ as given in~\eqref{Hilbert} on page~\pageref{Hilbert}. Initially, we assume that the measured signal has only the first-order harmonic contribution and therefore we can extract the complex amplitude $A_{1}$ as in~\eqref{etaABC} on page~\pageref{etaABC} but now with additional factor $e^{i k_{0} x_{1}}$. Then with this $A_{1}$, the complex amplitudes of the second-order double harmonic $B_{1}$ and the second-order non-harmonic $C_{1}$ are calculated. We obtain the complexified signal $\eta_{2}$ by subtracting the second-order contribution from $\eta_{1}$. By adding the complex conjugate, we obtain the wave signal $\eta_{2}$. By repeating the same procedure again for $N$ times, we now obtain the measured wave signal without the second-order contribution $\eta_{N + 1}$. An illustration for removing the second-order contribution is given in Algorithm~\ref{algorithm}. \index{extreme waves!experiments}
\begin{algorithm}		\label{algorithm}		
\begin{flushleft}
  [Removing the second-order contribution from a measured signal] \\
  \qquad $\omega_{0} = $ \texttt{3.7834 rad/s} \\
  \qquad $\eta_{1} \equiv $ \texttt{measured wave signal}\\
  \qquad \texttt{For} $j = 1$ to $N$ \\
  \begin{center}
  \begin{tabular}{lr}
    $\mathcal{H}[\eta_{j}]$ & \texttt{the Hilbert transform of $\eta_{j}$} \\
    $(\eta_{j})_{\mathbb{C}} = (\eta_{j} + i\mathcal{H}[\eta_{j}])$ & \texttt{the complexification of $\eta_{j}$} \\
    $A_{j} = (\eta_{j})_\mathbb{C}/(2 e^{-i \omega_{0} t})$ & \texttt{the complex amplitude $A$} \\
    $B_{j} = k_{0}c A_{j}^{2}/[2\omega_{0} - \Omega(2k_{0})]$ & \texttt{the complex amplitude $B$} \\
    $C_{j} = c|A_{j}|^{2}/[\Omega'(k_{0}) - \Omega'(0)]$ & \texttt{the complex amplitude $C$} \\
    $(\eta_{j + 1})_{\mathbb{C}} = (\eta_{j})_{\mathbb{C}} - \epsilon(B_{j} e^{-2i\omega_{0}t} + C_{j})$ &
    \texttt{the complexification of $\eta_{j + 1}$} \\
    $\eta_{j + 1} = (\eta_{j + 1})_{\mathbb{C}} $ + c.c. & \texttt{wave signal $\backslash \; 2^{\textmd{nd}}$ order effect}\\
  \end{tabular}
  \end{center}
  \qquad \texttt{end}
\end{flushleft}
\end{algorithm}
\index{extreme waves!experiments}

\begin{figure}[h]			
\begin{center}
\includegraphics[width = 0.45\textwidth]{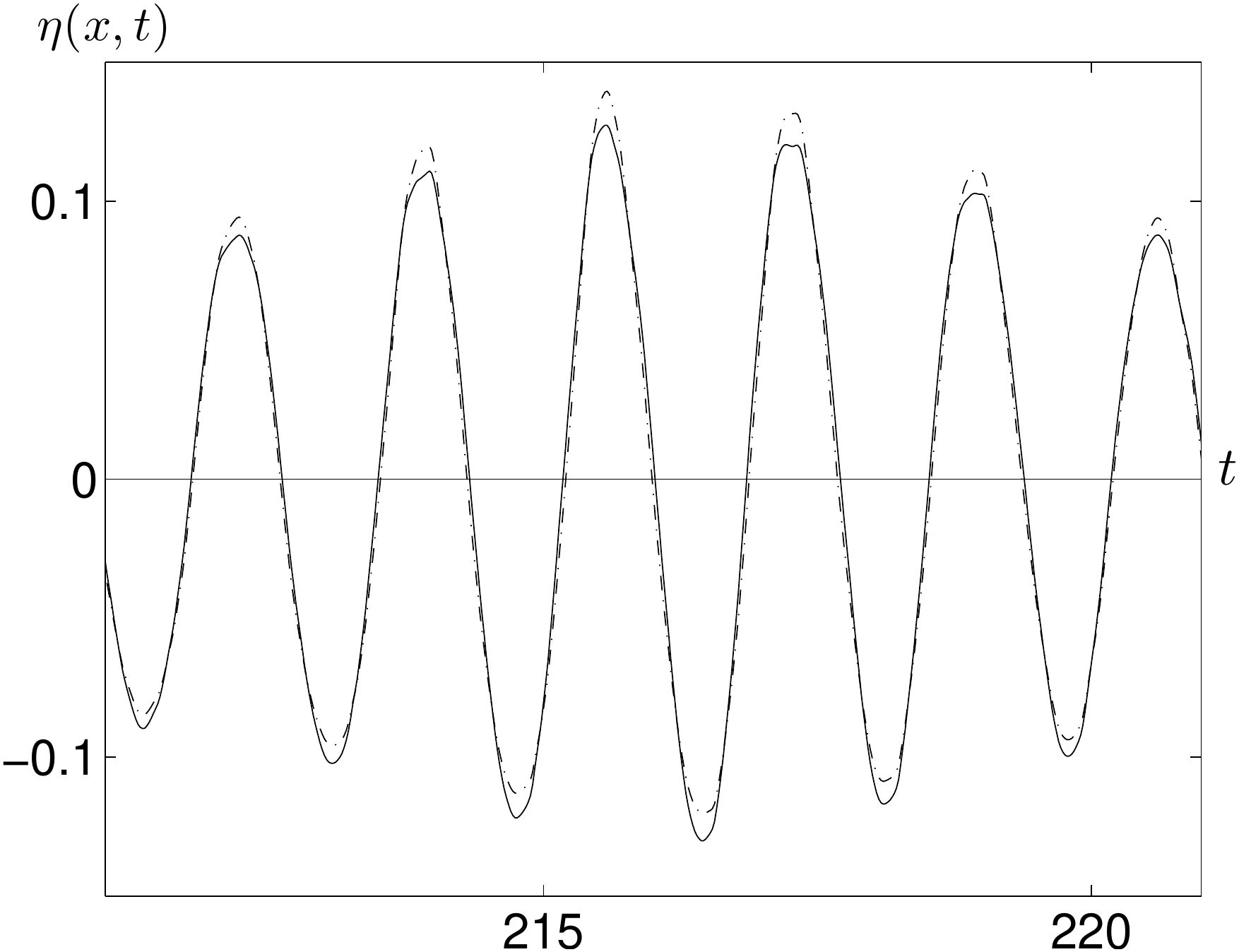}    \hspace{0.75cm}
\includegraphics[width = 0.45\textwidth]{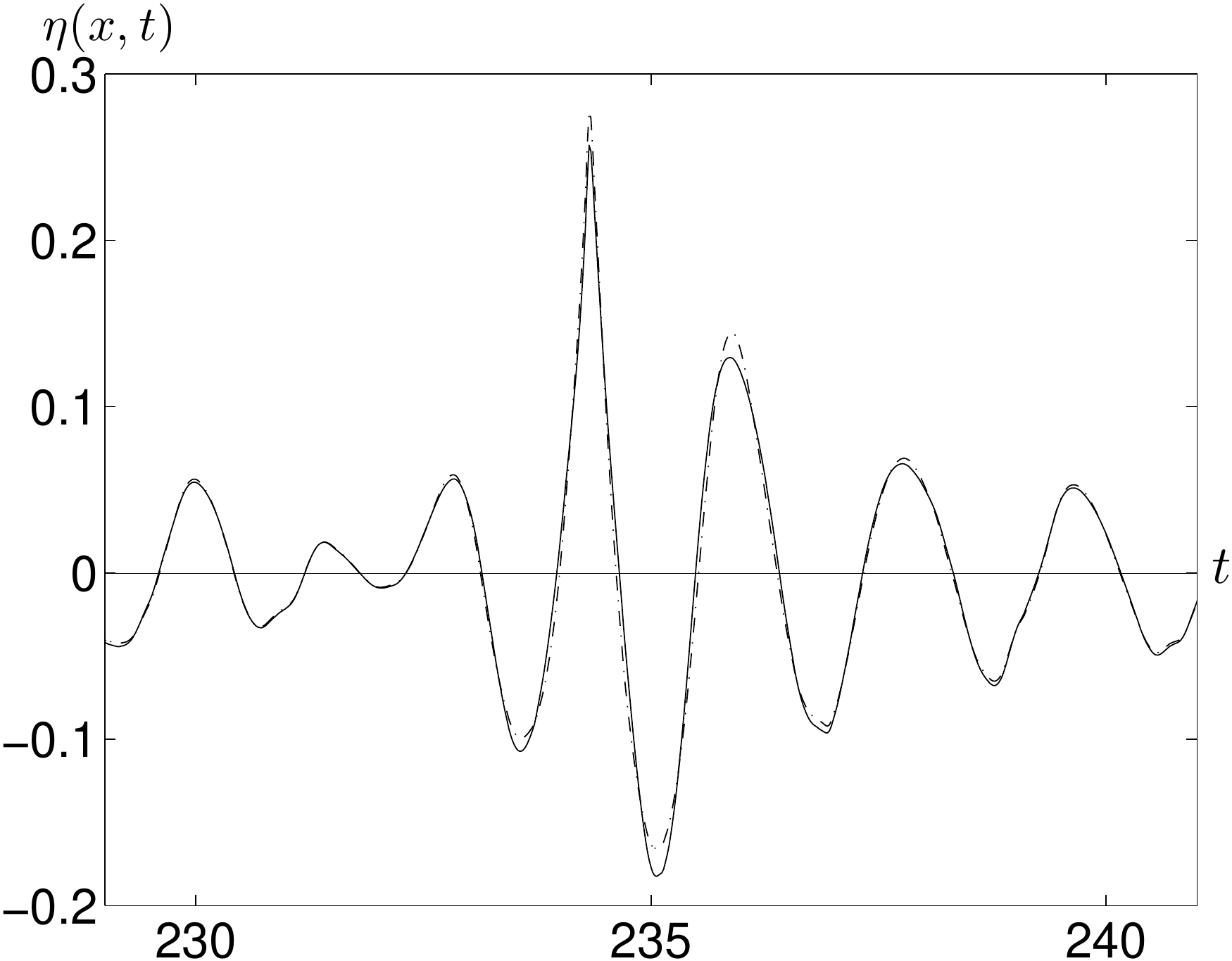}
\caption[Comparison of measured and experimental signals]{A comparison of a measured wave signal (dashed-dot) and the experimental signal (solid), that is the same wave signal without the second-order effect at 10~m (left) and 150~m (right) from the wavemaker. In this example, the norm of the difference between two final iterations is chosen less than $10^{-15}$, so it is sufficient to take the number of iteration as $N = 5$.}		  \label{2ordersignal}
\end{center}
\end{figure}

For the convergence criterion, we calculated the $L^{2}$-norm of the difference $\eta_{N + 1} - \eta_{N}$. The iteration stops for $\|\eta_{N + 1} - \eta_{N}\| < \delta_{0}$, \label{delta0} where $\delta_{0} > 0$ is a small parameter. Choosing $\delta_{0} = 10^{-6}$ gives the minimum number of iterations $N = 2$. In Figure~\ref{2ordersignal}, we choose $\delta_{0} = 10^{-15}$, so that the number of iterations $N = 5$ is needed to remove the second-order terms and to call the resulting signal the experimental signal. \index{experiments!experimental signal} \index{experiments!setting}

We remark that a second-order contribution gives the Stokes effect\index{Stokes effect}, which makes the crest steeper and the trough flatter: the second-order effect has a tendency to increase the crest and to decrease the trough. Therefore, the experimental signal in which the second-order terms removed has a lower crest and higher trough than the measured signal, see Figure~\ref{2ordersignal}. \index{extreme waves!experiments}
\index{experiments!setting} \index{extreme waves!experiments}

\subsection{Experimental wave parameters}
\index{experiments!wave parameters}

Once we have chosen the extreme position, we have to select a wave signal as input for the wavemaker. The selection of input is done using the information from the maximum temporal amplitude (MTA)\index{MTA} of the SFB that has been discussed in Subsection~\ref{MTAsubsection}. A snapshot showing the SFB wave profile, its envelope, and its MTA is given in Figure~\ref{MTASFB} of Section~\ref{SectionpropertiesSFB}. The MTA tells where the wave signal reaches an extreme condition. By taking the SFB solution backward to the wavemaker position,\index{wavemaker} we know theoretically the initial wave signal that, when fed to the wavemaker, will eventually become an extreme signal at the desired position. So, the movement of the wavemaker is specified according to this theoretical SFB signal. We say `theoretically' since we use the NLS equation\index{NLS equation} as a model for wave evolution. In practice and in reality, however, nature guides the wave evolution in a more complicated way, and many factors such as noise, generated currents, and reflected waves can cause deviations from the theoretical findings. \index{experiments!setting} \index{extreme waves!experiments}
\begin{table}[h!]			
\vspace*{-0.6cm}
\caption{Basic parameter values for the experiments}		  \label{basicdesign}
\begin{center}
\begin{tabular}{@{}lr@{}}
\toprule
Parameter & Value \\
\hline
carrier wave frequency $\omega_{0}$ (rad/s)                                			& 3.7284   			\\
carrier wave period $T_{\textmd{c}} = 2\pi/\omega_{0}$ (sec)               			& 1.6852   			\\
carrier wavenumber $k_{0} = K(\omega_{0})$ (m$^{-1}$)                      			& 1.4171   			\\
carrier wavelength $\lambda_{0} = 2\pi/k_{0}$ (m)                          			& 4.4337   			\\
extremal position $x_{\textmd{max}}$ (m)                                   		  	& 150      			\\
dispersive coefficient of the NLS equation $\beta$                         		  	& 1.013    			\\
nonlinear coefficient of the NLS equation $\gamma = \gamma_{\textmd{des}}$ \qquad 	& \qquad 230.2496 	\\
\bottomrule    
\end{tabular}
\end{center}
\end{table}

\vspace*{-0.6cm}
Table~\ref{basicdesign} shows the basic parameters of the carrier wave and the coefficients of the NLS equation with which the SFB solutions will be calculated. These parameters belong to one experiment that we will consider in more detail below. We designed a number of experiments with different modulation frequencies $\tilde{\nu}$ and different maximum amplitude $M$. Although experiments were also performed with $\tilde{\nu} = 1/\sqrt{2}$ and $\tilde{\nu} = \sqrt{3/2}$, better results were obtained for $\tilde{\nu} = 1$. Table~\ref{designM} gives additional values of the parameters that determine the SFB solution; we will refer to this set of parameters as test~A. Knowledge of the steepness at the extreme position can be used as a breaking criterion. Using breaking criteria based on the crest particle velocity exceeding the phase speed, the breaking limit of the two-dimensional Stokes wave is known to occur at a wave steepness\index{steepness} $s = ak = 0.443$ and crest angle 120\grad, where $a$ is wave amplitude and $k$ is wavenumber~\citep{6Stokes1880, 6Michell1893, 6Havelock18}. Hence, based on this information from the design, it is expected that the resulting waves will not break. \index{extreme waves!experiments}
\begin{table}[h!]			
\vspace*{-0.6cm}	
\caption{Design parameters for test~A}			  \label{designM}
\begin{center}
\begin{tabular}{@{}lr@{}}
\toprule 
Parameter & Value \\
\hline 
maximum amplitude $M$ (cm)                    & 21.3000 \\
asymptotic amplitude $2r_{0}$ (cm)            &  8.8227 \\
initial steepness $s_{0} = (2r_{0}) k_{0}$    &  0.1250 \\
extremal steepness $s_{M}$                    &  0.3019 \\
normalized modulation frequency $\tilde{\nu}$ &  1.0000 \\
modulation frequency $\nu$ (sec$^{-1}$)       &  0.3114 \\
modulation period $T = 2\pi/\nu$ (sec)        & 20.1800 \\
number of waves in one modulation period $N$  & 11.9747 \\
\bottomrule
\end{tabular}
\end{center}
\end{table}
\index{extreme waves!experiments} \index{experiments!setting}

\section{Qualitative comparisons}
\index{experiments!comparisons!qualitative}

The first observations after analyzing all experimental signals can be formulated as follows: \textsl{All experimental signals\index{experiments!experimental signal} show the typical development of the modulational instability\index{modulational instability} as described by the SFB solution \index{SFB!experiments} of the NLS equation.\index{NLS equation} Moreover, it is observed that the carrier wave frequency and the modulation frequency are conserved very accurately during downstream\index{downstream} evolution, keeping the same values as determined by the wavemaker motion.} \index{extreme waves!experiments}

\begin{figure}[htbp]			
\begin{center}
\includegraphics[height = 0.9\textheight]{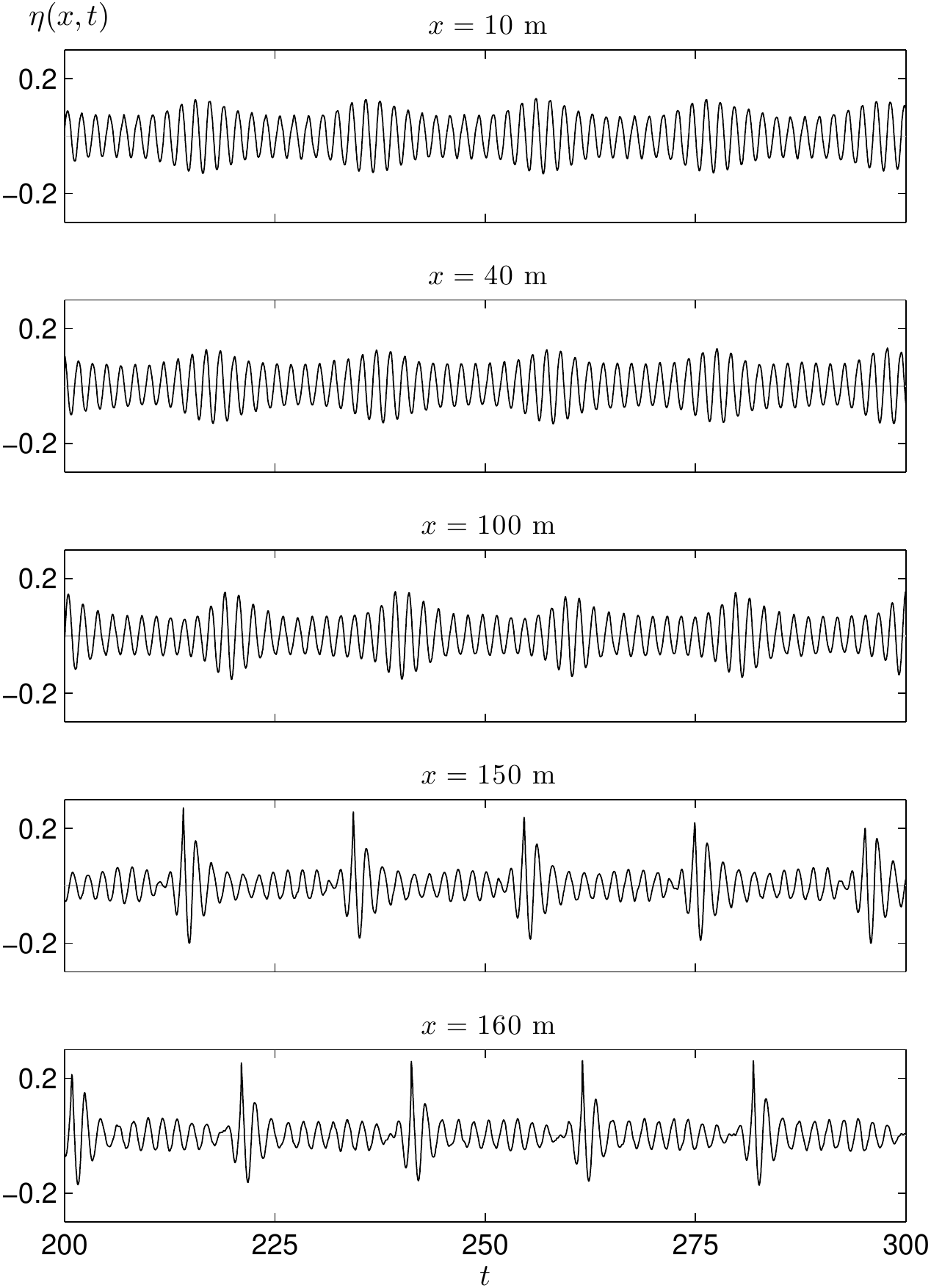} 	
\caption[Downstream evolution of the experimental signal]{Plots of the experimental wave signal at different positions. Wave amplitude increase is observed as the waves propagate further from the wavemaker towards the extremal position which is near 160~m.}    \index{experiments!comparisons!qualitative}    \label{Signalxjalan}
\end{center}
\end{figure}

Recall again the MTA plot of the theoretical SFB as given in Figure~\ref{MTASFB}. Although we do not have continuous data from the experiments, the experimental signals show the amplitude amplification in the first part of the downstream running waves. After that, a monotone decrease is observed. Therefore, the qualitative characteristic shape of the MTA is a robust phenomenon.\index{MTA} Figure~\ref{Signalxjalan} shows the experimental signals from test~A at several positions from the wavemaker. \index{extreme waves!experiments}

\subsection{Symmetry property} \label{symmetryproperty}
\index{experiments!symmetry property} \index{extreme waves!experiments}

In Figure~\ref{SFB_evolution} of Section~\ref{SectionSFBevol}, we have seen that an SFB wave signal maintains a symmetric structure\index{SFB!symmetric signal} within one modulation period as it propagates toward the extreme position. This is not surprising since from the theoretical SFB, expression~\eqref{SFBexpression} on page~\pageref{SFBexpression}, the complex amplitude $A$ is symmetric with respect to $\tau = 0$, namely $A(\xi,-\tau) = A(\xi,\tau)$. Furthermore, if we move with respect to the group velocity \index{group velocity} $V_{0}$, the real amplitude $a$ of expression~\eqref{physicaleta} on page~\pageref{physicaleta} is also symmetric with respect to $t = 0$, i.e., $a(x,-t) = a(x,t)$. The term symmetry in this context refers to the envelope, which is symmetric around the middle of the modulation period. Since the ratio of $\nu/\omega_{0}$ generally is not an integer, the carrier wave shifts a little bit from one modulation to the other. Consequently, the maximum amplitude will differ from one wave group to another and thus the signal is not really symmetric. \index{experiments!comparisons!qualitative} \index{extreme waves!experiments}

\begin{figure}[h]			
\begin{center}
\includegraphics[width = 0.45\textwidth, viewport = 29 218 551 631]{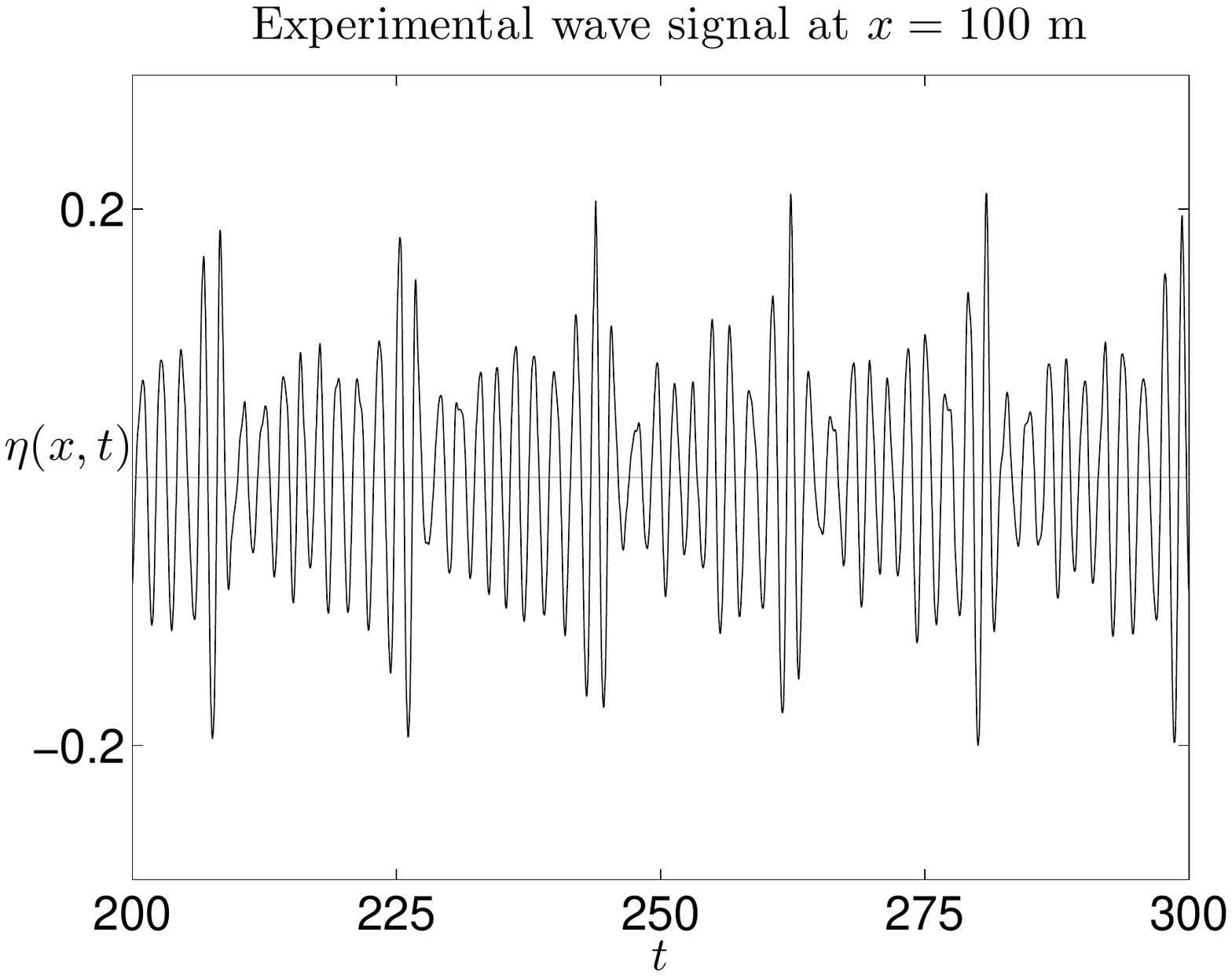}	    \hspace{0.75cm}
\includegraphics[width = 0.45\textwidth, viewport = 27 218 551 631]{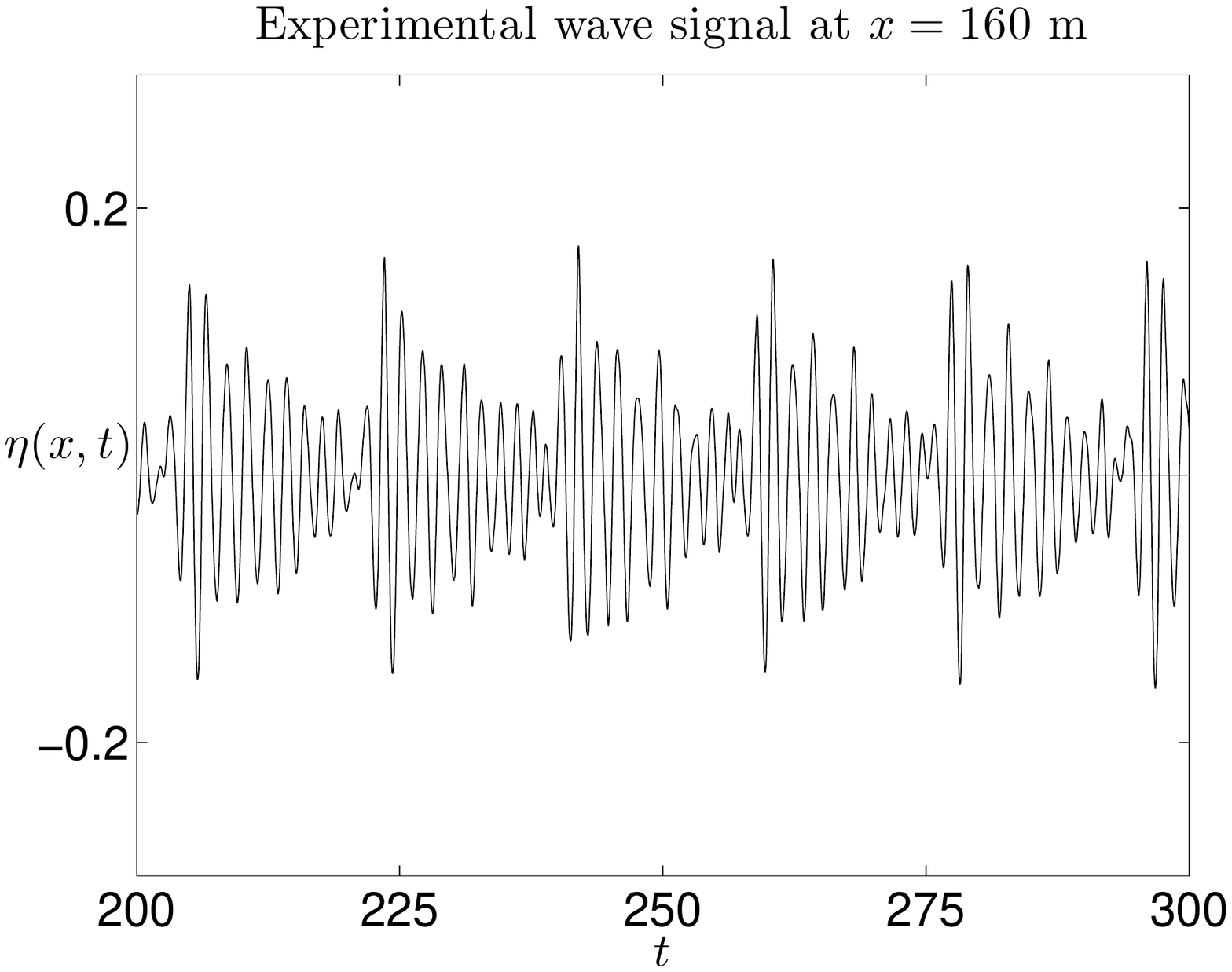}
\caption[Asymmetric structure in the experimental signal]{Plot of experimental wave signals from  test~B at two different positions $x = 100$~m and $x = 160$~m. This plot shows two important characteristics of experimental signals: the waves in between the extremal waves are not symmetric; before the wave group has passed the extreme position, the waves in one wave group are higher immediately after the highest wave, and smaller at the rear. When the wave group has passed the extreme position the waves after the largest wave are small and larger waves are found at the rear of the wave group.}		    \label{asymmetricsignals}
\end{center}
\end{figure}

\index{experiments!experimental signal} In contrast, however, the symmetry is lost in the experiments. We observe that the experimental wave signals show a deformed wave group\index{wave group(s)} structure with a steep front and a flat rear within one modulation period. This asymmetric structure is observed when the waves propagate sufficiently far from the wavemaker but before reaching the extreme position. This occurs in the region where waves experience amplitude amplification and an increase of the MTA. After the extreme position, the asymmetry becomes reversed: the experimental wave signals flatten at the front and steepen at the rear within one modulation period.\index{experiments!asymmetric signal} Figure~\ref{asymmetricsignals} shows an example of wave signal plots from test~B at two different positions with asymmetric structure. Test~B has design characteristics of maximum amplitude of $M = 21.30$~m and the normalized modulation frequency $\tilde{\nu} = \sqrt{3/2}$. \index{extreme waves!experiments}
\index{experiments!comparisons!qualitative} \index{extreme waves!experiments}

\subsection{Argand diagram representation} \label{Argandcomparison}
\index{experiments!Argand diagram}

\begin{figure}[h]			
\begin{center}
\includegraphics[width = 0.99\textwidth]{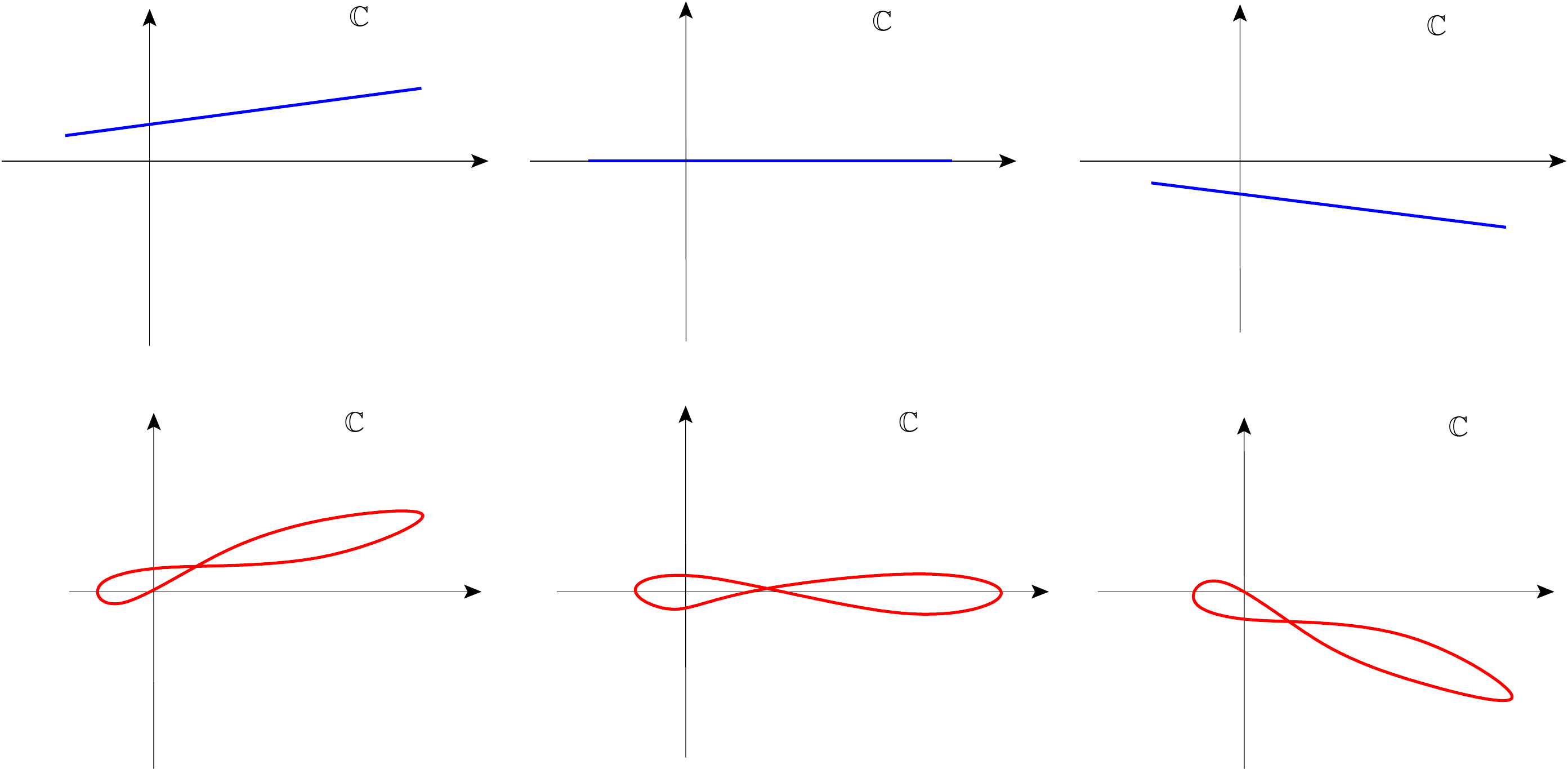}		
\caption[Sketch of a perturbed SFB in the Argand diagram]{(Top) Plots of the evolution corresponding to the SFB solution in the Argand diagram before, at, and after the extreme position. (Below) Qualitative sketches of a perturbed SFB around the extreme position showing twisted elliptical curves.}	 \label{Argandperturb}
\end{center}
\end{figure}

We have remarked and shown in Subsection~\ref{symmetryproperty} above that the symmetry within one modulation period of the envelope of the experimental wave signals is lost and that asymmetry is reversed when passing the extreme position. In this subsection, we describe the evolution of both theoretical and experimental wave signals in the Argand diagram (the complex plane). This comparison is new and gives a better understanding of the experimental signals. We have discussed in Subsection~\ref{evolargand} that the evolution of the SFB wave signal in the Argand diagram is a set of straight lines. In fact, returning back and forth in one period along the same curve is not a robust property. Any arbitrary perturbation will deform the path from a straight line into an elliptical curve, possibly even a twisted elliptical curve. This change will disturb the symmetry of the envelope wave signals. Figure~\ref{Argandperturb} shows a schematic perturbation of the straight lines in the Argand diagram. In this figure, the depicted plots are given near the extreme position. The clockwise rotation of the curves corresponds with the increasing position. For SFB, one pair of phase singularities occur at the extreme position, when the curve lies on the real axis. For the perturbed SFB, phase singularities occur before and after the extreme position. \index{extreme waves!experiments}
\index{experiments!Argand diagram}	\index{experiments!comparisons!qualitative}

\begin{figure}[h!]			
\begin{center}
\subfigure[]{\includegraphics[width = 0.275\textheight, viewport = 36 211 537 612]{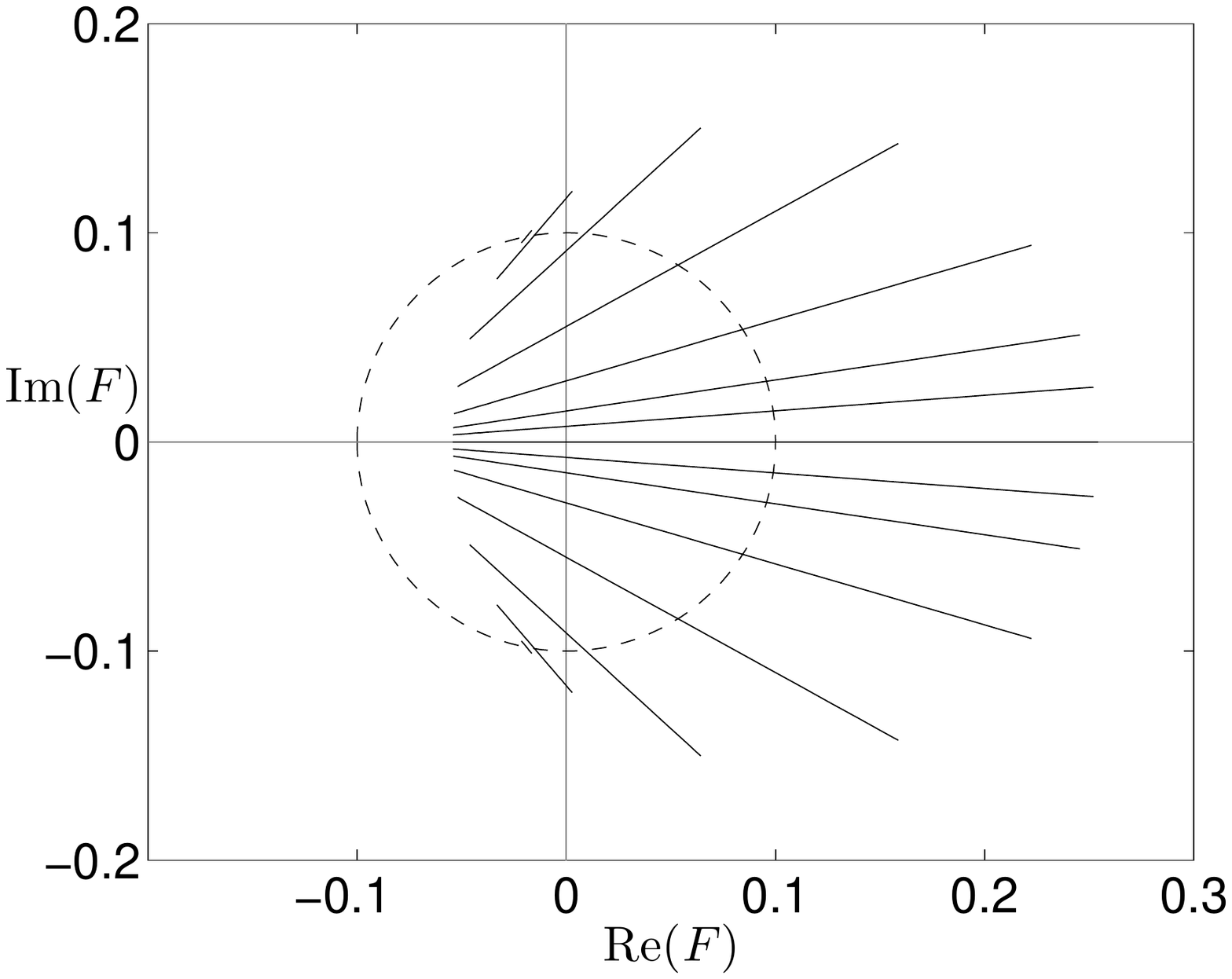}}   			\hspace{0.5cm}
\subfigure[]{\includegraphics[width = 0.275\textheight]{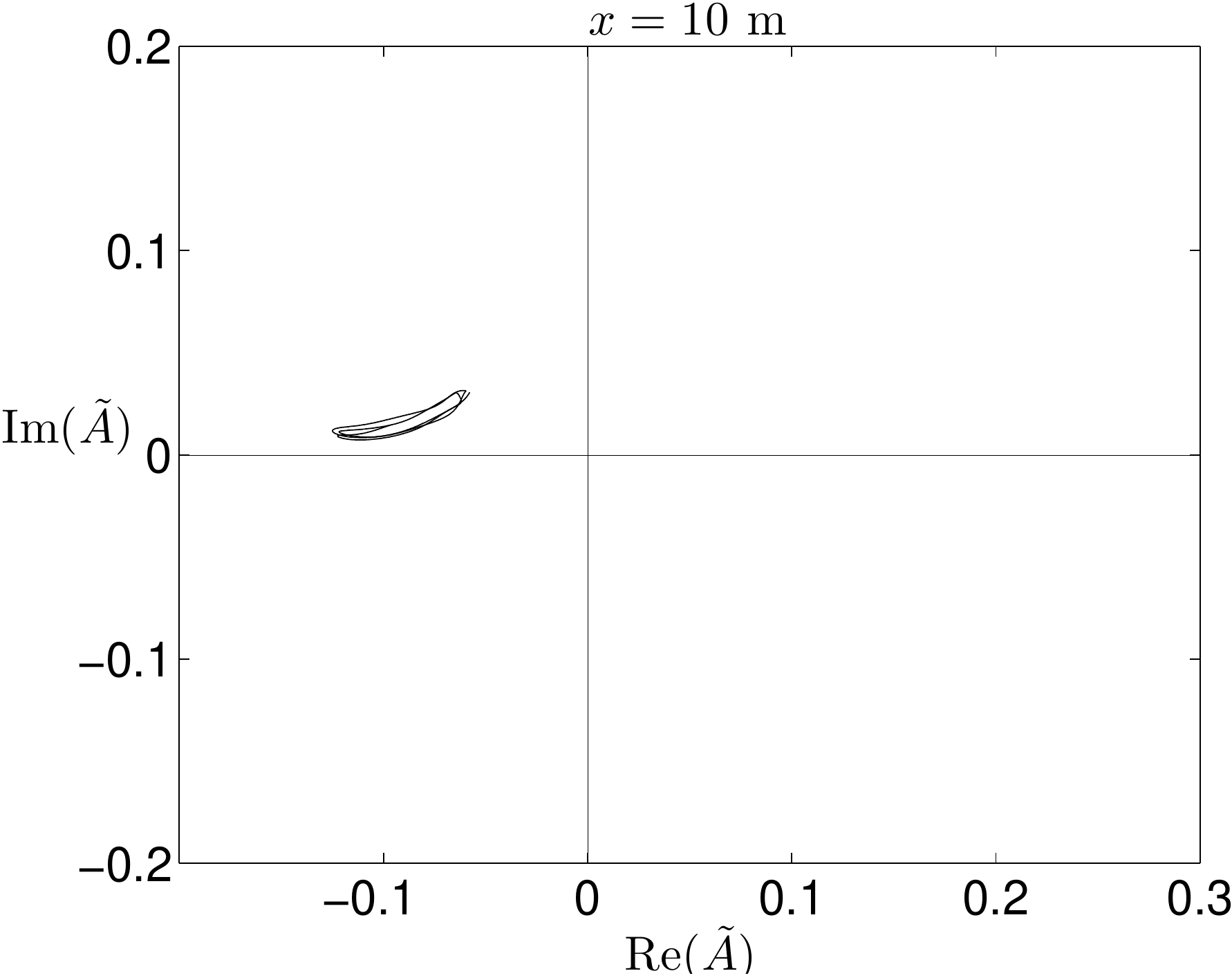}}   	\vspace{0.1cm} \\
\subfigure[]{\includegraphics[width = 0.275\textheight]{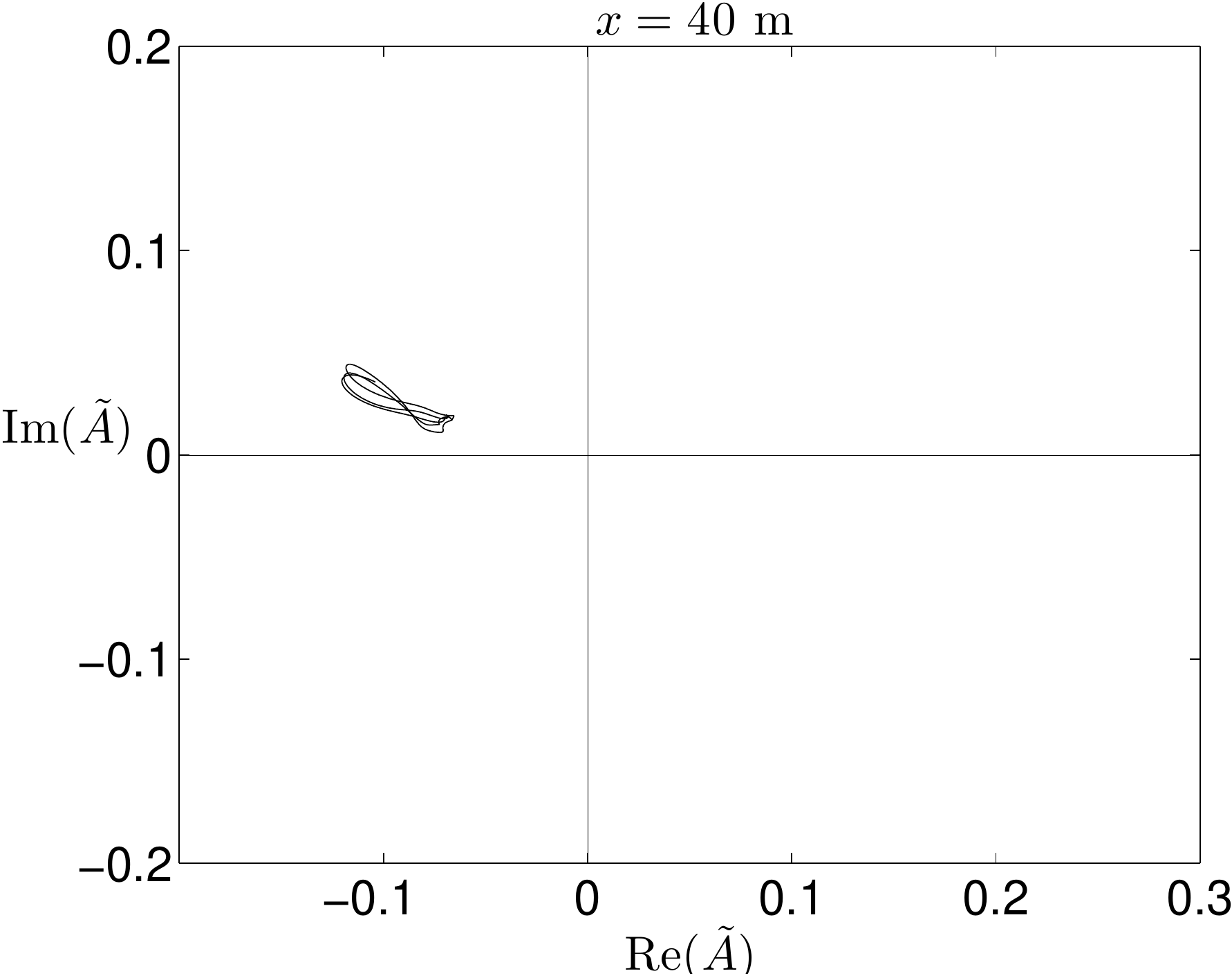}}   	\hspace{0.5cm}
\subfigure[]{\includegraphics[width = 0.275\textheight]{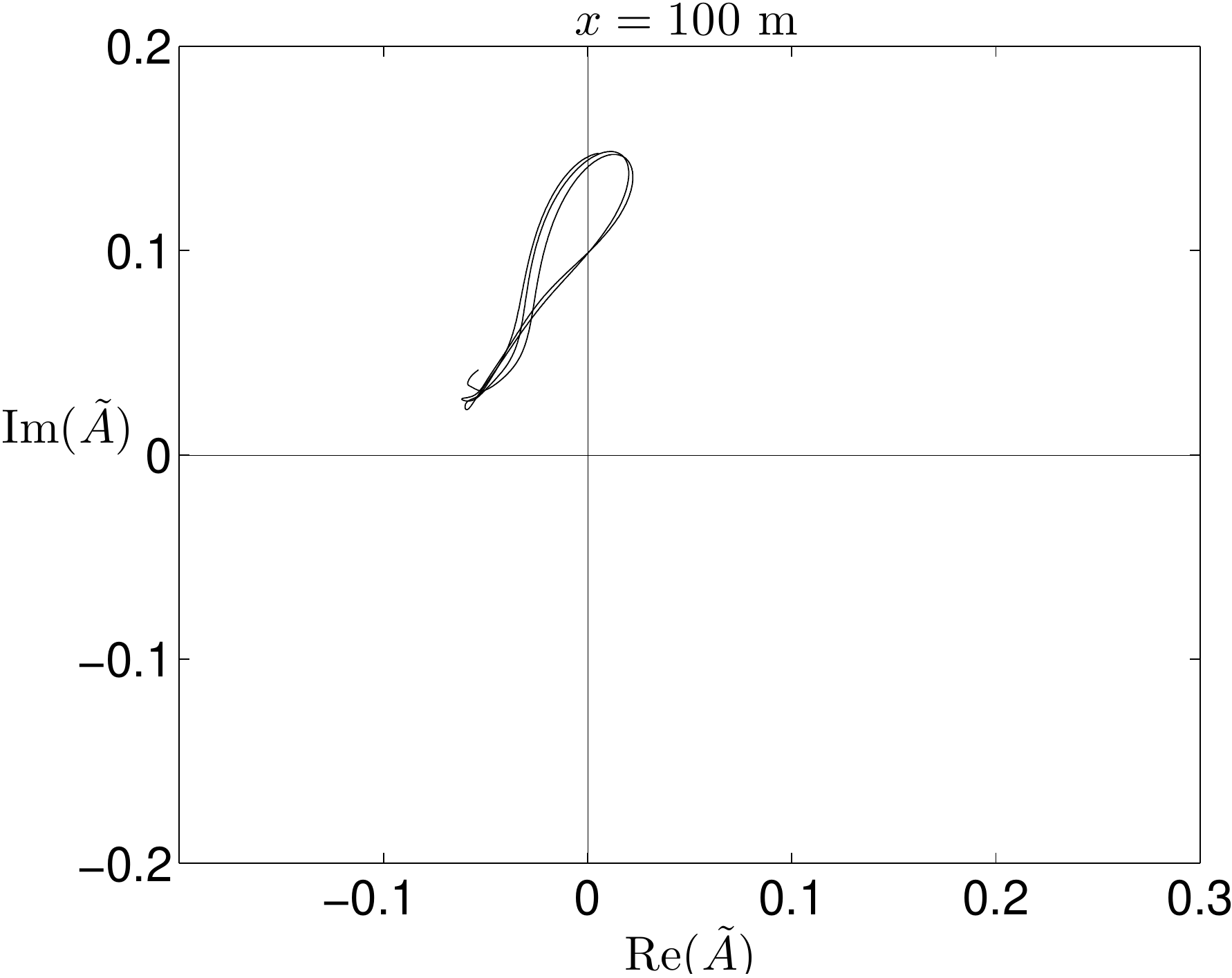}}   	\vspace{0.1cm} \\
\subfigure[]{\includegraphics[width = 0.275\textheight]{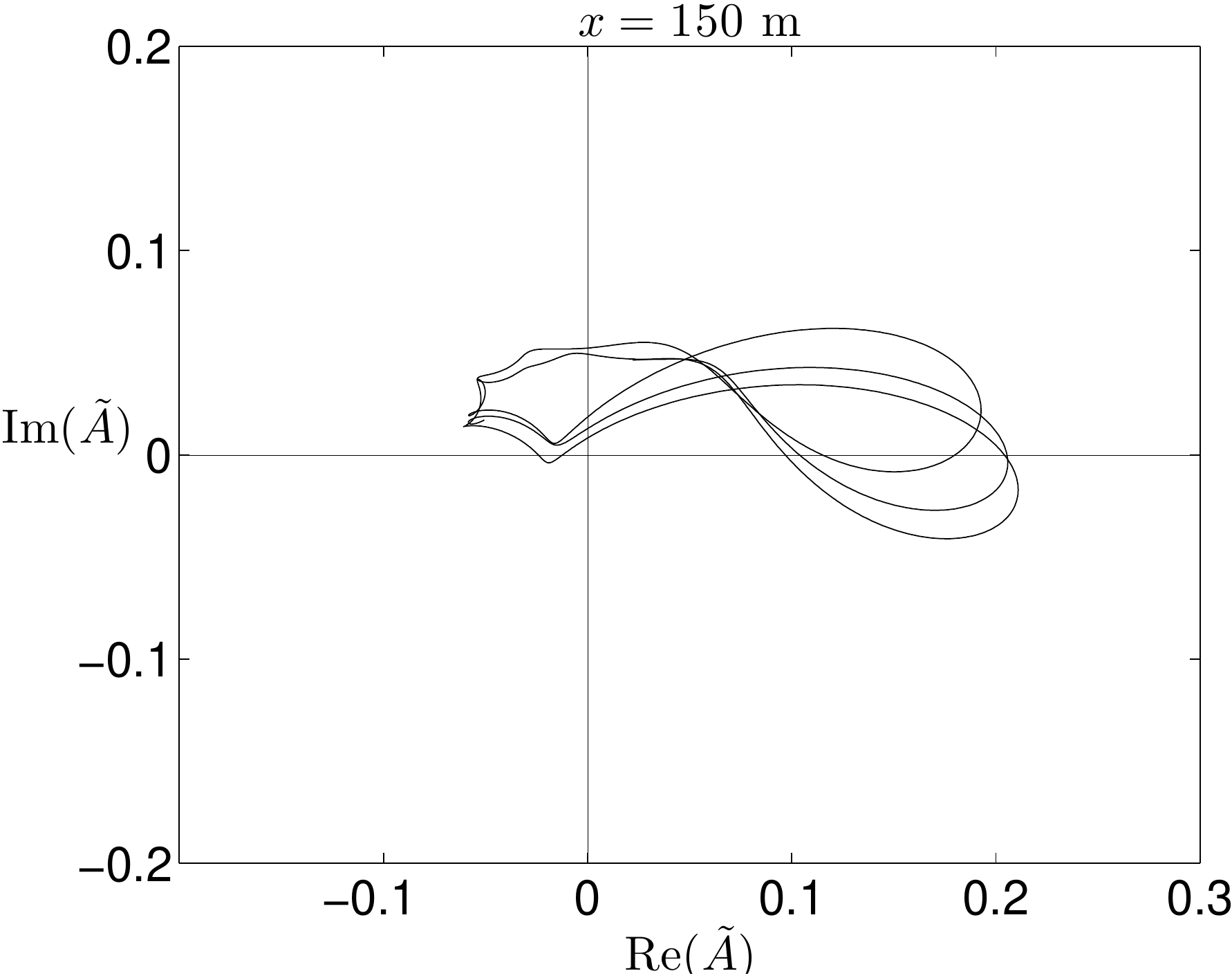}}   	\hspace{0.5cm}
\subfigure[]{\includegraphics[width = 0.275\textheight]{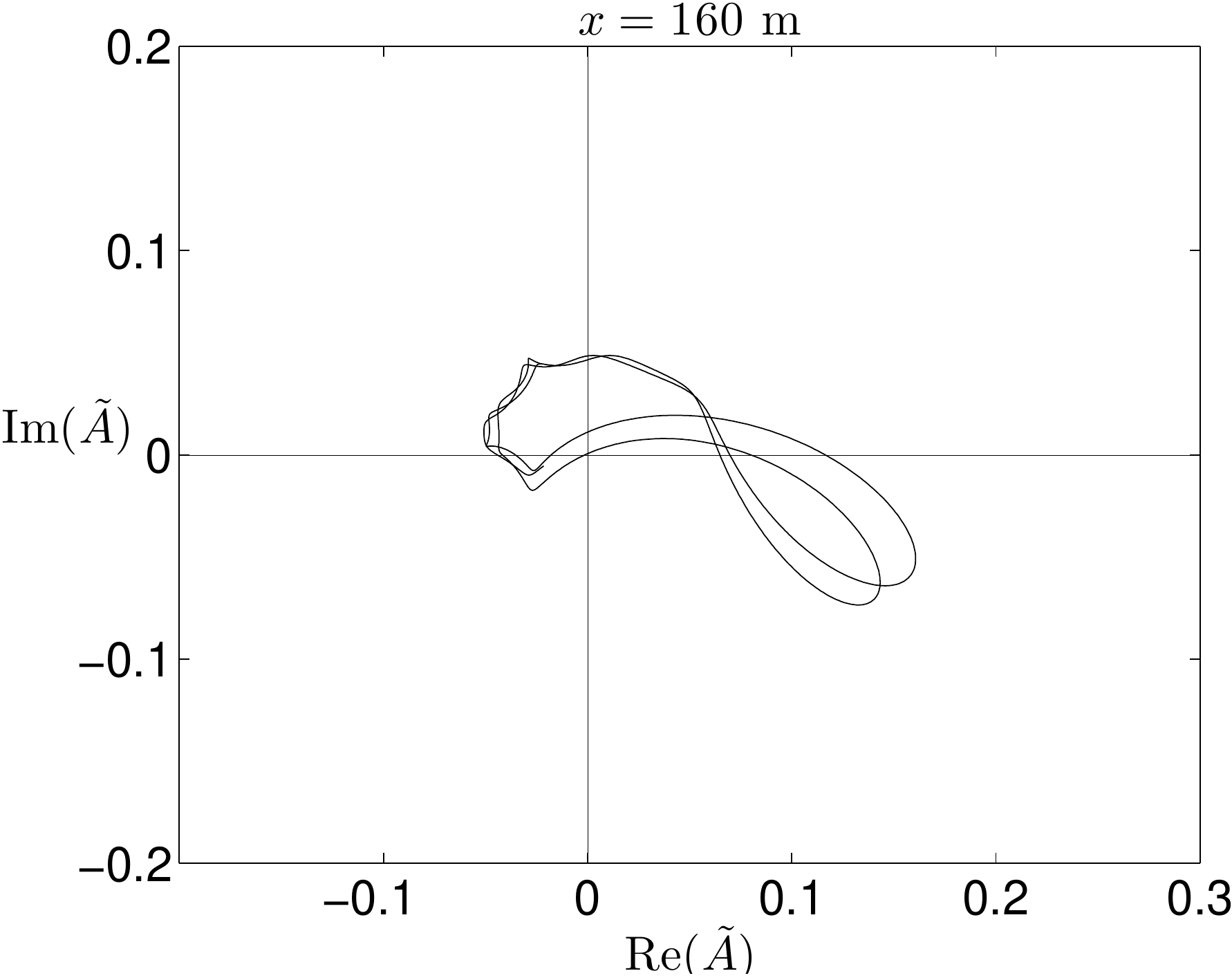}}
\caption[Comparison of evolutions in the Argand diagram]{Plots of the SFB (a) and experimental complex-valued amplitudes for test~A (b)--(f) in the Argand diagram after removing an oscillating part. The plots are parameterized in time at several positions. For the experiment, the following positions are the distance from the wavemaker (b) 10~m, (c) 40~m, (d) 100~m, (e) 150~m, and (f) 160~m.}				    \label{compareargand}
\end{center}
\end{figure}

We now compare the complex-valued amplitudes of the SFB and the experiments without the oscillating part from the plane-wave solution of the NLS equation. By writing the complex-valued amplitude of the SFB\index{SFB} signal as in~\eqref{solutionA}, $A(\xi,\tau) = A_{0}(\xi) F(\xi,\tau)$, we make plots of $F \in  \mathbb{C}$ for the theoretical result in the Argand diagram. For the experimental signal, the plots are parameterized in time at several positions. For this, we take its Hilbert transform\index{Hilbert transform} and then multiply with a complex exponential (different at different positions) with a phase to account for the increase in position. Figure~\ref{compareargand} shows the evolution in the Argand diagram for both the SFB and the experiment. \index{extreme waves!experiments}
\index{experiments!Argand diagram}		\index{experiments!comparisons!qualitative}

\begin{figure}[h]			
\begin{center}
\includegraphics[width = 0.45\textwidth]{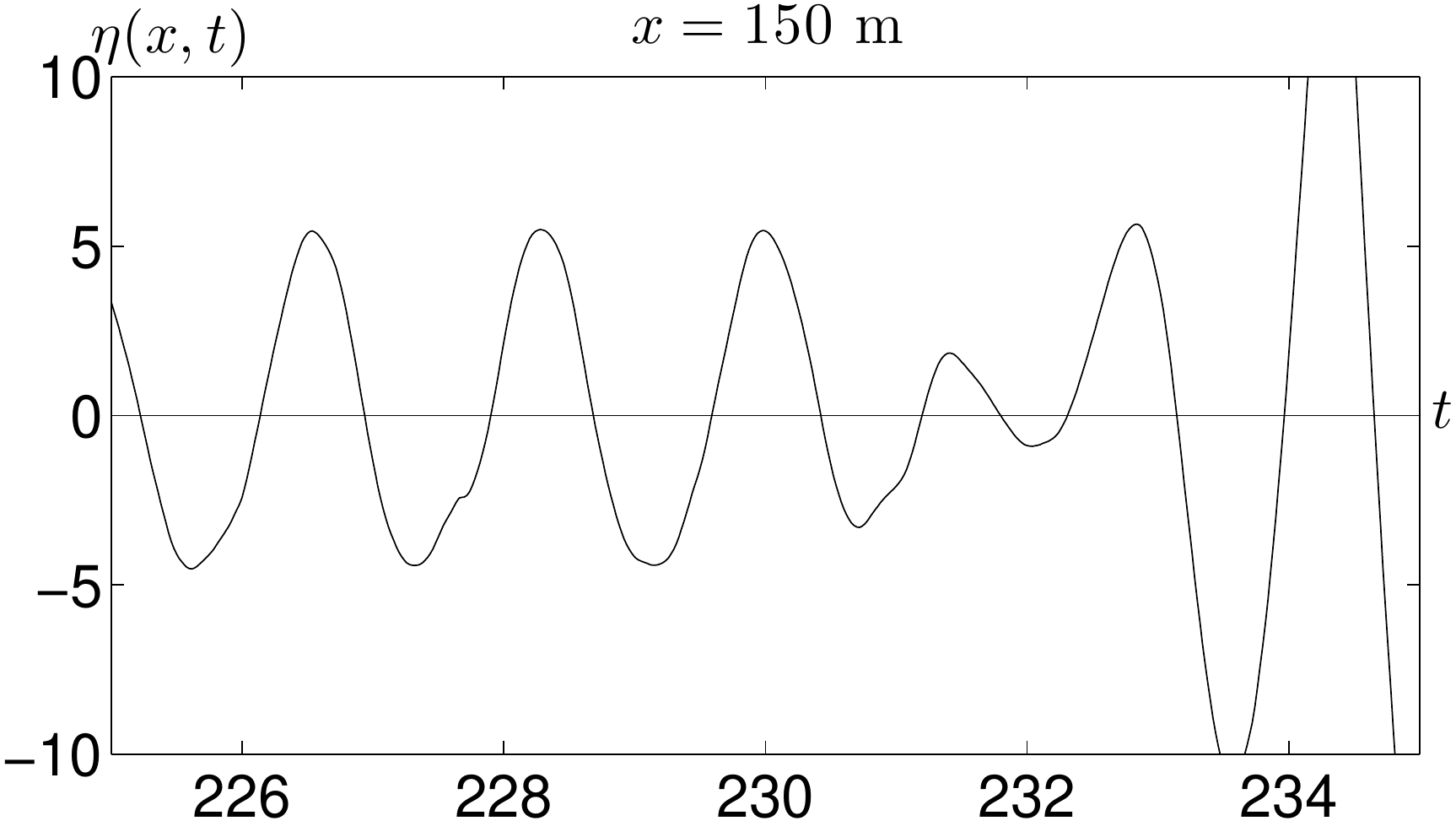}	    \hspace{0.75cm}
\includegraphics[width = 0.45\textwidth]{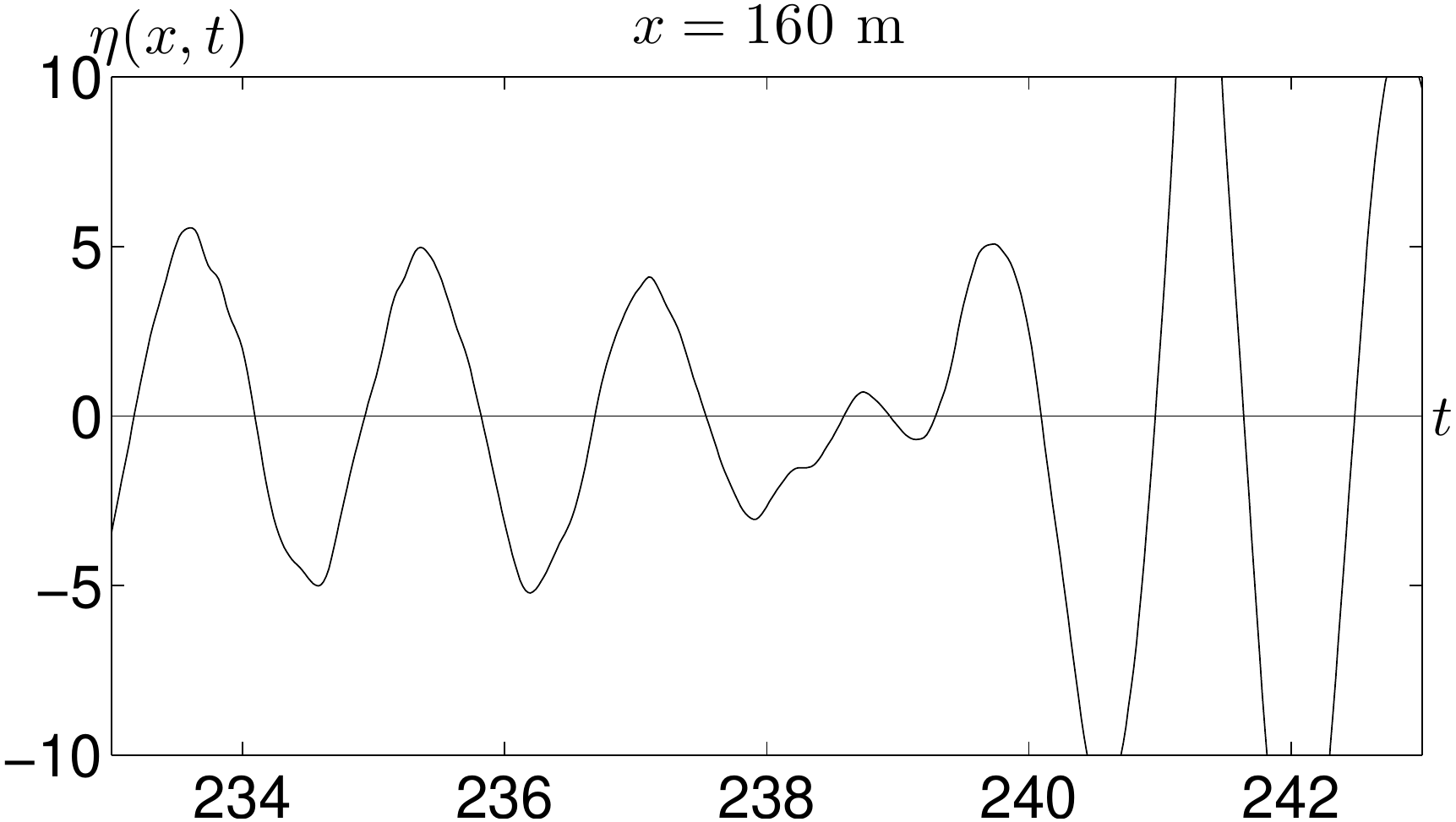}
\caption[Phase singularity in the experimental signal]{Magnification plots of the experimental signal at 150~m and 160~m from the wavemaker. The distance difference of 10~m and the group velocity of 1.32~m/s give the time difference of 7.8~sec in the occurrence of phase singularity. We observe that at a certain, time the wave period becomes as twice larger, indicating that we are at a phase singularity at which the local frequency is very large.} 		\label{zoomps}
\end{center}
\end{figure}

For the SFB, the evolution curves are a set of straight lines. At the extreme position, the line lies exactly on the real axes and crosses the origin twice. This is the reason why we have a pair of phase singularities in one modulation period only at the extreme position. The experimental results show different behavior of the evolution in the Argand diagram. They are not straight lines, but a set of twisted ellipse-like curves. This can be interpreted as a small perturbation of the straight lines of the SFB. \index{SFB!Argand diagram} As a consequence, the experiments do not show two phase singularities at one position, but rather one singularity at one position before the extreme position and one singularity at a place after the extreme position. This property is clearly related to the fact that the experimental signals are asymmetric. At one position, the singularity is immediately behind a large-amplitude wave while at another position the singularity is in front of that large-amplitude wave. Magnification plots of the experimental signal that show phase singularity are given in Figure~\ref{zoomps}. \index{experiments!phase singularity} \index{extreme waves!experiments}

\index{experiments!extreme waves} We discuss the special wave signals of both the theoretical SFB and the experimental results at or near the extreme position, the so-called extreme signals. \index{extreme signal} It should be noted that we do not know precisely the experimental extreme signal due to the limitations of measurement positions. Nevertheless, since the experiments show that the waves reach high amplitude and show extreme characteristics near 150--160~m from the wavemaker, we can use the wave signals at these positions for comparison with the theoretical SFB. \index{experiments!Argand diagram} \index{extreme waves!experiments}

We have discussed in Section~\ref{SectionSFBevol} the spatial evolution of the SFB to become the extreme signal. This extreme signal shows phase singularity when the amplitude vanishes, which occurs for $0 < \tilde{\nu} < \sqrt{3/2}$. In Section~\ref{WDwavegroups}, we have seen that the physical wave field shows wavefront dislocation, where waves merge and split. We have also shown that the Chu-Mei quotient is unbounded at a singular point and this is a generic property. Therefore, we expect that phase singularity also occurs in reality. With measurements at discrete positions only, we are not able to show completely that the experiments also show wavefront dislocation. However, the (almost) vanishing of the experimental signal at positions near 150~m, as in Figure~\ref{zoomps}, might be a strong indication of the appearance of these singularities. \index{experiments!Argand diagram} \index{experiments!comparisons!qualitative} \index{extreme waves!experiments}
\index{experiments!Argand diagram}		\index{experiments!comparisons!qualitative}

For the theoretical SFB at the extreme position, there is one pair of phase singularities in each modulation period. Due to the symmetry breaking in the experiments as described above, this property is lost in the experiments. As discussed earlier in this subsection, in the Argand diagram straight lines are deformed into elliptical curves under a small perturbation. This seems to describe what can be observed in the experiments. Indeed, phase singularities are observed in the sense of vanishingly small wave amplitudes at certain instants, one singularity before and another one after the experimental extreme position. This interpretation of the experiments is a strong indication that the generated wave fields indeed follow an evolution according to the theoretical SFB solution, albeit a bit perturbed. \index{experiments!Argand diagram} \index{experiments!comparisons!qualitative} \index{extreme waves!experiments}

\subsection{Phase plane representation}
\index{experiments!phase plane}

We now present phase plane representations for both the theoretical SFB and the experimental results. A phase plot is a plot of the complex-valued amplitude and its derivative parameterized by time. Since the complex-valued amplitude has real and imaginary parts, the phase plot will be a four-dimensional manifold. At the extreme position, the SFB is a real-valued function and therefore the phase plot can be shown in a plane. The experimental results, however, lead to a complexified signal that always contains a non-vanishing real and imaginary part. Since we do not know precisely the position where a phase singularity occurs in the experiments, it is rather difficult to make a comparison. However, we can still use the experimental signals close to or around the expected extreme position, which is near 150--160~m from the wavemaker. To compare the theoretical and the experimental signals, we took the absolute value of the complexified experimental signal. As a consequence, the phase plots are in the non-negative region right of the horizontal axis. As shown by the evolution curves in the Argand diagram, the experimental signals at these positions show that phase singularities are close by, since at some instant the amplitude almost vanishes. \index{experiments!phase plane} \index{experiments!comparisons!qualitative} \index{extreme waves!experiments}

\begin{figure}[h]			
\begin{center}
\subfigure[]{\includegraphics[width = 0.45\textwidth]{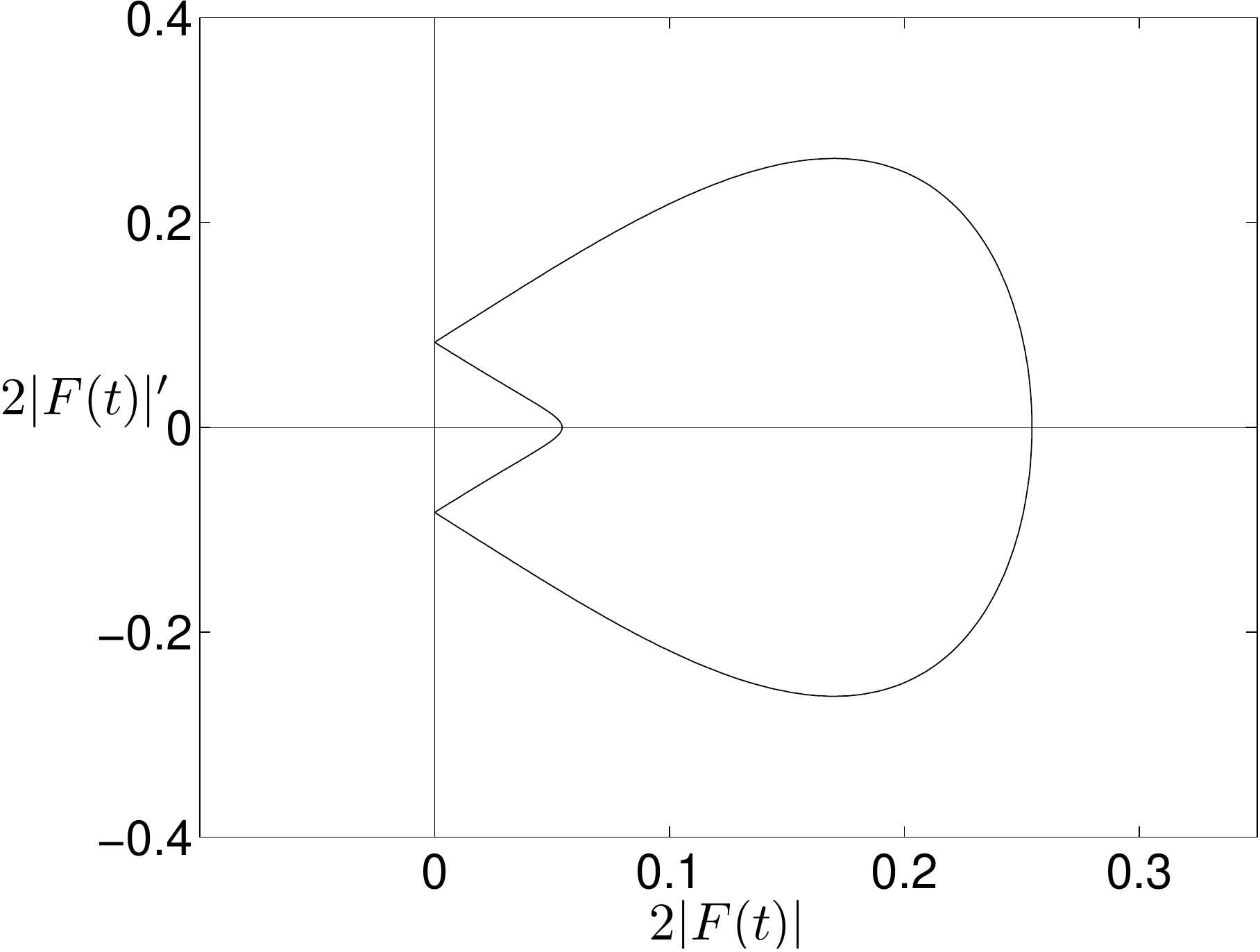}}	    	\hspace{0.75cm}
\subfigure[]{\includegraphics[width = 0.45\textwidth]{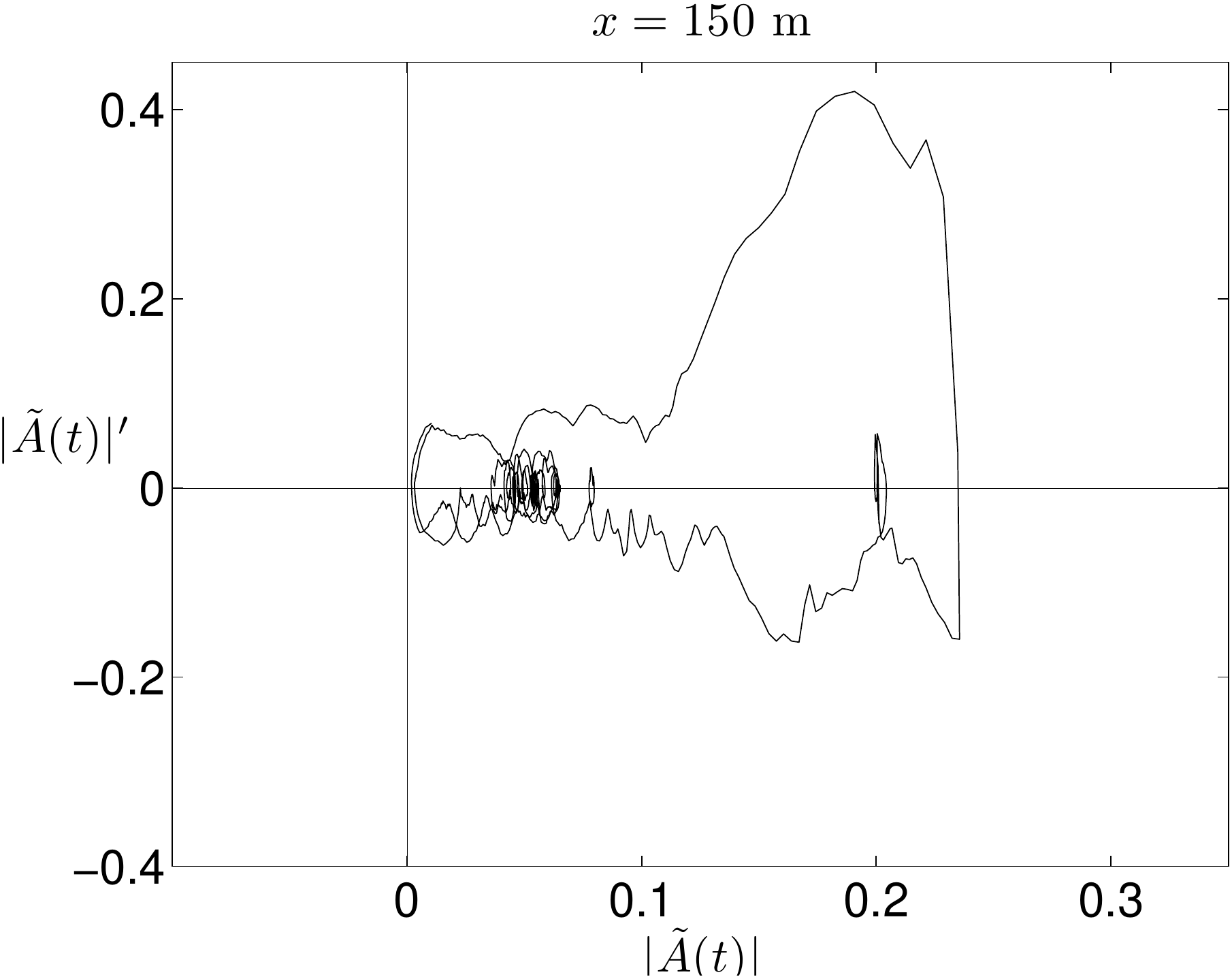}}
\caption[Phase curve comparison in the phase plane]{Phase plane plots of the SFB and the experimental envelope signals at and close by the extreme position. Observe that the SFB phase plot (a) is symmetric while the experimental signal phase plot (b) is asymmetric with respect to the horizontal axis.} \label{comparephaseplane} 			\vspace*{-0.2cm}
\end{center}
\end{figure}

Figure~\ref{comparephaseplane} shows the phase plane plots of the SFB and the experimental result. When looking at this plot of the experimental signal, it has to be kept in mind that calculation of the derivative of the absolute amplitude of the complexified experimental signal introduces additional inaccuracies when compared to the signal itself. Furthermore, the plot clearly shows the asymmetry within one modulation period. Despite the limited equivalence with the exact SFB representation, we like to remark that this kind of sensitive comparison of properties of theoretical and experimental signals appears to be novel.\index{experiments!phase plane} \index{extreme waves!experiments}
\index{experiments!phase plane}		\index{experiments!comparisons!qualitative}

\section{Quantitative comparisons}
\index{experiments!comparisons!quantitative}

In the previous section, we have described that some detailed characteristic properties of the SFB solutions are observable in the experimental signals. This is clear evidence that waves belonging to the SFB family were generated in the wave basin.\index{wave basin} Even though small perturbations developed during the propagation, most visibly the symmetry breaking, it can be concluded that the SFB family composes a robust class of surface wave fields when it concerns the gradual formation of wave groups\index{wave group(s)} that increase in amplitude when running towards the extreme position and decrease afterward. Above we made mainly qualitative comparisons between the theoretical SFB and the experimental signals. In this section, we will give some quantitative comparisons between these two signals. \index{experiments!comparisons!quantitative} \index{extreme waves!experiments} 

Since the theoretical SFB family possesses several parameters, it is not an obvious task to connect between the two wave signals. Indeed, in reality, these parameters could change during downstream\index{downstream} propagation in the wave basin, while for the theoretical model, these parameters are constant. Apart from this, differences between the model assumption and reality have to be taken into account. For instance, the theoretical model of the NLS equation deals with waves that propagate unidirectionally. In reality, however, there is an effect that waves reflect at the artificial beach. We assume that the effect of wave reflection is small and negligible. To make a quantitative comparison between the theoretical SFB and the experimental results, we have to find an SFB wave signal that resembles the experimental signal for the whole evolution. \index{experiments!comparisons!quantitative} \index{extreme waves!experiments} 

The comparisons presented in this section are based on choosing reasonable parameters that give a good agreement between the theoretical and the experimental signals. As mentioned already in the introduction of this chapter, we observed that the experimental signals maintain within experimental accuracy the values of the carrier wave frequency $\omega_{0}$ and the modulation frequency $\nu$. Then there remain three basic parameters corresponding to the SFB family: the asymptotic amplitude $2r_{0}$, the maximum amplitude $M$ and the extreme position $x_{\textmd{max}}$. By taking the origin of the wavemaker position as the origin of the $x$-axis, the physical wave field is denoted to be dependent on those parameters as $\eta_\text{SFB}(r_{0}, M, x_{\textmd{max}})(x,t)$. The MTA turns out to be a very useful and simple way to obtain a good view of the dependence on these parameters. \index{experiments!comparisons!quantitative} \index{extreme waves!experiments} This investigation is novel and contributes to understanding when choosing parameters for the experiments. 

\subsection{Sensitivity of MTA for parameter changes}
\index{experiments!comparisons!quantitative} \index{experiments!MTA} \index{MTA!sensitivity!SFB parameters}

We consider the SFB solutions with parameter values close to these of the experimental results and design. In this subsection, we investigate the sensitivity of the MTA by allowing one parameter to change but fixing the others. The wavemaker position is at $x = 0$ and since the closest measurement is at 10~m from the wavemaker, we will consider the MTA at this position. As an MTA reference, we take the SFB wave signal with $M = 25$~cm, $\tilde{\nu} \approx 0.9$ and the extreme position $x_{\textmd{max}} = 150$~m from the wavemaker. The corresponding parameter related to the asymptotic amplitude is $r_{0} = 4.9118$~cm and we denote this parameter as $r_\textmd{0ref}$. \label{ref} Earlier, test~A has design characteristics of $M = 21.3$~cm and $\tilde{\nu} = 1$. Since we want to keep the modulation frequency $\nu$ fixed, an increase in $M$ means a decrease in $\tilde{\nu}$.  The corresponding parameter related to the asymptotic amplitude for test~A is given by $r_{0} = 4.4114$~cm, see Table~\ref{designM}. \index{extreme waves!experiments} \index{experiments!MTA} \index{experiments!comparisons!quantitative}
\begin{figure}[h!]			
\begin{center}
\subfigure[For fixed     $M$]{\includegraphics[width = 0.45\textwidth]{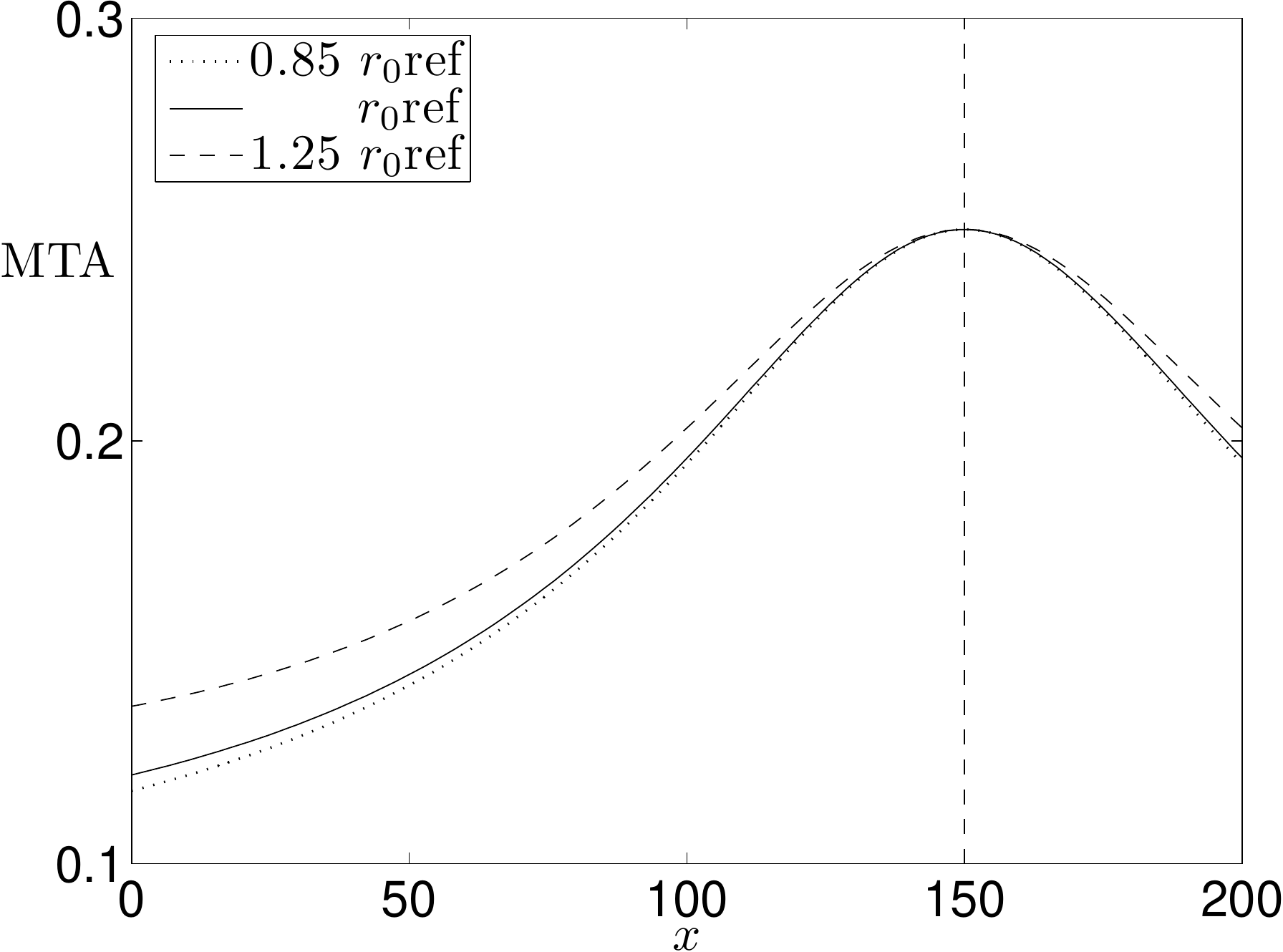}}    	\hspace{0.75cm}
\subfigure[For fixed $r_{0}$]{\includegraphics[width = 0.45\textwidth]{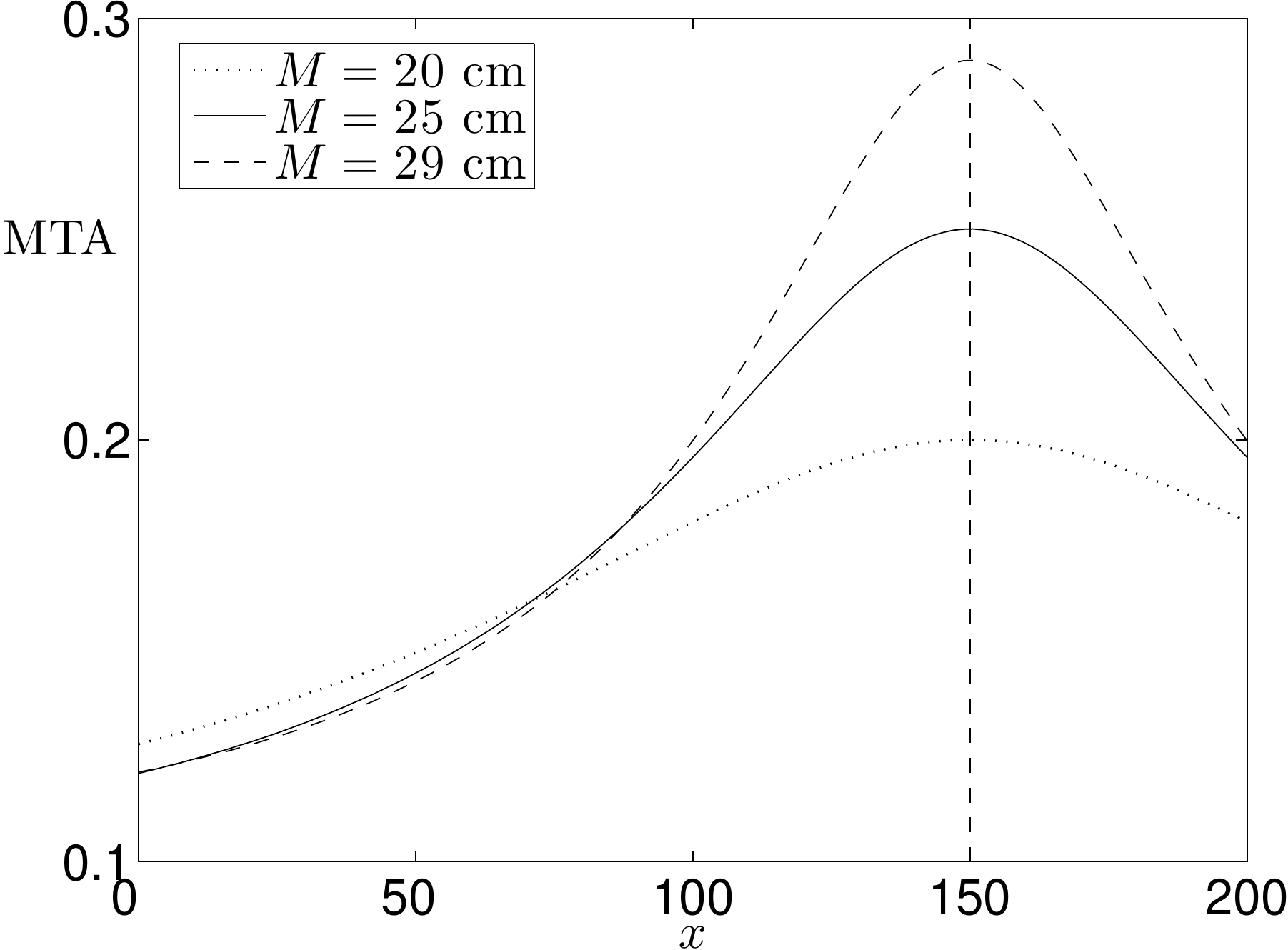}}
\caption[MTA of the SFB with variation in parameters]{Plots of the MTA of the SFB wave signal by allowing one parameter to change and fix the others. In all cases, the extreme position is kept at 150~m. (a) MTA plots for different values of $r_{0} = 0.85\, r_\textmd{0ref}$, $r_\textmd{0ref}$ and $1.25 \, r_\textmd{0ref}$ for fixed $M = 25$~cm. (b) MTA plots for different values of $M$, $M = 20, 25$, and 29~cm for fixed $r_{0} = r_\textmd{0ref} = 4.9118$~cm.}    \label{MTAversusr0M}
\end{center}
\end{figure}
\index{extreme waves!experiments} \index{experiments!MTA} \index{experiments!comparisons!quantitative}

The MTA plots are presented for restricted values of the amplitude since we are interested in a normalized modulation frequency that lies within the interval where the SFB shows wavefront dislocation, $0 < \tilde{\nu} < \sqrt{3/2}$. In fact, the experimental design was also restricted to this interval of modulation frequencies. We investigate which parameters give a significant change to the wave evolution. \index{experiments!MTA} \index{experiments!comparisons!quantitative} \index{extreme waves!experiments}

\begin{figure}[h!]			
\begin{center}
\subfigure[For fixed     $M$]{\includegraphics[width = 0.45\textwidth]{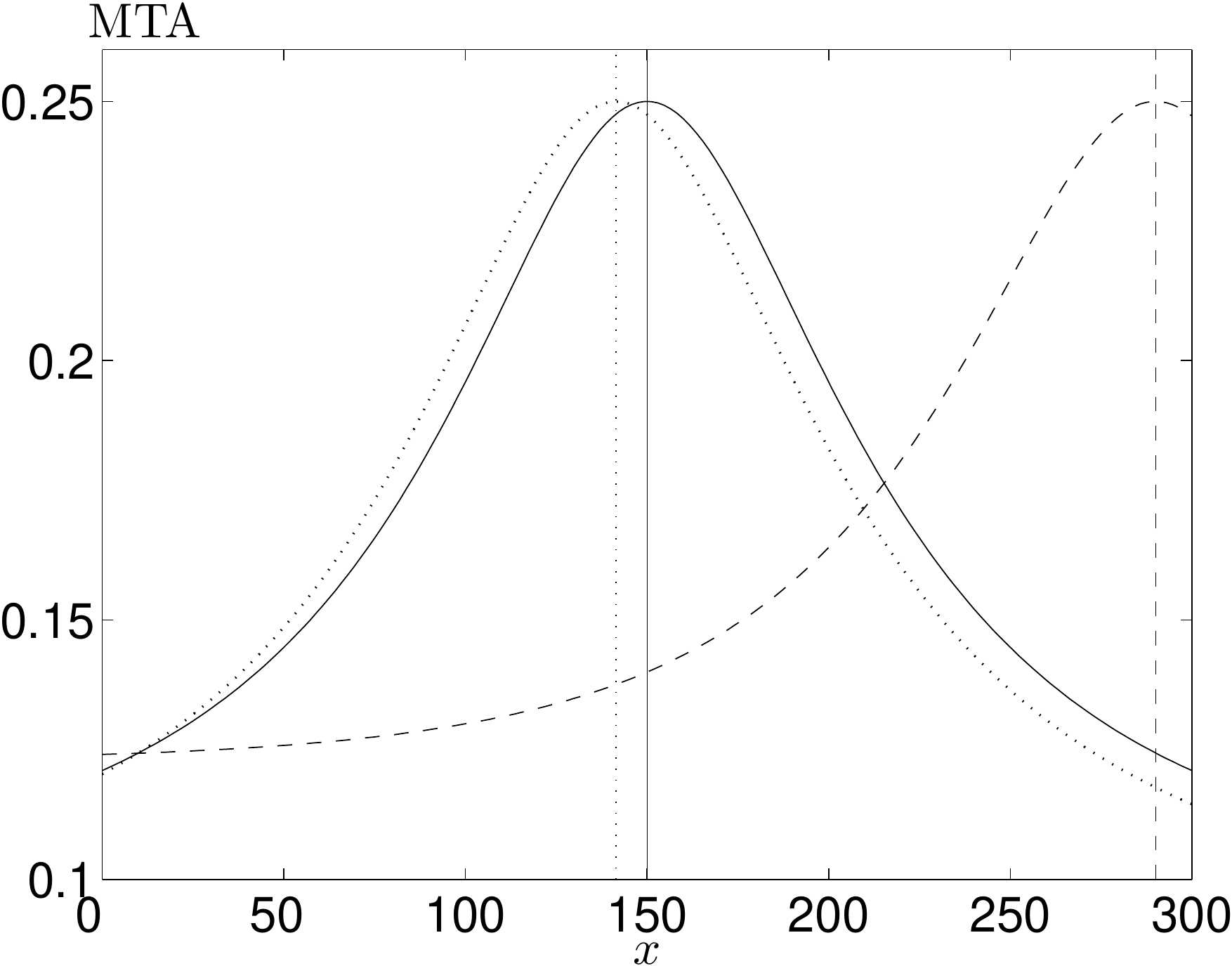}}    	\hspace{0.75cm}
\subfigure[For fixed $r_{0}$]{\includegraphics[width = 0.45\textwidth]{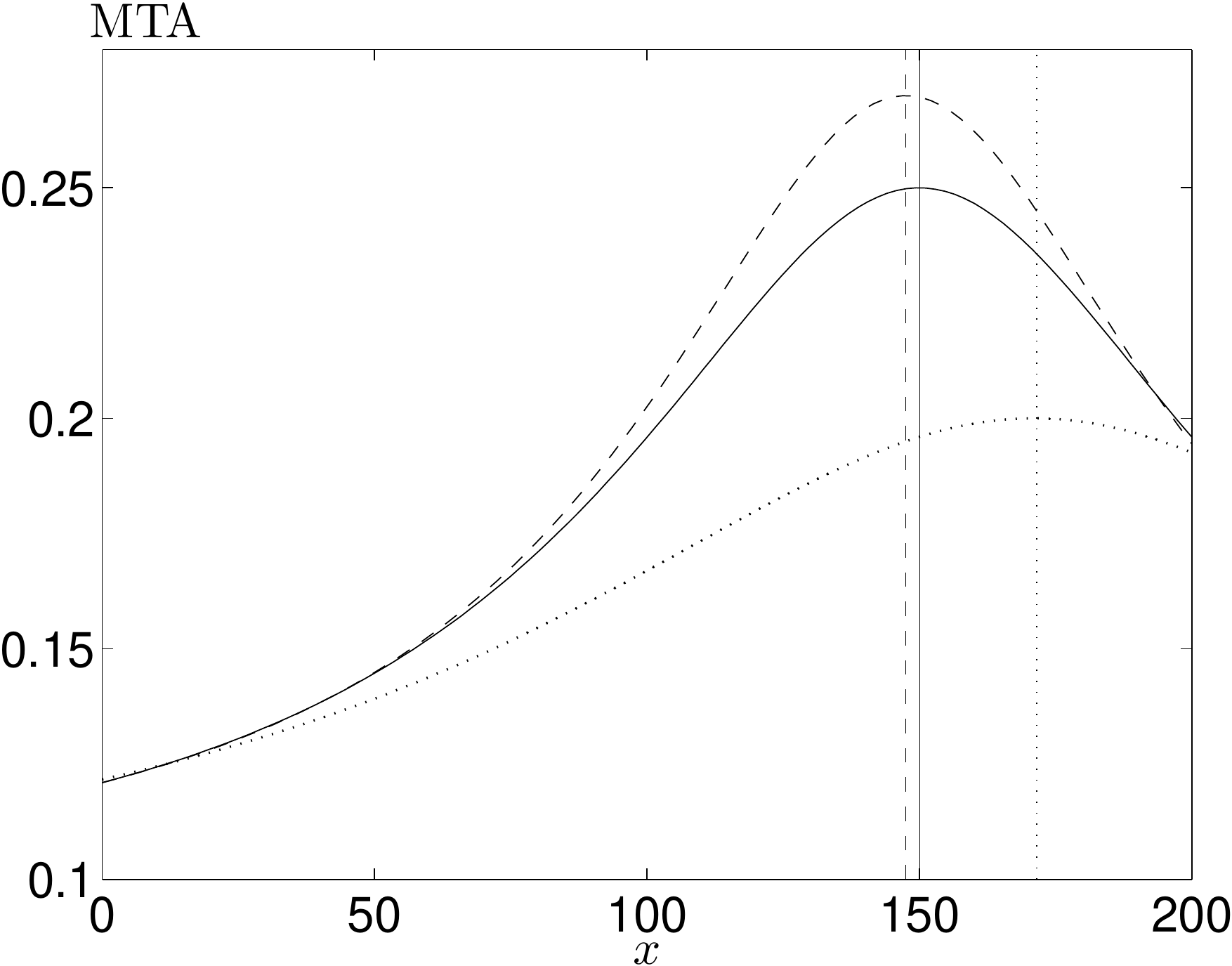}}
\caption[Shifted MTA of the SFB with variation in parameters]{Plots of the MTA after shifting the curves such that the maximum amplitude at 10~m from the wavemaker is equal: (a) for different values of $r_{0}$: $0.85\,r_\textmd{0ref}$ (dotted), $r_\textmd{0ref}$ (solid) and $1.25\,r_\textmd{0ref}$ (dashed) for given $M = 25$~cm. (b) for different values of $M$: 20~cm (dotted), 25~cm (solid) and 27~cm (dashed) for given $r_{0} = r_\textmd{0ref}$.}    \label{MTAgeser}
\end{center}
\vspace*{-0.03cm}
\end{figure}

Figure~\ref{MTAversusr0M} shows the MTA plots for different $r_{0}$ at given $M$ and vice versa for the fixed extreme position $x_{\textmd{max}} = 150$~m. The MTA that has parameters $M = 25$~cm and $r_\textmd{0ref}$ is now the MTA reference, denoted as MTA$_\textmd{ref}$. From Figure~\ref{MTAversusr0M} we observe that for different parameters $r_{0}$ or $M$, the maximum amplitude at 10~m from the wavemaker has different values. From Figure~\ref{MTAversusr0M}(a), we observe that for fixed $M$, the maximum wave amplitude at 10~m from the wavemaker for different $r_{0}$ ranges from 12.08~cm (for $0.85\, r_\textmd{0ref}$) to almost 14~cm (for $1.25\, r_\textmd{0ref}$). Figure~\ref{MTAversusr0M}(b) shows MTA plots for different $M$ with fixed $r_{0}$. In this plot, we present the lowest $M = 20$~cm, the MTA$_\textmd{ref}$ with $M = 25$~cm and the highest $M = 29$~cm. The maximum wave amplitude at 10~m from the wavemaker ranges from 12.34~cm, which corresponds to $M = 27$~cm (not shown in the plot) to 13.14~cm, which corresponds to $M = 20$~cm. In this case, the highest $M = 29$~cm does not give the lowest initial amplitude but the lowest $M$ gives the highest initial amplitude. If we require that all the MTA curves have the same maximum wave amplitude with MTA$_\textmd{ref}$ at 10~m from the wavemaker, we have to shift either to the left or to the right. It should be noted that requiring the maximum wave amplitude at 10~m from the wavemaker to be equal does not mean that the amplitude at the wavemaker itself is equal. As a consequence, the extreme position that is initially at 150~m from the wavemaker will be shifted as well. Figure~\ref{MTAgeser} shows the shifted MTA curves for different $r_{0}$ at given $M$ and vice versa. \index{extreme waves!experiments}
\index{experiments!MTA} \index{experiments!comparisons!quantitative}

From both cases, changing either $r_{0}$ or $M$, we observe that the extreme position is shifted from the initial reference at 150~m from the wavemaker. However, changes in $r_{0}$ give more significant shifts in the extreme position than changes in $M$. For changes in $r_{0}$, the range of possible extreme positions is almost 150~m. For changes in $M$, this range is less than 25~m. Therefore, this result indicates that the actual extreme position depends quite sensitive on the parameters, and in particular more sensitive on the parameter $r_{0}$ than on the maximum amplitude $M$. \vspace*{-0.028cm}

\subsection{Comparisons of the SFB and experimental signals}
\index{extreme waves!experiments} \index{experiments!comparisons!quantitative} \index{experiments!comparisons!SFB signal}

\begin{figure}[h]			
\begin{center}
\subfigure[]{\includegraphics[width = 0.45\textwidth, viewport = 28 218 551 631]{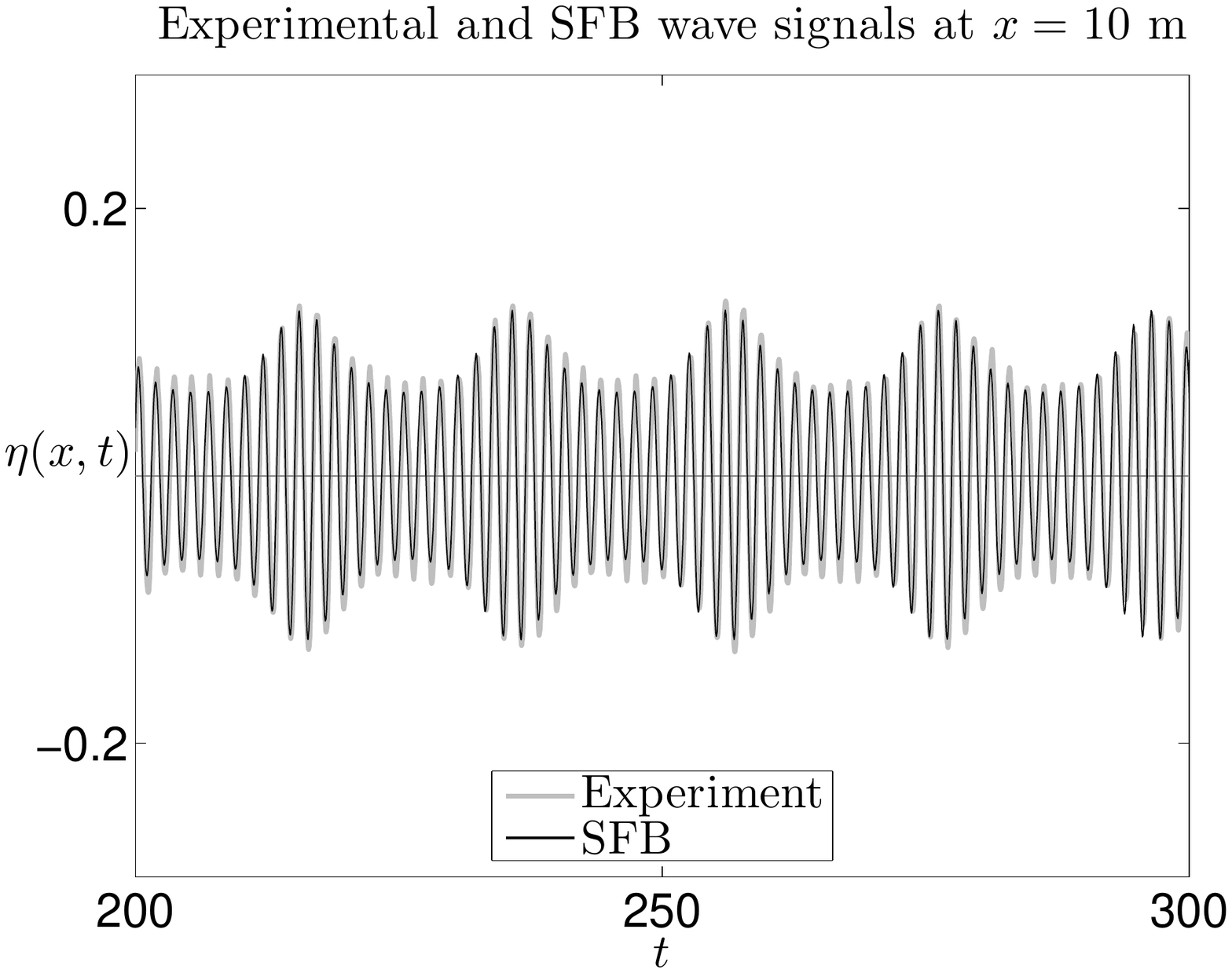}}	    \hspace{0.75cm}
\subfigure[]{\includegraphics[width = 0.45\textwidth, viewport = 28 218 551 631]{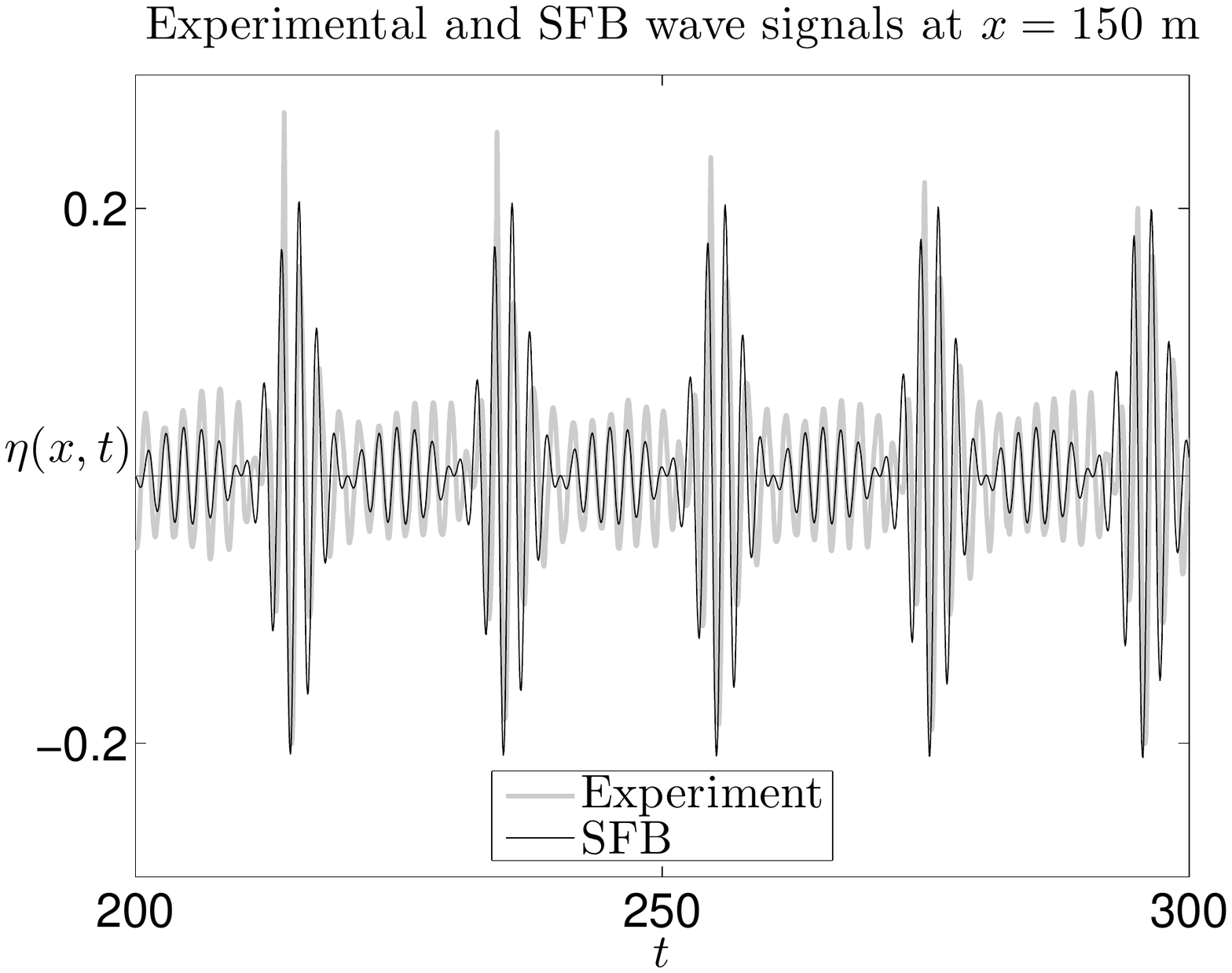}}
\caption[Wave signals comparison based on the design parameters]{Plot of an SFB and the experimental signals at $x = 10$~m (a) and $x = 150$~m (b). The SFB signal plot is based on the design with maximum amplitude $M = 21.30$~cm and $x_{\textmd{max}} = 150$~m.}			    \label{Signalcomparisongammarev}
\end{center}
\end{figure}

We compare the theoretical SFB signal based on the initial design with the corresponding result of the experimental signal (test~A). The design parameters read: carrier wave frequency $\omega_{0} = 3.7284$ rad/s, normalized modulation frequency $\tilde{\nu} = 1$, modulation period $T = 20.18$~sec,  maximum amplitude $M = 21.3$~cm, $r_{0} = 4.4114$~cm and extreme position $x_{\textmd{max}} = 150$~m. \index{experiments!comparisons!quantitative} Figure~\ref{Signalcomparisongammarev} shows the comparison at two different positions: at 10~m and 150~m from the wavemaker. The plots show that the initial signals do not have a good agreement since the experimental signal has a larger amplitude between 0.5--1.5~cm. After downstream evolution, it is no surprise that there exists no good agreement at all between the theoretical SFB and the experimental signals. The experimental signal shows a larger amplitude than the SFB signal with $M = 21.3$~cm. The initial maximum amplitude at 10~m from the wavemaker for the SFB signal is 12.41~m. Therefore the practical amplitude amplification factor\index{AAF!practical} for this SFB is 1.72.\footnote[1]{The practical amplitude amplification factor in this chapter is defined as the ratio of the maximum amplitude at the extreme position and the maximum amplitude at 10~m from the wavemaker. \par} \index{experiments!comparisons!quantitative} \index{extreme waves!experiments}
\index{experiments!comparisons!SFB signal}	\index{experiments!comparisons!quantitative}

\begin{figure}[t!]			
\begin{center}
\includegraphics[width = 0.406\textwidth]{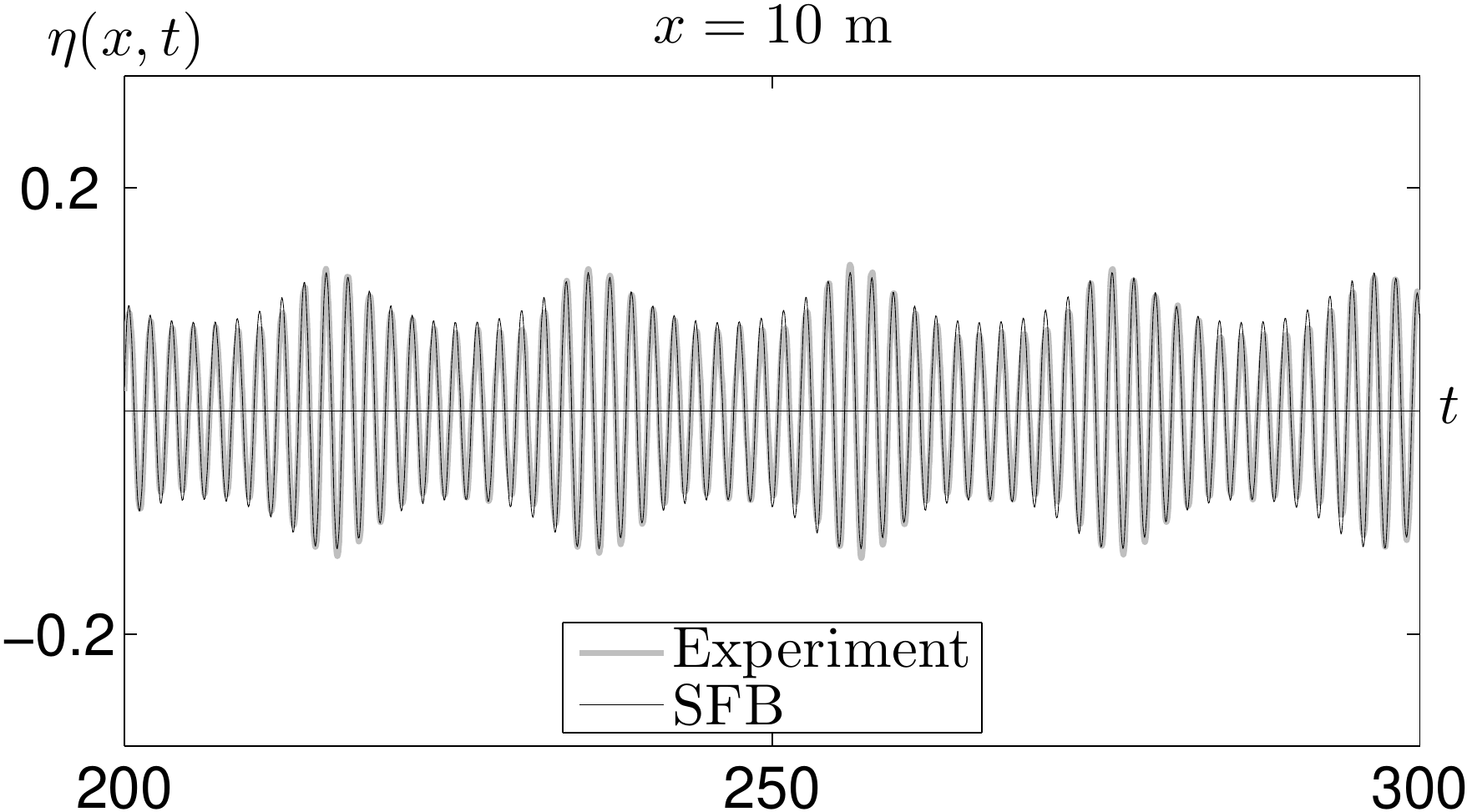} 			\vspace{0.2cm} \\
\includegraphics[width = 0.406\textwidth]{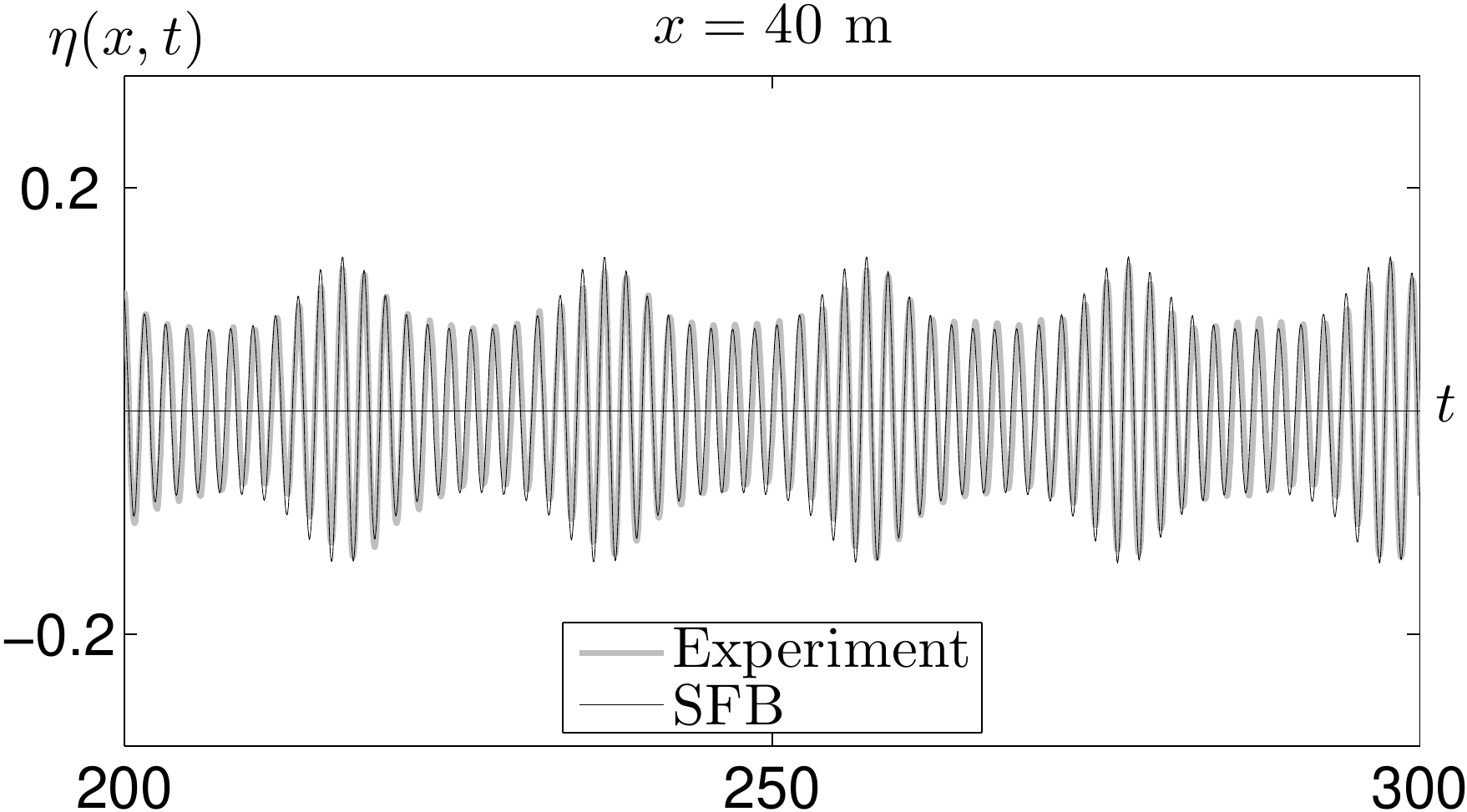} 			\vspace{0.2cm} \\
\includegraphics[width = 0.406\textwidth]{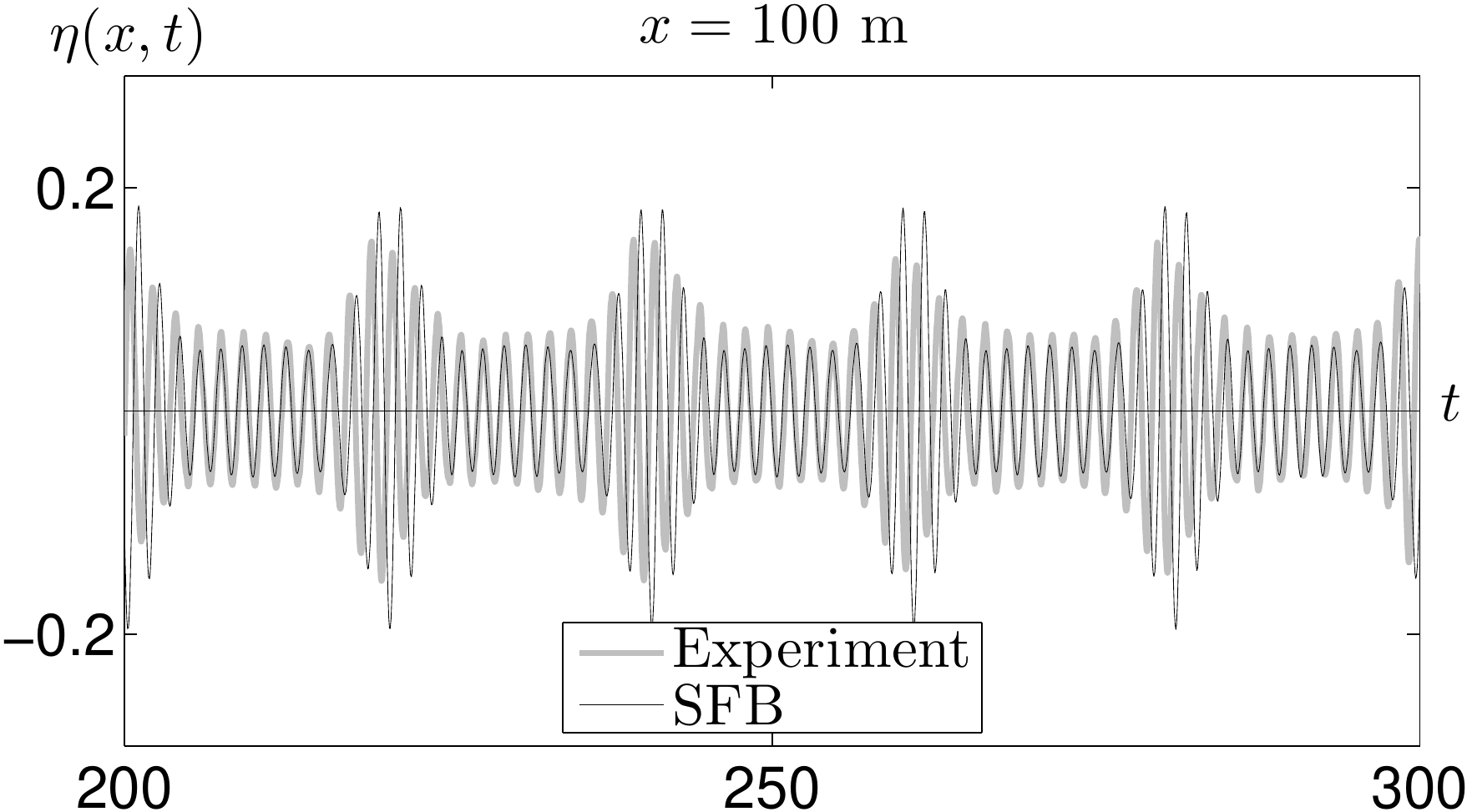} 		\vspace{0.2cm} \\
\includegraphics[width = 0.406\textwidth]{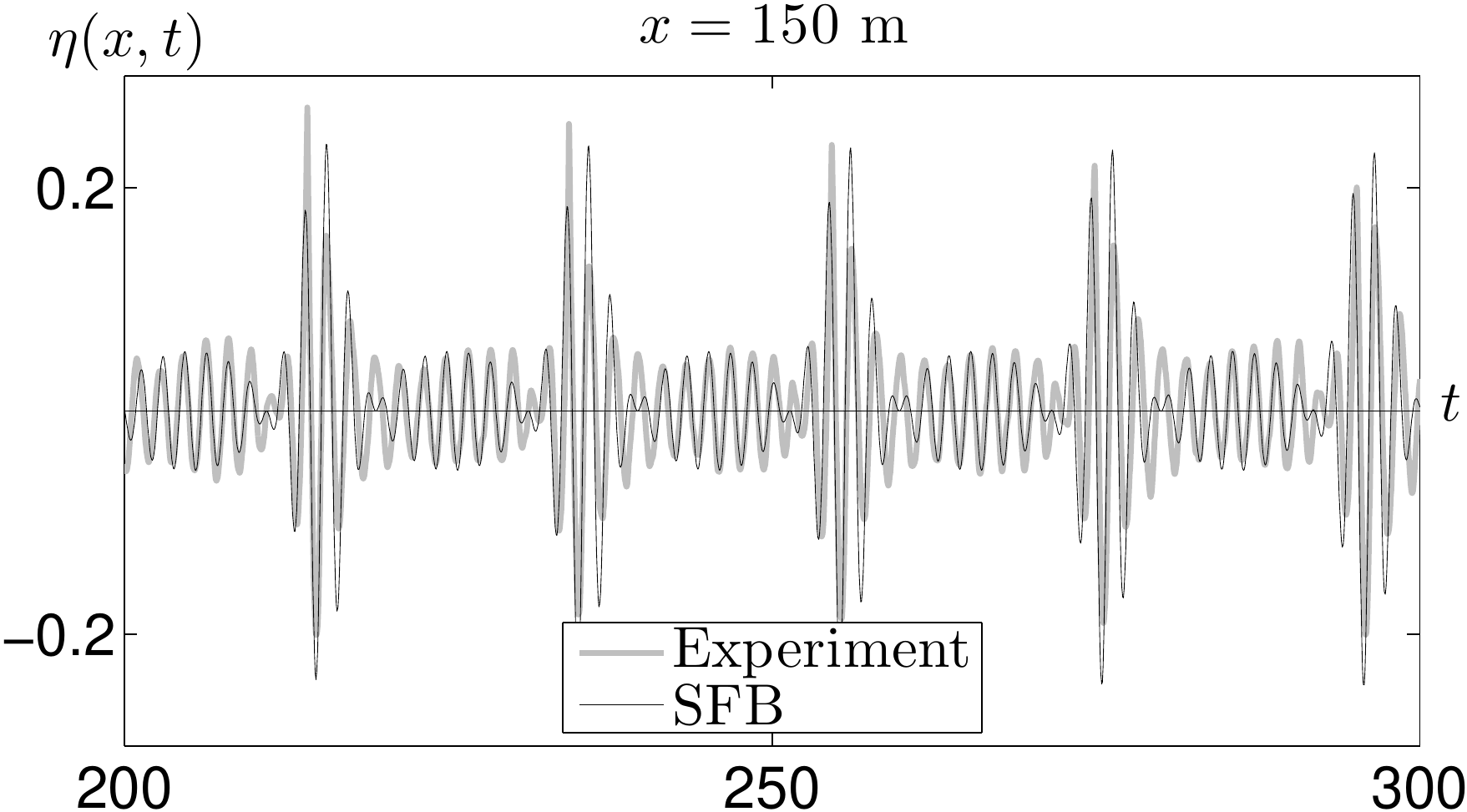} 		\vspace{0.2cm} \\
\includegraphics[width = 0.406\textwidth]{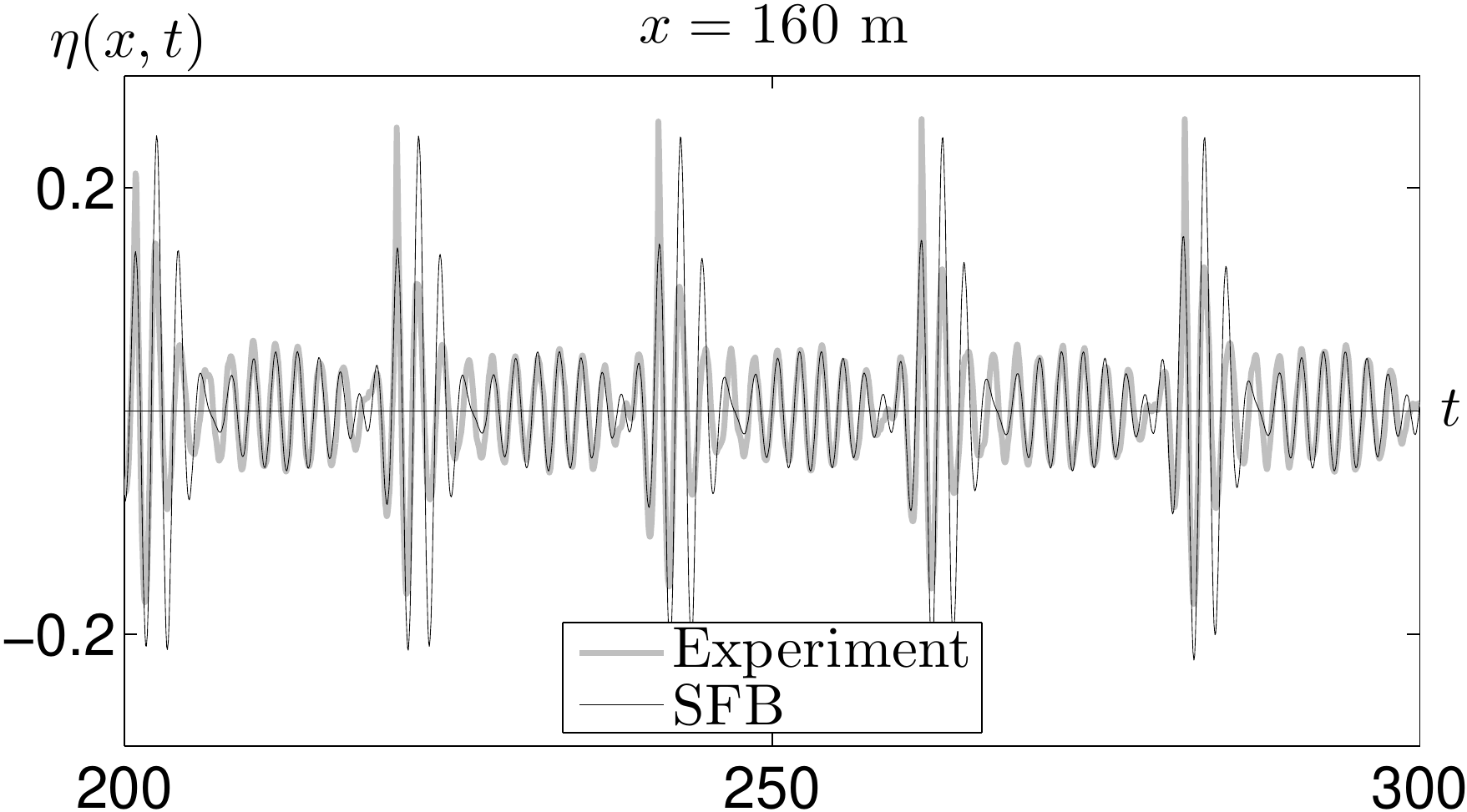}
\caption[Wave signals comparison with $M = 25$~cm]{Plot of an SFB and the experimental signals (test~A) at $x = 10$~m, $x = 40$~m, $x = 100$~m, $x = 150$~m and $x = 160$~m. The SFB wave signal has maximum amplitude $M = 25$~cm and $x_{\textmd{max}} = 150$~m from the wavemaker.}			    \label{SignalcomparisonM25}
\index{experiments!comparisons!SFB signal}    \index{experiments!comparisons!quantitative}
\end{center}
\end{figure}

We conclude that the experiment seems to produce waves with a larger amplitude than the initial design. Therefore we will compare the experimental signal with another SFB signal with larger maximum amplitude M. We choose $M = 25$~cm with an extreme position at 150~m from the wavemaker. Since we want to keep a fixed modulation period, choosing $M = 25$~cm gives a different normalized modulation frequency, now $\tilde{\nu} \approx 0.9$ and $r_{0} = 4.9118$~cm. This SFB signal has a maximum amplitude at 10~m from the wavemaker of 12.44~cm. Therefore, the practical amplitude amplification factor for this SFB is around 2.01. \index{experiments!comparisons!quantitative} Figure~\ref{SignalcomparisonM25} shows the comparison of downstream\index{downstream} evolution between the experimental signal from test~A and the SFB signal with $M = 25$~cm and $x_{\textmd{max}} = 150$~m from the wavemaker. This comparison shows much better agreement than the previous case, although at 100~m from the wavemaker the experimental signal has a larger amplitude than the SFB signal at a certain time interval and smaller amplitude than the SFB signal at another time interval. The choice of $M = 25$~cm (found by choosing various values in a small region around this value) is not a bad choice after all since the experimental signal has a reasonable comparison in maximum amplitude to the SFB signal at both 150~m and 160~m from the wavemaker. \index{extreme waves!experiments} \index{experiments!comparisons!quantitative}

We already remarked earlier the fact that the experimental signal shows an asymmetric structure after downstream\index{downstream} evolution, while we know that the theoretical SFB signal always maintains the symmetry structure during the evolution. Despite this difference between the two signals, we can conclude the following facts. First, the experimental signal still has good agreement with the theoretical SFB signal. Second, the experimental design predicts very well where the experimental extreme position is. Although it is not precise, from the time signals we can conclude that this extreme position is not too far from 150~m or 160~m from the wavemaker. This can also be concluded from the experimental signal that shows clearly the occurrence of phase singularity close to these positions, which is according to the properties of the SFB family it occurs for $0 < \tilde{\nu} < \sqrt{3/2}$. The fact that the experimental wave signal produces waves with a slightly larger amplitude far away from the wavemaker may be due to higher-order effects. \index{extreme waves!experiments} \index{experiments!comparisons!SFB signal} \index{experiments!comparisons!quantitative}

\subsection{Variations in a model parameter} \label{varmodelpar}
\index{extreme waves!experiments}	\index{experiments!comparisons!quantitative}	\index{experiments!comparisons!model parameter}

In the preceding subsections, we have shown the sensitivity of the theoretical SFB signal for changes in the parameters of the SFB: $r_{0}$, $M$, and $x_{\textmd{max}}$, and we compared members of this family with the experimental signal. \index{extreme waves!experiments} \index{experiments!comparisons!quantitative}

Of course, the SFB family, consisting of solutions of the NLS equation, \index{NLS equation} depends on the precise form of this equation, i.e., on the parameters in this equation. There are two essential parameters, the parameter $\beta$ which is the group velocity parameter, and the coefficient $\gamma$, the nonlinear coefficient. The parameter $\beta$ is well defined as group velocity dispersion, \index{dispersion!group velocity} \index{dispersion!group velocity|see{dispersion!coefficient}} related to the change in group velocity with wavenumber. The nonlinear parameter, however, is less well-defined, and various expressions can be found in the literature. The reason is that in the derivation of the NLS equation in Subsection~\ref{derivespatialNLS}, a choice has to be made about the third-order bound long-wave contribution, for which only its derivative is given in~\eqref{spatialNLSgauge}. Making a choice influences the resulting value of $\gamma$. This is related to the validity of a `zero mass flux'\index{zero mass flux} assumption of the water. Although we consider the unidirectional propagation of waves, there is a significant difference between the condition in the laboratory and the condition in the open sea or near the beach. An essentially different value appears when the NLS equation is derived for sea-waves when there is no closed basin that introduces return current. See for instance~\citep{6Mei83, 6Janssen06}. \index{experiments!comparisons!model parameter} \index{experiments!comparisons!quantitative} \index{extreme waves!experiments}

Another difficulty in finding the correct value of the nonlinear coefficient of the NLS equation that really describes the nonlinear effect in nature is as follows. We denote the nonlinear coefficient $\gamma$ presented in~\eqref{gamma} as $\gamma_\textmd{KdV}$ in the context of this chapter since it was derived from a KdV type of equation. This coefficient is obtained after applying the gauge transformation on page~\pageref{gaugetrans}. In fact, the original nonlinear coefficient $\gamma$ is the one before the gauge transformation is applied to the NLS equation. This coefficient now depends on the position and the time and is explicitly given by
\begin{equation}
\gamma(\xi,\tau) = \gamma_\textmd{KdV} + \frac{2k_{0}c}{\Omega'(k_{0})} \frac{\textmd{Re}[\alpha_{S}(\xi)]}{|A(\xi,\tau)|^{2}}.
\end{equation}
It is inconvenient for designing the experiments that this nonlinear coefficient is not constant. That is why we take a constant value for the design value of this nonlinear coefficient, as we will see in the following paragraph. Although the value we used for the design and comparison seems to be the correct one, as we shall indicate below, it seems worthwhile to investigate the dependence on this parameter also. This will be done in this subsection. We remark in addition that this coefficient, although it is in front of the nonlinear term in the NLS equation, depends strongly on dispersive contributions in the nonlinearity in the full (exact) equations, and therefore also on the nonlinear dispersion relation~\eqref{nondisrel}; the poor modeling of the dispersion in these nonlinear terms causes the problems and the different results that can be found in the literature. \index{extreme waves!experiments}
\begin{figure}[h!]			
\begin{center}
\subfigure[]{\includegraphics[width = 0.45\textwidth, viewport = 27 218 551 631]{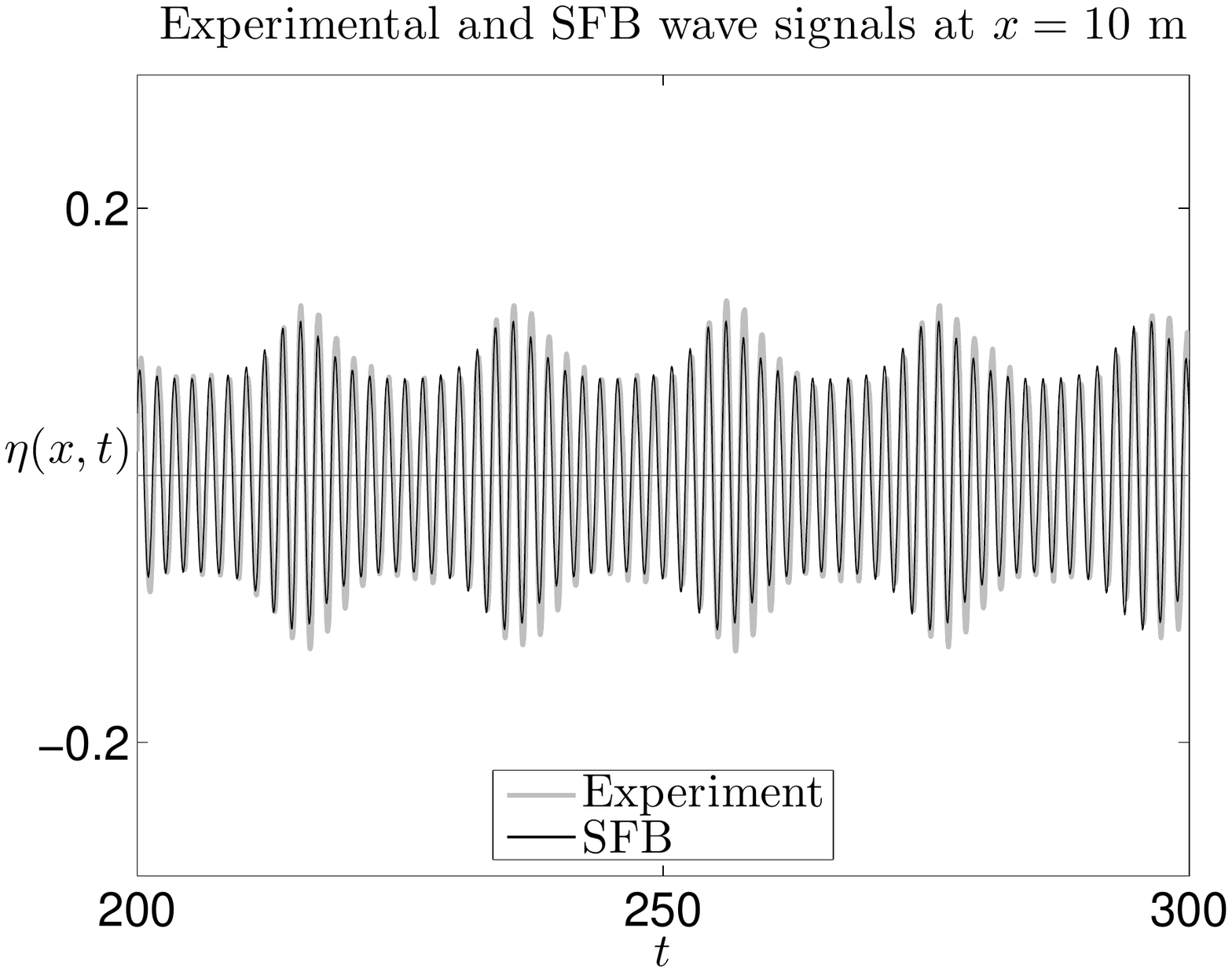}}	    \hspace{0.75cm}
\subfigure[]{\includegraphics[width = 0.45\textwidth, viewport = 28 218 551 633]{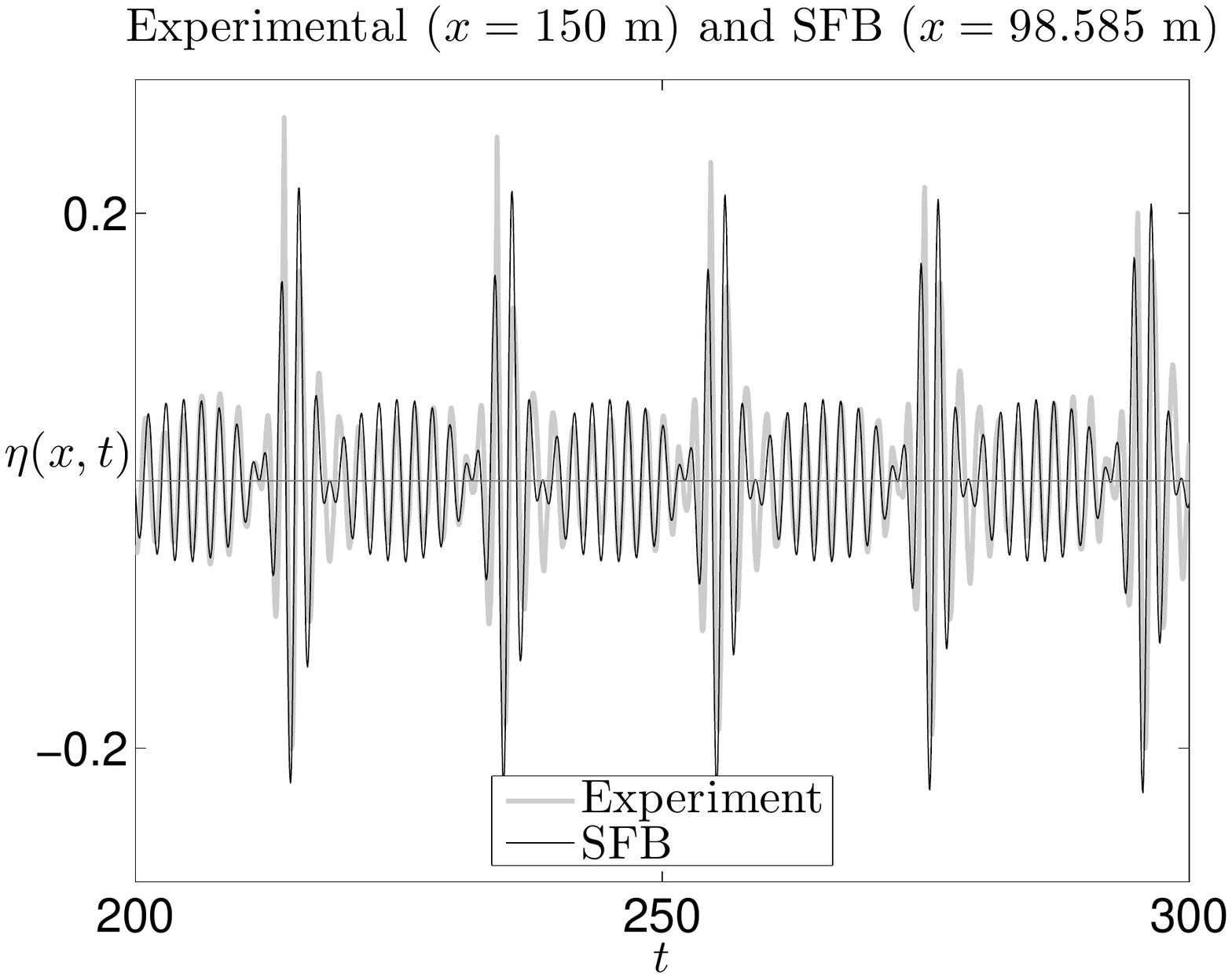}}
\caption[Wave signals comparison using the nonlinear coefficient $\gamma_{\textsc{do}}$]{Plot of the SFB signal and the experimental signal at $x = 10$~m (a) and the extreme position (b). With $\gamma = \gamma_{\textsc{do}}$ and fixing the asymptotic amplitude $2r_{0}$, the SFB wave signal now reaches an extreme at $x_{\textmd{max}} = 98.585$~m with maximum amplitude $M = 23.734$~cm.}		    \label{SignalcomparisongammaDO}
\end{center}
\end{figure}
\index{experiments!comparisons!model parameter}	\index{experiments!comparisons!quantitative}

Next, we discuss how changes of the nonlinear coefficient of the NLS equation\index{NLS equation!nonlinear coefficient} $\gamma$ influence the SFB signal, and in particular the extreme position $x_{\textmd{max}}$. The SFB in the previous comparisons has the design value of $\gamma$, denoted as $\gamma_{\textmd{des}}$ explicitly given by
\begin{equation}
\gamma_{\textmd{des}} = \gamma_{\textmd{KdV}} + \frac{2k_{0}^{2}}{\omega_{0} V_{0}},
\end{equation}
where $\gamma_{\textmd{KdV}}$ is the nonlinear coefficient based on derivation from the KdV equation with the exact dispersion relation given in Section~\ref{Bab2NDWE}. An explicit expression of $\gamma_{\textmd{KdV}}$ is given by~\eqref{gamma}. For $\omega_{0} = 3.7284$ rad/s, $\gamma_{\textmd{des}} = 230.2496$. \index{experiments!comparisons!model parameter} \index{experiments!comparisons!quantitative} \index{extreme waves!experiments}

Another value of this nonlinear coefficient has been proposed and will be denoted here as $\gamma_{\textsc{do}}$. This value of $\gamma$ takes into account the presence of the current after wave generation. An explicit expression of this coefficient can be found in~\citep{6Dingemans01, 6Klopman05, 6Huijsmans05} and is given by \index{extreme waves!experiments}
\begin{equation}
\gamma_{\textsc{do}} = \frac{\gamma_{1}(k_{0}) + k_{0} \alpha_{U} + \lambda(k_{0}) \alpha_{\zeta}}{V_{0}}
\end{equation}
where
\begin{align}
\gamma_{1}(k_{0}) &=  k_{0}^{2} \omega_{0} \frac{9 \tanh^{4}k_{0} - 10 \tanh^{2}k_{0} + 9}{4 \tanh^{4} k_{0}} \\
\lambda(k_{0})	  &=  \frac{1}{2}k_{0}^{2} \frac{1 - \tanh^{2} k_{0}}{\omega_{0}}
\end{align}
\begin{align}
\alpha_{\zeta} &= -\frac{1}{\Omega(k_{0})} \frac{4 k_{0} V_{0} - \omega_{0}}{[\Omega'(0)]^{2} - V_{0}^{2}} \\
\alpha_{U} 	   &=  \alpha_{\zeta} V_{0} - \frac{2k_{0}}{\omega_{0}}.
\end{align}
For given $\omega_{0} = 3.7834$ rad/s, the value is found to be $\gamma_{\textsc{do}} = 402.4946$ which, when compared to $\gamma_{\textmd{des}}$, leads to an almost two times larger value: $\gamma_{\textsc{do}} \approx 1.75 \gamma_{\textmd{des}}$. \index{experiments!comparisons!quantitative} \index{extreme waves!experiments}
\begin{figure}[h!]			
\begin{center}
\subfigure[]{\includegraphics[width = 0.45\textwidth]{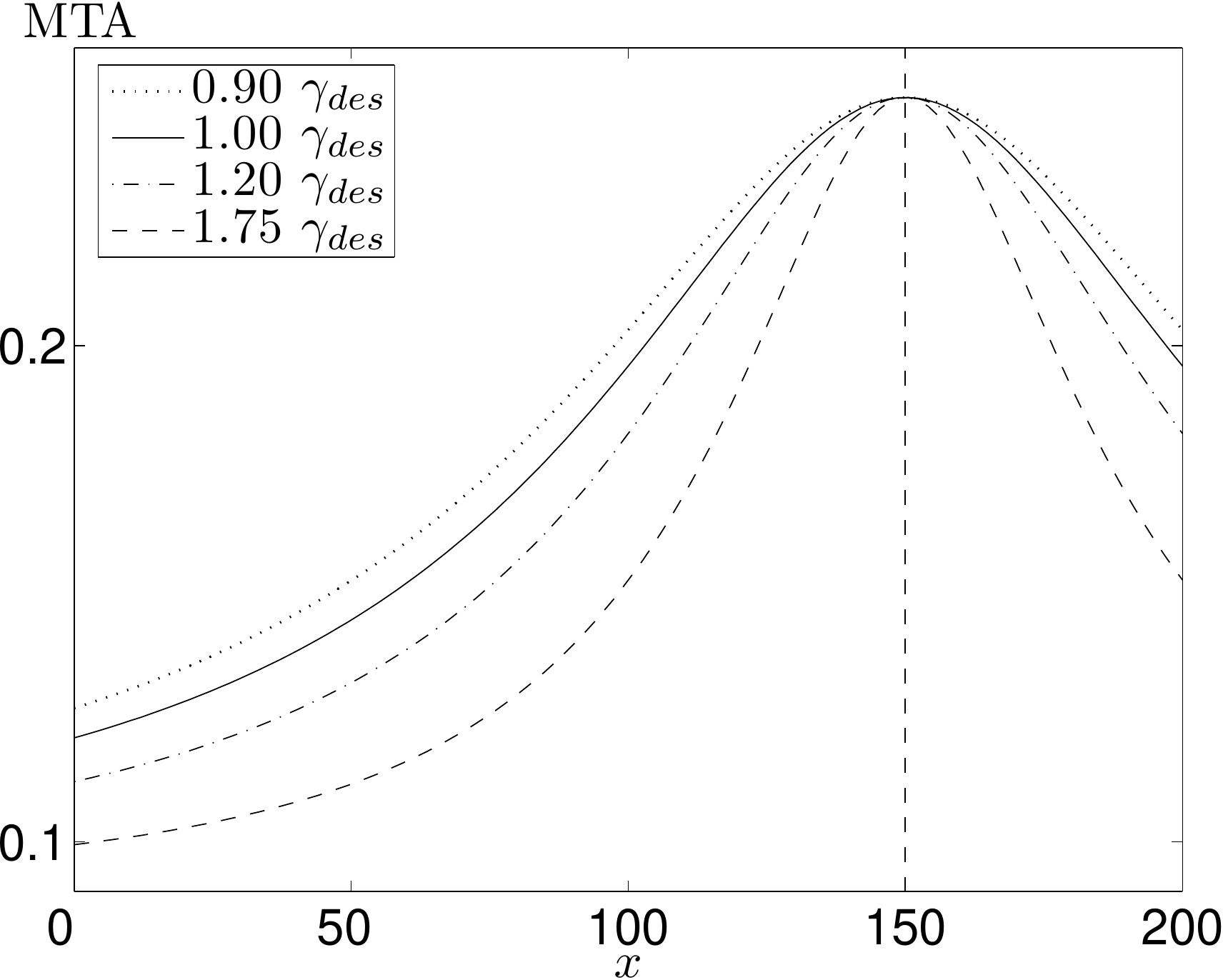}}    \hspace{0.75cm}
\subfigure[]{\includegraphics[width = 0.45\textwidth]{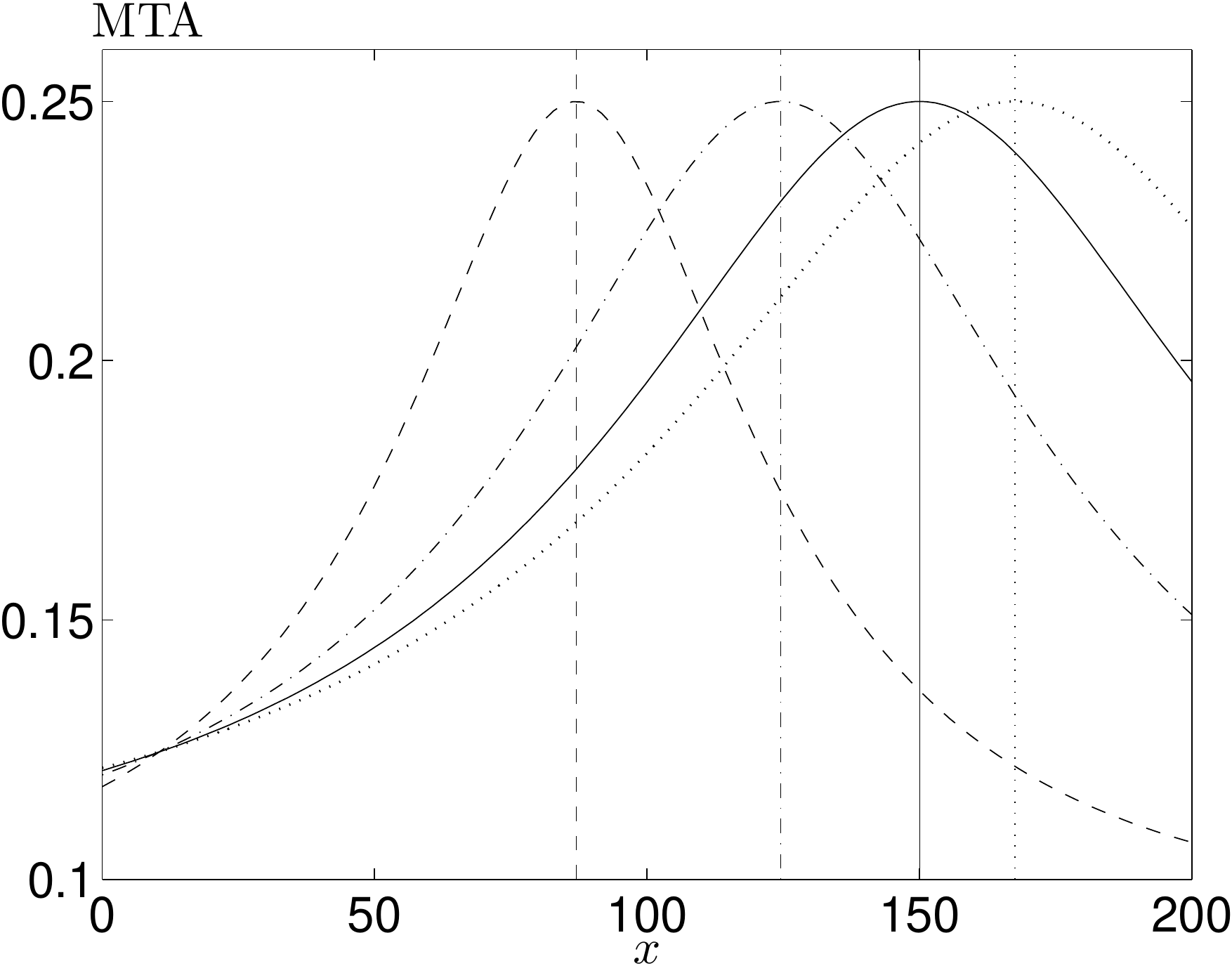}}
\caption[MTA of the SFB for variations in the nonlinear coefficient]{(a) Plot of the MTA corresponding to the theoretical SFB signal with $M = 25$~cm for different values of $\gamma_{\textmd{des}}$. (b) Plot of the shifted MTA curves such that the maximum amplitude at 10~m from the wavemaker is equal for different values of $\gamma$: $0.9\,\gamma_\textmd{des}$ (dotted), $\gamma_\textmd{des}$ (solid), $1.2\,\gamma_\textmd{des}$ (dash-dot) and   $1.75\,\gamma_\textmd{des}$ (dashed).}
\label{MTAgammarev}
\end{center}
\end{figure}
\index{extreme waves!experiments}	\index{experiments!comparisons!model parameter}		\index{experiments!comparisons!quantitative}

We will now compare the experimental signal with the theoretical SFB signal for $\gamma = \gamma_{\textsc{do}}$. We take the asymptotic amplitude $2 r_{0}$ to be fixed and allow the maximum amplitude $M$ and the extreme position $x_{\textmd{max}}$ to change. It is found that the normalized modulation frequency now becomes $\tilde{\nu} = 0.756$, the maximum amplitude $M = 23.734$~cm and the extreme position $x_{\textmd{max}} = 98.585$~m from the wavemaker. The maximum amplitude of the initial SFB signal at 10~m from the wavemaker is 11.63~cm and therefore the practical amplitude amplification factor is around 2.04. Figure~\ref{SignalcomparisongammaDO} shows the comparison between the theoretical SFB for $\gamma = \gamma_{\textsc{do}}$ and the experimental signals at 10~m and 150~m from the wavemaker for the experimental signal and at $x_{\textmd{max}}$ for the SFB signal. The comparison at 10~m from the wavemaker does not show a good agreement between the two signals since the experimental signal has a larger amplitude up to 1.5~cm. After downstream\index{downstream} evolution, both signals do not have a good agreement either, since the theoretical SFB signal reaches its extreme position at 98.585~m from the wavemaker, while the experimental signal reaches its largest amplitude at around 150~m or 160~m from the wavemaker. Based on the fact that the extreme position is shifted about 50~m to 60~m, it is unlikely that $\gamma_{\textsc{do}}$ is the correct nonlinear coefficient for the experiments since it does not describe accurately the experiments in the wave basin. \index{wave basin} \index{experiments!comparisons!model parameter} \index{experiments!comparisons!quantitative} \index{extreme waves!experiments}

In order to get a better understanding of how the nonlinear coefficient $\gamma$ influences the extreme position, we look at the corresponding MTA plots for fixed parameters except the value of $\gamma$. \index{MTA!sensitivity!model parameter} Figure~\ref{MTAgammarev}(a) shows the MTA plot corresponding to the SFB by allowing the nonlinear coefficient $\gamma$ to change. The solid curve that corresponds with $\gamma = \gamma_\textmd{des}$ is regarded as MTA$_\textmd{ref}$. It can be observed that for a fixed maximum amplitude $M$, the maximum amplitude close to the wavemaker changes according to the value of $\gamma$. Figure~\ref{MTAgammarev}(b) shows the plot of the same MTA after shifting such that the maximum wave amplitude at 10~m from the wavemaker is equal to MTA$_\textmd{ref}$. We observe that the extreme position is shifted significantly, particularly for a factor difference in $\gamma$ of 75\%. This indicates that the value of the nonlinear coefficient $\gamma$ plays an important role in designing the experiments on extreme wave generation. Therefore, the correct value of $\gamma$ deserves more attention in future research especially to design experiments using the theoretical SFB family. \index{experiments!comparisons!model parameter} \index{experiments!comparisons!quantitative} \index{extreme waves!experiments}

\section{Conclusion and remark}

In this chapter, we presented experimental results on extreme wave generation\index{extreme waves!generation} based on the theoretical SFB that has been conducted at the high-speed basin of MARIN. After removing the second-order contribution from the measured signal, we described both qualitative and quantitative comparisons between the theoretical and the experimental signals. It should be noted that all experimental results show an amplitude increase according to the Benjamin-Feir instability\index{Benjamin-Feir instability}\index{Benjamin-Feir instability|see{modulational instability}} of the SFB solution of the NLS equation. We also observed that both the carrier wave frequency and the modulation period are conserved during the propagation in the wave basin. A significant difference is that the experimental wave signal does not preserve the symmetry structure as the theoretical SFB does. \index{extreme waves!experiments}

We also observed from the evolution in the Argand diagram that the experimental signal resembles qualitatively the perturbed SFB. The theoretical SFB signal has one pair of singularities within one modulation period at the extreme position. In the Argand diagram, this is indicated by the curve that lies on the real axis. The evolution curves for the experimental signal in the Argand diagram are no longer straight lines, but twisted elliptical-like curves. It implies that at the extreme position, there is no phase singularity. However, the experimental signal still has two singularities within one modulation period, but now at two different positions close to the extreme position located between the singularity positions. \index{extreme waves!experiments}

A quantitative comparison was made between the theoretical SFB and the experimental signals. We found that the MTA plot is sensitive to the changes in the parameters of the SFB. We also compared the experimental signal with a family of the theoretical SFB with chosen parameters. With the choice of $M = 25$~cm, we have a reasonable comparison in maximum amplitude at both 150 and 160~m from the wavemaker. \index{extreme waves!experiments}

We investigated variations in one of the two model parameters. For the NLS equation, these are the group velocity dispersion $\beta$ and the nonlinear coefficient $\gamma$. The parameter $\beta$ is well-defined but $\gamma$ is less well-defined because we do not consider the gauge transformation in the derivation of the NLS equation carefully. Although another value of $\gamma$ has been proposed in the literature, the signal comparison showed that it does not describe accurately the experiments in the wave basin: it leads to a large shift of the extreme position of more than 50~m. \index{extreme waves!experiments}

Several experts in the field have suggested implementing the modified NLS equation\index{NLS equation!modified} since it contains higher-order terms~\citep{6Dysthe79, 6Trulsen96}. Further investigation on model equations, model parameters, design parameters, and some qualitative aspects of extreme waves gives many opportunities for future research. \index{extreme waves!experiments}

\newpage
{\renewcommand{\baselinestretch}{1} 
}

\setcounter{chapter}{6}
\chapter{Conclusion and recommendation} 	\label{7Conclude}

We have mentioned in the introduction of this thesis that we want to contribute to an understanding of the evolution and design of generation properties of large amplitude non-breaking waves, also known as extreme waves.\index{extreme waves} In this thesis, we have presented both theoretical and experimental studies of modeling and generation of extreme waves. In the following, we compare our contributions to the existing knowledge of extreme water waves. In the final section, we present possible topics for future research on this subject.

\section{Conclusion}

In the literature, studies of extreme wave modeling and nonlinear wave phenomena in general, are more often based on the temporal NLS equation\index{NLS equation!type!temporal} than on the spatial one. In this thesis, we focused on the signaling problem in which the evolution of the surface wave envelope is more directly described by the spatial NLS equation\index{NLS equation!type!spatial} than the temporal one. Both the spatial and the temporal NLS equations are derived under the assumption of a narrow-banded spectrum and give a quadratic contribution to the linear dispersion relation. The spatial NLS equation has better dispersive properties than the temporal NLS equation. For these reasons the entire content of this thesis is devoted to the spatial NLS equation, simply referred to as the NLS equation. Thus, the choice of the spatial NLS equation is especially relevant to extreme wave modeling in the wave basin of MARIN.

We considered several exact solutions of the NLS equation. In particular, we were interested in waves on finite background that do not vanish at infinity. The background is a plane-wave, which is a uniform monochromatic wave, and an exact solution of the NLS equation. We have shown that the plane-wave and the single soliton solutions are coherent since the phases of the spectrum are constant for all frequencies. On the other hand, waves on the finite background are not coherent, as we have shown in the case of the Soliton on Finite Background (SFB).\index{SFB} Therefore, we conclude that the SFB and other non-coherent solutions are more interesting to be investigated in order to understand the dynamics of the spectrum evolution.

We presented the description of waves on the finite background \index{waves on finite background} using the variational formulation of the displaced phase-amplitude\index{displaced phase-amplitude} representation. This representation seems to be novel and useful in understanding the dynamics that lead to extreme wave events. The dynamic evolution of the corresponding displaced amplitude has an analogy with the dynamics of an autonomous nonlinear oscillator in a certain effective potential energy. When the displaced-phase is restricted to be time-independent, the changing of phase with position physically corresponds to a change of the wavelength of the carrier wave of the wave groups. This turns out to be the only driving force responsible for the nonlinear amplitude amplification toward extreme wave events. Remarkably, the assumption of time-independent displaced phase leads to three (known) exact solutions of the NLS equation: the SFB, the Ma solution,\index{Ma solution} and the rational solution.\index{rational solution} Hence, the displaced phase-amplitude is a good representation for understanding the dynamics of waves on finite background.

We investigated the process that leads to amplitude amplification concerning waves with modulated initial evolution. This process is governed by the nonlinear modulational instability, \index{modulational instability} also known as the Benjamin-Feir instability in the context of water waves. According to this instability, a slightly modulated wave signal with one pair of sidebands\index{sideband(s)} corresponding to sufficiently long modulation length will grow exponentially in space during its evolution but eventually, the nonlinear effect will take over and binds the growth. The complete evolution of the wave signal according to the NLS equation is given by the SFB. The asymptotic behavior \index{asymptotic behavior!SFB} of the SFB wave signal is precisely the linearly modulated wave signal. In comparison to the Ma solution that has large amplitudes during its initial evolution, the SFB has moderate wave amplitudes during its initial evolution. Consequently, the SFB is a good candidate for extreme wave generation. SFB waves can reach a maximum amplification factor of three \index{AAF} for a very long modulation wavelength (when the modulation frequency $\nu \rightarrow 0$).

We studied in detail the corresponding physical wave field of the SFB and observed interesting physical phenomena. For a sufficiently long modulation wavelength, the physical wave field shows wavefront dislocation, \index{wavefront dislocation} which is related to phase singularity \index{phase singularity} and vanishing amplitude.\index{vanishing amplitude} To understand better these phenomena, we showed that unboundedness of the Chu-Mei quotient\index{Chu-Mei quotient} in the nonlinear dispersion relation is a generic property when the amplitude vanishes, which itself is a necessary condition for the occurrence of wavefront dislocation and phase singularity. This connection between the Chu-Mei quotient and the physical phenomena of the SFB has not been investigated in the literature. We used this observation to confirm that our experimental signals have similar properties with the theoretical SFB signal. Therefore, wavefront dislocation and phase singularity are not only interesting theoretically but also important characteristics for confirming the experimental results.

We examined the evolution in the Argand diagram \index{Argand diagram} in order to understand the dynamics of the SFB complex amplitude and also the experimental signals. The evolution curves are straight lines centered at $(-1,0)$ and the angle with the real axis depends on the time-independent displaced-phase. At the extreme position, the line lies at the real axis and crosses the origin twice during one sufficiently long modulation period. This explains why we have one pair of phase singularities/wavefront dislocations in one modulation period. A small perturbation to the SFB will deform the straight lines in the Argand diagram into twisted elliptical-like curves. As a consequence, the curves cross the origin at two different positions, and the extreme position is located in between. In addition, the symmetry structure of the SFB wave signal is not preserved anymore and the perturbed signal becomes asymmetric. Thus, the perturbed SFB signal has two phase singularities at two different positions with the extreme position is located in between. So, we conclude that the occurrence of phase singularity in the SFB will always be related to the extreme position.

We performed some experimental tests on extreme wave generation using a theoretical model based on the SFB of the NLS equation. \index{experiments!extreme waves} The comparisons using several mathematical concepts discussed in this thesis are new contributions to a better understanding of extreme wave characteristics. From the experimental results, we conclude that all experimental signals show amplitude increases according to the modulational instability process of the SFB wave signal of the NLS equation. We observed that both the carrier wave frequency $\omega_{0}$ and the modulation frequency $\nu$ are conserved very accurately during downstream\index{downstream} evolution. This gives an indication that the SFB is a good model for extreme wave generation and the modulational instability of the NLS equation is a robust property.

Furthermore, we found out that the experimental signals have an asymmetric structure, different from symmetric signals of the SFB. \index{experiments!asymmetric signal} From the comparisons of the evolution of the complex amplitudes in the Argand diagram, we conclude that the experimental signals resemble qualitatively with the perturbed SFB signal mentioned earlier. This conclusion is also supported by the fact that the experimental signals have two phase singularities at two different positions and the extreme position is located in between. Therefore, although the SFB is indeed a good model, the evolution curves in the Argand diagram are not robust quantities. \index{experiments!phase singularity}

We explored the concept and properties of maximum temporal amplitude (MTA),\index{MTA} in order to know qualitatively the sensitivity of this MTA and its extreme position toward parameter changes. The MTA is very useful in experimental design for determining the exact location of the initial signal such that the extreme position occurs at the desired location. By allowing to shift several MTA curves accordingly, we found out that the corresponding extreme position is more sensitive toward changes in the asymptotic amplitude $r_{0}$ rather than in the maximum amplitude $M$. Similar sensitivity is also observed for the variations in the nonlinear coefficient of the NLS equation. Therefore, if we want to have good accuracy in the precise location of the extreme position, we have to choose carefully the parameters in both the SFB and the NLS equation for the experimental design.

We also examined some theoretical aspects of a higher-order family of SFB solutions, in particular, SFB$_{2}$.\index{SFB$_{2}$} Since there is wave group interaction during downstream evolution, it is particularly interesting to investigate the dynamics of SFB$_{2}$. Several physical characteristics such as wavefront dislocation and phase singularity are also observed in SFB$_{2}$. Outstandingly, SFB$_{2}$ waves can reach a maximum amplification factor of five for a very long modulation wavelength. Therefore, SFB$_{2}$ is also a good candidate for extreme wave generation and indeed it deserves more further studies, as we will mention in the following section.

\section{Recommendation}

The following problems are recommended for future research in the field of nonlinear wave phenomena related to the work in this thesis.
\begin{itemize}[leftmargin=0.4cm]
\item We have presented the SFB (or SFB$_{1}$ in the context of Chapter~\ref{5HighOrder}) in the displaced phase-amplitude representation. Using a restriction to time-independent displaced-phase, the dynamics of SFB$_{1}$ at each position is given as the motion of a nonlinear autonomous oscillator in potential energy. For SFB$_{2}$, the dynamics can possibly be described as an interaction between two nonlinear oscillators. Further investigations are required to improve understanding of the characteristics of SFB$_{2}$. \index{SFB!SFB$_{2}$}

\item It is useful to improve the accuracy of model parameters in the NLS equation, particularly the nonlinear coefficient $\gamma$. This is important for a good comparison of other future experiments on extreme wave generation with solutions of the NLS equation. See Subsection~\ref{varmodelpar} on page~\pageref{varmodelpar} for the variations in this nonlinear coefficient.

\item Only unidirectional waves were considered here in a one-dimensional physical domain. An extension to the two-dimensional domain for studying the mechanism of extreme wave generation in the context of multidirectional waves would be interesting.\index{extreme waves!generation!multidirectional}
\end{itemize}

\appendix
\pagestyle{fancy}
\renewcommand{\chaptermark}[1]{\markboth{\large \appendixname \ \thechapter. \ #1}{}}
\renewcommand{\sectionmark}[1]{\markright{\normalsize \thesection. \ #1}}
\fancyhf{} 
\fancyfoot[LE,RO]{\thepage}%
\fancyhead[LE]{\large \bfseries \slshape \nouppercase{\leftmark}}
\fancyhead[RO]{\large \bfseries \slshape \nouppercase{\rightmark}}
\renewcommand{\headrulewidth}{0.1pt}
\fancypagestyle{plain}{%
\fancyhead{} 
\renewcommand{\headrulewidth}{0pt} 
}

\chapter{Spectrum of the single soliton solution} \label{Singlesolitonspectrum}
\index{single soliton!spectrum}

In this appendix, we present the spectrum of the single soliton solution of the NLS equation given by~\eqref{solitonspectrum} on page~\pageref{solitonspectrum}. To find the spectrum evolution, we apply the Fourier transform to the solution by multiplying with $e^{-i\omega \tau}$ and integrating with respect to $\tau$. Now it reads
\begin{equation}
\hat{A}(\xi,\omega) = \int_{-\infty}^{\infty} A_{0}(\xi) \sqrt{2} \frac{e^{-i\omega \tau}}{\cosh (\alpha \tau)} d \tau.
\end{equation}
Using the fact that the integration of an odd function $\sin (\omega \tau)$ over a symmetric interval vanishes and the integration of an even function $\cos (\omega \tau)$ results twice of the integral with half-interval, we have
\begin{equation}
\hat{A}(\xi,\omega) = 2 \sqrt{2} A_{0}(\xi) \int_{-\infty}^{0} \frac{\cos (\omega \tau)}{\cosh (\alpha \tau)} d \tau.
\end{equation}
The integration results to the following expression:
\begin{equation}
\hat{A}(\xi,\omega) = \frac{2 \sqrt{2} A_{0}(\xi)}{\alpha^{2} + \omega^{2}} \left[(\alpha + i \omega) _{2}F_{1}\left(\frac{1}{2}\left(1 - i\frac{\omega}{\alpha} \right), 1; \frac{1}{2}\left(3 - i\frac{\omega}{\alpha} \right); -1 \right) + \textmd{c.\,c.} \right],
\end{equation}
where $_{2}F_{1}(a,b;c;z)$ is frequently known as `the' hypergeometric function\index{hypergeometric function} or Gauss' hypergeometric function\index{Gauss' hypergeometric function|see{hypergeometric function}} since it is the first hypergeometric function to be studied back in the early 19$^\textmd{th}$ century {\small [\url{http://mathworld.wolfram.com/HypergeometricFunction.html}]}. The hypergeometric functions are solutions to the hypergeometric differential equation
\begin{equation}
z(1 - z)\frac{d^{2}y}{dz^{2}} + [c - (a + b + 1)z] \frac{dy}{dz} - a b y = 0.
\end{equation}
A generalized hypergeometric function $_{p}F_{q}(a_{1}, a_{2}, \dots, a_{p}; b_{1}, b_{2}, \dots, b_{q};z)$ is a function which can be defined in the form of a hypergeometric series, i.e., a series for which the ratio of successive terms can be written as
\begin{equation}
\frac{c_{k+1}}{c_{k}} = \frac{(k + a_{1})(k + a_{2}) \dots (k + a_{p})}{(k + b_{1})(k + b_{2}) \dots (k + b_{q})(k + 1)} z.
\end{equation}
For more information on this function, please consult ``Hypergeometric Functions", Chapter 15 in {\it Handbook of Mathematical Functions with Formulas, Graphs, and Mathematical Tables}, edited by M. Abramowitz and I. A. Stegun, 9th printing. New York: Dover, pp. 555--566, 1972.

So, the spectrum of the single soliton now reads
\begin{equation}
\hat{A}(\xi,\omega) = \frac{2 \sqrt{2} A_{0}(\xi)}{\alpha^{2} + \omega^{2}} \left[(\alpha + i \omega) \sum_{m =0}^{\infty} (-1)^{m} \frac{\alpha - i \omega}{\alpha(2m + 1) - i \omega} + \textmd{c\,c} \right].
\end{equation}
We can simplify this expression leading to~\eqref{solitonspectrum}, written as follows:
\begin{equation}
\hat{A}(\xi,\omega) = 4 \alpha \sqrt{2} A_{0}(\xi) \sum_{m =0}^{\infty} (-1)^{m} \left(\frac{2m + 1}{\alpha^{2}(2m + 1)^{2} + \omega^{2}} \right).
\end{equation}

\clearpage \thispagestyle{empty}
\chapter{Spectrum of the SFB} \label{SFBspectrum}
\index{SFB!spectrum}

In this appendix, we will show the derivation of the spectrum corresponding to the SFB solution of the NLS equation given by~(\ref{magnitude0spectrum}--\ref{magnitudespectrum}) on page~\pageref{magnitudespectrum}. The expressions are presented but not derived in Chapter~3 of {\it Solitons--Nonlinear Pulses and Beams} by N. N. Akhmediev and A. Ankiewicz, Chapman \& Hall, 1997. This appendix  helps readers who are curious about the derivation. To start with, we will show two useful trigonometric integrals and one useful trigonometric series. Later on, we will use these useful formulations to derive the spectrum formulation of the SFB.

\section{Proof of a useful trigonometric integral}
We show the first useful trigonometric integral is correct:
\begin{equation}
\int_{0}^{\pi} \frac{d\tau}{1 - a\,\cos \tau} = \frac{\pi}{\sqrt{1 - a^{2}}}. 	\label{trigoint1}
\end{equation}
Substituting $z = \tan\left(\frac{1}{2} \tau\right)$, we now have $d \tau = 2\,\cos^{2}\left(\frac{1}{2} \tau\right)\,dz = 2/(1 + z^{2})dz$ and $\cos \tau = 2/(1 + z^{2}) - 1$. The integral turns into
\begin{equation}
\int_{0}^{\pi} \frac{d \tau}{1 - a\,\cos \tau} = \int_{0}^{\infty} \frac{2\,dz}{(1 - a) + (1 + a)z^{2}}.
\end{equation}
Substituting $(1 + a)/(1 - a)\, z^{2} = \tan^{2} w$, we have $\sqrt{(1 + a)/(1 - a)}\, dz = (1 + \tan^{2} w)\, dw$ and the integral becomes
\begin{equation}
\int_{0}^{\pi} \frac{d \tau}{1 - a\,\cos \tau} = \frac{2}{\sqrt{1 - a^{2}}} \int_{0}^{\pi/2} \frac{1 + \tan^{2} w}{1 + \tan^{2} w}\, dw.
\end{equation}
The integral simplifies to $\pi/2$ and we obtain~\eqref{trigoint1}.

\section{Proof of a useful trigonometric series}
We show the following useful trigonometric series:
\begin{equation}
\sum_{n = 0}^{\infty} r^{n}\,\cos (n \tau) = \frac{1}{2}\left(1 + \frac{\cos s}{1 - \sin s \, \cos x} \right).	  \label{trigser}
\end{equation}
For $|r| < 1$ and $n \in \mathbb{N}_{0}$, we have the following series identity:
\begin{equation}
\sum_{n = 0}^{\infty} (r e^{i\tau})^{n} = \frac{1}{1 - r e^{i\tau}} = \frac{1 - r e^{-i \tau}}{1 - 2r \cos \tau + r^{2}}.
  \label{series1}
\end{equation}
Also, for $|r| < 1$ but $n \in \mathbb{Z \setminus N}$, we have another series identity:
\begin{equation}
\sum_{n = -\infty}^{0} \left(\frac{1}{r} e^{i\tau} \right)^{n} = \frac{1}{1 - r e^{-i\tau}} = \frac{1 - r e^{i \tau}}{1 - 2r \cos \tau + r^{2}}.  \label{series2}
\end{equation}
Taking the real part of both series~\eqref{series1} and~\eqref{series2}, we obtain
\begin{equation}
\sum_{n = 0}^{\infty} r^{n}\,\cos (n \tau) = \frac{1 - r\,\cos \tau}{1 - 2r\,\cos \tau + r^{2}} = \sum_{n = -\infty}^{0} \frac{1}{r^{n}}\, \cos(n \tau).
\end{equation}
Now consider $n \in \mathbb{N}_{0}$ and substitute $r = \tan \left(\frac{1}{2}s\right)$. Multiplying both the numerator and the denominator by $\cos^{2}\left(\frac{1}{2}s \right)$, the infinite series becomes
\begin{equation}
\sum_{n = 0}^{\infty} r^{n}\,\cos (n \tau) = \frac{\cos^{2}\left(\frac{1}{2}s \right) - \frac{1}{2}\sin s \, \cos \tau}{1 - \sin s\, \cos \tau}.
\end{equation}
Substituting $\cos^{2}\left(\frac{1}{2}s \right) = \frac{1}{2}(1 + \cos s)$, we obtain the desired useful trigonometric series~\eqref{trigser}.

\section{Proof of another useful trigonometric integral}
We show that the following trigonometric integral is correct:
\begin{equation}
\int_{0}^{\pi} \frac{\cos (n \tau)\, d \tau}{b - a\,\cos \tau} = \frac{\pi}{\sqrt{b^{2} - a^{2}}} \left(\frac{b - \sqrt{b^{2} - a^{2}}}{a} \right)^{n}, \qquad \; \textmd{for} \; n \in \mathbb{Z}.		  \label{trigoint2}
\end{equation}
Let
\begin{equation}
f_{n} = \int_{0}^{\pi} \frac{\cos (n \tau)\, dx}{b - a\,\cos \tau},		  \label{trigoint3}
\end{equation}
then we have the following series
\begin{equation}
\sum_{n = 0}^{\infty} r^{n}\,f_{n} = \int_{0}^{\pi} \left(\sum_{n = 0}^{\infty} r^{n}\,\cos (n \tau) \right) \frac{d \tau}{b - a\,\cos \tau}.	  \label{series3}
\end{equation}
By letting $a/b = \sin y$ and applying the trigonometric series~\eqref{trigser}, we obtain an integral series
\begin{eqnarray}
\sum_{n = 0}^{\infty} r^{n}\,f_{n} = \frac{1}{2b} \int_{0}^{\pi} \frac{d \tau}{1 - \sin y\, \cos \tau} + \frac{1}{2b} \int_{0}^{\pi} \frac{\cos s \, d \tau}{(1 - \sin y\, \cos \tau)(1 - \sin y\, \cos \tau)}.
\end{eqnarray}
The first integral of this expression can be evaluated using~\eqref{trigoint1}. After applying the partial fraction, we can also evaluate the second integral using~\eqref{trigoint1}. Thus the series~\eqref{series3} now becomes
\begin{equation}
\sum_{n = 0}^{\infty} r^{n}\,f_{n} = \frac{\pi}{2b}\left(\frac{1}{\cos y} + \cos s \frac{\tan s - \tan y}{\sin s - \sin y} \right).
\end{equation}
Let $r = \tan\left(\frac{1}{2}s\right)$, then $\sin s = 2r/(1 + r^{2})$, $\cos s = 1 - r^{2}/(1 + r^{2})$, and $\tan s = 2r/(1 - r^{2})$. Substituting these relations into the series we now have
\begin{equation}
\sum_{n = 0}^{\infty} r^{n} f_{n} = \sum_{n = 0}^{\infty} r^{n} \frac{\pi}{b \cos y} \left(\frac{\sin y}{1 + \cos y} \right)^{n}.
\end{equation}
For $n \in \mathbb{Z \setminus N}$, the following relation holds:
\begin{equation}
\sum_{n = -\infty}^{0} \frac{1}{r^{n}}\,f_{n} = \sum_{n = -\infty}^{0} \frac{1}{r^{n}}\frac{\pi}{b}\,\frac{1}{\cos y}\,\left(\frac{1 + \cos y}{\sin y}\right)^{n}.
\end{equation}
Therefore we have an expression for $f_{n}$:
\begin{equation}
f_{n} = \frac{\pi}{b}\frac{1}{\cos y} \left(\frac{\sin y}{1 + \cos y}\right)^{n},  \qquad \; \textmd{for} \; n \in \mathbb{N}_{0},
\end{equation}
and
\begin{equation}
f_{n} = \frac{\pi}{b}\frac{1}{\cos y} \left(\frac{1 + \cos y}{\sin y}\right)^{n},  \qquad \; \textmd{for} \; n \in \mathbb{Z \setminus N}.
\end{equation}
Substituting back the relation $\sin y = a/b$ and $\cos y = \sqrt{b^{2} - a^{2}}/b$, $f_{n}$ then turns into
\begin{equation}
f_{n} = \frac{\pi}{\sqrt{b^{2} - a^{2}}} \left(\frac{b - \sqrt{b^{2} - a^{2}}}{a} \right)^{n},  \qquad \; \textmd{for} \; n \in \mathbb{N}_{0}
\end{equation}
and
\begin{equation}
f_{n} = \frac{\pi}{\sqrt{b^{2} - a^{2}}} \left(\frac{b + \sqrt{b^{2} - a^{2}}}{a} \right)^{n},  \qquad \; \textmd{for} \; n \in \mathbb{Z \setminus N}.
\end{equation}
We can write these expressions into a single formula as follows:
\begin{equation}
f_{n} = \frac{\pi}{\sqrt{b^{2} - a^{2}}} \left(\frac{b - \sqrt{b^{2} - a^{2}}}{a} \right)^{n},  \qquad \; \textmd{for} \; n \in \mathbb{Z}.
\end{equation}
Thus, we have shown that the second trigonometric integral is correct.

\section{Derivation of the SFB spectrum}
\index{SFB!spectrum}

The spectrum is defined by the expression~\eqref{spectrumSFB}. Using the fact that the integrand is an even function with respect to $\tau$, we can write the spectrum explicitly as follows:
\begin{equation}
a_{n}(\xi) = \frac{A_{0}(\xi)}{\pi} \left( \int_{0}^{\pi} \frac{f(\xi)\, \cos(n \tau)}{\cosh(\sigma \xi) - \sqrt{1 - \frac{1}{2}\tilde{\nu}^{2}} \cos \tau}\, d\tau - \int_{0}^{\pi} \cos(n \tau)\, d\tau \right), \qquad n \in \mathbb{Z},
\end{equation}
where $f(\xi) = \tilde{\nu}^{2} \cosh(\sigma \xi) - \tilde{\sigma} \sinh(\sigma \xi)$. The expression $\int_{0}^{\pi} \cos (n \tau)\, d\tau$ vanishes for $n \neq 0$. Using the relation~\eqref{trigoint2}, we find the spectrum for $n \neq 0$ as given in~\eqref{magnitudespectrum}:
\begin{equation}
a_{n}(\xi) = A_{0}(\xi) \frac{\tilde{\nu}^{2}\cosh (\sigma \xi) - i \tilde{\sigma} \sinh(\sigma \xi)}{\sqrt{\cosh^{2}(\sigma \xi) - \left(1 - \frac{1}{2}\tilde{\nu}^{2} \right)}}\left(\frac{\cosh (\sigma \xi) - \sqrt{\cosh^{2}(\sigma \xi) - \left(1 - \frac{1}{2}\tilde{\nu}^{2} \right)}}{\sqrt{1 - \frac{1}{2}\tilde{\nu}^{2}}} \right)^{\!\!n}.
\end{equation}
For $n  = 0$, the integral $\int_{0}^{\pi} \cos (n \tau)\, d\tau = \pi$, and therefore the spectrum for $n = 0$ is given by
\begin{equation}
a_{0}(\xi) = A_{0}(\xi) \left(\frac{\tilde{\nu}^{2}\cosh(\sigma \xi) - i \tilde{\sigma} \sinh(\sigma \xi)}{\sqrt{\cosh^{2}(\sigma \xi) - \left( 1 - \frac{1}{2}\tilde{\nu}^{2}\right)}} - 1 \right).
\end{equation}
Therefore, we have obtained the spectrum of the SFB.

\clearpage \thispagestyle{empty}
\chapter{Quantities related to SFB$_{2}$} \label{ApendiksSFB2}

This appendix presents explicit expressions which are related to SFB$_{2}$ discussed in Chapter~\ref{5HighOrder}.

\section{Quantities related to the asymptotic behavior}
The phase difference $\phi_{20}$ mentioned in Subsection~\ref{SFB2asbe} on page~\pageref{SFB2asbe} reads
\begin{equation}
\tan \phi_{20} = \frac{3\tilde{\nu}_{2}(2\sqrt{2 - 4 \tilde{\nu}_{2}^{2}} - \sqrt{2 - \tilde{\nu}_{2}^{2}})}{13\tilde{\nu}_{2}^{2} - 5  + 2\sqrt{2 - 4 \tilde{\nu}_{2}^{2}}\,\sqrt{2 - \tilde{\nu}_{2}^{2}}}.
\end{equation}

The complex quantities mentioned in the same subsection have rather lengthy expressions. The quantity $w_{21}$ reads
\begin{equation}
w_{21} = \rho_{21}\,e^{\,i\phi_{21}}
\end{equation}
where
\begin{align}
\rho_{21} &= \frac{6\tilde{\nu}_{2}\sqrt{f_{1}(\tilde{\nu}_{2}) + f_{2}(\tilde{\nu}_{2})\,2\sqrt{2 - 4\tilde{\nu_{2}}^{2}}\,\sqrt{2 - \tilde{\nu}_{2}^{2}}}} {(4\tilde{\nu}_{2}^{2} - 5 + 2\sqrt{2 - 4\tilde{\nu}_{2}^{2}}\,\sqrt{2 - \tilde{\nu}_{2}^{2}})^{2}} \\
\tan \phi_{21} &= \frac{4(2 \tilde{\nu}_{2}^{2} - 3)\sqrt{2 - 4\tilde{\nu}_{2}^{2}} + (9 - 16\tilde{\nu}_{2}^{2})\sqrt{2 - \tilde{\nu}_{2}^{2}}} {\tilde{\nu}_{2}(16\tilde{\nu}_{2}^{2} - 11 + 2\sqrt{2 - 4\tilde{\nu}_{2}^{2}} \sqrt{2 - \tilde{\nu}_{2}^{2}})}
\end{align}
and
\begin{eqnarray}
f_{1}(\tilde{\nu}_{2}) \!\!\!\! &=& \!\!\!\! 4\tilde{\nu}_{2}^{4}(30\tilde{\nu}_{2}^{4} - 233\tilde{\nu}_{2}^{2} + 511) - 5(341\tilde{\nu}_{2}^{2} - 90)\\
f_{2}(\tilde{\nu}_{2}) \!\!\!\! &=& \!\!\!\! 48\tilde{\nu}_{2}^{6} - 217\tilde{\nu}_{2}^{4} + 296\tilde{\nu}_{2}^{2} - 108.
\end{eqnarray}

The quantity $\tilde{w}_{21}$ reads
\begin{equation}
\tilde{w}_{21} = \tilde{\rho}_{21} e^{\,i\tilde\phi_{21}}
\end{equation}
where
\begin{align}
\tilde{\rho}_{21} &= \frac{\sqrt{f_{3}(\tilde{\nu}_{2}) + f_{4}(\tilde{\nu}_{2}) \sqrt{2 - \tilde{\nu}_{2}^{2}} \sqrt{2 - 4\tilde{\nu}_{2}^{2}}}} {3\tilde{\nu}_{2}^{2}(2 - \tilde{\nu}_{2}^{2}) (1 - 2\tilde{\nu}_{2}^{2})} \\
\tan \tilde{\phi}_{21} &= \frac{3}{\tilde{\nu}_{2}} \frac{(2\tilde{\nu}_{2}^{2} - 1)\sqrt{2 - \tilde{\nu}_{2}^{2}} - (2\tilde{\nu}_{2}^{2} - 5) \sqrt{2 - 4\tilde{\nu}_{2}^{2}}}{13 - 14\tilde{\nu}_{2}^{2} - 7\sqrt{2 - \tilde{\nu}_{2}^{2}} \sqrt{2 - 4\tilde{\nu}_{2}^{2}}}
\end{align}
and
\begin{align}
f_{3}(\tilde{\nu}_{2}) &= 106\tilde{\nu}_{2}^{6} + 23\tilde{\nu}_{2}^{4} - 488\tilde{\nu}_{2}^{2} + 234\\
f_{4}(\tilde{\nu}_{2}) &=  62\tilde{\nu}_{2}^{4} + 17\tilde{\nu}_{2}^{2} - 45.
\end{align}
The quantity $\bar{w}_{21}$ reads
\begin{equation}
  \bar{w}_{21} = w_{21} + \tilde{w}_{21}.
\end{equation}

The quantity $w_{22}$ reads
\begin{equation}
w_{22} = \rho_{22} e^{\,i\phi_{22}}
\end{equation}
where
\begin{align}
\rho_{22} &= \frac{\sqrt{g_{0}(\tilde{\nu}_{2})}}{(4\tilde{\nu}_{2}^{2} - 5 + 2\sqrt{2 - 4\tilde{\nu}_{2}^{2}}\,\sqrt{2 - \tilde{\nu}_{2}^{2}})^{2}} \quad \\
\tan \phi_{22} &= \frac{4\tilde{\nu}_{2}^{2} - 5 + 2\sqrt{2 - 4\tilde{\nu}_{2}^{2}}\,\sqrt{2 - \tilde{\nu}_{2}^{2}} - 3\tilde{\nu}_{2}(\sqrt{2 - 4\tilde{\nu}_{2}^{2}} - \sqrt{2 - \tilde{\nu}_{2}^{2}})} {\tilde{\nu}_{2}^{2} [4 (\tilde{\nu}_{2}^{2} + 1) + 2\sqrt{2 - 4\tilde{\nu}_{2}^{2}} \sqrt{2 - \tilde{\nu}_{2}^{2}}]}
\end{align}
and
\begin{align}
g_{0}(\tilde{\nu}_{2}) &= g_{1}(\tilde{\nu}_{2}) + g_{2}(\tilde{\nu}_{2})\,\sqrt{2 - 4\tilde{\nu}_{2}^{2}}\,\sqrt{2 -\tilde{\nu}_{2}^{2}} \nonumber \\
& \quad \, + g_{3}(\tilde{\nu}_{2})\,\sqrt{2 - 4\tilde{\nu}_{2}^{2}} + g_{4}(\tilde{\nu}_{2})\,\sqrt{2 - \tilde{\nu}_{2}^{2}} \\
g_{1}(\tilde{\nu}_{2}) &= -288\tilde{\nu}_{2}^{6}(8\tilde{\nu}_{2}^{4} - 6\tilde{\nu}_{2}^{2} + 9) + 1139\tilde{\nu}_{2}^{4} - 44\tilde{\nu}_{2}^{2} + 41\\
g_{2}(\tilde{\nu}_{2}) &= -576\tilde{\nu}_{2}^{4}(2\tilde{\nu}_{2}^{2} -1)(\tilde{\nu}_{2}^{2} + 1) - (\tilde{\nu}_{2}^{2} + 10)\\
g_{3}(\tilde{\nu}_{2}) &=   18\tilde{\nu}_{2}(3 - 2\tilde{\nu}_{2}^{2})\\
g_{4}(\tilde{\nu}_{2}) &=   18\tilde{\nu}_{2}(4\tilde{\nu}_{2}^{2} - 3).
\end{align}

The quantity $\tilde{w}_{22}$ reads
\begin{equation}
\tilde{w}_{22} = \tilde{\rho}_{22} e^{\,i\tilde{\phi}_{22}}
\end{equation}
where
\begin{align}
\tilde{\rho}_{22} &= \frac{9 \tilde{\nu}_{2}^{2}(2 - \tilde{\nu}_{2}^{2})\sqrt{g_{5}(\tilde{\nu}_{2}) + g_{6}(\tilde{\nu}_{2}) \sqrt{2 - \tilde{\nu}_{2}^{2}} \sqrt{2 - 4\tilde{\nu}_{2}^{2}}}} {(4\tilde{\nu}_{2}^{2} - 5 + 2\sqrt{2 - \tilde{\nu}_{2}^{2}} \sqrt{2 - 4 \tilde{\nu}_{2}^{2}})^{3}} \\
\tan(\tilde{\phi}_{22}) &= 2\tilde{\nu}_{2}\sqrt{2 - 4 \tilde{\nu}_{2}^{2}} \frac{4\tilde{\nu}_{2}^{2} - 5 + 2\sqrt{2 - \tilde{\nu}_{2}^{2}}\sqrt{2 - 4\tilde{\nu}_{2}^{2}}} {16\tilde{\nu}_{2}^{2} - 11 + 8\sqrt{2 - \tilde{\nu}_{2}^{2}} \sqrt{2 - 4\tilde{\nu}_{2}^{2}}} \; \qquad%
\end{align}
and
\begin{align}
g_{5}(\tilde{\nu}_{2}) &=  377 - 8\tilde{\nu}_{2}^{2}(83 + 98\tilde{\nu}_{2}^{2} - 192\tilde{\nu}_{2}^{4} + 64\tilde{\nu}_{2}^{6}) \\
g_{6}(\tilde{\nu}_{2}) &= -16 (16\tilde{\nu}_{2}^{6} - 28\tilde{\nu}_{2}^{4} - 6 \tilde{\nu}_{2}^{2} - 11).
\end{align}
The quantity $\bar{w}_{22}$ reads:
\begin{equation}
\bar{w}_{22} = w_{22} + \tilde{w}_{22}.
\end{equation}

The quantity $w_{23}$ reads:
\begin{equation}
w_{23} = \rho_{23} e^{\,i\phi_{23}}
\end{equation}
where
\begin{align}
\rho_{23} &= \frac{3\sqrt{h_{1}(\tilde{\nu}_{2}) + h_{2}(\tilde{\nu}_{2})\sqrt{2 - \tilde{\nu}_{2}^{2}} \sqrt{2 - 4\tilde{\nu}_{2}^{2}}}} {(4\tilde{\nu}_{2}^{2} - 5 + 2\sqrt{2 - \tilde{\nu}_{2}^{2}} \sqrt{2 - 4 \tilde{\nu}_{2}^{2}})^{3}}\\
\tan \phi_{23} &= \frac{3\,h_{3}(\tilde{\nu}_{2})\,\tilde{\nu}_{2}\sqrt{2 - 4\tilde{\nu}_{2}^{2}} - 2\,h_{4}(\tilde{\nu}_{2})\,\tilde{\nu}_{2}\sqrt{2 - \tilde{\nu}_{2}^{2}}} {3\tilde{\nu}_{2}^{2}[(2 - \tilde{\nu}_{2}^{2})(5 - 16\tilde{\nu}_{2}^{2}) + 4(2\tilde{\nu}_{2}^{2} - 1) \sqrt{2 - \tilde{\nu}_{2}^{2}} \sqrt{2 - 4\tilde{\nu}_{2}^{2}}]}
\end{align}
and
\begin{align}
h_{1}(\tilde{\nu}_{2}) &= -128\tilde{\nu}_{2}^{10}(290\tilde{\nu}_{2}^{6} - 1614\tilde{\nu}_{2}^{4} + 4487\tilde{\nu}_{2}^{2} - 9286) \nonumber \\ 
& \quad \, - \tilde{\nu}_{2}^{2}(1693780\tilde{\nu}_{2}^{6} - 1348157\tilde{\nu}_{2}^{4} + 517080\tilde{\nu}_{2}^{2} - 74492) \\
h_{2}(\tilde{\nu}_{2}) &= -12\tilde{\nu}_{2}^{2}\left(256\tilde{\nu}_{2}^{12} - 6680\tilde{\nu}_{2}^{10} + 24904\tilde{\nu}_{2}^{8} - 33084\tilde{\nu}_{2}^{6} \right. \nonumber \\ & \quad \, \left. + \; 19788\tilde{\nu}_{2}^{4} - 6291\tilde{\nu}_{2}^{2} + 930 \right)\\
h_{3}(\tilde{\nu}_{2}) &= 32\tilde{\nu}_{2}^{6} -  76\tilde{\nu}_{2}^{4} +  35\tilde{\nu}_{2}^{2} - 10\\
h_{4}(\tilde{\nu}_{2}) &=  8\tilde{\nu}_{2}^{6} - 196\tilde{\nu}_{2}^{4} + 304\tilde{\nu}_{2}^{2} - 95.
\end{align}

\section{Coefficients of the cubic equation related to phase singularity}
\label{SFB2cubicPS}

In Subsection~\ref{SFB2PS}, we mention about coefficients of the cubic equations in $\cos(\nu_{2} \zeta_{2})$. Explicit expressions of these coefficients are given as follows:
\begin{align}
C_{0}(\tilde{\nu}_{2}) &= \frac{2\tilde{\nu}_{2}^{2}}{\tilde{\sigma}_{1} \tilde{\sigma}_{2}}(13 \tilde{\nu}_{2}^{2} - 5) + \frac{3 \sqrt{2}}{2\tilde{\sigma}_{1}} \tilde{\nu}_{2} (1 - \tilde{\nu}_{2}^{2}) \\
C_{1}(\tilde{\nu}_{2}) &= \frac{3}{2} + \frac{3 \sqrt{2}}{\tilde{\sigma}_{2}} \tilde{\nu}_{2} (1 - 4\tilde{\nu}_{2}^{2}) \\
C_{2}(\tilde{\nu}_{2}) &= \frac{3\sqrt{2}}{\tilde{\sigma}_{1}} \tilde{\nu}_{2} (\tilde{\nu}_{2}^{2} - 1).
\end{align}

\clearpage \thispagestyle{empty}
\chapter{Wave generation theory} \label{wavegeneration}
\index{wave generation theory}

\section{Introduction}

In this section, we will consider the problem of how to generate waves in a wave tank of a hydrodynamic laboratory. The wave tank in the context of this thesis is a facility with a wavemaker\index{wavemaker} on one side and a wave absorbing beach on the other side. We consider a tank with a flat bottom, and no water is flowing in or out of the tank. Typically, the situation is that waves are generated by a flap-type wavemaker at one side of a long tank; the motion of the flap `pushes' the waves to start propagating along the tank. This means that typically we are dealing with a signaling problem\index{signaling problem} or a boundary value problem\index{boundary value problem} (\textsc{bvp}), which is different from an initial value problem\index{initial value problem} (\textsc{ivp}) when one tries to find the evolution of waves from given surface elevation and velocities at an initial moment. \index{wave generation theory}

To illustrate this for the simplest possible case, consider the linear, non-dispersive second-order wave equation for waves in one spatial direction $x$ and time $t$: $\partial_{t}^{2}\eta = c^{2}\partial_{x}^{2}\eta$. Here, $\eta(x,t)$ denotes the surface wave elevation, and $c > 0$ is the constant propagation speed. The general solution is given by $\eta(x,t) = f(x - ct) + g(x + ct)$, for arbitrary functions $f$ and $g$. The term $f(x - ct)$ is the contribution of waves traveling to the right (in positive $x$-direction), and $g(x + ct)$ waves running to the left. For the \textsc{ivp}, specifying at an initial time (say at $t = 0$) the wave elevation $\eta(x,0)$ and the velocity $\partial_{t}\eta(x,0)$ determines the functions $f$ and $g$ uniquely. For the \textsc{bvp}, resembling the generation at $x = 0$, we prescribe the wave elevation at $x = 0$ for all positive time, assuming the initial elevation to be zero for positive $x$ (in the tank this means a flat surface prior to the start of the generation). If the signal is given by $s(t)$, vanishing for $t < 0$, the corresponding solution running into the tank is $\eta(x,t) = f(x - ct)$ which should be equal to $s(t)$ at $x = 0$, leading to $\eta (x,t) = s(t - x/c)$. Reversely, for a desired wave field $f(x - ct)$ running in the tank, the required surface elevation at $x = 0$ is given by $s(t) = f(-ct)$. This shows the characteristic property of the \textsc{bvp} for the signaling problem. \index{wave generation theory}

The actual equations for the water motion, and the precise incorporation of the flap motion, are much more difficult than shown in the simple example above. In particular, for the free motion of waves, there are two nontrivial effects: `dispersion'\index{dispersion} and `nonlinearity'\index{nonlinearity}. `Dispersion' means the propagation speed of waves depends on their wavelength (or frequency), described explicitly in the linear theory (i.e., for small surface elevations) by the linear dispersion relation\index{dispersion relation!linear} (\textsc{ldr}), see formula (\ref{ldr}). In fact, for a given frequency there is one normal mode that travels to the right as a harmonic wave, the propagating mode,\index{mode(s)!propagating} and there are solutions decaying exponentially for increasing distance (the evanescent modes).\index{mode(s)!evanescent} We will use the right propagating mode and the evanescent modes as building blocks to describe the generation of waves by the wavemaker. For each frequency in the spectrum, the Fourier amplitude of the flap motion is then related to the amplitude of the corresponding propagating and evanescent modes. If we assume these amplitudes to be sufficiently small, say of the `first-order' $\epsilon$, with $\epsilon$ a small quantity, we will be able to deal with nonlinear effects in a sequential way. This is needed because the second effect is that in reality, the equations are nonlinear. The quadratic nature of the `nonlinearity' implies that each two wave components will generate other components with an amplitude that is proportional to the product of the two amplitudes, the so-called `bound-wave' components\index{wave components!bound-wave} which have amplitudes of the order $\epsilon^{2}$. These are the so-called `second-order effects'. \index{second-order effects} \index{wave generation theory}

For instance, two harmonic waves of frequency $\omega_{1}$, $\omega_{2}$ and wavenumber $k_{1}$, $k_{2}$ (related by the \textsc{ldr}) will have a bound-wave with frequency $\omega_{1} + \omega_{2}$ and wavenumber $k_{1} + k_{2}$. Since the \textsc{ldr} is a concave function of the wavenumber, this last frequency--wavenumber combination does not satisfy the \textsc{ldr}, i.e., this is not a free-wave: it can only exist in the combination with the free-wave. `Free-wave' components \index{wave components!free-wave} mean that frequency and wavenumber satisfy the \textsc{ldr}. This second-order bound-wave that comes with a first-order free-wave has also its consequence for the wave generation. If the first-order free-wave component is compatible with the flap motion, the presence of the bound-wave component\index{wave components!bound-wave} will disturb the wave motion, such that the additional second-order free-wave will be generated as well. This is undesired, since the second-order free-wave component has a different propagation speed as the bound-wave component, thereby introducing a spatially inhomogeneous wave field. That is why we add to the flap motion the additional effects of second-order bound-waves, thereby preventing any second-order free-wave component to be generated. This process is called the `second-order steering' of the wavemaker motion.\index{wavemaker steering!second-order} \index{wave generation theory} \index{wavemaker}

This technique can be illustrated using a simple \textsc{ivp} for an ordinary differential equation as follows. Consider the nonlinear equation with a linear operator $\cal{L}$: \label{linop}
\begin{equation*}
\textmd{$\cal{L}$} \eta := \partial_{t}^{2} \eta + \omega_{0}^{2} \eta = \eta^{2},
\end{equation*}
for which we look for small solutions, say of order $\epsilon$, a small quantity. The series expansion technique then looks for a solution in the form
\begin{equation*}
\eta = \epsilon \eta^{(1)} + \epsilon^{2} \eta^{(2)} + \textmd{$\cal{O}$}(\epsilon^{3}).
\end{equation*}
Substitution in the equation and requiring each order of $\epsilon$ to vanish leads to a sequence of \textsc{ivp}s, the first two of which read
\begin{equation*}
\textmd{$\cal{L}$} \eta^{(1)} = 0; \qquad \textmd{$\cal{L}$} \eta^{(2)} = (\eta^{(1)})^{2}; \qquad \dots \; .
\end{equation*}
Observe that the equations for $\eta^{(1)}$ and $\eta^{(2)}$ are linear equations, homogeneous for $\eta^{(1)}$ and nonhomogeneous (with the known right-hand side after $\eta^{(1)}$ has been found)~for~$\eta^{(2)}$.	\index{wave generation theory}

\vspace*{-0.3cm} 
Suppose that the first-order solution we are interested in is $\eta^{(1)} = a e^{-i\omega_{0}t}$, already introducing the complex arithmetic that will be used in the sequel also. This solution is found for the initial values $\eta^{(1)}(0) = a$, $\partial_{t} \eta^{(1)} (0) = -i \omega _{0} a$. Then the equation for $\eta^{(2)}$, i.e., $\textmd{$\cal{L}$} \eta^{(2)} = a^{2} e^{-2i\omega_{0}t}$ has as particular solution: $\eta_{\textmd{p}}^{(2)} = A e^{-2i\omega_{0}t}$ with $A = -a^{2}/(3 \omega_{0}^{2})$. This particular solution is the equivalent of a bound-wave component\index{wave components!bound-wave} mentioned above: it comes inevitably with the first-order solution $\eta^{(1)}$. However, $\eta_{\textmd{p}}^{(2)}$ will change the initial condition; forcing it to remain unchanged could be done by adding a solution $\eta_{\textmd{h}}^{(2)}$ of the homogeneous equation: $\textmd{$\cal{L}$} \eta_{\textmd{h}}^{(2)} = 0$ that cancels the particular solution at $t = 0$, explicitly: $\eta_{\textmd{h}}^{(2)} = \frac{1}{2} A \left( e^{i\omega_{0}t} - 3 e^{i\omega_{0}t}\right)$. This homogeneous solution corresponds to the second-order free-wave component mentioned above. To avoid this solution to be present, the initial value has to be taken like
\begin{equation*}
\eta(0) = \epsilon a + \epsilon^{2}A;   \qquad 	  \partial_{t} \eta(0) = -i \epsilon \omega_{0}a - 2i\epsilon^{2} \omega_{0} A.
\end{equation*}
The second-order terms in $\epsilon$ in these initial conditions are similar to the second-order steering of the flap motion for the signaling problem.\footnote[1]{Just as in this example, the hierarchy of equations also continues for the \textsc{bvp}: there will also be third and higher-order contributions and bound and free-waves in each order. \index{wave components!free-wave} Higher-order steering than second-order has not been done until now, since the effects are smaller, although there are some exceptions. \par}

Besides the two difficult aspects of nature, i.e., dispersion and nonlinearity, the precise description of the signal is also quite involved, since the signal has to be described on a moving boundary, the flap, which complicates matters also. In the rest of this appendix, we will describe the major details of this procedure. The next section presents the \textsc{bvp} for the wave generation problem. Sections~\ref{ordesatu} and~\ref{ordedua} discuss the first and second-order wave generation theory, respectively. 	\index{wave generation theory}

The first-order wave generation theory for unidirectional regular waves corresponding to linearized Stokes theory has been known for almost a century and reference is made to the review by~\citet{8Svendsen85}, Chapter 6 of~\citet{8Dean91} and Chapter 7 of~\citet{8Hughes93}. Second-order wave generation theory has been studied since the 1960s, and many people have improved the theory thereafter. In this thesis, we only refer to relatively current publications by Sch\"{a}ffer for this theory. The full second-order wave generation theory for irregular waves is given by~\citet{8Schaffer96}. The complete second-order wave generation theory for multidirectional waves in a semi-infinite basin is given by~\citet{8Schaffer03}. Both papers include both superharmonic and subharmonic waves and cover wavemakers of the piston (translatory) and hinged (rotational) type. \index{subharmonic waves} \index{superharmonic waves}

\section{Governing equation}
\index{wave generation theory}

Let $\mathbf{u} = (u,w) = (\partial_{x} \phi, \partial_{z} \phi)$ define the velocity potential function\index{velocity potential function} $\phi = \phi(x,z,t)$ \label{phi} in a Cartesian coordinate system $(x,z)$. Let also $\eta = \eta(x,t)$ denotes the surface wave elevation, $\Xi = \Xi(z,t) = f(z)S(t)$ denotes the wavemaker position, $g$ denotes the gravitational acceleration, $h$ denotes the still water depth and $t$ denotes the time. The governing equation for the velocity potential is the Laplace equation\index{Laplace equation}
\begin{equation*}
\partial_{x}^{2} \phi + \partial_{z}^{2} \phi = 0,  \qquad \textmd{for}\; x \geq \Xi(z,t), \quad -h \leq z \leq \eta(x,t);
\end{equation*}
that results from the assumption that water (in a good approximation) is incompressible: $\nabla \cdot \mathbf{u} = 0$. The dynamic and kinematic free surface boundary conditions (\textsc{dfsbc} and \textsc{kfsbc}), the kinematic boundary condition at the wavemaker (\textsc{kwmbc}), and the bottom boundary condition (\textsc{bbc}) are given by
\begin{equation*}
\begin{array}{rrl}
\textsc{dfsbc}: & \partial_{t} \phi + \frac{1}{2}|\nabla \phi|^{2} + g \eta = 0 					& \qquad  \textrm{at} \; z = \eta(x,t); \\
\textsc{kfsbc}: & \partial_{t} \eta + \partial_{x} \eta \partial_{x} \phi - \partial_{z} \phi = 0 	& \qquad \textrm{at} \; z = \eta(x,t); \\
\textsc{kwmbc}: & \partial_{x} \phi - f(z)S'(t) - f'(z)S(t) \partial_{z} \phi = 0 					& \qquad \textrm{at} \; x = \Xi(z,t); \\
\textsc{bbc}:   & \partial_{z} \phi = 0 															& \qquad \textrm{at} \; z = -h.
\end{array}
\end{equation*}
The \textsc{dfsbc} is obtained from Bernoulli's equation\index{Bernoulli's equation}, the \textsc{kfsbc} and the \textsc{kwmbc} are derived by applying the material derivative to the surface elevation and wavemaker motion, respectively. The \textsc{bbc} is obtained from the fact that no water comes in nor goes out of the wave tank. Note that the \textsc{dfsbc} and \textsc{kfsbc} are nonlinear boundary conditions prescribed at a yet unknown and moving free surface $z = \eta(x,t)$. The elevation, potential, and wavemaker position are given by the following series expansions
\begin{eqnarray*}
\eta \!\!\!\! &=& \!\!\!\! \epsilon \eta^{(1)} + \epsilon^{2} \eta^{(2)} + \dots \\
\phi \!\!\!\! &=& \!\!\!\! \epsilon \phi^{(1)} + \epsilon^{2} \phi^{(2)} + \dots \\
   S \!\!\!\! &=& \!\!\!\! \epsilon    S^{(1)} + \epsilon^{2}    S^{(2)} + \dots,
\end{eqnarray*}
where $\epsilon$ is a small parameter, a measure of the surface elevation nonlinearity.

\begin{figure}[h]			
\begin{center}
\includegraphics[width = 0.4\textwidth, angle = -90, viewport = 30 62 333 774]{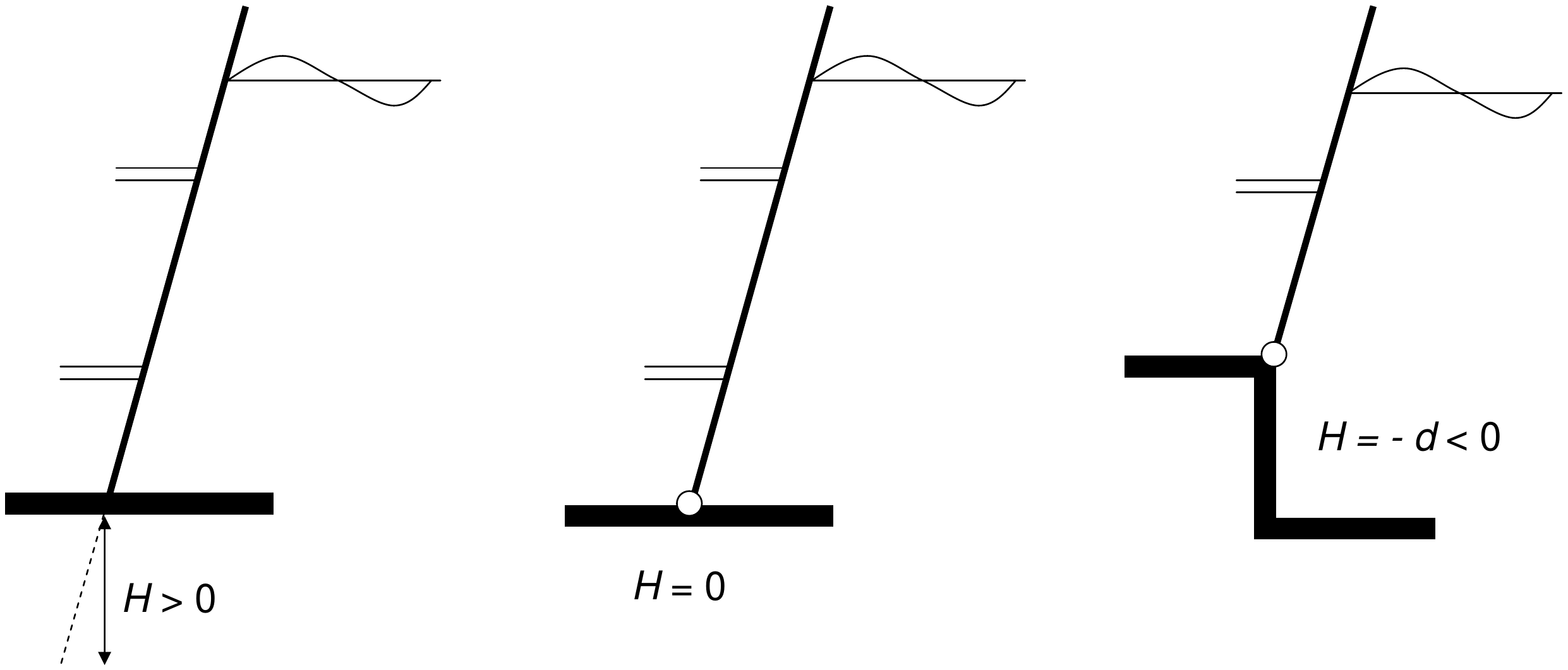}
\caption[Geometry of the flap type wavemaker]{The flap type of wavemaker with a different center of rotations: below the bottom (left), at the bottom (middle), and above the bottom (right).}		    \label{flapgeometry}
\end{center}
\vspace*{-0.72cm}
\end{figure}

The wavemaker\index{wavemaker} we will consider is a rotating flap, see Figure \ref{flapgeometry}. It is given by $\Xi(z,t) = f(z)S(t)$, where $f(z)$ describes the geometry of the wavemaker:\index{wavemaker}
\begin{equation}
f(z) = \left\{
\begin{array}{ll}
{\displaystyle
    1 + \frac{z}{h + H}}, & \; \hbox{for $-(h - d) \leq z \leq 0$;} \\
    0,                    & \; \hbox{for $-h \leq z < -(h - d)$.}   \\
\end{array}
\right. \label{wavemakertype}
\end{equation}
Note that $f(z)$ is given by design, and $S(t)$ is the wavemaker motion\index{wavemaker motion} that can be controlled externally to generate different types of waves. The center of rotation is at $z = -(h + H)$. If the center of rotation is at or below the bottom, then $d = 0$, and in fact, we do not have the last case of~\eqref{wavemakertype}. If the center of rotation is at a height $d$ above the bottom, then $d = -H$.

\section{First-order wave generation theory} \label{ordesatu}
\index{wave generation theory}

In this section, we solve a homogeneous \textsc{bvp} for the first-order wave generation theory. By prescribing the first-order wavemaker motion\index{wavemaker motion!first-order} as a linear superposition of monochromatic frequencies, we find the generated surface elevation also as a linear superposition of monochromatic modes. After applying the Taylor series expansion of the potential function $\phi$ around $x = 0$ and $z = 0$ as well as applying the series expansion method, the first-order potential function has to satisfy the Laplace equation\index{Laplace equation}
\begin{equation}
\partial_{x}^{2} \phi^{(1)} + \partial_{z}^{2} \phi^{(1)} = 0,  \qquad \textmd{for}\; x \geq 0, \quad -h \leq z \leq 0.		  \label{1stLaplace}
\end{equation}
We also obtain the \textsc{bvp} for the first-order wave generation theory at the lowest expansion order. It reads
\begin{equation}
\begin{array}{rcll}
g \eta^{(1)} + \partial_{t} \phi^{(1)} 			   \!\!\!\! &=& \!\!\!\! 0, \quad & \textmd{at} \; z = 0; \\
\partial_{t} \eta^{(1)} - \partial_{z} \phi^{(1)}  \!\!\!\! &=& \!\!\!\! 0, \quad & \textmd{at} \; z = 0; \\
\partial_{x} \phi^{(1)} - f(z) \frac{dS^{(1)}}{dt} \!\!\!\! &=& \!\!\!\! 0, \quad & \textmd{at} \; x = 0; \\
\partial_{z} \phi^{(1)} 						   \!\!\!\! &=& \!\!\!\! 0, \quad & \textmd{at} \; z = -h. \label{firstorderBCs}
\end{array}
\end{equation}
By combining the \textsc{dfsbc} and \textsc{kfsbc} at $z = 0$~\eqref{firstorderBCs}, we obtain the first-order homogeneous free surface boundary condition
\begin{equation}
g \partial_{z} \phi^{(1)} + \partial_{t}^{2} \phi^{(1)} = 0,  \qquad \textrm{at} \; z = 0.  	\label{combinedfsc}
\end{equation}
\index{wave generation theory}

\vspace*{-0.4cm}
We look for the so-called monochromatic waves:
\begin{equation*}
\phi^{(1)}(x,z,t) = \psi(z) e^{-i \theta(x,t)},
\end{equation*}
where $\theta(x,t) = k x - \omega t$. Then from the Laplace equation~\eqref{1stLaplace}, we have $\psi''(z) - k \psi(z) = 0$, for $-h \leq z \leq 0$. Applying the \textsc{bbc} leads to $\psi(z) = \alpha \cosh k(z + h)$, $\alpha \in \mathbb{C}$. From the combined free surface condition (\ref{combinedfsc}), we obtain  a relation between the wavenumber $k$ and frequency $\omega$, known as the \textit{linear dispersion relation} (\textsc{ldr}), explicitly given by\index{wave generation theory}
\begin{equation}
\omega^{2} = g k \tanh k h.  \label{ldr}
\end{equation}

Let us assume that the first-order wavemaker motion\index{wavemaker motion!first-order} $S^{(1)}(t)$ is given by a harmonic function with frequency $\omega_{n}$ and maximum stroke $|S_{n}|$ from an equilibrium position, represented in complex notation as 
\begin{equation*}
S^{(1)}(t) = \sum_{n = 1}^{\infty} -\frac{1}{2}i S_{n} e^{i \omega_{n}t} + \textmd{c.c.},
\end{equation*}
where c.c.~denotes the complex conjugate of the preceding term. Since this `first-order steering'\index{wavemaker steering!first-order} contains an infinite number of discrete frequencies $\omega_{n}$, it motivates us to write a general solution for the potential function by linear superposition of discrete spectrum. By choosing the arbitrary spectral coefficient $\alpha  = \frac{ig}{2\omega_{n}} \frac{C_{n}}{\cosh k_{n}h}$, the first potential function is found to be
\begin{equation*}
\phi^{(1)}(x,z,t) = \sum_{n = 1}^{\infty} \frac{ig}{2\omega_{n}} C_{n} \frac{\cosh k_{n}(z + h)}{\cosh k_{n} h} e^{-i\theta_{n}(x,t)} + \textmd{c.c.},
\end{equation*}
where $\theta_{n}(x,t) = k_{n} x - \omega_{n}t$, with wavenumber-frequency pairs $(k_{n},\omega_{n})$, $n \in \mathbb{Z}$ satisfying the \textsc{ldr}~\eqref{ldr}. For a continuous spectrum, the summation is replaced by an integral. Allowing the wavenumber to be complex-valued, the \textsc{ldr} becomes
\begin{equation*}
\omega_{n}^{2} = g k_{nj} \tanh k_{nj}h, \quad j \in \mathbb{N}_{0}.
\end{equation*}
For $j = 0$, the wavenumber is real and it corresponds to the propagating mode\index{mode(s)!propagating} of the surface wave elevation. For $j \in \mathbb{N}$, \label{bilasli} the wavenumbers are purely imaginary, and thus $i k_{nj} \in \mathbb{R}$. Since we are interested in the decaying solution, we choose $i k_{nj} > 0$ and hence the modes of these wavenumbers are called the evanescent modes. \index{mode(s)!evanescent} As a consequence, the first-order potential function can now be written as
\begin{equation}
\phi^{(1)}(x,z,t) = \sum_{n = 1}^{\infty} \sum_{j =  0}^{\infty} \frac{i g}{2 \omega_{n}} C_{nj} \frac{\cosh k_{nj}(z + h)}{\cosh k_{nj}h} e^{-i \theta_{nj}(x,t)} + \textmd{c.c.}, \label{1stpotential}
\end{equation}
where $\theta_{nj}(x,t) = k_{nj}x - \omega_{n}t$.\index{wave generation theory}

Furthermore, applying the \textsc{kwmbc} (\ref{firstorderBCs}), integrating along the water depth, and using the property that $\Big\{ \cosh k_{nj}(z + h), \cosh k_{nl}(z + h), \; j, l \in \mathbb{N}_{0} \Big\}$ is a set of orthogonal functions for $j \neq l$, we can find the surface wave complex-valued amplitude $C_{nj}$ as follows
\begin{align*}
C_{nj} &= \frac{\omega_{n}^{2} S_{n}}{g k_{nj}} \cosh k_{nj}h \frac{\displaystyle \int_{-h}^{0} f(z) \cosh k_{nj}(z + h)\,dz}{\displaystyle \int_{-h}^{0} \cosh^{2} k_{nj}(z + h)\,dz} \nonumber \\
       &= \frac{4 S_{n} \sinh k_{nj}h}{k_{nj}(h + H)}\,\frac{k_{nj}(h + H) \sinh k_{nj}h + \cosh k_{nj}d - \cosh k_{nj}h}{2k_{nj}h + \sinh (2k_{nj}h)}, \; j \in \mathbb{N}_{0}.
\end{align*}
Finally, the first-order surface elevation can be found from the \textsc{dfsbc} (\ref{firstorderBCs}) and is given as follows
\begin{equation*}
\eta^{(1)}(x,t) = \sum_{n = -\infty}^{\infty} \sum_{j =  0}^{\infty} \frac{1}{2} C_{nj} e^{-i \theta_{nj}(x,t)} + \textmd{c.c.}
\end{equation*}
This first-order theory can also be found in~\citep{8Dean91}.

\begin{figure}[htbp]			
\begin{center}
\includegraphics[width = 0.5\textwidth]{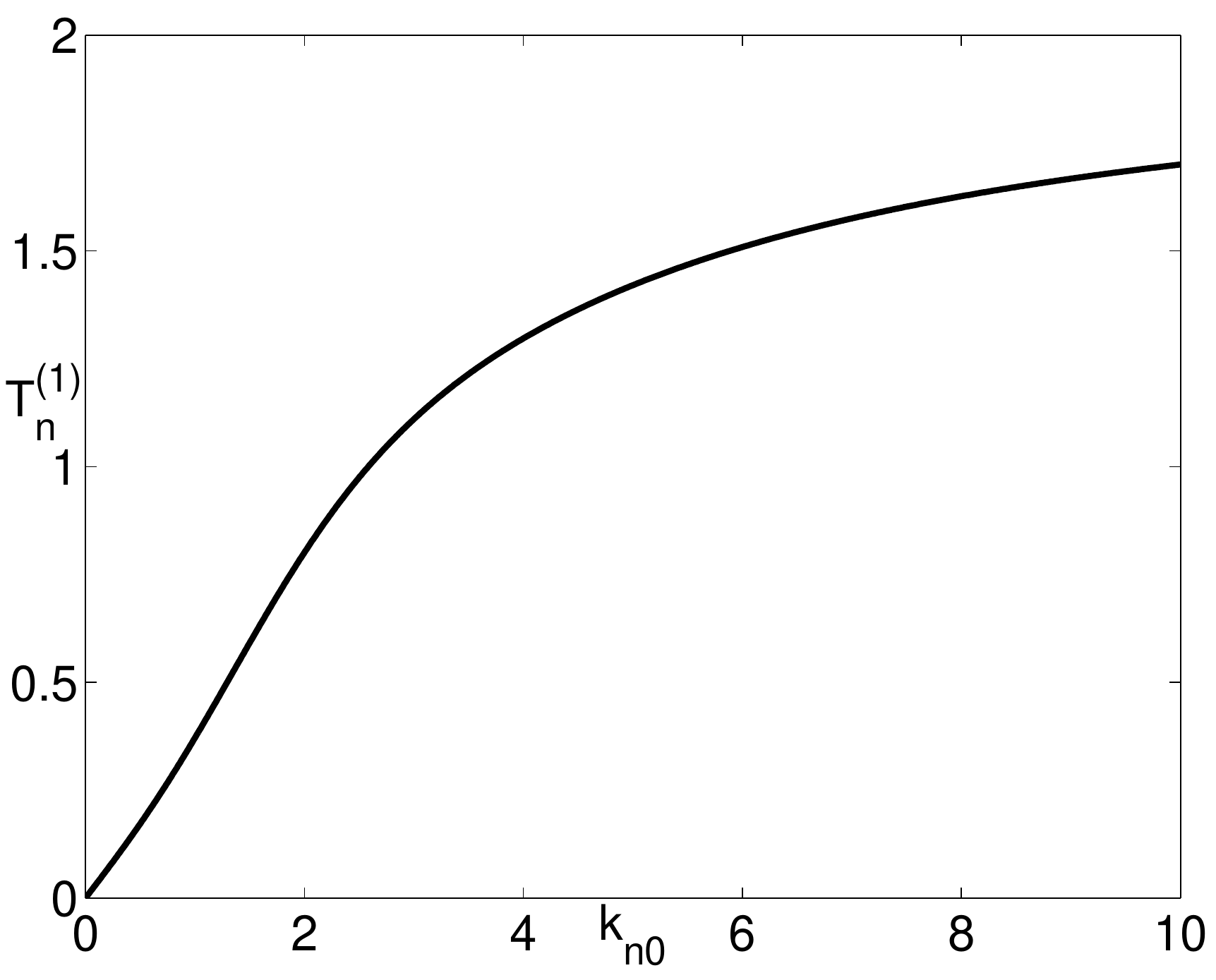}
\caption[First-order transfer function plot]{The first-order transfer function plot as a function of wavenumber $k_{n0}$, for the case that the water depth is $h = 1$ and the center of rotation is at $d = \frac{1}{3} h$ above the tank floor.} 		\label{gambarTF1}
\end{center}
\end{figure}

\begin{center}
\textsc{Remark 1.}
\end{center}
For `practical' purposes, it is useful to introduce the so-called \textit{transfer function}\index{transfer function} or \textit{frequency response}\index{frequency response} of a system. It is defined as the ratio of the output and the input of a system. In our wave generation problem, we have a system with a wavemaker motion as input and the surface wave amplitude as output. Therefore, the first-order transfer function $T_{n}^{(1)}$ is defined as the ratio between the surface wave amplitude of the propagating mode\index{mode(s)!propagating} $C_{n0}$ as output and the maximum stroke $|S_{n}|$ as input, explicitly given by
\begin{equation*}
T_{n}^{(1)} = 4 \; \frac{\sinh k_{n0}h}{k_{n0}(h + H)} \; \frac{k_{n0}(h + H) \sinh k_{n0}h + \cosh k_{n0}d - \cosh k_{n0}h}{2 k_{n0}h + \sinh 2k_{n0}h}.
\end{equation*}
Figure~\ref{gambarTF1} shows the first-order transfer function plot as a function of wavenumber $k_{n0}$ for a given water depth $h$ and the center of rotation $d$. For increasing $k_{n0}$, which also means increasing frequency $\omega_{n}$, the transfer function is monotonically increasing as well. It increases faster for a smaller value of $k_{n0}$ and slower for a larger value of $k_{n0}$, approaching the asymptotic limit of $T_{n}^{(1)} = 2$ for $k_{n0}h \rightarrow \infty$.

\section{Second-order wave generation theory} \label{ordedua}
\index{wave generation theory}

In this section, we solve a nonhomogeneous \textsc{bvp} for the second-order wave generation theory. Due to the nonhomogeneous boundary condition at the free surface, which causes interactions between each possible pair of first-order wave components, the resulting surface wave elevation has a second-order effect, known as the bound-wave component.\index{wave components!bound-wave} Furthermore, due to first-order wavemaker motion and the boundary condition at the wavemaker, the generated wave also has another second-order effect, namely the free-wave component. \index{wave components!free-wave} The latter component is undesired since it results in a spatially inhomogeneous wave field due to the different propagation velocities of bound-wave and free-wave components with the same frequency. Therefore, in order to prevent the free-wave component to be generated, we include an additional second-order bound-wave effect to the flap motion. This process is known as the `second-order steering' of the wavemaker motion. \index{wavemaker steering!second-order} More details about this theory, including an experimental verification can be found in~\citep{8Schaffer96}. For the history of wave generation theory, see also references in this paper. \index{wave generation theory}

Taking terms of the second-order in the series expansion, we obtain the \textsc{bvp} for the second-order wave generation theory. The second-order potential function also satisfies the Laplace equation \index{Laplace equation}
\begin{equation*}
\partial_{x}^{2} \phi^{(2)} + \partial_{z}^{2} \phi^{(2)} = 0,  \qquad \textmd{for}\; x \geq 0, \;\; -h \leq z \leq 0.
\end{equation*}
Almost all the second-order boundary conditions now become nonhomogeneous:
\begin{equation}
\begin{array}{rcll}
g \eta^{(2)} + \partial_{t} \phi^{(2)}
\!\!\!\! &=& \!\!\!\! - \Big(\eta^{(1)} \partial_{tz}^{2} \phi^{(1)} + \frac{1}{2} |\nabla \phi^{(1)}|^{2} \Big), & \quad \textmd{at} \; z = 0; \\
\partial_{t} \eta^{(2)} - \partial_{z} \phi^{(2)}
\!\!\!\! &=& \!\!\!\! \eta^{(1)} \partial_{z}^{2} \phi^{(1)} - \partial_{x} \eta^{(1)} \partial_{x} \phi^{(1)},   & \quad \textmd{at} \; z = 0; \\
\partial_{x} \phi^{(2)} - f(z) \frac{dS^{(2)}}{dt}
\!\!\!\! &=& \!\!\!\! S^{(1)}(t) \Big(f'(z) \partial_{z} \phi^{(1)} - f(z) \partial_{x}^{2} \phi^{(1)} \Big),     & \quad \textmd{at} \; x = 0; \\
\partial_{z} \phi^{(2)} \!\!\!\! &=& \!\!\!\! 0, & \quad \textmd{at} \; z = -h. \label{secondorderBCs}
\end{array}
\end{equation}
By combining the \textsc{dfsbc} and \textsc{kfsbc} of~\eqref{secondorderBCs} at $z = 0$, we have the second-order nonhomogeneous free surface boundary condition
\begin{equation}
g \partial_{z} \phi^{(2)} + \partial_{t}^{2} \phi^{(2)} = \textsc{rhs}_{1}, \quad \textrm{at} \; z = 0.  \label{FSDKBC}
\end{equation}
Using the first-order potential function~\eqref{1stpotential}, \textsc{rhs}$_{1}$ is explicitly given by
\begin{align*}
\textsc{rhs}_{1} &= \left. -\left(\frac{\partial}{\partial t} |\nabla \phi^{(1)}|^{2} + \eta^{(1)} \frac{\partial}{\partial z} \left[g \frac{\partial \phi^{(1)}}{\partial z} + \frac{\partial^{2} \phi^{(1)}}{\partial t^{2}} \right] \right) \right|_{z = 0} \nonumber \\
  				 &= \sum_{m,n = 1}^{\infty} \sum_{l,j =  0}^{\infty} \left(A_{mnlj}^{+} e^{-i(\theta_{ml} + \theta_{nj})} + A_{mnlj}^{-} e^{-i(\theta_{ml} - \theta_{nj}^{\ast})} \right) + \textmd{c.c.},
\end{align*}
where
\begin{align*}
\frac{A_{mnlj}^{+}}{C_{ml} C_{nj}} &= 
                     \frac{1}{4i} \!\! \left[(\omega_{m} + \omega_{n}) \! \left(g^{2} \frac{k_{ml} k_{nj}}{\omega_{m} \omega_{n}}
                   - \omega_{m} \omega_{n} \right)
                   + \frac{g^{2}}{2} \! \left(\frac{k_{ml}^{2}}{\omega_{m}} + \frac{k_{nj}^{2}}{\omega_{n}} \right)
                   - \frac{1}{2} (\omega_{m}^{3} + \omega_{n}^{3}) \right]\!\!, \\
\frac{A_{mnlj}^{-}}{C_{ml} C_{nj}^{\ast}} &= 
                     \frac{1}{4i} \!\! \left[(\omega_{m} - \omega_{n}) \! \left(g^{2} \frac{k_{ml} k_{nj}^{\ast}}{\omega_{m} \omega_{n}}
                   + \omega_{m} \omega_{n} \right)
                   + \frac{g^{2}}{2} \! \left(\frac{k_{ml}^{2}}{\omega_{m}} - \frac{k_{nj}^{2}}{\omega_{n}} \right)
                   - \frac{1}{2} (\omega_{m}^{3} - \omega_{n}^{3}) \right]\!\!.
\end{align*}
\index{wave generation theory}

In order to find the bound-wave component,\index{wave components!bound-wave} the free-wave component \index{wave components!free-wave}, and to apply the second-order steering wavemaker motion, we split the second-order \textsc{bvp}~\eqref{secondorderBCs} into three \textsc{bvp}s. For that purpose, the second-order potential function is split into three components as follows:
\begin{equation*}
\phi^{(2)}(x,z,t) = \phi^{(21)}(x,z,t) + \phi^{(22)}(x,z,t) + \phi^{(23)}(x,z,t).
\end{equation*}
Now the corresponding \textsc{bvp} for the first component of the potential function $\phi^{(21)}$ reads
\begin{equation}
\begin{array}{rcll}
g \partial_{z} \phi^{(21)} + \partial_{t}^{2} \phi^{(21)} \!\!\!\! &=& \!\!\!\! \textsc{rhs}_{1},  & \quad \textrm{at} \; z = 0;  \\
                                 \partial_{z} \phi^{(21)} \!\!\!\! &=& \!\!\!\! 0,                 & \quad \textrm{at} \; z = -h. \label{1stcomponent_bvp}
\end{array}
\end{equation}
The corresponding \textsc{bvp} for the second component of the potential function $\phi^{(22)}$ reads
\begin{equation}
\begin{array}{rcll}
g \partial_{z} \phi^{(22)} + \partial_{t}^{2} \phi^{(22)} \!\!\!\! &=& \!\!\!\! 0, 	 & \quad \textrm{at} \; z = 0; \\
  							     \partial_{x} \phi^{(22)} \!\!\!\! &=& \!\!\!\! S^{(1)}(t) \Big(f'(z) \partial_{z} \phi^{(1)} - f(z) \partial_{x}^{2} \phi^{(1)} \Big) - \partial_{x} \phi^{(21)},     				& \quad \textmd{at} \; x = 0;                   \label{2ndcomponent_bvp} \\
                                 \partial_{z} \phi^{(22)} \!\!\!\! &=& \!\!\!\! 0,	 & \quad \textrm{at} \; z = -h.
\end{array}
\end{equation}
And the \textsc{bvp} for the third component of the potential function $\phi^{(23)}$ reads
\begin{equation}
\begin{array}{rcll}
g \partial_{z} \phi^{(23)} + \partial_{t}^{2} \phi^{(23)} \!\!\!\! &=& \!\!\!\! 0, 						  & \quad \textrm{at} \; z = 0; \\
                                 \partial_{x} \phi^{(23)} \!\!\!\! &=& \!\!\!\! f(z) \frac{dS^{(2)}}{dt}, & \quad \textrm{at} \; x = 0; \\
                                 \partial_{z} \phi^{(23)} \!\!\!\! &=& \!\!\!\! 0,                        & \quad \textrm{at} \; z = -h. \label{3rdcomponent_bvp}
\end{array}
\end{equation}
\index{wave generation theory}

By taking the Ansatz for the first part of the second-order potential function $\phi^{(21)}$ as follows:
\begin{align*}
\phi^{(21)}(x,z,t) &= \sum_{m,n = 1}^{\infty} \sum_{l,j =  0}^{\infty} B_{mnlj}^{+} \frac{\cosh (k_{ml} + k_{nj})(z + h)}       {\cosh (k_{ml} + k_{nj})h}        e^{-i(\theta_{ml} + \theta_{nj})} \nonumber \\
& \qquad + B_{mnlj}^{-} \frac{\cosh (k_{ml} - k_{nj}^{\ast})(z + h)}{\cosh (k_{ml} - k_{nj}^{\ast})h} e^{-i(\theta_{ml} - \theta_{nj}^{\ast})} + \textmd{c.c.},
\end{align*}
then we can derive the corresponding coefficients to be
\begin{align*}
B_{mnlj}^{+} &= \frac{A_{mnlj}^{+}}{\Omega^{2}(k_{ml} + k_{nj})        - (\omega_{m} + \omega_{n})^{2}}, \\
B_{mnlj}^{-} &= \frac{A_{mnlj}^{-}}{\Omega^{2}(k_{ml} - k_{nj}^{\ast}) - (\omega_{m} - \omega_{n})^{2}}.
\end{align*}
\index{wave generation theory}

This first component of the second-order potential function will contribute the bound-wave component to the second-order surface wave elevation $\eta^{(2)}$. For $j = 0$, the wave component is a propagating mode\index{mode(s)!propagating} and for $j \in \mathbb{N}$, it consists of evanescent modes. \index{mode(s)!evanescent} Since the wavenumbers $k_{mj} + k_{nj}$ and $k_{mj} + k_{nj}^{\ast}$, $j \in \mathbb{N}_{0}$ do not satisfy the \textsc{ldr} with frequencies $\omega_{m} \pm \omega_{n}$, then the denominator part of $B_{mnj}^{\pm}$ will never vanish and thus the potential function is a bounded function. \index{wave generation theory}

Let the right-hand side of the boundary condition at the wavemaker for the second \textsc{bvp}~\eqref{2ndcomponent_bvp} be denoted by \textsc{rhs}$_{2}$, which is expressed as
\begin{equation*}
\textsc{rhs}_{2} = \sum_{m,n = 1}^{\infty} \sum_{l,j =  0}^{\infty} \left(F_{mnlj}^{+}(z) e^{i(\omega_{m} + \omega_{n})t} + F_{mnlj}^{-}(z) e^{i(\omega_{m} - \omega_{n})t} \right) + \textmd{c.c.},
\end{equation*}
where
\begin{align*}
F_{mnlj}^{+}(z) &= \frac{g}{8 \omega_{n}} \frac{S_{m} k_{nj} C_{nj}}{\cosh k_{nj}h} \left[f'(z) \sinh k_{nj}(z + h) + k_{nj} f(z) \cosh k_{nj}(z + h) \right] \nonumber \\
& \qquad + i (k_{ml} + k_{nj}) B_{mnlj}^{+} \frac{\cosh (k_{ml} + k_{nj})(z + h)}{\cosh (k_{ml} + k_{nj})h}, \\
F_{mnlj}^{-}(z) &= -\frac{g}{8 \omega_{n}} \frac{S_{m} k_{nj}^{\ast} C_{nj}^{\ast}}{\cosh k_{nj}^{\ast}h} \left[f'(z) \sinh k_{nj}^{\ast}(z + h) + k_{nj}^{\ast} f(z) \cosh k_{nj}^{\ast}(z + h) \right] \nonumber \\
& \qquad + i (k_{ml} - k_{nj}^{\ast}) B_{mnlj}^{-} \frac{\cosh (k_{ml} - k_{nj}^{\ast})(z + h)}{\cosh (k_{ml} - k_{nj}^{\ast})h}.
\end{align*}
Let the Ansatz for the second component of the second-order potential function $\phi^{(22)}$ be
\begin{align*}
\phi^{(22)}(x,z,t) &= \sum_{m,n = 1}^{\infty} \sum_{l,j =  0}^{\infty} \left(\frac{i\,g\,P_{mnlj}^{+} }{2(\omega_{m} + \omega_{n})}  \frac{\cosh K_{mnlj}^{+}(z + h)}{\cosh K_{mnlj}^{+}h} e^{-i(K_{mnj}^{+}x - (\omega_{m} + \omega_{n})t)} \right. \nonumber \\
& \qquad + \left. \frac{i\,g\,P_{mnlj}^{-}}{2(\omega_{m} - \omega_{n})} \frac{\cosh K_{mnlj}^{-}(z + h)}{\cosh K_{mnlj}^{-}h} e^{-i(K_{mnlj}^{-}x - (\omega_{m} - \omega_{n})t)} \right) + \textmd{c.c.},
\end{align*}
where the wavenumbers $K_{mnlj}^{\pm}$, $j \in \mathbb{N}_{0}$ and frequencies $\omega_{m} \pm \omega_{n}$ satisfy the \textsc{ldr}. Using the property that $\Big\{\cosh K_{mnlj}^{\pm}(z + h), \cosh K_{mnl'j'}^{\pm}(z + h), \; l, l', j, j' \in \mathbb{N}_{0} \Big\}$ is a set of orthogonal functions for $l \neq l'$ and $j \neq j'$, we find the coefficients $P_{mnlj}^{\pm}$ as follows:
\begin{align}
P_{mnlj}^{\pm} &= \frac{2(\omega_{m} \pm \omega_{n}) \cosh K_{mnlj}^{\pm}h}{g K_{mnlj}^{\pm}} \frac{\displaystyle \int_{-h}^{0} F_{mnlj}^{\pm}(z) \cosh K_{mnlj}^{\pm} (z + h)\,dz} {\displaystyle \int_{-h}^{0} \cosh^{2} K_{mnlj}^{\pm} (z + h)\,dz} \nonumber \\
&= 8 \frac{K_{mnlj}^{\pm} \sinh K_{mnlj}^{\pm}h}{\omega_{m} \pm \omega_{n}} \frac{\displaystyle \int_{-h}^{0} F_{mnlj}^{\pm}(z) \cosh K_{mnlj}^{\pm} (z + h)\,dz}{2 K_{mnlj}^{\pm} h + \sinh (2 K_{mnlj}^{\pm} h)}.			  \label{2ndcomponent_potential}
\end{align}
The second component of the second-order potential function $\phi^{(22)}$ will give contributions to the free-wave component of the second-order surface wave elevation $\eta^{(2)}$. This component arises due to the boundary condition at the wavemaker caused by the first-order wavemaker motion. Since the desired surface elevation is only the bound-wave component, we want to get rid of this term, especially the propagating mode. The evanescent modes vanish anyway after they evolve far away from the wavemaker. By prescribing the second-order wavemaker motion such that the propagating mode of the third component $\phi^{(23)}$ will cancel the same mode of the second one $\phi^{(22)}$, then far from the wavemaker, we have the desired bound-wave component only.\index{wave components!bound-wave} \index{wave generation theory}

Let the second-order wavemaker motion be given by \index{wavemaker motion!second-order}
\begin{equation*}
S^{(2)}(t) = \sum_{m,n = 1}^{\infty} -\frac{1}{2} i \left(S_{mn}^{+} e^{i (\omega_{m} + \omega_{n})t} + S_{mn}^{-} e^{i (\omega_{m} - \omega_{n})t} \right) + \textmd{c.c.}
\end{equation*}
Let also the Ansatz for the third component of the second-order potential function $\phi^{(23)}$ be
\begin{align*}
\phi^{(23)}(x,z,t) &= \sum_{m,n = 1}^{\infty} \sum_{l,j =  0}^{\infty} \left(\frac{i\,g\,Q_{mnlj}^{+}}{2(\omega_{m} + \omega_{n})} \frac{\cosh K_{mnlj}^{+}(z + h)}{\cosh K_{mnlj}^{+}h} e^{-i(K_{mnlj}^{+}x - (\omega_{m} + \omega_{n})t)} \right. \nonumber \\
& \qquad + \left. \frac{i\,g\,Q_{mnlj}^{-}}{2(\omega_{m} - \omega_{n})} \frac{\cosh K_{mnlj}^{-}(z + h)}{\cosh K_{mnlj}^{-}h} e^{-i(K_{mnlj}^{-}x - (\omega_{m} - \omega_{n})t)} \right) + \textmd{c.c.},
\end{align*}
where $\Omega(K_{mnlj}^{\pm}) = \omega_{m} \pm \omega_{n}$. Using the orthogonality property again, we find the coefficients $Q_{mnlj}^{\pm}$ as follows:
\begin{align}
Q_{mnlj}^{\pm} &= S_{mn}^{\pm} \sinh K_{mnlj}^{\pm}h \frac{\displaystyle \int_{-h}^{0} f(z) \cosh K_{mnlj}^{\pm} (z + h)\,dz} {\displaystyle \int_{-h}^{0} \cosh^{2} K_{mnlj}^{\pm} (z + h)\,dz} \nonumber \\
&= \frac{4 S_{mn}^{\pm} \sinh K_{mnlj}^{\pm}h}{K_{mnlj}^{\pm}(h + H)} \, \frac{K_{mnlj}^{\pm}(h + H) \sinh K_{mnlj}^{\pm}h + \cosh K_{mnlj}^{\pm}d - \cosh K_{mnlj}^{\pm}h} {2 K_{mnlj}^{\pm} h + \sinh (2 K_{mnlj}^{\pm} h)}. \label{3rdcomponent_potential}
\end{align}
To have the propagating mode of the free-wave from the second $\big(\phi^{(22)} \big)$ and the third $\big(\phi^{(23)} \big)$ components cancel each other, we must require $P_{mn00} + Q_{mn00} = 0$, which leads to the following second-order wavemaker motion, known as the `second-order steering':\index{wavemaker steering!second-order}
\begin{equation*}
S_{mn}^{\pm} = \frac{2 \big(K_{mn00}^{\pm} \big)^{2} I_{mn00}^{\pm} (h + H)} {\big(\omega_{m} \pm \omega_{n} \big)\big(\cosh K_{mn00}^{\pm}h - \cosh K_{mn00}^{\pm}d - K_{mn00}^{\pm}(h + l) \sinh K_{mn00}^{\pm}h\big)},
\end{equation*}
where
\begin{equation*}
I_{mn00}^{\pm} = \int_{-h}^{0} F_{mn00}^{\pm}(z) \cosh K_{mn00}^{\pm}(z + h)\,dz.
\end{equation*}
Therefore, with this choice of second-order wavemaker motion, the second-order potential function can be written as
\begin{equation*}
\phi^{(2)}(x,z,t) = \phi^{(2)}_{\textmd{\tiny propagating}}(x,z,t) + \phi^{(2)}_{\textmd{\tiny evanescent}}(x,z,t),
\end{equation*}
where \index{wave generation theory} \index{wave components!free-wave} \index{wave components!bound-wave} \index{mode(s)!propagating} \index{mode(s)!evanescent}
\begin{align*}
\phi^{(2)}_{\textmd{\tiny propagating}} &= \phi^{(21)}_{\textmd{\tiny bound-wave, propagating}}, \\
\phi^{(2)}_{\textmd{\tiny evanescent}}  &= (\phi^{(21)}_{\textmd{\tiny bound-wave}} + \phi^{(22)}_{\textmd{\tiny free-wave}}   +   \phi^{(23)}_{\textmd{\tiny free-wave}})_{\textmd{\tiny evanescent}}.
\end{align*}
Consequently, from the second-order \textsc{dfsbc}~\eqref{secondorderBCs}, we find the second-order surface wave elevation. It can be written as follows:
\begin{equation*}
\eta^{(2)}(x,t) = \eta^{(2)}_{\textmd{\tiny propagating}} + \phi^{(2)}_{\textmd{\tiny evanescent}},
\end{equation*}
where
\begin{equation*}
\eta^{(2)}_{\textmd{\tiny propagating}} = \sum_{m,n = 1}^{\infty} D_{mn00}^{+} e^{-i(\theta_{m0} + \theta_{n0})} + D_{mn00}^{-} e^{-i(\theta_{m0} - \theta_{n0})},
\end{equation*}
and
\begin{align*}
\eta^{(2)}_{\textmd{\tiny evanescent}} &= \sum_{m,n = 1}^{\infty} \sum_{j = 1, \atop l = 0}^{\infty} \left(D_{mnlj}^{+} e^{-i(\theta_{ml} + \theta_{nj})} + D_{mnlj}^{-} e^{-i(\theta_{ml} - \theta_{nj}^{\ast})} \right. \nonumber \\
& \qquad + \frac{1}{2} (P_{mnlj}^{+} + Q_{mnlj}^{+}) e^{-i(K_{mnlj}^{+}x - (\omega_{m} + \omega_{n})t)} \nonumber \\
& \qquad + \left. \frac{1}{2} (P_{mnlj}^{-} + Q_{mnlj}^{-}) e^{-i(K_{mnlj}^{-}x - (\omega_{m} - \omega_{n})t)} \right) +  \textmd{c.c.},
\end{align*}
where for $l,j \in \mathbb{N}_{0}$:
\begin{align*}
D_{mnlj}^{+} &= -\frac{1}{g} \left[i(\omega_{m} + \omega_{n})B_{mnlj}^{+} +  \frac{1}{8} \left(g^{2} \frac{k_{ml} k_{nj}}{\omega_{m} \omega_{n}} - \omega_{m} \omega_{n} - (\omega_{m}^{2} + \omega_{n}^{2}) \right) C_{ml} C_{nj} \right], \\
D_{mnlj}^{-} &= -\frac{1}{g} \left[i(\omega_{m} - \omega_{n})B_{mnlj}^{-} +  \frac{1}{8} \left(g^{2} \frac{k_{ml} k_{nj}^{\ast}}{\omega_{m} \omega_{n}} + \omega_{m} \omega_{n} - (\omega_{m}^{2} + \omega_{n}^{2}) \right) C_{ml} C_{nj}^{\ast} \right].
\end{align*}
\index{wave generation theory}

We have seen that the first-order surface wave elevation consists of a linear superposition of monochromatic frequencies. However, due to nonlinear effects, nonhomogeneous \textsc{bvp}, and interactions of the first-order wave components, the second-order surface elevation is composed of a superposition of bichromatic frequencies $\omega_{m} \pm \omega_{n}$. The components with frequency $\omega_{m} + \omega_{n}$ are called the `superharmonics' and those with frequency $|\omega_{m} - \omega_{n}|$ are called the `subharmonics'.\index{subharmonic waves} \index{superharmonic waves}

\begin{center}
\textsc{Remark 2.}
\end{center}
Similar to the first-order wave generation theory, we can define a second-order transfer function as well. The detailed formula for this transfer function can be found in~\citep{8Schaffer96}.	\index{wave generation theory}

\newpage
{\renewcommand{\baselinestretch}{1} \small \index{wave generation theory}

}

\backmatter
\renewcommand{\baselinestretch}{1}

{\pagestyle{plain}
\chapter*{\vspace*{-1cm} Summary}

In this thesis, we discuss mathematical aspects of extreme water wave generation in a hydrodynamic laboratory. The original problem comes from the Maritime Research Institute Netherlands (MARIN)\index{MARIN} to generate large amplitude and non-breaking waves to test ship and offshore construction. We choose the spatial nonlinear Schr\"odinger (NLS) equation\index{NLS equation} as a mathematical model for this problem and concentrate on the study of one family of exact solutions of this equation that describes extreme wave events in a wave basin. \addcontentsline{toc}{chapter}{Summary}

We derive the NLS equation using the multiple scale method, derive the phase-amplitude equations and introduce the Chu-Mei quotient\index{Chu-Mei quotient} from the nonlinear dispersion relation. We are interested in the modulational instability \index{modulational instability} of the nonlinear plane-wave solution of the NLS equation. The physical wave field of a plane-wave serves as the `finite' background of the extreme wave model. 

We discuss extensively the properties of waves on the finite background \index{waves on finite background} which are exact solutions of the NLS equation. Three types of such waves are known in the literature: the Soliton on Finite Background (SFB)\index{SFB}, the Ma solution,\index{Ma solution} and the rational solution\index{rational solution}. In particular, the SFB solution receives special attention since the corresponding physical wave signal is a good candidate for extreme wave generation. The asymptotic behavior\index{asymptotic behavior} of the SFB in the far distance corresponds to a modulated plane-wave solution of the NLS equation. For a very long modulation, the SFB has amplitude amplification up to a maximal factor of three.\index{AAF}

We introduce a transformation to displaced phase-amplitude \index{displaced phase-amplitude} variables with respect to a background of the monochromatic plane-wave solution. The transformation of the displaced phase is restricted to be time-independent. The change of phase with position physically corresponds to a change of the wavelength of the carrier wave of a wave group. This turns out to be the only driving force responsible for the nonlinear amplitude amplification toward extreme wave events. Remarkably, the assumption that the displaced-phase is time-independent leads to the waves on the finite background which are the three exact solutions of the NLS equation mentioned earlier.

We study the corresponding physical wave field of the SFB and observe that the interesting, purely linear, phenomena of vanishing amplitude, \index{vanishing amplitude} phase singularity, \index{phase singularity} and wavefront dislocation, \index{wavefront dislocation} occur simultaneously. We connect the unboundedness of the Chu-Mei quotient with the unboundedness of the local wavenumber\index{local wavenumber} and the local frequency \index{local frequency} at singular points. This unboundedness is a generic property and is responsible for the occurrence of phase singularity and wavefront dislocation.

We study some characteristics of higher-order waves on finite background,\index{waves on finite background!higher-order} particularly the SFB$_{2}$  solution.\footnote[1]{Index two now denotes the number of initial pairs of sidebands in the spectrum. \par} The corresponding physical wave field shows an interaction of two wave groups as they propagate downstream \index{downstream}. Theoretically, SFB$_{2}$\index{SFB!SFB$_{2}$} is also a good candidate for extreme wave generation since it has amplitude amplification up to a maximal factor of five. We present an explicit relation of the amplitude amplification factors between SFB$_{2}$ and SFB\index{SFB}. Vanishing amplitude, phase singularity, and wavefront dislocation also occur simultaneously in the physical wave field of SFB$_{2}$.

We designed a set of experiments\index{experiments} for extreme wave generation based on the theoretical prediction with SFB. These experiments were executed in the wave basin of MARIN. We compare the experimental results and the theoretical prediction qualitatively and quantitatively. All experimental signals show a pattern of modulational instability as described by SFB during the downstream \index{downstream} evolution. We observe that both the carrier wave frequency and the modulation frequency are conserved accurately during the evolution. We explain several differences between the theoretical SFB and the experimental signals using the evolution curves in the Argand diagram and the maximum temporal amplitude plots. The experimental signals have phase singularities at two different positions, with the extreme position located in between. We describe that the extreme position is sensitive for parameter changes in the SFB  family and also for the nonlinear coefficient of the NLS equation. We conclude that the SFB family provides suitable wave groups that can be used to generate extreme waves in the laboratory in a deterministic way.

\chapter*{\vspace*{-1cm} \addcontentsline{toc}{chapter}{Samenvatting (Dutch summary)} Samenvatting}

In dit proefschrift behandelen we wiskundige modellen voor het opwekken (genereren) van extreme golven in een waterloopkundig laboratorium. Het Maritiem Onderzoeks Instituut MARIN gebruikt deze modellen om hoge, niet brekende golven te genereren voor het testen van schepen en offshore constructies. Als wiskundig model is gekozen voor de niet-lineaire Schr\"odinger (NLS) vergelijking. We hebben een groep van exacte oplossingen van de vergelijking bestudeerd die extreme golven in een laboratorium beschrijven.

Wij leiden de NLS vergelijking af met behulp van de meer-schalen methode. Vervolgens leiden we de fase-amplitude vergelijkingen af en introduceren het Chu-Mei quoti\"ent van de niet-lineaire dispersie relatie. Wij zijn geinteresseerd in de modulatie instabiliteit van de niet-lineaire vlakke golf oplossing van de NLS vergelijking. Het fysische golfveld van een vlakke golf dient als de `eindige' achtergrond van het extreme golf model.

Wij bespreken uitgebreid de eigenschappen van golven op eindige achtergrond die nauwkeurige oplossingen van de NLS vergelijking zijn. Drie types van dergelijke golven zijn bekend in de literatuur: de Soliton op Eindige Achtergrond ({\sl SFB}), de Ma oplossing, en de rationele oplossing. In het bijzonder krijgt de SFB oplossing speciale aandacht aangezien het overeenkomstige fysische golfsignaal een goede kandidaat voor extreme golfgeneratie is. Het asymptotische gedrag van het SFB voor grote afstand is een gemoduleerde vlakke golfoplossing van de NLS vergelijking. Voor een zeer lange modulatie heeft SFB een amplitudevergroting tot een maximale factor van drie.

Wij introduceren een transformatie naar verschoven fase-amplitude variabelen met betrekking tot een achtergrond van de monochromatische vlakke golfoplossing. De transformatie van de verschoven fase wordt beperkt door deze tijdonafhankelijke te nemen. De verandering van fase met positie correspondeert fysisch met een verandering van de golflengte van de draaggolf van een golfgroep. Dit blijkt de enige drijvende kracht te zijn verantwoordelijk is voor de niet-lineaire amplitudetoename naar extreme golven. Opmerkelijk is dat de veronderstelde tijdonafhankelijkheid leidt tot de drie exacte oplossingen die bekend zijn van de eerder vermelde NLS vergelijking.


Wij bestuderen het bijbehorend fysisch golfveld van de SFB en constateren een gelijktijdig optreden van drie zuiver lineaire verschijnselen: verdwijnende amplitude, fase singulariteit, en golffront vertakking. In de singuliere punten leggen we een verband tussen de onbegrensdheid van het Chu-Mei quoti\"ent met die van het locale golfgetal en de locale frequentie. Dit onbegrensd zijn is een generieke eigenschap die verantwoordelijk is voor het optreden van fase singulariteit en golffront vertakking.

Wij onderzoeken eigenschappen van golven van hogere orde op eindige achtergrond, vooral de SFB$_{2}$ oplossing.\footnote[1]{Index twee geeft nu het aantal initi\"ele paren {\sl sidebands} in het spectrum aan. \par} Het bijhorende fysische golfpatroon toont een wisselwerking van twee golfgroepen die in de stromingsrichting voortbewegen. Theoretisch is de SFB$_{2}$ ook een goede kandidaat voor het opwekken van extreme golven omdat hij een amplitudegroei tot op het vijfvoudige veroorzaken kan. Wij geven een expliciete relatie voor het groei van de amplitude van SFB$_{2}$ en SFB. Verdwijnende amplitude, fase singulariteit, en golffront vertakking kunnen ook simultaan in het fysische golfveld van de SFB$_{2}$ voorkomen.

We hebben een aantal experimenten ontworpen voor het genereren van extreme golven gebaseerd op de theoretische voorspellingen met SFB. Deze experimenten werden uitgevoerd in het golfbasin van MARIN. Wij vergelijken de experimentele resultaten met theoretische voorspellingen, zowel kwalitatief als kwantitatief. Alle experimentele signalen tonen een patroon van modulatie instabiliteit zoals omschreven door de SFB tijdens een stroomafwaartse evolutie. Wij nemen waar dat zowel de frequentie van de draaggolf als die van de modulatie nauwkering behouden blijven tijdens de evolutie. Wij verklaren een aantal verschillen tussen de theoretische SFB en de experimentele signalen door gebruik te maken van evolutiekrommen in het Argand diagram en de informatie over de maximale tijdsafhankelijke amplitude ({\sl MTA}). De experimentele signalen hebben fasesingulariteiten op twee verschillende posities, met de extreme positie hier tussen. Wij beschrijven dat de extreme positie gevoelig is voor parameter veranderingen in de SFB familie en ook voor de niet-lineaire coefficient van de NLS vergelijking. Wij concluderen dat de SFB familie geschikte golfgroepen beschrijven die gebruikt kunnen worden om extreme golven in het laboratorium te genereren op een deterministische weijze.

\chapter*{\vspace*{-1cm} \addcontentsline{toc}{chapter}{Ringkasan (Indonesian summary)} Ringkasan}

Tesis ini membahas pembangkitan gelombang air ekstrim pada laboratorium hidrodinamika ditinjau dari segi matematis. Permasalahan asalnya diajukan oleh Institut Penelitian Kelautan Belanda ({\sl MARIN}) yang bermaksud membangkitkan gelombang dengan amplitudo tinggi namun tak pecah guna menguji konstruksi kapal dan bangunan lepas pantai. Persamaan Schr\"odinger tak linear tipe ruang ({\sl spatial NLS}) dipilih untuk memodelkan permasalahan ini dan penelitian dalam tesis ini dipusatkan pada satu kelas penyelesaian eksak dari persamaan tersebut, yang sekaligus menggambarkan kejadian gelombang ekstrim di suatu kolam pengujian gelombang.

Persamaan NLS diturunkan dengan menggunakan metode skala kelipatan ({\sl multiple-scale}), diturunkan juga persamaan fasa-amplitudo ({\sl phase-amplitude equations}) dan diperkenalkan suku Chu-Mei dari hubungan dispersi tak linear ({\sl nonlinear dispersion relation}). Ketidakstabilan modulasi ({\sl modulational instability}) dari penyelesaian gelombang datar ({\sl plane-wave}) tak linear yang berkaitan dengan persamaan NLS dipelajari dalam tesis ini. Medan gelombang fisik dari gelombang datar ({\sl plane-wave}) berperan sebagai latar terbatas dari model gelombang ekstrim tersebut.

Sifat-sifat gelombang pada latar terbatas ({\sl waves on finite background}) yang merupakan penyelesaian eksak dari persamaan NLS dibahas secara mendalam. Tiga jenis gelombang tersebut dapat ditemukan dalam kepustakaan: Soliton pada Latar Terbatas ({\sl SFB}), penyelesaian Ma, dan penyelesaian rasional. Dari ketiga penyelesaian ini, penyelesaian SFB mendapatkan perhatian khusus karena sinyal gelombang fisiknya merupakan kandidat yang cocok untuk pembangkitan gelombang ekstrim. Perilaku asimptotik dari SFB di kejauhan berkaitan dengan penyelesaian gelombang datar dari persamaan NLS yang termodulasi. Untuk modulasi gelombang yang sangat panjang, SFB mempunyai faktor kelipatan amplitudo ({\sl amplitude amplification}) sampai maksimal tiga kali.

Suatu transformasi pada peubah-peubah fase-amplitudo tergeser ({\sl displaced phase-amplitude}) terhadap suatu latar berupa penyelesaian gelombang datar dengan frekuensi tunggal diperkenalkan dalam tesis ini. Transformasi terhadap fase tergeser dibatasi sehingga tidak bergantung pada waktu. Perubahan fase terhadap posisi secara fisik berkaitan dengan perubahan panjang gelombang dari gelombang pembawa suatu kelompok gelombang. Hal ini ternyata menjadi satu-satunya gaya pemicu yang mengakibatkan kelipatan amplitudo tak linear pada peristiwa gelombang ekstrim. Sungguh menakjubkan bahwa fase tergeser yang diasumsikan tidak bergantung pada waktu menuntun pada gelombang-gelombang dengan latar terbatas yang tidak lain adalah ketiga penyelesaian pasti dari persamaan NLS di atas.

Sifat-sifat medan gelombang fisik dari SFB telah dipelajari dan terdapat fenomena linear yang menarik, yaitu terjadinya secara bersamaan lenyapnya amplitudo ({\sl vanishing amplitude}), singularitas fase ({\sl phase singularity}), dan dislokasi muka gelombang ({\sl wavefront dislocation}). Ketidakterbatasan suku Chu-Mei dikaitkan dengan ketidakterbatasan dari bilangan gelombang lokal ({\sl local wavenumber}) dan frekuensi lokal ({\sl local frequency}) pada titik-titik singular ({\sl singular points}). Ketidakterbatasan ini adalah sifat umum yang menentukan terjadinya singularitas fase serta dislokasi muka gelombang.

Beberapa sifat gelombang pada latar terbatas tingkat tinggi juga telah dipelajari, khususnya penyelesaian SFB$_{2}$.\footnote[1]{Indeks dua sekarang menyatakan jumlah pasangan awal pita samping ({\it sidebands}) pada spektrum. \par} Medan gelombang fisiknya menunjukan interaksi dua kelompok gelombang pada saat merambat ke arah hilir ({\sl downstream}). Secara teori, SFB$_{2}$ juga calon yang baik untuk pembangkitan gelombang ekstrim karena ia memiliki faktor kelipatan amplitudo sampai maksimal lima kali. Hubungan eksplisit dari faktor-faktor kelipatan amplitudo antara SFB$_{2}$ dan SFB juga disajikan dalam tesis ini. Lenyapnya amplitudo, singularitas fase, dan dislokasi gelombang muka juga terjadi secara bersamaan di medan gelombang fisik SFB$_{2}$.

Sejumlah percobaan telah dilaksanakan untuk membangkitkan gelombang ekstrim berdasarkan perkiraan teoretis dengan menggunakan SFB. Percobaan ini dilakukan pada kolam pengujian gelombang di MARIN. Hasil percobaan di laboratorium dan perkiraan teori telah dibandingkan baik dari segi kualitas maupun kuantitas. Semua sinyal percobaan menunjukan pola ketidakstabilan modulasi sebagaimana digambarkan oleh SFB selama perambatan ke arah hilir. Pada percobaan ini, dapat diamati bahwa frekuensi gelombang pembawa ({\sl carrier frequency}) dan frekuensi modulasi ({\sl modulation frequency}) dipertahankan dengan tepat selama perambatan gelombang. Beberapa perbedaan antara SFB teoretis dan sinyal percobaan dijelaskan dengan menggunakan kurva evolusi di diagram Argand dan grafik amplitudo maksimum terhadap waktu ({\sl MTA}). Sinyal percobaan mempunyai singularitas fase pada dua posisi yang berbeda, dengan posisi ekstrim terletak di antaranya. Posisi ekstrim ini ternyata cukup peka terhadap perubahan parameter pada kelas SFB dan terhadap koefisien tak linear dari persamaan NLS. Dari pembahasan dalam tesis ini, dapat disimpulkan bahwa kelas SFB memberikan kelompok gelombang yang cocok untuk digunakan pada pembangkitkan gelombang ekstrim di laboratorium secara deterministik.
\newpage
\thispagestyle{plain}}

\makeatletter
\renewenvironment{thebibliography}[1]
{\chapter*{\bibname 
		\@mkboth{\MakeUppercase\bibname}{\MakeUppercase\bibname}}%
	\list{\@biblabel{\@arabic\c@enumiv}}%
	{\settowidth\labelwidth{\@biblabel{#1}}%
		\leftmargin\labelwidth \advance\leftmargin\labelsep \@openbib@code
		\usecounter{enumiv}%
		\let\p@enumiv\@empty
		\renewcommand\theenumiv{\@arabic\c@enumiv}}%
	\sloppy \clubpenalty4000 \@clubpenalty \clubpenalty
	\widowpenalty4000%
	\sfcode`\.\@m%
	\setlength{\parindent}{2em}}%
{\def\@noitemerr
	{\@latex@warning{Empty `thebibliography' environment}}%
	\endlist}
\makeatother

\makeatletter
\def\cleardoublepage{\clearpage\if@twoside\ifodd\c@page\else
	\hbox{}\thispagestyle{empty}\newpage\fi\fi}
\makeatother
\renewcommand\bibname{Bibliography}

{\footnotesize \refstepcounter{chapter}
\addcontentsline{toc}{chapter}{\indexname}
\printindex}%

\newpage
\chapter*{\LARGE About the author}
\addcontentsline{toc}{chapter}{About the author}

{I was born in Bandung, West Java, Indonesia, on 1 April 1979, and also grew up there. I attended a senior high school at SMU/SMA Negeri 4 Bandung and finished the natural sciences program in 1997. From August 1997 until February 2001, I enrolled as an undergraduate student at the Department of Mathematics, Bandung Institute of Technology, also known as ITB. During my final project period, Dr. Andonowati was my undergraduate supervisor, resulting in an undergraduate degree thesis entitled `A two-dimensional flap-type wavemaker theory'. After that, I came to the Netherlands in August 2001 to pursue graduate studies at the Department of Applied Mathematics, University of Twente. Meanwhile, I also participated in the combined MSc--PhD program in the Applied Analysis and Mathematical Physics chair within the same department. I obtained my master's degree in June 2003 with the thesis entitled `Wave group evolution and interaction' and my PhD degree in December 2006, both under the supervision of Professor E. (Brenny) van Groesen. The result of my doctoral research entitled `Mathematical aspects of extreme water waves' is presented in this thesis.}

\newpage
\begin{figure}[h!]
\includepdf[width=\paperwidth, offset=-2.54cm -2.43cm]{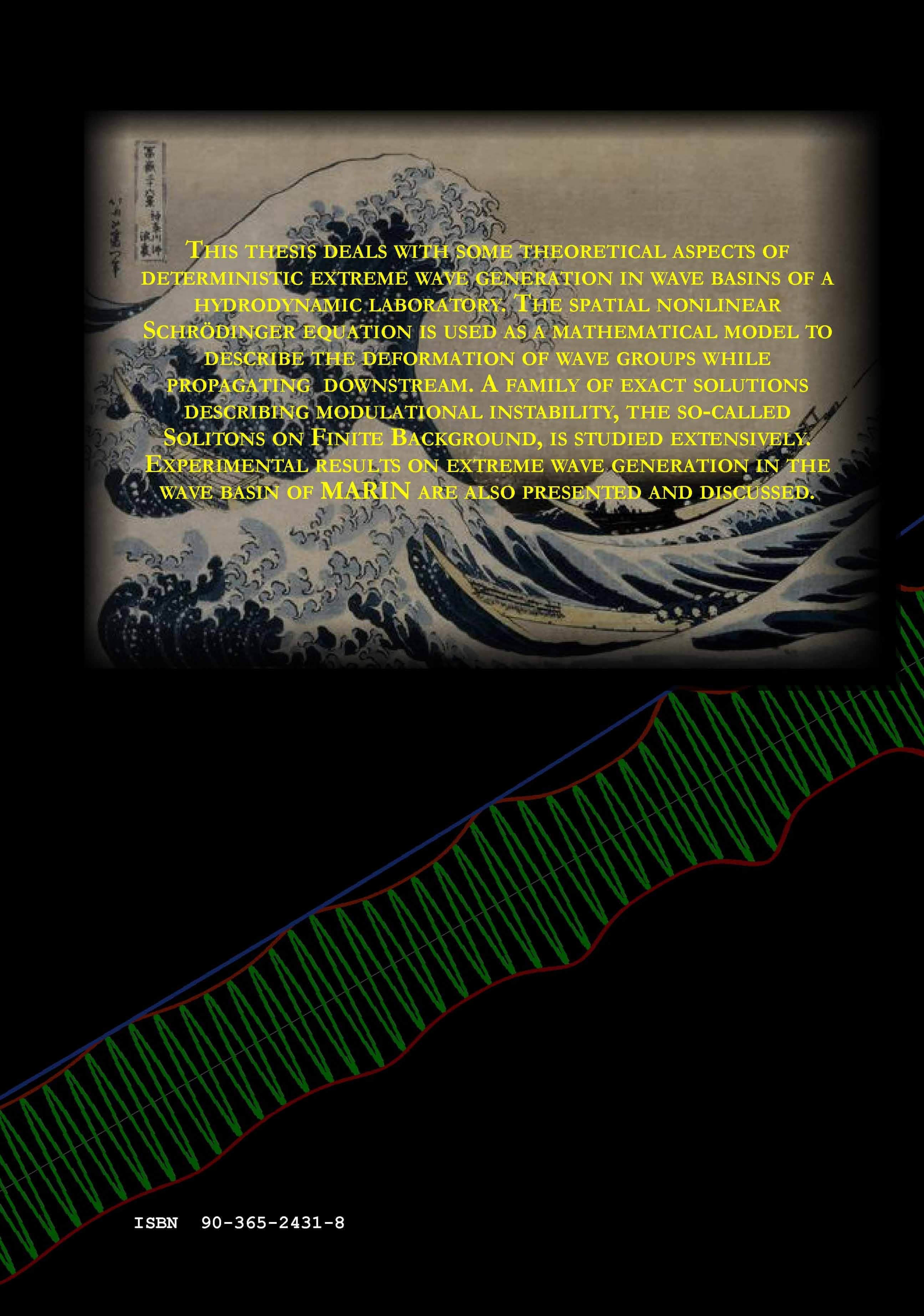}
\end{figure}

\small
\begin{thebibliography}{99}
\bibitem[Akhmediev and Korneev, 1986]{1Akhmediev86} \citep{1Akhmediev86} N. N. Akhmediev and V. I. Korneev. Modulation instability and periodic solution of the nonlinear Schr\"{o}dinger equation. \textit{Theor. Math. Phys.} \textbf{69}: 1089--1092, 1986. Translated from \textit{Teor. Mat. Fiz. (USSR)} \textbf{69}(2): 189--194, 1986.

\addcontentsline{toc}{section}{\bibname}
\bibitem[Andonowati and Van Groesen, 2003]{1Andonowati03} \citep{1Andonowati03} Andonowati and E. van Groesen. Optical pulse deformation in second order nonlinear media. \textit{J. Nonlinear Opt. Phys. Mat.} \textbf{12}(2): 221--234, 2003.

\bibitem[Andonowati et al., 2007]{1Andonowati07} \citep{1Andonowati07} Andonowati, N. Karjanto, and E. van Groesen. Extreme wave phenomena in down-stream running modulated waves. \textit{Appl. Math. Modelling} \textbf{31}(7): 1425--1443, 2007. \ arXiv:1710.10804 [physics.flu-dyn]

\bibitem[Benjamin and Feir, 1967]{1BenjaminFeir67} \citep{1BenjaminFeir67} T. B. Benjamin and J. E. Feir. The disintegration of wave trains in deep water. \textit{J. Fluid Mech.} \textbf{27}: 417--430, 1967.

\bibitem[Boccotti, 2000]{1Boccotti00} \citep{1Boccotti00} P. Boccotti. \textit{Wave Mechanics for Ocean Engineering}, Elsevier Science, New York, 2000.

\bibitem[Dankert et al., 2003]{1Dankert03} \citep{1Dankert03} H. Dankert, J. Horstmann, S. Lehner and W. Rosenthal. Detection of wave groups in SAR images and radar image sequences. \textit{IEEE Trans. Geosci. Remote Sens.} \textbf{41}(6): 1437--1446, 2003.

\bibitem[Dean, 1990]{1Dean90} \citep{1Dean90} R. G. Dean. Freak waves: a possible explanation. In A. T{\o}rum and O. T. Gudmestad, editors, \textit{Water Wave Kinematics}, pp~609--612, Kluwer Academic Publishers, Amsterdam, 1990.

\bibitem[Draper, 1965]{1Draper65} \citep{1Draper65} L. Draper. `Freak' ocean waves. \textit{Marine Observer} \textbf{35}: 193--195, 1965.

\bibitem[Dysthe and Trulsen, 1999]{1Dysthe99} \citep{1Dysthe99} K. B. Dysthe and K. Trulsen. Note on breather type solutions of the NLS as models for freak-waves. \textit{Phys. Scripta} \textbf{T82}: 48--52, 1999.

\bibitem[Fedele and Arena, 2005]{1Fedele05} \citep{1Fedele05} F. Fedele and F. Arena. Weakly nonlinear statistics of high random waves. \textit{Phys. Fluids} \textbf{17}: 026601, 2005.

\bibitem[Gibson et al., 2005]{1Gibson05} \citep{1Gibson05} R. Gibson, C. Swan, P. Tromans, L. Vanderscuren. Wave crest statistics calculated using a fully nonlinear spectral response surface method. In M. Olagnon and M. Prevosto, editors, \textit{Rogue Waves 2004}. Proceedings of a workshop in Brest, France (October 20--22, 2004), 10~pp, 2005.

\bibitem[Grimshaw and Saut, 2005]{1Grimshaw05} \citep{1Grimshaw05} R. Grimshaw and J.-C. Saut, organizers. \textit{Rogue Waves 2005}. Proceedings of a workshop in Edinburgh, United Kingdom (December 12--15, 2005), 2005. Available online at \url{http://www.icms.org.uk/meetings/2005/roguewaves/index.html}. Last accessed 6 November 2006. 

\bibitem[Haver, 2005]{1Haver05} \citep{1Haver05} S. Haver. A possible freak wave event measured at the Draupner jacket January~1, 1995. In M. Olagnon and M. Prevosto, editors, \textit{Rogue Waves 2004}.  Proceedings of a workshop in Brest, France (October 20--22, 2004), 8~pp, 2005.

\bibitem[Heller, 2005]{1Heller05} \citep{1Heller05} E. Heller. Freak waves: just bad luck, or avoidable? \textit{Europhysics News} \textbf{36}(5): 159--162, 2005.

\bibitem[Henderson et al., 1999]{1Henderson99} \citep{1Henderson99} K. L. Henderson, D. H. Peregrine and J. W. Dold. Unsteady water wave modulations: fully nonlinear solutions and comparison with the nonlinear Schr\"{o}dinger equation. \textit{Wave Motion} \textbf{29}: 341--461, 1999.

\bibitem[Janssen, 2003]{1Janssen03} \citep{1Janssen03} P. A. E. M. Janssen. Nonlinear four-wave interactions and freak waves. \textit{J. Phys. Ocean.} \textbf{33}: 863--884, 2003.

\bibitem[Kharif et al., 2000]{1Kharif00} \citep{1Kharif00} C. Kharif, E. Pelinovsky and T. Talipova. Formation de vagues g\'{e}antes en eau peu profonde. \textit{C. R. Acad. Sci. Paris, S\'{e}rie  II b, M\'{e}canique des fluides}, \textbf{328}(11): 801--807, 2000.

\bibitem[Kharif and Pelinovsky, 2003]{1Kharif03} \citep{1Kharif03} C. Kharif and E. Pelinovsky. Physical mechanisms of the rogue wave phenomenon. \textit{Eur. J. Mech. B: Fluids} \textbf{22}: 603--634, 2003.

\bibitem[Kokorina and Pelinovsky, 2002]{1Kokorina02} \citep{1Kokorina02} A. Kokorina and E. Pelinovsky. The applicability of the Korteweg-de Vries equation for description of the statistics of freak waves. \textit{J. Korean Soc. Coastal and Ocean Eng.} \textbf{14}(4): 308--318, 2002.

\bibitem[Liu and Pinho, 2004]{1Liu04} \citep{1Liu04} P. C. Liu and U. F. Pinho. Freak waves--more frequent than rare! \textit{Annal. Geophys.} \textbf{22}: 1839--1842, 2004.

\bibitem[Olagnon and Prevosto, 2005]{1Olagnon05} \citep{1Olagnon05} M. Olagnon and M. Prevosto, editors. \textit{Rogue Waves 2004}. Proceedings of a workshop in Brest, France (October 20--22, 2004), 308~pp, 2005.

\bibitem[Onorato et al., 2001]{1Onorato01} \citep{1Onorato01} M. Onorato, A. R. Osborne, M. Serio and S. Bertone. Freak waves in random oceanic sea states. \textit{Phys. Rev. Lett.} \textbf{86}: 5831--5834, 2001.

\bibitem[Osborne et al., 2000]{1Osborne00} \citep{1Osborne00} A. R. Osborne, M. Onorato, and M. Serio. The nonlinear dynamics of rogue waves and holes in deep-water gravity wave trains. \textit{Phys. Lett. A} \textbf{275}: 386--393, 2000.

\bibitem[Osborne, 2001]{1Osborne01} \citep{1Osborne01} A. R. Osborne. The random and deterministic dynamics of `rogue waves' in unidirectional, deep-water wave trains. \textit{Mar. Struct.} \textbf{14}: 275--293, 2001.

\bibitem[Pelinovsky et al., 2000]{1Pelinovsky00} \citep{1Pelinovsky00} E. Pelinovsky, T. Talipova and C. Kharif.  Nonlinear-dispersive mechanism of the freak wave formation in shallow water. \textit{Physica D} \textbf{147}: 83--94, 2000.

\bibitem[Pelinovsky et al., 2004]{1Pelinovsky04} \citep{1Pelinovsky04} E. Pelinovsky, T. Talipova, M. Ruderman, and R. Erd\'elyi. Freak waves described by the modified Korteweg-de Vries equation. \textit{Izvestia, Russian Academy of Engineering Sciences}, \textit{Applied Mathematics and Mechanics Series} \textbf{6}: 3--16, 2004.

\bibitem[Rosenthal, 2005]{1Rosenthal05} \citep{1Rosenthal05} W. Rosenthal. Result of the \textsl{MaxWave} project. In \textit{Proceedings of the 14th 'Aha Huliko'a Hawaiian Winter Workshop on Rogue Waves}, University of Hawaii, Honolulu, January 25--28, 2005.

\bibitem[Smith, 1976]{1Smith76} \citep{1Smith76} R. Smith. Giant waves. \textit{J. Fluid Mech.} \textbf{77}: 417--431, 1976.

\bibitem[Trulsen and Dysthe, 1997]{1Trulsen97} \citep{1Trulsen97} K. Trulsen and K. B. Dysthe. Freak waves--a three-dimensional wave simulation. In \textit{Proceedings of the Twenty-First (1996) Symposium on Naval Hydrodynamics}, Trondheim, Norway, June 24--28, pp~550--558, 1997.

\bibitem[Trulsen and Stansberg, 2001]{1Trulsen01} \citep{1Trulsen01} K. Trulsen and C. T. Stansberg. Spatial evolution of water surface waves: Numerical simulation and experiment of bichromatic waves. In \textit{Proceedings of the Eleventh (2001) International Offshore and Polar Engineering Conference}, Stavanger, Norway, June 17--22, pp~71--77, 2001.

\bibitem[Van Groesen et al., 2006]{1vanGroesen06} \citep{1vanGroesen06} E. van Groesen, Andonowati and N. Karjanto. Displaced phase-amplitude variables for waves on finite background. \textit{Phys. Lett. A} \textbf{354}: 312--319, 2006. \ arXiv:1906.00959 [nlin.PS]

\bibitem[Walker et al., 2004]{1Walker04} \citep{1Walker04} D. A. G. Walker, P. H. Taylor and R. E. Taylor. The shape of large surface waves on the open sea and the Draupner New Year wave. \textit{Appl. Ocean Res.} \textbf{26}: 73--83, 2004.

\bibitem[Wang et al., 2005]{1Wang05} \citep{1Wang05} D. W. Wang, D. A. Mitchell, W. J. Teague, E. Jarosz and M. S. Hulbert. Extreme waves under hurricane Ivan. \textit{Science} \textbf{309}: 896, 5 August 2005.

\bibitem[White and Fornberg, 1998]{1White98} \citep{1White98} B. S. White and B. Fornberg. On the chance of freak waves at sea. \textit{J. Fluid Mech.}, \textbf{355}: 113--138, 1998.
\end{thebibliography}

\small
\begin{thebibliography}{99}
\bibitem[Agrawal, 1995]{2Agrawal95} \citep{2Agrawal95} G. P. Agrawal. \textit{Nonlinear Fiber Optics}. Academic Press, New York, 1995.
\addcontentsline{toc}{section}{\bibname}

\bibitem[Benjamin, 1967]{2Benjamin67} \citep{2Benjamin67} T. B. Benjamin. Instability of periodic wave trains in nonlinear dispersive systems. \textit{Proc. R. Soc. London A} \textbf{229}(1456): 59--75, 1967.

\bibitem[Benjamin and Feir, 1967]{2BenjaminFeir67} \citep{2BenjaminFeir67} T. B. Benjamin and J. E. Feir. The disintegration of wave trains in deep water. \textit{J. Fluid Mech.} \textbf{27}: 417--430, 1967.

\bibitem[Benney and Newell, 1967]{2Benney67} \citep{2Benney67} D. J. Benney and A. C. Newell. The propagation of nonlinear wave envelopes \textit{J. Math. Phys.} \textbf{46}: 133--139, 1967.

\bibitem[Bespalov and Talanov, 1966]{2Bespalov66} \citep{2Bespalov66} V. I. Bespalov and V. I. Talanov. On the filament structure of a light beam in nonlinear liquids. \textit{JETP Lett.} \textbf{3}: 307--312, 1966. Translated from \textit{Zh. Eksperim. i Teor. Fiz.-Pis'ma Redakt.} \textbf{3}: 471--476, 1966.

\bibitem[Boyd and Chen, 2001]{2Boyd01} \citep{2Boyd01} J. P. Boyd and G.-Y. Chen. Weakly nonlinear wavepackets in the Korteweg-de Vries equation: the KdV/NLS connection. \textit{Math. Comput. Simulation} \textbf{55}: 317--328, 2001.

\bibitem[Bridges and Dias, 2004]{2Bridges04} \citep{2Bridges04} T. J. Bridges and F. Dias. On the enhancement of the Benjamin-Feir instability due to dissipation. Under consideration for publication in \textit{J. Fluid Mech.}, 2004.

\bibitem[Bullough and Caudrey, 1995]{2Bullough95} \citep{2Bullough95} R. K. Bulllough and P. J. Caudrey. Solitons and the Korteweg-de Vries equation: Integrable systems in 1834-1995. \textit{Acta Appl. Math. (Acta Applicandae Mathematicae)} \textbf{39}: 193--228, 1995.

\bibitem[Cahyono, 2002]{2Cahyono02} \citep{2Cahyono02} E. Cahyono. \textit{Analytical Wave Codes for Predicting Surface Waves in a Laboratory Basin}. Ph.D. thesis, Department of Applied Mathematics, University of Twente, The Netherlands, 2002.

\bibitem[Chu and Mei, 1970]{2Chu70} \citep{2Chu70} V. H. Chu and C. C. Mei. On slowly-varying Stokes waves. \textit{J. Fluid Mech.} \textbf{41}: 873--87, 1970.

\bibitem[Chu and Mei, 1971]{2Chu71} \citep{2Chu71} V. H. Chu and C. C. Mei. The nonlinear evolution of Stokes waves in deep water. \textit{J. Fluid Mech.} \textbf{47}: 337--351, 1971.

\bibitem[Craik, 2004]{2Craik04} \citep{2Craik04} A. D. D. Craik. The origins of water wave theory. \textit{Annu. Rev. Fluid Mech.} \textbf{36}: 1--28, 2004.

\bibitem[Craik, 2005]{2Craik05} \citep{2Craik05} A. D. D. Craik. George Gabriel Stokes on water wave theory. \textit{Annu. Rev. Fluid Mech.} \textbf{37}: 23--42, 2005.

\bibitem[Davey, 1972]{2Davey72} \citep{2Davey72} A. Davey. The propagation of a weak nonlinear wave. \textit{J. Fluid Mech.} \textbf{53}: 769--781, 1972.

\bibitem[Debnath, 1994]{2Debnath94} \citep{2Debnath94} L. Debnath. \textit{Nonlinear Water Waves}. Academic Press, San Diego, 1994.

\bibitem[Dingemans, 1997]{2Dingemans97} \citep{2Dingemans97} M. W. Dingemans. \textit{Water Wave Propagation over Uneven Bottoms}. World Scientific, Singapore, 1997.

\bibitem[Dingemans and Otta, 2001]{2Dingemans01} \citep{2Dingemans01} M. W. Dingemans and A. K. Otta. \textit{Nonlinear Modulation of Water Waves}. In P. L.-F. Liu, editor, volume \textbf{7} of \textit{Advances in Coastal and Ocean Engineering}, World Scientific, Singapore, 2001.

\bibitem[Djordjevi\'{c} and Redekopp, 1978]{2Djordjevic78} \citep{2Djordjevic78} V. D. Djordjevi\'{c} and L. G. Redekopp. On the development of packets of surface gravity waves moving over an uneven bottom. \textit{J. Appl. Math. Phys. (ZAMP)} \textbf{29}: 950--962, 1978.

\bibitem[Dodd et al., 1982]{2Dodd82} \citep{2Dodd82} R. K. Dodd, J. C. Eilbeck, J. D. Gibbon and H. C. Morris. \textit{Solitons and Nonlinear Wave Equations}. Academic Press, 1982.

\bibitem[Drazin and Johnson, 1989]{2Drazin89} \citep{2Drazin89} P. G. Drazin and R. S. Johnson. \textit{Solitons: an introduction}. Cambridge University Press, Cambridge, 1989.

\bibitem[Fermi et al., 1955]{2Fermi55} \citep{2Fermi55} E. Fermi, J. Pasta and S. M. Ulam. Studies of nonlinear problem. Los Alamos Scientific Laboratory Report No. LA-1940, 1955. Reprinted in A. C. Newell, editor, \textit{Nonlinear Wave Motion}, volume \textbf{15} of AMS Lectures in Applied Mathematics, pp 143--156, 1974.

\bibitem[Fornberg and Whitham, 1978]{2Fornberg78} \citep{2Fornberg78} B. Fornberg and G. B. Whitham. A numerical and theoretical study of certain nonlinear wave phenomena. \textit{Phil. Trans. R. Soc. London. A, Math. Phys. Sci.}, \textbf{289}: 373--404, 1978.

\bibitem[Gel'fand and Fomin, 1963]{2Gelfand63} \citep{2Gelfand63} I. M. Gel'fand and S. V. Fomin. \textit{Calculus of Variations}. Prentice Hall, Englewood Cliffs, New Jersey, 1963.

\bibitem[Ginzburg and Landau, 1950]{2Ginzburg50} \citep{2Ginzburg50} V. L. Ginzburg and L. D. Landau. On the theory of superconductivity. \textit{Zh. Eksp. Teor. Fiz.} (USSR), \textbf{20}: 1064--1082 1950. English translation in L.D. Landau. \textit{Men of Physics}, edited by D. ter Haar, volume \textbf{1}, Pergamon Press, Oxford, pp 138--167, 1965.

\bibitem[Goldman and Sirovich, 1994]{2Goldman94} \citep{2Goldman94} D. Goldman and L. Sirovich. The one dimensional complex Ginzburg-Landau equation in the low dissipation limit. \textit{Nonlinearity}, \textbf{7}: 417--439, 1994.

\bibitem[Hasegawa and Kodama, 1995]{2Hasegawa95} \citep{2Hasegawa95} A. Hasegawa and Y. Kodama. \textit{Solitons in Optical Communications}, volume \textbf{7} of \textit{Oxford Series in Optical and Imaging Sciences}. Clarendon Press, Oxford, 1995.

\bibitem[Hasegawa and Tappert, 1973]{2Hasegawa73} \citep{2Hasegawa73} A. Hasegawa and F. D. Tappert. Transmission of stationary nonlinear optical pulses in dispersive dielectric fibers. I. Anomalous dispersion. II. Normal dispersion. \textit{Appl. Phys. Lett.} \textbf{23}: 142--144, 171--172, 1973.

\bibitem[Hasimoto and Ono, 1972]{2Hasimoto72} \citep{2Hasimoto72} H. Hasimoto and H. Ono.  Nonlinear modulation of gravity waves. \textit{J. Phys. Soc. Japan} \textbf{33}: 805--811, 1972.

\bibitem[Hunt, 2005]{2Hunt05} \citep{2Hunt05} J. C. R. Hunt. Nonlinear and wave theory contributions of T. Brooke Benjamin (1929-1995). \textit{Annu. Rev. Fluid Mech.} \textbf{38}: 1--25, 2005.

\bibitem[Ichikawa et al., 1972]{2Ichikawa72} \citep{2Ichikawa72} Y. H. Ichikawa, T. Imamura, and T. Taniuti. Nonlinear wave modulation in collisionless plasmas. \textit{J. Phys. Soc. Japan} \textbf{33}: 189--197, 1972.

\bibitem[Infeld and Rowlands, 1990]{2Infeld90} \citep{2Infeld90} E. Infeld and G. Rowlands. \textit{Nonlinear Waves, Solitons, and Chaos}, Cambridge University Press, 1990.

\bibitem[Janssen, 1981]{2Janssen81} \citep{2Janssen81} P. A. E. M. Janssen. Modulational instability and the Fermi-Pasta-Ulam recurrence. \textit{Phys. Fluids} \textbf{24}(1): 23--26, 1981.

\bibitem[Jeffrey and Kawahara, 1982]{2Jeffrey82} \citep{2Jeffrey82} A. Jeffrey and T. Kawahara. \textit{Asymptotic Methods in Nonlinear Wave Theory}. Applicable Mathematics Series, Pitman, Boston, 1982.

\bibitem[Johnson, 1976]{2Johnson76} \citep{2Johnson76} R. S. Johnson. On the modulation of water waves on shear flows. \textit{Proc. R. Soc. London. Ser. A} \textbf{347}(1651): 537--546, 1976.

\bibitem[Johnson, 1997]{2Johnson97} \citep{2Johnson97} R. S. Johnson. \textit{A modern introduction to the mathematical theory of water waves}, volume \textbf{18} of \textit{Cambridge Texts in Applied Mathematics}, Cambridge University Press, Cambridge, 1997.

\bibitem[Kadomtsev and Karpman, 1971]{2Kadomtsev71} \citep{2Kadomtsev71} B. B. Kadomtsev and V. I. Karpman. Nonlinear waves. \textit{Sov. Phys. Uspekhi}, \textbf{14}(1): 40--60, 1971. Translated from \textit{Usp. Fiz. Nauk} \textbf{103}: 193--232, 1971.

\bibitem[Karpman and Kruskal, 1969]{2Karpman69} \citep{2Karpman69} V. I. Karpman and E. M. Kruskal. Modulated waves in a nonlinear dispersive media. \textit{Sov. Phys. JETP}, \textbf{28}: 277--281, 1969.

\bibitem[Karpman, 1975]{2Karpman75} \citep{2Karpman75} V. I. Karpman. \textit{Nonlinear Waves in Dispersive Media}. Pergamon Press, Oxford, 1975.

\bibitem[Kelley, 1965]{2Kelley65} \citep{2Kelley65} P. L. Kelley. Self-focusing of optical beams. \textit{Phys. Rev. Lett.} \textbf{15}: 1005--1008, 1965.

\bibitem[Kevorkian, 1961]{2Kevorkian61} \citep{2Kevorkian61} J. K. Kevorkian. \textit{The Uniformly Valid Asymptotic Approximations to the Solutions of Certain Nonlinear Ordinary Differential Equations}. PhD thesis, California Institute of Technology, 1961.

\bibitem[Kivshar and Luther-Davies, 1998]{2Kivshar98} \citep{2Kivshar98} Y. S. Kivshar and B. Luther-Davies. Dark optical solitons: Physics and applications. \textit{Phys. Reports} \textbf{298}: 81--197, 1998.

\bibitem[Lake et al., 1977]{2Lake77} \citep{2Lake77} B. M. Lake, H. C. Yuen, H. Rundgaldier and W. E. Ferguson. Nonlinear deep water waves: Theory and experiment, Part 2, Evolution of a continuous wave train. \textit{J. Fluid Mech.} \textbf{83}: 49--74, 1977.

\bibitem[Lamb, 1980]{2Lamb80} \citep{2Lamb80} G. L. Lamb, Jr. \textit{Elements of Soliton Theory}, John Wiley \& Sons, New York, 1980.

\bibitem[Lighthill, 1965]{2Lighthill65} \citep{2Lighthill65} M. J. Lighthill. Contributions to the theory of waves in nonlinear dispersive systems. \textit{Journal of the Institute of Mathematics and Its Applications  (J. Inst. Math. Applics.)} \textbf{1}: 269--306, 1965.

\bibitem[Lighthill, 1967]{2Lighthill67} \citep{2Lighthill67} M. J. Lighthill. Some special cases treated by the Whitham theory. \textit{Proc. R. Soc. London. Ser. A} \textbf{299}(1456): 28--53, 1967.

\bibitem[L'vov, 1994]{2Lvov94} \citep{2Lvov94} V. S. L'vov. \textit{Wave Turbulence Under Parametric Excitation--Applications to Magnets}. In V. E. Zakharov, series editor, \textit{Springer Series in Nonlinear Dynamics}.  Springer-Verlag, Berlin, 1994.

\bibitem[Madelung, 1927]{2Madelung27} \citep{2Madelung27} E. Madelung. Quantentheorie in Hydrodynamischer Form (Quantum theory in hydrodynamic form). \textit{Zeitschrift f\"{u}r Physik (Journal of Physics)} \textbf{40}: 322--326, 1927.

\bibitem[Mei, 1983]{2Mei83} \citep{2Mei83} C. C. Mei. \textit{The Applied Dynamics of Ocean Surface Waves}. John Wiley \& Sons, New York, 1983.

\bibitem[Miles, 1981]{2Miles81} \citep{2Miles81} J. W. Miles. The Korteweg-de Vries equation: a historical essay. \textit{J. Fluid Mech.} \textbf{106}: 131--147, 1981.

\bibitem[Nayfeh, 1973]{2Nayfeh73} \citep{2Nayfeh73} A. H. Nayfeh. \textit{Perturbation Methods}, Wiley-Interscience, New York, 1973.

\bibitem[Newell, 1974]{2Newell74} \citep{2Newell74} A. C. Newell. Envelope equations. In A. C. Newell, editor, \textit{Nonlinear Wave Motion}, volume \textbf{15} of \textit{Lectures in Applied Mathematics}.  American Mathematical Society, 1974.

\bibitem[Newell, 1985]{2Newell85} \citep{2Newell85} A. C. Newell. \textit{Solitons in Mathematical Physics}, volume \textbf{48} of \textit{Regional Conference Series in Applied Mathematics}. Society for Industrial and Applied Mathematics, 1985.

\bibitem[Noether, 1918]{2Noether18} \citep{2Noether18} E. Noether. Invariante Variationsprobleme. \textit{Nachr. v.d. Ges. d. Wiss. zu G\"{o}ttingen}, 235--257, 1918. English translation: M. A. Tavel. \textit{Transport Theory and Stat. Phys.} \textbf{1}(3): 186--207, 1971.

\bibitem[Onorato et al., 2001]{2Onorato01} \citep{2Onorato01} M. Onorato, A. R. Osborne, M. Serio and S. Bertone. Freak waves in random oceanic sea states. \textit{Phys. Rev. Lett.} \textbf{86}: 5831--5834, 2001.

\bibitem[Osborne et al., 2000]{2Osborne00} \citep{2Osborne00} A. R. Osborne, M. Onorato, and M. Serio. The nonlinear dynamics of rogue waves and holes in deep-water gravity wave trains. \textit{Phys. Lett. A} \textbf{275}: 386--393, 2000.

\bibitem[Ostrovski\u{\i}, 1967]{2Ostrovsky67} \citep{2Ostrovsky67} L. A. Ostrovski\u{\i}. Propagation of wave packets and space-time self-focusing in a nonlinear medium. \textit{Sov. Phys. JETP} \textbf{24}: 797--800, 1967.

\bibitem[Remoissenet, 1999]{2Remoissenet99} \citep{2Remoissenet99} M. Remoissenet. \textit{Waves Called Solitons: Concepts and Experiments}. Springer, third edition, 1999.

\bibitem[Segur et al., 2005]{2Segur05} \citep{2Segur05} H. Segur, D. Henderson, J. Carter, J. Hammack, C.-M. Li, D. Pheiff and K. Socha. Stabilizing the Benjamin-Feir instability. \textit{J. Fluid Mech.} \textbf{539}: 229--271, 2005.

\bibitem[Scott et al., 1973]{2Scott73} \citep{2Scott73} A. C. Scott, F. Y. H. Chu and D. W. McLaughlin. The soliton: a new concept in applied science \textit{Proc. IEEE} \textbf{61}(10): 1443--1483, 1973.

\bibitem[Scott, 2003]{2Scott03} \citep{2Scott03} A. C. Scott. \textit{Nonlinear Science--Emergence and Dynamics of Coherent Structures}. Oxford University Press, second edition, 2003.

\bibitem[Scott, 2005]{2Scott05} \citep{2Scott05} A. C. Scott. Solitons, a brief history. In A. C. Scott, editor, \textit{Encyclopedia of Nonlinear Science}. Routledge Taylof \& Francis Group, New York, 2005.

\bibitem[Stuart and DiPrima, 1978]{2Stuart78} \citep{2Stuart78} J. T. Stuart and R. C. DiPrima. The Eckhaus and Benjamin-Feir resonance mechanisms. \textit{Proc. R. Soc. London A} \textbf{362}(1708): 27--41, 1978.

\bibitem[Sulem and Sulem, 1999]{2Sulem99} \citep{2Sulem99} C. Sulem and P.-L. Sulem. \textit{The Nonlinear Schr\"{o}dinger Equation---Self-Focusing and Wave Collapse}, volume \textbf{139} of \textit{Applied Mathematical Sciences}, Springer-Verlag, New York, 1999.

\bibitem[Talanov, 1965]{2Talanov65} \citep{2Talanov65} V. I. Talanov. Self focusing of wave beams in nonlinear media. \textit{JETP Lett.} \textbf{2}: 138--141, 1965. Translated from \textit{Pis'ma Zh. Eksp. Teor. Fiz.} \textbf{2}: 223--236, 1965.

\bibitem[Taniuti and Washimi, 1968]{2Taniuti68} \citep{2Taniuti68} T. Taniuti and H. Washimi. Self-trapping and instability of hydromagnetics waves along the magnetic field in a cold plasma. \textit{Phys. Rev. Lett.} \textbf{21}: 209--212, 1968.

\bibitem[Tappert and Varma, 1970]{2Tappert70} \citep{2Tappert70} F. Tappert and C. M. Varma. Asymptotic theory of self-trapping of heat pulses in solids. \textit{Phys. Rev. Lett.} \textbf{25}: 1108--1111, 1970.

\bibitem[Tracy et al., 1988]{2Tracy88} \citep{2Tracy88} E. R. Tracy, J. W. Larson, A. R. Osborne and L. Bergamasco. On the nonlinear Schr\"{o}dinger limit of the Korteweg-de Vries equation. \textit{Physica D} \textbf{32}(1): 83--106, 1988.

\bibitem[Tracy et al., 1991]{2Tracy91} \citep{2Tracy91} E. R. Tracy, J. W. Larson, A. R. Osborne and L. Bergamasco. The relationship between the spectral theories for the periodic Korteweg-de Vries and nonlinear Schr\"{o}dinger equations. In A. R. Osborne, editor, \textit{Nonlinear Topics of Ocean Physics}, Fermi Summer School, Course LIX, pp. 769--825, North-Holland, Amsterdam, 1991.

\bibitem[Van Groesen, 1998]{2vanGroesen98} \citep{2vanGroesen98} E. van Groesen. Wave groups in uni-directional surface-wave models. \textit{J. Engng. Math.} \textbf{34}(1--2): 215--226, 1998.

\bibitem[Van Groesen and Andonowati, 2006]{2BrennyAan06} \citep{2BrennyAan06} E. van Groesen and Andonowati. Finite energy wave signals of extremal amplitude in the spatial NLS-dynamics. \textit{Phys. Lett. A} \textbf{357}: 86--91, 2006.

\bibitem[Van Saarloos and Hohenberg, 1992]{2vanSaarloos92} \citep{2vanSaarloos92} W. van Saarloos and P. C. Hohenberg. Fronts, pulses, sources and sinks in generalized complex Ginzburg-Landau equations. \textit{Physica D} \textbf{56}: 303--367, 1992.

\bibitem[Whitham, 1967]{2Whitham67} \citep{2Whitham67} G. B. Whitham. Nonlinear dispersion of water waves. \textit{J. Fluid Mech.} \textbf{27}(2): 399--412, 1967.

\bibitem[Whitham, 1974]{2Whitham74} \citep{2Whitham74} G. B. Whitham. \textit{Linear and Nonlinear Waves}. John Wiley \& Sons, New York, 1974.

\bibitem[Yuen and Ferguson, 1978a]{2Yuen78a} \citep{2Yuen78a} H. C. Yuen and W. E. Ferguson, Jr. Relationship between Benjamin-Feir instability and recurrence in the nonlinear Schr\"{o}dinger equation. \textit{Phys. Fluids} \textbf{21}(8): 1275--1278, 1978.

\bibitem[Yuen and Ferguson, 1978b]{2Yuen78b} \citep{2Yuen78b} H. C. Yuen and W. E. Ferguson, Jr.  Fermi-Pasta-Ulam recurrence in the two-space dimensional nonlinear Schr\"{o}dinger equation. \textit{Phys. Fluids} \textbf{21}(11): 2116--2118, 1978.

\bibitem[Yuen and Lake, 1975]{2Yuen75} \citep{2Yuen75} H. C. Yuen and B. M. Lake. Nonlinear deep water waves: Theory and experiment. \textit{Phys. Fluids} \textbf{18}: 956--960, 1975.

\bibitem[Yuen and Lake, 1982]{2Yuen82} \citep{2Yuen82} H. C. Yuen and B. M. Lake. Nonlinear dynamics of deep-water gravity waves. \textit{Adv. Appl. Mech.} \textbf{22}: 67--229, 1982.

\bibitem[Zakharov, 1967]{2Zakharov67} \citep{2Zakharov67} V. E. Zakharov. The wave stability in nonlinear media. \textit{Sov. Phys. JETP} \textbf{24}: 455--459, 1967.

\bibitem[Zakharov, 1968]{2Zakharov68} \citep{2Zakharov68} V. E. Zakharov. Stability of periodic waves of finite amplitude on the surface of a deep fluid. \textit{J. Appl. Mech. Tech. Phys.} \textbf{9}: 190--194, 1968. Translated from \textit{Zh. Prikl. Mekh. Tekh. Fiz.} \textbf{9}(2): 86--94, 1968.

\bibitem[Zakharov and Shabat, 1972]{2ZakharovShabat72} \citep{2ZakharovShabat72} V. E. Zakharov and A. B Shabat. Exact theory of two-dimensional self-focusing and one-dimensional self-modulation of waves in nonlinear media. \textit{Sov. Phys. JETP} \textbf{34}(1): 62--69, 1972. Translated from \textit{Zh. Eksp. Teor. Fiz.} \textbf{61}: 118--134, 1971.

\bibitem[Zakharov, 1972]{2Zakharov72} \citep{2Zakharov72} V. E. Zakharov. Collapse of Langmuir waves. \textit{Sov. Phys. JETP} \textbf{35}(5): 908--914, 1972. Translated from \textit{Zh. Eksp. Teor. Fiz.} \textbf{62}: 1745--1759, 1972.
\end{thebibliography}

\small
\begin{thebibliography}{99}
\bibitem[Ablowitz et al., 1974]{3Ablowitz74} \citep{3Ablowitz74} M. J. Ablowitz, D. J. Kaup, A. C. Newell, and H. Segur. The inverse scattering transform-Fourier analysis for nonlinear problems. \textit{Stud. Appl. Math.} \textbf{53}(4): 249--315, 1974.

\addcontentsline{toc}{section}{\bibname}
\bibitem[Ablowitz and Herbst, 1990]{3Ablowitz90} \citep{3Ablowitz90} M. J. Ablowitz and B. M. Herbst. On homoclinic structure and numerically induced chaos for the nonlinear Schr\"{o}dinger equation. \textit{SIAM J. Appl. Math.} \textbf{50}(2): 339--351, 1990.

\bibitem[Akhmediev et al., 1985]{3Akhmediev85} \citep{3Akhmediev85} N. N. Akhmediev, V. M. Eleonski\u{\i}, and N. E. Kulagin. Generation of periodic trains of picosecond pulses in an optical fiber: exact solutions. \textit{Sov. Phys. JETP} \textbf{62}(5): 894--899, 1985.

\bibitem[Akhmediev and Korneev, 1986]{3Akhmediev86} \citep{3Akhmediev86} N. N. Akhmediev and V. I. Korneev. Modulation instability and periodic solutions of the nonlinear Schr\"{o}dinger equation. \textit{Theoret. Math. Phys.} \textbf{69}: 1089--1092, 1986. Translated from \textit{Teoret. Mat. Fiz.} \textbf{62}(2): 189--194, 1986.

\bibitem[Akhmediev et al., 1987]{3Akhmediev87} \citep{3Akhmediev87} N. N. Akhmediev, V. M. Eleonski\u{\i}, and N. E. Kulagin. Exact first-order solutions of the nonlinear Schr\"{o}dinger equation. \textit{Theoret. Math. Phys.} \textbf{72}(2): 809--818, 1987. Translated from \textit{Teoret. Mat. Fiz.} \textbf{72}(2): 183--196, 1987.

\bibitem[Akhmediev and Ankiewicz, 1997]{3Akhmediev97} \citep{3Akhmediev97} N. N. Akhmediev and A. Ankiewicz. \textit{Solitons---Nonlinear Pulses and Beams}, volume \textbf{5} of \textit{Optical and Quantum Electronic Series}. Chapman \& Hall, first edition, 1997.

\bibitem[Andonowati and Van Groesen, 2003]{3Andonowati03} \citep{3Andonowati03} Andonowati and E. van Groesen. Optical pulse deformation in second order nonlinear media. \textit{J. Nonlinear Opt. Phys.} \textbf{12}(2): 221--234, 2003.

\bibitem[Andonowati et al., 2007]{3Andonowati07} \citep{3Andonowati07} Andonowati, N. Karjanto, and E. van Groesen. Extreme wave phenomena in down-stream running modulated waves. \textit{Appl. Math. Modelling} \textbf{31}(7): 1425--1443, 2007. \ arXiv:1710.10804 [physics.flu-dyn] 

\bibitem[Benjamin and Feir, 1967]{3BenjaminFeir67} \citep{3BenjaminFeir67} T. B. Benjamin and J. E. Feir. The disintegration of wave trains in deep water. \textit{J. Fluid Mech.} \textbf{27}: 417--430, 1967.

\bibitem[Drazin and Johnson, 1989]{3Drazin89} \citep{3Drazin89} P. G. Drazin and R. S. Johnson. \textit{Solitons: an Introduction}. Cambridge University Press, 1989.

\bibitem[Dysthe and Trulsen, 1999]{3Dysthe99} \citep{3Dysthe99} K. B. Dysthe and K. Trulsen. Note on breather type solutions of the NLS as models for freak-waves. \textit{Phys. Scripta} \textbf{T82}: 48--52, 1999.

\bibitem[Dysthe, 2000]{3Dysthe00} \citep{3Dysthe00} K. B. Dysthe. Modelling a ``rogue wave''--speculations or a realistic possibility? In M. Olagnon and G. A. Athanassoulis, editors, \textit{Proceedings of the Rogue Waves 2000}, Ifremeer, Brest, France, November 2000.

\bibitem[Grimshaw et al., 2001]{3Grimshaw01} \citep{3Grimshaw01} R. Grimshaw, D. Pelinovsky, E. Pelinovsky, and T. Talipova. Wave group dynamics in weakly nonlinear long-wave models. \textit{Physica D} \textbf{159}: 35--57, 2001.

\bibitem[Ma, 1979]{3Ma79} \citep{3Ma79} Y.-C. Ma. The perturbed plane-wave solutions of the cubic Schr\"{o}dinger equation. \textit{Stud. Appl. Math.} \textbf{60}(1): 43--58, 1979.

\bibitem[Nye and Berry, 1974]{3NyeBerry74} \citep{3NyeBerry74} J.F. Nye and M.V. Berry. Dislocation in wave trains. \textit{Proc. R. Soc. Lond. A} \textbf{336}: 165--190, 1974.

\bibitem[Onorato et al., 2000]{3Onorato00} \citep{3Onorato00} M. Onorato, A. Osborne, M. Serio and T. Damiani. Occurrence of freak waves from envelope equations in random ocean wave simulations. In M. Olagnon, editor, \textit{Proceedings Rogue Waves 2000}, Brest, France, 2000.

\bibitem[Osborne et al., 2000]{3Osborne00} \citep{3Osborne00} A. R. Osborne, M. Onorato, and M. Serio. The nonlinear dynamics of rogue waves and holes in deep-water gravity wave trains. \textit{Phys. Lett. A} \textbf{275}: 386--393, 2000.

\bibitem[Osborne, 2001]{3Osborne01} \citep{3Osborne01} A. R. Osborne. The random and deterministic dynamics of `rogue waves' in unidirectional, deep-water wave trains. \textit{Mar. Struct.} \textbf{14}: 275--293, 2001.

\bibitem[Peregrine, 1983]{3Peregrine83} \citep{3Peregrine83} D. H. Peregrine. Water waves, nonlinear Schr\"{o}dinger equations and their solutions. \textit{J. Austral. Math. Soc. Ser. B} \textbf{25}(1): 16--43, 1983.

\bibitem[Tajiri and Arai, 2000]{3Tajiri00} \citep{3Tajiri00} M. Tajiri and T. Arai. Periodic soliton solutions to the Davey-Stewartson equation. \textit{Proc. Inst. Math. Natl. Acad. Sci. Ukr.} \textbf{30}(1): 210--217, 2000.

\bibitem[Van Groesen et al., 2005]{3vanGroesen05} \citep{3vanGroesen05} E. van Groesen, Andonowati and N. Karjanto. Deterministic aspect of nonlinear modulation instability. In M. Olagnon and M. Prevosto, editors, \textit{Rogue Waves 2004}. Proceedings of a workshop in Brest, France (October 20-22, 2004), 12~pp, 2005. \ arXiv:1110.5120 [physics.flu-dyn]

\bibitem[Van Groesen et al., 2006]{3vanGroesen06} \citep{3vanGroesen06} E. van Groesen, Andonowati and N. Karjanto. Displaced phase-amplitude variables for waves on finite background. \textit{Phys. Lett. A} \textbf{354}: 312--319, 2006. \ arXiv:1906.00959 [nlin.PS]
\end{thebibliography}

\small
\begin{thebibliography}{99}
\bibitem[Balistreri et al., 2000]{4Balistreri00} \citep{4Balistreri00} M. L. M. Balistreri, J. P. Korterik, L. Kuipers and N. F. van Hulst. Local observation of phase singularities in optical fields in waveguide structures. {\it Phys. Rev. Lett.} {\bf 85}: 294--297, 2000.

\addcontentsline{toc}{section}{\bibname}
\bibitem[Basistiy et al., 1995]{4Basistiy95} \citep{4Basistiy95} I. V. Basistiy, M. S. Soskin, and M. V. Vasnetsov. Optical wavefront dislocations and their properties. {\it Opt. Comm.} \textbf{119}: 604--612, 1995.

\bibitem[Berry et al., 1980]{4Berry80} \citep{4Berry80} M. V. Berry, R. G. Chambers, M. D. Large, C. Upstill and J. C. Walmsley. Wavefront dislocations in the Aharonov-Bohm effect and its water wave analogue. {\it Eur. J. Phys.} {\bf 1}: 154--162, 1980.

\bibitem[Berry, 1981]{4Berry81} \citep{4Berry81} M. V. Berry. Singularities in waves and rays. In R. Balian, M. Kl\'{e}man, and J.-P. Poirier, editors, {\it Les Houches 1980, Session XXXV, Physics of Defects}, North-Holland, Amsterdam,  p~453--459, 1981.

\bibitem[Berry, 1998]{4Berry98} \citep{4Berry98} M. V. Berry. Much ado about nothing: optical dislocation lines (phase singularities, zeros, vortices...). In M. S. Soskin, editor, {\it Proceedings of International Conference on Singular Optics}, The International Society of Optical Engineering (SPIE), p~1--10, 1998.

\bibitem[Berry and Dennis, 2000]{4Berry00} \citep{4Berry00} M. V. Berry and M. R. Dennis. Phase singularities in isotropic random waves. {\it Proc. R. Soc. Lond. A} {\bf 456}: 2059--2079, 2000.

\bibitem[Berry and Dennis, 2001]{4Berry01} \citep{4Berry01} M. V. Berry and M. R. Dennis. Knotted and linked phase singularities in monochromatic waves. {\it Proc. R. Soc. Lond. A} {\bf 457}: 2251--2263, 2001.

\bibitem[Berry et al., 2004]{4Berry04} \citep{4Berry04} M. V. Berry, M. R. Dennis and M. S. Soskin. The plurality of optical singularities. {\it J. Opt. A: Pure Appl. Opt.} {\bf 6}: S155--S156, 2004.

\bibitem[Chu and Mei, 1970]{4Chu70} \citep{4Chu70} V. H. Chu and C. C. Mei. On slowly-varying Stokes waves. \textit{J. Fluid Mech.} \textbf{41}: 873--887, 1970.

\bibitem[Chu and Mei, 1971]{4Chu71} \citep{4Chu71} V. H. Chu and C. C. Mei. The nonlinear evolution of Stokes waves in deep water. \textit{J. Fluid Mech.} \textbf{47}: 337--351, 1971.

\bibitem[Coullet et al., 1989]{4Coullet89} \citep{4Coullet89} P. Coullet, L. Gil and F. Rocca. Optical vortices. \textit{Opt. Comm.} \textbf{73}: 403--408, 1989.

\bibitem[Dennis, 2001]{4Dennis01} \citep{4Dennis01} M. R. Dennis, {\it Topological Singularities in Wave Fields}. PhD thesis, University of Bristol, H. H. Wills Physics Laboratory, November 2001.

\bibitem[Dieter, 1988]{4Dieter88} \citep{4Dieter88} G. E. Dieter. {\it Mechanical Metallurgy}. McGraw-Hill, London, SI Metric edition, 1988.

\bibitem[Fornberg and Whitham, 1978]{4Fornberg78} \citep{4Fornberg78} V. Fornberg and G. B. Whitham. A numerical and theoretical study of certain nonlinear wave phenomena. {\it Phil. Trans. R. Soc. Lond. A} {\bf 289}: 373--404, 1978.

\bibitem[Honeycombe, 1984]{4Honeycombe84} \citep{4Honeycombe84} R. W. K. Honeycombe. {\it The Plastic Deformation of Metals}. Edward Arnold, London, second edition, 1984.

\bibitem[Hull and Bacon, 2001]{4Hull01} \citep{4Hull01} D. Hull and D. J. Bacon. {\it Introduction to Dislocations}. Butterworth Heinemann, Oxford, fourth edition, 2001.

\bibitem[Infeld and Rowlands, 1990]{4Infeld90} \citep{4Infeld90} E. Infeld and G. Rowlands. {\it Nonlinear waves, solitons and chaos}. Cambridge University Press, Cambridge, pp~117--119, 1990.

\bibitem[Karpman, 1967]{4Karpman67} \citep{4Karpman67}  V. I. Karpman. Self-modulation of nonlinear plane waves in dispersive media. {\it JETP Lett.} {\bf 6}: 277--279, 1967.

\bibitem[Karpman and Krushkal', 1969]{4Karpman69} \citep{4Karpman69} V. I. Karpman and E. M. Krushkal'. Modulated waves in nonlinear dispersive media. {\it Sov. Phys. JETP} {\bf 28}: 277--281, 1969.

\bibitem[Kr\'{a}sa, 1981]{4Krasa81} \citep{4Krasa81} J. Kr\'{a}sa. Dislocation in turbulent ionisation waves. {\it J. Phys. D: Appl. Phys.} {\bf 14}: 1241--1246, 1981.

\bibitem[Nabarro, 1979-2004]{4Nabarro7904} \citep{4Nabarro7904} F. R. N. Nabarro, editor, {\it Dislocations in Solids} {\bf 1--12}, 1979--2004.

\bibitem[Nye and Berry, 1974]{4Nye&Berry74} \citep{4Nye&Berry74} J. F. Nye and M. V. Berry. Dislocation in wave trains. {\it Proc. R. Soc. Lond. A} {\bf 336}: 165--190, 1974.

\bibitem[Nye, 1981]{4Nye81} \citep{4Nye81} J. F. Nye. The motion and structure of dislocations in wavefronts. {\it Proc. R. Soc. Lond. A} {\bf 378}: 219--239, 1981.

\bibitem[Nye, 1997]{4Nye97} \citep{4Nye97} J. F. Nye. Line singularities in wave fields. {\it Proc. R. Soc. Lond. A} {\bf 355}: 2065--2069, 1997.

\bibitem[Nye, 1999]{4Nye99} \citep{4Nye99} J. F. Nye. {\it Natural Focussing and Fine Structure of Light: Caustics and Wave Dislocations}. IOP, Bristol, 1999.

\bibitem[Orowan, 1934]{4Orowan34} \citep{4Orowan34} E. Orowan. Zur Kristallplastizit\"{a}t. I:   Tieftemperaturplastizit\"{a}t und Beckersche Formel. II:  Die dynamische Auffassung der Kristallplastizit\"{a}t. III: \"{U}ber die Mechanismus des Gleitvorganges. {\it Zeitschrift f\"{u}r Physik (Z. Phys.)} {\bf 89}: 605--659, 1934.

\bibitem[Polanyi, 1934]{4Polanyi34} \citep{4Polanyi34} M. Polanyi. \"{U}ber eine Art Gitterst\"{o}rung, die einen Kristall plastisch machen k\"{o}nnte. {\it Z. Phys.} {\bf 89}: 660--664, 1934.

\bibitem[Read, 1953]{4Read53} \citep{4Read53} W. T. Read, Jr. {\it Dislocations in Crystals}. McGraw-Hill, New York, 1953.

\bibitem[Soskin and Vasnetsov, 2001]{4Soskin01} \citep{4Soskin01} M. S. Soskin and M. V. Vasnetsov. Singular optics. {\it Prog. Opt.} {\bf 42}: 219--276, 2001.

\bibitem[Tanaka, 1995]{4Tanaka95} \citep{4Tanaka95} M. Tanaka. Dissapearance of waves in modulated train of surface gravity waves. \textit{Structure and Dynamics of Nonlinear Waves in Fluids}. In A. Mielke and K. Kirchg\"{a}ssner, editors, Proceedings of the IUTAM/ISIMM Symposium held in Hannover August 1994, volume \textbf{7} of \textit{Advanced Series in Nonlinear Dynamics}, World Scientific, Singapore, pp.~392--398, 1995.

\bibitem[Taylor, 1934]{4Taylor34} \citep{4Taylor34} G. I. Taylor. The mechanism of plastic deformation of crystals. Part I: Theoretical. Part II: Comparison with observations. {\it Proc. R. Soc. London} {\bf 145A}: 362--404, 1934.

\bibitem[Trulsen, 1998]{4Trulsen98} \citep{4Trulsen98} K. Trulsen. Crest pairing predicted by modulation theory. {\it J. Geophys. Res.} {\bf 103}(C2): 3143--3147, 1998.

\bibitem[Whitham, 1967]{4Whitham67} \citep{4Whitham67} G. B. Whitham. Nonlinear dispersion of water waves. {\it J. Fluid Mech.} {\bf 27}: 399--412, 1967.

\bibitem[Whitham, 1974]{4Whitham74} \citep{4Whitham74} G. B. Whitham. {\it Linear and Nonlinear Waves}. John Wiley \& Sons, New York, 1974.

\bibitem[Wright, 1979]{4Wright79} \citep{4Wright79} F. J. Wright. Wavefront dislocations and their analysis using catastrophe theory. In W. G\"{u}ttinger and H. Eikemeier, editors {\it Structural Stability in Physics}, pp.~141--156, 1979.
\end{thebibliography}

\small
\begin{thebibliography}{99}
\bibitem[Ablowitz and Herbst, 1990]{5Ablowitz90} \citep{5Ablowitz90} M. J. Ablowitz and B. M. Herbst. On homoclinic structure and numerically induced chaos for the nonlinear Schr\"{o}dinger equation. \textit{SIAM J. Appl. Math.} \textbf{50}(2): 339--351, 1990.

\addcontentsline{toc}{section}{\bibname}
\bibitem[Akhmediev et al., 1985]{5Akhmediev85} \citep{5Akhmediev85} N. N. Akhmediev, V. M. Eleonski\u{\i}, and N. E. Kulagin. Generation of periodic trains of picosecond pulses in an optical fiber: exact solutions. \textit{Sov. Phys. JETP} \textbf{62}(5): 894--899, 1985.

\bibitem[Akhmediev and Ankiewicz, 1997]{5Akhmediev97} \citep{5Akhmediev97} N. N. Akhmediev and A. Ankiewicz. \textit{Solitons---Nonlinear Pulses and Beams}, volume \textbf{5} of \textit{Optical and Quantum Electronic Series}. Chapman \& Hall, first edition, 1997.

\bibitem[Calini and Schober, 2002]{5Calini02} \citep{5Calini02} A. Calini and C. M. Schober. Homoclinic chaos increases the likelihood of rogue wave formation.  \textit{Phys. Lett. A} \textbf{298}: 335--349, 2002.

\bibitem[Matveev and Salle, 1991]{5Matveev91} \citep{5Matveev91} V. B. Matveev and M. A. Salle. \textit{Darboux Transformations and Solitons}. Springer-Verlag, 1991.

\bibitem[Rogers and Schief, 2002]{5Rogers02} \citep{5Rogers02} C. Rogers and W. K. Schief. \textit{B\"{a}cklund and Darboux Transformations: Geometry and Modern Applications in Soliton Theory}. Cambridge University Press, 2002.
\end{thebibliography}

\small
\begin{thebibliography}{99}
\bibitem[Chaplin, 1996]{6Chaplin96} \citep{6Chaplin96} J. R. Chaplin. On frequency-focusing unidirectional waves. \textit{Int. J. Offshore Polar Eng.} \textbf{6}(2): 131--137, 1996.  \addcontentsline{toc}{section}{\bibname}

\bibitem[Dingemans and Otta, 2001]{6Dingemans01} \citep{6Dingemans01} M. W. Dingemans and A. K. Otta. \textit{Nonlinear Modulation of Water Waves}, volume \textbf{7} of \textit{Advances in Coastal and Ocean Engineering}. World Scientific, Singapore, 2001.

\bibitem[Dysthe, 1979]{6Dysthe79} \citep{6Dysthe79} K. B. Dysthe. Note on a modification to the nonlinear Schr\"{o}dinger equation for application to deep water waves. \textit{Proc. Royal Soc. London A}, \textbf{369}: 105--114, 1979.

\bibitem[Havelock, 1918]{6Havelock18} \citep{6Havelock18} T. Havelock. Periodic, irrotational waves of finite height. \textit{Proc. Roy. Soc. London}, \textbf{95A}: 38--51, 1918.

\bibitem[Huijsmans et al., 2005]{6Huijsmans05} \citep{6Huijsmans05} R. H. M. Huijsmans, G. Klopman, N. Karjanto and Andonowati. Experiments on extreme wave generation using the Soliton on Finite Background. In M. Olagnon and M. Prevosto, editors, \textit{Rogue Waves 2004}. Proceedings of a workshop in Brest, France (October 20-22, 2004), 10 pp, 2005. \ arXiv:1110.5119 [physics.flu-dyn]

\bibitem[Janssen, 2006]{6Janssen06} \citep{6Janssen06} T. T. Janssen. {\it Nonlinear Surface Waves Over Topography}. PhD thesis, Department of Civil Engineering, Technical University of Delft, The Netherlands, 2006.

\bibitem[Klopman, 2005]{6Klopman05} \citep{6Klopman05} G. Klopman. Heuristic derivation of the spatial NLS equation for signalling problem in wave flumes. \textit{Concept Report}, unpublished, University of Twente, 2005.

\bibitem[Longuet-Higgins, 1974]{6LonguetHiggins74} \citep{6LonguetHiggins74} M. S. Longuet-Higgins. Breaking waves--in deep and shallow water. \textit{Proc. 10th Symp. on Naval Hydrodynamics}, pp. 597--605, US Government Printing Office, Cambridge, Massachusetts, 1974.

\bibitem[Mei, 1983]{6Mei83} \citep{6Mei83} C. C. Mei. \textit{The Applied Dynamics of Ocean Surface Waves}. John Wiley \& Sons, New York, 1983.

\bibitem[Michell, 1893]{6Michell1893} \citep{6Michell1893} J. H. Michell. The highest waves in water. \textit{Philos. Mag.}, Ser. 5, \textbf{36}: 430--437, 1893.

\bibitem[Onorato, 2004]{6Onorato04} \citep{6Onorato04} M. Onorato, personal communication, 2004.

\bibitem[Stokes, 1880]{6Stokes1880} \citep{6Stokes1880} G. Stokes. Supplement to a paper on the theory of oscillatory waves. \textit{Math. Phys. Papers}, \textbf{1}: 314--326, 1880.

\bibitem[Trulsen and Dysthe, 1996]{6Trulsen96} \citep{6Trulsen96} K. Trulsen and K. B. Dysthe. A modified nonlinear Schr\"{o}dinger equation for broader bandwith gravity waves on deep water. \textit{Wave Motion}, \textbf{24}: 281--289, 1996.
\end{thebibliography}

\begin{thebibliography}{99}
\bibitem[Dean and Dalrymple, 1991]{8Dean91} \citep{8Dean91} R. G. Dean and R. A. Dalrymple. \textit{Water Wave Mechanics for Engineers and Scientists}, volume \textbf{2} of \textit{Advanced Series of Ocean Engineering}. World Scientific, Singapore, 1991.

\bibitem[Hughes, 1993]{8Hughes93} \citep{8Hughes93} S. A. Hughes. \textit{Physical Models and Laboratory Techniques in Coastal Engineering}, volume \textbf{7} of \textit{Advanced Series of Ocean Engineering}. World Scientific, Singapore, 1993.

\bibitem[Sch\"{a}ffer, 1996]{8Schaffer96} \citep{8Schaffer96} H. A. Sch\"{a}ffer. Second-order wavemaker theory for irregular waves. \textit{Ocean Engng.} \textbf{23}(1):47--88, 1996.

\bibitem[Sch\"{a}ffer and Steenberg, 2003]{8Schaffer03} \citep{8Schaffer03} H. A. Sch\"{a}ffer and C. M. Steenberg. Second-order wavemaker theory for multidirectional waves. \textit{Ocean Engng.} \textbf{30}:1203--1231, 2003.

\bibitem[Svendsen, 1985]{8Svendsen85} \citep{8Svendsen85} I. A. Svendsen. Physical modelling of water waves. In R. A. Dalrymple, editor, \textit{Physical Modelling in Ocean Engineering}, A. A. Balkema, Rotterdam, 1985.
\end{thebibliography}

\newpage
{\small
\begin{thebibliography}{99}
\bibitem[Ablowitz et al., 1974]{Ablowitz74} \citep{Ablowitz74} M. J. Ablowitz, D. J. Kaup, A. C. Newell, and H. Segur. The inverse scattering transform-Fourier analysis for nonlinear problems. \textit{Studies in Applied Mathematics} \textbf{53}(4): 249--315, 1974. \addcontentsline{toc}{chapter}{\bibname}

\bibitem[Ablowitz and Herbst, 1990]{Ablowitz90} \citep{Ablowitz90} M. J. Ablowitz and B. M. Herbst. On homoclinic structure and numerically induced chaos for the nonlinear Schr\"{o}dinger equation. \textit{(Society of Industrial and Applied Mathematics) SIAM Journal on Applied Mathematics} \textbf{50}(2): 339--351, 1990.

\bibitem[Abramowitz and Stegun, 1972]{Abramowitz72} \citep{Abramowitz72} M. Abramowitz and I. A. Stegun, (Editors). \textit{Handbook of Mathematical Functions with Formulas, Graphs, and Mathematical Tables}, ninth printing. Dover, New York, 1972.

\bibitem[Agrawal, 1995]{Agrawal95} \citep{Agrawal95} G. P. Agrawal. \textit{Nonlinear Fiber Optics}, second edition. Academic Press, San Diego, California, 1995. 

\bibitem[Akhmediev et al., 1985]{Akhmediev85} \citep{Akhmediev85} N. N. Akhmediev, V. M. Eleonski\u{\i}, and N. E. Kulagin. Generation of periodic trains of picosecond pulses in an optical fiber: exact solutions. \textit{Soviet Physics JETP (Journal of Experimental and Theoretical Physics)} \textbf{62}(5): 894--899, 1985. Translated from Russian, the original version appeared in \textit{Журнал Экспериментальной и Теоретической Физики (ЖЭТФ), Zhurnal \'Eksperimental'no\u{\i} i Teoretichesko\u{\i} Fiziki (ZhETF)} \textbf{89}: 1542--1551, 1985.

\bibitem[Akhmediev and Korneev, 1986]{Akhmediev86} \citep{Akhmediev86} N. N. Akhmediev and V. I. Korneev. Modulation instability and periodic solutions of the nonlinear Schr\"{o}dinger equation. \textit{Theoretical and Mathematical Physics} \textbf{69}(2): 1089--1092, 1986. Translated from Russian, the original version appeared in \textit{Теоретическая и математическая физика, Teoreticheskaya i Matematicheskaya Fizika} \textbf{69}(2): 189--194, 1986.

\bibitem[Akhmediev et al., 1987]{Akhmediev87} \citep{Akhmediev87} N. N. Akhmediev, V. M. Eleonski\u{\i}, and N. E. Kulagin. Exact first-order solutions of the nonlinear Schr\"{o}dinger equation. \textit{Theoretical and Mathematical Physics} \textbf{72}(2): 809--818, 1987. Translated from Russian, the original version appeared in \textit{Теоретическая и математическая физика, Teoreticheskaya i Matematicheskaya Fizika} \textbf{72}(2): 183--196, 1987.

\bibitem[Akhmediev and Ankiewicz, 1997]{Akhmediev97} \citep{Akhmediev97} N. N. Akhmediev and A. Ankiewicz. \textit{Solitons---Nonlinear Pulses and Beams}, volume \textbf{5} of \textit{Optical and Quantum Electronic Series}. Chapman \& Hall, London, UK, 1997.

\bibitem[Andonowati and Van Groesen, 2003]{Andonowati03} \citep{Andonowati03} Andonowati and E. van Groesen. Optical pulse deformation in second order nonlinear media. \textit{Journal of Nonlinear Optical Physics and Materials} \textbf{12}(2): 221--234, 2003.

\bibitem[Andonowati et al., 2007]{Andonowati07} \citep{Andonowati07} Andonowati, N. Karjanto, and E. van Groesen. Extreme wave phenomena in down-stream running modulated waves. \textit{Applied Mathematical Modelling} \textbf{31}(7): 1425--1443, 2007. \ arXiv:1710.10804 [physics.flu-dyn] 

\bibitem[Balistreri et al., 2000]{Balistreri00} \citep{Balistreri00} M. L. M. Balistreri, J. P. Korterik, L. Kuipers, and N. F. van Hulst. Local observation of phase singularities in optical fields in waveguide structures. \textit{Physical Review Letters} \textbf{85}(2): 294--297, 2000.

\bibitem[Basistiy et al., 1995]{Basistiy95} \citep{Basistiy95} I. V. Basistiy, M. S. Soskin, and M. V. Vasnetsov. Optical wavefront dislocations and their properties. {\it Optics Communications} \textbf{119}(5--6): 604--612, 1995.

\bibitem[Benjamin, 1967]{Benjamin67} \citep{Benjamin67} T. B. Benjamin. Instability of periodic wave trains in nonlinear dispersive systems. \textit{Proceedings of the Royal Society of London. Series A. Mathematical and Physical Sciences} \textbf{229}(1456): 59--75, 1967.

\bibitem[Benjamin and Feir, 1967]{BenjaminFeir67} \citep{BenjaminFeir67} T. B. Benjamin and J. E. Feir. The disintegration of wave trains in deep water \ Part 1. Theory. \textit{Journal of Fluid Mechanics} \textbf{27}(3): 417--430, 1967.

\bibitem[Benney and Newell, 1967]{Benney67} \citep{Benney67} D. J. Benney and A. C. Newell. The propagation of nonlinear wave envelopes. \textit{Journal of Mathematics and Physics} \textbf{46}(1--4): 133--139, 1967.

\bibitem[Berry et al., 1980]{Berry80} \citep{Berry80} M. V. Berry, R. G. Chambers, M. D. Large, C. Upstill, and J. C. Walmsley. Wavefront dislocations in the Aharonov-Bohm effect and its water wave analogue. \textit{European Journal of Physics} \textbf{1}(3): 154--162, 1980.

\bibitem[Berry, 1981]{Berry81} \citep{Berry81} M. V. Berry. Singularities in waves and rays. In R. Balian, M. Kl\'{e}man, and J.-P. Poirier (Editors), \textit{Les Houches 1980, Session XXXV, Physics of Defects}, North-Holland, Amsterdam, the Netherlands, pp.~453--459, 1981.

\bibitem[Berry, 1998]{Berry98} \citep{Berry98} M. V. Berry. Much ado about nothing: optical dislocation lines (phase singularities, zeros, vortices...). In M. S. Soskin (Editor), \textit{Proceedings of International Conference on Singular Optics}, The International Society of Optical Engineering (SPIE, Society of Photo-Optical Instrumentation Engineers), Bellingham, Washington, pp.~1--10, 1998.

\bibitem[Berry and Dennis, 2000]{Berry00} \citep{Berry00} M. V. Berry and M. R. Dennis. Phase singularities in isotropic random waves. \textit{Proceedings of the Royal Society of London. Series A. Mathematical, Physical and Engineering Sciences} \textbf{456}(2001): 2059--2079, 2000.

\bibitem[Berry and Dennis, 2001]{Berry01} \citep{Berry01} M. V. Berry and M. R. Dennis. Knotted and linked phase singularities in monochromatic waves. \textit{Proceedings of the Royal Society of London. Series A. Mathematical, Physical and Engineering Sciences} \textbf{457}(2013): 2251--2263, 2001.

\bibitem[Berry et al., 2004]{Berry04} \citep{Berry04} M. V. Berry, M. R. Dennis, and M. S. Soskin. The plurality of optical singularities. \textit{Journal of Optics A: Pure and Applied Optics} {\bf 6}(5): S155--S156, 2004.

\bibitem[Bespalov and Talanov, 1966]{Bespalov66} \citep{Bespalov66} V. I. Bespalov and V. I. Talanov. On the filament structure of a light beam in nonlinear liquids. \textit{JETP Letters} \textbf{3}: 307--312, 1966. Translated from Russian, the original version appeared in \textit{Письма в Журнал экспериментальной и теоретической физики (Письма в ЖЭТФ), Pis'ma v Zhurnal \'Eksperimental'no\u{\i} i Teoretichesko\u{\i} Fiziki (Pis'ma v ZhETF)} \textbf{3}: 471--476, 1966.

\bibitem[Boccotti, 2000]{Boccotti00} \citep{Boccotti00} P. Boccotti. \textit{Wave Mechanics for Ocean Engineering}, Elsevier Science, Amsterdam, the Netherlands, 2000.

\bibitem[Boyd and Chen, 2001]{Boyd01} \citep{Boyd01} J. P. Boyd and G.-Y. Chen. Weakly nonlinear wavepackets in the Korteweg-de Vries equation: the KdV/NLS connection. \textit{Mathematics and Computers in Simulation} \textbf{55}: 317--328, 2001.

\bibitem[Bridges and Dias, 2007]{Bridges07} \citep{Bridges07} T. J. Bridges and F. Dias. Enhancement of the Benjamin-Feir instability with dissipation. \textit{Physics of Fluid} \textbf{19}(10): 104104, 2007.

\bibitem[Bullough and Caudrey, 1995]{Bullough95} \citep{Bullough95} R. K. Bulllough and P. J. Caudrey. Solitons and the Korteweg-de Vries equation: Integrable systems in 1834--1995. \textit{Acta Applicandae Mathematicae} \textbf{39}: 193--228, 1995.

\bibitem[Cahyono, 2002]{Cahyono02} \citep{Cahyono02} E. Cahyono. \textit{Analytical Wave Codes for Predicting Surface Waves in a Laboratory Basin}. Ph.D. thesis, Department of Applied Mathematics, University of Twente, the Netherlands, 2002.

\bibitem[Calini and Schober, 2002]{Calini02} \citep{Calini02} A. Calini and C. M. Schober. Homoclinic chaos increases the likelihood of rogue wave formation.  \textit{Physics Letters A} \textbf{298}(5--6): 335--349, 2002.

\bibitem[Chaplin, 1996]{Chaplin} \citep{Chaplin} J. R. Chaplin. On frequency-focusing unidirectional waves. \textit{International Journal of Offshore and Polar Engineering} \textbf{6}(02): 131--137, 1996.

\bibitem[Chu and Mei, 1970]{Chu70} \citep{Chu70} V. H. Chu and C. C. Mei. On slowly-varying Stokes waves. \textit{Journal of Fluid Mechanics} \textbf{41}(4): 873--887, 1970.

\bibitem[Chu and Mei, 1971]{Chu71} \citep{Chu71} V. H. Chu and C. C. Mei. The nonlinear evolution of Stokes waves in deep water. \textit{Journal of Fluid Mechanics} \textbf{47}(2): 337--351, 1971.

\bibitem[Coullet et al., 1989]{Coullet89} \citep{Coullet89} P. Coullet, L. Gil, and F. Rocca. Optical vortices. \textit{Optics Communications} \textbf{73}(5): 403--408, 1989.

\bibitem[Craik, 2004]{Craik04} \citep{Craik04} A. D. D. Craik. The origins of water wave theory. \textit{Annual Review of Fluid Mechanics} \textbf{36}: 1--28, 2004.

\bibitem[Craik, 2005]{Craik05} \citep{Craik05} A. D. D. Craik. George Gabriel Stokes on water wave theory. \textit{Annual Review of Fluid Mechanics} \textbf{37}: 23--42, 2005.

\bibitem[Dankert et al., 2003]{Dankert03} \citep{Dankert03} H. Dankert, J. Horstmann, S. Lehner, and W. Rosenthal. Detection of wave groups in SAR (synthetic-aperture radar) images and radar image sequences. \textit{(Institute of Electrical and Electronics Engineers) IEEE Transactions on Geoscience and Remote Sensing} \textbf{41}(6): 1437--1446, 2003.

\bibitem[Davey, 1972]{Davey72} \citep{Davey72} A. Davey. The propagation of a weak nonlinear wave. \textit{Journal of Fluid Mechanics} \textbf{53}(4): 769--781, 1972.

\bibitem[Dean, 1990]{Dean90} \citep{Dean90} R. G. Dean. Freak waves: a possible explanation. In A. T{\o}rum and O. T. Gudmestad (Editors), \textit{Water Wave Kinematics}, pp.~609--612, Kluwer Academic Publishers, Amsterdam, the Netherlands, 1990.

\bibitem[Dean and Dalrymple, 1991]{Dean} \citep{Dean} R. G. Dean and R. A. Dalrymple. \textit{Water Wave Mechanics for Engineers and Scientists}, volume \textbf{2} of \textit{Advanced Series of Ocean Engineering}. World Scientific, Singapore, 1991.

\bibitem[Debnath, 1994]{Debnath94} \citep{Debnath94} L. Debnath. \textit{Nonlinear Water Waves}. Academic Press, San Diego, California, 1994.

\bibitem[Dennis, 2001]{Dennis01} \citep{Dennis01} M. R. Dennis, \textit{Topological Singularities in Wave Fields}. PhD thesis, University of Bristol, H. H. Wills Physics Laboratory, UK, November 2001.

\bibitem[Dieter, 1988]{Dieter88} \citep{Dieter88} G. E. Dieter. \textit{Mechanical Metallurgy}. SI Metric edition, McGraw-Hill, London, UK, 1988.

\bibitem[Dingemans, 1997]{Dingemans97} \citep{Dingemans97} M. W. Dingemans. \textit{Water Wave Propagation over Uneven Bottoms}. World Scientific, Singapore, 1997.

\bibitem[Dingemans and Otta, 2001]{Dingemans01} \citep{Dingemans01} M. W. Dingemans and A. K. Otta. \textit{Nonlinear Modulation of Water Waves}. In P. L.-F. Liu (Editor), volume \textbf{7} of \textit{Advances in Coastal and Ocean Engineering}, World Scientific, Singapore, 2001.

\bibitem[Djordjevi\'{c} and Redekopp, 1978]{Djordjevic78} \citep{Djordjevic78} V. D. Djordjevi\'{c} and L. G. Redekopp. On the development of packets of surface gravity waves moving over an uneven bottom. \textit{Journal of Applied Mathematics and Physics (Zeitschrift f\"ur Angewandte Mathematik und Physik, ZAMP)} \textbf{29}(6): 950--962, 1978.

\bibitem[Dodd et al., 1982]{Dodd82} \citep{Dodd82} R. K. Dodd, J. C. Eilbeck, J. D. Gibbon, and H. C. Morris. \textit{Solitons and Nonlinear Wave Equations}. Academic Press, London, UK, 1982.


\bibitem[Draper, 1966]{Draper66} \citep{Draper66} L. Draper. `Freak' ocean waves. \textit{Weather} \textbf{21}(1): 2--4, 1966.

\bibitem[Drazin and Johnson, 1989]{Drazin89} \citep{Drazin89} P. G. Drazin and R. S. Johnson. \textit{Solitons: an introduction}. Cambridge University Press, Cambridge, UK, 1989.

\bibitem[Dysthe, 1979]{Dysthe} \citep{Dysthe} K. B. Dysthe. Note on a modification to the nonlinear Schr\"{o}dinger equation for application to deep water waves. \textit{Proceedings of the Royal Society of London. Series A. Mathematical and Physical Sciences}, \textbf{369}(1736): 105--114, 1979.

\bibitem[Dysthe and Trulsen, 1999]{Dysthe99} \citep{Dysthe99} K. B. Dysthe and K. Trulsen. Note on breather type solutions of the NLS as models for freak-waves. \textit{Physica Scripta} \textbf{T82}: 48--52, 1999.

\bibitem[Dysthe, 2000]{Dysthe00} \citep{Dysthe00} K. B. Dysthe. Modelling a ``rogue wave''--speculations or a realistic possibility? In M. Olagnon and G. A. Athanassoulis (Editors), \textit{Proceedings of the Rogue Waves 2000}, Ifremeer, Brest, France, November 2000.


\bibitem[Fedele and Arena, 2005]{Fedele05} \citep{Fedele05} F. Fedele and F. Arena. Weakly nonlinear statistics of high random waves. \textit{Physics of Fluids} \textbf{17}(2): 026601, 2005.

\bibitem[Fermi et al., 1955]{Fermi55} \citep{Fermi55} E. Fermi, J. Pasta, and S. M. Ulam. Studies of nonlinear problem. Los Alamos Scientific Laboratory Report No. LA-1940, 1955. Reprinted in A. C. Newell (Editor), \textit{Nonlinear Wave Motion}, volume \textbf{15} of AMS Lectures in Applied Mathematics, pp.~143--156, 1974.

\bibitem[Fornberg and Whitham, 1978]{Fornberg78} \citep{Fornberg78} V. Fornberg and G. B. Whitham. A numerical and theoretical study of certain nonlinear wave phenomena. \textit{Philosophical Transactions of the Royal Society of London. Series A, Mathematical and Physical Sciences} {\bf 289}(1361): 373--404, 1978.

\bibitem[Gel'fand and Fomin, 1963]{Gelfand63} \citep{Gelfand63} I. M. Gel'fand and S. V. Fomin. \textit{Calculus of Variations}. Prentice Hall, Englewood Cliffs, New Jersey, 1963.

\bibitem[Gibson et al., 2005]{Gibson05} \citep{Gibson05} R. Gibson, C. Swan, P. Tromans, L. Vanderscuren. Wave crest statistics calculated using a fully nonlinear spectral response surface method. In M. Olagnon and M. Prevosto (Editors), \textit{Rogue Waves 2004}. Proceedings of a workshop in Brest, France (October 20--22, 2004), 10~pp., 2005.

\bibitem[Ginzburg and Landau, 1950]{Ginzburg50} \citep{Ginzburg50} V. L. Ginzburg and L. D. Landau. On the theory of superconductivity. \textit{Журнал Экспериментальной и Теоретической Физики (ЖЭТФ), Zhurnal \'Eksperimental'no\u{\i} i Teoretichesko\u{\i} Fiziki (ZhETF), Journal of Experimental and Theoretical Physics (JETP)} \textbf{20}: 1064--1082, 1950. English translation in L. D. Landau. \textit{Men of Physics}, edited by D. ter Haar, volume \textbf{1}, Pergamon Press, Oxford, pp.~138--167, 1965. Also in Chapter 4 of \textit{On Superconductivity and Superfluidity--A Scientific Autobiography}, edited by  V. L. Ginzburg, Springer, Berlin, Heildelberg, pp.~113--137, 2009.

\bibitem[Goldman and Sirovich, 1994]{Goldman94} \citep{Goldman94} D. Goldman and L. Sirovich. The one dimensional complex Ginzburg-Landau equation in the low dissipation limit. \textit{Nonlinearity}, \textbf{7}(2): 417--439, 1994.

\bibitem[Grimshaw et al., 2001]{Grimshaw01} \citep{Grimshaw01} R. Grimshaw, D. Pelinovsky, E. Pelinovsky, and T. Talipova. Wave group dynamics in weakly nonlinear long-wave models. \textit{Physica D: Nonlinear Phenomena} \textbf{159}(1--2): 35--57, 2001.

\bibitem[Grimshaw and Saut, 2005]{Grimshaw05} \citep{Grimshaw05} R. Grimshaw and J.-C. Saut, organizers. \textit{Rogue Waves 2005}. Proceedings of a workshop in Edinburgh, United Kingdom (December 12--15, 2005), 2005. Available online at \url{http://www.icms.org.uk/meetings/2005/roguewaves/index.html}. Last accessed 6 November 2006. 

\bibitem[Hasegawa and Kodama, 1995]{Hasegawa95} \citep{Hasegawa95} A. Hasegawa and Y. Kodama. \textit{Solitons in Optical Communications}, volume \textbf{7} of \textit{Oxford Series in Optical and Imaging Sciences}. Clarendon Press, Oxford, UK, 1995.


\bibitem[Hasegawa and Tappert, 1973a]{Hasegawa73a} \citep{Hasegawa73a} A. Hasegawa and F. D. Tappert. Transmission of stationary nonlinear optical pulses in dispersive dielectric fibers. I. Anomalous dispersion. \textit{Applied Physics Letters} \textbf{23}(3): 142--144, 1973.

\bibitem[Hasegawa and Tappert, 1973b]{Hasegawa73b} \citep{Hasegawa73b} A. Hasegawa and F. D. Tappert. Transmission of stationary nonlinear optical pulses in dispersive dielectric fibers. II. Normal dispersion. \textit{Applied Physics Letters} \textbf{23}(4): 171--172, 1973.

\bibitem[Hasimoto and Ono, 1972]{Hasimoto72} \citep{Hasimoto72} H. Hasimoto and H. Ono. Nonlinear modulation of gravity waves. \textit{Journal of the Physical Society of Japan} \textbf{33}(3): 805--811, 1972.

\bibitem[Havelock, 1918]{Havelock18} \citep{Havelock18} T. Havelock. Periodic, irrotational waves of finite height. \textit{Proceedings of the Royal Society of London. Series A, Containing Papers of a Mathematical and Physical Character}, \textbf{95}(665): 38--51, 1918.

\bibitem[Haver, 2005]{Haver05} \citep{Haver05} S. Haver. A possible freak wave event measured at the Draupner jacket January~1, 1995. In M. Olagnon and M. Prevosto (Editors), \textit{Rogue Waves 2004}. Proceedings of a workshop in Brest, France (October 20--22, 2004), 8~pp., 2005.

\bibitem[Heller, 2005]{Heller05} \citep{Heller05} E. Heller. Freak waves: just bad luck, or avoidable? \textit{Europhysics News} \textbf{36}(5): 159--162, 2005.

\bibitem[Henderson et al., 1999]{Henderson99} \citep{Henderson99} K. L. Henderson, D. H. Peregrine, and J. W. Dold. Unsteady water wave modulations: fully nonlinear solutions and comparison with the nonlinear Schr\"{o}dinger equation. \textit{Wave Motion} \textbf{29}(4): 341--461, 1999.

\bibitem[Honeycombe, 1984]{Honeycombe84} \citep{Honeycombe84} R. W. K. Honeycombe. {\it The Plastic Deformation of Metals}, second edition. Edward Arnold, London, UK, 1984.

\bibitem[Hughes, 1993]{Hughes} \citep{Hughes} S. A. Hughes. \textit{Physical Models and Laboratory Techniques in Coastal Engineering}, volume \textbf{7} of \textit{Advanced Series of Ocean Engineering}. World Scientific, Singapore, 1993.

\bibitem[Huijsmans et al., 2005]{Huijsmans} \citep{Huijsmans} R. H. M. Huijsmans, G. Klopman, N. Karjanto, and Andonowati. Experiments on extreme wave generation using the Soliton on Finite Background. In M. Olagnon and M. Prevosto (Editors), \textit{Rogue Waves 2004}. Proceedings of a workshop in Brest, France (October 20-22, 2004), 10~pp., 2005. \ arXiv:1110.5119 [physics.flu-dyn]

\bibitem[Hull and Bacon, 2001]{Hull01} \citep{Hull01} D. Hull and D. J. Bacon. {\it Introduction to Dislocations}, fourth edition. Butterworth Heinemann, Oxford, UK, 2001.

\bibitem[Hunt, 2005]{Hunt05} \citep{Hunt05} J. C. R. Hunt. Nonlinear and wave theory contributions of T. Brooke Benjamin (1929-1995). \textit{Annual Review of Fluid Mechanics} \textbf{38}: 1--25, 2005.

\bibitem[Ichikawa et al., 1972]{Ichikawa72} \citep{Ichikawa72} Y. H. Ichikawa, T. Imamura, and T. Taniuti. Nonlinear wave modulation in collisionless plasmas. \textit{Journal of the Physical Society of Japan} \textbf{33}(1): 189--197, 1972.

\bibitem[Infeld and Rowlands, 1990]{Infeld90} \citep{Infeld90} E. Infeld and G. Rowlands. \textit{Nonlinear Waves, Solitons, and Chaos}, Cambridge University Press, Cambridge, UK, 1990.

\bibitem[Janssen, 1981]{Janssen81} \citep{Janssen81} P. A. E. M. Janssen. Modulational instability and the Fermi-Pasta-Ulam recurrence. \textit{The Physics of Fluids} \textbf{24}(1): 23--26, 1981.

\bibitem[Janssen, 2003]{Janssen03} \citep{Janssen03} P. A. E. M. Janssen. Nonlinear four-wave interactions and freak waves. \textit{Journal of Physical Oceanography} \textbf{33}(4): 863--884, 2003.

\bibitem[Janssen, 2006]{Janssen06} \citep{Janssen06} T. T. Janssen. {\it Nonlinear Surface Waves Over Topography}. PhD thesis, Department of Civil Engineering, Technical University of Delft, the Netherlands, 2006.

\bibitem[Jeffrey and Kawahara, 1982]{Jeffrey82} \citep{Jeffrey82} A. Jeffrey and T. Kawahara. \textit{Asymptotic Methods in Nonlinear Wave Theory}. Applicable Mathematics Series, Pitman, Boston, Massachusetts, 1982.

\bibitem[Johnson, 1976]{Johnson76} \citep{Johnson76} R. S. Johnson. On the modulation of water waves on shear flows. \textit{Proceedings of the Royal Society of London. Series A, Mathematical and Physical Sciences} \textbf{347}(1651): 537--546, 1976.

\bibitem[Johnson, 1997]{Johnson97} \citep{Johnson97} R. S. Johnson. \textit{A Modern Introduction to the Mathematical Theory of Water Waves}, volume \textbf{18} of \textit{Cambridge Texts in Applied Mathematics}, Cambridge University Press, Cambridge, UK, 1997.

\bibitem[Kadomtsev and Karpman, 1971]{Kadomtsev71} \citep{Kadomtsev71} B. B. Kadomtsev and V. I. Karpman. Nonlinear waves. \textit{Soviet Physics--Uspekhi (Advances in Physical Sciences)}, \textbf{14}(1): 40--60, 1971. Translated from Russian, the original version appeared in \textit{Успехи физических наук, Uspekhi Fizicheskikh Nauk} \textbf{103}: 193--232, 1971.

\bibitem[Karpman and Krushkal', 1969]{Karpman69} \citep{Karpman69} V. I. Karpman and E. M. Krushkal'. Modulated waves in nonlinear dispersive media. \textit{Soviet Physics JETP} {\bf 28}(2): 277--281, 1969. Translated from Russian, the original version appeared in \textit{Журнал Экспериментальной и Теоретической Физики (ЖЭТФ), Zhurnal \'Eksperimental'no\u{\i} i Teoretichesko\u{\i} Fiziki (ZhETF)} \textbf{55}(2): 530--538, 1969.

\bibitem[Karpman, 1967]{Karpman67} \citep{Karpman67}  V. I. Karpman. Self-modulation of nonlinear plane waves in dispersive media. \textit{JETP Letters} \textbf{6}: 277--279, 1967. Translated from Russian, the original version appeared in \textit{Письма в Журнал экспериментальной и теоретической физики (Письма в ЖЭТФ), Pis'ma v Zhurnal \'Eksperimental'no\u{\i} i Teoretichesko\u{\i} Fiziki (Pis'ma v ZhETF)} \textbf{6}(8): 829--832, 1967.

\bibitem[Karpman, 1975]{Karpman75} \citep{Karpman75} V. I. Karpman. \textit{Nonlinear Waves in Dispersive Media}. Pergamon Press, Oxford, UK, 1975.

\bibitem[Kelley, 1965]{Kelley65} \citep{Kelley65} P. L. Kelley. Self-focusing of optical beams. \textit{Physical Review Letters} \textbf{15}(26): 1005--1008, 1965.

\bibitem[Kevorkian, 1961]{Kevorkian61} \citep{Kevorkian61} J. K. Kevorkian. \textit{The Uniformly Valid Asymptotic Approximations to the Solutions of Certain Nonlinear Ordinary Differential Equations}. PhD thesis, California Institute of Technology, California, 1961.

\bibitem[Kharif et al., 2000]{Kharif00} \citep{Kharif00} C. Kharif, E. Pelinovsky, and T. Talipova. Formation de vagues g\'{e}antes en eau peu profonde. (Freak wave generation in shallow water.) \textit{Comptes Rendus de l'Acad\'emie des Sciences Paris--S\'{e}rie  II b, M\'{e}canique des fluides (Proceedings of the French Academy of Sciences Paris--Section II b, Fluid mechanics)}, \textbf{328}(11): 801--807, 2000.

\bibitem[Kharif and Pelinovsky, 2003]{Kharif03} \citep{Kharif03} C. Kharif and E. Pelinovsky. Physical mechanisms of the rogue wave phenomenon. \textit{European Journal of Mechanics. B: Fluids} \textbf{22}(6): 603--634, 2003.

\bibitem[Klopman, 2005]{Klopman} \citep{Klopman} G. Klopman. Heuristic derivation of the spatial NLS equation for signalling problem in wave flumes. \textit{Concept Report}, unpublished, University of Twente, 2005.

\bibitem[Kokorina and Pelinovsky, 2002]{Kokorina02} \citep{Kokorina02} A. Kokorina and E. Pelinovsky. The applicability of the Korteweg-de Vries equation for description of the statistics of freak waves. \textit{Journal of the Korean Society of Coastal and Ocean Engineers} \textbf{14}(4): 308--318, 2002.

\bibitem[Kivshar and Luther-Davies, 1998]{Kivshar98} \citep{Kivshar98} Y. S. Kivshar and B. Luther-Davies. Dark optical solitons: physics and applications. \textit{Physics Reports} \textbf{298}(2--3): 81--197, 1998.

\bibitem[Kr\'{a}sa, 1981]{Krasa81} \citep{Krasa81} J. Kr\'{a}sa. Dislocation in turbulent ionisation waves. \textit{Journal of Physics D: Applied Physics} \textbf{14}: 1241--1246, 1981.

\bibitem[Lake et al., 1977]{Lake77} \citep{Lake77} B. M. Lake, H. C. Yuen, H. Rundgaldier, and W. E. Ferguson. Nonlinear deep water waves: theory and experiment. Part 2. Evolution of a continuous wave train. \textit{Journal of Fluid Mechanics} \textbf{83}(1): 49--74, 1977.

\bibitem[Lamb, 1980]{Lamb80} \citep{Lamb80} G. L. Lamb, Jr. \textit{Elements of Soliton Theory}, John Wiley \& Sons, New York, 1980.

\bibitem[Lighthill, 1965]{Lighthill65} \citep{Lighthill65} M. J. Lighthill. Contributions to the theory of waves in nonlinear dispersive systems. \textit{Institute of Mathematics and Its Applications (IMA) Journal of Applied Mathematics} \textbf{1}(3): 269--306, 1965.

\bibitem[Lighthill, 1967]{Lighthill67} \citep{Lighthill67} M. J. Lighthill. Some special cases treated by the Whitham theory. \textit{Proceedings of the Royal Society of London. Series A, Mathematical and Physical Sciences} \textbf{299}(1456): 28--53, 1967.

\bibitem[Liu and Pinho, 2004]{Liu04} \citep{Liu04} P. C. Liu and U. F. Pinho. Freak waves--more frequent than rare! \textit{Annales Geophysicae} \textbf{22}(5): 1839--1842, 2004.

\bibitem[Longuet-Higgins, 1974]{LonguetHiggins} \citep{LonguetHiggins} M. S. Longuet-Higgins. Breaking waves in deep and shallow water. \textit{Proceedings of the 10th Symposium on Naval Hydrodynamics}, pp.~597--605, MIT, Cambridge, Massachusetts, 1974.

\bibitem[L'vov, 1994]{Lvov94} \citep{Lvov94} V. S. L'vov. \textit{Wave Turbulence Under Parametric Excitation--Applications to Magnets}. In V. E. Zakharov (Series Editor), \textit{Springer Series in Nonlinear Dynamics}.  Springer-Verlag, Berlin, 1994.

\bibitem[Ma, 1979]{Ma79} \citep{Ma79} Y.-C. Ma. The perturbed plane-wave solutions of the cubic Schr\"{o}dinger equation. \textit{Studies in Applied Mathematics} \textbf{60}(1): 43--58, 1979.

\bibitem[Madelung, 1927]{Madelung27} \citep{Madelung27} E. Madelung. Quantentheorie in Hydrodynamischer Form (Quantum theory in hydrodynamic form). \textit{Zeitschrift f\"{u}r Physik (Journal of Physics)} \textbf{40}(3--4): 322--326, 1927.


\bibitem[Matveev and Salle, 1991]{Matveev91} \citep{Matveev91} V. B. Matveev and M. A. Salle. \textit{Darboux Transformations and Solitons}. Springer-Verlag, Berlin Heildelberg, Germany, 1991.

\bibitem[Mei, 1983]{Mei83} \citep{Mei83} C. C. Mei. \textit{The Applied Dynamics of Ocean Surface Waves}. John Wiley \& Sons, New York, 1983.

\bibitem[Michell, 1893]{Michell1893} \citep{Michell1893} J. H. Michell M. A. XLIV. The highest waves in water. \textit{The London, Edinburgh, and Dublin Philosophical Magazine and Journal of Science} \textbf{36}(222): 430--437, 1893.

\bibitem[Miles, 1981]{Miles81} \citep{Miles81} J. W. Miles. The Korteweg-de Vries equation: a historical essay. \textit{Journal of Fluid Mechanics} \textbf{106}: 131--147, 1981.

\bibitem[Nabarro, 1979-2004]{Nabarro7904} \citep{Nabarro7904} F. R. N. Nabarro (Editor), {\it Dislocations in Solids} {\bf 1--12}, 1979--2004.

\bibitem[Nayfeh, 1973]{Nayfeh73} \citep{Nayfeh73} A. H. Nayfeh. \textit{Perturbation Methods}, Wiley-Interscience, New York, 1973.

\bibitem[Newell, 1974]{Newell74} \citep{Newell74} A. C. Newell. Envelope equations. In A. C. Newell (Editor), \textit{Nonlinear Wave Motion}, volume \textbf{15} of \textit{Lectures in Applied Mathematics}.  American Mathematical Society, Providence, Rhode Island, 1974.

\bibitem[Newell, 1985]{Newell85} \citep{Newell85} A. C. Newell. \textit{Solitons in Mathematical Physics}, volume \textbf{48} of \textit{Regional Conference Series in Applied Mathematics}. Society for Industrial and Applied Mathematics, Philadelphia, Pennsylvania, 1985.

\bibitem[Noether, 1918]{Noether18} \citep{Noether18} E. Noether. Invariante Variationsprobleme. (Invariant Variation Problems.) \textit{Nachrichten von der Gessellschaft der Wissenschaften zu G\"{o}ttingen--Mathematisch-Physikalische Klasse}, 235--257, 1918. English translation: M. A. Tavel. \textit{Transport Theory and Statistical Physics} \textbf{1}(3): 186--207, 1971. \ arXiv:physics/0503066 [physics.hist-ph]

\bibitem[Nye and Berry, 1974]{NyeBerry74} \citep{NyeBerry74} J. F. Nye and M. V. Berry. Dislocation in wave trains. \textit{Proceedings of the Royal Society of London. Series A, Mathematical and Physical Sciences} \textbf{336}(1605): 165--190, 1974.

\bibitem[Nye, 1981]{Nye81} \citep{Nye81} J. F. Nye. The motion and structure of dislocations in wavefronts. \textit{Proceedings of the Royal Society of London. Series A, Mathematical and Physical Sciences} \textbf{378}(1773): 219--239, 1981.

\bibitem[Nye, 1997]{Nye97} \citep{Nye97} J. F. Nye. Line singularities in wave fields. \textit{Philosophical Transactions: Mathematical, Physical and Engineering Sciences} \textbf{355}(1731): 2065--2069, 1997.

\bibitem[Nye, 1999]{Nye99} \citep{Nye99} J. F. Nye. \textit{Natural Focussing and Fine Structure of Light: Caustics and Wave Dislocations}. Insitute of Physics, Bristol, UK, 1999.


\bibitem[Olagnon and Prevosto, 2005]{Olagnon05} \citep{Olagnon05} M. Olagnon and M. Prevosto (Editors), \textit{Rogue Waves 2004}. Proceedings of a workshop in Brest, France (October 20--22, 2004), 308~pp., 2005.

\bibitem[Onorato et al., 2000]{Onorato00} \citep{Onorato00} M. Onorato, A. Osborne, M. Serio, and T. Damiani. Occurrence of freak waves from envelope equations in random ocean wave simulations. In M. Olagnon (Editor), \textit{Proceedings Rogue Waves 2000}, Brest, France, 2000.

\bibitem[Onorato et al., 2001]{Onorato01} \citep{Onorato01} M. Onorato, A. R. Osborne, M. Serio, and S. Bertone. Freak waves in random oceanic sea states. \textit{Physical Review Letters} \textbf{86}(25): 5831--5834, 2001.

\bibitem[Onorato, 2004]{Onorato04} \citep{Onorato04} M. Onorato, personal communication, 2004.


\bibitem[Orowan, 1934a]{Orowan34a} \citep{Orowan34a} E. Orowan. Zur Kristallplastizit\"{a}t. I: Tieftemperaturplastizit\"{a}t und Beckersche Formel. (On crystal plasticity. Part I: Low temperature plasticity and Becker's formula.) \textit{Zeitschrift f\"{u}r Physik (Journal of Physics)} \textbf{89}: 605--613, 1934.

\bibitem[Orowan, 1934b]{Orowan34b} \citep{Orowan34b} E. Orowan. Zur Kristallplastizit\"{a}t. II: Die dynamische Auffassung der Kristallplastizit\"{a}t. (On crystal plasticity. Part II: The dynamic view of crystal plasticity.) \textit{Zeitschrift f\"{u}r Physik (Journal of Physics)} \textbf{89}: 614--633, 1934.

\bibitem[Orowan, 1934c]{Orowan34c} \citep{Orowan34c} E. Orowan. Zur Kristallplastizit\"{a}t. III: \"{U}ber die Mechanismus des Gleitvorganges. (On crystal plasticity. Part III: About the mechanism of the sliding process.) \textit{Zeitschrift f\"{u}r Physik (Journal of Physics)} \textbf{89}: 634--659, 1934.

\bibitem[Osborne et al., 2000]{Osborne00} \citep{Osborne00} A. R. Osborne, M. Onorato, and M. Serio. The nonlinear dynamics of rogue waves and holes in deep-water gravity wave trains. \textit{Physics Letters A} \textbf{275}(5--6): 386--393, 2000.

\bibitem[Osborne, 2001]{Osborne01} \citep{Osborne01} A. R. Osborne. The random and deterministic dynamics of `rogue waves' in unidirectional, deep-water wave trains. \textit{Marine Structures} \textbf{14}(3): 275--293, 2001.

\bibitem[Ostrovski\u{\i}, 1967]{Ostrovsky67} \citep{Ostrovsky67} L. A. Ostrovski\u{\i}. Propagation of wave packets and space-time self-focusing in a nonlinear medium. \textit{Soviet Physics JETP} \textbf{24}(4): 797--800, 1967. Translated from Russian, the original version appeared in \textit{Журнал Экспериментальной и Теоретической Физики (ЖЭТФ), Zhurnal \'Eksperimental'no\u{\i} i Teoretichesko\u{\i} Fiziki (ZhETF)} \textbf{51}(4): 1189--1194, 1966.

\bibitem[Pelinovsky et al., 2000]{Pelinovsky00} \citep{Pelinovsky00} E. Pelinovsky, T. Talipova, and C. Kharif. Nonlinear-dispersive mechanism of the freak wave formation in shallow water. \textit{Physica D: Nonlinear Phenomena} \textbf{147}(1--2): 83--94, 2000.

\bibitem[Pelinovsky et al., 2004]{Pelinovsky04} \citep{Pelinovsky04} E. Pelinovsky, T. Talipova, M. Ruderman, and R. Erdelyi. Freak waves described by the modified Korteweg-de Vries equation. \textit{Izvestia, Russian Academy of Engineering Sciences}, \textit{Applied Mathematics and Mechanics Series} \textbf{6}: 3--16, 2004.

\bibitem[Peregrine, 1983]{Peregrine83} \citep{Peregrine83} D. H. Peregrine. Water waves, nonlinear Schr\"{o}dinger equations and their solutions. \textit{Journal of the Australian Mathematical Society Series B/The ANZIAM  (Australia and New Zealand Industrial and Applied Mathematics) Journal} \textbf{25}(1): 16--43, 1983.

\bibitem[Polanyi, 1934]{Polanyi34} \citep{Polanyi34} M. Polanyi. \"{U}ber eine Art Gitterst\"{o}rung, die einen Kristall plastisch machen k\"{o}nnte. (About a kind of lattice defect that could make a crystal plastic.) \textit{Zeitschrift f\"ur Physik (Journal of Physics)} \textbf{89}(9--10): 660--664, 1934.


\bibitem[Read, 1953]{Read53} \citep{Read53} W. T. Read, Jr. \textit{Dislocations in Crystals}. McGraw-Hill, New York, 1953.

\bibitem[Remoissenet, 1999]{Remoissenet99} \citep{Remoissenet99} M. Remoissenet. \textit{Waves Called Solitons: Concepts and Experiments}, third edition. Springer, Berlin Heidelberg, Germany, 1999.

\bibitem[Rogers and Schief, 2002]{Rogers02} \citep{Rogers02} C. Rogers and W. K. Schief. \textit{B\"{a}cklund and Darboux Transformations: Geometry and Modern Applications in Soliton Theory}. Cambridge University Press, Cambridge, UK, 2002.

\bibitem[Rosenthal, 2005]{Rosenthal05} \citep{Rosenthal05} W. Rosenthal. Result of the \textsl{MaxWave} project. In \textit{Proceedings of the 14th `Aha Huliko`a Hawaiian Winter Workshop on Rogue Waves}, 7~pp., University of Hawaii, Honolulu, HI, January 25--28, 2005. Available online at \url{http://www.soest.hawaii.edu/PubServices/AhaHulikoa.html}. Last accessed 6 November 2006.

\bibitem[Sch\"{a}ffer, 1996]{Schaffer96} \citep{Schaffer96} H. A. Sch\"{a}ffer. Second-order wavemaker theory for irregular waves. \textit{Ocean Engineering} \textbf{23}(1):47--88, 1996.

\bibitem[Sch\"{a}ffer and Steenberg, 2003]{Schaffer03} \citep{Schaffer03} H. A. Sch\"{a}ffer and C. M. Steenberg. Second-order wavemaker theory for multidirectional waves. \textit{Ocean Engineering} \textbf{30}(10): 1203--1231, 2003.

\bibitem[Segur et al., 2005]{Segur05} \citep{Segur05} H. Segur, D. Henderson, J. Carter, J. Hammack, C.-M. Li, D. Pheiff, and K. Socha. Stabilizing the Benjamin-Feir instability. \textit{Journal of Fluid Mechanics} \textbf{539}: 229--271, 2005.

\bibitem[Scott et al., 1973]{Scott73} \citep{Scott73} A. C. Scott, F. Y. H. Chu, and D. W. McLaughlin. The soliton: A new concept in applied science \textit{Proceedings of the IEEE} \textbf{61}(10): 1443--1483, 1973.

\bibitem[Scott, 2003]{Scott03} \citep{Scott03} A. C. Scott. \textit{Nonlinear Science--Emergence and Dynamics of Coherent Structures}, second edition. Oxford University Press, Oxford, UK, 2003.

\bibitem[Scott, 2005]{Scott05} \citep{Scott05} A. C. Scott. Solitons, a brief history. In A. C. Scott (Editor), \textit{Encyclopedia of Nonlinear Science}. Routledge Taylof \& Francis Group, New York, 2005.

\bibitem[Smith, 1976]{Smith76} \citep{Smith76} R. Smith. Giant waves. \textit{Journal of Fluid Mechanics} \textbf{77}(3): 417--431, 1976.

\bibitem[Soskin and Vasnetsov, 2001]{Soskin01} \citep{Soskin01} M. S. Soskin and M. V. Vasnetsov. Singular optics. \textit{Progress in Optics} \textbf{42}(4): 219--276, 2001.

\bibitem[Stuart and DiPrima, 1978]{Stuart78} \citep{Stuart78} J. T. Stuart and R. C. DiPrima. The Eckhaus and Benjamin-Feir resonance mechanisms. \textit{ Proceedings of the Royal Society of London. Series A, Mathematical and Physical Sciences} \textbf{362}(1708): 27--41, 1978.

\bibitem[Stokes, 1880]{Stokes1880} \citep{Stokes1880} G. Stokes. Supplement to a paper on the theory of oscillatory waves. \textit{Mathematical and Physical Papers} \textbf{1}: 314--326, 1880.

\bibitem[Sulem and Sulem, 1999]{Sulem99} \citep{Sulem99} C. Sulem and P.-L. Sulem. \textit{The Nonlinear Schr\"{o}dinger Equation---Self-Focusing and Wave Collapse}, volume \textbf{139} of \textit{Applied Mathematical Sciences}, Springer-Verlag, New York, 1999.

\bibitem[Svendsen, 1985]{Svendsen} \citep{Svendsen} I. A. Svendsen. Physical modelling of water waves. In R. A. Dalrymple (Editor), \textit{Physical Modelling in Ocean Engineering}, A. A. Balkema, Rotterdam, the Netherlands, 1985.

\bibitem[Talanov, 1965]{Talanov65} \citep{Talanov65} V. I. Talanov. Self focusing of wave beams in nonlinear media. \textit{JETP Letters} \textbf{2}(5): 138--141, 1965. Translated from Russian, the original version appeared in \textit{Письма в Журнал экспериментальной и теоретической физики (Письма в ЖЭТФ), Pis'ma v Zhurnal \'Eksperimental'no\u{\i} i Teoretichesko\u{\i} Fiziki (ZhETF)} \textbf{2}: 223--236, 1965.

\bibitem[Tajiri and Arai, 2000]{Tajiri00} \citep{Tajiri00} M. Tajiri and T. Arai. Periodic soliton solutions to the Davey-Stewartson equation. \textit{Proceedings of Institute of Mathematics of National Academy of Science of Ukraine} \textbf{30}(1): 210--217, 2000.

\bibitem[Tanaka, 1995]{Tanaka95} \citep{Tanaka95} M. Tanaka. Dissapearance of waves in modulated train of surface gravity waves. \textit{Structure and Dynamics of Nonlinear Waves in Fluids}. In A. Mielke and K. Kirchg\"{a}ssner (Editors), Proceedings of the IUTAM/ISIMM (International Union for Theoretical and Applied Mechanics/Institut Sup\'erieur d'Informatique et de Math\'ematiques de Monastir, The Higher Institute of Informatics and Mathematics of the University of Monastir) Symposium held in Hannover August 1994, volume \textbf{7} of \textit{Advanced Series in Nonlinear Dynamics}, World Scientific, Singapore, pp.~392--398, 1995.

\bibitem[Taniuti and Washimi, 1968]{Taniuti68} \citep{Taniuti68} T. Taniuti and H. Washimi. Self-trapping and instability of hydromagnetics waves along the magnetic field in a cold plasma. \textit{Physical Review Letters} \textbf{21}(4): 209--212, 1968.

\bibitem[Tappert and Varma, 1970]{Tappert70} \citep{Tappert70} F. Tappert and C. M. Varma. Asymptotic theory of self-trapping of heat pulses in solids. \textit{Physical Review Letters} \textbf{25}(16): 1108--1111, 1970.


\bibitem[Taylor, 1934a]{Taylor34a} \citep{Taylor34a} G. I. Taylor. The mechanism of plastic deformation of crystals. Part I: Theoretical. \textit{Proceedings of the Royal Society of London. Series A, Containing Papers of a Mathematical and Physical Character} \textbf{145}(855): 362--387, 1934.

\bibitem[Taylor, 1934b]{Taylor34b} \citep{Taylor34b} G. I. Taylor. The mechanism of plastic deformation of crystals. Part II: Comparison with observations. \textit{Proceedings of the Royal Society of London. Series A, Containing Papers of a Mathematical and Physical Character} \textbf{145}(855): 388--404, 1934.

\bibitem[Tracy et al., 1988]{Tracy88} \citep{Tracy88} E. R. Tracy, J. W. Larson, A. R. Osborne, and L. Bergamasco. On the nonlinear Schr\"{o}dinger limit of the Korteweg-de Vries equation. \textit{Physica D: Nonlinear Phenomena} \textbf{32}(1): 83--106, 1988.

\bibitem[Tracy et al., 1991]{Tracy91} \citep{Tracy91} E. R. Tracy, J. W. Larson, A. R. Osborne, and L. Bergamasco. The relationship between the spectral theories for the periodic Korteweg-de Vries and nonlinear Schr\"{o}dinger equations. In A. R. Osborne (Editor), \textit{Nonlinear Topics of Ocean Physics}, Fermi Summer School, Course LIX, pp. 769--825, North-Holland, Amsterdam, the Netherlands, 1991.

\bibitem[Trulsen and Dysthe, 1996]{Trulsen} \citep{Trulsen} K. Trulsen and K. B. Dysthe. A modified nonlinear Schr\"{o}dinger equation for broader bandwith gravity waves on deep water. \textit{Wave Motion}, \textbf{24}(3): 281--289, 1996.

\bibitem[Trulsen and Dysthe, 1997]{Trulsen97} \citep{Trulsen97} K. Trulsen and K. B. Dysthe. Freak waves--a three-dimensional wave simulation. In \textit{Proceedings of the Twenty-First (1996) Symposium on Naval Hydrodynamics}, Trondheim, Norway, June 24--28, pp.~550--558, 1997.

\bibitem[Trulsen, 1998]{Trulsen98} \citep{Trulsen98} K. Trulsen. Crest pairing predicted by modulation theory. \textit{Journal of Geophysical Research: Oceans} \textbf{103}(C2): 3143--3147, 1998.

\bibitem[Trulsen and Stansberg, 2001]{Trulsen01} \citep{Trulsen01} K. Trulsen and C. T. Stansberg. Spatial evolution of water surface waves: Numerical simulation and experiment of bichromatic waves. In \textit{Proceedings of the Eleventh (2001) International Offshore and Polar Engineering Conference}, Stavanger, Norway, June 17--22, pp.~71--77, 2001.


\bibitem[Van Groesen, 1998]{vanGroesen98} \citep{vanGroesen98} E. van Groesen. Wave groups in uni-directional surface-wave models. \textit{Journal of Engineering Mathematics} \textbf{34}(1--2): 215--226, 1998.

\bibitem[Van Groesen et al., 2005]{vanGroesen05} \citep{vanGroesen05} E. van Groesen, Andonowati, and N. Karjanto. Deterministic aspect of nonlinear modulation instability. In M. Olagnon and M. Prevosto (Editors), \textit{Rogue Waves 2004}. Proceedings of a workshop in Brest, France (October 20-22, 2004), 12~pp., 2005. \ arXiv:1110.5120 [physics.flu-dyn]

\bibitem[Van Groesen et al., 2006]{vanGroesen06} \citep{vanGroesen06} E. van Groesen, Andonowati, and N. Karjanto. Displaced phase-amplitude variables for waves on finite background. \textit{Physics Letters A} \textbf{354}(4): 312--319, 2006. \ arXiv:1906.00959 [nlin.PS]

\bibitem[Van Groesen and Andonowati, 2006]{BrennyAan06} \citep{BrennyAan06} E. van Groesen and Andonowati. Finite energy wave signals of extremal amplitude in the spatial NLS-dynamics. \textit{Physics Letters A} \textbf{357}(2): 86--91, 2006.

\bibitem[Van Saarloos and Hohenberg, 1992]{vanSaarloos92} \citep{vanSaarloos92} W. van Saarloos and P. C. Hohenberg. Fronts, pulses, sources and sinks in generalized complex Ginzburg-Landau equations. \textit{Physica D: Nonlinear Phenomena} \textbf{56}(4): 303--367, 1992.

\bibitem[Walker et al., 2004]{Walker04} \citep{Walker04} D. A. G. Walker, P. H. Taylor, and R. E. Taylor. The shape of large surface waves on the open sea and the Draupner New Year wave. \textit{Applied Ocean Research} \textbf{26}(3--4): 73--83, 2004.

\bibitem[Wang et al., 2005]{Wang05} \citep{Wang05} D. W. Wang, D. A. Mitchell, W. J. Teague, E. Jarosz, and M. S. Hulbert. Extreme waves under hurricane Ivan. \textit{Science} \textbf{309}(5736): 896, 2005.

\bibitem[White and Fornberg, 1998]{White98} \citep{White98} B. S. White and B. Fornberg. On the chance of freak waves at sea. \textit{Journal of Fluid Mechanics} \textbf{355}: 113--138, 1998.

\bibitem[Whitham, 1967]{Whitham67} \citep{Whitham67} G. B. Whitham. Non-linear dispersion of water waves. \textit{Journal of Fluid Mechanics} \textbf{27}(2): 399--412, 1967.

\bibitem[Whitham, 1974]{Whitham74} \citep{Whitham74} G. B. Whitham. \textit{Linear and Nonlinear Waves}. John Wiley \& Sons, New York, 1974.

\bibitem[Wright, 1979]{Wright79} \citep{Wright79} F. J. Wright. Wavefront dislocations and their analysis using catastrophe theory. In W. G\"{u}ttinger and H. Eikemeier (Editors), \textit{Structural Stability in Physics: Proceedings of Two International Symposia on Applications of Catastrophe Theory and Topological Concepts in Physics T\"ubingen, Federal Republic of Germany, May 2--6 and December 11--14, 1978}, pp.~141--156, Springer, Berlin Heilderberg, Germany, 1979.


\bibitem[Yuen and Ferguson, 1978a]{Yuen78a} \citep{Yuen78a} H. C. Yuen and W. E. Ferguson,~Jr. \ Relationship between Benjamin-Feir instability and recurrence in the nonlinear Schr\"{o}dinger equation. \textit{The Physics of Fluids} \textbf{21}(8): 1275--1278, 1978.

\bibitem[Yuen and Ferguson, 1978b]{Yuen78b} \citep{Yuen78b} H. C. Yuen and W. E. Ferguson,~Jr. \ Fermi-Pasta-Ulam recurrence in the two-space dimensional nonlinear Schr\"{o}dinger equation. \textit{The Physics of Fluids} \textbf{21}(11): 2116--2118, 1978.

\bibitem[Yuen and Lake, 1975]{Yuen75} \citep{Yuen75} H. C. Yuen and B. M. Lake. Nonlinear deep water waves: Theory and experiment. \textit{The Physics of Fluids} \textbf{18}(8): 956--960, 1975.

\bibitem[Yuen and Lake, 1982]{Yuen82} \citep{Yuen82} H. C. Yuen and B. M. Lake. Nonlinear dynamics of deep-water gravity waves. \textit{Advances in Applied Mechanics} \textbf{22}: 67--229, 1982.

\bibitem[Zakharov, 1967]{Zakharov67} \citep{Zakharov67} V. E. Zakharov. The wave stability in nonlinear dispersive media. \textit{Soviet Physics JETP} \textbf{24}(4): 740--744, 1967. Translated from Russian, the original version appeared in \textit{Журнал Экспериментальной и Теоретической Физики (ЖЭТФ), Zhurnal \'Eksperimental'no\u{\i} i Teoretichesko\u{\i} Fiziki (ZhETF)} \textbf{51}(4): 1107--1114, 1966.

\bibitem[Zakharov, 1968]{Zakharov68} \citep{Zakharov68} V. E. Zakharov. Stability of periodic waves of finite amplitude on the surface of a deep fluid. \textit{Journal of Applied Mechanics and Technical Physics} \textbf{9}(2): 190--194, 1968. Translated from Russian, the original version appeared in \textit{Прикладная механика и техническая физика (ПМТФ), Zhurnal Prikladno\u{\i} Mekhaniki i Tekhnichesko\u{\i} Fiziki} \textbf{9}(2): 86--94, 1968.

\bibitem[Zakharov and Shabat, 1972]{ZakharovShabat72} \citep{ZakharovShabat72} V. E. Zakharov and A. B Shabat. Exact theory of two-dimensional self-focusing and one-dimensional self-modulation of waves in nonlinear media. \textit{Soviet Physics JETP} \textbf{34}(1): 62--69, 1972. Translated from Russian, the original version appeared in \textit{Журнал Экспериментальной и Теоретической Физики (ЖЭТФ), Zhurnal \'Eksperimental'no\u{\i} i Teoretichesko\u{\i} Fiziki (ZhETF)} \textbf{61}(1): 118--134, 1971.

\bibitem[Zakharov, 1972]{Zakharov72} \citep{Zakharov72} V. E. Zakharov. Collapse of Langmuir waves. \textit{Soviet Physics JETP} \textbf{35}(5): 908--914, 1972. Translated from Russian, the original version appeared in \textit{Журнал Экспериментальной и Теоретической Физики (ЖЭТФ), Zhurnal \'Eksperimental'no\u{\i} i Teoretichesko\u{\i} Fiziki (ZhETF)} \textbf{62}(5): 1745--1759, 1972.
\end{thebibliography}
}
\end{document}